\documentclass[letterpaper]{article}
\pdfoutput=1
\usepackage[margin=1in]{geometry}
\usepackage[utf8]{inputenc}
\usepackage[T1]{fontenc}
\usepackage{microtype}
\usepackage{amsfonts}
\usepackage{amssymb}
\usepackage{amsmath}
\usepackage{amsthm}
\usepackage{fullpage}
\usepackage{booktabs}
\usepackage[inline]{enumitem}
\usepackage[hidelinks,bookmarksopen=true]{hyperref}
\usepackage[capitalize]{cleveref}
\usepackage{mathtools}
\usepackage{tikz}
\usepackage[font=small,labelfont=bf]{caption}
\usepackage{titlesec}
\usepackage{thmtools}
\usepackage{wrapfig}
\usepackage{tablefootnote}
\usepackage{thm-restate}
\usepackage{soul}
\usepackage{fixme}
\usepackage[norefs,nocites,nomsgs]{refcheck}  %

\input{settings}

\newcommand{\bigO}{\mathcal{O}}
\newcommand{\Oh}{\bigO}
\DeclareMathOperator{\polylog}{polylog}

\newcommand{\SubstringComplexitySym}{\delta}
\newcommand{\LZSizeSym}{z}
\newcommand{\RLBWTSizeSym}{r}
\newcommand{\GrammarSizeSym}{g}
\newcommand{\AttractorSizeSym}{\gamma}

\newcommand{\SubstringComplexity}[1]{\SubstringComplexitySym(#1)}
\newcommand{\LZSize}[1]{\LZSizeSym(#1)}
\newcommand{\RLBWTSize}[1]{\RLBWTSizeSym(#1)}
\newcommand{\SmallestAttractor}[1]{\AttractorSizeSym^{*}(#1)}
\newcommand{\SmallestGrammar}[1]{\GrammarSizeSym^{*}(#1)}

\newcommand{\LCE}{\mathrm{LCE}}
\newcommand{\lcp}{\mathrm{lcp}}
\newcommand{\lcs}{\mathrm{lcs}}
\newcommand{\per}{\mathrm{per}}
\newcommand{\rot}{\mathrm{rot}}
\newcommand{\revstr}[1]{\overline{#1}}
\newcommand{\CompRepr}[3]{\mathrm{comp}_{#1}(#2, #3)}
\newcommand{\Successor}[2]{\mathrm{succ}_{#1}(#2)}
\newcommand{\ArrayRank}[2]{\mathsf{rank}(#1,#2)}
\newcommand{\IntervalRepr}[1]{\mathcal{I}(#1)}
\newcommand{\SubstrCount}[2]{\mathsf{d}_{#1}(#2)}

\newcommand{\OccTwo}[2]{\mathrm{Occ}(#1, #2)}
\newcommand{\OccThree}[3]{\mathrm{Occ}_{#1}(#2, #3)}
\newcommand{\RangeBegTwo}[2]{\mathrm{RangeBeg}(#1, #2)}
\newcommand{\RangeBegThree}[3]{\mathrm{RangeBeg}_{#1}(#2, #3)}
\newcommand{\RangeEndTwo}[2]{\mathrm{RangeEnd}(#1, #2)}
\newcommand{\RangeEndThree}[3]{\mathrm{RangeEnd}_{#1}(#2, #3)}
\newcommand{\PosBeg}[3]{\mathrm{Pos}^{\rm beg}_{#1}(#2, #3)}
\newcommand{\PosEnd}[3]{\mathrm{Pos}^{\rm end}_{#1}(#2, #3)}
\newcommand{\DeltaBeg}[3]{\delta^{\rm beg}_{#1}(#2, #3)}
\newcommand{\DeltaEnd}[3]{\delta^{\rm end}_{#1}(#2, #3)}

\newcommand{\RootPos}[4]{\mathrm{root}_{#1}(#2,#3,#4)}
\newcommand{\RootPat}[3]{\mathrm{root}_{#1}(#2,#3)}
\newcommand{\HeadPos}[4]{\mathrm{head}_{#1}(#2,#3,#4)}
\newcommand{\HeadPat}[3]{\mathrm{head}_{#1}(#2,#3)}
\newcommand{\TailPos}[4]{\mathrm{tail}_{#1}(#2,#3,#4)}
\newcommand{\TailPat}[3]{\mathrm{tail}_{#1}(#2,#3)}
\newcommand{\ExpPos}[4]{\mathrm{exp}_{#1}(#2,#3,#4)}
\newcommand{\ExpPat}[3]{\mathrm{exp}_{#1}(#2,#3)}
\newcommand{\TypePos}[3]{\mathrm{type}(#1,#2,#3)}
\newcommand{\TypePat}[2]{\mathrm{type}(#1,#2)}
\newcommand{\RunEndFullPos}[4]{e^{\rm full}_{#1}(#2,#3,#4)}
\newcommand{\RunEndFullPat}[3]{e^{\rm full}_{#1}(#2,#3)}
\newcommand{\RunEndPos}[3]{e(#1,#2,#3)}
\newcommand{\RunEndPat}[2]{e(#1,#2)}
\newcommand{\RunBeg}[3]{b(#1,#2,#3)}
\newcommand{\ExpCutPos}[5]{\mathrm{exp}^{\rm cut}_{#1}(#2,#3,#4,#5)}
\newcommand{\ExpCutPat}[4]{\mathrm{exp}^{\rm cut}_{#1}(#2,#3,#4)}
\newcommand{\RunEndCutPos}[5]{e^{\rm cut}_{#1}(#2,#3,#4,#5)}
\newcommand{\RunEndCutPat}[4]{e^{\rm cut}_{#1}(#2,#3,#4)}

\newcommand{\PosLowMinus}[3]{\mathrm{Pos}_{#1,\ell}^{\rm low-}(#2,#3)}
\newcommand{\PosLowPlus}[3]{\mathrm{Pos}_{#1,\ell}^{\rm low+}(#2,#3)}
\newcommand{\PosLowAll}[3]{\mathrm{Pos}_{#1,\ell}^{\rm low}(#2,#3)}
\newcommand{\PosMidMinus}[3]{\mathrm{Pos}_{#1,\ell}^{\rm mid-}(#2,#3)}
\newcommand{\PosMidPlus}[3]{\mathrm{Pos}_{#1,\ell}^{\rm mid+}(#2,#3)}
\newcommand{\PosMidAll}[3]{\mathrm{Pos}_{#1,\ell}^{\rm mid}(#2,#3)}
\newcommand{\PosHighMinus}[3]{\mathrm{Pos}_{#1,\ell}^{\rm high-}(#2,#3)}
\newcommand{\PosHighPlus}[3]{\mathrm{Pos}_{#1,\ell}^{\rm high+}(#2,#3)}
\newcommand{\PosHighAll}[3]{\mathrm{Pos}_{#1,\ell}^{\rm high}(#2,#3)}
\newcommand{\DeltaLowMinus}[3]{\delta_{#1,\ell}^{\rm low-}(#2,#3)}
\newcommand{\DeltaMidMinus}[3]{\delta_{#1,\ell}^{\rm mid-}(#2,#3)}
\newcommand{\DeltaHighMinus}[3]{\delta_{#1,\ell}^{\rm high-}(#2,#3)}
\newcommand{\DeltaLowPlus}[3]{\delta_{#1,\ell}^{\rm low+}(#2,#3)}
\newcommand{\DeltaMidPlus}[3]{\delta_{#1,\ell}^{\rm mid+}(#2,#3)}
\newcommand{\DeltaHighPlus}[3]{\delta_{#1,\ell}^{\rm high+}(#2,#3)}

\newcommand{\RangeCountFourSide}[5]{\mathsf{weight\mbox{-}count}_{#1}(#2, #3, #4, #5)}
\newcommand{\RangeCountThreeSide}[4]{\mathsf{weight\mbox{-}count}_{#1}(#2, #3, #4)}
\newcommand{\IncRangeCountThreeSide}[4]{\mathsf{weight\mbox{-}count}^{\preceq}_{#1}(#2, #3, #4)}
\newcommand{\RangeCountTwoSide}[3]{\mathsf{weight\mbox{-}count}_{#1}(#2, #3)}
\newcommand{\RangeSelect}[4]{\mathsf{weight\mbox{-}select}_{#1}(#2, #3, #4)}
\newcommand{\RangeMinFourSide}[5]{\mathsf{r\mbox{-}min}_{#1}(#2, #3, #4, #5)}
\newcommand{\RangeMinTwoSide}[3]{\mathsf{r\mbox{-}min}_{#1}(#2, #3)}

\newcommand{\ModCountTwoSide}[5]{\mathsf{mod\mbox{-}count}_{#1,#2}(#3, #4, #5)}
\newcommand{\ModCountOneSide}[4]{\mathsf{mod\mbox{-}count}_{#1,#2}(#3, #4)}
\newcommand{\ModCountZeroSide}[3]{\mathsf{mod\mbox{-}count}_{#1,#2}(#3)}
\newcommand{\ModSelect}[4]{\mathsf{mod\mbox{-}select}_{#1,#2}(#3, #4)}

\newcommand{\CompSaCore}[1]{\mathrm{CompSACore}(#1)}
\newcommand{\CompSaPeriodic}[1]{\mathrm{CompSAPeriodic}(#1)}
\newcommand{\CompSaNonperiodic}[1]{\mathrm{CompSANonperiodic}(#1)}
\newcommand{\CompSa}[1]{\mathrm{CompSA}(#1)}

\newcommand{\Pat}{P}
\newcommand{\Text}{T}
\newcommand{\Textinf}{\Text^{\infty}}
\newcommand{\Textlen}{n}

\newcommand{\RName}{\mathsf{R}}
\newcommand{\RMinusName}{\RName^{-}}
\newcommand{\RPlusName}{\RName^{+}}
\newcommand{\RTwo}[2]{\RName(#1, #2)}
\newcommand{\RFour}[4]{\RName_{#1,#2}(#3, #4)}
\newcommand{\RFive}[5]{\RName_{#1,#2,#3}(#4, #5)}
\newcommand{\RSix}[6]{\RName_{#1,#2,#3,#4}(#5, #6)}
\newcommand{\RMinusTwo}[2]{\RMinusName(#1, #2)}
\newcommand{\RMinusFour}[4]{\RMinusName_{#1,#2}(#3, #4)}
\newcommand{\RMinusFive}[5]{\RMinusName_{#1,#2,#3}(#4, #5)}
\newcommand{\RMinusSix}[6]{\RMinusName_{#1,#2,#3,#4}(#5, #6)}
\newcommand{\RPlusTwo}[2]{\RPlusName(#1, #2)}
\newcommand{\RPlusFour}[4]{\RPlusName_{#1,#2}(#3, #4)}
\newcommand{\RPlusFive}[5]{\RPlusName_{#1,#2,#3}(#4, #5)}
\newcommand{\RPlusSix}[6]{\RPlusName_{#1,#2,#3,#4}(#5, #6)}

\newcommand{\RPrimName}{\RName'}
\newcommand{\RPrimMinusName}{\RPrimName^{-}}
\newcommand{\RPrimPlusName}{\RPrimName^{+}}
\newcommand{\RPrimTwo}[2]{\RPrimName(#1, #2)}
\newcommand{\RPrimMinusTwo}[2]{\RPrimMinusName(#1, #2)}
\newcommand{\RPrimPlusTwo}[2]{\RPrimPlusName(#1, #2)}
\newcommand{\RPrimMinusFour}[4]{\RPrimMinusName_{#1, #2}(#3, #4)}
\newcommand{\RPrimPlusFour}[4]{\RPrimPlusName_{#1, #2}(#3, #4)}

\newcommand{\DistinguishingPrefs}[3]{\mathcal{D}(#1, #2, #3)}
\newcommand{\Cover}[2]{\mathsf{C}(#1,#2)}
\newcommand{\SSScompgen}[1]{\SSS_{{\rm comp},#1}}

\newcommand{\StrStrPoints}[3]{\mathrm{StrStrPoints}_{#1}(#2, #3)}
\newcommand{\IntStrPoints}[3]{\mathrm{IntStrPoints}_{#1}(#2, #3)}
\newcommand{\WInts}[3]{\mathrm{WeightedIntervals}_{#1}(#2, #3)}
\newcommand{\PairsMinus}[4]{\mathrm{Seed}^{-}_{#1, #2}(#3, #4)}

\newcommand{\SSS}{\mathsf{S}}
\newcommand{\SSScomp}{\SSS_{\rm comp}}
\newcommand{\Rcomp}{\RName_{\rm comp}}
\newcommand{\IntegerAlphabet}{[0 \dd \sigma)}
\newcommand{\Z}{\mathbb{Z}}
\newcommand{\Zz}{\Z_{\ge 0}}
\newcommand{\Zn}{\Zz}
\newcommand{\Zp}{\Z_{>0}}

\newcommand{\X}{\mathcal{X}}
\newcommand{\Y}{\mathcal{Y}}
\newcommand{\Pts}{\mathcal{P}}
\newcommand{\Ints}{\mathcal{I}}

\newcommand{\SA}{\mathrm{SA}}
\newcommand{\ISA}{\mathrm{ISA}}
\newcommand{\ArrSSSComp}[1]{A_{{\rm comp}, #1}}
\newcommand{\ArrRuns}[1]{A_{{\rm runs}, #1}}
\newcommand{\ArrRoot}[1]{A_{{\rm root},#1}}
\newcommand{\ArrPtrFirst}[1]{A_{{\rm ptr},#1}}
\newcommand{\ArrPtrSecond}[1]{A'_{{\rm ptr},#1}}
\newcommand{\ArrStr}{A_{\rm str}}
\newcommand{\ArrRange}{A_{\rm range}}
\newcommand{\ArrMinOcc}{A_{\rm minocc}}

\newcommand{\sm}{\setminus}
\newcommand{\sub}{\subseteq}
\newcommand{\dd}{\mathinner{.\,.}}
\newcommand{\emptystring}{\varepsilon}

\newcommand{\floor}[1]{\lfloor #1 \rfloor}
\newcommand{\ceil}[1]{\lceil #1 \rceil}

\newcommand{\G}{\mathcal{G}} %
\newcommand{\Bnd}{\mathsf{B}} %
\newcommand{\Cov}{\mathsf{C}} %
\newcommand{\N}{\mathcal{N}}
\renewcommand{\S}{\mathcal{S}}
\newcommand{\Tr}{\mathcal{T}}
\newcommand{\rhs}{\mathsf{rhs}}
\newcommand{\Symb}{\mathcal{A}}
\newcommand{\Act}{\mathcal{B}}
\newcommand{\Pres}{\mathcal{S}}
\newcommand{\Left}{\mathcal{L}}
\newcommand{\Right}{\mathcal{R}}
\newcommand{\rle}{\mathsf{rle}}
\newcommand{\pc}{\mathsf{pc}}

\newcommand{\PP}{\mathsf{P}}
\newcommand{\LMR}{\mathsf{LR}}
\newcommand{\RMR}{\mathsf{RR}}
\newcommand{\LML}{\mathsf{LS}}
\newcommand{\RML}{\mathsf{RS}}
\newcommand{\LMB}{\mathsf{LB}}
\newcommand{\RMB}{\mathsf{RB}}
\newcommand{\T}{T}
\newcommand{\hT}{\hat{\T}}
\newcommand{\hG}{\hat{\G}}
\newcommand{\hA}{\hat{A}}
\newcommand{\hB}{\hat{B}}
\newcommand{\hC}{\hat{C}}
\newcommand{\hS}{\hat{S}}
\newcommand{\height}{\mathsf{height}}
\newcommand{\symb}{\mathsf{s}}

\begin{document}

\title{Collapsing the Hierarchy of Compressed Data Structures:\\
  Suffix Arrays in Optimal Compressed Space}

\author{
  \large Dominik Kempa\thanks{Supported by the
    Simons Foundation Junior Faculty Fellowship.}\\[-0.3ex]
  \normalsize Department of Computer Science,\\[-0.3ex]
  \normalsize Stony Brook University,\\[-0.3ex]
  \normalsize Stony Brook, NY, USA\\[-0.3ex]
  \normalsize \texttt{kempa@cs.stonybrook.edu}
  \and
  \large Tomasz Kociumaka\\[-0.3ex]
  \normalsize Max Planck Institute for Informatics,\\[-0.3ex]
  \normalsize Saarland Informatics Campus,\\[-0.3ex]
  \normalsize Saarbrücken, Germany\\[-0.3ex]
  \normalsize \texttt{tomasz.kociumaka@mpi-inf.mpg.de}
}

\date{\vspace{-0.5cm}}
\maketitle

\begin{abstract}
  The last two decades have witnessed a dramatic increase in the
  amount of highly repetitive datasets consisting of sequential data
  (strings, texts).  Processing these massive amounts of data using
  conventional data structures is infeasible. This fueled the
  development of \emph{compressed text indexes}, which 
  efficiently answer various queries on a given text, typically in polylogarithmic time,
  while occupying space proportional to the compressed representation of the
  text. There exist
  numerous structures supporting queries ranging from simple ``local''
  queries, such as random access, through more complex ones, including
  longest common extension (LCE) queries, to the most powerful
  queries, such as the suffix array (SA) functionality.  Alongside
  the rich repertoire of queries followed a detailed study of the
  trade-off between the size and functionality of compressed indexes
  (see: Navarro; ACM Comput.\ Surv.\ 2021).  It is widely
  accepted that this hierarchy of structures tells a simple story: the
  more powerful the queries, the more space is needed. On the one
  hand, random access, the most basic query, can be supported using
  $\bigO(\delta \log \tfrac{n \log \sigma}{\delta \log n})$ space
  (where $n$ is the length of the text, $\sigma$ is the alphabet size,
  and $\delta$ is the text's substring complexity), which is known to be
  the asymptotically smallest space sufficient to represent any string
  with parameters $n$, $\sigma$, and $\delta$, (Kociumaka, Navarro, and Prezza; IEEE
  Trans.\ Inf.\ Theory 2023).  The other end of the hierarchy is
  occupied by indexes supporting the suffix array queries.
  The currently smallest one takes $\bigO(r \log\tfrac{n}{r})$ space,
  where $r \geq \delta$ is the number of runs in the Burrows--Wheeler
  Transform of the text (Gagie, Navarro, and Prezza; J.\ ACM 2020).

  We present a new compressed index, referred to as $\delta$-SA, that supports the powerful SA
  functionality and needs only $\bigO(\delta \log \tfrac{n \log
  \sigma}{\delta \log n})$ space.  This collapses the hierarchy of
  compressed data structures into a single point: The space
  \emph{required} to represent the text is simultaneously
  \emph{sufficient} to efficiently support the full SA functionality.
  Since suffix array queries are the most widely utilized queries in
  string processing and data compression, our result immediately
  improves the space complexity of dozens of algorithms, which can now
  be executed in $\delta$-optimal compressed space.  The $\delta$-SA supports
  both suffix array and inverse suffix array queries in
  $\bigO(\log^{4+\epsilon} n)$ time (where $\epsilon>0$ is any
  predefined constant).
  
  Our second main result is an $\bigO(\delta \polylog n)$ time construction
  of the $\delta$-SA from the Lempel--Ziv (LZ77) parsing
  of the text. This is the first algorithm that builds an
  SA index in \emph{compressed time}, i.e., time nearly
  linear in the compressed input size. For highly
  repetitive texts, this is up to exponentially faster than the
  previously best algorithm, which builds an $\bigO(r
  \log\tfrac{n}{r})$-size index in $\bigO(\sqrt{\delta n} \polylog n)$~time.
  
  To obtain our results, we develop numerous new techniques
  of independent interest. This includes deterministic restricted
  recompression, $\delta$-compressed string synchronizing sets, and
  their construction in compressed time. We also improve many
  other auxiliary data structures; e.g., we show the first
  $\bigO(\delta \log \tfrac{n \log \sigma}{\delta \log n})$-size index
  for LCE queries along with its efficient construction from the LZ77
  parsing.
\end{abstract}

\thispagestyle{empty}
\pagenumbering{arabic}
\newpage

\section{Introduction}\label{sec:intro}

The last few decades witnessed explosive growth in the amount of data humanity generates and needs to process.
Many rapidly expanding datasets consist of sequential (textual) data,
such as source code in version control systems~\cite{NavarroIndexes}, 
results of web crawls~\cite{fm2010}, versioned
document collections (such as Wikipedia)~\cite{resolution}, and, perhaps most notably,
biological sequences~\cite{Przeworski2000,BergerDY16}. The sizes of these datasets
already reach petabytes~\cite{hernaez2019genomic} and are predicted
to still get orders of magnitudes larger~\cite{estimate}. One of the
key characteristics of this data, and what turns searching such
datasets into the ultimate needle-in-a-haystack scenario, is that none
of it can be discarded: in computational biology, the presence or lack
of disease can depend on a single
mutation~\cite{Przeworski2000,estimate}, whereas in source code
repositories, a bug could be a result of a~single~typo.

What comes to the rescue is that these datasets are extremely redundant,
e.g., genomic databases are known to be up to 99.9\%
repetitive~\cite{Przeworski2000}. Researchers have therefore turned
their attention to techniques from lossless data compression. 
Compressing alone is not enough, however,
as this renders the text unreadable. The solution lies in incorporating
techniques from data compression directly into the design of
\emph{compressed algorithms} and \emph{compressed data structures}:
\begin{itemize}
\item To date, compressed algorithms have been developed for numerous
  problems, ranging from
  exact~\cite{Gaw11,Gaw13,Jez2015,AbboudBBK17,GanardiG22}
  and approximate string matching~\cite{AbboudBBK17,BringmannWK19,CKW20}, via
  computing edit distance~\cite{HermelinLLW13,Tiskin15,AbboudBBK17,GaneshKLS22}, 
  to fundamental linear algebra operations (such as inner product,
  matrix-vector multiplication, and matrix multiplication) ubiquitous
  in machine
  learning~\cite{AbboudBBK20,FerraginaMGKNST22,FranciscoGKLN22}.
\item Same can be said about data structures.  One can keep the data
  in compressed form and, at the price of a moderate (typically polylogarithmic) increase 
  in space complexity, efficiently support various queries on the original (uncompressed) text.
  The currently supported queries range
  from the fundamental local queries like random
  access~\cite{BLRSRW15,balancing,blocktree}, through less local
  rank and select~\cite{PereiraNB17,Prezza19,blocktree} or
  longest common extension (LCE)
  queries~\cite{NishimotoMFCS,tomohiro-lce,dynstr}, to the most
  powerful and complex queries like pattern
  matching~\cite{ClaudeN11,GagieGKNP12,
    GagieGKNP14,ClaudeNP21,Diaz-DominguezN21,ChristiansenEKN21,%
    KociumakaNO22} and full suffix array
  functionality~\cite{Gagie2020}.  The suffix array queries that,
  given a rank $i \in [1 \dd \Textlen]$, ask for the starting position of the $i$th
  lexicographically smallest suffix of the length-$\Textlen$ text, are known to
  be particularly powerful, as they form the backbone of dozens of
  string processing and data compression
  algorithms~\cite{gusfield,bwtbook}.
\end{itemize}

As the field matured, numerous ways to classify and compare
different compressed structures
emerged \cite{attractors,NavarroMeasures,delta}, resulting in
hierarchies of structures ordered according to their size and
functionality.  As expected, structures supporting the most
basic queries (such as random access) occupy the low-space
regime~\cite{delta}, while the most powerful indexes supporting suffix
array functionality, such as~\cite{Gagie2020}, require the most space.
The natural question was thus raised: How much space is \ul{required}
to efficiently support each functionality? Kociumaka, Navarro, and
Prezza~\cite{delta} recently proved that, letting $\SubstringComplexitySym$ be the
\emph{substring complexity} of the text, $\Textlen$ be its
length, and $\sigma$ be the size of the alphabet, a text can be
represented in $\bigO(\SubstringComplexitySym \log \tfrac{\Textlen \log \sigma}{\SubstringComplexitySym \log \Textlen})$
space, and this bound is asymptotically
tight as a function of $\SubstringComplexitySym$, $\Textlen$, and~$\sigma$.
Simultaneously, they showed that it is possible to
support random access and pattern-matching queries in the same space
(see also~\cite{KociumakaNO22} for improvements of pattern matching query time).
Given this situation, we thus ask:

\begin{center}
  \emph{What is the space required to efficiently support the most
    powerful queries,\\ such as the suffix array functionality?}
\end{center}

\paragraph{Our Results}

In this paper, we establish the following two main results:
\begin{enumerate}
\item We develop the first data structure, referred to as $\delta$-SA, that uses only
  $\bigO(\SubstringComplexitySym \log \tfrac{\Textlen \log \sigma}{\SubstringComplexitySym \log \Textlen})$ space and
  supports efficient suffix array queries (more precisely, it answers
  $\SA$ and inverse $\SA$ queries in $\bigO(\log^{4 + \epsilon} \Textlen)$
  time, where $\epsilon > 0$ is any given constant). 
  This collapses the existing rich hierarchy of compressed data structures
  (see~\cite{attractors,Gagie2020,resolution,NavarroMeasures,delta})
  into a single point: In view of our result, $\bigO(\SubstringComplexitySym \log \tfrac{n
    \log \sigma}{\SubstringComplexitySym \log \Textlen})$ is the fundamental space complexity for
  compressed text indexing since, on the one hand, such
  space is \emph{required} to represent the string~\cite{delta} (and
  this bound holds for all combinations of $\Textlen$, $\sigma$, and $\SubstringComplexitySym$) and, on the
  other hand, it is already \emph{sufficient} to support the powerful
  $\SA$ queries. Since the suffix array queries constitute the fundamental building block of
  string processing algorithms~\cite{gusfield,bwtbook}, our result immediately implies that dozens
  of algorithms can be executed in this \emph{$\delta$-optimal}
  space.
\item We present an algorithm that constructs the $\delta$-SA
  in $\bigO(\SubstringComplexitySym\polylog \Textlen)$ time from the Lempel--Ziv (LZ77) parsing
  of text~\cite{LZ77}.
  This is the first construction
  of an SA index running in \emph{compressed time}, i.e., in time
  nearly-linear in the compressed input size. The relevance of this
  result lies in the fact that LZ77 can be very efficiently approximated (using,
  e.g.,~\cite{KosolobovVNP20}) and then converted (in compressed
  time) into the canonical greedy form (see~\cref{prop:realLZ}). At the same time,
  LZ77 is already strong enough to compress any string into the $\delta$-optimal size
  $\bigO(\SubstringComplexitySym \log \tfrac{n
  \log \sigma}{\SubstringComplexitySym \log \Textlen})$~\cite{delta}. This makes LZ77 the
  perfect input in any pipeline of algorithms running in compressed
  time.  The only similar prior construction is the
  $\bigO(\SubstringComplexitySym\polylog \Textlen)$-time construction of run-length-encoded
  Burrows--Wheeler Transform (BWT) from the LZ77 parsing~\cite{resolution}. The similarity
  lies in the fact that RLBWT is one of the components of the $r$-index of
  Gagie, Navarro, and Prezza~\cite{Gagie2020}, which is also capable of answering SA
  queries. Our construction, however, is much stronger
  than~\cite{resolution}:
  \begin{itemize}
  \item Our algorithm builds a fully-functional SA index (the $\delta$-SA), whereas the
    construction from~\cite{resolution} builds only the run-length-encoded
    BWT~\cite{bwt}, which is just a single component of
    the index of~\cite{Gagie2020}. To date, the fastest algorithm building the complete index of~\cite{Gagie2020}
    based on the run-length-encoded BWT required
    $\bigO(\sqrt{\RLBWTSizeSym n} \polylog \Textlen) =
    \bigO(\sqrt{\SubstringComplexitySym n} \polylog \Textlen)$
    time~\cite{GT23}.
  \item The $\delta$-SA uses the $\delta$-optimal space,
    while the index of~\cite{Gagie2020} uses more space, i.e.,
    $\bigO(\RLBWTSizeSym \log \tfrac{n}{\RLBWTSizeSym})$, where
    $\RLBWTSizeSym \ge \SubstringComplexitySym$ is the number of
    runs in the BWT~\cite{bwt}.
  \item Our algorithm is deterministic, whereas~\cite{resolution}
    only provides a Las-Vegas randomized procedure.
  \end{itemize}
\end{enumerate}

On the way to our main results, we also achieve several auxiliary
goals of independent interest. In particular, we describe the first data structure efficiently
answering longest common extension (LCE) queries using the $\delta$-optimal
space of $\bigO(\SubstringComplexitySym \log \tfrac{\Textlen \log \sigma}{\SubstringComplexitySym \log \Textlen})$.
Moreover, we show how to deterministically construct it from
the LZ77 parsing in $\bigO(\SubstringComplexitySym\polylog \Textlen)$ time
(\cref{th:lce}). We also obtain the first analogous construction of a
data structure that supports random-access queries in $\bigO(\SubstringComplexitySym
\log \tfrac{\Textlen \log \sigma}{\SubstringComplexitySym \log \Textlen})$ space
(\cref{th:random-access});
the previous such indexes~\cite{delta,KociumakaNO22} only had 
$\Omega(\Textlen)$-time randomized construction algorithms.

One of the biggest technical hurdles to obtaining the above results is to
simultaneously achieve
\begin{enumerate}[label=(\alph*)]
\item $\delta$-optimal space,
\item polylogarithmic worst-case query time, and
\item construction in compressed time (preferably deterministic)
\end{enumerate}
\emph{for every component of the structure}. Satisfying any two out of
three would already constitute an improvement compared to the
state-of-the-art SA indexes and their construction
algorithms~\cite{Gagie2020,resolution}. We nevertheless show that simultaneously satisfying all
three is possible.  Our main new techniques to achieve this are: \begin{enumerate*}[label=(\arabic*)]
\item deterministic restricted recompression, and \item $\SubstringComplexitySym$-compressed
string synchronizing sets.\end{enumerate*}  Restricted recompression is a technique
proposed in~\cite{IPM} that, as shown in~\cite{delta}, allows constructing an RLSLP
(i.e., a run-length grammar; see \cref{sec:recompression-prelim})
representing the text in $\delta$-optimal space $\bigO(\SubstringComplexitySym \log \tfrac{n
  \log \sigma}{\SubstringComplexitySym \log \Textlen})$. The analysis
in~\cite{delta}, however, is probabilistic, and hence yields
only a Las-Vegas randomized algorithm that takes $\bigO(\Textlen)$ expected time. 
Here (\cref{sec:deterministic-recompression}), we improve it not only by proposing an
explicit construction, resulting in the first $\bigO(\Textlen)$-time deterministic
algorithm, but we also show how to achieve $\bigO(\SubstringComplexitySym\polylog \Textlen)$-time
construction from the LZ77 parsing.  The second
main new technical contribution is developing $\SubstringComplexitySym$-compressed string
synchronizing sets. String synchronizing sets~\cite{sss} are a
powerful symmetry-breaking mechanism with numerous applications,
including algorithms for longest common substrings computation~\cite{Charalampopoulos21},
indexing packed strings~\cite{sss,Dinklage0HKK20,breaking}, dynamic suffix
array~\cite{dynsa}, converting between compressed
representations~\cite{resolution}, and quantum string
algorithms~\cite{AkmalJ22,JinN23}.  
For a given parameter $\tau\in [1\dd \Textlen]$, this technique selects $\Oh(\Textlen/\tau)$ \emph{synchronizing positions}
so that positions with matching contexts are treated consistently 
and (except for highly periodic regions of the text) there is at least one synchronizing position among any $\tau$ consecutive positions.
In~\cite{resolution}, LZ77-compressed synchronizing sets (i.e., synchronizing sets represented by synchronizing positions located close to LZ77 phrase boundaries~\cite{LZ77}) were used to obtain an
$\bigO(\SubstringComplexitySym \polylog \Textlen)$-time algorithm for converting
the LZ77 parsing into the run-length compressed BWT~\cite{bwt}.  Their
technique, however, cannot be utilized here due to three major
obstacles: First,~\cite{resolution} uses $\Omega(\LZSizeSym \log \Textlen)$ space (where $\LZSizeSym \ge \SubstringComplexitySym$ is the size
of the LZ77 parsing), and
hence does not meet the $\delta$-optimal space bound requirement. Second, the algorithm
in~\cite{resolution} is able to infer some suffix ordering only
for a batch of suffixes. In other words, it is an offline solution to
the problem stated in this paper. Obtaining an online solution (i.e.,
a data structure) requires a different approach. Finally, the synchronizing set
construction used in~\cite{resolution} is Las-Vegas randomized
and hence does not satisfy our goal of achieving deterministic construction.
To overcome the first obstacle, 
rather than storing the synchronizing positions around the LZ77 phrase boundaries,
 we store them in what we call a \emph{cover}: the
set of positions in the text covered by the leftmost occurrences of
substrings of some fixed length (see~\cref{sec:cover}). This lets us
bound the number of stored synchronizing positions in terms of 
the substring complexity $\SubstringComplexitySym$ (see~\cref{sec:sss}). 
The bulk of our paper is devoted to overcoming the second obstacle. We show how to
combine weighted range counting and selection queries~\cite{chazelle} with
the ``range refinement'' technique inspired by dynamic suffix
arrays~\cite{dynsa}, which gradually shrinks the range of $\SA$ to
contain only suffixes prefixed with a desired length-$2^k$ string, to
obtain the SA functionality. This requires substantial modifications
compared to~\cite{dynsa} since dynamic suffix arrays are not
compressed (they use $\bigO(\Textlen \polylog \Textlen)$ space). Finally, to
address the third challenge, we utilize a novel construction of
synchronizing sets using restricted recompression, binding our last
problem to the first technique. This is similar to~\cite{dynsa},
except that here we avoid the $\bigO(\log^{*} \Textlen)$ space increase since our
structure is static. Instead, we carefully design a potential function
that guides the recompression algorithm so that the outcome is deterministically 
as good as it would have been in expectation if we used randomization.
 We give a more detailed overview of our
techniques in \cref{sec:technical-overview}. Our final result is
summarized as follows.\footnote{We did not aim to optimize the $\polylog \Textlen$ factor in the construction algorithm.
In particular, we utilized existing procedures whose running time has not been optimized either.}

\begin{restatable}[$\delta$-SA]{theorem}{thmain}\label{th:main}
  Given the LZ77 parsing of $\Text \in \IntegerAlphabet^{\Textlen}$ and any constant $\epsilon \in (0,
  1)$, we can in $\bigO(\SubstringComplexitySym \log^7 \Textlen)$
  time construct a data structure of size
  $\bigO(\SubstringComplexitySym \log \tfrac{\Textlen \log \sigma}{\SubstringComplexitySym \log \Textlen})$
  (where $\SubstringComplexitySym$ is the substring complexity of $\Text$) that, given
  any position $i \in [1 \dd \Textlen]$, returns the values $\SA[i]$ and $\SA^{-1}[i]$ in
  $\bigO(\log^{4 + \epsilon} \Textlen)$ time.
  The construction algorithm is deterministic, and the running times are worst-case.
\end{restatable}

\paragraph{Related Work}
In this paper, we focus on algorithms and data structures working for
\emph{highly repetitive} strings~$\Text$, which can be defined as those for
which either of the values: $\LZSize{\Text}$ (the size of the LZ77
parsing~\cite{LZ77}), $\SmallestAttractor{\Text}$ (the size of the smallest
string attractor~\cite{attractors}), $\SmallestGrammar{\Text}$ (the size of the
smallest context-free grammar~\cite{KiefferY00,Rytter03,charikar}), $\RLBWTSize{\Text}$ (the number of runs
in the BWT~\cite{bwt}), or $\SubstringComplexity{\Text}$ (the substring
complexity~\cite{delta}) are significantly smaller than $\Textlen$ (the list
of such measures goes on~\cite{macro,collage,kreft2010navarro}; see~\cite{NavarroMeasures} for a survey).  We can use
either of them, since a series of
papers~\cite{Rytter03,charikar,attractors,GNPlatin18,resolution,KempaS22,delta}
demonstrates that the ratio between any two of these values is
$\bigO(\polylog \Textlen)$ for every text $\Text$ of length $\Textlen$. 
The redundancy captured by these measures is present in modern massive datasets. A lot of the
earlier work on small-space data structures, however, focused on
reducing the sizes of structures relative to the size of text, i.e.,
$\bigO(\Textlen \log \sigma)$ bits (assuming $\Text \in \IntegerAlphabet^{\Textlen}$) or, a step
further, on achieving some variant of the $k$th order entropy bound
$\bigO(\Textlen H_k(\Text)) + o(\Textlen \log \sigma)$, i.e., removing the
``statistical'' redundancy caused by skewed frequencies of individual symbols or short substrings of length $o(\log_\sigma n)$.
Some of the most popular structures in this setting include those answering
rank and select queries, such as wavelet trees~\cite{wt}, or those with
pattern matching and suffix array or suffix tree functionality, such as
the FM-index~\cite{FerraginaM05}, the compressed suffix array
(CSA)~\cite{GrossiV05}, or the compressed suffix tree (CST)~\cite{cst}.
Many of these structures have subsequently been implemented and are
now available via libraries, such as
\href{https://github.com/simongog/sdsl-lite}{\texttt{sdsl}}~\cite{sdsl}.
The construction of these indexes has also received a lot of attention
and nowadays, most of them can be constructed very
efficiently~\cite{HonSS03,Belazzougui14,MunroNN17,Kempa19}.
Recently, new $\Oh(n\log\sigma)$-bit indexes with CSA and CST capabilities have been
proposed that also admit $o(\Textlen)$-time construction~\cite{breaking} if $\log \sigma = o(\sqrt{\log n})$.
We refer to~\cite{NavarroM07,Navarro14,BelazzouguiN14,navarrobook} for
further details.

\paragraph{Organization of the Paper}

After introducing the basic notation and tools in \cref{sec:prelim},
we present a technical overview in~\cref{sec:technical-overview}.  In
\cref{sec:cover}, we then describe the notion of string covers, which
we subsequently combine with the new deterministic restricted recompression in
\cref{sec:recompression}. Next, we describe the auxiliary
structures supporting range queries on various grids (\cref{sec:range-queries})
and the so-called modular constraint queries (\cref{sec:mod-queries}). Finally,
in the main \cref{sec:sa}, we present our new index, the $\delta$-SA, which answers SA and inverse SA
queries.

\section{Preliminaries}\label{sec:prelim}

\begin{wrapfigure}{R}{0.36\textwidth}
  \vspace{-.5cm}
  \begin{tikzpicture}[yscale=0.35]
    \foreach \x [count=\i] in {a, aababa, aababababaababa,
      aba, abaababa, abaababababaababa, ababa, ababaababa,
      abababaababa, ababababaababa, ba, baababa,
      baababababaababa, baba, babaababa, babaababababaababa,
      bababaababa, babababaababa, bbabaababababaababa}
        \draw (1.9, -\i) node[right]
          {$\texttt{\x}$};
    \draw(1.9,0) node[right] {\scriptsize $\Text[\SA[i]\dd \Textlen]$};
    \foreach \x [count=\i] in {b, b, b, b, b, b, a, b, b,
                               a, a, a, a, a, a, b, a, a,\$}
      \draw (0.7, -\i) node {\footnotesize $\i$};
    \draw(0.7,0) node{\scriptsize $i$};
    \foreach \x [count=\i] in {19,14,5,17,12,3,15,10,
                               8,6,18,13,4,16,11,2,9,7,1}
      \draw (1.4, -\i) node {$\x\vphantom{\textbf{\underline{7}}}$};
    \draw(1.4,0) node{\scriptsize $\SA[i]$};
  \end{tikzpicture}

  \vspace{-1.5ex}
  \caption{A list of all sorted suffixes of $\Text =
    \texttt{bbabaababababaababa}$ along with
    the suffix array.}\label{fig:example}
  \vspace{-0.2cm}
\end{wrapfigure}

\paragraph{Basic definitions}

A \emph{string} is a finite sequence of characters from a given
\emph{alphabet} $\Sigma$.  The length of a string $S$ is denoted $|S|$. For $i
\in [1\dd |S|]$,\footnote{ For $i,j\in \mathbb{Z}$, denote $[i\dd j] =
\{k \in \mathbb{Z} : i \le k \le j\}$, $[i
\dd j)=\{k \in \mathbb{Z} : i \le k <
j\}$, and $(i\dd j]={\{k \in \mathbb{Z}: i
< k \le j\}}$.  } the $i$th character of $S$ is denoted $S[i]$.
A~\emph{substring} of $S$ is a string of the form $S[i \dd j) =
S[i]S[{i+1}]\cdots S[{j-1}]$ for some $1\le i \le j \le |S|+1$. 
Substrings the form $S[1\dd j)$ and $S[i\dd |S|]$ are called
\emph{prefixes} and \emph{suffixes}, respectively. We use $\revstr{S}$
to denote the \emph{reverse} of $S$, i.e., $S[|S|]\cdots S[2]S[1]$.
We denote the \emph{concatenation} of two strings $U$ and $V$, that
is, $U[1]\cdots U[|U|]V[1]\cdots V[|V|]$, by $UV$ or $U\cdot V$.
Furthermore, $S^k = \bigodot_{i=1}^k S$ is the concatenation of $k \in
\Zz$ copies of $S$; note that $S^0 = \emptystring$ is the \emph{empty
string}. A nonempty string $S$ is said to be \emph{primitive} if it
cannot be written as $S = U^k$, where $k \geq 2$.  An integer $p\in
[1\dd |S|]$ is a \emph{period} of $S$ if $S[i] = S[i + p]$ holds for
every $i \in [1 \dd |S|-p]$. We denote the shortest period of $S$ as
$\per(S)$.  For every $S \in \Sigma^{+}$, we define the
infinite power $S^{\infty}$ so that $S^{\infty}[i] = S[1 + (i-1) \bmod |S|]$
for $i \in \Z$.  In particular, $S = S^{\infty}[1 \dd |S|]$.  
The \emph{rotation} operation
$\rot(\cdot)$, given a string $S\in \Sigma^+$, moves the last
character of $S$ to the front so that $\rot(S) = S[|S|] \cdot S[1 \dd
|S| - 1]$.  The inverse operation $\rot^{-1}(\cdot)$ moves the first
character of $S$ to the back so that $\rot^{-1}(S) = S[2 \dd |S|]
\cdot S[1]$.  For an integer $s \in \Z$, the operation
$\rot^{s}(\cdot)$ denotes the $|s|$-time composition of $\rot(\cdot)$
(if $s \ge 0$) or $\rot^{-1}(\cdot)$ (if $s \le 0$).  Strings $S, S'$
are \emph{cyclically equivalent} if $S' = \rot^{s}(S)$ for some $s \in
\Z$.  By $\lcp(U, V)$ (resp.\ $\lcs(U, V)$) we denote the length of
the longest common prefix (resp.\ suffix) of $U$ and $V$. For any
string $S \in \Sigma^{*}$ and any $j_1, j_2 \in [1 \dd |S|]$, we
denote $\LCE_{S}(j_1, j_2) = \lcp(S[j_1 \dd |S|], S[j_2 \dd |S|])$.

We use $\preceq$ to denote the order on $\Sigma$, extended to the
\emph{lexicographic} order on $\Sigma^*$ so that $U,V\in \Sigma^*$
satisfy $U \preceq V$ if and only if either \begin{enumerate*}[label=(\alph*)] \item $U$ is a prefix of $V$, or \item
$U[1 \dd i) = V[1 \dd i)$ and $U[i]\prec V[i]$ holds for some $i\in
[1\dd \min(|U|,|V|)]$. \end{enumerate*}

\begin{definition}\label{def:occ}
  For any $\Text\in \Sigma^{\Textlen}$, $\Pat \in \Sigma^*$, and
  integer $\ell \geq 0$, we define\vspace{-.4cm}
  \begin{align*}
    \OccThree{\ell}{\Pat}{\Text}
      &= \{j' \in [1 \dd \Textlen] : \lcp(\Pat, \Text[j' \dd \Textlen]) \geq
         \min(|\Pat|, \ell)\},\\
    \RangeBegThree{\ell}{\Pat}{\Text}
      &= |\{j' \in [1 \dd \Textlen] : \Text[j' \dd \Textlen] \prec \Pat\text{ and }
         j'\notin \OccThree{\ell}{\Pat}{\Text}\}|,\\
    \RangeEndThree{\ell}{\Pat}{\Text}
      &= \RangeBegThree{\ell}{\Pat}{\Text} + |\OccThree{\ell}{\Pat}{\Text}|.
  \end{align*}
  When $\ell = |\Pat|$, we simply write $\OccTwo{\Pat}{\Text}$,
  $\RangeBegTwo{\Pat}{\Text}$, and $\RangeEndTwo{\Pat}{\Text}$. Moreover, by $\OccThree{\ell}{j}{\Text}$,
  $\RangeBegThree{\ell}{j}{\Text}$, and $\RangeEndThree{\ell}{j}{\Text}$ we mean, respectively,
  $\OccThree{\ell}{\Text[j \dd \Textlen]}{\Text}$, $\RangeBegThree{\ell}{\Text[j \dd \Textlen]}{\Text}$, and
  $\RangeEndThree{\ell}{\Text[j \dd \Textlen]}{\Text}$.
\end{definition}

\begin{remark}
  Note that the above generalizes for the notation for ranges used
  in~\cite{breaking} (where the parameter $\ell$ was not used) and
  from~\cite{dynsa} (where it was only defined for patterns of the
  form $\Pat = \Text[j \dd \Textlen]$). Here, we obtain all these notations as
  special cases.
\end{remark}

\paragraph{Suffix array}
For any $\Text \in \Sigma^{\Textlen}$ (where $\Textlen \geq 1$), the \emph{suffix
array} $\SA_{\Text}[1 \dd \Textlen]$ of $\Text$ is a permutation of $[1 \dd \Textlen]$
such that $\Text[\SA_{\Text}[1] \dd \Textlen] \prec \Text[\SA_{\Text}[2] \dd \Textlen] \prec \cdots
\prec \Text[\SA_{\Text}[\Textlen] \dd \Textlen]$, i.e., $\SA_{\Text}[i]$ is the starting position
of the lexicographically $i$th suffix of $\Text$; see \cref{fig:example}
for an example.  The \emph{inverse suffix array} $\ISA_{\Text}[1 \dd \Textlen]$ (also
denoted $\SA^{-1}_{\Text}[1 \dd \Textlen]$)
is the inverse permutation of $\SA_{\Text}$, i.e., $\ISA_{\Text}[j] = i$ holds if
and only if $\SA_{\Text}[i] = j$. Intuitively, $\ISA_{\Text}[j]$ stores the
lexicographic \emph{rank} of $\Text[j \dd \Textlen]$ among the suffixes of~$\Text$. Note
that if $\Text \neq \emptystring$, then $\OccThree{\ell}{\Pat}{\Text} =
\{\SA_{\Text}[i] : i \in (\RangeBegThree{\ell}{\Pat}{\Text} \dd
\RangeEndThree{\ell}{\Pat}{\Text}]\}$ holds for every $\Pat \in \Sigma^*$
and $\ell \geq 0$. Whenever $\Text$ is clear from the context, we drop
the subscript in $\SA_{\Text}$ and $\ISA_{\Text}$.

\paragraph{Substring complexity}
For a string $\Text\in \Sigma^{\Textlen}$ and $\ell \in \Zp$, we denote the number of
length-$\ell$ substrings by $\SubstrCount{\ell}{\Text} = |\{\Text[i \dd i + \ell) : i \in [1\dd \Textlen-\ell+1]\}|$; note that $\SubstrCount{\ell}{\Text}=0$ if $\ell > \Textlen$.
The \emph{substring complexity} of $\Text$ is defined as $\SubstringComplexity{\Text} = 
\max_{\ell=1}^{\Textlen}\frac{1}{\ell}\SubstrCount{\ell}{\Text}$~\cite{delta}.
On the one hand, as shown in~\cite{delta}, the measure $\SubstringComplexitySym$ is an asymptotic lower
bound for nearly all known compression algorithms and repetitiveness
measures, including LZ77~\cite{LZ77}, run-length-compressed
BWT~\cite{bwt}, grammar compression~\cite{charikar}, and string
attractors~\cite{attractors}. On the other hand,~\cite{delta} also
shows that, if $\Sigma = \IntegerAlphabet$, then it is possible to represent $\Text$
using $\bigO(\SubstringComplexity{\Text} \log
\tfrac{\Textlen \log \sigma}{\SubstringComplexity{\Text} \log \Textlen})$ space (or more
precisely, $\bigO(\SubstringComplexity{\Text}
\log \tfrac{\Textlen \log \sigma}{\SubstringComplexity{\Text} \log \Textlen} \log \Textlen)$
bits), and this bound is asymptotically tight as a function of $\Textlen$, $\sigma$, and $\SubstringComplexity{\Text}$.
In other words, there is no combination of values 
$\Textlen$, $\sigma$, and $\SubstringComplexity{\Text}$ such that every string 
$\Text\in \IntegerAlphabet^{\Textlen}$ can be encoded using $o(\SubstringComplexity{\Text} \log
\tfrac{\Textlen \log \sigma}{\SubstringComplexity{\Text} \log \Textlen} \log \Textlen)$ bits.

\paragraph{Lempel--Ziv compression}

A fragment $\Text[i \dd i + \ell)$ of $\Text$ is a \emph{previous factor} if
it has an earlier occurrence in $\Text$, i.e., $\LCE_{\Text}(i,i')\ge \ell$
holds for some $i'\in [1 \dd i)$.  An \emph{LZ77-like factorization}
of $\Text$ is a factorization $\Text = F_1 \cdots F_f$ into non-empty
\emph{phrases} such that each phrase $F_j$ with $|F_j| > 1$ is a
previous factor. In the underlying \emph{LZ77-like representation},
every phrase $F_j = \Text[i \dd i + \ell)$ that is a previous factor is
encoded as $(i', \ell)$, where $i' \in [1 \dd i)$ satisfies $\LCE_{\Text}(i,
i')\geq \ell$ (and is chosen arbitrarily in case of multiple
possibilities); if $F_j = \Text[i]$ is not a previous factor, we encode it
as $(\Text[i], 0)$.

The LZ77 factorization~\cite{LZ77} (or the LZ77 parsing) of a string
$\Text$ is then just an LZ77-like factorization constructed by greedily
parsing $\Text$ from left to right into the longest possible phrases.  More
precisely, the $j$th phrase $F_j$ is the longest previous factor
starting at position $1 + |F_1 \cdots F_{j - 1}|$; if no previous
factor starts there, then $F_j$ consists of a single character. 
This greedy construction yields the smallest LZ77-like factorization of~$\Text$ \cite[Theorem~1]{LZ76}.
We denote the number of phrases in the LZ77 parsing of $\Text$ by $\LZSize{\Text}$.
For example, the text $\Text = \texttt{bbabaababababaababa}$ of
\cref{fig:example} has LZ77 factorization $\texttt{b}\cdot
\texttt{b}\cdot \texttt{a}\cdot \texttt{ba} \cdot \texttt{aba}\cdot
\texttt{bababa} \cdot \texttt{ababa}$ with $\LZSize{\Text} = 7$ phrases, and its
LZ77 representation is $ (\texttt{b},0), (1,1), (\texttt{a},0), (2,2),
(3,3), (7,6), (10,5). $

\begin{proposition}\label{prop:realLZ}
  Given a string $\Text$ of length $\Textlen$, represented using an LZ77-like
  parsing consisting of $f$ phrases, the LZ77 parsing of $\Text$ can be
  constructed in $\bigO(f \log^4 \Textlen)$ time.
\end{proposition}
\begin{proof}
  We use \cite[Theorem~6.11]{resolutionfull} to build a data structure
  that, for any pattern $P$ represented by its arbitrary occurrence in
  $\Text$, returns the leftmost occurrence of $P$ in $\Text$.  Then, we
  process $\Text$ from left to right constructing the LZ77 parsing of $\Text$.
  Suppose that we have already parsed a prefix $\Text[1 \dd i)$. We binary
  search for the maximum length $\ell$ such that the leftmost
  occurrence of $\Text[i \dd i+\ell)$ is $\Text[i' \dd i' + \ell)$ for some
  $i' \in [1 \dd i)$. By definition of the LZ77 parsing, the next phrase is either
  $\Text[i]$ (if $\ell = 0$) or $\Text[i \dd i + \ell)$ (otherwise).
  The construction time of \cite[Theorem~6.11]{resolutionfull} is
  $\bigO(f \log^4 \Textlen)$, whereas the query time is $\bigO(\log^3 \Textlen)$.  For
  each phrase of the LZ77 parsing, we make $\bigO(\log \Textlen)$ queries,
  which take $\bigO(\log^4 \Textlen)$ time in total. Since $\LZSize{\Text} \leq f$, the
  overall running time is $\big(f \log^4 \Textlen)$.
\end{proof}

\paragraph{String Synchronizing Sets}

\begin{definition}[$\tau$-synchronizing set~\cite{sss}]\label{def:sss}
  Let $\Text\in \Sigma^{\Textlen}$ be a string and let $\tau \in
  [1 \dd \lfloor\frac{\Textlen}{2}\rfloor]$ be a parameter. A set $\SSS
  \subseteq [1 \dd \Textlen - 2\tau + 1]$ is called a
  \emph{$\tau$-synchronizing set} of $\Text$ if it satisfies the
  following \emph{consistency} and \emph{density} conditions:
  \begin{enumerate}
  \item\label{def:sss-consistency}
    If $\Text[i \dd i + 2\tau) = \Text[j\dd j + 2\tau)$, then $i \in \SSS$
    holds if and only if $j \in \SSS$
    (for $i, j \in [1 \dd \Textlen - 2\tau + 1]$),
  \item\label{def:sss-density}
    $\SSS\cap[i \dd i + \tau) = \emptyset$ if and only if
    $i \in \RTwo{\tau}{\Text}$ (for $i \in [1 \dd \Textlen - 3\tau + 2]$),
    where
    \[
      \RTwo{\tau}{\Text} := \{i \in [1 \dd \Textlen - 3\tau + 2] :
      \per(\Text[i \dd i + 3\tau - 2]) \leq \tfrac{1}{3}\tau\}.
    \]
  \end{enumerate}
\end{definition}

\begin{remark}\label{rm:sss-size}
  In most applications, we want to minimize $|\SSS|$. Note, however, that
  the density condition imposes a lower bound
  $|\SSS| = \Omega(\frac{\Textlen}{\tau})$ for strings of length
  $\Textlen \ge 3\tau-1$ that do not contain substrings of length $3\tau - 1$
  with period at most $\frac{1}{3}\tau$.  Thus, we cannot hope to achieve an
  upper bound improving in the worst case upon the following one.
\end{remark}

\begin{theorem}[{\cite[Proposition~8.10]
      {sss}}]\label{th:sss-existence-and-construction}
  For every string $\Text$ of length $\Textlen$ and parameter $\tau \in
  [1 \dd \lfloor\frac{\Textlen}{2}\rfloor]$, there exists a $\tau$-synchronizing
  set $\SSS$ of size $|\SSS| =
  \bigO\left(\frac{\Textlen}{\tau}\right)$.
  Moreover, if $\Text \in \IntegerAlphabet^{\Textlen}$, where
  $\sigma = \Textlen^{\bigO(1)}$, such $\SSS$ can be deterministically
  constructed in $\bigO(\Textlen)$ time.
\end{theorem}

\paragraph{Model of computation}

We use the standard word RAM model of computation~\cite{Hagerup98}
with $w$-bit \emph{machine words}, where $w \ge \log \Textlen$, and all
standard bit-wise and arithmetic operations taking $\bigO(1)$ time.
Unless explicitly stated otherwise, we measure the space complexity in machine words.

\section{Technical Overview}\label{sec:technical-overview}

Let $\Text \in \Sigma^{\Textlen}$, where $\Sigma =\IntegerAlphabet$.
Assume that $\Text[\Textlen]$ is a symbol that
does not occur in $\Text[1 \dd \Textlen)$.  
Moreover, let $\epsilon \in (0, 1)$ be a constant.
In this section, we give an overview of the $\delta$-SA,
which is a compressed text index that
\begin{enumerate*}[label=(\alph*)]
  \item takes
  $\bigO(\SubstringComplexity{\Text} \log \tfrac{\Textlen \log
  \sigma}{\SubstringComplexity{\Text} \log \Textlen})$ space,
  \item answers
  $\SA$ and $\ISA$ queries on $\Text$ in $\bigO(\log^{4 + \epsilon}
  \Textlen)$ time, and 
  \item can be constructed from the LZ77 parsing of
  $\Text$ in $\bigO(\SubstringComplexity{\Text} \log^7 \Textlen)$ time.
\end{enumerate*} 

\subsection{SA and ISA Queries}\label{sec:to-queries}

\paragraph{The Basic Idea}

First, we observe (\cref{lm:occ-equivalence}) that the uniqueness of
$\Text[\Textlen]$ implies that $\OccThree{\ell}{j}{\Text} = \{j' \in [1
\dd \Textlen] : \Textinf[j' \dd j' + \ell) = \Textinf[j \dd j +
\ell)\}$ holds for  every $j \in [1 \dd \Textlen]$ and $\ell \geq 0$
(cf. \cref{def:occ}).  The main idea of the query algorithms
is as follows:

\begin{itemize}
\item To calculate $\ISA[j]$ given $j \in [1 \dd \Textlen]$,
   we compute the following three values for subsequent $k\in [4\dd \ceil{\log n}]$:
   the ranks $b=\RangeBegThree{2^k}{j}{\Text}$ and $e=\RangeEndThree{2^k}{j}{\Text}$
   such that $\OccThree{2^k}{j}{\Text} = \{\SA[i] : i \in (b \dd e]\}$ (as discussed in \cref{sec:prelim})
   as well as an arbitrary position $j' \in \OccThree{2^k}{j}{\Text}$ satisfying $j' = \min
   \OccThree{2^{k+1}}{j'}{\Text}$.
   For $k=4$, these values are computed from scratch;
   subsequently, we rely on the output of the preceding step.
   After completing the final step, we return $\ISA[j] :=
   \RangeEndThree{\ell}{j}{\Text}$, where $\ell = 2^{\lceil \log
  \Textlen \rceil} \geq n$. 
   This is the correct answer because $\OccThree{\ell}{j}{\Text} =
  \OccThree{n}{j}{\Text} = \{j\}$.
\item To calculate $\SA[i]$ given $i \in [1 \dd \Textlen]$, we proceed
  similarly, that is, we compute, for $k\in [4\dd \ceil{\log n}]$,
  the ranks $b=\RangeBegThree{2^k}{\SA[i]}{\Text}$ and $e=\RangeEndThree{2^k}{\SA[i]}{\Text}$
  as well as an arbitrary position $j' \in \OccThree{2^k}{\SA[i]}{\Text}$ satisfying $j' = \min
   \OccThree{2^{k+1}}{j'}{\Text}$.
  After completing the final step, we return the position $j'$
  satisfying $j' \in \OccThree{\ell}{\SA[i]}{\Text}$ for $\ell = 2^{\lceil \log
  \Textlen \rceil} \geq n$.
  This is the correct answer because  $\OccThree{\ell}{\SA[i]}{\Text} =
  \OccThree{n}{\SA[i]}{\Text} = \{\SA[i]\}$.
  Note that the individual steps of computing $\SA[i]$
  are different from those for $\ISA[j]$ since we are given the rank $i$ rather than the position $\SA[i]$.
\end{itemize}

This basic framework is similar to~\cite{dynsa}. The major difference,
however, lies in the implementation of the ``refinement'' procedure:
While~\cite{dynsa} uses $\widetilde{\Theta}(\Textlen)$
space\footnote{The $\widetilde{\Oh}(\cdot)$ notation hides factors
polylogarithmic in the (uncompressed) input size $n$. In other words,
for any function $f$, we have $\widetilde{\Oh}(f)=\Oh(f\polylog
n)$. Similarly, $\widetilde{\Omega}(f)=\Omega(f/\polylog n)$ and
$\widetilde{\Theta}(f)=\widetilde{\Oh}(f)\cap
\widetilde{\Omega}(f)$.}, here we can only store
$\widetilde{\bigO}(\SubstringComplexity{\Text})$ words. Since
this space allowance can be up to exponentially smaller,
a much more complex approach is required.

To implement the initial step in both queries, it suffices to store
all length-$16$ substrings of $\Textinf$ in the lexicographic order, each
augmented with the endpoints of the corresponding $\SA$ range and the
position of the leftmost occurrence in $\Textinf[1 \dd)$
(\cref{sec:sa-final-ds}). Since $\Textinf$ contains fewer than $\SubstrCount{16}{\Text}+16 \leq
16\SubstringComplexity{\Text}+16 \le 32\SubstringComplexity{\Text}$ length-$16$ substrings, the resulting arrays need
$\bigO(\SubstringComplexity{\Text})$ space. They are also easy
to obtain from the LZ77 parsing: It suffices to consider all
length-$16$ substrings overlapping phrase boundaries; for each of them,
the leftmost occurrence and the number of occurrences can be determined using 
existing compressed text indexes~\cite[Theorems~6.11 and~6.21]{resolutionfull} 
(note that we use these indexes solely within our construction procedure; they are not included in the $\delta$-SA).
The key difficulty is thus the refinement procedure.

\begin{definition}\label{def:pat-posless}
  For any $\ell \geq 1$ and $\Pat \in \Sigma^{+}$, we define
  \vspace{-1ex}
  \begin{align*}
    \PosBeg{\ell}{\Pat}{\Text}
      &= \{j' \in \OccThree{\ell}{\Pat}{\Text} : \Text[j' \dd \Textlen] \prec \Pat\text{ and }
         j'\notin \OccThree{2\ell}{\Pat}{\Text}\},\\
    \PosEnd{\ell}{\Pat}{\Text}
      &= \{j' \in \OccThree{\ell}{\Pat}{\Text} : \Text[j' \dd \Textlen] \succ \Pat\text{ and }
      j'\notin \OccThree{2\ell}{\Pat}{\Text}\}.
  \vspace{-1ex}
  \end{align*}
  We denote $\DeltaBeg{\ell}{\Pat}{\Text} :=
  |\PosBeg{\ell}{\Pat}{\Text}|$ and $\DeltaEnd{\ell}{\Pat}{\Text}: =
  |\PosEnd{\ell}{\Pat}{\Text}|$.  For any $j \in [1 \dd \Textlen]$, we
  then let $\PosBeg{\ell}{j}{\Text} := \PosBeg{\ell}{\Text[j \dd
  \Textlen]}{\Text}$ and $\PosEnd{\ell}{j}{\Text} :=
  \PosEnd{\ell}{\Text[j \dd \Textlen]}{\Text}$.  The values
  $\DeltaBeg{\ell}{j}{\Text}$ and $\DeltaEnd{\ell}{j}{\Text}$ are
  defined analogously.
\end{definition}

Let us fix $k \in [4 \dd \lceil \log \Textlen \rceil)$ and denote $\ell =
2^k$.  We observe that every $\Pat$ satisfies
$\RangeBegThree{2\ell}{\Pat}{\Text} =
\RangeBegThree{\ell}{\Pat}{\Text} + \DeltaBeg{\ell}{\Pat}{\Text}$ and
$\RangeEndThree{2\ell}{\Pat}{\Text} =
\RangeEndThree{\ell}{\Pat}{\Text} - \DeltaEnd{\ell}{\Pat}{\Text}$
(\cref{lm:pat-posless}). In particular, every $j \in [1 \dd
\Textlen]$ satisfies $\RangeBegThree{2\ell}{j}{\Text} =
\RangeBegThree{\ell}{j}{\Text} + \DeltaBeg{\ell}{j}{\Text}$ and
$\RangeEndThree{2\ell}{j}{\Text} = \RangeEndThree{\ell}{j}{\Text} -
\DeltaEnd{\ell}{j}{\Text}$. Thus, to refine the suffix array range, it
suffices to compute any two values among $\DeltaBeg{\ell}{j}{\Text},
\DeltaEnd{\ell}{j}{\Text}, |\OccThree{2\ell}{j}{\Text}|$ (during an
$\ISA$ query) or among $\DeltaBeg{\ell}{\SA[i]}{\Text},
\DeltaEnd{\ell}{\SA[i]}{\Text},\allowbreak |\OccThree{2\ell}{\SA[i]}{\Text}|$
(during an $\SA$ query).

Denote $\tau = \lfloor \tfrac{\ell}{3} \rfloor$ and let $\RName$ be a
shorthand for $\RTwo{\tau}{\Text}=\{i \in [1 \dd \Textlen - 3\tau + 2] :
\per(\Text[i \dd i + 3\tau - 2]) \leq \tfrac{1}{3}\tau\}$. The refinement step during the
computation of $\ISA[j]$ (resp.\ $\SA[i]$) works differently depending
on whether $j \in \RName$ (resp.\ $\SA[i] \in \RName$), in which case
we call $j$ (resp.\ $\SA[i]$) \emph{periodic}. Otherwise, the position is called
\emph{nonperiodic}.  To distinguish these two cases, we store the set $\RName
\cap \Cov$, where $\Cov$ is a $14\tau$-cover of $\Text$, i.e., a
subset of $[1 \dd \Textlen]$ including all positions covered by the
leftmost occurrences of length-$14\tau$ substrings of $\Text$ (\cref{def:cover}).
As $\RName \cap \Cov$ might be
large, we store its \emph{interval representation} $\IntervalRepr{\RName \cap \Cov}$,
that is, we express $\RName \cap \Cov$ as a union of disjoint
integer intervals (\cref{def:interval-representation}).

\begin{description}[style=sameline,itemsep=1ex,font={\normalfont\itshape}]
\item[Observation 1: There exists a $14\tau$-cover $\Cov$ such that $|\IntervalRepr{\Cov}| =
  \bigO(\SubstringComplexity{\Text})$ and $\IntervalRepr{\Cov}$ admits a fast
  construction algorithm from the LZ77 parsing of $\Text$.] Let $\Cov$
  be a union of $(\Textlen - 28\tau \dd \Textlen]$ as well as
  intervals $[x \dd x + 28\tau)$ over all $x \in [1 \dd \Textlen -
  28\tau]$ such that $x \equiv 1 \pmod{14\tau}$ and $x = \min
  \OccThree{28\tau}{x}{\Text}$.  The set $\Cov$ is a $14\tau$-cover
  since the leftmost occurrence of every length-$14\tau$ substring of
  $\Text$ can be extended into an interval $[x \dd x + 28\tau)$ for
  $x$ as above; see \cref{lm:c-cover}. Note that $\Cov$ is a subset of
  positions covered by the leftmost occurrences of \emph{all}
  length-$28\tau$ substring of $\Text$.  By~\cite{delta}, we thus have
  $|\Cov| \leq 84\tau\SubstringComplexity{\Text}$ (see also
  \cref{lm:c-cover}).  On the other hand, $\Cov$ is a union of
  length-$28\tau$ intervals. Thus, $|\IntervalRepr{\Cov}| = \bigO(|\Cov|/\tau) =
  \bigO(\SubstringComplexity{\Text})$. To construct $\IntervalRepr{\Cov}$, it
  suffices to check at most two positions around each LZ77 phrase
  boundary (\cref{pr:cover-construction}). Using an index for finding
  leftmost occurrences~\cite[Theorem~6.11]{resolutionfull}, we can thus build
  $\IntervalRepr{\Cov}$ in $\bigO(\LZSize{\Text} \polylog \Textlen) =
  \bigO(\SubstringComplexity{\Text} \polylog \Textlen)$~time.

\item[Observation 2: The above $\Cov$ satisfies
  $|\IntervalRepr{\RName \cap \Cov}| =
  \bigO(\SubstringComplexity{\Text})$ and
  $\IntervalRepr{\RName \cap \Cov}$
  also admits fast construction.] First, we observe that any two
  maximal blocks of consecutive positions in $\RName$ are separated by
  a gap of size $\Omega(\tau)$ (\cref{lm:gap}). This implies that,
  using the above $\Cov$, the interval representation of $\Cov \cap
  \RName$ is of size $\bigO(\SubstringComplexity{\Text})$
  (\cref{lm:IR-comp-R-size}). Above, we noted that $\IntervalRepr{\Cov}$ can be
  constructed from the LZ77 parsing in
  $\bigO(\SubstringComplexity{\Text}\polylog \Textlen)$ time. It
  remains to observe that, given $\IntervalRepr{\Cov}$, constructing
  $\IntervalRepr{\RName \cap \Cov}$
  reduces to computing the shortest periods via so-called 2-period
  queries (\cref{lm:sa-core-construction}).  Consequently, by
  utilizing the index for 2-period queries \cite[Theorem 6.7]{resolutionfull}, we
  can construct $\IntervalRepr{\RName \cap \Cov}$ in
  $\bigO(\SubstringComplexity{\Text}\polylog \Textlen)$ time; see
  \cref{pr:sa-core-construction}.

\item[Observation 3: Using $\IntervalRepr{\RName\cap \Cov}$, we can efficiently
  check if $j \in \RName$ (resp.\ ${\SA[i] \in \RName}$).] Recall that,
  at the beginning of the refinement step, we have some $j' \in
  \OccThree{\ell}{j}{\Text}$ (resp.\ $j' \in
  \OccThree{\ell}{\SA[i]}{\Text}$). By $3\tau - 1 \leq \ell$ and the
  definition of $\RName$, we thus have $j \in \RName$ (resp.\ $\SA[i]
  \in \RName$) if and only if $j' \in \RName$. Moreover, since $j'$
  satisfies $j' = \min \OccThree{2\ell}{j'}{\Text}$, and any
  $14\tau$-cover is also a $2\ell$-cover (\cref{lm:cover}), it thus
  follows that $j' \in \Cov$. Thus, $j' \in \Cov \cap \RName$ if and
  only if $j' \in \RName$. Consequently, to check if $j' \in \RName$,
  it suffices to check $j'$ belongs to any interval contained in 
  $\IntervalRepr{\RName \cap \Cov}$.
  Provided that the intervals in $\IntervalRepr{\RName \cap \Cov}$
  are ordered left-to-right, this takes $\bigO(\log |\IntervalRepr{\RName \cap \Cov}|)=\bigO(\log \Textlen)$ time (\cref{pr:sa-core-nav}).
\end{description}

The above is a simplified analysis, and reaching
$\bigO(\SubstringComplexity{\Text})$ space for each $k\in [4\dd \lceil
\log \Textlen \rceil)$ is still not enough to achieve the $\delta$-optimal
bound of $\bigO(\SubstringComplexity{\Text} \log \tfrac{\Textlen \log
\sigma}{\SubstringComplexity{\Text} \log \Textlen})$ for the entire
structure. In our complete analysis, we prove a tighter upper bound:
$|\IntervalRepr{\RName \cap \Cov}| =
\bigO(\tfrac{1}{\ell}\SubstrCount{38\ell}{\Text}+1)$; see
\cref{sec:sa-core}.

\paragraph{The Nonperiodic Positions}

Assume $j \in [1 \dd \Textlen] \setminus \RName$ (resp.\ $\SA[i] \in
[1 \dd \Textlen] \setminus \RName$). We first focus on computing
$\ISA[j]$. Recall that we are given $b =
\RangeBegThree{\ell}{j}{\Text}$, $e = \RangeEndThree{\ell}{j}{\Text}$,
and some $j' \in \OccThree{\ell}{j}{\Text}$ satisfying $j' = \min
\OccThree{2\ell}{j'}{\Text}$ as input. The refinement step for
nonperiodic positions first computes the position $j'' = \min
\OccThree{2\ell}{j}{\Text}$ (this condition implies $j'' =
\min \OccThree{4\ell}{j''}{\Text}$), and then the values
$\DeltaBeg{\ell}{j}{\Text}$ and $|\OccThree{2\ell}{j}{\Text}|$.  By
the above discussion, this is sufficient to infer
$\RangeBegThree{2\ell}{j}{\Text}$ and
$\RangeEndThree{2\ell}{j}{\Text}$. 

Let $\SSS$ be a
$\tau$-synchronizing set of $\Text$ (\cref{def:sss}). Observe that, by $3\tau \leq \ell
< \Textlen$, the uniqueness of $\Text[\Textlen]$ in $\Text$ yields 
$\per(\Text[\Textlen - 3\tau + 2 \dd \Textlen]) >
\tfrac{1}{3}\tau$. Thus, $\Textlen - 3\tau + 2 \not\in \RName$, and
hence the density condition (\cref{def:sss}\eqref{def:sss-density}) 
implies $\SSS \cap [\Textlen - 3\tau + 2 \dd \Textlen - 2\tau + 2) \neq
\emptyset$.  Consequently, $\max \SSS \geq \Textlen - 3\tau + 2$,
and, for every $p \in [1 \dd \Textlen - 3\tau +
2]$, we can define $\Successor{\SSS}{p} = \min\{s \in \SSS : s \geq
p\}$.
If $j > \Textlen - 3\tau - 2$, the uniqueness of
$\Text[\Textlen]$ in $\Text$ implies $\OccThree{\ell}{j}{\Text}=
\{j\}$ (\cref{pr:sa-nonperiodic-occ-min}).
Thus, we henceforth assume $j \in [1 \dd \Textlen - 3\tau +2]\setminus \RName$.

\begin{description}[style=sameline,itemsep=1ex,font={\normalfont\itshape}]
\item[Observation 1: The set $\OccThree{2\ell}{j}{\Text}$ can be
  characterized using $\SSS$ and range queries.] First, note that $j
  \in [1 \dd \Textlen - 3\tau + 2] \setminus \RName$ and the density
  condition (\cref{def:sss}\eqref{def:sss-density}) yield
  $\Successor{\SSS}{j} - j < \tau$. On the other hand, by $3\tau - 1
  \leq 2\ell$ and the consistency condition
  (\cref{def:sss}\eqref{def:sss-consistency}), every $p \in
  \OccThree{2\ell}{j}{\Text}$ satisfies $\Successor{\SSS}{p} - p =
  \Successor{\SSS}{j} - j$. Thus, letting $x_1 = \revstr{\Textinf[j
  \dd \Successor{\SSS}{j})}$, and $y_1 = \Textinf[\Successor{\SSS}{j}
  \dd j + 2\ell)$, the set $\OccThree{2\ell}{j}{\Text}$ consists of
  all positions of the form $s - |x_1|$, where $s \in \SSS$ and $s$ is
  preceded by $\revstr{x_1}$ and followed by $y_1$ in $\Textinf$. 
  In other words, $\OccThree{2\ell}{j}{\Text}=\{s-|x_1| : s\in \SSS \text{ and } \Textinf[s-|x_1|\dd s+|y_1|)=\revstr{x_1}\cdot y_1\}$.
  If
  we further denote $c = \max\Sigma$, $x_2 = x_1c^{\infty}$, and $y_2
  = y_1c^\infty$, we have $\OccThree{2\ell}{j}{\Text} = \{s - |x_1| :
  s \in \SSS,\,\allowbreak x_1 \preceq \revstr{\Textinf[s - 7\tau \dd
  s)} \prec x_2,\allowbreak \text{ and }y_1 \preceq \Textinf[s \dd s +
  7\tau) \prec y_2\}$ due to $|x_1|, |y_1| \leq 7\tau$.  Thus, letting
  $\Pts = \{(\revstr{\Textinf[s - 7\tau \dd s)}, \Textinf[s \dd s +
  7\tau), s) : s \in \SSS\}$ be a set of labeled points,
  the set $\OccThree{2\ell}{j}{\Text}$ shifted by $|x_1|$ forward
  consists of the labels of the points in the range $[x_1 \dd x_2) \times [y_1 \dd y_2)$.
\end{description}

By the above, computing $\min \OccThree{2\ell}{j}{\Text}$ reduces to an orthogonal
range minimum query, returning the minimum label of a point occurring
in the rectangle $[x_1 \dd x_2) \times [y_1 \dd y_2)$
(\cref{lm:sa-nonperiodic-pos-occ-min}). Implementing such queries,
however, is challenging.  First, the coordinates of points in $\Pts$ are
substrings of $\Textinf$. Comparing them reduces to $\LCE$ and random
access queries, and hence it suffices to only store labels of points
in~$\Pts$, i.e., the set $\SSS$ (\cref{pr:str-str}).  The smallest
prior structure for $\LCE$ queries~\cite{tomohiro-lce}, however, does
not match our space bound. To avoid this issue, we develop the first structure that uses
$\delta$-optimal space $\bigO(\SubstringComplexity{\Text} \log \tfrac{\Textlen \log
\sigma}{\SubstringComplexity{\Text} \log \Textlen})$.  In
addition, we describe its $\bigO(\SubstringComplexity{\Text}\polylog
\Textlen)$-time deterministic construction from the LZ77
parsing~\cite{LZ77} (see \cref{th:lce}). We also describe the first
deterministic construction of the $\delta$-optimal-space structure for random
access queries (\cref{th:random-access}). We elaborate more on these indexes in
\cref{sec:to-recompression}.

The challenge thus reduces to storing and querying $\SSS$.  The plain
representation is too large since it is not possible to reduce
$|\SSS|$ below $\Theta(\tfrac{n}{\tau})$ in the worst case
(\cref{rm:sss-size}). We thus need to store $\SSS$ in a compressed
form. Our starting point is the technique introduced
in~\cite{resolution}, which compresses $\SSS$ by only keeping elements
of $\SSS$ that are within distance $\Theta(\tau)$ from
LZ77 phrase boundaries. These \emph{LZ77-compressed $\tau$-synchronizing sets},
however, do not meet the $\delta$-optimal space bound and come only with
Las-Vegas randomized construction~\cite{resolution}. To solve the
space issue, we again employ a $14\tau$-cover $\Cov$ of $\Text$. 
Denote $\SSScomp = \SSS \cap \Cov$ and $\Pts_{\rm comp} =
\{(\revstr{\Textinf[s - 7\tau \dd s)}, \Textinf[s \dd s + 7\tau), s) :
s \in \SSScomp\}$.

\begin{description}[style=sameline,itemsep=1ex,font={\normalfont\itshape}]
\item[Observation 2: We can compute $x_1$, $x_2$, $y_1$, and $y_2$
  using $\SSScomp$.] Observe that, since $|x_1| + |y_1| = 2\ell$ and
  all strings in question occur near $j$, the difficulty lies in
  computing $|x_1|$.  Recall that we are given some $j' \in
  \OccThree{\ell}{j}{\Text}$ satisfying $j' = \min
  \OccThree{2\ell}{j'}{\Text}$ as input.  By $3\tau - 1 \leq \ell$,
  the consistency of~$\SSS$
  (\cref{def:sss}\eqref{def:sss-consistency}) yields
  $|x_1| = \Successor{\SSS}{j} - j = \Successor{\SSS}{j'} - j'$.  On the other
  hand, since $j' = \min \OccThree{2\ell}{j'}{\Text}$ and $\Cov$ is
  also a $2\ell$-cover (\cref{lm:cover}), it follows that $[j' \dd j'
  + 2\ell) \cap [1 \dd \Textlen] \subseteq \Cov$. Thus
  $\Successor{\SSS}{j} - j < \tau$ implies $\Successor{\SSS}{j'}\in \SSScomp$
  (\cref{lm:sa-nonperiodic-succ}).  Consequently, it suffices to store
  the sorted set $\SSScomp$. Given $j'$, we can then quickly determine
  $\Successor{\SSScomp}{j'}-j'= \Successor{\SSS}{j'} - j' = \Successor{\SSS}{j} - j=|x_1|$.

\item[Observation 3: Orthogonal range minimum queries on $\Pts_{\rm comp}$ and $\Pts$ are equivalent.] Let $x_1, x_2, y_1, y_2 \in \Sigma^{*}$ and
  let $s_{\rm comp}$ (resp.\ $s$) denote the output of the range
  minimum query in $[x_1 \dd x_2) \times [y_1 \dd y_2)$ on $\Pts_{\rm
  comp}$ (resp.\ $\Pts$). Observe that $\SSScomp \subseteq \SSS$
  implies $s \leq s_{\rm comp}$. To show the opposite inequality, let
  $s_{\min} = \min\{ i \in [1 \dd \Textlen] : \Textinf[i - 7\tau \dd i
  + 7\tau) = \Textinf[s - 7\tau \dd s + 7\tau)\} \leq s$. Then, either
  $s_{\min} \in [1\dd 7\tau]\cup (\Textlen-7\tau\dd \Textlen]$ or $s_{\min} - 7\tau = \min
  \OccThree{14\tau}{s_{\min} - 7\tau}{\Text}$. In both cases,
  $s_{\min} \in \Cov$. Moreover, by the consistency of $\SSS$
  (\cref{def:sss}\eqref{def:sss-consistency}), $s_{\min} \in
  \SSS$. Thus, $s_{\min} \in \SSScomp$.  Finally, $\Textinf[s_{\min} -
  7\tau \dd s_{\min} + 7\tau) = \Textinf[s - 7\tau \dd s + 7\tau)$
  implies that $(\revstr{\Textinf[s_{\min} - 7\tau \dd s_{\min})},
  \Textinf[s_{\min} \dd s_{\min} + 7\tau)) \in [x_1 \dd x_2) \times
  [y_1 \dd y_2)$ if and only if $(\revstr{\Textinf[s - 7\tau \dd s)},
  \Textinf[s \dd s + 7\tau)) \in [x_1 \dd x_2) \times [y_1 \dd y_2)$.
  Thus, $s_{\rm comp} \leq s_{\min} \leq s$.
\end{description}

By the above, it suffices to use $\Pts_{\rm comp}$ during the
computation of $\min \OccThree{2\ell}{j}{\Text}$
(\cref{lm:sa-nonperiodic-pos-occ-min}). The computation of
$\DeltaBeg{\ell}{j}{\Text}$ and $|\OccThree{2\ell}{j}{\Text}|$ uses
similar ideas, except range minimum queries are replaced with range
counting (\cref{lm:sa-nonperiodic-posbeg-posend-size,lm:sa-nonperiodic-pos-occ-size}). For this, we
augment each point with a \emph{weight} storing the frequency of the
corresponding substring.  The correctness of this follows by the local
consistency of $\SSS$ (\cref{lm:sa-nonperiodic-count}). To avoid
double counting, we also need to ensure that no two points in $\Pts_{\rm comp}$
coincide. To simultaneously still allow range minimum queries, we thus
leave only points with the smallest labels; see \cref{def:str-str}.

Let us now return to computing $\SA[i]$. Observe that, to compute $\min
\OccThree{2\ell}{j}{\Text}$, $\DeltaBeg{\ell}{j}{\Text}$, and
$|\OccThree{2\ell}{j}{\Text}|$, we needed some occurrences of strings
$x_1$ and $y_1$ satisfying $|x_1| = \Successor{\SSS}{j} - j$ and
$\Textinf[j \dd j + 2\ell) = \revstr{x_1}y_1$. The input of the refinement
procedure in the computation of $\SA[i]$, however, does not include any
element of $\OccThree{2\ell}{\SA[i]}{\Text}$.  Consequently, our query
procedure performs an additional step that computes some position $p \in
\OccThree{2\ell}{\SA[i]}{\Text}$. Such a position can be retrieved
from $\Pts_{\rm comp}$ using the inverse of range counting, i.e.,
range selection queries (see \cref{lm:sa-nonperiodic-occ-elem}).  The
rest of the query proceeds similarly to the computation of
$\ISA[j]$ for $j = \SA[i]$, except we can now use $p$ instead of $j$
since $p \in \OccThree{2\ell}{\SA[i]}{\Text}$ implies that
$\OccThree{2\ell}{\SA[i]}{\Text} = \OccThree{2\ell}{p}{\Text}$
(\cref{lm:occ-equivalence}).

The remaining challenge is ensuring that $\SSScomp$ is small.  Unlike
for the compressed version of $\RName$, where upper bounds on $|\Cov|$
and $|\IntervalRepr{\Cov}|$ already impose an upper
bound on $|\IntervalRepr{\RName \cap \Cov}|$, the size of $\SSS \cap \Cov$ can be
large if $\SSS$ is not constructed carefully. In \cref{sec:recompression}, we
develop a deterministic construction that ensures that the total size
of $\SSScomp$ across all levels $k\in [4\dd \lceil\log
\Textlen\rceil)$ of the data structure is
$\bigO(\SubstringComplexity{\Text} \log \tfrac{\Textlen \log
\sigma}{\SubstringComplexity{\Text} \log \Textlen})$
(see \cref{sec:to-recompression}).

\paragraph{The Periodic Positions}

Let us now assume $j \in \RName$ (resp.\ $\SA[i] \in \RName$). One of
the key challenges, compared to previous work using the ``refinement''
framework~\cite{dynsa} is as follows.  The basic property of every $p
\in \RName$ (extending easily to blocks of such positions), dictating
the rest of the query algorithm, is its \emph{type}, defined either as
$-1$ or $+1$, depending on whether the symbol following the periodic
substring is larger or smaller than the symbol that would extend the
period (see \cref{sec:sa-periodic-prelim}). Dealing with positions of each type
is straightforward if $\tilde{\Theta}(\Textlen)$ space is available:
We separately store all maximal blocks of
positions in $\RName$ of each type~\cite{dynsa}. In compressed space,
however, we are much more constrained. For example, $j\in \RName$
and $j'\in \OccThree{2\ell}{j}{T}$ may have different types, so we cannot
distinguish the type simply based on the occurrence of a periodic
fragment. This requires numerous new and more general combinatorial
properties, allowing separate processing of elements of
$\OccThree{2\ell}{j}{\Text}$
(resp.\ $\OccThree{2\ell}{\SA[i]}{\Text}$) depending on whether they
are \emph{partially periodic} (i.e., the length of their periodic
prefix is less than $2\ell$) or \emph{fully periodic} (otherwise); see
\cref{sec:sa-periodic}.

\vspace{-1.5ex}
\subsection{Deterministic Restricted Recompression}\label{sec:to-recompression}

\emph{Restricted recompression}~\cite{IPM} is a general technique for
constructing a run-length grammar (RLSLP) of a given text (see
\cref{sec:recompression-prelim}).  Utilizing this technique,
Kociumaka, Navarro, and Prezza~\cite{delta} proved that every
$\Text \in \IntegerAlphabet^{\Textlen}$ can be represented using an
RLSLP of size $\bigO(\SubstringComplexity{\Text} \log \tfrac{\Textlen
\log \sigma}{\SubstringComplexity{\Text} \log \Textlen})$ and height
$\Oh(\log \Textlen)$.  They also showed that
$\bigO(\SubstringComplexity{\Text} \log \tfrac{\Textlen \log
\sigma}{\SubstringComplexity{\Text} \log \Textlen})$ is the
asymptotically optimal space to represent a string, for all
combinations of $\Textlen$, $\sigma$, and
$\SubstringComplexity{\Text}$.  Consequently, random access to $\Text$
can be efficiently supported in the $\delta$-optimal space. Finally, they
developed an $\bigO(\Textlen)$-expected-time Las-Vegas randomized
construction of such RLSLP.\@ At the heart of their construction is
the problem of approximating the directed max-cut of graphs derived
from partially compressed representations of $\Text$.
In~\cite{delta}, it is proved that a uniformly random partition at
every level of the grammar is sufficient to achieve the $\delta$-optimal total size
in expectation. In this paper, we describe an explicit partitioning
technique (\cref{def:deterministic}) resulting in the same bound on
the size of the RLSLP
(\cref{sec:deterministic-recompression-analysis}).  The unique
component of our construction is the use of a \emph{cover hierarchy}
(\cref{sec:cover}), allowing us to account for the effects of
partitioning at the current level of the grammar on the properties of
the grammar at all future levels.  In addition, we develop an
$\bigO(\SubstringComplexity{\Text}\polylog \Textlen)$-time
deterministic construction algorithm of our RLSLP from the LZ77
parsing of $\Text$~\cite{LZ77}.  

Equipped with this RLSLP, it is relatively easy to answer random access
and LCE queries on $\Text$ in $\Oh(\log \Textlen)$ time (\cref{th:random-access,th:lce}). 
Moreover, in $\bigO(\SubstringComplexity{\Text}\polylog \Textlen)$
time, we can derive from the RLSLP a sequence of string synchronizing sets
such that, after pairwise intersection with the cover
hierarchy guiding the RLSLP construction, we obtain their representation of total size
$\bigO(\SubstringComplexity{\Text} \log \tfrac{\Textlen \log
\sigma}{\SubstringComplexity{\Text} \log \Textlen})$
(\cref{pr:comp-sss-construction}).

\section{Optimal Compressed Space String Covers}\label{sec:cover}

\begin{definition}[Cover]\label{def:cover}
  Let $\Text \in \Sigma^{\Textlen}$ and $\ell \in \Zp$.  A set $\Cov
  \sub [1 \dd \Textlen]$ is called an \emph{$\ell$-cover} of $\Text$
  if $(\max(0, \Textlen - \ell) \dd \Textlen] \subseteq \Cov$ and 
  $[i \dd i + \ell) \sub \Cov$ holds for
  every $i \in [1 \dd \Textlen - \ell]$ satisfying $i = \min
  \OccThree{\ell}{i}{\Text}$.
  In other words, $\Cov$ must contain the last $\min(\ell, \Textlen)$
  positions of $\Text$, as well as all positions covered by the
  leftmost occurrences of length-$\ell$ substrings of $\Text$.
\end{definition}

\begin{definition}[Interval representation]\label{def:interval-representation}
  The \emph{interval representation} of a finite set $\mathsf{P}\sub
  \Zp$ is the unique sequence $\IntervalRepr{\mathsf{P}} = (a_i, t_i)_{i \in [1
  \dd m]}$ such that $a_1 < a_1 + t_1 < a_2 < a_2 + t_2 < \dots <
  a_m < a_m + t_m$ and $\mathsf{P}=\bigcup_{i \in [1 \dd m]}[a_i \dd
  a_i + t_i)$.
\end{definition}

\begin{definition}[Cover hierarchy]\label{def:cover-hierarchy}
  A \emph{cover hierarchy} of a string $\Text\in \Sigma^{+}$ is an
  ascending sequence of sets $\Cov=(\Cov_\ell)_{\ell\in \Zp}$ such
  that $\Cov_\ell$ is an $\ell$-cover of $\Text$ and
  $|\IntervalRepr{\Cov_\ell}|\le \max(1, \frac{|\Cov_\ell|}{\ell})$.
\end{definition}

\begin{lemma}\label{lm:cover}
  Let $\Text \in \Sigma^{\Textlen}$ and $\ell \in\Zp$. If $\Cov$ is an
  $\ell$-cover of $\Text$, then it is also an $\ell'$-cover of $\Text$
  for every $\ell' \in [1 \dd \ell)$.
\end{lemma}
\begin{proof}
  Let $\ell' \in [1 \dd \ell)$.  First, note that $(\max(0, \Textlen -
  \ell') \dd \Textlen] \subseteq (\max(0, \Textlen - \ell) \dd
  \Textlen] \subseteq \Cov$.  Next, let us consider $i \in [1 \dd
  \Textlen - \ell']$ such that $i = \min \OccThree{\ell'}{i}{\Text}$.
  If $i + \ell > \Textlen$, then $[i \dd i + \ell') \subseteq (\max(0,
  \Textlen - \ell) \dd \Textlen] \subseteq \Cov$ follows as above. We
  can thus assume $i \in [1 \dd \Textlen - \ell]$.  Observe that this
  implies $i = \min \OccThree{\ell}{i}{\Text}$. Otherwise, there would
  exist $i' \in [1 \dd i)$ such that $\Text[i' \dd i' + \ell) =
  \Text[i \dd i + \ell)$.  By $\ell' < \ell$, we then have $\Text[i'
  \dd i' + \ell') = \Text[i \dd i + \ell')$, contradicting $i = \min
  \OccThree{\ell'}{i}{\Text}$.  Thus, $i = \min
  \OccThree{\ell}{i}{\Text}$, and hence $[i \dd i + \ell) \subseteq
  \Cov$.  By $\ell' < \ell$, we thus obtain $[i \dd i + \ell')
  \subseteq \Cov$.
\end{proof}

\begin{lemma}\label{lm:cover-equivalence}
  Let $\Text \in \Sigma^{\Textlen}$ and $\ell \in\Zp$.
  If $\Cov$ is an $\ell$-cover of $\Text$,
  then $[i \dd i + \ell) \cap [1 \dd \Textlen] \subseteq \Cov$
  holds for every $i \in [1 \dd \Textlen]$
  satisfying $i = \min \OccThree{\ell}{i}{\Text}$ .
\end{lemma}
\begin{proof}
  Let $i \in [1 \dd \Textlen]$ be such that $i = \min
  \OccThree{\ell}{i}{\Text}$. If $i + \ell > \Textlen$, then by
  \cref{def:cover}, we immediately obtain $[i \dd i + \ell) \cap [1
  \dd n] \subseteq (\max(0, \Textlen - \ell) \dd \Textlen]
  \subseteq \Cov$.  Otherwise (i.e., $i \in [1 \dd n - \ell]$), it
  holds $[i \dd i + \ell) \cap [1 \dd \Textlen] = [i \dd i + \ell)$,
  and hence the claim follows by the second property in
  \cref{def:cover}.
\end{proof}

\begin{construction}\label{cons:cover}
  Let $\Text \in \Sigma^{\Textlen}$ and $\ell \in\Zp$.
  Denoting $k = 2^{\lceil \log \ell \rceil}$, we define the set 
   \[\Cover{\ell}{\Text} :=   \bigcup_{i \in \mathsf{I}} [i \dd i + 2k)\; \cup \; (\max(0, \Textlen-2k)\dd \Textlen].\]
  where $\mathsf{I} = \{i \in [1 \dd \Textlen-2k] : i \bmod k =1\text{
  and } i = \min \OccThree{2k}{i}{\Text}\}$.
\end{construction}

\begin{lemma}\label{lm:c-cover}
  For every text $\Text\in \Sigma^{+}$,
  the family $(\Cover{\ell}{\Text})_{\ell \in \Zp}$ forms a cover hierarchy.
  Moreover, each set in this hierarchy satisfies $|\Cover{\ell}{\Text}| \le \SubstrCount{8\ell}{\Text}+8\ell \le 16\ell \SubstringComplexity{\Text}$
  and $|\IntervalRepr{\Cover{\ell}{\Text}}| \le \frac{\SubstrCount{8\ell}{\Text}}{\ell}+8 \le 16\SubstringComplexity{\Text}$.
\end{lemma}
\begin{proof}
  Denote $\Textlen = |\Text|$.
  Let us first assume that $\ell \le \frac{1}{2}\Textlen$ is a power
  of two and prove that $\Cover{\ell}{\Text}$ is an $\ell$-cover.
  First, we note that $(\max(0, \Textlen - \ell) \dd \Textlen]
  \subseteq (\max(0, \Textlen - 2k), \dd \Textlen] \subseteq
  \Cover{\ell}{\Text}$.  Let us consider any $i \in [1 \dd \Textlen -
  \ell]$ such that $i = \min \OccThree{\ell}{i}{\Text}$.  If $i +
  2\ell > \Textlen$, then $[i \dd i + \ell) \sub (\max(0, \Textlen -
  2\ell) \dd \Textlen] \sub (\max(0, \Textlen - 2k) \dd \Textlen]
  \sub \Cover{\ell}{\Text}$.  We can thus assume $i \in [1 \dd n -
  2\ell]$.  Define $j$ as the largest integer such that $j \leq i$
  and $j \bmod \ell = 1$. Then, $i - j < \ell$ and hence $[i \dd i +
  \ell) \subseteq [j \dd j + 2\ell)$. Moreover, we claim that $j =
  \min \OccThree{2\ell}{j}{\Text}$.  Otherwise, there would exist $j'
  < j$ such that $\Text[j' \dd j' + 2\ell) = \Text[j \dd j + 2\ell)$.
  This would imply that $i' = j' + (i - j) = i - (j - j'') < i$
  satisfies $\Text[i' \dd i' + \ell) = \Text[i \dd i + \ell)$,
  contradicting $i = \min \OccThree{\ell}{i}{\Text}$.  Thus, $j\in
  \mathsf{I}$ and $[i\dd i+\ell)\sub [j\dd j+2\ell)\subseteq
  \Cover{\ell}{\Text}$.  Hence, $\Cover{\ell}{\Text}$ is an
  $\ell$-cover of $\Text$.

  Next, we shall bound $|\Cover{\ell}{\Text}|$. 
  Observe that if $i\in \Cover{\ell}{\Text}\cap (2\ell\dd \Textlen-2\ell]$, then there is a position $j\in (i-2\ell\dd i]$
  such that $j=\min\OccThree{2\ell}{j}{\Text}$.
  Consequently, $i-2\ell = \min\OccThree{4\ell}{i-2\ell}{\Text}$.
  Thus, $|\Cover{\ell}{\Text}|\le \SubstrCount{4\ell}{\Text}+4\ell \le 8\ell \SubstringComplexity{\Text}$.

  Furthermore, observe that $\Cover{\ell}{\Text}$ is a union of length-$2\ell$ intervals,
  and hence $|\IntervalRepr{\Cover{\ell}{\Text}}|\le \frac{1}{2\ell}|\Cover{\ell}{\Text}|\le \frac{1}{\ell}|\Cover{\ell}{\Text}|$.
  Moreover, the sequence $\Cover{\ell}{\Text}$ is ascending when restricted to powers of two not exceeding $\frac12\Textlen$:
  Each position $i\in \Cover{\ell}{\Text}\cap[1\dd \Textlen-2\ell]$ is covered by the leftmost occurrence of a length-$2\ell$ substring of $\Text$, and 
  thus it must be contained in any $2\ell$-cover.
  Furthermore, each position $i\in (\Textlen-2\ell\dd \Textlen]$ is contained in $\Cover{2\ell}{\Text}$ by construction.

  It remains to consider the case of arbitrary $\ell\in \Zp$.
  If $\ell > \frac12\Textlen$, we have $\Cover{\ell}{\Text}=[1\dd \Textlen]$.
  This set is trivially an $\ell$-cover, it satisfies $|\IntervalRepr{\Cover{\ell}{\Text}}|=1$ and $\Cover{\ell}{\Text}=\Textlen < 2\ell$.
  If $\ell \le \frac12\Textlen$ is not a power of two, then we have $\Cover{\ell}{\Text}:=\Cover{k}{\Text}$, where $k=2^{\ceil{\log \ell}}$.
  This set is an $\ell$-cover by \cref{lm:cover}. Furthermore, $|\Cover{\ell}{\Text}|=|\Cover{k}{\Text}|\le \SubstrCount{4k}{\Text}+4k
  \le \SubstrCount{8\ell}{\Text}+8\ell \le 16\ell \cdot \SubstringComplexity{\Text}$
  and $|\IntervalRepr{\Cover{\ell}{\Text}}|=|\IntervalRepr{\Cover{k}{\Text}}|
  \le \max(1, \frac1k|\Cover{k}{\Text}|)\le \max(1, \frac1\ell|\Cover{\ell}{\Text}|) \le 8+\frac{1}{\ell}\SubstrCount{8\ell}{\Text} \le 16\SubstringComplexity{\Text}$.
\end{proof}

\begin{proposition}\label{pr:cover-construction}
  Let $\Text \in \Sigma^{+}$ and $\ell \in \Zz$ Assume
  that, for every substring $Q$ of $\Text$ (specified with its starting
  position and the length), we can in $\bigO(t_{\rm minocc})$ time
  compute $\min \OccTwo{Q}{\Text}$.  Given the LZ77 parsing of
  $\Text$, we can construct the 
  interval representation $\IntervalRepr{\Cover{\ell}{\Text}}$ of the cover 
  of \cref{cons:cover} in $\bigO(\LZSize{\Text} \cdot t_{\rm minocc})$ time.
\end{proposition}
\begin{proof}
  Since $\Cover{\ell}{\Text}=\Cover{k}{\Text}$ holds for $k=2^{\lceil \log \ell \rceil}$,
  we can assume without loss of generality that $\ell$ is a power of two.
  
  Let $\mathsf{I}$ and $\Cover{\ell}{\Text}$ be defined as in
  \cref{lm:c-cover}. Observe that to compute $\IntervalRepr{\Cover{\ell}{\Text}}$, it suffices to enumerate the set $\mathsf{I}$
  left-to-right. The elements of the sequence are then easily
  constructed.

  The set $\mathsf{I}$ can be enumerated as follows.  Let $b_j$,
  where $j \in [1 \dd \LZSize{\Text}]$, denote the first position of the $j$th
  leftmost phrase in the input parsing. Observe that for every $i \in
  \mathsf{I}$, there exists an index $j \in [1 \dd \LZSize{\Text}]$ such that $i <
  b_j < i + 2\ell$, since otherwise $\Text[i \dd i + 2\ell)$ would be
  entirely contained inside some phrase and hence have an earlier
  occurrence in $\Text$, contradicting $i = \min
  \OccThree{2\ell}{i}{\Text}$.  For such $j$ we therefore have $i \in (b_j -
  2\ell \dd b_j)$. Note now that by taking into account the constraint $i
  \bmod i = 1$ in the definition of $\mathsf{I}$, we have $|(b_j - 2\ell
  \dd b_j) \cap \mathsf{I}| \leq 2$. Consequently, to enumerate
  $\mathsf{I}$ it suffices for every $j \in [1 \dd \LZSize{\Text}]$ to inspect at
  most two candidate positions: the first at $i_1 = \ell \lceil
  \tfrac{b_j - 2\ell}{\ell} \rceil + 1$ and second at $i_2 = i_1 + \ell$ (if
  $i_2 < b_j$). For each candidate $i$, testing reduces to checking if
  it holds $i = \min \OccThree{2\ell}{i}{\Text}$. The latter check can be
  implemented in $\bigO(t_{\rm minocc})$ time by 
  computing $i' = \min \OccTwo{\Text[i \dd i + 2\ell)}{\Text}$ and checking
  if $i' = i$.  Over all $j \in [1 \dd \LZSize{\Text}]$, we thus spend $\bigO(\LZSize{\Text}
  \cdot t_{\rm minocc})$ time. Note that some candidates $i$ may
  repeat across different $j \in [1 \dd \LZSize{\Text}]$, but it is easy to detect
  and eliminate such candidates by storing the last two candidate
  positions.
\end{proof}

By  \cite[Theorem~6.11]{resolutionfull}, the queries specified in the statement of \cref{pr:cover-construction}
can be answered in $\Oh(\log^3 \Textlen)$ time after $\Oh(\LZSize{\Text} \log^4 \Textlen)$-time preprocessing.
Thus, we get the following

\begin{corollary}\label{cor:cover}
  Given the LZ77 parsing of a string $\Text\in \Sigma^{\Textlen}$
  and an integer $\ell\in \Zp$, the interval representation $\IntervalRepr{\Cover{\ell}{\Text}}$
  of the cover of \cref{cons:cover} can be built in $\Oh(\LZSize{\Text}\cdot \log^4 \Textlen)$ time.
\end{corollary}

\begin{definition}[Compressed representation]\label{def:comp}
  Let $\Text \in \Sigma^{\Textlen}$ and ${\sf P} \subseteq [1 \dd \Textlen]$.
  For every $\ell \geq 1$, we define the \emph{compressed representation}
  of ${\sf P}$ as $\CompRepr{\ell}{{\sf P}}{\Text} := {\sf P} \cap
  \Cover{\ell}{\Text}$, where $\Cover{\ell}{\Text}$ is the $\ell$-cover
  of $\Text$ defined in \cref{cons:cover}.
\end{definition}

\section{From LZ Parsing to Restricted Recompression}\label{sec:recompression}

As shown in~\cite{delta}, every string $\T\in [0\dd \sigma)^n$ can be represented using a run-length context-free grammar of size $\Oh(\SubstringComplexity{\T}\cdot \log\frac{n \log \sigma}{\SubstringComplexity{\T}\log n})$.
In this section, we revise the original construction in order to achieve a deterministic $\Oh(\SubstringComplexity{\T}\log^7 n)$-time algorithm that builds such a grammar from the LZ77 representation of $\T$. 
The underlying modifications are also crucial to make sure that the synchronizing sets defined in~\cite{IPM} (using a similar run-length context-free grammar)
admit compressed representations of $\Oh(\SubstringComplexity{\T}\cdot \log\frac{n \log \sigma}{\SubstringComplexity{\T}\log n})$ size in total.

\subsection{Preliminaries}\label{sec:recompression-prelim}
For a context-free grammar $\G$, we denote by $\Sigma_\G$ and $\N_\G$ the set of non-terminals and the set of terminals, respectively. The set of \emph{symbols} is $\S_\G:=\Sigma_\G\cup \N_\G$.
A \emph{straight-line grammar} (SLG) is a context-free grammar $\G$ such that:
\begin{itemize}
\item each non-terminal $A\in \N_\G$ has a unique production $A\to \rhs_\G(A)$, where $\rhs_\G(A)\in \S_\G^*$,
\item the set of symbols $\S_\G$ admits a partial order $\prec$ such that $B \prec A$ if $B$ appears in $\rhs(A)$.
\end{itemize}
The \emph{expansion} function $\exp_\G: \S_G \to \Sigma_\G^*$ is defined as follows:
\[\exp_\G(A) = \begin{cases}
  A & \text{if $A\in \Sigma_\G$},\\
  \exp(A_1)\exp(A_2)\cdots \exp(A_a) & \text{if $A\in \N_\G$ with $\rhs_\G(A)=A_1A_2\cdots A_a$.}
\end{cases}\]
In particular, the expansion $\exp_\G(S)$ of a starting symbol $S\in \S$ is the unique string \emph{represented} by~$\G$.
Moreover, $\exp_\G$ is lifted to $\exp_\G : \S_\G^*\to \Sigma_\G^*$ by setting $\exp_\G(A_1\cdots A_a)=\exp_\G(A_1)\cdots \exp_\G(A_a)$ for $A_1\cdots A_a\in \S_\G^*$. When the grammar $\G$ is clear from context, we omit the superscript.

For every symbol $A\in \S$, we also define $\LML(A)$ and $\RML(A)$ as the leftmost and the rightmost character of $\exp(A)$,
respectively, with $\LML(A)=\RML(A)=\emptystring$ if $\exp(A)=\emptystring$.
These values can be obtained in $\Oh(|\G|)$ time by processing symbols of $\G$ according to the underlying order $\prec$.

The \emph{parse tree} $\Tr(A)$ of a symbol $A\in \S$ is a rooted ordered tree with each node $\nu$ associated to a symbol $\symb(\nu)\in \S$. 
The root of $\Tr(A)$ is a node $\rho$ with $\symb(\rho)=A$.
If $A \in \Sigma$, then $\rho$ has no children. 
If $A\in \N$ and $\rhs(A)=A_1\cdots A_a$, then $\rho$ has $a$ children,
and the subtree rooted at the $i$th child is (a copy of) $\Tr(A_i)$.
The \emph{height} $\height(A)$ of a symbol $A\in \S$ is defined as the height of its parse tree $\Tr(A)$.
In other words, $\height(A)=0$ if $A\in \Sigma$ and $\height(A)=1+\max_{i=1}^a \height(A_i)$ if $\rhs(A)=A_1\cdots A_a$. The parse tree $\Tr_\G$ of an SLG $\G$ is defined as the parse tree $\Tr(S)$ of the starting symbol $S$,
and the height of $\G$ is defined as the height of $S$.

Each node $\nu$ of $\Tr(A)$ is associated with a fragment $\exp(\nu)$ of $\exp(A)$ matching $\exp(\symb(\nu))$.
For the root $\rho$, we define $\exp(\rho)=\exp(A)[1\dd |\exp(A)|]$ to be the whole $\exp(A)$.
Moreover, if $\exp(\nu)=\exp(A)[\ell\dd r)$, $\rhs(\symb(\nu))=A_1\cdots A_a$,
and $\nu_1,\ldots,\nu_a$ are the children of $\nu$, then $\exp(\nu_i)=\exp(A)[r_{i-1}\dd r_{i})$,
where $r_i = \sum_{j=1}^{i} |\exp(A_j)|$ for $0\le i \le a$.
This way, the fragments $\exp(\nu_i)$ form a partition of $\exp(\nu)$,
and $\exp(\nu_i)$ matches $\exp(\symb(\nu_i))$ (as claimed).

Without loss of generality, we assume that each symbol $A\in \S$ appears as $\symb(\nu)$ for a node $\nu$ of $\Tr_\G$;
the remaining symbols can be removed from $\G$ without affecting the string generated by~$\G$.

\paragraph*{Straight-Line Programs}
We say that a straight-line grammar $\G$ is in \emph{Chomsky normal form} (CNF)
if $|\rhs(A)|=2$ holds for each $A\in \N$. Such a grammar is also called a \emph{straight-line program} (SLP).
An SLP $\G$ of size $g$ (with $g$ symbols) representing a text $\T$ of length $n$
can be stored $\Oh(g)$ space ($\Oh(g \log n)$ bits) with each non-terminal $A\in \N$ 
storing $\rhs(A)$ and $|\exp(A)|$. 
This representation allows for efficiently traversing the parse tree~$\Tr_\G$: given a node $\nu$ represented as a pair $(s(\nu),\exp(\nu))$, it is possible to retrieve in constant time an analogous representation of a child $\nu_i$ of $\nu$ given its index $i\in\{1,2\}$ (among the children of $\nu$) or an arbitrary position $\T[j]$ contained in $\exp(\nu_i)$.

\paragraph*{AVL Grammars}
Rytter~\cite{Rytter03} and Charikar et al.~\cite{charikar} provided efficient algorithms that convert the LZ77 parsing of a string into an SLP generating it.
The original constructions work only for the non-self-referential version of LZ77, but this restriction has been lifted in subsequent works.
\begin{theorem}[{\cite[Theorem 5.1]{Gaw11,resolutionfull}}]\label{thm:rytter}
  Given an LZ77-like parsing of a string $\T[1\dd n]$ into $f$ phrases,
  an SLP $\G$ of height $\Oh(\log n)$ and size $\Oh(f \log n)$ generating $\T$ can be constructed in $\Oh(f\log n)$ time.
\end{theorem}

The construction in \cite{Rytter03,resolutionfull} is based on the notion of \emph{AVL grammars},
which are SLPs satisfying the following extra condition: if $\rhs(A)=BC$ for $A\in \N$, then $|\height(B)-\height(C)|\le 1$.
This guarantees~\cite[Lemma 1]{Rytter03} that $\height(A) = \Oh(\log |\exp(A)|)$ holds for every $A\in \S$.
The algorithm of~\cite{Rytter03,resolutionfull} builds $\G$ incrementally: each step involves adding a symbol $A$ with a desired expansion $\exp(A)$,
as well as a bounded number of auxiliary symbols. In the last step, the starting symbol $S$ with $\exp(S)=\T$ is added.
Each step is of one of three kinds:
\begin{enumerate}[label={\rm(\alph*)}]
  \item\label{it:new} A new terminal symbol can be added to $\G$ in $\Oh(1)$ time (along with no auxiliary symbols).
  \item\label{it:concat} Given two symbols $B,C\in \S$, a new symbol $A$ with $\exp(A)=\exp(B)\exp(C)$ can be added to $\G$ in $\Oh(1+\log|\exp(A)|)$ time along with $\Oh(\log|\exp(A)|)$ auxiliary symbols~\cite[Lemma 2]{Rytter03}.
  \item\label{it:extract} Given a symbol $A\in \S$ and two positions $1\le i \le j \le |\exp(A)|$, 
  a new symbol $B$ with $\exp(B)=\exp(A)[i\dd j]$ can be added to $\G$ in $\Oh(1+\log |\exp(A)|)$ time
  along with $\Oh(\log |\exp(A)|)$ auxiliary symbols~\cite[Lemma 3 and Theorem 2]{Rytter03}.
\end{enumerate}

\paragraph*{Run-Length Straight-Line Programs}
A \emph{run-length straight-line program} (RLSLP) is a straight-line grammar $\G$ whose non-terminals can be classified into \emph{pairs} 
with $\rhs(A)= BC$ for symbols $B,C\in \S$ such that $B \ne C$,
and \emph{powers} with $\rhs(A) = B^k$ for a symbol $B\in \S$ and an integer $k \ge 2$. 
Analogously to an SLP, an RLSLP of size $g$ (with $g$ symbols) representing a text $\T$ of length $n$
can be stored in $\Oh(g)$ space ($\Oh(g \log n)$ bits)  allowing efficient traversal of the parse tree $\Tr_\G$.

\subsection{Run-Length Grammar Construction via Restricted Recompression}\label{sec:deterministic-recompression}
Both recompression and restricted recompression, given a string $\T\in \Sigma^+$,
construct a sequence of strings $(\T_k)_{k=0}^\infty$ over the alphabet $\Symb$
 defined as the least fixed point of the following equation:
 \[\Symb = \Sigma \cup (\Symb \times \Symb)\cup (\Symb \times \mathbb{Z}_{\ge 2} ).\]
Symbols in $\Symb \sm \Sigma$ are non-terminals
with $\rhs((A_1,A_2))=A_1A_2$ for $(A_1,A_2)\in \Symb \times \Symb$ and $\rhs((A_1,m))= A_1^m$
for $(A_1,m)\in \Symb\times \mathbb{Z}_{\ge 2}$.
With any symbol in $\Symb$ designated as the start symbol, this yields a run-length straight-line program (RLSLP).
Intuitively, $\Symb$ forms a \emph{universal} RLSLP:
for every RLSLP with symbols $\Pres$ and terminals $\Sigma\sub \Pres$, there is a unique homomorphism $f:\Pres \to \Symb$ such that $f(A)=A$ if $A\in \Sigma$ and $\rhs(f(A))=f(A_1)\cdots f(A_a)$ if $\rhs_\G(A)=A_1\cdots A_a$.
As a result, $\Symb$ provides a convenient formalism to argue about procedures generating RLSLPs.

The main property of strings $(\T_k)_{k=0}^\infty$ generated using (restricted) recompression is that $\exp(\T_k)=\T$ holds for all $k\in \Zz$.
The subsequent strings $\T_k$, starting from $\T_0=\T$, are obtained by alternate applications 
of the following two functions which
decompose a string of symbols into \emph{blocks} and then \emph{collapse} blocks
into appropriate symbols. In \cref{it:rle}, all blocks of length at least $2$ are maximal blocks of the same symbol in $\Act$,
and they are collapsed to symbols in $\Act\times \mathbb{Z}_{\ge 2}$.
In \cref{it:pair}, all blocks consisting of a symbol in $\Left$ followed by a symbol in $\Right$ are collapsed to a symbol 
in $\Left \times \Right$.
We provide efficient algorithms that implement both transformations, with the input and the output strings represented using their LZ77 parsings.
Previous work~\cite{IPM,delta} relied on straightforward linear-time implementations in the uncompressed settings.

\begin{definition}[Restricted run-length encoding~\cite{IPM,delta}]\label{it:rle}
Given $\T\in \Symb^+$ and $\Act \sub \Symb$, we define $\rle_{\Act}(\T)\in \Symb^+$
to be the string obtained as follows by decomposing $\T$ into blocks and collapsing these blocks:
\begin{enumerate}
  \item For $i\in [1\dd |\T|)$, place a \emph{block boundary} between $\T[i]$ and $\T[i+1]$
  unless $\T[i]=\T[i+1]\in \Act$.
  \item Replace each block $\T[i\dd i+m)= A^m$ of length $m\ge 2$ with a symbol $(A,m)\in \Symb$.
\end{enumerate}
\end{definition}

\begin{lemma}\label{lem:rle}
  Given the LZ77 parsing of a string $\T\in \Symb^+$ and a set $\Act\sub \Symb$,
  the LZ77 parsing of $\hT:=\rle_\Act(\T)$ can be constructed in $\Oh(|\Act|+\LZSize{\T}\log^5 n)$ time.
\end{lemma}
\begin{proof}
  First, we use \cref{thm:rytter} to build an SLP $\G$ of size $|\G|=\Oh(f\log n)$ generating $\T$.
  Next, we construct a new grammar $\hG$ (over the same alphabet) of size $\Oh(|\G|)$ generating $\hT$. 
  For each symbol $A$ of $\G$, we construct a non-terminal $\hA$ in $\hG$, as well as strings $\LMR(A),\RMR(A)$ of length $0$ or $1$ such that:
  \begin{itemize}
    \item $\rle_\Act(\exp_\G(A))=\LMR(A)\cdot \exp_{\hG}(\hA)\cdot \RMR(A)$;
    \item $\LMR(A)\ne\emptystring$ if and only if $\LML(A)\in \Act$;
    \item $\RMR(A)\ne\emptystring$ if and only if $\RML(A)\in \Act$ and $|\rle_\Act(\exp_\G(A))|>1$.
  \end{itemize}
  It is easy to verify that the following construction satisfies these invariants.
  For every terminal $A$ of~$\G$, we consider two cases:
  \begin{itemize}
    \item If $A\in \Act$, then $\LMR(A)=A$, $\rhs_{\hG}(\hA)=\emptystring$, and $\RMR(A)=\emptystring$.
    \item Otherwise, $\LMR(A)=\emptystring$,  $\rhs_{\hG}(\hA)=A$, and $\RMR(A)=\emptystring$.
  \end{itemize}
  For every non-terminal $A$ with $\rhs_\G(A)=BC$, we consider multiple possibilities:
  \begin{itemize}
    \item If $\RML(B)\ne \LML(C)$ or $\RML(B)=\LML(C)\notin \Act$, then $\rle_\Act(\exp_\G(A))=\rle_\Act(\exp_\G(B))\cdot \rle_\Act(\exp_\G(C))$. Thus, we proceed as follows:
    \begin{itemize}
      \item If $|\rle_\Act(\exp_\G(C))|>1$, then $\LMR(A)=\LMR(B)$, $\rhs_{\hG}(\hA)=\hB \cdot \RMR(B)\cdot \LMR(C)\cdot \hC$, and $\RMR(A)=\RMR(C)$.
      \item If $|\rle_\Act(\exp_\G(C))|=1$, then $\LMR(A)=\LMR(B)$, $\rhs_{\hG}(\hA)= \hB \cdot \RMR(B)$, and $\RMR(A)=\LMR(C)$.
    \end{itemize}
    \item In the remaining cases, $\RML(B)=\LML(C)=X$ for some $X\in \Act$. Thus, $\rle_\Act(\exp_\G(A))$ is obtained from $\rle_\Act(\exp_\G(B))\cdot \rle_\Act(\exp_\G(C))$ by merging the last run in $\exp_\G(B)$ and the first run in $\exp_\G(C)$, both of which are powers of $X$.
    To formalize this operation, we identify $X$ with $(X,1)$ and define $(X,i)+(X,j):=(X,i+j)$.
    \begin{itemize}
    \item If $|\rle_\Act(\exp_\G(B))|>1$ and $|\rle_\Act(\exp_\G(C))|>1$, then $\LMR(A)=\LMR(B)$, $\rhs_{\hG}(\hA)= \hB \cdot (\RMR(B)+\LMR(C))\cdot \hC$, and $\RMR(A)=\RMR(C)$.
    \item If $|\rle_\Act(\exp_\G(B))|>1$ and $|\rle_\Act(\exp_\G(C))|=1$, then $\LMR(A)=\LMR(B)$, $\rhs_{\hG}(\hA)= \hB$, and $\RMR(A)=\RMR(B)+\LMR(C)$.
    \item If $|\rle_\Act(\exp_\G(B))|=1$ and $|\rle_\Act(\exp_\G(C))|>1$, then $\LMR(A)=\LMR(B)+\LMR(C)$, $\rhs_{\hG}(\hA)= \hC$, and $\RMR(A)=\RMR(C)$.
    \item If $|\rle_\Act(\exp_\G(B))|=1$ and $|\rle_\Act(\exp_\G(C))|=1$, then $\LMR(A)=\LMR(B)+\LMR(C)$, $\rhs_{\hG}(\hA)= \emptystring$, and $\RMR(A)=\emptystring$.
  \end{itemize}
\end{itemize}
Additionally, we add to $\hG$ a new starting symbol $\hS'$ such that $\rhs_{\hG}(\hS')=\LMR(S)\cdot \hS\cdot \RMR(S)$, where $S$ is the starting symbol of $\G$. 
Thus, the grammar $\hG$ is of size $|\hG|=\Oh(|\G|)$ and generates $\exp_{\hG}(\hS')=\rle_\Act(\exp_\G(S))=\rle_\Act(\T)=\hT$.
We construct an LZ77-like parsing of $\hT$ of size $\hG$ by traversing the parse tree of $\hG$ and creating a new previous factor whenever the symbol $\symb(\nu)$ has already been encountered, 
This parsing can be converted into the LZ77 parsing of $\hT$ in $\Oh(|\hG|\log^4 n)=\Oh(\LZSize{\T}\log^5 n)$ time using \cref{prop:realLZ}.
\end{proof}

\begin{definition}[Restricted pair compression~\cite{IPM,delta}]\label{it:pair}
  Given $\T\in \Symb^+$ and disjoint sets $\Left,\Right \sub \Symb$, we define $\pc_{\Left,\Right}(\T)\in \Symb^+$
  to be the string obtained as follows by decomposing $\T$ into blocks and collapsing these blocks:
  \begin{enumerate}
    \item For $i\in [1\dd |\T|)$, place a \emph{block boundary} between $\T[i]$ and $\T[i+1]$
    unless $\T[i]\in \Left$ and $\T[i+1]\in \Right$.
    \item Replace each block $\T[i\dd i+1]$ of length $2$ with a symbol $(\T[i],\T[i+1])\in \Symb$.
  \end{enumerate}
\end{definition}

\begin{lemma}\label{lem:pc}
  Given the LZ77 parsing of a string $\T\in \Symb^n$ and disjoint sets $\Left,\Right\sub \Symb$,
  the LZ77 parsing of $\hT:=\pc_{\Left,\Right}(\T)$ can be constructed in $\Oh(|\Left|+|\Right|+\LZSize{\T}\log^5 n)$ time.
\end{lemma}
\begin{proof}
  First, we use \cref{thm:rytter} to build an SLP $\G$ of size $\Oh(f\log n)$ generating $\T$.
  We construct a new grammar $\hG$ (over the same alphabet) of size $\Oh(|\G|)$ generating $\hT$. 
  For each symbol $X$ of $\G$, we construct a non-terminal $\hA$ in $\hG$, as well as strings $\LMB(A),\RMB(A)$ of length $0$ or $1$ such that:
  \begin{itemize}
    \item $\pc_{\Left,\Right}(\exp_\G(A))=\LMB(A)\cdot \exp_{\hG}(\hA)\cdot \RMB(A)$;
    \item $\LMB(A)\ne \emptystring$ if and only if $\LML(A)\in \Right$;
    \item $\RMB(A)\ne \emptystring$ if and only if $\RML(A)\in \Left$.
  \end{itemize}  
  It is easy to verify that the following construction satisfies these invariants.
  For every terminal $A$ of~$\G$, we consider two cases:
  \begin{itemize}
    \item If $A\in \Left$, then $\LMB(A)=\emptystring$, $\rhs_{\hG}(\hA)= \emptystring$, and $\RMB(A)=A$.
    \item If $A\in \Right$, then $\LMB(A)=A$, $\rhs_{\hG}(\hA)= \emptystring$, and $\RMB(A)=\emptystring$.
    \item Otherwise, $\LMB(A)=\emptystring$, $\rhs_{\hG}(\hA)= A$, and $\RMB(A)=\emptystring$.
  \end{itemize}
  For every non-terminal $A$ with $\exp_\G(A)=BC$, we consider two possibilities:
  \begin{itemize}
    \item If $\RMB(B)=\emptystring$ or $\LMB(C)=\emptystring$, then $\pc_{\Left,\Right}(\exp_\G(A))=\pc_{\Left,\Right}(\exp_\G(B))\cdot \pc_{\Left,\Right}(\exp_\G(C))$. Thus, we set $\LMB(A)=\LMB(B)$, $\rhs_{\hG}(\hA)= \hB \cdot \RMB(B)\cdot \LMB(C)\cdot \hC$, and $\RMB(A)=\RMB(C)$.
    \item Otherwise, $\pc_{\Left,\Right}(\exp_\G(A))$ is obtained from $\pc_{\Left,\Right}(\exp_\G(B))\cdot \pc_{\Left,\Right}(\exp_\G(C))$ by merging the last symbol of $B$ and the first symbol of $C$ into a single block $(\RMB(B),\LMB(C))$.
    Consequently, we set $\LMB(A)=\LMB(B)$, $\rhs_{\hG}(\hA)= \hB \cdot (\RMB(B),\LMB(C))\cdot \hC$, and $\RMB(A)=\RMB(C)$.
\end{itemize}
Additionally, we add to $\hG$ a new starting symbol $\hS'$ such that $\hS'\to \LMB(S)\cdot \hS\cdot \RMB(S)$, where $S$ is the starting symbol of $\G$.
Thus, the grammar $\hG$ is of size $|\hG|=\Oh(|\G|)$ and generates $\exp_{\hG}(\hS')=\pc_{\Left,\Right}(\exp_\G(S))=\pc_{\Left,\Right}(\T)=\hT$.
We construct an LZ77-like parsing of $\hT$ of size $\hG$ by traversing the parse tree of $\hG$ and creating a new previous factor whenever the symbol $\symb(\nu)$ has already been encountered, 
This parsing can be converted into the LZ77 parsing of $\hT$ in $\Oh(|\hG|\log^4 n)=\Oh(\LZSize{\T}\log^5 n)$ time using \cref{prop:realLZ}.
\end{proof}

We are now ready to formally define the sequence $(\T_k)_{k=0}^\infty$
constructed through restricted recompression.
\begin{construction}[Restricted recompression~\cite{IPM,delta}]\label{constr:Jez}
Given a string $\T\in \Sigma^+$, the strings $\T_k$ for $k\in \Zz$ are constructed as follows,
based on $\ell_k := (\tfrac{8}{7})^{\ceil{\frac{k}{2}}-1}$ and $\Symb_k := \{A\in \Symb : |\exp(A)|\le \ell_k\}$:
\begin{itemize}
  \item If $k=0$, then $\T_k = \T$.
  \item If $k>0$ is odd, then $\T_{k}=\rle_{\Symb_k}(\T_{k-1})$.
  \item If $k>0$ is even, then $\T_{k}=\pc_{\Left_k,\Right_k}(\T_{k-1})$, where $\Left_k,\Right_k\sub \Symb_k$ are disjoint.
\end{itemize}
\end{construction}

It is easy to see that $\exp(\T_k)=\T$ indeed holds for all $k\in \Zz$.
The main challenge is to appropriately select the subsets $\Left_k,\Right_k$.
The first (easier) goal is to make sure that $|\T_k|=1$ holds for some $k\in \Zz$ so that an RLSLP generating $\T$ can be obtained by setting $\T_k[1]$ as the starting symbol of the RLSLP derived from by $\Symb$.
While this RLSLP contains infinitely many symbols, we can remove symbols that do not occur in any string~$\T_k$.
Formally, for each $k\in \Zz$, we define the family $\Pres_k:=\{\T_k[j] : j\in [1\dd |\T_k|]\}\sub \Symb$
of symbols occurring in~$\T_k$. Then, the actual symbols present in the generated RLSLP can be 
expressed as $\Pres := \bigcup_{k=0}^\infty \Pres_k$.

The main goal behind our selection of the subsets $\Left_k,\Right_k$ is to make sure that $|\Pres| = \Oh(\SubstringComplexity{\T}\cdot \frac{n\log \sigma}{\SubstringComplexity{\T}\log n})$. As shown in~\cite{delta}, this holds in expectation if $\Left_k$ and $\Right_k$ form a random partition of $\Symb_k$.
In this work, we develop an alternative deterministic construction parameterized by a cover hierarchy $\Cov$ of the string $\T$ (cf. \cref{def:cover-hierarchy})

\begin{construction}\label{def:deterministic}
Fix a text $\T\in \Sigma^n$ and its cover hierarchy $\Cov$.
For every $k\in \Zz$, denote $\alpha_k = \floor{16\ell_k}$ and $m_k = 2\alpha_k+\floor{\ell_{k+1}}$,
For integers $k,h\in \Zp$, let us define 
\begin{align*}
  J_{k,h} &:= \{j\in [1\dd |\T_{k-1}|) : |\exp(\T_{k-1}[1\dd j])|\in \Cov_{m_h}\},\\
  w_{k,h}(A,B) &:= |\{j \in J_{k,h} : \T_{k-1}[j\dd j+1]=AB\}|,\\
  w_k(A,B) &:= \sum_{h=k}^\infty (\tfrac34)^{\floor{\frac{h}{2}}-\floor{\frac{k}{2}}} w_{k,h}(A,B).
\end{align*}
The subsets $\Left_k,\Right_k\sub \Symb_k$ are chosen so that $\sum_{(A,B)\in \Left_k\times \Right_k} w_k(A,B) \ge \tfrac14 \sum_{(A,B)\in \Symb_k\times \Symb_k} w_k(A,B)$.
\end{construction}

\subsubsection{Analysis of the Grammar Size}\label{sec:deterministic-recompression-analysis}
Our argument relies on several properties of restricted recompression proved in~\cite{IPM,delta}.

\begin{fact}[{\cite[Fact V.7]{delta}}]\label{fct:cons}
  For every $k\in \Zz$, if $\exp(x)=\exp(x')$ holds for two fragments of $\T_k$,
  then $x= x'$.
 \end{fact}
  
\begin{corollary}[{\cite[Corollary V.8]{delta}}]\label{cor:distinct}
  For every odd $k\in \Zz$, there is no $j\in [1\dd |\T_k|)$ such that $\T_k[j]=\T_k[j+1]\in \Symb_{k+1}$.
\end{corollary}

Recall that $\exp(\T_k)=\T$ for every $k\in \Zz$. Hence, for every $j\in [1\dd |\T_k|]$,
we can associate $\T_k[j]$ with a fragment $\T(|\exp(\T_k[1\dd j))|\dd |\exp(\T_k[1\dd j])|]= \exp(\T_k[j])$;
these fragments are called \emph{phrases} (of $\T$) induced by $\T_k$.
We also define a set $\Bnd_k$ of \emph{phrase boundaries} induced by $\T_k$:
\[\Bnd_k = \{|\exp(\T_{k}[1\dd j])| : j\in [1\dd |\T_k|)\}.\]

\begin{fact}[see~\cite{IPM}]\label{fct:recompr}
  For every $k\in \Zz$ and every symbol $A\in \Pres_k$, the string $\exp(A)$ has length at most $2{\ell_k}$ or period at most $\ell_k$.
  \end{fact}
  \begin{proof}
    We proceed by induction on $k$. 
    Let $A\in \Pres_k$. If $k=0$, then $|\exp(A)|=1 \le 2\cdot \frac{7}{8} =2 \ell_0$.
    Thus, we may assume $k>0$.
    If $A\in \Pres_{k-1}$, then the inductive assumption shows that $\exp(A)$ 
    is of length at most $2{\ell_{k-1}} \le 2{\ell_k}$
    or period most $\ell_{k-1}\le \ell_k$. 
    Otherwise, we have two possibilities.
    If $k$ is odd, then $A = (B,m)\in \Act_{k}\times \mathbb{Z}_{\ge 2}$,
    and thus the period $\exp(A)$ is at most $|\exp(B)|\le \ell_k$.
    If $k$ is even, on the other hand, then $A = (B,C)\in \Act_{k}^2$,
    so $|\exp(A)|=|\exp(B)|+|\exp(C)|\le 2{\ell_{k}}$.
  \end{proof}

The following lemma captures the ``local consistency'' property of our construction:
the presence of phrase boundaries is determined by a small context.

\begin{lemma}[{\cite[Lemma V.9]{delta}}]\label{lem:recompr1}
Let $\alpha \in \mathbb{Z}_{\ge 1}$ and let $i,i'\in [\alpha \dd n-\alpha]$ be such that $\T(i-\alpha\dd i+\alpha]= \T(i'-\alpha\dd i'+\alpha]$.
For every $k\in \Zz$, if $\alpha \ge \floor{16\ell_k}$, then $i\in \Bnd_k \Longleftrightarrow i'\in \Bnd_k$.
\end{lemma}

As a first step towards proving $|\Pres|=\Oh(\SubstringComplexity{\T} \log \frac{n \log \sigma}{\SubstringComplexity{\T} \log n})$,
we apply \cref{lem:recompr1} to bound $|\Pres_k\sm \Pres_{k+1}|$ for a given $k\in \Zz$. 

\begin{lemma}\label{lem:bkpk}
  For every $k\in \Zz$, we have $|\Pres_k\sm \Pres_{k+1}|\le 2|\Bnd_k\cap \Cov_{m_k}|$.
\end{lemma}
\begin{proof}
  First, suppose that $|\Bnd_k\cap \Cov_{m_k}|=0$. Consider $p = \min \Bnd_k$.
  Since $p\notin \Cov_{m_k}$, we must have $p>m_k$ and $p-\alpha_k+1 \ne \OccThree{m_k}{p-\alpha_k+1}{\T}$.
  Consequently, there is a position $p'\in [\alpha_k\dd p)$
  such that $\T(p-\alpha_k \dd p-\alpha_k+m_k]=\T(p'-\alpha_k\dd p'-\alpha_k+m_k]$.
  In particular, $\T(p-\alpha_k \dd p+\alpha_k]=\T(p'-\alpha_k\dd p'+\alpha_k]$,
  so \cref{lem:recompr1} yields $p'\in \Bnd_k$, contradicting the choice of~$p$.
  Thus, $\Bnd_k=\emptyset$, which means that $\Pres_{k+1}=\Pres_k=\{\T_k[1]\}$.

  Next, we prove that $|\Pres_k\sm \Pres_{k+1}|\le 1+|\Bnd_k\cap \Cov_{m_k}|$.
  Let $\T_k[j]$ be the leftmost occurrence in $\T_k$ of $A\in \Pres_{k}\sm \Pres_{k+1}$.  
  Moreover, let $p=|\exp(\T_k[1\dd j))|$ and $q=|\exp(\T_k[1\dd j])|$
  so that $\T(p\dd q]=\exp(A)$ is the phrase induced by $\T_k[j]$. 
  By \cref{constr:Jez}, we have $A\in \Symb_{k+1}$, and therefore $q-p \le \ell_{k+1}$.

  We shall prove that $j=1$ or $p\in \Bnd_k\cap \Cov_{m_k}$.
  This will complete the proof of $|\Pres_k\sm \Pres_{k+1}|\le 1+|\Bnd_k\cap \Cov_{m_k}|$ because 
  distinct symbols $A$ yield distinct positions $j$ and $p$.
  For a proof by contradiction, suppose that $j\in (1\dd |\T_k|]$ 
  yet $p\notin \Bnd_k\cap \Cov_{m_k}$.
  Since $p\in \Bnd_k$ holds due to $j>1$, we derive $p\notin \Cov_{m_k}$.
  Hence, $p > m_k$ and $p-\alpha_k+1 \ne \OccThree{m_k}{p-\alpha_k+1}{\T}$.
  Consequently, there is a position $p'\in [\alpha_k\dd p)$
  such that $\T(p-\alpha_k \dd p-\alpha_k+m_k]=\T(p'-\alpha_k\dd p'-\alpha_k+m_k]$.
  In particular, $\T(p-\alpha_k \dd p+\alpha_k]=\T(p'-\alpha_k\dd p'+\alpha_k]$,
  so \cref{lem:recompr1} yields $p'\in \Bnd_k$.
  Similarly, due to $q-p=|\exp(A)|\le \floor{\ell_{k+1}}=m_k-2\alpha_k$,
  we have $\T(q-\alpha_k \dd q+\alpha_k]=\T(q'-\alpha_k\dd q'+\alpha_k]$
  for $q':=p'+|\exp(A)|$, and therefore $q'\in \Bnd_k$ holds due to $q\in \Bnd_k$.
  \cref{lem:recompr1} further implies $\Bnd_k\cap(p'\dd q') = \emptyset = \Bnd_k\cap (p\dd q)$.
  Consequently, $\T(p'\dd q']$ is a phrase induced by $\T_k$,
  and, since $p'<p$, it corresponds to $\T_k[j']$ for some $j'<j$.
  By \cref{fct:cons}, we have $\T_k[j']=\T_k[j]=A$, which contradicts the choice of $\T_k[j]$
  as the leftmost occurrence of $A$ in $\T_k$.

  Overall, we have $|\Pres_k\sm \Pres_{k+1}|=0$ if $|\Bnd_k\cap \Cov_{m_k}|=0$
  and $|\Pres_k\sm \Pres_{k+1}|\le 1+|\Bnd_k\cap \Cov_{m_k}|$ otherwise.
  Combining these two claims, we derive $|\Pres_k\sm \Pres_{k+1}|\le 2|\Bnd_k\cap \Cov_{m_k}|$.
\end{proof}

Next, we use our choice of $\Left_k$ and $\Right_k$ to bound the sizes $|\Bnd_h\cap \Cov_{m_h}|$ in terms of the sizes $|\Cov_{m_h}|$.
\begin{lemma}\label{lem:inductivebound}
For every $k\in \Zz$, we have 
\[\sum_{h=0}^{k-1}|\Bnd_h\cap \Cov_{m_h}| + \sum_{h=k}^\infty (\tfrac34)^{\floor{\frac{h}{2}}-\floor{\frac{k}{2}}} |\Bnd_k\cap \Cov_{m_h}| \le \sum_{h=0}^{k-1} \left(\tfrac{2|\Cov_{m_h}|}{m_h}+\tfrac{4|\Cov_{m_h}|}{\ell_{h+1}}\right) + \sum_{h=k}^\infty (\tfrac34)^{\floor{\frac{h}{2}}-\floor{\frac{k}{2}}}\left(\tfrac{2|\Cov_{m_h}|}{m_h}+\tfrac{4|\Cov_{m_h}|}{\ell_{k+1}}\right).\]
\end{lemma}
\begin{proof}
Denote \[L_k := \sum_{h=0}^{k-1}|\Bnd_h\cap \Cov_{m_h}| + \sum_{h=k}^\infty (\tfrac34)^{\floor{\frac{h}{2}}-\floor{\frac{k}{2}}} |\Bnd_k\cap \Cov_{m_h}|.\]
We proceed by induction on $k$. For $k = 0$, due to $\ell_1 = 1$, we have
\begin{align*}
  L_0 
&= \sum_{h=0}^\infty (\tfrac34)^{\floor{\frac{h}{2}}} |\Bnd_0\cap \Cov_{m_h}| \\
&\le \sum_{h=0}^\infty (\tfrac34)^{\floor{\frac{h}{2}}} |\Cov_{m_h}| \\
&< \sum_{h=0}^\infty (\tfrac34)^{\floor{\frac{h}{2}}} \left(\tfrac{2|\Cov_{m_h}|}{m_h}+\tfrac{4|\Cov_{m_h}|}{\ell_{1}}\right).
\end{align*}
If $k$ is odd, then $\Bnd_{k}\sub \Bnd_{k-1}$, $\floor{\frac{k}{2}}=\floor{\frac{k-1}{2}}$, and $\ell_{k+1}=\ell_k$. Therefore,
\begin{align*}
  L_k 
&=\sum_{h=0}^{k-1}|\Bnd_h\cap \Cov_{m_h}| + \sum_{h=k}^\infty (\tfrac34)^{\floor{\frac{h}{2}}-\floor{\frac{k}{2}}} |\Bnd_k\cap \Cov_{m_h}|\\
&\le \sum_{h=0}^{k-1}|\Bnd_h\cap \Cov_{m_h}| + \sum_{h=k}^\infty (\tfrac34)^{\floor{\frac{h}{2}}-\floor{\frac{k-1}{2}}} |\Bnd_{k-1}\cap \Cov_{m_h}| \\
&=\sum_{h=0}^{k-2}|\Bnd_h\cap \Cov_{m_h}| + \sum_{h=k-1}^\infty (\tfrac34)^{\floor{\frac{h}{2}}-\floor{\frac{k-1}{2}}} |\Bnd_{k-1}\cap \Cov_{m_h}| \\
&\le \sum_{h=0}^{k-2} \left(\tfrac{2|\Cov_{m_h}|}{m_h}+\tfrac{4|\Cov_{m_h}|}{\ell_{h+1}}\right) + \sum_{h=k-1}^\infty (\tfrac34)^{\floor{\frac{h}{2}}-\floor{\frac{k-1}{2}}}\left(\tfrac{2|\Cov_{m_h}|}{m_h}+\tfrac{4|\Cov_{m_h}|}{\ell_{k}}\right) \\
& = \sum_{h=0}^{k-1} \left(\tfrac{2|\Cov_{m_h}|}{m_h}+\tfrac{4|\Cov_{m_h}|}{\ell_{h+1}}\right) + \sum_{h=k}^\infty (\tfrac34)^{\floor{\frac{h}{2}}-\floor{\frac{k}{2}}}\left(\tfrac{2|\Cov_{m_h}|}{m_h}+\tfrac{4|\Cov_{m_h}|}{\ell_{k+1}}\right).
\end{align*}
It remains to consider the case when $k$ is even.
\begin{claim}
  For every $h\in \Zz$, the set $J'_{k,h} := \{j \in J_{k,h} : \T_{k-1}[j]\notin \Symb_{k} \text{ or } \T_{k-1}[j+1]\notin \Symb_k\}$ satisfies
  $|J'_{k,h}|<\frac{2}{m_h}|\Cov_{m_h}|+\frac{2}{\ell_k}|\Cov_{m_h}|$.
\end{claim}
\begin{proof}
Let us first consider the case when $m_h \ge n$ so that $\Cov_{m_h} = [1\dd n]$.
If $j\in J'_{k,h}$, then $\T_{k-1}[j]\notin \Symb_k$ or $\T_{k-1}[j+1]\notin \Symb_k$.
The level-$(k-1)$ phrase corresponding to this pausing symbol is of length greater than $\ell_k$, and there are fewer than $\frac{n}{\ell_k}$ such phrases.
Thus, $|J'_{k,h}|< \frac{2n}{\ell_k} < \frac{2}{m_h}|\Cov_{m_h}|+\frac{2}{\ell_k}|\Cov_{m_h}|$.

Next, consider $m_h < n$. 
Denote $\bar{\Bnd}_{k-1}=\Bnd_{k-1}\cup\{0,n\}$.
Let us fix an interval $I\sub \Cov_{m_h}$ and define $J=\{j\in [1\dd |\T_{k-1}|) : |\exp(\T_{k-1}[1\dd j-1])|\in I\text{ and }|\exp(\T_{k-1}[1\dd j+1])|\in I\}$.
Observe that $|J|=\max(0, |\bar{\Bnd}_{k-1}\cap I|-2)$: every $j\in J$ corresponds to a position $|\exp(\T_{k-1}[1\dd j])|\in \bar{\Bnd}_{k-1}\cap I$ which is neither the leftmost nor the rightmost one in $\bar{\Bnd}_{k-1}\cap I$. If $j\in J'_{k,h}$, then $\T_{k-1}[j]\notin \Symb_k$ or $\T_{k-1}[j+1]\notin \Symb_k$.
The level-$(k-1)$ phrase $\T(p\dd q]$ corresponding to this pausing symbol satisfies $[p\dd q]\cap I$ and, by definition of $\Symb_k$,
it is of length $q-p> \ell_k$. There are fewer than $\frac{|I|-1}{\ell_k}$ phrases $\T(p\dd q]$ satisfying both conditions, 
and thus $|J\cap J'_{k,h}| < \frac{2|I|-2}{\ell_k}$.
Taking into account the leftmost and the rightmost element of $\bar{\Bnd}_{k-1}\cap I$,
we conclude that $|\{j\in J'_{k,h} : |\exp(\T_{k-1}[1\dd j])|\in I\}| < 2 + \frac{2|I|}{\ell_k}$.
Since $|\IntervalRepr{\Cov_{m_h}}|\le \frac{1}{m_h}|\Cov_{m_h}|$, we conclude that $|J'_{k,h}| <  \frac{2|\Cov_{m_h}|}{m_h}+\frac{2|\Cov_{m_h}|}{\ell_{k}}$.
\end{proof}
Therefore,
\begin{align*}
  L_k 
  &=\sum_{h=0}^{k-1}|\Bnd_h\cap \Cov_{m_h}| + \sum_{h=k}^\infty (\tfrac34)^{\floor{\frac{h}{2}}-\floor{\frac{k}{2}}} |\Bnd_k\cap \Cov_{m_h}|\\
  &=\sum_{h=0}^{k-1}|\Bnd_h\cap \Cov_{m_h}| + \sum_{h=k}^\infty (\tfrac34)^{\floor{\frac{h}{2}}-\floor{\frac{k}{2}}} (|\Bnd_{k-1}\cap \Cov_{m_h}|-|\{j\in J_{k,h} : \T_{k-1}[j\dd j+1]\in \Left_k \Right_k\}|)\\
  &=\sum_{h=0}^{k-1}|\Bnd_h\cap \Cov_{m_h}| + \sum_{h=k}^\infty (\tfrac34)^{\floor{\frac{h}{2}}-\floor{\frac{k}{2}}} \left(|\Bnd_{k-1}\cap \Cov_{m_h}|-\sum_{(A,B)\in \Left_k\times \Right_k}w_{k,h}(A,B)\right) \\
  &=\sum_{h=0}^{k-1}|\Bnd_h\cap \Cov_{m_h}| + \sum_{h=k}^\infty (\tfrac34)^{\floor{\frac{h}{2}}-\floor{\frac{k}{2}}} |\Bnd_{k-1}\cap \Cov_{m_h}|-\sum_{(A,B)\in \Left_k\times \Right_k}w_{k}(A,B) \\
  &\le \sum_{h=0}^{k-1}|\Bnd_h\cap \Cov_{m_h}| + \sum_{h=k}^\infty (\tfrac34)^{\floor{\frac{h}{2}}-\floor{\frac{k}{2}}} |\Bnd_{k-1}\cap \Cov_{m_h}|-\tfrac14\sum_{(A,B)\in \Symb_k\times \Symb_k}w_{k}(A,B) \\
  &=\sum_{h=0}^{k-1}|\Bnd_h\cap \Cov_{m_h}| + \sum_{h=k}^\infty (\tfrac34)^{\floor{\frac{h}{2}}-\floor{\frac{k}{2}}} \left(|\Bnd_{k-1}\cap \Cov_{m_h}|-\tfrac14\sum_{(A,B)\in \Symb_k\times \Symb_k}w_{k,h}(A,B)\right) \\
  &=\sum_{h=0}^{k-1}|\Bnd_h\cap \Cov_{m_h}| + \sum_{h=k}^\infty (\tfrac34)^{\floor{\frac{h}{2}}-\floor{\frac{k}{2}}} (|\Bnd_{k-1}\cap \Cov_{m_h}|-\tfrac14|\{j\in J_{k,h} : \T_{k-1}[j\dd j+1]\in \Symb_k \Symb_k\}|)\\
  &=\sum_{h=0}^{k-1}|\Bnd_h\cap \Cov_{m_h}| + \sum_{h=k}^\infty (\tfrac34)^{\floor{\frac{h}{2}}-\floor{\frac{k}{2}}} (\tfrac34|\Bnd_{k-1}\cap \Cov_{m_h}|+\tfrac14|J'_{k,h}|)\\
&=\sum_{h=0}^{k-2}|\Bnd_h\cap \Cov_{m_h}| + \sum_{h=k-1}^\infty (\tfrac34)^{\floor{\frac{h}{2}}-\floor{\frac{k-1}{2}}} |\Bnd_{k-1}\cap \Cov_{m_h}| +\tfrac14 \sum_{h=k}^\infty (\tfrac34)^{\floor{\frac{h}{2}}-\floor{\frac{k}{2}}} |J'_{k,h}| \\
&\le \sum_{h=0}^{k-2} \left(\tfrac{2|\Cov_{m_h}|}{m_h}+\tfrac{4|\Cov_{m_h}|}{\ell_{h+1}}\right) + \sum_{h=k-1}^\infty (\tfrac34)^{\floor{\frac{h}{2}}-\floor{\frac{k-1}{2}}}\left(\tfrac{2|\Cov_{m_h}|}{m_h}+\tfrac{4|\Cov_{m_h}|}{\ell_{k}}\right) +  \tfrac14 \sum_{h=k}^\infty (\tfrac34)^{\floor{\frac{h}{2}}-\floor{\frac{k}{2}}} \left(\tfrac{2|\Cov_{m_h}|}{m_h}+\tfrac{2|\Cov_{m_h}|}{\ell_{k}}\right)\\
& = \sum_{h=0}^{k-1} \left(\tfrac{2|\Cov_{m_h}|}{m_h}+\tfrac{4|\Cov_{m_h}|}{\ell_{h+1}}\right) + \tfrac34\sum_{h=k}^\infty (\tfrac34)^{\floor{\frac{h}{2}}-\floor{\frac{k}{2}}}\left(\tfrac{2|\Cov_{m_h}|}{m_h}+\tfrac{4|\Cov_{m_h}|}{\ell_{k}}\right) + \tfrac14 \sum_{h=k}^\infty (\tfrac34)^{\floor{\frac{h}{2}}-\floor{\frac{k}{2}}} \left(\tfrac{2|\Cov_{m_h}|}{m_h}+\tfrac{2|\Cov_{m_h}|}{\ell_{k}}\right) \\
& = \sum_{h=0}^{k-1} \left(\tfrac{2|\Cov_{m_h}|}{m_h}+\tfrac{4|\Cov_{m_h}|}{\ell_{h+1}}\right) + \sum_{h=k}^\infty (\tfrac34)^{\floor{\frac{h}{2}}-\floor{\frac{k}{2}}}\left(\tfrac{2|\Cov_{m_h}|}{m_h}+\tfrac{7|\Cov_{m_h}|}{2\ell_{k}}\right)\\
& = \sum_{h=0}^{k-1} \left(\tfrac{2|\Cov_{m_h}|}{m_h}+\tfrac{4|\Cov_{m_h}|}{\ell_{h+1}}\right) + \sum_{h=k}^\infty (\tfrac34)^{\floor{\frac{h}{2}}-\floor{\frac{k}{2}}}\left(\tfrac{2|\Cov_{m_h}|}{m_h}+\tfrac{4|\Cov_{m_h}|}{\ell_{k+1}}\right).
\end{align*}
This completes the proof.
\end{proof}

The following result is used to conclude that $|\T_k|=1$ for sufficiently large $k=\Oh(\log n)$.
\begin{lemma}\label{lem:height}
Let $\kappa = 2\ceil{\log_{8/7} n}$. For every integer $k\ge \kappa$, we have $|\Bnd_k| < (\frac34)^{\floor{\frac{k}{2}}-\floor{\frac{\kappa}{2}}}n$.
\end{lemma}
\begin{proof}
We proceed by induction on $n$. The base case of $k=\kappa$ is trivial because $\Bnd_\kappa \sub [1\dd n)$.
If $k>\kappa$ is odd, then $|\Bnd_k| \le |\Bnd_{k-1}| < (\frac34)^{\floor{\frac{k-1}{2}}-\floor{\frac{\kappa}{2}}}n=(\frac34)^{\floor{\frac{k}{2}}-\floor{\frac{\kappa}{2}}}n$.
If $k>\kappa$ is even, then we note that $\ell_{k}\ge n$ and $\Cov_{m_h} = [1\dd n]$ for all $h\ge k$. Consequently, $J_{k,h}=[1\dd |\T_{k-1}|)$ and, for every $A,B\in \Symb$, we have 
\begin{align*}
  w_k(A,B)&=\sum_{k=h}^\infty (\tfrac34)^{\floor{\frac{h}{2}}-\floor{\frac{k}{2}}} |\{j\in J_{k,h}: \T_{k-1}[j\dd j+1]=AB\}|\\
  &=\sum_{k=h}^\infty (\tfrac34)^{\floor{\frac{h}{2}}-\floor{\frac{k}{2}}} |\{j\in [1\dd |\T_{k-1}|): \T_{k-1}[j\dd j+1]=AB\}|\\ 
  &= 8 |\{j\in [1\dd |\T_{k-1}|): \T_{k-1}[j\dd j+1]=AB\}|
\end{align*}
Consequently,
 \begin{multline*} |\Bnd_k| \le |\Bnd_{k-1}| - |\{j\in [1\dd |\T_{k-1}|): \T_{k-1}[j\dd j+1]\in \Left_k\cdot \Right_k\}|\\
   = |\Bnd_{k-1}|-\tfrac18\sum_{(A,B)\in \Left_k\times \Right_k} w_k(A,B) \le |\Bnd_{k-1}|-\tfrac1{32}\sum_{(A,B)\in \Symb_k\times \Symb_k} w_k(A,B)= |\Bnd_{k-1}|-\tfrac{1}{4}|\Bnd_{k-1}| \\= \tfrac34|\Bnd_{k-1}| 
   < \tfrac34\cdot (\tfrac34)^{\floor{\frac{k-1}{2}}-\floor{\frac{\kappa}{2}}}n=(\tfrac34)^{\floor{\frac{k}{2}}-\floor{\frac{\kappa}{2}}}n.\qedhere
\end{multline*}
\end{proof}
\begin{corollary}\label{cor:height}
  If $k \ge 2\ceil{\log_{8/7} n}+2\ceil{\log_{4/3} n}$, then $|\T_k|=1$.
\end{corollary}

We conclude the analysis of $|\Pres|$ by combining \cref{lem:bkpk,lem:inductivebound}.
\begin{theorem}\label{thm:grsize}
  We have $|\Pres|=\Oh(\sum_{k=0}^\infty \frac{1}{m_k}\cdot |\Cov_{m_k}|)$.
  \end{theorem}
\begin{proof}
  By \cref{cor:height}, there exists $k=\Oh(\log n)$ such that $|\T_k|=1$ (and thus $\T_h = \T_k$ for all $h \ge k$).
  Applying \cref{lem:bkpk,lem:inductivebound} for this $k$, we conclude that 
  \begin{align*}
    |\Pres| &= |\Pres_{k}|+\sum_{h=0}^{k-1} |\Pres_{h}\sm \Pres_{h+1}|\\
  &\le 1 + 2\sum_{h=0}^{k-1} |\Bnd_h\cap \Cov_{m_h}| \\
  &\le 1+ 2\sum_{h=0}^{k-1} \left(\tfrac{2|\Cov_{m_h}|}{m_h}+\tfrac{4|\Cov_{m_h}|}{\ell_{h+1}}\right) + 2\sum_{h=k}^\infty (\tfrac34)^{\floor{\frac{h}{2}}-\floor{\frac{k}{2}}}\cdot\left(\tfrac{2|\Cov_{m_h}|}{m_h}+\tfrac{4|\Cov_{m_h}|}{\ell_{k+1}}\right)\\
    &\le 1+ 2\sum_{h=0}^{k-1} \left(\tfrac{2|\Cov_{m_h}|}{m_h}+\tfrac{4|\Cov_{m_h}|}{\ell_{h+1}}\right) + 2\sum_{h=k}^\infty (\tfrac78)^{\floor{\frac{h}{2}}-\floor{\frac{k}{2}}}\cdot\left(\tfrac{2|\Cov_{m_h}|}{m_h}+\tfrac{4|\Cov_{m_h}|}{\ell_{k+1}}\right)\\
  &\le 1+2\sum_{h=0}^{\infty} \left(\tfrac{2|\Cov_{m_h}|}{m_h}+\tfrac{4|\Cov_{m_h}|}{\ell_{h+1}}\right)\\
  &\le 1+268\sum_{h=0}^{\infty} \tfrac{|\Cov_{m_h}|}{m_h}\\
  &= \Oh\left(\sum_{h=0}^\infty \tfrac{|\Cov_{m_h}|}{m_h}\right).
  \end{align*}
  Here, the last inequality follows from the fact that $m_h = \floor{16\ell_k}+\floor{\ell_{k+1}}\le 33\ell_{k+1}$.
  \end{proof}

Finally, we provide a concrete bound for the cover hierarchy of \cref{cons:cover}.%
\begin{lemma}\label{lem:level}
The cover hierarchy $\Cov$ of \cref{cons:cover} satisfies 
$\sum_{k=0}^\infty \frac{1}{m_k}|\Cov_{m_k}|=\Oh(\SubstringComplexity{\T} \log \tfrac{n \log \sigma}{\SubstringComplexity{\T} \log n})$.
\end{lemma}
\begin{proof}
  Define $\mu = \floor{2\log_{8/7}\frac{\log \SubstringComplexity{\T}}{264\log \sigma}}$ and $\nu = \ceil{2\log_{8/7}\frac{n}{\SubstringComplexity{\T}}}$
  so that 
  \begin{align*}
    m_{\mu} &\le 33\ell_{\mu+1} = 33(\tfrac{8}{7})^{\floor{\mu/2}} \le 33(\tfrac{8}{7})^{\log_{8/7}\frac{\log \SubstringComplexity{\T}}{264\log \sigma}}
  =\tfrac{\log\SubstringComplexity{\T}}{8\log \sigma},\\
  \ell_{\nu+2} &= (\tfrac{8}{7})^{\ceil{\nu/2}} \ge (\tfrac{8}{7})^{\log_{8/7}\frac{n}{\SubstringComplexity{\T}}} = \tfrac{n}{\SubstringComplexity{\T}}.
\end{align*}
For $k\in [0\dd \mu]$, we observe that $|\Cov_{m_k}|\le \SubstrCount{8m_k}{\T}+8m_k \le \sigma^{8m_k}+8m_k \le 2\sigma^{8m_k}$.
Thus,
\[
  \sum_{k=0}^\mu 
  \tfrac{|\Cov_{m_k}|}{m_k}\le \sum_{k=0}^\mu |\Cov_{m_k}| \le \sum_{k=0}^{\mu} 2\sigma^{8m_{k}} \le 2\sigma^{8m_\mu}\le 2\SubstringComplexity{\T}.
\]
For $k\in (\mu\dd \nu]$, we observe that $|\Cov_{m_k}|\le \SubstrCount{8m_k}{\T}+8m_k \le 16\SubstringComplexity{\T} m_k$.
Thus,
\[
  \sum_{k=\mu+1}^\nu 
  \tfrac{|\Cov_{m_k}|}{m_k}\le \sum_{k=\mu+1}^\nu 16\SubstringComplexity{\T} = 16\SubstringComplexity{\T} (\nu-\mu) = \Oh(\SubstringComplexity{\T} \log\tfrac{n \log\sigma}{\SubstringComplexity{\T} \log n})=\Oh(\SubstringComplexity{\T} \log\tfrac{n \log\sigma}{\SubstringComplexity{\T} \log n}).
\]
For $k\in (\nu\dd \infty)$, we observe that $|\Cov_{m_k}|\le n$.
Thus,
\[
  \sum_{k=\nu+1}^\infty
  \tfrac{|\Cov_{m_k}|}{m_k}\le \sum_{k=\nu+1}^\infty \tfrac{n}{\ell_{k+1}} \le \tfrac{2n}{\ell_{\nu+2}}\sum_{i=0}^\infty (\tfrac{7}{8})^i  \le 8\SubstringComplexity{\T}.\qedhere
\]
\end{proof}

\subsection{Efficient Construction}\label{sec:alg}
Our next big goal is to develop an efficient implementation of \cref{constr:Jez}.
We start with an auxiliary tools needed to find the sets $\Left_k$ and $\Right_k$ according to \cref{def:deterministic}.
The first of them is a folklore linear-time approximation algorithm for the directed max-cut problem.
\begin{lemma}\label{lem:maxcut}
There exists a linear-time algorithm that, given a weighted directed graph $G=(V,E,w)$ without self-loops,
constructions a partition $V=L\cup R$ such that $w(L,R)\ge \frac14 w(V,V)$, where, for any $A,B\sub V$, 
we denote $w(A,B) = \sum_{e\in E\cap (A\times B)} w(e)$.
\end{lemma}
\begin{proof}
  First, we preprocess $G$ so that each $v\in V$ stores both incoming and outgoing arcs. 
  We maintain a partition $V = L \cup M \cup R$ into three disjoint classes.
  Initially, $M=V$ and, until $M\ne \emptyset$, we pick an arbitrary vertex $v\in M$
  and move $v$ to $L$ or $R$, depending on whether $2w(v,R)+w(v,M) \ge 2w(L,v)+w(M,v)$ or not.
  This decision can be implemented in $\Oh(1+\deg(v))$ time, which yields a total running time of $\Oh(|V|+|E|)$.
  
  As for correctness, we shall prove that $\Phi:=4w(L,R) + 2w(L,M) +2w(M,R)+w(M,M)$
  cannot decrease throughout the algorithm.
  Consider the effect of moving $v$ from $M$ to $L$ on the four terms of~$\Phi$ (recall that there are no self-loops incident to $v$):
  \begin{itemize}
    \item $w(L,R)$ increases by $w(v,R)$;
    \item $w(L,M)$ increases by $w(v,M)$ and decreases by $w(L,v)$;
    \item $w(M,R)$ decreases by $w(v,R)$;
    \item $w(M,M)$ decreases by $w(v,M)$ and decreases by $w(M,v)$.
  \end{itemize}
  Overall, $\Phi$ increases by 
  \begin{multline*}4w(v,R)+2(w(v,M)-w(L,v))-2w(v,R)-(w(v,M)+w(M,v))\\
  =2w(v,R)+w(v,M)-2w(L,v)-w(M,v),\end{multline*}
  and this quantity is nonnegative when the algorithm decides to move $v$ to $L$. 
  
  Similarly, if $v$ is moved from $M$ to $R$, then 
  \begin{itemize}
    \item $w(L,R)$ increases by $w(L,v)$;
    \item $w(L,M)$ decreases by $w(L,v)$;
    \item $w(M,R)$ increases by $w(M,v)$ and decreases by $w(v,R)$;
    \item $w(M,M)$ decreases by $w(v,M)$ and decreases by $w(M,v)$.
  \end{itemize}
  Overall, $\Phi$ increases by 
  \begin{multline*}4w(L,v)-2w(L,v)+2(w(M,v)-w(v,R))-(w(v,M)+w(M,v))\\
  =2w(L,v)+w(M,v)-2w(v,R)-w(v,M),\end{multline*}
  and this quantity is positive when the algorithm decides to move $v$ to $R$.
  Upon the end of the algorithm, we have $\Phi=4w(L,R)$ due to $M=\emptyset$,
  whereas, initially, $\Phi = w(M,M) = w(V,V)$ due to $M=V$.
  Since $\Phi$ is nondecreasing, we conclude that $4w(L,R)\ge w(V,V)$ holds as claimed. 
  \end{proof}

The following lemma and its corollary are needed to evaluate the $w_{k,h}$ functions of \cref{def:deterministic}.
  \begin{lemma}\label{lem:cnt}
    Given a text $\T\in \Sigma^n$, represented by a straight-line grammar of size $g$, and the interval representation of a set $\PP\sub [1\dd n]$,
    in $\Oh((g+|\IntervalRepr{\PP}|)\log n)$ time one can compute, for every length-1 substring $a$ of $\T$, the value $|\{i\in \PP: \T[i]=a\}|$.
  \end{lemma}
  \begin{proof}
    First, we construct an AVL grammar $\G$ of size $\Oh(g \log n)$ that generates $\T$; see~\cite{Rytter03}.
    Next, for every interval $I=[b\dd e)$ in the representation of $\PP$, we extend the grammar with a symbol $X_I$ whose expansion is $\T[b\dd e)$.
    Finally, we add a new start symbol to $\G$ whose production is $\bigodot_{I}X_I$, where $I$ iterates over the intervals representing $\PP$.
    This grammar can be constructed in $\Oh((g+|\IntervalRepr{\PP}|)\log n)$ time.

    It is easy to see that, for every length-$1$ substring $a$ of $\T$, the value $|\{i\in \PP: \T[i]=a\}|$ equals the number of occurrences of $a$ in the string produced by $\G$. This corresponds to the number of paths from the start symbol to $a$ in the DAG representing $\G$;
    such values can be computed in $\Oh(|\G|)=\Oh((g+|\IntervalRepr{\PP}|)\log n)$ time.
  \end{proof}
  \begin{corollary}\label{cor:cnt}
    Given a text $\T\in \Sigma^n$, represented by an SLP of size $g$, and the interval representation of a set $\PP\sub [1\dd n)$,
    in $\Oh((g+|\IntervalRepr{\PP}|)\log n)$ time one can compute, for every length-2 substring $ab$ of $\T$, the value $|\{i\in \PP: \T[i\dd i+1]=ab\}|$.
  \end{corollary}
  \begin{proof}
    We construct a text $\hT\in (\Sigma^2)^{n-1}$ such that $\hT[i] = \T[i\dd i+1]$.
    For this, we transform the input grammar $\G$ into a grammar $\hG$.
    For each symbol $A$ with $\rhs_\G(A)=BC$, we create a symbol $\hA$ with $\rhs_{\hG}(\hA)= \hB \cdot (\RML(B),\LML(C))\cdot \hC$.
    If the starting symbol of $\G$ is $S$, then the starting symbol of $\hG$ is $\hS$.
    It is easy to see that $\exp(\hS)=\hT$ and that $\{i\in \PP: \T[i\dd i+1]=ab\}=\{i\in \PP : \hT[i]=(a,b)\}$.
    Thus, it suffices to use \cref{lem:cnt}.
  \end{proof}

  We are now ready to provide the implementation of the construction algorithm.

  \begin{theorem}\label{thm:rlslp}
    Suppose that we are given the LZ77 parsing of a text $\T\in [0\dd \sigma)^n$
    and, for every $h$ with $m_h \le n$, the interval representation $\IntervalRepr{\Cov_{m_h}}$ of size $\Oh(\frac{1}{m_h}|\Cov_{m_h}|)$.
    Then, the RLSLP of \cref{constr:Jez,def:deterministic} can be constructed in $\Oh(C\log^6 n)$ time,
    where $C=\sum_{h=0}^\infty \frac{1}{m_h}|\Cov_{m_h}|$.
  \end{theorem}
  \begin{proof}
    Our main goal is to build strings $(\T_k)_{k=0}^\infty$ for subsequent integers $k\in \Zz$.
    For each iteration, the output consists of the set $\Pres_{\le k} := \bigcup_{h=0}^{k} \Pres_h$
    as well as the LZ77 parsing of $\T_k$, where each symbol is stored as a pointer to an element of $\Pres_{\le k}$.
    Moreover, for every non-terminal $A\in \Pres_{\le k}$, if $A=(B,C)\in \Symb^2$, then $B,C\in \Pres_{\le k}$ are represented as pointers to elements of $\Pres_{\le k}$.
    Similarly, if $A=(B,m)\in \Pres_{\le k}\cap  \Symb\times \mathbb{Z}_{\ge 2}$, then $B$ is represented as a pointer to an element of $\Pres_{\le k}$.
    Each symbol $A\in \Pres_{\le k}$ also stores $|\exp(A)|$, i.e., the length of its expansion.

    In the base case of $k=0$, the string $\T_0=\T$ is already represented by its LZ77-parsing.
    A linear scan over this parsing lets us retrieve the set $\Pres_{\le 0}$ of symbols that occur in $\T$
    and alter the parsing so that symbols are stored using pointers to $\Pres_{\le 0}$.

    Once the algorithm reaches $|\T_k|=1$, which happens for $k=\Oh(\log n)$ by \cref{cor:height},
    the set $\Pres_{\le k}$ forms an RLSLP $\G$ generating $\T$. Moreover, \cref{thm:grsize} guarantees
    that $|\Pres_{\le k}|=|\Pres|=\Oh(C)$.

    By \cref{fct:cons}, if we change $\G$ so that all symbols in $\Pres_{\le h}$ become terminals,
    then $\G$ would generate $\T_h$ instead. Consequently, $\LZSize{\T_h}=\Oh(C)$ holds for all $h\in \Zz$.

    It remains to implement the iteration, where we derive $\T_{k}$ from $\T_{k-1}$ for $k\in \Zp$.
    If $k$ is odd, then our algorithm applies \cref{lem:rle} with $\Act = \Symb_{k}\cap \Pres_{k-1}$.
    This subroutine costs $\Oh(\LZSize{\T_{k-1}}\cdot \log^5 n)=\Oh(C\log^5 n)$ time and produces the LZ77 representation of $\T_k$.
    A left-to-right scan over this representation lets us determine $\Pres_{k}\sm \Pres_{k-1}$ (which is equal to $\Pres_{\le k}\sm \Pres_{\le (k-1)}$
    by \cref{fct:cons}). For each symbol in $A\in \Pres_{k}\sm \Pres_{k-1}$, we add an appropriate entry to $\Pres_{\le k}$
    and replace its representation (of the form $(B,m)\in \Pres_{\le k-1}\times \mathbb{Z}_{\ge 2}$)
    with a pointer to $\Pres_{\le k}$.

    The algorithm for even $k$ is more difficult because we first need to find sets $\Left_k,\Right_k\sub \Symb_{k}\cap \Pres_{k-1}$ satisfying \cref{def:deterministic}, that is, $w_{k}(\Left_k,\Right_k) \ge \frac14 w_{k}(\Symb_k,\Symb_k)$.
    For this, we convert the LZ77-representation of $\T_{k-1}$ into an AVL grammar $\G_{k}$ using \cref{thm:rytter};
    the size of this grammar is $\Oh(\LZSize{\T_{k-1}}\log n)\sub \Oh(C\log n)$ and the grammar is constructed in $\Oh(C\log n)$ time.
    For each symbol $X\in \T_{\G_{k}}$, we compute the value $|\exp(\exp_{\G_k}(X))|$.
    Since $\G_k$ is of logarithmic height, given any $i\in [0\dd n]$, we can compute the rank of $i$ in $\Bnd_{k-1}$ in $\Oh(\log n)$ time.
    In particular, for every $h\in \Zz$, this lets us convert the interval representation of $\Cov_{m_h}$ into 
    the interval representation of $J_{k,h}$ of size $|\IntervalRepr{J_{k,h}}|=|\IntervalRepr{\Cov_{m_h}}|=\Oh(\frac{1}{m_h}|\Cov_{m_h}|)=\Oh(C)$.
    We execute this step for all $h\in [k\dd \kappa]$, where $\kappa=2\ceil{\log_{8/7} n}$ is defined in \cref{lem:height} and satisfies $m_{\kappa}\ge n$.
    Now, we can use \cref{cor:cnt} to compute all the non-zero values of the $w_{k,h}$ function.
    Since $|\G_k|=\Oh(C\log n)$, this step costs $\Oh(C\log n + C\log^2)=\Oh(C\log^2 n)$ time for each $h\in [0\dd \kappa]$
    and $\Oh(C\log^3 n)$ time in total.
    To retrieve $w_k(A,B)$, we note that $w_{k,h}(A,B)=w_{k,\kappa}(A,B)$ holds for $h\ge \kappa$ due to $\Cov_{m_h} = [1\dd n]=\Cov_{m_\kappa}$.
    Hence, since $\kappa$ is even,
    \[\sum_{h=k}^\infty (\tfrac34)^{\floor{\frac{h}{2}}-\floor{\frac{k}{2}}}w_{k,h}(A,B)=\sum_{h=k}^{\kappa -1}  (\tfrac34)^{\floor{\frac{h}{2}}-\floor{\frac{k}{2}}}w_{k,h}(A,B)+8\cdot  (\tfrac34)^{\floor{\frac{\kappa}{2}}-\floor{\frac{k}{2}}}w_{k,\kappa}(A,B).\]
    Consequently, we can retrieve $w_{k}(A,B)$ by adding up the values $w_{k,h}(A,B)$ for $h\in [k\dd \kappa]$ with appropriate coefficients
    (which are integers when scaled up by a factor of $2^\kappa=n^{\Oh(1)}$).
    Thus, it takes $\Oh(C\log^3 n)$ time to construct a multigraph whose vertices are $\Act_{k}\cap \Pres_{k-1}$
    and with edges $(A,B)$ with (integer) weights $2^\kappa\cdot w_k(A,B)$ (we do not add the edge if $AB$ is not a substring of $\T_{k-1}$; in that case, its weight would be $0$ anyway). By \cref{cor:distinct}, this graph does not contain loops.
    Therefore, we can use  \cref{lem:maxcut} to obtain the desired sets $\Left_k,\Right_k\sub \Symb_{k}\cap \Act_{k-1}$ satisfying 
    $w_{k}(\Left_k,\Right_k) \ge \frac14 w_{k}(\Symb_k,\Symb_k)$. The overall cost of this subroutine is $\Oh(C\log^3 n)$.

    Finally, our algorithm applies \cref{lem:pc}.
    This subroutine costs $\Oh(\LZSize{\T_{k-1}}\cdot \log^4 n)=\Oh(C\log^5 n)$ time and produces the LZ77 representation of $\T_k$.
    A left-to-right scan over this representation lets us determine $\Pres_{k}\sm \Pres_{k-1}$ (which is equal to $\Pres_{\le k}\sm \Pres_{\le (k-1)}$
    by \cref{fct:cons}). For each symbol in $A\in \Pres_{k}\sm \Pres_{k-1}$, we add an appropriate entry to $\Pres_{\le k}$
    and replace its representation (of the form $(B,C)\in \Pres_{\le k-1}^2$) with a pointer to $\Pres_{\le k}$.

    The total cost of the algorithm (across all iterations) is $\Oh(C\log^6 n)$ due to \cref{cor:height}.
  \end{proof}

  \begin{corollary}\label{cor:rlslp}
    Given the LZ77 parsing of a text $\T\in [0\dd \sigma)^n$,
    the RLSLP of \cref{constr:Jez,def:deterministic} can be constructed in $\Oh(\SubstringComplexity{\T}\cdot \log^7 n)$ time
    assuming that the cover hierarchy is chosen according to \cref{cons:cover}.
    Moreover, the size of this RLSLP is $\Oh(\SubstringComplexity{\T}\log \frac{n\log \sigma}{\SubstringComplexity{\T}\log n})$.
  \end{corollary}
  \begin{proof}
    First, we use \cref{cor:cover} to compute $\IntervalRepr{\Cov_{m_h}}$ for all $h\in \Zz$ such that $m_h\le n$;
    this step costs $\Oh(\LZSize{\T}\log^4 n)=\Oh(\SubstringComplexity{\T}\log^5 n)$ time for each $h$ and $\Oh(\SubstringComplexity{\T}\log^6 n)$ time in total.
    Next, we use the algorithm of \cref{thm:rlslp}. Note that $C=\Oh(\SubstringComplexity{\T}\log \frac{n\log \sigma}{\SubstringComplexity{\T}\log n})$ holds due to \cref{lem:level},
    so the total running time is $\Oh(C\log^6 n)=\Oh(\SubstringComplexity{\T}\log^7 n)$.
  \end{proof}

  \subsection{Consequences}
\subsubsection{Optimal Compressed Space Random Access}\label{sec:random-access}

\begin{theorem}\label{th:random-access}
  For any text $\T\in [0\dd \sigma)^n$, there exists a data structure of
  size $\bigO(\SubstringComplexity{\T} \log \tfrac{n \log \sigma}{\SubstringComplexity{\T} \log n})$
  that, given any $i
  \in [1 \dd n]$, returns $\T[i]$ in $\bigO(\log n)$ time. Moreover,
  given the LZ77-parsing of $\T$, it can be constructed in $\bigO(\SubstringComplexity{\T}\log^7 n)$ time.
\end{theorem}
\begin{proof}
  Our data structure consists of the RLSLP constructed in $\bigO(\SubstringComplexity{\T}\log^7 n)$ time using \cref{cor:rlslp}.
  At the query time, we traverse the parse tree of $\T$ from the root to the leaf representing $\T[i]$.
  The query time is proportional to the RLSLP height, which is $\Oh(\log n)$ by \cref{cor:height}.
\end{proof}

\subsubsection{Optimal Compressed Space LCE Queries}\label{sec:lce-queries}

\begin{theorem}\label{th:lce}
  For any text $\T\in [0\dd \sigma)^n$, there exists a data structure of
  size $\bigO(\SubstringComplexity{\T} \log \tfrac{n \log \sigma}{\SubstringComplexity{\T} \log n})$
  answering $\LCE_{\T}$ and $\LCE_{\revstr{\T}}$ queries in $\bigO(\log n)$
  time.
  Moreover, given the LZ77-parsing of $\T$, it can be constructed in $\bigO(\SubstringComplexity{\T}\log^7 n)$ time.
\end{theorem}
\begin{proof}
  Our data structure consists of the RLSLP constructed in $\bigO(\SubstringComplexity{\T}\log^7 n)$ time using \cref{cor:rlslp}.
  In the following, we describe an $\Oh(\log n)$-time algorithm answering an $\LCE_{\T}(i,i')$ query;
  the procedure for an $\LCE_{\revstr{\T}}$ query is symmetric. 
  We assume $i\ne i'$ without loss of generality;
  otherwise, $\LCE_{\T}(i,i)=n-i+1$ can be computed trivially.

  The query procedure maintains two nodes $\nu,\nu'$ of the parse tree $\Tr$ whose expansions $\exp(\nu)=\T[\ell\dd r)$
  and $\exp(\nu')=\T[\ell'\dd r')$ satisfy $\T[i\dd \ell)=\T[i'\dd \ell')$.
  Technically, each of these nodes is associated with a stack storing (the expansions and symbols of) all its ancestors. 

  We initialize $\nu$ as the highest node such that $\exp(\nu)$ starts at position $i$,
  and $\nu'$ as the highest node such that $\exp(\nu')$ starts at position $i'$.
  For this, we follow the paths from the root to the leaves representing $\T[i]$ and $\T[i']$, respectively.
  At each step, we proceed as follows:
  \begin{enumerate}
    \item\label{it:lce:1} If $|\exp(\nu)|=|\exp(\nu')|=1$ and $\symb(\nu)\ne \symb(\nu')$, return $\ell-i$.
    \item\label{it:lce:2} If $|\exp(\nu)| > |\exp(\nu')|$, we replace $\nu$ by its leftmost child.
    \item\label{it:lce:3} If $|\exp(\nu')| > |\exp(\nu)|$, we replace $\nu'$ by its leftmost child.
    \item\label{it:lce:4} If $\symb(\nu)\ne \symb(\nu')$, we replace both $\nu$ and $\nu'$ by their leftmost children.
    \item\label{it:lce:5} Suppose that $\nu$ is the $j$th among the $d$ children of its parent
    and $\nu'$ is the $j'$th among the $d'$ children of its parent. Let $\lambda = \min(d-j,d'-j')$.
    If $\lambda \le 1$, we replace $\nu$ and $\nu'$ the highest nodes whose expansions start at position $r$
    and $r'$, respectively.
    \item\label{it:lce:6} Otherwise, we replace $\nu$ by its $(j+\lambda)$th sibling and $\nu'$ by its $(j'+\lambda)$th sibling.
  \end{enumerate}

  Let us justify the correctness of the algorithm.
  In case~\ref{it:lce:1}, we have $\T[i\dd \ell)=\T[i'\dd \ell')$ yet $\T[\ell]=\symb(v)\ne \symb(\nu')= \T[\ell']$.
  Hence, $\LCE_{\T}(i,i')=\ell-i=\ell'-i'$ is computed correctly.
  In cases~\ref{it:lce:2}--\ref{it:lce:4}, the invariant is still satisfied because the leftmost 
  positions in $\exp(\nu)$ and $\exp(\nu')$ do not change. Moreover, whenever we replace $\nu$ or $\nu'$ with its leftmost child,
  we have $|\exp(\nu)|>1$ and $|\exp(\nu')|>1$, respectively, so the child exists.
  In the remaining cases, we have $\symb(\nu)=\symb(\nu')$ and thus $\T[i\dd r)=\T[i'\dd r')$.
  Thus, the invariant still holds in case~\ref{it:lce:5}, where we replace $\nu$ and $\nu'$ by the highest nodes whose expansions start
  at positions $r$ and $r'$, respectively.
  If $\lambda>1$, then $d,d'\ge 3$, and thus all siblings of $\nu$ and $\nu'$ have the same symbol.
  Thus, the invariant holds because we shift $\nu$ and $\nu'$ by $\lambda$ siblings to the right.

  It remains to analyze the query time. For this, we say that nodes $(\nu,\nu')$ form a \emph{matching pair}
  if  $\symb(\nu)=\symb(\nu')$ and the expansions $\exp(\nu)=\T[\ell\dd r)$ and $\exp(\nu)=\T[\ell\dd r)$ satisfy $\T[i\dd \ell)=\T[i'\dd \ell')$.
  Denote by $N$ and $N'$ the set of nodes $\nu$ and $\nu'$ participating in a matching pair; these pairs form a perfect matching between $N$ and $N'$.
  We claim that the query algorithm satisfies the following additional invariant:   None of the ancestors of $\nu$ belongs to $N$ and none of the ancestors of $\nu'$ belong to $N'$.  
  The invariant is clearly satisfied at the beginning because all the ancestors of $\nu$ and $\nu$ have their expansions starting before position $i$
  and $i'$, respectively.
  In cases~\ref{it:lce:2} and~\ref{it:lce:4}, the node $\nu$ does for a matching pair with any ancestor of $\nu'$ (because they belong to $N'$),
  with the node $\nu$ itself (because $\symb(\nu)\ne \symb(\nu')$), nor with any descendant of $\nu'$ (because their expansions are shorter than $\exp(\nu)$). Thus, $\nu$ does not belong to $N$. By symmetry, the same holds in cases~\ref{it:lce:3} and~\ref{it:lce:4} when we replace $\nu'$ by its leftmost child.
  In cases~\ref{it:lce:5} and~\ref{it:lce:6}, all the ancestors of the new nodes $\nu$ and $\nu'$ are also the ancestors of the old nodes $\nu$ and $\nu'$.

  Let $\hat{N}\sub N$ consist of nodes of $N$ whose ancestors do not belong to $N$, and define $\hat{N}'$ analogously.
  Our algorithm visits only nodes in $\hat{N}$ and $\hat{N}'$ and their ancestors, as well as the ancestors of the two leaves whose mismatch causes the algorithm to return.
  Moreover, if $\hat{N}$ or $\hat{N}'$ contains three or more siblings,
  then Case~\ref{it:lce:6} guarantees that we visit at most two of them. 
  We claim that $\hat{N}$ and $\hat{N}'$ have $\Oh(\log n)$ parents in total. 
  Specifically, we claim that each level of the parsing contributes nodes with at most two parents.
  For this, let us fix a level and let $\T_k[j]$ and $\T_k[\jmath]$ be the leftmost and the rightmost symbol of $\T_k$
  that corresponds to a node in $\hat{N}$.
  By definition of matching pairs, there are analogous symbols $\T_k[j']$ and $\T_k[\jmath']$ corresponding to nodes on $\hat{N}'$.
  By \cref{fct:cons}, the fragments $\T_k[j\dd \jmath]$ and $\T_k[j'\dd \jmath']$ match.
  The block boundaries of \cref{constr:Jez} depend only on the adjacent two symbols,
  so the block boundaries strictly within $\T_k[j\dd \jmath]$ and $\T_k[j'\dd \jmath']$ are placed analogously.
  Hence, the only symbols of $\T_k[j\dd \jmath]$ corresponding to nodes in $\hat{N}$ might be located before
  the leftmost block boundary within $\T_k[j\dd \jmath]$ or after the rightmost block boundary within $\T_k[j\dd \jmath]$.
  These symbols belong to two blocks, and thus the corresponding nodes have two parents.
  By \cref{cor:height}, the parsing has $\Oh(\log n)$ levels, so the total query time is $\Oh(\log n)$.
\end{proof}

\subsubsection{Optimal Compressed Space Synchronizing Sets}\label{sec:sss}

\newcommand{\RUNS}{\mathsf{RUNS}}

\begin{definition}[$\tau$-runs]
For a string $\T\in \Sigma^+$ and an integer $\tau\in [1\dd n]$, we define the set $\RUNS_{\tau}(\T)$
of $\tau$-runs in $\T$ that consists of all fragments $\T[p\dd q]$ of length at least $\tau$ that 
satisfy $\per(\T[p\dd q])\le \frac13\tau$ yet cannot be extended (in any direction) while preserving the shortest period.
\end{definition}

\begin{construction}[\cite{IPM}]\label{cons:sss}
  For $\tau \in [1\dd \floor{\tfrac{n}{2}}]$, let $k=\max\{h\in \Zz\,:\, \tau\ge \alpha_h\}$.
  A position $i\in [1\dd n-2\tau+1]$ is included in $\SSS(\tau,\T)$ if at least one of the following conditions holds:
    \begin{enumerate}[label={\rm (\arabic*)}]
      \item\label{it:sss:reg} $i+\tau-1 \in \Bnd_k$ and $\T[i\dd i+2\tau)$ is not contained in any $\tau$-run;
\item\label{it:sss:beg} $i=p-1$ for some $\tau$-run $\T[p\dd q]\in \RUNS_\tau(\T)$;
\item\label{it:sss:end} $i=q-2\tau+2$ for some $\tau$-run $\T[p\dd q]\in \RUNS_\tau(\T)$.
    \end{enumerate}
\end{construction}

\begin{lemma}\label{lem:sss}
The set $\SSS(\tau,\T)$ obtained using \cref{cons:sss} is a $\tau$-synchronizing set.
Moreover, $|\{i\in \SSS(\tau,\T) : i+\tau-1\in \Cov_{m_k}\}|= \Oh(|\Bnd_k\cap \Cov_{m_k}|+\frac{1}{m_k}|\Cov_{m_k}|)$,
where $k=\max\left(\{0\}\cup \{h\in \Zp\,:\, \tau\ge \alpha_h\}\right)$.
\end{lemma}
\begin{proof}
First, suppose that $i,i'\in [1\dd n-2\tau+1]$ satisfy $\T[i\dd i+2\tau)=\T[i'\dd i'+2\tau)$ and $i\in \SSS(\tau,\T)$.
We will show that if $i$ satisfies one of the conditions~\ref{it:sss:reg}--\ref{it:sss:end}, then $i'$ satisfies the same condition.
If $i$ satisfies condition~\ref{it:sss:reg}, then $\per(\T[i'\dd i'+2\tau))=\per(\T[i\dd i+2\tau))>\frac13\tau$,
so $\T[i'\dd i'+2\tau)$ is not contained in any $\tau$-run. At the same time, due to 
$\T(i+\tau-1-\alpha_k\dd i+\tau-1+2\alpha_k]=\T(i'+\tau-1-\alpha_k\dd i'+\tau-1+2\alpha_k]$,
by \cref{lem:recompr1}, $i+\tau-1\in \Bnd_k$ implies $i'+\tau-1\in \Bnd_k$. Consequently, $i'$ also satisfies condition~\ref{it:sss:reg}.
If $i$ satisfies condition~\ref{it:sss:beg}, then
$\per(\T[i+1 \dd i+\tau])=\per(\T[i'+1 \dd i'+\tau])\le\tfrac13\tau<\per(\T[i \dd i+\tau])=\per(\T[i' \dd i'+\tau])$.
Hence, $\T[i'+1\dd i'+\tau]$ can be extended to a $\tau$-run $\T[p'\dd q']$ that starts at position $p'=i'+1$, and thus $i'$ satisfies condition~\ref{it:sss:beg}.
Similarly, if $i$ satisfies condition~\ref{it:sss:end}, then
$\per(\T[i'+\tau-1 \dd i'+2\tau-2])=\per(\T[i+\tau-1\dd i+2\tau-2])\le\tfrac13\tau<\per(\T[i+\tau-1\dd i+2\tau-1])=\per(\T[i'+\tau-1 \dd i'+2\tau-1])$.
Hence, $\T[i'+\tau-1\dd i'+2\tau-2]$ can be extended to a $\tau$-run $\T[p'\dd q']$ that ends at position $q'=i'+2\tau-2$,
and thus $i'$ satisfies condition~\ref{it:sss:end}.

For a proof of the density condition, consider a position $i\in [1\dd n-3\tau+2]$
with $[i\dd i+\tau)\cap \SSS(\tau,\T) = \emptyset$.
We start by identifying a $\tau$-run $\T[p\dd q]$ with $p\le i+\tau$ and $q\ge i+2\tau-2$.
First, suppose that there exists a position $b\in [i+\tau-1\dd i+2\tau-1)\cap \Bnd_k$.
Since $b-\tau+1\in [i\dd i+\tau)$ has not been added to $\SSS(\tau,\T)$, the fragment $\T[b-\tau+1\dd b+\tau]$
must be contained in a $\tau$-run that satisfies $p\le b-\tau \le i+\tau-1$ and $q\ge b+\tau \ge i+2\tau-1$.

Next, suppose that $[i+\tau\dd i+2\tau)\cap \Bnd_k = \emptyset$.
Then, $\T[i+\tau-1\dd i+2\tau]$ is contained in a single phrase induced by $\T_k$,
and the length of this phrase is at least $\tau+1$.
If $k=0$, this contradicts the fact that all level-$0$ phrases are of length $1$.
Otherwise, $\tau+1\ge \alpha_k + 1 > 2\ell_k$, so \cref{fct:recompr} yields $\per(\T[i+\tau-1\dd i+2\tau-1])\le \ell_k \le  \frac13\tau$.
The $\tau$-run $\T[p\dd q]$ extending $\T[i+\tau-1\dd i+2\tau]$ satisfies $p\le i+\tau-1$ and $q\ge i+2\tau$.

Note that positions $i=p-1$ and $i=q-2\tau+2$ satisfy conditions~\ref{it:sss:beg} and~\ref{it:sss:end}, respectively. Due to $[i\dd i+\tau)\cap \SSS(\tau,\T) = \emptyset$, this implies $p\le i$ and $q\ge  i+3\tau-2$, which means that $\per(\T[i\dd i+3\tau-1))\le \frac13\tau$ holds as claimed.

For the converse implication, note that if $s\in [i\dd i+\tau)\cap \SSS(\tau,\T)$,
then $\per(\T[i\dd i+3\tau-1))\ge \per(\T[s\dd s+2\tau)) > \frac13\tau$ because $\T[s\dd s+2\tau)$ is not contained in any $\tau$-run
(the latter observation is trivial if $s$ satisfies condition~\ref{it:sss:reg};
in the remaining two cases, it follows from the upper bound of $\frac23\tau$ on the overlap of two $\tau$-runs).

As for the size of $\SSS(\tau,\T)$, observe that if $i\in \SSS(\tau,\T)$, then $i+\tau-1\in \Bnd_k$
or there is a $\tau$-run $\T[p\dd q]$ with $p=i+1$ or $q=i+2\tau-2$.
The contribution of the first term is $|\Bnd_k\cap \Cov_{m_k}|$.
The latter two cases contribute $\Oh(|\IntervalRepr{\Cov_{m_k}}|+\frac{1}{m_k}|\Cov_{m_k}|)=\Oh(\frac{1}{m_k}|\Cov_{m_k}|)$ due to  the upper bound of $\frac23\tau$ on the overlap of two $\tau$-runs.
\end{proof}

\begin{lemma}\label{lem:sss:construction}
  Given the LZ77 representation of a text $\T\in [0\dd \sigma)^n$, and integer $\tau\in [1\dd \floor{\tfrac{n}{2}}]$,
  and the interval representation of the set $\Cov_{m_k}$ such that $k=\max\left(\{0\}\cup \{h\in \Zp\,:\, \tau\ge \alpha_h\}\right)$,
  the set $\{i\in \SSS(\tau,\T) : i+\tau-1\in \Cov_{m_k}\}$ can be constructed in $\Oh(\SubstringComplexity{\T}\cdot \log^7 n + (|\Bnd_k\cap \Cov_{m_k}|+\frac{1}{m_k}|\Cov_{m_k}|)\log^3 n)$ time.
\end{lemma}
\begin{proof}
  We use the $\LCE$ queries of \cref{th:lce} as well as the period queries of \cite[Theorem~6.7]{resolutionfull}:
  Given any fragment $x$ of $\T$, in $\Oh(\log^3 n)$ time,
  we can compute $\per(x)$ or report that $\per(x)>\tfrac12|x|$.
  The total preprocessing time is $\Oh(\SubstringComplexity{\T}\cdot \log^7 n)$.

  We separately process each interval $I$ in the interval representation of the set $\Cov_{m_k}$.
  Let $I=[\ell\dd r]$.
  Our first goal is to compute all runs in $\RUNS_\tau(\T)$ that have at least $\tau$ characters in common with $\T[\ell\dd r+2\tau)$.
  We partition $\T[\ell\dd r+2\tau)$ into blocks of length $\floor{\tfrac13\tau}$ (leaving up to $\floor{\tfrac13\tau}$ trailing characters behind).
  For any two consecutive blocks, we apply a 2-period query to determine the shortest period of their concatenation
  (provided that it does not exceed $\floor{\tfrac13\tau}$). If the period does not exceed  $\floor{\tfrac13\tau}$,
  we use $\LCE$ queries of \cref{th:lce} to maximally extend the fragment while preserving its shortest period.
  If the extended fragment has at least $\tau$ characters in common with $\T[\ell\dd r+2\tau)$, we save it as one of the desired $\tau$-runs.
  For each of the computed runs, we report the appropriate positions on the output.
  This step costs $\Oh(\ceil{|I|/\tau}\log^3 n)$ time per interval and $\Oh((|\IntervalRepr{\Cov_{m_k}}|+\frac{1}{m_k}|\Cov_{m_k}|)\log^3 n)=\Oh(\frac{1}{m_k}|\Cov_{m_k}|\log^3 n)$ time in total.

  Next, we construct the set $\{i : i+\tau-1\in \Bnd_k\cap \Cov_{m_k}\}$ by traversing the parse tree of $\T$.
  For each position $i$ in this set, we use a period query to check if $\T[i\dd i+2\tau)$ is contained in any run and, if not, add $i$ to the output.
  This step costs $\Oh(|\Bnd_k\cap \Cov_{m_k}|\log^3 n)$ time.
\end{proof}

\begin{proposition}\label{pr:comp-sss-construction}
  Let $\T\in [0\dd \sigma)^n$ and $c \in \Zp$. Given the LZ77 parsing
  of $\T$, in $\bigO(c\cdot \SubstringComplexity{\T} \log^7 n)$ time we can
  construct a collection $\{\SSScompgen{i}\}_{i \in [4 \dd \lceil
  \log n \rceil)}$ such that:
  \begin{itemize}
  \item For every $i \in [4 \dd \lceil \log n \rceil)$, letting
    $\tau_i = \lfloor \tfrac{2^i}{3} \rfloor$, it holds $\SSScompgen{i} = \CompRepr{c\tau_i}{\SSS_i}{\T} = \SSS_i \cap
    \Cover{c\tau_i}{\T}$ (\cref{def:comp}), where $\SSS_i$ is a
    $\tau_i$-synchronizing set of $\T$ and $\Cover{c\tau_i}{\T}$ is as
    in \cref{cons:cover},
  \item It holds $\sum_{i \in [4 \dd \lceil \log n \rceil)}
    |\SSScompgen{i}| = \bigO(c\SubstringComplexity{\T} \log \tfrac{n \log
    \sigma}{\SubstringComplexity{\T} \log n})$.
  \end{itemize}
\end{proposition}
\begin{proof}
  For $\ell\in \Zp$, define $\Cov_\ell:=\bigcup_{i\in \Cover{c\ell}{\T}}[i\dd i+\ell)\cap [1\dd n]$.
  We shall prove that $(\Cov_\ell)_{\ell\in \Zp}$ forms a cover hierarchy of $\T$.
  Since $\Cover{\ell}{\T} \sub \Cover{c\ell}{\T} \sub \Cov_\ell$, the set $\Cov_\ell$ is an $\ell$-cover.
  Moreover, $|\IntervalRepr{\Cov_\ell}| \le |\IntervalRepr{\Cover{c\ell}{\T}}|\le \max(1, \frac{1}{c\ell}|\Cover{c\ell}{\T}|)
  \le \max(1, \frac{1}{c\ell}|\Cov_\ell|)\le \max(1, \frac{1}{\ell}|\Cov_\ell|)$.
  It remains to prove that the family $\Cov_\ell$ is monotone.
  For this, consider $\ell,\ell'\in \Zp$ with $\ell \le \ell'$ and a position $j\in \Cov_\ell$.
  By definition of $\Cov_\ell$, there exists $i\in \Cover{c\ell}{\T}$ such that $j\in [i\dd i+\ell)$.
  Since $\Cover{c\ell}{\T}\sub \Cover{c\ell'}{\T}$ and $[i\dd i+\ell)\sub [i\dd i+\ell')$, we conclude that $j\in \Cov_{\ell'}$ holds as claimed.
  Observe now that by a simple generalization of \cref{lem:level}, it holds
  $\sum_{k=0}^\infty \frac{1}{m_k}|\Cov_{m_k}| = \Oh(c \SubstringComplexity{\T}\log \frac{n\log \sigma}{\SubstringComplexity{\T}\log n})$.

  The algorithm proceeds as follows:
  \begin{itemize}
  \item Note that, for each $\ell \in \Zp$,
    the interval representation $\IntervalRepr{\Cov_{\ell}}$ can be constructed
    in $\bigO(\LZSize{\T}\log^4 n)$ time using \cref{cor:cover} to build
    $\IntervalRepr{\Cover{c\ell}{\T}}$.  We apply \cref{thm:rlslp} on top of
    this construction. By the above, this takes $\bigO(c\cdot
    \SubstringComplexity{\T}\log^7 n)$ time.
  \item We preprocess the LZ77 parsing of $\T$ using
    \cref{lem:sss:construction} in $\bigO(\SubstringComplexity{\T} \log^7 n)$ time.
    Consider $i \in [4 \dd \lceil \log n \rceil)$ and let
    $k = \max\left(\{0\}\cup \{h\in \Zp\,:\, \tau_i\ge
    \alpha_h\}\right)$.  First, using \cref{lem:sss:construction}, in
    $\bigO(|\Bnd_k \cap \Cov_{m_k}| + \frac{1}{m_k}|\Cov_{m_k}| \log^3 n)$ time we compute the
    set $\{i \in \SSS(\tau_i, \T) : i + \tau - 1 \in \Cov_{m_k}\}$.
    By definition of $\Cov_{\tau_i}$, $j\in \Cover{c\tau_i}{\T}$
    implies $[j\dd j+\tau_i)\sub \Cov_{\tau_i}\sub \Cov_{m_k}$ (where
    $\Cov_{\tau_i}\sub \Cov_{m_k}$ holds since $\tau_i \le
    \alpha_{k+1}\le m_k$).  In particular, $j \in \Cover{c\tau_i}{\T}$
    implies $j + \tau - 1 \in \Cov_{m_k}$. Consequently, $\SSS(\tau_i,
    \T) \cap \Cover{c\tau_i}{\T} \subseteq \{i\in\SSS(\tau_i,\T) : i +
    \tau - 1 \in \Cov_{m_k}\}$. Using \cref{cor:cover}, in
    $\bigO(\LZSize{\T} \log^4 n)$ we thus construct $\IntervalRepr{\Cover{c\tau_i}{\T}}$
    and filter the set $\{i\in\SSS(\tau_i,\T) : i + \tau - 1 \in
    \Cov_{m_k}\}$ to obtain $\SSS(\tau_i, \T) \cap \Cover{c\tau_i}{\T}$.
    By $\sum_{k=0}^\infty \frac{1}{m_k}|\Cov_{m_k}| = \Oh(c \SubstringComplexity{\T}\log \frac{n\log \sigma}{\SubstringComplexity{\T}\log n})$
    and the summation in the proof of \cref{thm:grsize},
    the application of \cref{lem:sss:construction} over all $i$
    takes $\bigO(c \SubstringComplexity{\T} \log \frac{n \log \sigma}{\SubstringComplexity{\T}\log n} \log^3 n)$ time.
  \end{itemize}
  In total, we thus spend $\bigO(c \SubstringComplexity{\T} \log^7 n)$ time.  By
  \cref{lem:sss}, the summation in the proof of \cref{thm:grsize}, and the upper bound $\sum_{k=0}^\infty
  \frac{1}{m_k}|\Cov_{m_k}| = \Oh(c \SubstringComplexity{\T}\log \frac{n\log
  \sigma}{\SubstringComplexity{\T}\log n})$, the total size of the sets
  $\SSS(\tau_i,\T) \cap \Cover{c\tau_i}{\T}$ is $\Oh(c \SubstringComplexity{\T}\log
  \frac{n\log \sigma}{\SubstringComplexity{\T}\log n})$.
\end{proof}

\section{Weighted Range Queries}\label{sec:range-queries}

Let $\X$ and $\Y$ be some linearly ordered sets (we denote the order
on both sets using $\prec$ or $\preceq$).  Let $\Pts \sub \X \times \Y
\times \Zp \times \Z$ be a finite set of points, where each point is
associated with some positive integer \emph{weight} and some integer
\emph{label}. Unless explicitly stated otherwise, we do not place any
restriction on $\Pts$, and only require that $\Pts$ is a set (not a
multiset). In particular, $\Pts$ can in general contain two points
with equal coordinates as long as they differ either on the weight or
on the label.

\begin{description}[style=sameline,itemsep=1ex]
\item[Weighted range counting:] Let $x_l, x_u {\in} \X$ and $y_l, y_u
  {\in} \Y$. We define
  \begin{itemize}
  \item $\RangeCountFourSide{\Pts}{x_l}{x_u}{y_l}{y_u} :=
    \sum_{(x,y,w,\ell) \in \mathcal{R}_4} w$,
  \item $\RangeCountThreeSide{\Pts}{x_l}{x_u}{y_u} :=
    \sum_{(x,y,w,\ell) \in \mathcal{R}} w$,
  \item $\IncRangeCountThreeSide{\Pts}{x_l}{x_u}{y_u} :=
    \sum_{(x,y,w,\ell) \in \mathcal{R}^{\preceq}_3} w$,
  \item $\RangeCountTwoSide{\Pts}{x_l}{x_u} :=
    \sum_{(x,y,w,\ell) \in \mathcal{R}_2} w$,
  \end{itemize}
  where
  \begin{itemize}
  \item $\mathcal{R}_4 = \{(x,y,w,\ell) \in \Pts :
    x_l \preceq x \prec x\text{ and }y_l \preceq y \prec y\}$,
  \item $\mathcal{R}_3 = \{(x,y,w,\ell) \in \Pts :
    x_l \preceq x \prec x_u\text{ and }\allowbreak y \prec y_u\}$,
  \item $\mathcal{R}^{\preceq}_3 = \{(x,y,w,\ell) \in \Pts :
    x_l \preceq x \prec x_u\text{ and }\allowbreak y \preceq y_u\}$,
  \item $\mathcal{R}_2 = \{(x,y,w,\ell) \in \Pts :
    x_l \preceq x \prec x_u\}$.
  \end{itemize}

\item[Weighted range selection:] Let $x_l, x_u \in \X$ and $r \in [1
  \dd \RangeCountTwoSide{\Pts}{x_l}{x_u}]$. We define
  \[\RangeSelect{\Pts}{x_l}{x_u}{r} := \{\ell \in \Z : (x,y,w, \ell) \in
  \Pts,\ x_l \preceq x \prec x_u,\text{ and }\allowbreak y = y_u\},\]
  where $y_u \in \Y$ is such that $r \in
  (\RangeCountThreeSide{\Pts}{x_l}{x_u}{y_u} \dd
  \IncRangeCountThreeSide{\Pts}{x_l}{x_u}{y_u}]$. A weighted range selection
  query asks to return any element of the set
  $\RangeSelect{\Pts}{x_l}{x_u}{r}$.
\item[Range minimum:] Let $x_l, x_u \in \X$ and $y_l, y_u \in \Y$ be
  such that $\RangeCountFourSide{\Pts}{x_l}{x_u}{y_l}{y_u} > 0$. We define
  \begin{itemize}
  \item $\RangeMinFourSide{\Pts}{x_l}{x_u}{y_l}{y_u} :=
    \min_{(x,y,w,\ell) \in \mathcal{R}_4} \ell$,
  \item $\RangeMinTwoSide{\Pts}{x_l}{x_u} :=
    \min_{(x,y,w,\ell) \in \mathcal{R}_2} \ell$,
  \end{itemize}
  where
  \begin{itemize}
  \item $\mathcal{R}_4 = \{(x,y,w,\ell) \in \Pts :
    x_l \preceq x \prec x_u\text
    { and }\allowbreak y_l \preceq y \prec y_u\}$,
  \item $\mathcal{R}_2 = \{(x,y,w,\ell) \in \Pts :
    x_l \preceq x \prec x_u\}$.
  \end{itemize}
\end{description}

\subsection{Integer-Integer Coordinates}\label{sec:int-int}

\begin{proposition}\label{pr:int-int}
  Let $\epsilon > 0$ be a fixed constant. Given a set $\Pts$ of $m$
  points with coordinates in $\Zz$, such that for every $(x,y,w,\ell),
  (x',y',w',\ell') \in \Pts$, $(x,y) = (x',y')$ implies $w' = w$ and
  $\ell' = \ell$, we can in $\bigO(m \log m)$ time construct a
  structure of size $\bigO(m)$ supporting the following queries on
  $\Pts$:
  \begin{itemize}
  \item Weighted range counting queries in $\bigO(\log^{2 + \epsilon}
    m)$ time,
  \item Weighted range selection queries in $\bigO(\log^{3 + \epsilon}
    m)$ time,
  \item Range minimum queries in $\bigO(\log^{1 + \epsilon} m)$ time.
  \end{itemize}
\end{proposition}
\begin{proof}

  We use the following definitions. Let $m = |\Pts|$.  Let $X_{\Pts}[1
  \dd m_x]$ (resp.\ $Y_{\Pts}[1 \dd m_y]$) be a sorted array
  containing all different first (resp.\ second) coordinates of points
  in $\Pts$, i.e., such that $\{X_{\Pts}[i]\}_{i \in [1 \dd m_x]} =
  \{x : (x,y,w,\ell) \in \Pts\}$ (resp.\ $\{Y_{\Pts}[i]\}_{i \in [1
  \dd m_y]} = \{y : (x,y,w,\ell) \in \Pts\}$).  For any array of
  integers $A[1 \dd k]$ and any $a \in \Z$, we define $\ArrayRank{A}{a}
  = |\{i \in [1 \dd k]: A[i] < a\}|$.  Let $\Pts_{\ell}$ be the set
  $\Pts$ with coordinates reduced to rank space and weights removed,
  i.e., $\Pts_{\ell} = \{(\ArrayRank{X_{\Pts}}{x} + 1,
  \ArrayRank{Y_{\Pts}}{y} + 1,\ell) : (x,y,w,\ell) \in \Pts\}$.  Let
  $\Pts_{w}$ be the set $\Pts$ with coordinates reduced to rank space
  and labels removed, i.e., $\Pts_{w} = \{(\ArrayRank{X_{\Pts}}{x} + 1,
  \ArrayRank{Y_{\Pts}}{y} + 1, w) : (x,y,w,\ell) \in \Pts\}$.  Note that
  by the assumption on $\Pts$, we have $|\Pts_{\ell}| = |\Pts_{w}| =
  m$.  Let $P_{\rm sort}[1 \dd m]$ be an array containing all elements
  of $\Pts_{\ell}$ sorted by the second coordinate and, in case of
  ties, by the first coordinate.

  \paragraph{Components}

  The structure consists of five components:
  \begin{enumerate}
  \item The array $P_{\rm sort}[1 \dd m]$ in plain form using
    $\bigO(m)$ space.
  \item The array $X_{\Pts}[1 \dd m_x]$ in plain form using
    $\bigO(m_x) = \bigO(m)$ space.
  \item The array $Y_{\Pts}[1 \dd m_y]$ in plain form using
    $\bigO(m_y) = \bigO(m)$ space.
  \item The data structure for weighted range counting
    from~\cite{chazelle} (where these queries are referred to as
    semigroup range searching) for the set $\Pts_{w}$. We use the
    variant using $\bigO(m)$ space that answers queries in
    $\bigO(\log^{2 + \epsilon} m)$ time~\cite[Table~1]{chazelle}.
    Note that as required by~\cite{chazelle}, the coordinates of
    $\Pts_w$ are in the rank space (i.e., $[1 \dd |\Pts_w|]$) and
    there are no two points equal on both coordinates.
  \item The data structure for range minimum queries
    from~\cite{chazelle} for the set $\Pts_{\ell}$. We use the variant
    using $\bigO(m)$ space that answers queries in $\bigO(\log^{1 +
    \epsilon} m)$ time~\cite[Table~1]{chazelle}.  Note that here
    again, as required by~\cite{chazelle}, the coordinates of
    $\Pts_{\ell}$ are in the rank space, and there are no two points
    equal on both coordinates.

  \end{enumerate}

  \paragraph{Implementation of queries}

  The weighted range counting queries are answered as follows.  Let
  $x_l, x_u \in \Zz$ and $y_u \in \Zz$. To compute
  $\RangeCountThreeSide{\Pts}{x_l}{x_u}{y_u}$, we first reduce the query
  coordinates to rank space by performing binary search over
  $X_{\Pts}$ and $Y_{\Pts}$ to compute $x'_l = \ArrayRank{X_{\Pts}}{x_l}
  + 1$, $x'_u = \ArrayRank{Y_{\Pts}}{x_u} + 1$, and $y'_u =
  \ArrayRank{Y_{\Pts}}{y_u} + 1$. Observe, that then it holds
  $\RangeCountThreeSide{\Pts}{x_l}{x_u}{y_u} = \sum_{(x,y,w) \in \mathcal{R}}
  w$, where $\mathcal{R} = \{(x,y,w) \in \Pts_w : x'_l \leq x <
  x'_u\text{ and }y < y'_u\}$.  Thus, using the structure
  from~\cite{chazelle} on $\Pts_w$, the computation of
  $\RangeCountThreeSide{\Pts}{x_l}{x_u}{y_u}$ takes $\bigO(\log^{2 +
  \epsilon} m)$ time. The computation of
  $\IncRangeCountThreeSide{\Pts}{x_l}{x_u}{y_u}$ is performed analogously,
  except we first increment $y_u$. Lastly, the computation of
  $\RangeCountTwoSide{\Pts}{x_l}{x_y}$ is reduced to
  $\RangeCountThreeSide{\Pts}{x_l}{x_u}{y_u}$, where $y_u = Y_{\Pts}[m_y] + 1$,
  i.e., we can immediately set $y'_u = m_y + 1$.

  To answer range selection queries with arguments $x_l, x_u \,{\in}\,
  \Zz$, and $r \in [1 \dd \RangeCountTwoSide{\Pts}{x_l}{x_u}]$, we first
  analogously reduce $x_l$ and $x_u$ in $\bigO(\log m)$ time to rank
  space coordinates $x'_l$ and $x'_u$. We then perform a binary search
  in $P_{\rm sort}[1 \dd m]$ with weighted range counting queries on
  $\Pts_w$ as an oracle to find any $(x,y_u,\ell) \in \Pts_{\ell}$
  such that $r \in (\sum_{(x,y,w) \in \mathcal{R}_1}w \dd
  \sum_{(x,y,w) \in \mathcal{R}_2}w]$, where $\mathcal{R}_1 =
  \{(x,y,w) \in \Pts_w : x'_l \leq x < x'_u\text{ and }y < y_u\}$ and
  $\mathcal{R}_2 = \{(x,y,w) \in \Pts_w : x'_l \leq x < x'_u\text{ and
  }y \leq y_u\}$.  We then continue the binary search to locate any
  $(x',y',\ell') \in \Pts_{\ell}$ such that $y' = y_u$ and $x' \in
  [x'_l \dd x'_u)$, and return $\ell' \in
  \RangeSelect{\Pts}{x_l}{x_u}{r}$ as the answer. In total, the selection
  query takes $\bigO(\log^{3 + \epsilon} m)$ time.

  To answer range minimum queries with arguments $x_l, x_u \in \Zz$
  and $y_l, y_u \in \Zz$, we first reduce all coordinates in
  $\bigO(\log m)$ time to rank space coordinates $x'_l, x'_u, y'_l,
  y'_u$. Note that then it holds $\RangeMinFourSide{\Pts}{x_l}{x_u}{y_l}{y_u} =
  \min_{(x,y,\ell) \in \mathcal{R}} \ell$, where $\mathcal{R} =
  \{(x,y,\ell) \in \Pts_{\ell} : x'_l \leq x < x'_u\text{ and }y'_l
  \leq y < y'_u\}$.  Thus, using the structure from~\cite{chazelle} on
  $\Pts_{\ell}$, the computation of $\RangeMinFourSide{\Pts}{x_l}{x_u}{y_l}{y_u}$
  takes $\bigO(\log^{1 + \epsilon} m)$ time.
  The computation of $\RangeMinTwoSide{\Pts}{x_l}{x_y}$ is reduced to
  $\RangeMinFourSide{\Pts}{x_l}{x_u}{y_l}{y_u}$, where
  $y_l = Y_{\Pts}[1]$ and $y_u = Y_{\Pts}[m_y] + 1$,
  i.e., we can immediately set $y'_l = 1$ and $y'_u = m_y + 1$.

  \paragraph{Construction algorithm}

  Given the set $\Pts$, we first construct the arrays $X_{\Pts}$ and
  $Y_{\Pts}$ by scanning $\Pts$, and then sorting the resulting arrays
  and removing the duplicates. In total, this takes $\bigO(m \log m)$
  time.  We then construct $\Pts_w$ and $\Pts_{\ell}$. To this end, we
  need to reduce the coordinates of every point in $\Pts$ to rank
  space. Overall, the construction $\bigO(m \log m)$ time. Next, we
  construct the structure for weighted range counting queries on
  $\Pts_w$ and range minimum queries on $\Pts_{\ell}$ in $\bigO(m \log
  m)$ time using the algorithm described in~\cite{chazelle}. Finally,
  the array $P_{\rm sort}$ is computed from $\Pts_{\ell}$ in $\bigO(m
  \log m)$ time.
\end{proof}

\begin{proposition}\label{pr:int-int-2}
  Let $\epsilon > 0$ be a fixed constant. Given a set $\Pts$ of $m$
  points with coordinates in $\Zz$, we can in $\bigO(m \log m)$ time
  construct a structure of size $\bigO(m)$ supporting the following
  queries on $\Pts$:
  \begin{itemize}
  \item Weighted range counting queries in $\bigO(\log^{2 + \epsilon}
    m)$ time,
  \item Weighted range selection queries in $\bigO(\log^{3 + \epsilon}
    m)$ time,
  \item Range minimum queries in $\bigO(\log^{1 + \epsilon} m)$ time.
  \end{itemize}
\end{proposition}
\begin{proof}
  Observe, that if there exists $p_1 = (x_1,y_2,w_1,\ell_1)$ and $p_2
  = (x_2,y_2,w_2,\ell_2)$ in $\Pts$ such that $p_1 \neq p_2$ and
  $(x_1, y_1) = (x_2, y_2)$ then, letting $\Pts' = \Pts \setminus
  \{p_1, p_2\} \cup \{(x_1, y_1, w_1 + w_2, \min(\ell_1, \ell_2))\}$,
  the answers to range counting and range minimum or $\Pts$ and
  $\Pts'$ are the same, and for every $x_l, x_u \in \Zz$ and $r \in [1
  \dd \RangeCountTwoSide{\Pts}{x_l}{x_u}]$, $\RangeSelect{\Pts}{x_l}{x_r}{r}
  \neq \emptyset$ holds if and only if $\RangeSelect{\Pts'}{x_l}{x_r}{r}
  \neq \emptyset$, which implies that we can use $\Pts'$ instead of
  $\Pts$ to execute range selection queries, since they return an
  arbitrary element of $\RangeSelect{\Pts}{x_l}{x_r}{r}$.  Let thus
  $\Pts_{\rm unique}$ denote a set obtained by repeatedly merging
  pairs points in $\Pts$ that are equal on both coordinates, as
  described above, until there are no such pairs. Formally, $\Pts_{\rm
  unique}$ is such that
  \begin{enumerate}
  \item For every $(x,y,w,\ell) \in \Pts$, there exists
    $(x',y',w',\ell') \in \Pts_{\rm unique}$ with $(x',y') = (x,y)$.
  \item For every $(x,y,w,\ell) \in \Pts_{\rm unique}$, it holds $w =
    \sum_{(x',y',w',\ell') \in \mathcal{R}} w'$ and $\ell =
    \min_{(x',y',w',\ell') \in \mathcal{R}} \ell'$, where $\mathcal{R}
    = \{(x',y',w',\ell') \in \Pts : (x, y) = (x', y')\}$.
  \end{enumerate}

  The data structure consists of a single component, namely, the
  structure from \cref{pr:int-int} for $\Pts_{\rm unique}$. Letting
  $m' = |\Pts_{\rm unique}|$, we have $m' \leq m$, and hence it needs
  $\bigO(m') = \bigO(m)$ space.

  Queries are answered using the structure from \cref{pr:int-int}.
  The correctness follows by the above discussion, and the time
  complexities follow by $m' \leq m$ and by \cref{pr:int-int}.

  To construct $\Pts_{\rm unique}$, we first sort in $\bigO(m \log m)$
  time the input set $\Pts$ by the first coordinate and, in case of
  ties, by the second one. The set $\Pts_{\rm unique}$ is then easily
  obtained in $\bigO(m)$ time. In $\bigO(m' \log m') = \bigO(m \log
  m)$ we then construct the structure from \cref{pr:int-int} for
  $\Pts_{\rm unique}$.
\end{proof}

\subsection{String-String Coordinates}\label{sec:str-str}

\begin{definition}\label{def:str-str}
  Let $\Text \in \Sigma^{\Textlen}$ and $P \subseteq [1 \dd \Textlen]$. For every $q \geq
  1$, we define
  \[
    \StrStrPoints{q}{P}{\Text} := \{(\revstr{\Textinf[i - q \dd i)},
    \Textinf[i \dd i + q), c(i), m(i)) : i \in P\},
  \]
  where $c(i)$ and $m(i)$ are defined as follows:
  \begin{itemize}
  \item $c(i) = |\{i' \in [1 \dd \Textlen] : \Textinf[i' - q \dd i' + q) =
    \Textinf[i - q \dd i + q)\}|$,
  \item $m(i) = \min \{i' \in [1 \dd \Textlen] : \Textinf[i' - q \dd i' + q) =
    \Textinf[i - q \dd i + q)\}$.
  \end{itemize}
\end{definition}

\begin{proposition}\label{pr:str-str}
  Let $\Text \in \Sigma^{\Textlen}$, $c = \max \Sigma$, $q \geq 1$, and $P
  \subseteq [1 \dd \Textlen]$ be a set of $|P| = p$ positions in $\Text$.  Let
  $\epsilon > 0$ be a fixed constant and assume that we can compare
  any two substrings of $\Textinf$ or $\revstr{\Textinf}$ (specified with
  their starting positions and lengths) in $\bigO(t_{\rm cmp})$
  time. There exists a data structure of size $\bigO(p)$ that,
  denoting $\Pts = \StrStrPoints{q}{P}{\Text}$ (\cref{def:str-str}), provides
  support for the following queries:
  \begin{enumerate}[leftmargin=3.5ex]
  \item\label{pr:str-str-it-1} Given any $i \in [1 \dd \Textlen]$ and $q_l,
    q_r \geq 0$, return $\RangeCountThreeSide{\Pts}{x_l}{x_u}{y_l}$ and
    $\RangeCountThreeSide{\Pts}{x_l}{x_u}{y_u}$ in $\bigO(\log^{2 +
      \epsilon} \Textlen {+} t_{\rm cmp} \log \Textlen)$ time, where $\revstr{x_l}
    \,{=}\, \Textinf[i - q_l \dd i)$, $x_u \,{=}\, x_l c^{\infty}$, $y_l
    \,{=}\, \Textinf[i \dd i + q_r)$, $y_u \,{=}\, y_l c^{\infty}$.
  \item\label{pr:str-str-it-2} Given any $i \in [1 \dd \Textlen]$, $q_l \geq
    0$, and $r \in [1 \dd \RangeCountTwoSide{\Pts}{x_l}{x_u}]$ (where
    $\revstr{x_l} = \Textinf[i - q_l \dd i)$ and $x_u = x_l c^{\infty}$),
    return any $j \in \RangeSelect{\Pts}{x_l}{x_u}{r}$ in
    $\bigO(\log^{3 + \epsilon} \Textlen + t_{\rm cmp} \log \Textlen)$ time.
  \item\label{pr:str-str-it-3} Given any $i \in [1 \dd \Textlen]$ and $q_l,
    q_r \geq 0$, return $\RangeMinFourSide{\Pts}{x_l}{x_u}{y_l}{y_u}$ in
    $\bigO(\log^{1 + \epsilon} \Textlen \,{+}\,t_{\rm cmp} \log \Textlen)$ time,
    where $\revstr{x_l} \,{=}\, \Textinf[i - q_l \dd i)$, $x_u \,{=}\,
    x_l c^{\infty}$, $y_l \,{=}\, \Textinf[i \dd i + q_r)$, $y_u
    \,{=}\, y_l c^{\infty}$.
  \end{enumerate}
  Furthermore, assuming that for any substring $S$ of $\Text$
  (specified as above), we can compute $|\OccTwo{S}{\Text}|$ and $\min
  \OccTwo{S}{\Text}$ in $\bigO(t_{\rm count})$ and $\bigO(t_{\rm minocc})$
  time (respectively), then given the values $q$, $\epsilon$, and the
  set $P$, we can construct the above data structure in $\bigO(p \cdot
  (t_{\rm cmp} \log \Textlen + t_{\rm count} + t_{\rm minocc}))$ time.
\end{proposition}
\begin{proof}

  We use the following definitions. For every $i \in [1 \dd \Textlen]$,
  denote $L(i) = \revstr{\Textinf[i - q \dd i)}$ and $R(i) = \Textinf[i \dd
  i + q)$.  Let $P_L[1 \dd p]$ (resp.\ $P_R[1 \dd p]$) be an array
  containing all positions in $P$ ordered according to the reversed
  length-$q$ left context (resp.\ according to the length-$q$ right
  context) in $\Textinf$, i.e., such that $\{P_L[i]\}_{i \in [1 \dd p]} =
  P$ (resp.\ $\{P_R[i]\}_{i \in [1 \dd p]} = P$) and that for every
  $i, i' \in [1 \dd p]$, $i < i'$ implies $L(P_L[i]) \preceq
  L(P_L[i'])$ (resp.\ $R(P_R[i]) \preceq R(P_R[i'])$). Let also $S_L[1
  \dd p]$ (resp.\ $S_R[1 \dd p]$) be an array of strings corresponding
  to $P_L$ and $P_R$, i.e., defined by $S_L[i] = L(P_L[i])$
  (resp.\ $S_R[i] = R(P_R[i])$). For any array $A[1 \dd k]$ of strings
  over alphabet $\Sigma$, and any string $Q \in \Sigma^{*}$, we define
  $\ArrayRank{A}{Q} = |\{i \in [1 \dd k]: A[i] \prec Q\}|$. We then
  define $\Pts_{\rm int}$ as $\Pts$ with coordinates reduced from
  strings to integers, i.e., $\Pts_{\rm int} = \{(\ArrayRank{S_L}{x} +
  1, \ArrayRank{S_R}{y} + 1, w, \ell) : (x,y,w,\ell) \in \Pts\}$.

  \paragraph{Components}

  The structure consists of three components:
  \begin{enumerate}
  \item The array $P_L[1 \dd p]$ in plain form using $\bigO(p)$ space.
  \item The array $P_R[1 \dd p]$ in plain form using $\bigO(p)$ space.
  \item The data structure from \cref{pr:int-int-2} for $\Pts_{\rm
    int}$ using $\bigO(|\Pts_{\rm int}|) = \bigO(|\Pts|) = \bigO(p)$
    space.
  \end{enumerate}

  \paragraph{Implementation of queries}

  \begin{enumerate}
  \item First, we compute $y'_l = \ArrayRank{S_R}{y_l} + 1$ and $y'_u =
    \ArrayRank{S_R}{y_u} + 1$.  To do this, we perform binary search
    over array $P_R$ with the help of the oracle from the claim to
    compare substrings of $\Textinf$.  This lets us obtain $y'_l$. The
    computation of $y'_u$ is slightly different since $y_u = y_l
    c^{\infty}$ is not a substring of $\Textinf$ or $\revstr{\Textinf}$.  To
    compute $y'_u$, we utilize the fact that for every substring $S$
    of $\Textinf$ it holds $S \prec y_l c^{\infty}$ if and only if $S
    \prec y_l$ or $y_l$ is a prefix of $S$.  The first condition is
    easy to check using the oracle.  The second is equivalent to
    checking that $|y_l| \leq |S|$ and $y_l = S[1 \dd |y_l|]$, which
    is again the query answered by the oracle. After computing $y'_l$
    and $y'_u$, we analogously compute $x'_l = \ArrayRank{S_L}{x_l} + 1$
    and $x'_u = \ArrayRank{S_L}{x_u} + 1$ by binary search over $P_L$.
    In total, this takes $\bigO(t_{\rm cmp} \log \Textlen)$ time.  Observe
    now that it holds $\RangeCountThreeSide{\Pts}{x_l}{x_u}{y_l} =
    \RangeCountThreeSide{\Pts_{\rm int}}{x'_l}{x'_u}{y'_l}$ and
    $\RangeCountThreeSide{\Pts}{x_l}{x_u}{y_u} = \RangeCountThreeSide{\Pts_{\rm
    int}}{x'_l}{x'_u}{y'_u}$.  Thus, using the structure from
    \cref{pr:int-int-2}, the rest of the query takes $\bigO(\log^{2 +
    \epsilon} \Textlen)$ time.  In total, we spend $\bigO(\log^{2 +
    \epsilon} \Textlen + t_{\rm cmp} \log \Textlen)$ time.
  \item First, as explained above, we compute $x'_l =
    \ArrayRank{S_L}{x_l} + 1$ and $x'_u = \ArrayRank{S_L}{x_u} + 1$ in
    $\bigO(t_{\rm cmp} \log \Textlen)$ time.  Observe that then
    $\RangeSelect{\Pts}{x_l}{x_u}{r} = \RangeSelect{\Pts_{\rm
    int}}{x'_l}{x'_u}{r}$.  Thus, using the structure from
    \cref{pr:int-int-2}, the rest of the query takes $\bigO(\log^{3 +
    \epsilon} \Textlen)$ time. In total, we spend $\bigO(\log^{3 + \epsilon} \Textlen
    + t_{\rm cmp} \log \Textlen)$ time.
  \item First, compute in $\bigO(t_{\rm cmp} \log \Textlen)$ time the values
    $x'_l = \ArrayRank{S_L}{x_l} + 1$, $x'_u = \ArrayRank{S_L}{x_u} + 1$,
    $y'_l = \ArrayRank{S_R}{y_l} + 1$, and $y'_u = \ArrayRank{S_R}{y_u} +
    1$.  Then, it holds $\RangeMinFourSide{\Pts}{x_l}{x_u}{y_l}{y_r} =
    \RangeMinFourSide{\Pts_{\rm int}}{x'_l}{x'_u}{y'_l}{y'_u}$. Thus, using the
    structure from \cref{pr:int-int-2}, the rest of the query takes
    $\bigO(\log^{1 + \epsilon} \Textlen)$ time. In total, we spend
    $\bigO(\log^{1 + \epsilon} \Textlen + t_{\rm cmp} \log \Textlen)$ time.
  \end{enumerate}

  \paragraph{Construction algorithm}

  First, we compute the arrays $P_L[1 \dd p]$ and $P_R[1 \dd p]$.  In
  each case, we first initialize the array to contain all positions in
  $P$, and then perform comparison based sorting using either $L(j)$
  or $R(j)$ as the key of each position $j$ and the oracle to compare
  any two keys. In total, this takes $\bigO(t_{\rm cmp} \cdot p \log
  p)$ time. Next, we construct the set $\Pts_{\rm int}$. For each
  position $i \in P$, we first construct a tuple $(\ArrayRank{S_L}{L(i)}
  + 1, \ArrayRank{S_R}{R(i)} + 1, c(i), m(i))$, where $c(i)$ and $m(i)$ are
  as in \cref{def:str-str}. The first two elements in the tuple are
  computed using binary search over $P_L$ and $P_R$, and the oracle
  for substring comparisons in $\bigO(t_{\rm cmp} \log \Textlen)$. The third
  and fourth elements are computed as follows.  First, we check if
  $\Textinf[i - q \dd i + q)$ overlaps $\Text[\Textlen]$ (which holds if and only
  if $i - q < 1$ or $i + q > \Textlen$). If so, then $c(i) = 1$ and $m(i) = i$.
  Otherwise, we have $c(i) = |\OccTwo{\Text[i - q \dd i + q)}{\Text}|$ and $m(i)
  = \min \OccTwo{\Text[i - q \dd i + q)}{\Text}$, which we compute in
  $\bigO(t_{\rm count} + t_{\rm minocc})$ time.  Over all points, we
  spend $\bigO(p \cdot (t_{\rm cmp} \log \Text + t_{\rm count} + t_{\rm
  minocc}))$ time. We collect all the resulting tuples in the array,
  which we then sort lexicographically in $\bigO(p \log p)$ time.
  Finally, we scan the array one last time and eliminate
  duplicates. The resulting array contains $\Pts_{\rm int}$. In
  $\bigO(p \log p)$ time we then construct the structure from
  \cref{pr:int-int-2}. In total, we spend $\bigO(p \cdot (t_{\rm cmp}
  \log \Textlen + t_{\rm count} + t_{\rm minocc}))$ time.
\end{proof}

\subsection{Integer-String Coordinates}\label{sec:int-str}

\begin{definition}\label{def:int-str}
  Let $\Text \in \Sigma^{\Textlen}$ and $P \subseteq [1 \dd \Textlen] \times \Zz$. For
  every $q \geq 1$, we define
  \[
    \IntStrPoints{q}{P}{\Text} := \{(h,
    \Textinf[i \dd i + q), c(i), m(i)) : (i, h) \in P\},
  \]
  where $c(i)$ and $m(i)$ are defined as follows:
  \begin{itemize}
  \item $c(i) = |\{i' \in [1 \dd \Textlen] : \Textinf[i' - q \dd i' + q) =
    \Textinf[i - q \dd i + q)\}|$,
  \item $m(i) = \min \{i' \in [1 \dd \Textlen] : \Textinf[i' - q \dd i' + q)
    = \Textinf[i - q \dd i + q)\}$.
  \end{itemize}
\end{definition}

\begin{proposition}\label{pr:int-str}
  Let $\Text \in \Sigma^{\Textlen}$, $c = \max \Sigma$, $q \geq 1$, and $P
  \subseteq [1 \dd \Textlen] \times \Zz$ be a set of $|P| = p$ pairs.  Let
  $\epsilon > 0$ be a fixed constant and assume that we can compare
  any two substrings of $\Textinf$ or $\revstr{\Textinf}$ (specified with
  their starting positions and lengths) in $\bigO(t_{\rm cmp})$
  time. There exists a data structure of size $\bigO(p)$ that,
  denoting $\Pts = \IntStrPoints{q}{P}{\Text}$ (\cref{def:int-str}), provides
  support for the following queries:
  \begin{enumerate}[leftmargin=3.5ex]
  \item\label{pr:int-str-it-1} Given any $i \in [1 \dd \Textlen]$, $x_l \geq
    0$, and $q_r \geq 0$, return
    $\RangeCountThreeSide{\Pts}{x_l}{\Textlen}{y_l}$ and
    $\RangeCountThreeSide{\Pts}{x_l}{\Textlen}{y_u}$ in $\bigO(\log^{2 + \epsilon} \Textlen
    {+} t_{\rm cmp} \log \Textlen)$ time, where $y_l \,{=}\, \Textinf[i \dd i +
    q_r)$ and $y_u \,{=}\, y_l c^{\infty}$. Given any $i \in [1 \dd \Textlen]$
    and $x_l \geq 0$, return the value $\RangeCountTwoSide{\Pts}{x_l}{\Textlen}$ in
    $\bigO(\log^{2 + \epsilon} \Textlen)$ time. 
  \item\label{pr:int-str-it-2} Given any $x_l \geq
    0$ and $r \in [1 \dd \RangeCountTwoSide{\Pts}{x_l}{\Textlen}]$, return an
    element $j \in \RangeSelect{\Pts}{x_l}{\Textlen}{r}$ in $\bigO(\log^{3 +
    \epsilon} \Textlen)$ time.
  \item\label{pr:int-str-it-3} Given any $i \in [1 \dd \Textlen]$, $x_l \geq
    0$, and $q_r \geq 0$, return $\RangeMinFourSide{\Pts}{x_l}{\Textlen}{y_l}{y_u}$ in
    $\bigO(\log^{1 + \epsilon} \Textlen \,{+}\,t_{\rm cmp} \log \Textlen)$ time,
    where $y_l \,{=}\, \Textinf[i \dd i + q_r)$ and $y_u \,{=}\, y_l
    c^{\infty}$. Given any $x_l \geq 0$, return $\RangeMinTwoSide{\Pts}{x_l}{\Textlen}$
    in $\bigO(\log^{1 + \epsilon} \Textlen)$ time.
  \end{enumerate}
  Furthermore, assuming that for any substring $S$ of $\Text$
  (specified as above), we can compute $|\OccTwo{S}{\Text}|$ and $\min
  \OccTwo{S}{\Text}$ in $\bigO(t_{\rm count})$ and $\bigO(t_{\rm minocc})$
  time (respectively), then given the values $q$, $\epsilon$, and the
  set $P$, we can construct the above data structure in $\bigO(p \cdot
  (t_{\rm cmp} \log \Textlen + t_{\rm count} + t_{\rm minocc}))$ time.
\end{proposition}
\begin{proof}

  The data structure and its construction is similar to
  \cref{pr:str-str} and hence here we only describe the key
  differences.  We use the following definitions.  Denote $P' = \{i
  \in [1 \dd \Textlen] : (i, h) \in P\}$ and $p' = |P'|$.  Note that $p'
  \leq p$ and $|\Pts| \leq p$, but the inequalities can be strict.
  Let $P_L[1 \dd p']$ (resp.\ $P_R[1 \dd p']$) be an array containing
  all positions in $P'$ ordered according to the left (resp.\ right)
  length-$q$ contexts in $\Textinf$. Let also $S_L[1 \dd p']$
  (resp.\ $S_R[1 \dd p']$) be an array of strings defined by $S_L[i] =
  L(P_L[i])$ (resp.\ $S_R[i] = R(P_R[i])$), where $L(\cdot)$
  (resp.\ $R(\cdot)$) is as in the proof of \cref{pr:str-str}.  We
  define $\Pts_{\rm int}$ as $\Pts$ with second coordinates reduced
  from strings to integers, i.e., $\Pts_{\rm int} = \{(x,
  \ArrayRank{S_R}{y} + 1, w, \ell) : (x,y,w,\ell) \in \Pts\}$ (where
  $\ArrayRank{\cdot}{\cdot}$ is defined as in the proof of
  \cref{pr:str-str}).

  \paragraph{Components}

  The structure consists of two components:
  \begin{enumerate}
  \item The array $P_R[1 \dd p']$ in plain form using $\bigO(p') =
    \bigO(p)$ space.
  \item The data structure from \cref{pr:int-int-2} for $\Pts_{\rm
    int}$ using $\bigO(|\Pts_{\rm int}|) = \bigO(|\Pts|) = \bigO(p)$
    space.
  \end{enumerate}

  \paragraph{Implementation of queries}

  \begin{enumerate}
  \item First, in $\bigO(t_{\rm cmp} \log \Textlen)$ time we compute $y'_l =
    \ArrayRank{S_R}{y_l} + 1$ and $y'_u = \ArrayRank{S_R}{y_u} + 1$ as in
    the proof of \cref{pr:str-str}. Then, it holds
    $\RangeCountThreeSide{\Pts}{x_l}{\Textlen}{y_l} =
    \RangeCountThreeSide{\Pts_{\rm int}}{x_l}{\Textlen}{y'_l}$ and
    $\RangeCountThreeSide{\Pts}{x_l}{\Textlen}{y_u} =
    \RangeCountThreeSide{\Pts_{\rm int}}{x_l}{\Textlen}{y'_u}$. Thus, using
    \cref{pr:int-int-2}, the rest of the query takes $\bigO(\log^{2 +
    \epsilon} \Textlen)$ time.  In total, we spend $\bigO(\log^{2 +
    \epsilon} \Textlen + t_{\rm cmp} \log \Textlen)$ time.
    To compute $\RangeCountTwoSide{\Pts}{x_l}{\Textlen}$ given $x_l \geq 0$, we
    observe that $\RangeCountTwoSide{\Pts}{x_l}{\Textlen} =
    \RangeCountTwoSide{\Pts_{\rm int}}{x_l}{\Textlen}$. Thus, using
    \cref{pr:int-int-2}, the query takes $\bigO(\log^{2 + \epsilon} \Textlen)$
    time.
  \item Note that $\RangeSelect{\Pts}{x_l}{\Textlen}{r} = \RangeSelect{\Pts_{\rm
    int}}{x_l}{\Textlen}{r}$. Thus, using the structure from
    \cref{pr:int-int-2}, the query takes $\bigO(\log^{3 + \epsilon}
    \Textlen)$ time.
  \item First, compute in $\bigO(t_{\rm cmp} \log \Textlen)$ time the values
    $y'_l = \ArrayRank{S_R}{y_l} + 1$ and $y'_u = \ArrayRank{S_R}{y_u} +
    1$.  Then, it holds $\RangeMinFourSide{\Pts}{x_l}{\Textlen}{y_l}{y_r} =
    \RangeMinFourSide{\Pts_{\rm int}}{x_l}{\Textlen}{y'_l}{y'_u}$.
    Thus, using the structure from
    \cref{pr:int-int-2}, the rest of the query takes
    $\bigO(\log^{1 + \epsilon} \Textlen)$ time.
    In total, we spend
    $\bigO(\log^{1+\epsilon} \Textlen + t_{\rm cmp} \log \Textlen)$ time.
    To compute $\RangeMinTwoSide{\Pts}{x_l}{\Textlen}$ given $x_l \geq 0$, we
    observe that $\RangeMinTwoSide{\Pts}{x_l}{\Textlen} =
    \RangeMinTwoSide{\Pts_{\rm int}}{x_l}{\Textlen}$. Thus, using
    \cref{pr:int-int-2}, the query takes $\bigO(\log^{1 + \epsilon} \Textlen)$
    time.
  \end{enumerate}

  \paragraph{Construction algorithm}

  We begin by constructing an array containing elements of $P'$. Given
  $P$, this is easily done in $\bigO(p \log p)$ time.  Next, we
  compute the arrays $P_L[1 \dd p']$ and $P_R[1 \dd p']$ as in the
  proof of \cref{pr:str-str} in $\bigO(t_{\rm cmp} \cdot p' \log p')$
  time. We then construct the set $\Pts_{\rm int}$. To this end, for
  each $(i, h) \in P$, we construct a tuple $(\ArrayRank{S_L}{L(i)} + 1,
  \ArrayRank{S_R}{R(i)} + 1, i, c(i), m(i))$, where $c(i)$ and $m(i)$ are as
  in \cref{def:int-str}. The computation is performed as in the proof
  of \cref{pr:str-str} and takes $\bigO(p \cdot (t_{\rm cmp} \log \Textlen +
  t_{\rm count} + t_{\rm minocc}))$ time. We then collect all the
  resulting tuples in the array, which we sort lexicographically in
  $\bigO(p \log p)$ time and with another scan remove the
  duplicates. The resulting array contains the set $\Pts_{\rm
  int}$. In $\bigO(p \log p)$ time we then construct the structure
  from \cref{pr:int-int-2}. In total, we spend
  $\bigO(p \cdot (t_{\rm cmp} \log \Textlen + t_{\rm count} + t_{\rm minocc}))$
  time.
\end{proof}

\section{Weighted Modular Constraint Queries}\label{sec:mod-queries}

Let $\Ints \sub \Zn \times \Zp \times \Z$ be a finite set, where each
tuple $(e,w,\ell) \in \Ints$ represents an interval $[0 \dd e]$ with
an associated nonnegative \emph{endpoint} $e$, a positive
\emph{weight} $w$, and an integer \emph{label} $\ell$. We assume that
no two intervals share the same label.

\begin{description}[style=sameline,itemsep=1ex]
\item[Weighted modular constraint counting:]
  Let $h \in \Zp$, $r \in [0 \dd h)$, and $k_1, k_2 \in \Zn$ be such
  that $k_1 \leq k_2$. We define
  \begin{itemize}
  \item $\ModCountTwoSide{\Ints}{h}{r}{k_1}{k_2} :=
    \sum_{(e,w,\ell) \in \Ints} w \cdot |\{j \in [0 \dd e] :
    j \bmod h = r\text{ and }k_1 < \lfloor \tfrac{j}{h} \rfloor \leq k_2\}|$,
  \item $\ModCountOneSide{\Ints}{h}{r}{k_2} :=
    \sum_{(e,w,\ell) \in \Ints} w \cdot |\{j \in [0 \dd e] :
    j \bmod h = r\text{ and } \lfloor \tfrac{j}{h} \rfloor \leq k_2\}|$,
  \item $\ModCountZeroSide{\Ints}{h}{r} :=
    \sum_{(e,w,\ell) \in \Ints} w \cdot |\{j \,{\in}\, [0 \dd e] :
    j \bmod h = r\}|$.
  \end{itemize}
\item[Weighted modular constraint selection:]
  Let $h \in \Zp$, $r \in [0 \dd h)$, and $c \in [1 \dd
  \ModCountZeroSide{\Ints}{h}{r}]$. We define
  $\ModSelect{\Ints}{h}{r}{c} := k$, where $k \in \Zp$ is the unique
  nonnegative integer satisfying $c \in
  (\ModCountOneSide{\Ints}{h}{r}{k-1} \dd
  \ModCountOneSide{\Ints}{h}{r}{k}]$.
\end{description}

\begin{definition}\label{def:intervals}
  Let $\Text \in \Sigma^{\Textlen}$ and $P \subseteq [1 \dd \Textlen] \times \Zn$. For
  every $q \geq 1$, we define
  \[
    \WInts{q}{P}{\Text} := \{(e,c(i),m(i)) : (i,e) \in P\},
  \]
  where $c(i)$ and $m(i)$ are defined as follows:
  \begin{itemize}
  \item $c(i) = |\{i' \in [1 \dd \Textlen] : \Textinf[i' - q \dd i' + q) =
    \Textinf[i - q \dd i + q)\}|$,
  \item $m(i) = \min \{i' \in [1 \dd \Textlen] : \Textinf[i' - q \dd i' + q) =
    \Textinf[i - q \dd i + q)\}$.
  \end{itemize}
\end{definition}

\begin{lemma}\label{lm:mod-queries-properties}
  Let $\Text \in \Sigma^{\Textlen}$, $q \geq 1$, and $P \subseteq [1 \dd \Textlen] \times
  \Zn$ be such that labels in $\Ints = \WInts{q}{P}{\Text}$ are
  unique.  For every $h \in \Zp$, $r \in [0 \dd h)$, and $k_1, k_2,
  k_3 \in \Zn$ such that $k_1 \leq k_2 \leq k_3$, it holds:
  \begin{enumerate}
  \item\label{lm:mod-queries-properties-it-1}
    $\ModCountTwoSide{\Ints}{h}{r}{k_1}{k_3} =
    \ModCountOneSide{\Ints}{h}{r}{k_3} -
    \ModCountOneSide{\Ints}{h}{r}{k_1}$,
  \item\label{lm:mod-queries-properties-it-2}
    $\ModCountTwoSide{\Ints}{h}{r}{k_1}{k_3} =
    \ModCountTwoSide{\Ints}{h}{r}{k_1}{k_2} +
    \ModCountTwoSide{\Ints}{h}{r}{k_2}{k_3}$.
  \end{enumerate}
\end{lemma}
\begin{proof}
  By \cref{def:intervals}, it follows that
  \begin{align*}
    \ModCountTwoSide{\Ints}{h}{r}{k_1}{k_3}
      &= \textstyle\sum_{(e,w,\ell) \in \Ints}w \cdot
         |\{j \in [0 \dd e] : j \bmod h = r\text{ and }
         k_1 < \lfloor \tfrac{j}{h} \rfloor \leq k_3\}|\\
      &= \textstyle\sum_{(e,w,\ell) \in \Ints} w \cdot (
         |\{j \in [0 \dd e] : j \bmod h = r\text{ and }
         \lfloor \tfrac{j}{h} \rfloor \leq k_3\}| -\\
      &  \hspace{3.0cm}
         |\{j \in [0 \dd e] : j\bmod h = r\text{ and }
         \lfloor \tfrac{j}{h} \rfloor \leq k_1\}|)\\
      &= \textstyle\sum_{(e,w,\ell) \in \Ints}w \cdot
         |\{j \in [0 \dd e] : j\bmod h = r\text{ and }
         \lfloor \tfrac{j}{h} \rfloor \leq k_3\}| -\\
      &  \hspace{0.47cm}
         \textstyle\sum_{(e,w,\ell) \in \Ints}w \cdot
         |\{j \in [0 \dd e] : j\bmod h = r\text{ and }
         \lfloor \tfrac{j}{h} \rfloor \leq k_1\}|\\
      &= \ModCountOneSide{\Ints}{h}{r}{k_3} -
         \ModCountOneSide{\Ints}{h}{r}{k_1}.
  \end{align*}
  By \cref{lm:mod-queries-properties-it-1}, we thus have
  \begin{align*}
    \ModCountTwoSide{\Ints}{h}{r}{k_1}{k_3}
      &= (\ModCountOneSide{\Ints}{h}{r}{k_2} -
         \ModCountOneSide{\Ints}{h}{r}{k_1}) +\\
      &  \hspace{0.5cm}
         (\ModCountOneSide{\Ints}{h}{r}{k_3} -
         \ModCountOneSide{\Ints}{h}{r}{k_2})\\
      &= \ModCountTwoSide{\Ints}{h}{r}{k_1}{k_2} +
         \ModCountTwoSide{\Ints}{h}{r}{k_2}{k_3}.
      \qedhere
  \end{align*}
\end{proof}

\begin{proposition}\label{pr:mod-queries}
  Let $\Text \in \Sigma^{\Textlen}$, $q \geq 1$, $h \in \Zp$, and $P \subseteq [1
  \dd \Textlen] \times [0 \dd \Textlen]$ be a set of $|P| = p$ pairs such that the
  labels in $\Ints = \WInts{q}{P}{\Text}$ are unique. Let $\epsilon > 0$
  be a fixed constant.  There exists a data structure of size
  $\bigO(p)$ that provides support for the following queries:
  \begin{enumerate}
  \item Given any $r \in [0 \dd h)$ and $k_1, k_2 \in \Zn$ satisfying
    $k_1 \leq k_2$, return the values
    $\ModCountTwoSide{\Ints}{h}{r}{k_1}{k_2}$,
    $\ModCountOneSide{\Ints}{h}{r}{k_1}$,
    and $\ModCountZeroSide{\Ints}{h}{r}$ in
    $\bigO(\log^{2 + \epsilon} \Textlen)$ time.
  \item Given any $r \in [0 \dd h)$ and $c \in [1 \dd
    \ModCountZeroSide{\Ints}{h}{r}]$, return the value
    $\ModSelect{\Ints}{h}{r}{c}$ in $\bigO(\log^{3 + \epsilon} \Textlen)$
    time.
  \end{enumerate}
  Furthermore, assuming that we can compare any two substrings of
  $\Textinf$ of $\revstr{\Textinf}$ (specified with their starting positions
  and lengths) in $\bigO(t_{\rm cmp})$ time and that for any substring
  $S$ of $\Textinf$ (specified as above), we can compute $|\OccTwo{S}{\Text}|$
  and $\min \OccTwo{S}{\Text}$ in $\bigO(t_{\rm count})$ and $\bigO(t_{\rm
  minocc})$ time (respectively), then given the values $q$, $h$, and
  $\epsilon$, and the set $P$, we can construct the above data
  structure in $\bigO(p \cdot (t_{\rm cmp} \log \Textlen + t_{\rm count} +
  t_{\rm minocc}))$ time.
\end{proposition}
\begin{proof}
  Let $\Ints = \{(e_1, w_1, \ell_1), \dots, (e_{p'}, w_{p'},
  \ell_{p'})\}$, where $p' = |\Ints|$ and $e_1 \leq e_2 \leq \dots
  \leq e_{p'}$. We also define $(e_0, w_0) = (0, 0)$.  Note that $p'
  \leq p$. Note also that by \cref{def:intervals}, for every $(i, e),
  (i', e') \in P$, $i = i'$ implies that labels of the corresponding
  points in $\Ints$ are equal. Thus, the assumption about unique
  labels implies $p \leq \Textlen$, and hence $p' \leq \Textlen$.  For any $i \in [0
  \dd p']$, denote $x_i = e_i \bmod h$ and $y_i = \lfloor
  \tfrac{e_i}{h} \rfloor$. Let $Y[0 \dd p']$, $S_{\rm pref}[0 \dd
  p']$, and $S_{\rm suf}[0 \dd p']$ be integer arrays defined such
  that for $i \in [0 \dd p']$,
  \begin{itemize}
  \item $Y[i] = y_i$,
  \item $S_{\rm pref}[i] = \sum_{j=0}^{i} w_j \cdot y_j$,
  \item $S_{\rm suf}[i] = \sum_{j=i+1}^{p'} w_j$.
  \end{itemize}
  Finally, let $\Pts = \{(x_i, y_i, w_i, \ell_i) : i \in [1 \dd
    p']\}$.

  \paragraph{Components}

  The structure consists of two components:
  \begin{enumerate}
  \item The arrays $Y[0 \dd p']$, $S_{\rm pref}[0 \dd p']$, and
    $S_{\rm suf}[0 \dd p']$ in plain form using $\bigO(p') = \bigO(p)$
    space.
  \item The data structure from \cref{pr:int-int-2} for the set
    $\Pts$. It needs $\bigO(p') = \bigO(p)$ space.
  \end{enumerate}

  \paragraph{Implementation of queries}

  \begin{enumerate}
  \item To compute $\ModCountOneSide{\Ints}{h}{r}{k_1}$, we proceed
    as follows.  First, using binary search over the array $Y$, in
    $\bigO(\log p') = \bigO(\log \Textlen)$ time we compute $j = \max \{i \in
    [0 \dd p'] : Y[i] \leq k_1\}$. Letting $\mathcal{J} = \{i \in [1
      \dd p'] : r \leq x_i < h\text{ and }0 \leq y_i < k_1 + 1\}$, we
    then have:
    \begin{align*}
      \ModCountOneSide{\Ints}{h}{r}{k_1}
        &= \textstyle\sum_{i=0}^{p'} w_i \cdot \min(k_1, y_i) +
           \textstyle\sum_{i \in \mathcal{J}} w_i\\
        &= \textstyle\sum_{i=0}^{j} w_i \cdot y_i +
           \textstyle\sum_{i=j+1}^{p'} w_i \cdot k_1 +
           \textstyle\sum_{i \in \mathcal{J}} w_i\\
        &= S_{\rm pref}[j] + k_1 \cdot S_{\rm suf}[j] +
           \RangeCountFourSide{\Pts}{r}{h}{0}{k_1+1}.
    \end{align*}
    The last term is computed in $\bigO(\log^{2 + \epsilon} p') =
    \bigO(\log^{2 + \epsilon} \Textlen)$ time using \cref{pr:int-int-2}.  In
    total, we thus spend $\bigO(\log^{2 + \epsilon} \Textlen)$ time.  The
    computation of $\ModCountTwoSide{\Ints}{h}{r}{k_1}{k_2}$ is
    reduced to $\ModCountOneSide{\Ints}{h}{r}{k_2} -
    \ModCountOneSide{\Ints}{h}{r}{k_1}$
    (\cref{lm:mod-queries-properties}\eqref{lm:mod-queries-properties-it-1}),
    and thus also takes $\bigO(\log^{2 + \epsilon} \Textlen)$ time.  To
    compute $\ModCountZeroSide{\Ints}{h}{r}$, we proceed as above,
    except we immediately set $j = p'$.
  \item The value $\ModSelect{\Ints}{h}{r}{c}$ is computed using
    binary search (in the range $[0 \dd \Textlen]$) and modular rank queries
    (implemented as above). Thus, the query takes $\bigO(\log^{3 +
    \epsilon} \Textlen)$ time.
  \end{enumerate}

  \paragraph{Construction algorithm}

  First, for each pair $(i, e) \in P$, we construct a tuple $(e, c(i),
  m(i))$, where $c(i)$ and $m(i)$ are as in \cref{def:intervals}.  The
  computation is performed as in the proof of \cref{pr:str-str} and
  takes $\bigO(p \cdot (t_{\rm cmp} \log \Textlen + t_{\rm count} + t_{\rm
  minocc}))$ time. We then collect all the resulting tuples in the
  array, which we sort lexicographically in $\bigO(p \log p)$ time and
  with another scan remove the duplicates. The resulting array
  contains the set $\Ints$ (sorted by the first coordinate) and is of
  size $p'$.  Using this array we compute the array $S$ in $\bigO(p')$
  time.  Lastly, in $\bigO(p')$ time we construct the set $\Pts$, and
  then in $\bigO(p' \log p') = \bigO(p \log p)$ time we construct the
  structure from \cref{pr:int-int-2} for $\Pts$. In total, we spend
  $\bigO(p \cdot (t_{\rm cmp} \log \Textlen + t_{\rm count} + t_{\rm
    minocc}))$ time.
\end{proof}

\section{Optimal Compressed Space SA and ISA Queries}\label{sec:sa}

For the duration of this section, we fix some $\Text \in
\Sigma^{\Textlen}$ (where $\Sigma = \IntegerAlphabet$), such that
$\Text[\Textlen]$ does not occur in $\Text[1 \dd \Textlen)$.  Let
$\epsilon \in (0, 1)$ be any fixed constant. We describe a data
structure of size $\bigO(\SubstringComplexity{\Text} \log
\tfrac{\Textlen \log \sigma}{\SubstringComplexity{\Text} \log
\Textlen})$ that given any $j \in [1 \dd \Textlen]$ (resp.\ $i \in [1
\dd \Textlen]$) returns the value $\ISA[j]$ (resp.\ $\SA[i])$ in
$\bigO(\log^{4 + \epsilon} \Textlen)$ time. Moreover, we show that
given an LZ77 parsing of $\Text$, we can construct our data structure
in $\bigO(\SubstringComplexity{\Text} \log^7 \Textlen)$ time.

\subsection{Preliminaries}\label{sec:sa-prelim}

\begin{definition}[$\tau$-periodic and $\tau$-nonperiodic patterns]\label{def:periodic-pattern}
  Let $\Pat \in \Sigma^{m}$ and $\tau \geq 1$. We say that $\Pat$ is
  \emph{$\tau$-periodic} if it holds $m \geq 3\tau - 1$ and
  $\per(\Pat[1 \dd 3\tau - 1]) \leq \tfrac{1}{3}\tau$. Otherwise, it
  is called \emph{$\tau$-nonperiodic}.
\end{definition}

\begin{definition}[Inverted lexicographic order]\label{def:inv}
  By $\preceq_{\rm inv}$ we denote the inverted order on
  elements of $\Sigma$, i.e., $a \preceq_{\rm inv} b$ if and only if
  $b \preceq a$ for $a, b \in \Sigma$. We then extend $\preceq_{\rm inv}$ to
  \emph{inverted lexicographic order} analogously to how $\preceq$ is extended
  to lexicographic order in \cref{sec:prelim}.
\end{definition}

\begin{remark}\label{rm:lex-order-inv}
  Note that $\preceq_{\rm inv}$ in \cref{def:inv} is not the same as $\succeq$.
  For example, assuming $\Sigma = \{\texttt{a}, \texttt{b}, \texttt{c}\}$ with
  $\texttt{a} \prec \texttt{b} \prec \texttt{c}$, it holds
  $\texttt{ab} \preceq \texttt{abc}$ and $\texttt{ab} \preceq_{\rm inv} \texttt{abc}$.
\end{remark}

We now present equivalent characterizations of $\RangeBegThree{\ell}{\Pat}{\Text}$, $\PosBeg{\ell}{\Pat}{\Text}$, and $\PosEnd{\ell}{\Pat}{\Text}$. 
Although less intuitive than the original \cref{def:occ,def:pat-posless}, they are useful in many contexts.
\begin{lemma}\label{lm:equiv}
  Let $\Pat \in \Sigma^{*}$ be such that no suffix of $\Text$ is a proper prefix of $\Pat$.
  Then, the following holds for every integer $\ell\ge 0$.
  \begin{align*} 
    \RangeBegThree{\ell}{\Pat}{\Text} &=|\{j' \in [1 \dd \Textlen] : \Text[j' \dd \Textlen] \prec \Pat\text{ and }
  \lcp(\Pat, \Text[j' \dd \Textlen]) < \ell\}|,\\
  \PosBeg{\ell}{\Pat}{\Text}
  &= \{j' \in [1 \dd \Textlen] : \Text[j' \dd \Textlen] \prec \Pat\text{ and }
     \lcp(\Pat, \Text[j' \dd \Textlen]) \in [\ell \dd 2\ell)\},\\
\PosEnd{\ell}{\Pat}{\Text}
  &= \{j' \in [1 \dd \Textlen] : \Text[j' \dd \Textlen] \prec_{\rm inv}\Pat\text{ and }
     \lcp(\Pat, \Text[j' \dd \Textlen]) \in [\ell \dd 2\ell)\}.
\end{align*}
\end{lemma}
\begin{proof}
Recall that $\RangeBegThree{\ell}{\Pat}{\Text}=|\{j'\in [1\dd \Textlen] : \Text[j'\dd \Textlen] \prec \Pat\text{ and }j'\notin \OccThree{\ell}{\Pat}{\Text}\}|$
and consider a position $j'$ contributing to the right-hand side. By definition of $\OccThree{\ell}{\Pat}{\Text}$, we have 
$\lcp(\Pat, \Text[j'\dd \Textlen])< \min(|\Pat|,\ell)<\ell$. Consequently,
$\RangeBegThree{\ell}{\Pat}{\Text} \le |\{j' \in [1 \dd \Textlen] : \Text[j' \dd \Textlen] \prec \Pat\text{ and }
\lcp(\Pat, \Text[j' \dd \Textlen]) < \ell\}|$.
For a proof of the converse inequality, consider $j'\in [1\dd \Textlen]$ such that $\Text[j' \dd \Textlen] \prec \Pat$ and $\lcp(\Pat, \Text[j' \dd \Textlen]) < \ell$.
The former condition implies that $\Pat$ is not a prefix of $\Text[j' \dd \Textlen]$, and thus $\lcp(\Pat, \Text[j' \dd \Textlen]) < |\Pat|$.
Together with the latter condition, this yields $\lcp(\Pat, \Text[j'\dd \Textlen])< \min(|\Pat|,\ell)$ and, in turn, $j'\notin \OccThree{\ell}{\Pat}{\Text}$.
Consequently, $|\{j' \in [1 \dd \Textlen] : \Text[j' \dd \Textlen] \prec \Pat\text{ and }
\lcp(\Pat, \Text[j' \dd \Textlen]) < \ell\}| \le \RangeBegThree{\ell}{\Pat}{\Text}$.

Recall that $\PosBeg{\ell}{\Pat}{\Text}=\{j'\in \OccThree{\ell}{\Pat}{\Text} : \Text[j'\dd \Textlen] \prec \Pat\text{ and }j'\notin \OccThree{2\ell}{\Pat}{\Text}\}$
and consider a position $j'\in \PosBeg{\ell}{\Pat}{\Text}$. Due to $j'\in \OccThree{\ell}{\Pat}{\Text}\setminus \OccThree{2\ell}{\Pat}{\Text}$,
we have $\min(|\Pat|,\ell) \le \lcp(\Pat,\Text[j'\dd \Textlen])< \min(|\Pat|,2\ell)$. 
In particular, $\min(|\Pat|,\ell) < \min(|\Pat|,2\ell)$ implies $|\Pat|\ge \ell$, so we derive 
$\ell = \min(|\Pat|,\ell) \le \lcp(\Pat,\Text[j'\dd \Textlen])< \min(|\Pat|,2\ell)\le 2\ell$.
Consequently, $\PosBeg{\ell}{\Pat}{\Text} \sub \{j' \in [1 \dd \Textlen] : \Text[j' \dd \Textlen] \prec \Pat\text{ and } \lcp(\Pat, \Text[j' \dd \Textlen]) \in [\ell \dd 2\ell)\}$.
For a proof of the converse inclusion, consider $j'\in [1\dd \Textlen]$ such that $\Text[j' \dd \Textlen] \prec \Pat$ and $\lcp(\Pat, \Text[j' \dd \Textlen]) \in [\ell\dd 2\ell)$.
In particular, $\lcp(\Pat, \Text[j' \dd \Textlen]) \ge \ell \ge \min(|\Pat|,\ell)$, so $j'\in \OccThree{\ell}{\Pat}{\Text}$.
At the same time, $\Text[j' \dd \Textlen] \prec \Pat$ implies that $\Pat$ is not a prefix of $\Text[j' \dd \Textlen]$,
so $\lcp(\Pat, \Text[j' \dd \Textlen]) <|\Pat|$. Combined with $\lcp(\Pat, \Text[j' \dd \Textlen]) < 2\ell$, this implies $j'\notin \OccThree{2\ell}{\Pat}{\Text}$.
Consequently, $\{j' \in [1 \dd \Textlen] : \Text[j' \dd \Textlen] \prec \Pat\text{ and }
\lcp(\Pat, \Text[j' \dd \Textlen]) \in [\ell \dd 2\ell)\}\sub \PosBeg{\ell}{\Pat}{\Text}$.

Recall that $\PosEnd{\ell}{\Pat}{\Text}=\{j'\in \OccThree{\ell}{\Pat}{\Text} : \Text[j'\dd \Textlen] \succ \Pat\text{ and }j'\notin \OccThree{2\ell}{\Pat}{\Text}\}$
and consider a position $j'\in \PosEnd{\ell}{\Pat}{\Text}$. Due to $j'\in \OccThree{\ell}{\Pat}{\Text}\setminus \OccThree{2\ell}{\Pat}{\Text}$,
we have $\min(|\Pat|,\ell) \le \lcp(\Pat,\Text[j'\dd \Textlen])< \min(|\Pat|,2\ell)$. 
In particular, $\min(|\Pat|,\ell) < \min(|\Pat|,2\ell)$ implies $|\Pat|\ge \ell$, so we derive 
$\ell = \min(|\Pat|,\ell) \le \lcp(\Pat,\Text[j'\dd \Textlen])< \min(|\Pat|,2\ell)\le 2\ell$.
Moreover, $\lcp(\Pat,\Text[j'\dd \Textlen])< |\Pat|$ means that $\Pat$ is not a prefix of $\Text[j'\dd \Textlen]$,
and thus $\Text[j'\dd \Textlen] \succ \Pat$ implies $\Text[j'\dd \Textlen] \prec_{\rm inv} \Pat$.
Consequently, $\PosEnd{\ell}{\Pat}{\Text} \sub \{j' \in [1 \dd \Textlen] : \Text[j' \dd \Textlen] \prec_{\rm inv} \Pat\text{ and } \lcp(\Pat, \Text[j' \dd \Textlen]) \in [\ell \dd 2\ell)\}$.
For a proof of the converse inclusion, consider $j'\in [1\dd \Textlen]$ such that $\Text[j' \dd \Textlen] \prec_{\rm inv} \Pat$ and $\lcp(\Pat, \Text[j' \dd \Textlen]) \in [\ell\dd 2\ell)$.
In particular, $\lcp(\Pat, \Text[j' \dd \Textlen]) \ge \ell \ge \min(|\Pat|,\ell)$, so $j'\in \OccThree{\ell}{\Pat}{\Text}$.
At the same time, $\Text[j' \dd \Textlen] \prec_{\rm inv} \Pat$ implies that $\Pat$ is not a prefix of $\Text[j' \dd \Textlen]$,
so $\lcp(\Pat, \Text[j' \dd \Textlen]) <|\Pat|$. Combined with $\lcp(\Pat, \Text[j' \dd \Textlen]) < 2\ell$, this implies $j'\notin \OccThree{2\ell}{\Pat}{\Text}$.
Moreover, $\Text[j' \dd \Textlen] \prec_{\rm inv} \Pat$ implies $\Text[j' \dd \Textlen] \succ \Pat$
because $\Text[j' \dd \Textlen]$ is not a proper prefix of $\Pat$ (by the assumption in the lemma statement).
Consequently, $\{j' \in [1 \dd \Textlen] : \Text[j' \dd \Textlen] \prec_{\rm inv} \Pat\text{ and }
\lcp(\Pat, \Text[j' \dd \Textlen]) \in [\ell \dd 2\ell)\}\sub \PosEnd{\ell}{\Pat}{\Text}$.
\end{proof}

\begin{remark}\label{rm:posbeg-posend-equiv}
  We characterize $\PosEnd{\ell}{\Pat}{\Text}$ with
  the help of $\prec_{\rm inv}$, since due to the asymmetry of
  lexicographic order, $\{j' \in [1 \dd \Textlen] : \Text[j' \dd \Textlen] \succ
  \Pat\text{ and }\lcp(\Pat, \Text[j' \dd \Textlen]) \in [\ell \dd 2\ell)\}$ may
  contain elements of $\OccTwo{\Pat}{\Text}$.
\end{remark}

\begin{lemma}\label{lm:pat-posless}
  Let $\ell \geq 1$ and $\Pat \in \Sigma^{+}$. Then,
  $\RangeBegThree{2\ell}{\Pat}{\Text} = \RangeBegThree{\ell}{\Pat}{\Text} +
  \DeltaBeg{\ell}{\Pat}{\Text}$.
\end{lemma}
\begin{proof}
  By combining \cref{def:pat-posless,lm:equiv}, we have $\RangeBegThree{2\ell}{\Pat}{\Text} =
  |\{j' \in [1 \dd \Textlen] : \Text[j' \dd \Textlen] \prec \Pat\text{ and
  }\lcp(\Pat, \Text[j' \dd \Textlen]) < 2\ell\}| = |\{j' \in [1 \dd \Textlen] : \Text[j'
  \dd \Textlen] \prec \Pat\text{ and }\lcp(\Pat, \Text[j' \dd \Textlen]) < \ell\}|
  + |\{j' \in [1 \dd \Textlen] : \Text[j' \dd \Textlen] \prec \Pat\text{ and
  }\lcp(\Pat, \Text[j' \dd \Textlen]) \in [\ell \dd 2\ell)\}| =
  \RangeBegThree{\ell}{\Pat}{\Text} + \DeltaBeg{\ell}{\Pat}{\Text}$.
\end{proof}

\begin{lemma}\label{lm:pos-posless}
  Let $\ell \geq 1$ and $j \in [1 \dd \Textlen]$. Then,
  $\RangeBegThree{2\ell}{j}{\Text} = \RangeBegThree{\ell}{j}{\Text} + \DeltaBeg{\ell}{j}{\Text}$.
\end{lemma}
\begin{proof}
  Denote $\Pat = \Text[j \dd \Textlen]$. Note that, by definition, it holds
  $\RangeBegThree{\ell}{j}{\Text} = \RangeBegThree{\ell}{\Pat}{\Text}$, $\RangeBegThree{2\ell}{j}{\Text} =
  \RangeBegThree{2\ell}{\Pat}{\Text}$ (\cref{def:occ}), and $\DeltaBeg{\ell}{j}{\Text}
  = \DeltaBeg{\ell}{\Pat}{\Text}$ (\cref{def:pat-posless}). Thus, the
  claim follows by \cref{lm:pat-posless}.
\end{proof}

\begin{lemma}\label{lm:sa-prelim}
  Let $\Pat \in \Sigma^{*}$ and $j \in [1 \dd \Textlen]$.
  Denote $m = |\Pat|$. Then:
  \begin{enumerate}
  \item\label{lm:sa-prelim-it-1}
    For $0 \leq \ell_1$, $\Pat[1 \dd \min(m,\ell_1)] \preceq \Text[j \dd
    \Textlen]$ if and only if $\Text[j \dd \Textlen] \succeq \Pat$ or $\lcp(\Pat, \Text[j \dd
    \Textlen]) \geq \ell_1$.
  \item\label{lm:sa-prelim-it-2}
    For $0 < \ell_2$, $\Text[j \dd \Textlen] \prec \Pat[1 \dd \min(m,\ell_2)]$
    if and only if $\Text[j \dd \Textlen] \prec \Pat$ and $\lcp(\Pat, \Text[j \dd \Textlen])
    < \ell_2$.
  \item\label{lm:sa-prelim-it-3}
    For $0 \leq \ell_1 < \ell_2$, $\Pat[1 \dd \min(m,\ell_1)] \preceq
    \Text[j \dd \Textlen] \prec \Pat[1 \dd \min(m,\ell_2)]$ if and only if $\Text[j \dd
    \Textlen] \prec \Pat$ and $\lcp(\Pat, \Text[j \dd \Textlen]) \in [\ell_1 \dd
    \ell_2)$.
  \end{enumerate}
\end{lemma}
\begin{proof}
  1. Denote $\Pat_1 = \Pat[1 \dd \min(m, \ell_1)]$ and $\ell' =
  \lcp(\Pat, \Text[j \dd \Textlen])$. Let us first assume $\Pat_1 \preceq \Text[j
  \dd \Textlen]$. If $m \leq \ell_1$, then $|\Pat_1| = \min(m, \ell_1) =
  m$, and hence $\Pat_1 = \Pat$. From the assumption we thus have
  $\Pat = \Pat_1 \preceq \Text[j \dd \Textlen]$. Let us thus assume $m >
  \ell_1$. Then, $|\Pat_1| = \ell_1$, and hence $\Pat_1 = \Pat[1 \dd
  \ell_1]$. The assumption $\Pat_1 \preceq \Text[j \dd \Textlen]$ implies
  that either $\Pat_1$ is a prefix of $\Text[j \dd \Textlen]$, or it holds $1 +
  \ell' \leq \ell_1$, $j + \ell' \leq \Textlen$, $\Pat_1[1 \dd \ell'] = \Text[j
  \dd j + \ell')$, and $\Pat_1[1 + \ell'] \prec \Text[j + \ell']$.
  Consider two cases.
  \begin{itemize}
  \item In the first case, we immediately obtain $\ell' = |\Pat_1| =
    \ell_1$.
  \item In the second case, since $\Pat_1$ is a prefix of $\Pat$, we
    obtain $\Pat[1 \dd \ell'] = \Pat_1[1 \dd \ell'] = \Text[j \dd j +
    \ell')$ and $\Pat[1 + \ell'] = \Pat_1[1 + \ell'] \prec \Text[j +
    \ell']$. This implies $\Pat \prec \Text[j \dd \Textlen]$.
  \end{itemize}

  Let us now assume that it holds $\Text[j \dd \Textlen] \succeq \Pat$ or
  $\ell' \geq \ell_1$. We consider both cases as follows:
  \begin{itemize}
  \item If $\Pat \preceq \Text[j \dd \Textlen]$, then $\Pat_1 \preceq \Pat
    \preceq \Text[j \dd \Textlen]$ follows, since $\Pat_1$ is a prefix of
    $\Pat$.
  \item Let us now assume $\ell' \geq \ell_1$. First, note that $m
    \geq \ell' \geq \ell_1$. Thus, $|\Pat_1| = \min(m, \ell_1) =
    \ell_1$, i.e., $\Pat_1 = \Pat[1 \dd \ell_1]$.  On the other hand,
    the assumption $\ell' \geq \ell_1$ implies $\Pat[1 \dd \ell_1] =
    \Text[j \dd j + \ell_1)$.  We thus obtain $\Pat_1 = \Pat[1 \dd \ell_1]
    = \Text[j \dd j + \ell_1) \preceq \Text[j \dd \Textlen]$.
  \end{itemize}

  2. Denote $\Pat_2 = \Pat[1 \dd \min(m, \ell_2)]$ and $\ell' =
  \lcp(\Pat, \Text[j \dd \Textlen])$.  Let us first assume $\Text[j \dd \Textlen] \prec
  \Pat_2$. We prove the two claims as follows:
  \begin{itemize}
  \item Since $\Pat_2$ is a prefix of $\Pat$, we immediately obtain
    $\Text[j \dd \Textlen] \prec \Pat_2 \preceq \Pat$.
  \item To show $\ell' < \ell_2$, suppose that we have $\ell' \geq
    \ell_2$. First, note that letting $\Pat_3 = \Pat[1 \dd \ell']$ we
    have $\Pat_3 = \Text[j \dd j + \ell') \preceq \Text[j \dd \Textlen]$.  On the
    other hand, by $|\Pat_2| = \min(m, \ell_2) \leq \ell_2 \leq
    \ell'$, we then also have $\Pat_2 \preceq \Pat_3$. Combining the
    two we thus obtain $\Pat_2 \preceq \Pat_3 \preceq \Text[j \dd \Textlen]$, a
    contradiction.
  \end{itemize}

  Let us now assume that $\Text[j \dd \Textlen] \prec \Pat$ and $\ell' <
  \ell_2$.  The assumption $\Text[j \dd \Textlen] \prec \Pat$ implies that
  either $\Text[j \dd \Textlen]$ is a proper prefix of $\Pat$, or $j + \ell'
  \leq \Textlen$, $1 + \ell' \leq m$, $\Text[j \dd j + \ell') = \Pat[1 \dd
  \ell']$, and $\Text[j + \ell'] \prec \Pat[1 + \ell']$. Consider two
  cases:
  \begin{itemize}
  \item First, assume that $\Text[j \dd \Textlen]$ is a proper prefix of
    $\Pat$.  This implies that $\ell' = \Textlen - j + 1$ and $m > \ell'$.
    On the other hand, we assumed $\ell' < \ell_2$. Thus, $|\Pat_2| =
    \min(m, \ell_2) \geq \ell' + 1$, and hence $\Text[j \dd \Textlen] = \Pat[1
    \dd \ell'] \prec \Pat[1 \dd \ell' + 1] \preceq \Pat_2$.
  \item Let us now assume that $j + \ell' \leq \Textlen$, $1 + \ell' \leq
    m$, $\Text[j \dd j + \ell') = \Pat[1 \dd \ell']$, and $\Text[j + \ell']
    \prec \Pat[1 + \ell']$. We first observe that combining $1 + \ell'
    \leq m$ with the assumption $\ell' < \ell_2$ yields $|\Pat_2| =
    \min(m, \ell_2) \geq \ell' + 1$. In particular, $\Pat[1 \dd \ell'
    + 1] \preceq \Pat_2$.  On the other hand, by the assumption, we
    have $\Text[j \dd \Textlen] \prec \Pat[1 \dd \ell' + 1]$.  Putting the two
    observations together we thus obtain $\Text[j \dd \Textlen] \prec \Pat_2$.
  \end{itemize}

  3. Denote $\Pat_1 = \Pat[1 \dd \min(m, \ell_1)]$, $\Pat_2 = \Pat[1
  \dd \min(m, \ell_2)]$, and $\ell' = \lcp(\Pat, \Text[j \dd \Textlen])$.
  Let us first assume $\Pat_1 \preceq \Text[j \dd \Textlen] \prec \Pat_2$.  By
  \cref{lm:sa-prelim}\eqref{lm:sa-prelim-it-2}, it then follows that
  $\Text[j \dd \Textlen] \prec \Pat$ and $\ell' < \ell_2$. By subsequently
  applying \cref{lm:sa-prelim}\eqref{lm:sa-prelim-it-1}, we thus
  obtain that it must additionally hold $\ell' \geq \ell_1$. Putting
  everything together, we obtain the claim.  Let us now assume that
  $\Text[j \dd \Textlen] \prec \Pat$ and $\ell' \in [\ell_1 \dd \ell_2)$. The
  claim then follows immediately by
  \cref{lm:sa-prelim}\eqref{lm:sa-prelim-it-1} and
  \cref{lm:sa-prelim}\eqref{lm:sa-prelim-it-2}.
\end{proof}

\begin{lemma}\label{lm:sa-prelim-inf}
  Let $\Pat_1, \Pat_2 \in \Sigma^{*}$ be such that no nonempty
  suffix of $\Text$ is a proper prefix of $\Pat_1$ or $\Pat_2$. Let $j \in
  [1 \dd \Textlen]$.  Then:
  \begin{enumerate}
  \item\label{lm:sa-prelim-inf-it-1}
    For every $\ell \geq |\Pat_1|$, $\Pat_1 \preceq \Text[j \dd \Textlen]$
    if and only if  $\Pat_1 \preceq \Textinf[j \dd j + \ell)$.
  \item\label{lm:sa-prelim-inf-it-2}
    For every $\ell \geq |\Pat_2|$, $\Text[j \dd \Textlen] \prec \Pat_2$
    if and only if $\Textinf[j \dd j + \ell) \prec \Pat_2$.
  \item\label{lm:sa-prelim-inf-it-3}
    For every $\ell \geq \max(|\Pat_1|, |\Pat_2|)$, $\Pat_1 \preceq
    \Text[j \dd \Textlen] \prec \Pat_2$ if and only if $\Pat_1 \preceq
    \Textinf[j \dd j + \ell) \prec \Pat_2$.
  \end{enumerate}
\end{lemma}
\begin{proof}
  1. Assume $\Pat_1 \preceq \Text[j \dd \Textlen]$. Then, either $\Pat_1$ is a
  prefix of $\Text[j \dd \Textlen]$, or there exists $\ell' \geq 0$ such that
  $\ell' < |\Pat_1|$, $j + \ell' \leq \Textlen$, $\Pat_1[1 \dd \ell'] = \Text[j
  \dd j + \ell')$, and $\Pat_1[1 + \ell'] \prec \Text[j + \ell']$.  We
  consider each case separately:
  \begin{itemize}
  \item If $\Pat_1$ is a prefix of $\Text[j \dd \Textlen]$, then $j + |\Pat_1|
    \leq \Textlen + 1$, and hence $\Pat_1 = \Text[j \dd j + |\Pat_1|) =
    \Textinf[j \dd j + |\Pat_1|) \preceq \Textinf[j \dd j +
    \ell)$.
  \item Let us now assume that there exists $\ell' \geq 0$ such that
    $\ell' < |\Pat_1|$, $j + \ell' \leq \Textlen$, $\Pat_1[1 \dd \ell'] =
    \Text[j \dd j + \ell')$, and $\Pat_1[1 + \ell'] \prec \Text[j +
    \ell']$. By definition of the lexicographic order, this implies
    $\Pat_1 \prec \Text[j \dd j + \ell')$, and hence we obtain $\Pat_1
    \prec \Text[j \dd j + \ell') = \Textinf[j \dd j + \ell') \preceq
    \Textinf[j \dd j + \ell)$.
  \end{itemize}

  Let us now assume that for some $\ell \geq |\Pat_1|$, it holds
  $\Pat_1 \preceq \Textinf[j \dd j + \ell)$. Consider two cases:
  \begin{itemize}
  \item First, assume $j + \ell \leq \Textlen + 1$. Then, we obtain $\Pat_1
    \preceq \Textinf[j \dd j + \ell) = \Text[j \dd j + \ell) \preceq \Text[j
    \dd \Textlen]$.
  \item Let us now assume $j + \ell > \Textlen + 1$. Denote $\ell' =
    \lcp(\Pat_1, \Textinf[j \dd j + \ell))$.  Since we assumed that
    no nonempty suffix of $\Text$ is a proper prefix of $\Pat_1$, it must
    hold $j + \ell' \leq \Textlen$. By definition of the lexicographic
    order, $\Pat_1 \preceq \Textinf[j \dd j + \ell)$ then implies
    $\Pat_1[1 + \ell'] \prec \Text[j + \ell']$. Consequently, $\Pat_1
    \prec \Text[j \dd j + \ell'] \preceq \Text[j \dd \Textlen]$.
  \end{itemize}

  2. Assume $\Text[j \dd \Textlen] \prec \Pat_2$. Observe that $\Text[j \dd \Textlen]$
  not a prefix of $\Pat_2$, since otherwise we either have $\Text[j \dd
  \Textlen] = \Pat_2$, or some nonempty suffix of $\Text$ is a proper prefix
  of $\Pat_2$. Consequently, letting $\ell' = \lcp(\Text[j \dd \Textlen],
  \Pat_2)$, it holds $j + \ell' \leq \Textlen$, $\ell' < |\Pat_2|$, and
  $\Text[j + \ell'] \prec \Pat_2[1 + \ell']$.  Since $\Text[j \dd j + \ell']$
  is a prefix of $\Textinf[j \dd j + \ell)$, it thus follows by
  definition of the lexicographic order that $\Textinf[j \dd j +
  \ell) \prec \Pat_2[1 \dd 1 + \ell'] \preceq \Pat_2$.

  Let us now assume $\Textinf[j \dd j + \ell) \prec \Pat_2$.  Denote
  $\ell' = \lcp(\Textinf[j \dd j + \ell), \Pat_2)$. Observe that it
  holds $\ell' < |\Pat_2|$ and $j + \ell' \leq \Textlen$. The inequality
  $\ell' < |\Pat_2|$ holds since otherwise we would have $\Textinf[j
  \dd j + \ell) \succeq \Pat_2$. The inequality $j + \ell' \leq \Textlen$
  then holds since otherwise $\Text[j \dd \Textlen]$ would be a nonempty suffix
  of $\Text$ that is a proper prefix of $\Pat_2$. Consequently, by $\ell'
  < |\Pat_2| \leq \ell$, we must have $\Text[j + \ell'] \prec \Pat_2[1 +
  \ell']$. By definition of the lexicographic order, this implies
  $\Text[j \dd \Textlen] \prec \Pat_2[1 \dd 1 + \ell'] \preceq \Pat_2$.

  3. The proof follows by combining the above two equivalences.
\end{proof}

\begin{lemma}\label{lm:pat-occ-equivalence}
  Let $\Pat \in \Sigma^{*}$. Let $j, j' \in [1 \dd \Textlen]$
  and $k \geq 0$ be such that $\Textinf[j \dd j + k) = \Textinf[j'
  \dd j' + k)$. Then, $j \in \OccThree{k}{\Pat}{\Text}$ if and only if $j' \in
  \OccThree{k}{\Pat}{\Text}$.
\end{lemma}
\begin{proof}
  Let $j \in \OccThree{k}{\Pat}{\Text}$. We will prove that $j' \in
  \OccThree{k}{\Pat}{\Text}$ (the proof of the opposite implication follows by
  symmetry). If $j = j'$, then the claim follows immediately.  Let us
  thus assume $j \neq j'$. By the uniqueness of $\Text[\Textlen]$ in $\Text$, the
  assumption $\Textinf[j \dd j + k) = \Textinf[j' \dd j' + k)$
  implies $\lcp(\Text[j \dd \Textlen], \Text[j' \dd \Textlen]) = k'$, where $k' \geq k$.
  The assumption $j \in \OccThree{k}{\Pat}{\Text}$ implies $\lcp(\Text[j \dd \Textlen],
  \Pat) = k''$, where $k'' \geq \min(|\Pat|, k)$.  We therefore obtain
  $\lcp(\Text[j' \dd \Textlen], \Pat) \geq \min(\lcp(\Text[j' \dd \Textlen], \Text[j \dd
  \Textlen]),\allowbreak \lcp(\Text[j \dd \Textlen], \Pat)) = \min(k', k'') \geq
  \min(k, \min(|\Pat|, k)) = \min(|\Pat|, k)$. Thus, $j' \in
  \OccThree{k}{\Pat}{\Text}$.
\end{proof}

\begin{lemma}\label{lm:occ-equivalence}
  For every $j, j' \in [1 \dd \Textlen]$ and every $k \geq 0$,
  the following conditions are equivalent:
  \begin{itemize}
  \item $\Textinf[j \dd j + k) = \Textinf[j' \dd j' + k)$,
  \item $\OccThree{k}{j}{\Text} = \OccThree{k}{j'}{\Text}$,
  \item $j \in \OccThree{k}{j'}{\Text}$,
  \item $j = j'$ or $\LCE_{\Text}(j, j') \geq k$.
  \end{itemize}
\end{lemma}
\begin{proof}
  Denote $\Pat = \Text[j \dd \Textlen]$, $\Pat' = \Text[j' \dd \Textlen]$, $m = |\Pat|$,
  and $m' = |\Pat'|$.  Recall (\cref{def:occ}) that $\OccThree{k}{j}{\Text}$
  is defined as $\OccThree{k}{j}{\Text} = \OccThree{k}{\Pat}{\Text} = \{j'' \in [1 \dd
  \Textlen] : \lcp(\Pat, \Text[j'' \dd \Textlen]) \geq \min(m, k)\}$.
  Analogously, $\OccThree{k}{j'}{\Text} = \OccThree{k}{\Pat'}{\Text} = \{j'' \in [1
  \dd \Textlen] : \lcp(\Pat', \Text[j'' \dd \Textlen]) \geq \min(m', k)\}$.  The
  four implications are proved as follows.
  \begin{itemize}
  \item Assume $\Textinf[j \dd j + k) = \Textinf[j' \dd j' +
    k)$. If $j = j'$, then $\Pat = \Pat'$, and we immediately obtain
    $\OccThree{k}{j}{\Text} = \OccThree{k}{\Pat}{\Text} = \OccThree{k}{\Pat'}{\Text} =
    \OccThree{k}{j'}{\Text}$. Let us thus assume $j \neq j'$.  By the
    uniqueness of $\Text[\Textlen]$ in $\Text$, we then have $\LCE_{\Text}(j, j') \geq
    k$. In other words, $\lcp(\Pat, \Pat') = k'$, where $k' \geq
    k$. In particular, $m \geq k$ and $m' \geq k$. Consider any $j''
    \in \OccThree{k}{j}{\Text}$. Then, $\lcp(\Pat, \Text[j'' \dd \Textlen]) \geq
    \min(m, k) = k$.  We thus have $\lcp(\Pat', \Text[j'' \dd \Textlen]) \geq
    \min(\lcp(\Pat', \Pat), \lcp(\Pat, \Text[j'' \dd \Textlen])) = \min(k', k)
    = k = \min(m', k)$.  We have thus proved $j'' \in
    \OccThree{k}{j'}{\Text}$. Hence, $\OccThree{k}{j}{\Text} \subseteq \OccThree{k}{j'}{\Text}$. The
    inclusion $\OccThree{k}{j'}{\Text} \subseteq \OccThree{k}{j}{\Text}$ follows by
    symmetry. Thus, $\OccThree{k}{j}{\Text} = \OccThree{k}{j'}{\Text}$.
  \item Assume $\OccThree{k}{j}{\Text} = \OccThree{k}{j'}{\Text}$. By definition of
    $\Pat$, it holds $\lcp(\Text[j \dd \Textlen], \Pat) = m \geq \min(m,
    k)$. Thus, $j \in \OccThree{k}{j}{\Text}$. By the assumption, this implies
    $j \in \OccThree{k}{j'}{\Text}$.
  \item Assume $j \in \OccThree{k}{j'}{\Text}$. If $j = j'$, the claim follows
    immediately. Let us thus assume $j \neq j'$.  By $j \in
    \OccThree{k}{j'}{\Text}$, we then have $\lcp(\Pat', \Text[j \dd \Textlen]) \geq
    \min(m', k)$. Equivalently, it holds $\LCE_{\Text}(j', j) \geq
    \min(m', k)$. Suppose that $m' \leq k$. Then, $\min(m', k) = m'$,
    and the assumption $\LCE_{\Text}(j', j) \geq m'$ implies that $\Text[j \dd
    j + m' - 1] = \Text[j' \dd j' + m' - 1] = \Text[j' \dd \Textlen]$. In
    particular, $\Text[\Textlen] = \Text[j + m' - 1] = \Text[j + (\Textlen - j' + 1) - 1] =
    \Text[\Textlen - (j' - j)]$. Since $j \neq j'$, this contradicts the
    uniqueness of $\Text[\Textlen]$ in $\Text$. We thus have $m' >
    k$. Consequently, $\LCE_{\Text}(j', j) \geq \min(m', k) = k$. We have
    thus proved that it holds $j = j'$ or $\LCE_{\Text}(j, j') \geq k$.
  \item Assume $j = j'$ or $\LCE_{\Text}(j, j') \geq k$. If $j = j'$, then
    we immediately obtain $\Textinf[j \dd j + k) = \Textinf[j' \dd
    j' + k)$. Let us thus assume $\LCE_{\Text}(j, j') \geq k$. This
    implies $\Textinf[j \dd j + k) = \Text[j \dd j + k) = \Text[j' \dd j' +
    k) = \Textinf[j' \dd j' + k)$. In both cases, we thus obtain
    $\Textinf[j \dd j + k) = \Textinf[j' \dd j' + k)$. \qedhere
  \end{itemize}
\end{proof}

\subsection{The Index Core}\label{sec:sa-core}

\subsubsection{Preliminaries}\label{sec:sa-core-prelim}

\begin{lemma}\label{lm:d-sum}
  For every $c \geq 1$, it holds
  \[
    \sum_{i =0}^\infty \tfrac{1}{2^i}\SubstrCount{c2^i}{\Text}
    = \bigO(c\SubstringComplexity{\Text} \log \tfrac{c\Textlen \log \sigma}{\SubstringComplexity{\Text} \log \Textlen}).
  \]
\end{lemma}
\begin{proof}
  Let $\mu = \lfloor \log \tfrac{\log \SubstringComplexity{\Text}}{c \log \sigma} \rfloor$
  and $\nu = \lceil \log \tfrac{\Textlen}{\SubstringComplexity{\Text}} \rceil$, so that
  $2^{\mu} \leq \tfrac{\log \SubstringComplexity{\Text}}{c \log \sigma}$ and
  $2^{\nu} \geq \tfrac{\Textlen}{\SubstringComplexity{\Text}}$.
  \begin{itemize}
  \item For $i \in [0 \dd \mu]$, we observe that $\SubstrCount{c2^i}{\Text}
    \leq \sigma^{c2^i}$, and hence
    \begin{align*}
      \sum_{i=0}^{\mu} \tfrac{1}{2^i} \SubstrCount{c2^i}{\Text}
        \leq \sum_{i=0}^{\mu} \SubstrCount{c2^i}{\Text}
        \leq \sum_{i=0}^{\mu} \sigma^{c2^i}
        \leq 2\sigma^{c2^{\mu}}
        \leq 2\SubstringComplexity{\Text}.
    \end{align*}
  \item For $i \in (\mu \dd \nu]$, we observe that
    $\SubstrCount{c2^i}{\Text} \leq c\cdot2^{i}\cdot \SubstringComplexity{\Text}$, and hence
    \begin{align*}
      \sum_{i=\mu+1}^{\nu} \tfrac{1}{2^i}\SubstrCount{c2^i}{\Text}
        &\leq \sum_{i=\mu+1}^{\nu} c\SubstringComplexity{\Text}
         \leq c\SubstringComplexity{\Text} (\nu - \mu)\\
        &=    \bigO(c\SubstringComplexity{\Text} \log \tfrac{c\Textlen \log \sigma}{\SubstringComplexity{\Text} \log \SubstringComplexity{\Text}})
         =    \bigO(c\SubstringComplexity{\Text} \log \tfrac{c\Textlen \log \sigma}{\SubstringComplexity{\Text} \log \Textlen}).
    \end{align*}
  \item For $i \in (\nu \dd \infty)$, we observe that
    $\SubstrCount{c2^i}{\Text} \leq \Textlen$, and hence
    \begin{align*}
      \sum_{i=\nu+1}^{\infty}
      \tfrac{1}{2^i}\SubstrCount{c2^i}{\Text}
        \leq \sum_{i=\nu+1}^{\infty} \tfrac{\Textlen}{2^i}
        \leq \tfrac{\Textlen}{2^{\nu+1}} \sum_{i=0}^{\infty}
             \left(\tfrac{1}{2} \right)^i
        \leq \SubstringComplexity{\Text}.
    \end{align*}
  \end{itemize}
  Thus, in total, we obtain $\sum_{i =0}^\infty
  \tfrac{1}{2^i} \SubstrCount{c2^i}{\Text} = \bigO(c\SubstringComplexity{\Text} \log \tfrac{c\Textlen \log
  \sigma}{\SubstringComplexity{\Text} \log \Textlen})$.
\end{proof}

\begin{lemma}\label{lm:gap}
  Let $\tau \in [1 \dd \floor{\frac{\Textlen}{2}}]$.  Let $j, j',
  j'' \in [1 \dd \Textlen]$ be such that $j,
  j'' \in \RTwo{\tau}{\Text}$, $j' \not\in \RTwo{\tau}{\Text}$, and $j
  < j' < j''$. Then, it holds $j'' - j \geq 2\tau$.
\end{lemma}
\begin{proof}
  Suppose that $j'' - j < 2\tau$. Denote $p_1 = \per(\Text[j \dd j +
  3\tau - 1))$ and $p_2 = \per(\Text[j'' \dd j'' + 3\tau - 1))$. By
  $j, j'' \in \RTwo{\tau}{\Text}$, we have $p_1,
  p_2 \leq \tfrac{1}{3}\tau$. Let us first assume $p_1 \neq p_2$.  By
  $p_1, p_2 \leq \tfrac{1}{3}\tau$, each of the substrings
  $\Text[j \dd j + 3\tau - 1)$ and $\Text[j'' \dd j'' + 3\tau - 1)$
  can be extended into
  a \emph{run}~\cite{KolpakovK99,michael1989detecting} (i.e., a
  maximal substring $\T[x \dd y)$ of $\Text$ satisfying $\per(\T[x \dd
  y)) \leq \tfrac{1}{2}(y-x)$). Let $\Text[x_1 \dd y_1)$ (resp.\
  $\Text[x_2 \dd y_2)$) be the run extending the substring
  $\Text[j \dd j + 3\tau - 1)$ (resp.\ $\Text[j'' \dd j'' + 3\tau -
  1)$).  Note that since $p_1 \neq p_2$, these two runs are
  different. It is known that $\Text[x_1 \dd y_1)$ and $\Text[x_2 \dd
  y_2)$ cannot overlap by $p_1 + p_2$ or more
  (e.g.,~\cite[Fact~2.2.4]{phdtomek}).  However, $p_1 +
  p_2 \leq \tfrac{2}{3}\tau$, and $j'' - j < 2\tau$ implies that they
  overlap by at least $\tau$ symbols, a contradiction. We thus must
  have $p_1 = p_2$. Note, however, that by definition of a period,
  this implies that $\Text[j \dd j'' + 3\tau - 1)$ has period $p_1$.
  In particular, $\per(\Text[j' \dd j' + 3\tau -
  1)) \leq \tfrac{1}{3}\tau$, which contradicts
  $j' \not\in \RTwo{\tau}{\Text}$. We therefore must have $j'' -
  j \geq 2\tau$.
\end{proof}

\begin{lemma}\label{lm:IR-comp-R-size}
  Let $\tau \in [1 \dd \lfloor \tfrac{\Textlen}{2}
  \rfloor]$. For every $c \geq 1$, it holds
  \[
    |\IntervalRepr{\CompRepr{c\tau}{\RTwo{\tau}{\Text}}{\Text}}| \leq
    \tfrac{5}{2\tau} \left(\SubstrCount{8c\tau}{\Text} + 8c\tau \right) \le 40c\SubstringComplexity{\Text}.
  \]
\end{lemma}
\begin{proof}
  Denote $(s_k,t_k)_{k \in [1 \dd m']} =
  \IntervalRepr{\CompRepr{c\tau}{\RTwo{\tau}{\Text}}{\Text}}$.
  Recall that $\CompRepr{c\tau}{\RTwo{\tau}{\Text}}{\Text} =
  \RTwo{\tau}{\Text} \cap \Cover{c\tau}{\Text}$ (\cref{def:comp}).  Consider any maximal
  subinterval $[j \dd j + \ell)$ of $\Cover{c\tau}{\Text}$, i.e., such
  that $[j \dd j + \ell) \subseteq \Cover{c\tau}{\Text}$ and $\{j - 1, j
  + \ell\} \cap \Cover{c\tau}{\Text} = \emptyset$. By definition, for
  every $k \in [1 \dd m]$, $s_k \in (j \dd j + \ell)$ implies $s_k -
  1 \not\in \RTwo{\tau}{\Text}$.  On the other hand, we have by definition
  that $s_k \in \RTwo{\tau}{\Text}$. Therefore, letting $P = \{s_k : k \in
  [1 \dd m]\}$, for every $j' \in (j \dd j + \ell)$, $j' \in P$
  implies $j' \in \RTwo{\tau}{\Text}$ and $j' - 1 \not\in \RTwo{\tau}{\Text}$.
  By \cref{lm:gap}, we thus have that for every $j',
  j'' \in (j \dd j + \ell) \cap P$, it holds $|j'' - j'| \geq
  2\tau$. Consequently, $|(j \dd j + \ell) \cap
  P| \leq \lceil \tfrac{\ell - 1}{2\tau} \rceil \leq 1
  + \lfloor \tfrac{\ell - 1}{2\tau} \rfloor$ and thus $|[j \dd j
  + \ell) \cap P| \leq 2 + \lfloor \tfrac{\ell}{2\tau} \rfloor$.  Let
  us now denote $(x_k,y_k)_{k \in [1 \dd m]} = \IntervalRepr{\Cover{c\tau}{\Text}}$.
  By \cref{lm:c-cover} and the above discussion, it holds
  \begin{align*}
    |\IntervalRepr{\CompRepr{c\tau}{\RTwo{\tau}{\Text}}{\Text}}|
         &= \sum_{i=1}^{m} \left|[x_i \dd x_i + y_i) \cap P\right|
       \leq \sum_{i=1}^{m} \left(2 + \left\lfloor\tfrac{y_i}{2\tau}
                          \right\rfloor\right)
       \leq 2m + \tfrac{1}{2\tau} \sum_{i=1}^{m} y_i\\
      &\leq 2\left(\tfrac{\SubstrCount{8c\tau}{\Text}}{c\tau} + 8 \right) +
            \tfrac{1}{2\tau} \Big(\SubstrCount{8c\tau}{\Text} + 8c\tau\Big)\\
      & = \left(\tfrac{2}{c\tau}+\tfrac{1}{2\tau}\right) \cdot \Big(\SubstrCount{8c\tau}{\Text} + 8c\tau \Big)\\
      &\leq \tfrac{5}{2\tau} \left(\SubstrCount{8c\tau}{\Text} + 8c\tau \right).
  \end{align*}

  To show the second part of the claim, recall that for every $q \in \Zp$, we have $\SubstrCount{q}{\Text}+q \leq q\cdot \SubstringComplexity{\Text}+q\le 2q\cdot \SubstringComplexity{\Text}$.  Thus, 
  $\tfrac{5}{2\tau} \left(\SubstrCount{8c\tau}{\Text} + 8c\tau \right) \le \tfrac{5}{2\tau}\cdot 16c\tau\cdot \SubstringComplexity{\Text}
  \le 40c\SubstringComplexity{\Text}$.
\end{proof}

\begin{lemma}\label{lm:sa-core-construction}
  Let $\tau \in
  [3 \dd \lfloor \tfrac{\Textlen}{2} \rfloor]$.  Let $i \in [1 \dd \Textlen]$,
  and $b \leq 2\tau$ be such that $[i \dd i + b) \subseteq [1 \dd \Textlen
  - 3\tau + 2]$. Let $p = \per(\Text[i + b \dd i + 3\tau - 1))$.  If $p
  > \tfrac{1}{3}\tau$, then $[i \dd i + b) \cap \RTwo{\tau}{\Text}
  = \emptyset$.  Otherwise, letting $x \in [i \dd i + b]$ and $y \in
  [i + 3\tau - 1 \dd i + b + 3\tau - 2]$ be such that $\per(\Text[x \dd
  y)) = p$ and $y - x$ is maximized, it holds
  \vspace{2ex}
  \[
    [i \dd i + b) \cap \RTwo{\tau}{\Text} =
    \begin{cases}
      \emptyset & \text{if $y - x < 3\tau - 1$},\\
      [x \dd y - 3\tau + 1] & \text{otherwise}.
    \end{cases}
    \vspace{2ex}
  \]
\end{lemma}
\begin{proof}
  We show the first implication by contraposition. Assume that $[i \dd
  i + b) \cap \RTwo{\tau}{\Text} \neq \emptyset$, i.e., there exists $i' \in
  [i \dd i + b)$ such that $i' \in \RTwo{\tau}{\Text}$. We then have
  $\per(\Text[i' \dd i' + 3\tau - 1)) \leq \tfrac{1}{3}\tau$.  By $i' < i
  + b$ and $i + 3\tau - 1 \leq i' + 3\tau - 1$, $\Text[i + b \dd i + 3\tau
  - 1)$ is a substring of $\Text[i' \dd i' + 3\tau - 1)$ and thus
  $p \leq \tfrac{1}{3}\tau$.

  Let us thus assume $p \leq \tfrac{1}{3}\tau$. We first show, again
  by contraposition, that $y - x < 3\tau - 1$ implies $[i \dd i +
  b) \cap \RTwo{\tau}{\Text}$. Let $i'$ be as above. Note that by
  $\per(\Text[i' \dd i' + 3\tau - 1)) \leq \tfrac{1}{3}\tau$ we then have
  $x \leq i'$ and $y' \geq i'+ 3\tau - 1$. Consequently, $y - x \geq
  3\tau - 1$.  This concludes the proof of the first case. Let us now
  assume $y - x \geq 3\tau - 1$. We aim to show that $[i \dd i +
  b) \cap \RTwo{\tau}{\Text} = [x \dd y - 3\tau + 1]$.
  \begin{itemize}
  \item Let $i' \in [x \dd y - 3\tau + 1]$.  By $x \geq i$, we have
    $i' \geq i$. On the other hand, by $y \leq i + b + 3\tau - 2$, we
    have $i' \leq y - 3\tau + 1 \leq i + b - 1$.  Thus, $i' \in [i \dd
    i + b)$. To show $i' \in \RTwo{\tau}{\Text}$, note that by the assumption
    $i' \leq y - 3\tau + 1$, or equivalently, $i' + 3\tau - 1 \leq y$,
    we have that $\Text[i' \dd i' + 3\tau - 1)$ is a substring of $\Text[x \dd
    y)$.  Thus, $\per(\Text[i' \dd i' + 3\tau - 1)) \leq \per(\Text[x \dd y))
    = p \leq \tfrac{1}{3}\tau$, i.e., $i' \in \RTwo{\tau}{\Text}$.
  \item Let us now assume $i' \in [i \dd i + b) \cap
    \RTwo{\tau}{\Text}$. Denote $p' = \per(\Text[i' \dd i' + 3\tau - 1))$. By
    $i' \in \RTwo{\tau}{\Text}$, we have $p' \leq \tfrac{1}{3}\tau$. As
    observed above, $i' \in [i \dd i + b)$ implies that $\Text[i + b \dd i
    + 3\tau - 1)$ is a substring of $\Text[i' \dd i' + 3\tau - 1)$. This
    implies that $\Text[i' \dd i' + 3\tau - 1)$ in addition to $p$, also
    has period $p'$. Clearly, we have $p' \leq p$. We claim that $p' =
    p$.  For the proof by contradiction, suppose $p' < p$. Note that
    $p + p' \leq 2 \lfloor \tfrac{1}{3}\tau \rfloor \leq \tau - 1 \leq
    3\tau - 1 - b$ (where the last inequality follows by $b \leq
    2\tau$). Thus, by the weak periodicity
    lemma~\cite{periodicitylemma}, $\Text[i + b \dd i + 3\tau - 1)$ also
    has period $d = \gcd(p', p)$. By $p' < p$ and $d \mid p'$, this
    implies $d < p$, contradicting $p = \per(\Text[i + b \dd i + 3\tau -
    1))$.  Thus, $p' = p$. This implies $x \leq i'$ and $y \geq i' +
    3\tau - 1$ and hence $i' \in [x \dd y - 3\tau + 1]$. \qedhere
  \end{itemize}
\end{proof}

\begin{lemma}\label{lm:sa-core-nav}
  Let $\tau \in
  [1 \dd \lfloor \tfrac{\Textlen}{2} \rfloor]$. Let $t, t' \in \Zp$ be such
  that $t \leq t'$. Let $j \in [1 \dd \Textlen]$ and $j' \in
  \OccThree{3\tau - 1}{j}{\Text}$ be such that $j' = \min \OccThree{t}{j'}{\Text}$. Then,
  $j \in \RTwo{\tau}{\Text}$ holds if and only if $j' \in
  \CompRepr{t'}{\RTwo{\tau}{\Text}}{\Text}$ (\cref{def:comp}).
\end{lemma}
\begin{proof}
  Assume $j \in \RTwo{\tau}{\Text}$. This implies $\per(\Text[j \dd j + 3\tau -
  1)) \leq \tfrac{1}{3}\tau$.  Thus, by the uniqueness of $\Text[\Textlen]$ in
  $\Text$, the substring $\Text[j \dd j + 3\tau - 1)$ does not contain the
  symbol $\Text[\Textlen]$ and hence $j \leq \Textlen - 3\tau + 1$.  By
  $j' \in \OccThree{3\tau - 1}{j}{\Text}$, we thus obtain $j' \leq \Textlen - 3\tau
  - 1$.  Consequently, $\Text[j \dd j + 3\tau - 1) = \Text[j' \dd j' + 3\tau -
  1)$.  This implies $\per(\Text[j' \dd j' + 3\tau - 1)) = \per(\Text[j \dd j
  + 3\tau - 1)) \leq \tfrac{1}{3}\tau$, i.e., $j' \in \RTwo{\tau}{\Text}$.
  Recall now that $\CompRepr{t'}{\RTwo{\tau}{\Text}}{\Text} =
  \RTwo{\tau}{\Text} \cap \Cover{t'}{\Text}$ (see \cref{cons:cover,def:comp}). By $t \leq
  t'$ and \cref{lm:cover}, $\Cover{t'}{\Text}$ is a $t$-cover of
  $\Text$. Thus, $j' = \min \OccThree{t}{j'}{\Text}$ implies that $[j' \dd j' +
  t) \subseteq \Cover{t'}{\Text}$ and hence $j' \in \Cover{t'}{\Text}$.
  Combining with $j' \in \RTwo{\tau}{\Text}$, we thus obtain $j' \in
  \RTwo{\tau}{\Text} \cap \Cover{t'}{\Text} = \CompRepr{t'}{\RTwo{\tau}{\Text}}{\Text}$.

  Let us now assume $j' \in \CompRepr{t'}{\RTwo{\tau}{\Text}}{\Text}$.  By
  definition, this implies $j' \in \RTwo{\tau}{\Text} \subseteq [1 \dd \Textlen -
  3\tau + 2]$. By $\Textinf[j' \dd j' + 3\tau - 1) =
  \Textinf[j \dd j + 3\tau - 1)$ (following from $j' \in
  \OccThree{3\tau - 1}{j}{\Text}$) and the uniqueness of $\Text[\Textlen]$ in $\Text$, we have $j \in
  [1 \dd \Textlen - 3\tau + 2]$.  Thus, $\Text[j \dd j + 3\tau - 1) = \Text[j' \dd
  j' + 3\tau - 1)$ and hence $\per(\Text[j \dd j + 3\tau - 1))
  = \per(\Text[j' \dd j' + 3\tau - 1))
  \leq \tfrac{1}{3}\tau$. Consequently, $j \in \RTwo{\tau}{\Text}$.
\end{proof}

\subsubsection{The Data Structure}\label{sec:sa-core-ds}

\paragraph{Definitions}

For every $k \in [4 \dd \lceil \log \Textlen \rceil)$, denote $\ell_k = 2^k$,
$\tau_k = \lfloor \tfrac{\ell_k}{3} \rfloor$, and let
$\ArrRuns{k}[1 \dd n_{{\rm runs},k}]$ be an array containing the
sequence $\IntervalRepr{\CompRepr{14\tau_k}{\RTwo{\tau_k}{\Text}}{\Text}}$
(\cref{def:interval-representation,def:comp}) sorted lexicographically.

\paragraph{Components}

The index core, denoted $\CompSaCore{\Text}$, consists of three
components:
\begin{enumerate}
\item The structure from \cref{th:random-access}. It needs
  $\bigO(\SubstringComplexity{\Text} \log \tfrac{\Textlen \log \sigma}{\SubstringComplexity{\Text} \log \Textlen})$ space.
\item The structure from \cref{th:lce}. It needs the same space as the first component.
\item For $k \in [4 \dd \lceil \log \Textlen \rceil)$, we store the array
  $\ArrRuns{k}[1 \dd n_{{\rm runs},k}]$ in plain form using
  $\bigO(n_{{\rm runs},k})$ space. To bound the total space usage,
  first note that by \cref{lm:IR-comp-R-size}, for every $k \in
  [4 \dd \lceil \log \Textlen \rceil)$, it holds $n_{{\rm
  runs},k} \leq \tfrac{2}{5\tau_k}\left(\SubstrCount{112\tau_k}{\Text} + 112\tau_k\right)$.  By
  utilizing that $3\tau_k \leq \ell_k \le 4\tau_k$ and because $\SubstrCount{q}{\Text} +q$
  is a non-decreasing function of $q\in \Zp$, we thus have
  $n_{{\rm runs},k} \leq \tfrac{10}{\ell_k}\left(\SubstrCount{38\ell_k}{\Text} + 38\ell_k\right)$.
  Combining this with \cref{lm:d-sum}, we therefore obtain
  \begin{align*}
    \textstyle\sum_{k = 4}^{\lceil \log \Textlen \rceil-1} n_{{\rm runs},k}
      &\leq  10\textstyle\sum_{k=4}^{\infty}
             \tfrac{1}{\ell_k}\SubstrCount{38\ell_k}{\Text} + 380\lceil \log \Textlen \rceil
      =     \bigO(\SubstringComplexity{\Text} \log \tfrac{\Textlen \log \sigma}{\SubstringComplexity{\Text} \log \Textlen})
  \end{align*}
  where the in the last inequality we utilized that
  $\SubstringComplexity{\Text} \log \tfrac{\Textlen \log \sigma}{\SubstringComplexity{\Text} \log \Textlen} = \Omega(\log \Textlen)$.
  The total space needed by all arrays is thus
  $\bigO(\SubstringComplexity{\Text} \log \tfrac{\Textlen \log \sigma}{\SubstringComplexity{\Text} \log \Textlen})$.
\end{enumerate}

In total, $\CompSaCore{\Text}$ needs
$\bigO(\SubstringComplexity{\Text} \log \tfrac{\Textlen \log \sigma}{\SubstringComplexity{\Text} \log \Textlen})$ space.

\subsubsection{Basic Navigation Primitives}\label{sec:sa-core-nav}

\begin{proposition}\label{pr:sa-core-nav}
  Let $k \in [4 \dd \lceil \log \Textlen \rceil)$, $\ell = 2^k$, $\tau =
  \lfloor \tfrac{\ell}{3} \rfloor$, and $j \in [1 \dd \Textlen]$.
  Given $\CompSaCore{\Text}$, the value $k$, the position $j$, and any
  $j' \in \OccThree{\ell}{j}{\Text}$ satisfying $j'
  = \min \OccThree{2\ell}{j'}{\Text}$, we can in $\bigO(\log \Textlen)$ time
  determine if $j \in \RTwo{\tau}{\Text}$.
\end{proposition}
\begin{proof}
  First, note that by $3\tau - 1 \leq \ell$, we have
  $\OccThree{\ell}{j}{\Text} \subseteq \OccThree{3\tau - 1}{j}{\Text}$, and thus
  $j' \in \OccThree{3\tau - 1}{j}{\Text}$. On the other hand, we have
  $2\ell \leq 14\tau$ (since for $\tau
  = \lfloor \tfrac{\ell}{3} \rfloor$ and $\ell \geq 16$, it holds
  $2\ell \leq 7\tau$). Thus, by \cref{lm:sa-core-nav},
  $j \in \RTwo{\tau}{\Text}$ holds if and only if
  $j' \in \CompRepr{14\tau}{\RTwo{\tau}{\Text}}{\Text}$. We thus proceed as
  follows. If $n_{{\rm runs},k} = 0$, then
  $\CompRepr{14\tau}{\RTwo{\tau}{\Text}}{\Text} = \emptyset$ and hence we return
  that $j \not\in \RTwo{\tau}{\Text}$.  Let us thus assume $n_{{\rm runs},k} >
  0$. If $\ArrRuns{k}[1] > j'$, then by definition of $\ArrRuns{k}$, we
  have $j' \not\in \CompRepr{14\tau}{\RTwo{\tau}{\Text}}{\Text}$ and thus again
  $j \not\in \RTwo{\tau}{\Text}$. Otherwise, using binary search in
  $\bigO(\log \Textlen)$ time we compute the largest
  $i \in [1 \dd n_{{\rm runs},k}]$ such that letting $(p_i,t_i) = \ArrRuns{k}[i]$, it holds
  $p_i \leq j'$. If $j' \in [p_i \dd p_i + t_i)$ then we have
  $j' \in \CompRepr{14\tau}{\RTwo{\tau}{\Text}}{\Text}$ and hence
  $j \in \RTwo{\tau}{\Text}$. Otherwise we return $j \not\in \RTwo{\tau}{\Text}$.
\end{proof}

\subsubsection{Construction Algorithm}\label{sec:sa-core-construction}

\begin{proposition}\label{pr:sa-core-construction}
  Given the LZ77 parsing of $\Text$, we can construct $\CompSaCore{\Text}$
  in $\bigO(\SubstringComplexity{\Text} \log^7 \Textlen)$ time.
\end{proposition}
\begin{proof}
  We construct the components of $\CompSaCore{\Text}$
  (\cref{sec:sa-core-ds}) as follows:
  \begin{enumerate}

  \item First, using \cref{th:random-access}, we construct the structure
    from \cref{th:random-access} in $\bigO(\SubstringComplexity{\Text} \log^7 \Textlen)$ time.

  \item Next, using \cref{th:lce}, we construct the structure
    from \cref{th:lce} in $\bigO(\SubstringComplexity{\Text} \log^7 \Textlen)$ time.

  \item We construct arrays $\ArrRuns{i}$ as follows.
    First, using~\cite[Theorem~6.7]{resolutionfull}, in
    $\bigO(\LZSize{\Text} \log^2 \Textlen)$ time we construct a structure that, given
    any substring $S$ of $\Text$ (specified with its starting position
    and the length) in $\bigO(\log^3 \Textlen)$ time checks if
    $\per(S) \leq \tfrac{|S|}{2}$, and if so, returns $\per(S)$.  Next,
    using~\cite[Theorem~6.11]{resolutionfull}, in $\bigO(\LZSize{\Text} \log^4
    \Textlen)$ time we construct a structure that, given any substring $S$ of
    $\Text$ (represented as above) in $\bigO(\log^3 \Textlen)$ time returns
    $\min \OccTwo{S}{\Text}$. For $i \in [4 \dd \lceil \log \Textlen \rceil)$ we then
    execute the following algorithm:
    \begin{enumerate}
    \item Using \cref{pr:cover-construction} in $\bigO(\LZSize{\Text} \log^3 \Textlen)$
      time we compute $(s_k,t_k)_{k \in [1 \dd m]}
      = \IntervalRepr{\Cover{14\tau_i}{\Text}}$.
    \item We then separately process each element of $(s_k,t_k)_{k \in
      [1 \dd m]}$ in the order of increasing $k$ (note that this
      corresponds to maximal blocks of $\Cover{14\tau_i}{\Text}$ in the
      left-to-right order).  For every $k \in [1 \dd m]$, we split
      $[s_k \dd s_k + t_k)$ into $\lceil \tfrac{t_k}{2\tau_i} \rceil$
      subblocks of size $2\tau_i$ (except possibly the last one), and
      for each subblock $[j \dd j + b)$, we determine
      using \cref{lm:sa-core-construction} if $[j \dd j +
      b) \cap \RTwo{\tau_i}{\Text} \neq \emptyset$ and if so, compute positions
      $p, p' \in [j \dd j + b]$ such that $[j \dd j +
      b) \cap \RTwo{\tau_i}{\Text} = [p \dd p')$. All subblocks within each
      block are processed left-to-right.  Observe, that under this
      assumption, it is easy to generate subsequent elements of the
      array $\ArrRuns{i}$ from the computed indexes $p, p'$ for each
      subblock. It remains to explain how to efficiently implement the
      computation of $p, p'$.
      \begin{enumerate}
      \item First, in $\bigO(\log^3 \Textlen)$ time, we check if $\per(\Text[j +
        b \dd j + 3\tau_i - 1)) \leq \tfrac{3\tau_i - 1 - b}{2}$, and if
        so, we compute $\per(\Text[j + b \dd j + 3\tau_i - 1))$.  Observe
        that since $b \leq 2\tau$, it holds $3\tau_i - 1 - b \geq \tau_i
        - 1$.  Thus, $\per(\Text[j + b \dd j + 3\tau_i - 1))
        > \tfrac{3\tau_i - 1 - b}{2}$ implies $\per(\Text[j + b \dd j +
        3\tau_i - 1)) > \tfrac{\tau_i - 1}{2} \geq \tfrac{1}{3}\tau_i$.
        By \cref{lm:sa-core-construction}, we then have $[j \dd j +
        b) \cap \RTwo{\tau_i}{\Text} = \emptyset$ and hence we can set $p = p'
         = j$. Let us thus assume that $p = \per(\Text[j + b \dd j + 3\tau_i
        - 1))$ satisfies $p \leq \frac{3\tau_i - 1 - b}{2}$, and we
        obtained $p$.
      \item Next, we check if $p > \tfrac{1}{3}\tau_i$. If so,
        by \cref{lm:sa-core-construction} we again have $[j \dd j +
        b) \cap \RTwo{\tau_i}{\Text} = \emptyset$ and hence we return $p = p'
        = j$. Let us thus assume $p \leq \tfrac{1}{3}\tau_i$.
      \item Next, we determine the positions $x$ and $y$ by first
        computing in $\bigO(\log \Textlen)$ time the values $\delta_{\rm left}
        = \LCE_{\revstr{\Text}}(\Textlen - j - b + 2, \Textlen - j - b + 2 - p)$ and
        $\delta_{\rm right} = \LCE_{\Text}(j + 3\tau_i - 1, j + 3\tau_i - 1
        - p)$.  We then have $x = j + b - \delta_{\rm left}$ and $y = j
        + 3\tau_i - 1 + \delta_{\rm right}$.  If $y - x < 3\tau_i - 1$,
        then by \cref{lm:sa-core-construction}, we have $[j \dd j +
        b) \cap \RTwo{\tau_i}{\Text} = \emptyset$, and thus we set $p = p' =
        j$.  Otherwise, by \cref{lm:sa-core-construction}, we set $p =
        x$ and $p' = y - 3\tau_i + 2$.
      \end{enumerate}
      Recall that block $[s_k \dd s_k + t_k)$ is split into
      $\lceil \tfrac{t_k}{2\tau_i} \rceil$ subblocks.  The above
      procedure thus takes $\bigO((1 + \frac{t_k}{\tau_i}) \cdot \log^3
      \Textlen)$ time.  By \cref{lm:c-cover}, we have $m \leq 16\SubstringComplexity{\Text}
      = \bigO(\SubstringComplexity{\Text})$ and $\sum_{k \in [1 \dd m]} t_k \leq 16 \cdot
      14\tau_i \cdot \SubstringComplexity{\Text} = \bigO(\tau_i\SubstringComplexity{\Text})$.  Over all
      $k \in [1 \dd m]$ we thus spend $\bigO(\sum_{k \in [1 \dd m]} (1
      + \frac{t_k}{\tau_i}) \cdot \log^3 \Textlen) = \bigO(m \cdot \log^3 \Textlen
      + \tfrac{\tau_i\SubstringComplexity{\Text}}{\tau_i} \cdot \log^3 \Textlen)
      = \bigO(\SubstringComplexity{\Text} \log^3 \Textlen)$ time.
    \end{enumerate}
    In total, the construction of all arrays $\ArrRuns{i}$ takes
    $\bigO((\LZSize{\Text} + \SubstringComplexity{\Text}) \log^3 \Textlen) = \bigO(\SubstringComplexity{\Text} \log^4 \Textlen)$
    time.
    \qedhere
  \end{enumerate}
\end{proof}

\subsection{The Nonperiodic Patterns and Positions}\label{sec:sa-nonperiodic}

\subsubsection{Preliminaries}\label{sec:sa-nonperiodic-prelim}

\begin{definition}\label{def:dist-prefix}
  Let $\tau \in [1 \dd \lfloor \tfrac{\Textlen}{2} \rfloor]$.  For
  every $Q \subseteq [1 \dd \Textlen - 2\tau + 1]$ satisfying
  $Q \neq \emptyset$ and $\max Q \geq \Textlen - 3\tau + 2$, and every
  $j \in [1 \dd \Textlen - 3\tau + 2]$, we denote $\Successor{Q}{j}
  = \min \{j' \in Q : j' \geq j'\}$. We then denote
  $\DistinguishingPrefs{\tau}{\Text}{Q}
  := \{\Text[j \dd \Successor{Q}{j} + 2\tau) : j \in [1 \dd \Textlen -
  3\tau + 2] \setminus \RTwo{\tau}{\Text}\}$.
\end{definition}

\begin{lemma}\label{lm:nonperiodic-pat-lce}
  Let $\tau \geq 1$ and $\Pat \in \Sigma^{+}$ be a $\tau$-nonperiodic
  pattern. For every $\Pat' \in \Sigma^{+}$, $\lcp(\Pat, \Pat') \geq
  3\tau - 1$ implies that $\Pat'$ is $\tau$-nonperiodic.
\end{lemma}
\begin{proof}
  By \cref{def:periodic-pattern}, $\Pat$ being $\tau$-nonperiodic
  implies that either $|\Pat| < 3\tau - 1$, or $|\Pat| \geq 3\tau - 1$
  and $\per(\Pat[1 \dd 3\tau - 1]) > \tfrac{1}{3}\tau$. From
  $\lcp(\Pat, \Pat') \geq 3\tau - 1$, we obtain $|\Pat| \geq 3\tau -
  1$. Thus, it holds $\per(\Pat[1 \dd 3\tau - 1])
  > \tfrac{1}{3}\tau$. Consequently, by $\lcp(\Pat, \Pat') \geq 3\tau
  - 1$, we obtain $\per(\Pat'[1 \dd 3\tau - 1]) = \per(\Pat[1 \dd
  3\tau - 1]) > \tfrac{1}{3}\tau$, and hence $\Pat'$ is
  $\tau$-nonperiodic.
\end{proof}

\begin{lemma}\label{lm:sss-nonempty}
  Let $\tau \geq 1$ be such that $3\tau - 1 \leq \Textlen$ and $\SSS$
  be a $\tau$-synchronizing set of $\Text$. Let also $\ell$ be such
  that $3\tau - 1 \leq \ell$ and $\SSScomp
  = \CompRepr{\ell}{\SSS}{\Text}$ (\cref{def:comp}). Then,
  \begin{enumerate}
  \item It holds $\SSS \neq \emptyset$ and $\max \SSS \geq \Textlen -
    3\tau + 2$,
  \item It holds $\SSScomp \neq \emptyset$ and
    $\max \SSScomp \geq \Textlen - 3\tau + 2$.
  \end{enumerate}
\end{lemma}
\begin{proof}
  1. By the uniqueness of $\Text[\Textlen]$ in $\Text$, it follows
  that $\per(\Text[\Textlen - 3\tau + 2 \dd \Textlen])
  > \tfrac{1}{3}\tau$, i.e., $\Textlen - 3\tau +
  2 \not\in \RTwo{\tau}{\Text}$. By the density property of $\SSS$
  (\cref{def:sss}), we thus have $\SSS \cap [\Textlen - 3\tau +
  2 \dd \Textlen - 2\tau + 2) \neq \emptyset$.  This immediately
  implies both claims.

  2. By the above, $\SSS \neq \emptyset$. Let $s = \max \SSS$. We will
  prove that $s \in \SSScomp$. Since $s \geq \Textlen - 3\tau + 2$,
  this will immediately imply both claims. Recall that
  by \cref{def:comp}, $\SSScomp
  = \SSS \cap \Cover{\ell}{\Text}$. Thus, to show $s \in \SSScomp$, we
  need to prove $s \in \Cover{\ell}{\Text}$. By $s \geq \Textlen -
  3\tau + 2$ and $\ell \geq 3\tau - 1$, it follows that
  $\Textinf[s \dd s + \ell)$ contains the symbol $\Text[\Textlen]$. By
  the uniqueness of $\Text[\Textlen]$ in $\Text$, it thus follows that
  the position $s$ satisfies $s = \min \OccThree{\ell}{s}{\Text}$.
  Since $\Cover{\ell}{\Text}$ is an $\ell$-cover of $\Text$
  (\cref{def:comp}), it thus follows by \cref{lm:cover-equivalence}
  that $s \in \Cover{\ell}{\Text}$.
\end{proof}

\begin{lemma}\label{lm:nonperiodic-pos-lce}
  Let $\tau \in
  [1 \dd \lfloor \tfrac{\Textlen}{2} \rfloor]$, $\SSS$ be a
  $\tau$-synchronizing set of $\Text$, and $j \in [1 \dd \Textlen]$.
  \begin{enumerate}
  \item\label{lm:nonperiodic-pos-lce-it-1}
    Let $\Pat$ be a $\tau$-nonperiodic pattern satisfying $|\Pat| \geq
    3\tau - 1$. Then, $j \in \OccThree{3\tau - 1}{\Pat}{\Text}$ implies that:
    \begin{itemize}
    \item $j \in [1 \dd \Textlen - 3\tau + 2] \setminus \RTwo{\tau}{\Text}$,
    \item $\Text[j \dd \Successor{\SSS}{j} + 2\tau)$ is a
      prefix of $\Pat$.
    \end{itemize}
  \item\label{lm:nonperiodic-pos-lce-it-2}
    Let $j' \in [1 \dd \Textlen - 3\tau +
    2] \setminus \RTwo{\tau}{\Text}$. Then, $j \in \OccThree{3\tau - 1}{j'}{\Text}$
    implies that:
    \begin{itemize}
    \item  $j \in [1 \dd \Textlen - 3\tau + 2] \setminus \RTwo{\tau}{\Text}$,
    \item $\Successor{\SSS}{j} - j
      = \Successor{\SSS}{j'} - j'$.
    \end{itemize}
  \end{enumerate}
\end{lemma}
\begin{proof}

  1. Denote $m = |\Pat|$. By definition of $\OccThree{3\tau -
  1}{\Pat}{\Text}$ and the assumption $m \geq 3\tau - 1$, it holds
  $\lcp(\Text[j \dd \Textlen], \Pat) \geq \min(m, 3\tau - 1) = 3\tau -
  1$.  Consequently, we must have $j + 3\tau - 1 \leq \Textlen + 1$,
  or equivalently, $j \in [1 \dd \Textlen - 3\tau + 2]$. By
  \cref{lm:nonperiodic-pat-lce}, we then obtain that $\T[j \dd \Textlen]$
  is $\tau$-nonperiodic. Thus, $j \in [1 \dd \Textlen - 3\tau + 2]
  \setminus \RTwo{\tau}{\Text}$.

  We now show the second claim. First, note that $\Successor{\SSS}{j}$
  is well-defined by \cref{lm:sss-nonempty} (note that $3\tau -
  1 \leq \Textlen$ follows by $j \in [1 \dd \Textlen - 3\tau + 2]$).
  To show the main claim, note that by the density property of $\SSS$
  (\cref{def:sss}) and $j \in [1 \dd \Textlen - 3\tau +
  2] \setminus \RTwo{\tau}{\Text}$, we have $\SSS \cap [j \dd j
  + \tau) \neq \emptyset$. Thus, $\Successor{\SSS}{j} - j < \tau$, and
  thus letting $D = \Text[j \dd \Successor{\SSS}{j} + 2\tau)$, it
  holds $|D| = (\Successor{\SSS}{j} - j) + 2\tau \leq 3\tau - 1$.
  Consequently, by $\lcp(\Text[j \dd \Textlen], \Pat) \geq 3\tau - 1$,
  we obtain that $D$ is a prefix of $\Pat$.

  2. Denote $\Pat = \Text[j' \dd \Textlen]$ and $m = |\Pat'| \geq
  3\tau - 1$.  By $j' \not\in \RTwo{\tau}{\Text}$, $\Pat$ is
  $\tau$-nonperiodic. Recall that, by definition,
  $\OccThree{3\tau}{j'}{\Text} = \OccThree{3\tau -
  1}{\Pat}{\Text}$. Thus, by $j \in \OccThree{3\tau - 1}{\Pat}{\Text}$
  and \cref{lm:nonperiodic-pos-lce}\eqref{lm:nonperiodic-pos-lce-it-1},
  we obtain that $j \in [1 \dd \Textlen - 3\tau +
  2] \setminus \RTwo{\tau}{\Text}$.

  We now show the second claim. The notation $\Successor{\SSS}{j}$ and
  $\Successor{\SSS}{j'}$ is well-defined by an analogous argument as
  above. Denote $s = \Successor{\SSS}{j}$ and $s'
  = \Successor{\SSS}{j'}$. By the above, we have $j, j' \in
  [1 \dd \Textlen - 3\tau + 2] \setminus \RTwo{\tau}{\Text}$. Thus,
  by \cref{def:sss}, $\SSS \cap [j \dd j + \tau) \neq \emptyset$ and
  $\SSS \cap [j' \dd j' + \tau) \neq \emptyset$. On the other hand,
  note that $\LCE_{\Text}(j, j') \geq 3\tau - 1$. By the consistency
  condition of $\SSS$, we thus have that for every $\delta \in
  [0 \dd \tau)$, $j + \delta \in \SSS$ holds if and only if $j'
  + \delta \in \SSS$. Applying this for $\delta \in [0 \dd s - j]$, we
  thus obtain that $\SSS \cap [j' \dd j' + (s - j)] = \{j' + (s -
  j)\}$.  Consequently, $s' = j' + (s - j)$ which is equivalent to $s
  - j = s' - j'$, i.e., the claim.
\end{proof}

\subsubsection{The Data Structure}\label{sec:sa-nonperiodic-ds}

\paragraph{Definitions}

For every $k \in [4 \dd \lceil \log \Textlen \rceil)$, we denote $\ell_k =
2^k$ and $\tau_k = \lfloor \tfrac{\ell_k}{3} \rfloor$. Note that
$\ell_k \in [16 \dd \Textlen)$. By $\{\SSScompgen{k}\}_{k \in
[4 \dd \lceil \log \Textlen \rceil)}$, we denote the collection obtained
using \cref{pr:comp-sss-construction} for $c = 14$.  For every $k \in
[4 \dd \lceil \log \Textlen \rceil)$, by $\SSS_{k}$ we denote the
$\tau_k$-synchronizing set of $\Text$ satisfying $\SSScompgen{k}
= \CompRepr{14\tau_k}{\SSS_k}{\Text}$ (\cref{def:comp}). Such $\SSS_k$ exists
by \cref{pr:comp-sss-construction}. We also let $n_k =
|\SSScompgen{k}|$ and by $\ArrSSSComp{k}[1 \dd n_k]$ denote an array
containing the elements of $\SSScompgen{k}$ in sorted order.

\paragraph{Components}

The data structure, denoted $\CompSaNonperiodic{\Text}$, to handle
nonperiodic positions consists of three components:

\begin{enumerate}
\item The index core $\CompSaCore{\Text}$ (\cref{sec:sa-core-ds}). It
  needs $\bigO(\SubstringComplexity{\Text} \log \tfrac{\Textlen \log \sigma}{\SubstringComplexity{\Text} \log \Textlen})$
  space.
\item For $k \in [4 \dd \lceil \log \Textlen \rceil)$, we store
  the array $\ArrSSSComp{k}[1 \dd n_{k}]$ in plain form using
  $\bigO(n_{k})$ space. By \cref{pr:comp-sss-construction}, the total
  space is
  $
    \bigO(\textstyle\sum_{k \in [4 \dd \lceil \log \Textlen \rceil)} n_k) =
    \bigO(\SubstringComplexity{\Text} \log \tfrac{\Textlen \log \sigma}{\SubstringComplexity{\Text} \log \Textlen})
  $.
\item For $k \in [4 \dd \lceil \log \Textlen \rceil)$, we store the
  structure from \cref{pr:str-str} for $P = \SSScompgen{k}$ and $q =
  7\tau_k$. By \cref{pr:comp-sss-construction,pr:str-str}, in total
  they need $\bigO(\sum_{k \in [4 \dd \lceil \log \Textlen \rceil)} n_k)
  = \bigO(\SubstringComplexity{\Text} \log \tfrac{\Textlen \log \sigma}{\SubstringComplexity{\Text} \log \Textlen})$ space.
\end{enumerate}

In total, $\CompSaNonperiodic{\Text}$ needs
$\bigO(\SubstringComplexity{\Text} \log \tfrac{\Textlen \log \sigma}{\SubstringComplexity{\Text} \log \Textlen})$ space.

\subsubsection{Basic Combinatorial Properties}\label{sec:sa-nonperiodic-basic}

\begin{lemma}\label{lm:sa-nonperiodic-count}
  Let $\tau \in [1 \dd \lfloor \tfrac{\Textlen}{2} \rfloor]$. Let $\SSS$
  be a $\tau$-synchronizing set of $\Text$ and let $\SSScomp =
  \CompRepr{14\tau}{\SSS}{\Text}$ (\cref{def:comp}).
  Let $\Pts = \StrStrPoints{7\tau}{\SSScomp}{\Text}$
  (\cref{def:str-str}). For any strings $x_l, x_u, y_l, y_u$, it holds:
  \begin{enumerate}[leftmargin=4.5ex]
  \item\label{lm:sa-nonperiodic-count-it-1}
    $\RangeCountFourSide{\Pts}{x_l}{x_u}{y_l}{y_u} =
      |\{s \in \SSS :
      x_l \,{\preceq}\, \revstr{\Textinf[s {-} 7\tau {\dd} s)} \,{\prec}\, x_u
      \text{ and }
      y_l \,{\preceq}\, \Textinf[s {\dd} s {+} 7\tau) \,{\prec}\, y_u\}|$,
  \item\label{lm:sa-nonperiodic-count-it-2}
    $\IncRangeCountThreeSide{\Pts}{x_l}{x_u}{y_u} =
      |\{s \in \SSS :
      x_l \,{\preceq}\, \revstr{\Textinf[s {-} 7\tau {\dd} s)} \,{\prec}\, x_u
      \text{ and }
      \Textinf[s {\dd} s {+} 7\tau) \,{\preceq} y_u\}|$,
  \item\label{lm:sa-nonperiodic-count-it-3}
    $\RangeCountTwoSide{\Pts}{x_l}{x_u} =
      |\{s \in \SSS :
      x_l \,{\preceq}\, \revstr{\Textinf[s {-} 7\tau {\dd} s)}
      \,{\prec}\, x_u\}|$,
  \end{enumerate}
\end{lemma}
\begin{proof}

  1. The proof consists of three steps labeled (a) through (c).

  (a) Denote
  \begin{align*}
    Q &= \{\Textinf[s - 7\tau \dd s + 7\tau) : s \in \SSScomp,\,
         x_l \preceq \revstr{\Textinf[s - 7\tau \dd s)}\prec x_u
         \text{ and }
         y_l \preceq \Textinf[s \dd s + 7\tau) \prec y_u\},\\
    A &= \{i \in [1 \dd \Textlen] : \Textinf[i - 7\tau \dd i + 7\tau) \in Q\}.
  \end{align*}
  In the first step, we prove that $\RangeCountFourSide{\Pts}{x_l}{x_u}{y_l}{y_u}
  = |A|$.  Let $\mathcal{R} = \{(x,y,w,\ell) \in \Pts : x_l \preceq x
  \prec x_u\text{ and } y_l \preceq y \prec y_u\}$. We begin by proving that for
  every $p = (x_p, y_p, w_p, \ell_p) \in \mathcal{R}$, it holds
  $\Textinf[\ell_p - 7\tau \dd \ell_p + 7\tau) \in Q$. By $p \in \Pts$
  (see \cref{def:str-str}), there exists $i_p \in \SSScomp$ such that
  $x = \revstr{\Textinf[i_p - 7\tau \dd i_p)}$, $y = \Textinf[i_p \dd i_p +
  7\tau)$, and $\ell_p$ satisfies $\Textinf[\ell_p - 7\tau \dd \ell_p +
  7\tau) = \Textinf[i_p - 7\tau \dd i_p + 7\tau)$. By $p \in
  \mathcal{R}$, we thus have $x_l \preceq x \prec x_u$, i.e., $x_l
  \preceq \revstr{\Textinf[i_p - 7\tau \dd i_p)} \prec x_u$ and $y_l \preceq y \prec
  y_u$, i.e., $y_l \preceq \Textinf[i_p \dd i_p + 7\tau) \prec y_u$. We have thus
  proved that $\Textinf[i_p - 7\tau \dd i_p + 7\tau) \in Q$. As observed
  above, however, it holds $\Textinf[\ell_p - 7\tau \dd \ell_p + 7\tau) =
  \Textinf[i_p - 7\tau \dd i_p + 7\tau)$.  Thus, $\Textinf[\ell_p - 7\tau
  \dd \ell_p + 7\tau) \in Q$. Let $g : \mathcal{R} \rightarrow Q$ be a
  function defined so that for every $p = (x, y, w, \ell) \in
  \mathcal{R}$, it holds $g(p) = \Textinf[\ell - 7\tau \dd \ell +
  7\tau)$. We prove that $g$ is a bijection.
  \begin{itemize}
  \item Let $p_1, p_2 \in \mathcal{R}$ be such that $g(p_1) =
    g(p_2)$. We will show that $p_1 = p_2$.  For $k \in \{1, 2\}$,
    denote $p_k = (x_k, y_k, w_k, \ell_k)$ and let $i_k \in \SSScomp$
    be such that $x_k = \revstr{\Textinf[i_k - 7\tau \dd i_k)}$, $y_k =
    \Textinf[i_k \dd i_k + 7\tau)$, and $\Textinf[\ell_k - 7\tau \dd \ell_k
    + 7\tau) = \Textinf[i_k - 7\tau \dd i_k + 7\tau)$. Let also $S_k =
    \Textinf[\ell_k - 7\tau \dd \ell_k + 7\tau) = \Textinf[i_k - 7\tau \dd
    i_k + 7\tau)$.  Note that then $g(p_k) = S_k$. By the assumption,
    $S_1 = g(p_1) = g(p_2) = S_2$. Consequently, $\ell_1 = \min\{j \in
    [1 \dd \Textlen] : \Textinf[j - 7\tau \dd j + 7\tau) = S_1\} = \min\{j \in
    [1 \dd \Textlen] : \Textinf[j - 7\tau \dd j + 7\tau) = S_2\} =
    \ell_2$. Similarly, $w_1 = |\{j \in [1 \dd \Textlen] : \Textinf[j - 7\tau
    \dd j + 7\tau) = S_1\}| = |\{j \in [1 \dd \Textlen] : \Textinf[j - 7\tau \dd
    j + 7\tau) = S_2\}| = w_2$. We also have $x_1 = \revstr{\Textinf[i_1
    - 7\tau \dd i_1)} = \revstr{S_1[1 \dd 7\tau]} = \revstr{S_2[1 \dd
    7\tau]} = \revstr{\Textinf[i_2 - 7\tau \dd i_2)} = x_2$ and $y_1 =
    \Textinf[i_1 \dd i_1 + 7\tau) = S_1(7\tau \dd 14\tau] = S_2(7\tau \dd
    14\tau] = \Textinf[i_2 \dd i_2 + 7\tau) = y_2$. We have thus proved
    $p_1 = p_2$.
  \item Let $X \in Q$. Then, there exists $s \in \SSScomp$ such that
    $x_l \preceq \revstr{\Textinf[s - 7\tau \dd s)} \prec x_u$, $y_l \preceq \Textinf[s
    \dd s + 7\tau) \prec y_u$, and $\Textinf[s - 7\tau \dd s + 7\tau) =
    X$. Note that by definition of $\Pts$, $s \in \SSScomp$ implies
    that there exists $p = (x,y,w,\ell) \in \Pts$ such that $x =
    \revstr{\Textinf[s - 7\tau \dd s)}$, $y = \Textinf[s \dd s + 7\tau)$,
    and $\ell$ satisfies $\Textinf[\ell - 7\tau \dd \ell + 7\tau) =
    \Textinf[s - 7\tau \dd s + 7\tau) = X$. By the above observations, we
    thus have $x_l \preceq x \prec x_u$ and $y_l \preceq y \prec y_u$. Thus, $p
    \in \mathcal{R}$. Since, as noted earlier, we have $\Textinf[\ell -
    7\tau \dd \ell + 7\tau) = X$, we thus obtain $g(p) = \Textinf[\ell -
    7\tau \dd \ell + 7\tau) = X$.
  \end{itemize}
  We have thus proved that $g$ is a bijection. Observe now that, by
  definition (see \cref{sec:range-queries}), it holds
  $\RangeCountFourSide{\Pts}{x_l}{x_u}{y_l}{y_u} = \sum_{(x,y,w,\ell) \in
  \mathcal{R}} w$. Observe also (see \cref{def:str-str}) that for
  every $(x,y,w,\ell) \in \Pts$, it holds $w = |\{j \in [1 \dd \Textlen] :
  \Textinf[j - 7\tau \dd j + 7\tau) = \Textinf[\ell - 7\tau \dd \ell +
  7\tau)\}|$. Thus, $\RangeCountFourSide{\Pts}{x_l}{x_u}{y_l}{y_u} =
  \sum_{(x,y,w,\ell) \in \mathcal{R}} |\{j \in [1 \dd \Textlen] : \Textinf[j -
  7\tau \dd j + 7\tau) = \Textinf[\ell - 7\tau \dd \ell + 7\tau)\}|$.
  Observe that since $g$ is bijection, for any $(x_1, y_1, w_1,
  \ell_1), (x_2, y_2, w_2, \ell_2) \in \mathcal{R}$, $(x_1, y_1, w_1,
  \ell_1) \neq (x_2, y_2, w_2, \ell_2)$ implies $\Textinf[\ell_1 - 7\tau
  \dd \ell_1 + 7\tau) \neq \Textinf[\ell_2 - 7\tau \dd \ell_2 +
  7\tau)$. Thus, in the above expression, no position $j \in [1 \dd
  \Textlen]$ is accounted twice (i.e., for different elements of
  $\mathcal{R}$). On the other hand, $g$ being a bijection also
  implies that for every $j \in [1 \dd \Textlen]$, $\Textinf[j - 7\tau \dd j +
  7\tau) \in Q$ implies that there exists some $p = (x, y, w, \ell)
  \in \mathcal{R}$ such that $\Textinf[j - 7\tau \dd j + 7\tau) =
  \Textinf[\ell - 7\tau \dd \ell + 7\tau)$.  Consequently,
  $\RangeCountFourSide{\Pts}{x_l}{x_u}{y_l}{y_u} = \sum_{(x,y,w,\ell) \in
  \mathcal{R}}|\{j \in [1 \dd \Textlen] : \Textinf[j - 7\tau \dd j + 7\tau) =
  \Textinf[\ell - 7\tau \dd \ell + 7\tau)\}| = |\{j \in [1 \dd \Textlen] :
  \Textinf[j - 7\tau \dd j + 7\tau) \in Q\}| = |A|$.

  (b) Denote
  \begin{align*}
    A' &= \{s \in \SSS :
          x_l \preceq \revstr{\Textinf[s - 7\tau \dd s)} \prec x_u
          \text{ and }
          y_l \preceq \Textinf[s \dd s + 7\tau) \prec y_u\}.
  \end{align*}
  In the second step, we prove that $A = A'$.
  \begin{itemize}
  \item Let $i \in A$. Then, there exists $s \in \SSScomp$ such that
    $\Textinf[i - 7\tau \dd i + 7\tau) = \Textinf[s - 7\tau \dd s + 7\tau)$,
    $x_l \preceq \revstr{\Textinf[s - 7\tau \dd s)} \prec x_u$, and
    $y_l \preceq \Textinf[s \dd s + 7\tau) \prec y_u$. Thus, we also have $x_l
    \preceq \revstr{\Textinf[i - 7\tau \dd i)} \prec x_u$ and $y_l \preceq \Textinf[i
    \dd i + 7\tau) \prec y_u$. It therefore remains to show $i \in
    \SSS$. First, note that by $s \in \SSScomp \subseteq \SSS
    \subseteq [1 \dd \Textlen - 2\tau + 1]$, we have $s \in [1 \dd \Textlen - 2\tau
    + 1]$. Thus, by the uniqueness of $\Text[\Textlen]$ in $\Text$ and by $\Textinf[i
    \dd i + 2\tau) = \Textinf[s \dd s + 2\tau)$, we have $i \in [1 \dd \Textlen
    - 2\tau + 1]$. Consequently, by the consistency of $\SSS$
    (\cref{def:sss}), we have $i \in \SSS$.  Thus, $i \in A'$.
  \item Let $s \in A'$. Then, $s \in \SSS$, $x_l \preceq
    \revstr{\Textinf[s - 7\tau \dd s)} \prec x_u$, and $y_l \preceq \Textinf[s \dd s +
    7\tau) \prec y_u$. Denote $X = \Textinf[s - 7\tau \dd s + 7\tau)$ and
    $i_{\rm left} = \min\{i \in [1 \dd \Textlen] : \Textinf[i - 7\tau \dd i +
    7\tau) = X\}$.  We will prove that it holds $i_{\rm left} \in
    \SSScomp$. First, note that $s + 2\tau \leq \Textlen + 1$ and $i_{\rm
    left} \leq s$. Thus, $\Text[s \dd s + 2\tau) = \Text[i_{\rm left} \dd
    i_{\rm left} + 2\tau)$.  By the consistency of $\SSS$
    (\cref{def:sss}), we thus have $i_{\rm left} \in \SSS$. To show
    $i_{\rm left} \in \SSScomp$, we need to additionally prove that
    $i_{\rm left} \in \Cover{14\tau}{\Text}$, i.e., that there exists $j
    \in [1 \dd \Textlen]$ such that $j = \min \OccThree{14\tau}{j}{\Text}$ and
    $i_{\rm left} \in [j \dd j + 14\tau)$. We consider two cases.
    \begin{itemize}
    \item If $i_{\rm left} \leq 14\tau$, then it suffices to take $j = 1$.
    \item Otherwise, let $j = i_{\rm left} - 7\tau$.  By definition of
      $i_{\rm left}$, for every $i \in [1 \dd i_{\rm left})$, we have
      $\Textinf[i - 7\tau \dd i + 7\tau) \neq X$.  This implies that for
      every $i \in [1 \dd i_{\rm left} - 7\tau)$, it holds $\Textinf[i
      \dd i + 14\tau) \neq X$. Thus, by $\Textinf[j \dd j + 14\tau) = X$,
      we obtain $j = \min \OccThree{14\tau}{j}{\Text}$ and hence $i_{\rm
      left} \in [i_{\rm left} - 7\tau \dd i_{\rm left} + 7\tau) = [j
      \dd j + 14\tau)$.
    \end{itemize}
    We have thus proved that there exists $i \in \SSScomp$ such that
    $\Textinf[s - 7\tau \dd s + 7\tau) = \Textinf[i - 7\tau \dd i + 7\tau)$.
    By our assumption on $s$, it holds $x_l \preceq
    \revstr{\Textinf[i - 7\tau \dd i)} \prec x_u$ and
    $y_l \preceq \Textinf[i \dd i + 7\tau) \prec y_u$.
    Thus, $s \in A$.
  \end{itemize}

  (c) Combining (a) and (b), we obtain
  $\RangeCountFourSide{\Pts}{x_l}{x_u}{y_l}{y_u} = |A| = |A'|$, i.e., the claim.

  2. The proof as analogous, except we replace the condition $y_l \preceq \Textinf[s
  \dd s + 7\tau) \prec y_u$ in the definition of $Q$ and $A'$ with
  the condition $\Textinf[s \dd s + 7\tau) \preceq y_u$.

  3. The proof is analogous, except we remove the condition $y_l \preceq \Textinf[s
  \dd s + 7\tau) \prec y_u$ in the definition of $Q$ and $A'$.
\end{proof}

\begin{lemma}\label{lm:sa-nonperiodic-min}
  Let $\tau \in [1 \dd \lfloor \tfrac{\Textlen}{2} \rfloor]$. Let $\SSS$
  be a $\tau$-synchronizing set of $\Text$ and let $\SSScomp =
  \CompRepr{14\tau}{\SSS}{\Text}$. Let $\Pts = \StrStrPoints{7\tau}{\SSScomp}{\Text}$
  (\cref{def:str-str}). For any $x_l, x_u, y_l, y_u$ such that
  $\RangeCountFourSide{\Pts}{x_l}{x_u}{y_l}{y_u} > 0$, it holds
  \[\RangeMinFourSide{\Pts}{x_l}{x_u}{y_l}{y_u} = \min\{s \in \SSS :
  x_l \preceq \revstr{\Textinf[s - 7\tau \dd s)} \prec x_u\text{ and }
  y_l \preceq \Textinf[s \dd s + 7\tau) \prec y_u\}.\]
\end{lemma}
\begin{proof}

  The proof consists of two steps labeled (a) through (b).

  (a) Let $Q$ and $A$ be defined as in the proof
  of \cref{lm:sa-nonperiodic-count}\eqref{lm:sa-nonperiodic-count-it-1},
  i.e., $Q = \{\Textinf[s - 7\tau \dd s + 7\tau) : s \in \SSScomp,\,
  x_l \preceq \revstr{\Textinf[s - 7\tau \dd s)}\prec x_u \text{ and
  }y_l \preceq \Textinf[s \dd s + 7\tau) \prec y_u\}$ and $A = \{i \in
  [1 \dd \Textlen] : \Textinf[i - 7\tau \dd i + 7\tau) \in Q\}$.  Denote
  $\mathcal{R} = \{(x,y,w,\ell) \in \Pts : x_l \preceq x \prec
  x_u\text{ and } y_l \preceq y \prec y_u\}$. In the first step, we
  prove that $\RangeMinFourSide{\Pts}{x_l}{x_u}{y_l}{y_u} = \min A$.
  \begin{itemize}
  \item Let $i = \min A$. By definition of $A$, we then have $\Textinf[i
    - 7\tau \dd i + 7\tau) \in Q$. Denote $S = \Textinf[i - 7\tau \dd i +
    7\tau)$. Recall from the proof
    of \cref{lm:sa-nonperiodic-count}\eqref{lm:sa-nonperiodic-count-it-1},
    that the function $g : \mathcal{R} \rightarrow Q$ defined such
    that for every $p = (x, y, w, \ell) \in \mathcal{R}$, $g(p)
    = \Textinf[\ell - 7\tau \dd \ell + 7\tau)$, is a bijection. Let us
    thus consider $p_S = g^{-1}(S) \in \mathcal{R}$. Denote $p_S =
    (x_S, y_S, w_S, \ell_S)$. Observe, that we then have $\Textinf[\ell_S
    - 7\tau \dd \ell_S + 7\tau) = g(p_S) = S = \Textinf[i - 7\tau \dd i +
    7\tau)$.  By \cref{def:str-str}, it then holds $\ell_S
    = \min\{i' \in [1 \dd \Textlen] : \Textinf[i' - 7\tau \dd i' + 7\tau) =
    S\}$. Consequently, $\ell_S \leq i$ and thus
    $\RangeMinFourSide{\Pts}{x_l}{x_u}{y_l}{y_u}
    = \min_{(x,y,w,\ell) \in \mathcal{R}} \ell \leq \ell_S \leq i
    = \min A$.
  \item Let $p_{\min} = (x_{\min}, y_{\min},
    w_{\min}, \ell_{\min}) \in \mathcal{R}$ be such that
    $\RangeMinFourSide{\Pts}{x_l}{x_u}{y_l}{y_u}
    = \min_{(x,y,w,\ell) \in \mathcal{R}} \ell = \ell_{\min}$. Let $S
    = g(p_{\min}) = \Textinf[\ell_{\min} - 7\tau \dd \ell_{\min} +
    7\tau)$, where $g : \mathcal{R} \rightarrow Q$ is the bijection
    defined above. We then have $S \in Q$. Consequently, by
    $\ell_{\min} \in [1 \dd \Textlen]$ and $\Textinf[\ell_{\min} -
    7\tau \dd \ell_{\min} + 7\tau) = S$, we have $\ell_{\min} \in
    A$. Thus, $\min A \leq \ell_{\min}
    = \RangeMinFourSide{\Pts}{x_l}{x_u}{y_l}{y_u}$.
  \end{itemize}

  (b) Let $A'$ be defined as in the proof
  of \cref{lm:sa-nonperiodic-count}\eqref{lm:sa-nonperiodic-count-it-1},
  i.e., $A' = \{s \in \SSS : x_l \preceq \revstr{\Textinf[s - 7\tau \dd
  s)} \prec x_u\text{ and }y_l \preceq \Textinf[s \dd s + 7\tau) \prec
  y_u\}$. In the proof
  of \cref{lm:sa-nonperiodic-count}\eqref{lm:sa-nonperiodic-count-it-1},
  we showed that $A = A'$.  Combining this with
  $\RangeMinFourSide{\Pts}{x_l}{x_u}{y_l}{y_u} = \min A$ shown above, we obtain
  $\RangeMinFourSide{\Pts}{x_l}{x_u}{y_l}{y_u} = \min A = \min A'
  = \min \{s \in \SSS : x_l \preceq \revstr{\Textinf[s - 7\tau \dd
  s)} \prec x_u\text{ and }y_l \preceq \Textinf[s \dd s + 7\tau) \prec
  y_u\}$, i.e., the claim.
\end{proof}

\subsubsection{Basic Navigation Primitives}\label{sec:sa-nonperiodic-nav}

\begin{lemma}\label{lm:sa-nonperiodic-succ}
  Let $\tau \in [1 \dd \lfloor \tfrac{\Textlen}{2} \rfloor]$ and
  $\ell, \ell', \ell' \in \Zp$ be such that $3\tau -
  1 \leq \ell \leq \ell' \leq \ell''$.  Let $\SSS$ be a
  $\tau$-synchronizing set of $\Text$ and let $\SSScomp
  = \CompRepr{\ell''}{\SSS}{\Text}$.  Let $j \in [1 \dd \Textlen - 3\tau
  + 2] \setminus \RTwo{\tau}{\Text}$ and
  $j' \in \OccThree{\ell}{j}{\Text}$ be a position satisfying $j'
  = \min \OccThree{\ell'}{j'}{\Text}$. Then, $\Successor{\SSS}{j} - j
  = \Successor{\SSScomp}{j'} - j'$.
\end{lemma}
\begin{proof}
  We begin by observing that $j \in [1 \dd \Textlen - 3\tau + 2]$
  implies $3\tau - 1 \leq \Textlen$. It thus follows
  by \cref{lm:sss-nonempty} that $\SSS \neq \emptyset$ and
  $\max \SSS \geq \Textlen - 3\tau + 2$.  Consequently,
  $\Successor{\SSS}{j}$ is well-defined. Next, we prove that $j' \in
  [1 \dd \Textlen - 3\tau + 2]$ and $\Text[j \dd j + 3\tau - 1)
  = \Text[j' \dd j' + 3\tau - 1)$.  By $3\tau - 1 \leq \ell$, we have
  $j' \in \OccThree{\ell}{j}{\Text} \subseteq \OccThree{3\tau -
  1}{j}{\Text}$. Consider two cases:
  \begin{itemize}
  \item If $j = j'$, then the claim follows immediately.
  \item If $j \neq j'$, then by \cref{lm:occ-equivalence},
    $j' \in \OccThree{3\tau - 1}{j}{\Text}$ implies $\LCE_{\Text}(j,
    j') \geq 3\tau - 1$, i.e., $j' \in [1 \dd \Textlen - 3\tau + 2]$
    and $\Text[j \dd j + 3\tau - 1) = \Text[j' \dd j' + 3\tau - 1)$.
  \end{itemize}
  In particular, we have $j' \in
  [1 \dd \Textlen - 3\tau + 2]$.  By \cref{lm:sss-nonempty}, this
  implies $\SSScomp \neq \emptyset$ and $\max \SSScomp \geq \Textlen -
  3\tau + 2$.  Hence, $\Successor{\SSScomp}{j'}$ is well-defined.

  We show the main claim in two steps:
  \begin{enumerate}
  \item By the consistency of $\SSS$ (\cref{def:sss}), it follows that for
    every $t \in [0 \dd \tau)$, $j' + t \in \SSS$ holds if and only if
    $j + t \in \SSS$.  Observe, however, that by the density condition
    (\cref{def:sss}), we have $\SSS \cap [j \dd j
    + \tau) \neq \emptyset$. Consequently, letting $s_j
    = \Successor{\SSS}{j}$, it holds $s_j - j < \tau$. By the above
    property for $t \in [0 \dd s_j - j]$ we thus obtain that $s_{j'} -
    j' = s_j - j$, where $s_{j'} = \Successor{\SSS}{j'}$.

  \item Denote $s' = \Successor{\SSScomp}{j'}$.  Observe that $j'
    = \min \OccThree{\ell''}{j'}{\Text}$. Otherwise, there would exist
    $j'' \in [1 \dd j')$ such that $\Textinf[j'' \dd j'' + \ell'')
    = \Textinf[j' \dd j' + \ell'')$.  By $\ell' \leq \ell''$, we would
    then have $\Textinf[j'' \dd j'' + \ell') = \Textinf[j' \dd j'
    + \ell')$, which would contradict $j'
    = \min \OccThree{\ell'}{j'}{\Text}$.  Recalling from \cref{def:comp}
    that $\SSScomp = \SSS \cap \Cover{\ell''}{\Text}$, it follows that
    $[j' \dd j' + \ell'') \subseteq \Cover{\ell''}{\Text}$, and thus for
    every $q \in [0 \dd \ell'')$, $j' + q \in \SSS$ holds if and only if
    $j' + q \in \SSScomp$.  Applying this for $q \in [0 \dd s_{j'} -
    j']$ and recalling that $s_{j'} - j' < \tau$, we obtain
    $\Successor{\SSScomp}{j'} - j' = \Successor{\SSS}{j'} - j'$, i.e.,
    $s' - j' = s_{j'} - j'$. Combining this with $s_j - j = s_{j'} -
    j'$, we thus have $s_j - j = s_{j'} - j' = s' - j'$.
    \qedhere
  \end{enumerate}
\end{proof}

\begin{proposition}\label{pr:sa-nonperiodic-succ}
  Let $k \in [4 \dd \lceil \log \Textlen \rceil)$, $\ell = 2^k$, $\tau
  = \lfloor \tfrac{\ell}{3} \rfloor$, and $j \in [1 \dd
  \Textlen - 3\tau + 2] \setminus \RTwo{\tau}{\Text}$. Given the structure $\CompSaNonperiodic{\Text}$, the
  value $k$, the position $j$, and any $j' \in \OccThree{\ell}{j}{\Text}$
  satisfying $j' = \min \OccThree{2\ell}{j'}{\Text}$ as input, we can compute
  $\Successor{\SSS_k}{j}$ in $\bigO(\log \Textlen)$ time.
\end{proposition}
\begin{proof}
  First, using binary search over the array $\ArrSSSComp{k}$, we
  compute $s' = \Successor{\SSScompgen{k}}{j'}$ in $\bigO(\log \Textlen)$
  time.  Observe now that by $\ell \geq 16$ and $\tau
  = \lfloor \tfrac{\ell}{3} \rfloor$, it holds $2\ell \leq
  14\tau$. By \cref{lm:sa-nonperiodic-succ} (with $\ell = \ell$,
  $\ell' = 2\ell$, and $\ell'' = 14\tau$), we then obtain
  $\Successor{\SSS_k}{j} = j + (s' - j')$.
\end{proof}

\subsubsection{Computing the Size of Posbeg and Posend}

\paragraph{Combinatorial Properties}

\begin{lemma}\label{lm:sa-nonperiodic-pat-posbeg-posend}
  Let $\ell \in [16 \dd \Textlen)$ and $\tau = \lfloor \tfrac{\ell}{3} \rfloor$.
  Let $\SSS$ be a $\tau$-synchronizing set of $\Text$.
  Let $\Pat \in \Sigma^{m}$ be a $\tau$-nonperiodic pattern such that
  $m \geq 3\tau - 1$, no nonempty suffix of $\Text$ is a proper prefix of
  $\Pat$, and $\Pat$ is prefixed with some $D \in \DistinguishingPrefs{\tau}{\Text}{\SSS}$ (\cref{def:dist-prefix}).
  Let $c = \max \Sigma$ and $x, x', y, y'$ be such that:
  \begin{itemize}
  \item $\revstr{x}$ is a prefix of $D$ and $|\revstr{x}| + 2\tau =
    |D|$,
  \item $x' = x c^{\infty}$,
  \item $\revstr{x}y = \Pat[1 \dd \min(m, \ell)]$,
  \item $\revstr{x}y' = \Pat[1 \dd \min(m, 2\ell)]$.
  \end{itemize}
  Then, it holds:
  \begin{enumerate}
  \item\label{lm:sa-nonperiodic-pat-posbeg-posend-it-1}
    $\PosBeg{\ell}{\Pat}{\Text} = \{s - |x| : s \in \SSS,\
    x \preceq \revstr{\Textinf[s {-} 7\tau \dd s)} \prec x',\text{ and }
    y \preceq \Textinf[s \dd s {+} 7\tau) \prec y'\}$,
  \item\label{lm:sa-nonperiodic-pat-posbeg-posend-it-2}
    $\PosEnd{\ell}{\Pat}{\Text} = \{s - |x| : s \in \SSS,\
    x \preceq \revstr{\Textinf[s {-} 7\tau \dd s)} \prec x',\text{ and }
    y \preceq_{\rm inv} \Textinf[s \dd s {+} 7\tau) \prec_{\rm inv} y'\}$.
  \end{enumerate}
\end{lemma}
\begin{proof}

  1. Let $j \in \PosBeg{\ell}{\Pat}{\Text}$. Denote $s = j + |x|$ We
  need to prove (a) $s \in \SSS$, (b) $x \preceq \revstr{\Textinf[s -
  7\tau \dd s)} \prec x'$, and (c) $y \preceq \Textinf[s \dd s + 7\tau)
  \prec y'$.
  \begin{enumerate}[label=(\alph*)]
  \item By \cref{lm:equiv}, $\Text[j \dd \Textlen] \prec \Pat$ and $\lcp(\Pat,
    \Text[j \dd \Textlen]) \in [\ell \dd 2\ell)$.  On the other hand, by $D \in
    \DistinguishingPrefs{\tau}{\Text}{\SSS}$ there exists $j' \in [1 \dd \Textlen - 3\tau + 2] \setminus \RTwo{\tau}{\Text}$ such
    that $j' \in \OccTwo{D}{\Text}$ and $\Successor{\SSS}{j'} = j' + |D| -
    2\tau = j' + |x|$. Thus, $j' + |x| \in \SSS$.  Since $j' \in [1
    \dd \Textlen - 3\tau + 2] \setminus \RTwo{\tau}{\Text}$ implies $[j' \dd j' + \tau) \cap
    \SSS \neq \emptyset$, we thus obtain $|x| < \tau$ and hence $|D| =
    |x| + 2\tau \leq 3\tau - 1 \leq \ell$. By $\lcp(\Pat, \Text[j \dd \Textlen])
    \geq \ell$ and $D$ being a prefix of $\Pat$, we therefore have $j
    \in \OccTwo{D}{\Text}$. Consequently, $\Text[j' + |x| \dd j' + |x| + 2\tau)
    = \Text[j + |x| \dd j + |x| + 2\tau)$, and hence by the consistency
    of $\SSS$, it holds $s = j + |x| \in \SSS$.
  \item By the above, $j \in \OccTwo{D}{\Text}$.  Since $\revstr{x}$ is a
    prefix of $D$, we thus have $\Text[j \dd j + |x|) = \revstr{x}$, or
    equivalently, $\revstr{\Text[j \dd j + |x|)} = x$, which we can write
    as $\revstr{\Text[s - |x| \dd s)} = x$.  Since $|x| < \tau$, this
    implies that $x$ is a prefix of $\revstr{\Textinf[s - 7\tau \dd
    s)}$. By $x' = x c^{\infty}$, we thus have $x \preceq
    \revstr{\Textinf[s {-} 7\tau \dd s)} \prec x'$.
  \item Above we observed that $\revstr{x}$ is a prefix of both $\Pat$
    and $\Text[j \dd \Textlen]$.  Thus, recalling that $j + |x| = s$ and letting
    $\Pat'$ be such that $\revstr{x}\Pat' = \Pat$, the conditions
    $\Text[j \dd \Textlen] \prec \Pat$ and $\lcp(\Pat, \Text[j \dd \Textlen]) \in [\ell
    \dd 2\ell)$ imply $\Text[s \dd \Textlen] \prec \Pat'$ and $\lcp(\Pat', \Text[s
    \dd \Textlen]) \in [\ell - |x| \dd 2\ell - |x|)$. Denote $m' = |\Pat'| =
    m - |x|$.  By \cref{lm:sa-prelim}\eqref{lm:sa-prelim-it-3} with
    $\ell_1 = \ell - |x|$ and $\ell_2 = 2\ell - |x|$, it holds $\Pat_1
    \preceq \Text[s \dd \Textlen] \prec \Pat_2$, where $\Pat_1 = \Pat'[1 \dd
    \min(m', \ell_1)] = \Pat'[1 \dd \min(m, \ell) - |x|]$ and $\Pat_2
    = \Pat'[1 \dd \min(m', \ell_2)] = \Pat'[1 \dd \min(m, 2\ell) -
    |x|]$. Note that $\ell_1 < \ell_2 \leq 7\tau$.  By
    \cref{lm:sa-prelim-inf}\eqref{lm:sa-prelim-inf-it-3}, we thus
    obtain $\Pat_1 \preceq \Textinf[s \dd s + 7\tau) \prec \Pat_2$.  It
    remains to observe that $\revstr{x}\Pat_1$ and $\revstr{x}\Pat_2$
    are prefixes of $\Pat$, and it holds $|\revstr{x}\Pat_1| =
    |\revstr{x}| + \min(m, \ell) - |x| = \min(m, \ell)$ and
    $|\revstr{x}\Pat_2| = |\revstr{x}| + \min(m, 2\ell) - |x| =
    \min(m, 2\ell)$. Thus, $\Pat_1 = y$ and $\Pat_2 = y'$, and hence
    we obtain $y \preceq \Textinf[s \dd s + 7\tau) \prec y'$.
  \end{enumerate}

  Let us now consider any $s \in \SSS$ satisfying $x \preceq
  \revstr{\Textinf[s - 7\tau \dd s)} \prec x'$ and $y \preceq \Textinf[s \dd
  s + 7\tau) \prec y'$. Denote $j = s - |x|$. We will prove that $j
  \in \PosBeg{\ell}{\Pat}{\Text}$. By \cref{lm:equiv}, this is
  equivalent to (a) $j \in [1 \dd \Textlen]$ and (b) $\Text[j \dd \Textlen] \prec \Pat$
  and $\lcp(\Pat, \Text[j \dd \Textlen]) \geq [\ell \dd 2\ell)$.
  \begin{enumerate}[label=(\alph*)]
  \item Recall that $x' = x c^{\infty}$. The inequality $x \preceq
    \revstr{\Textinf[s - 7\tau \dd s)} \prec x'$ thus implies that $x$ is
    a prefix of $\revstr{\Textinf[s - 7\tau \dd s)}$, or equivalently,
    $\Textinf[s - |x| \dd s) = \revstr{x}$. Suppose that $s - |x| <
    1$. Then, there exists a nonempty prefix of $\revstr{x}$ that is a
    suffix of $\Text$. Note, however, that $|x| < |D| \leq m$, i.e.,
    $\revstr{x}$ is a proper prefix of $\Pat$. Thus, this contradicts
    the assumption about $\Text$ not having any nonempty suffix that is a
    proper prefix of $\Pat$. We thus have $s - |x| \geq 1$, and hence
    $j \in [1 \dd \Textlen]$.
  \item Recall that $|y| = \min(m, \ell) - |x| \leq \ell$ and $|y'| =
    \min(m, 2\ell) - |x| \leq 2\ell$. By $2\ell \leq 7\tau$ and
    \cref{lm:sa-prelim-inf}\eqref{lm:sa-prelim-inf-it-3}, we thus
    obtain $y \preceq \Text[s \dd \Textlen] \prec y'$. Next, note that letting
    $\Pat'$ be such that $\revstr{x}\Pat' = \Pat$, it holds $y =
    \Pat'[1 \dd \min(m', \ell_1)]$ and $y' = \Pat'[1 \dd \min(m,
    \ell_2)]$, where $m' = |\Pat'| = m - |x|$, $\ell_1 = \ell - |x|$
    and $\ell_2 = 2\ell - |x|$.  By $\ell_1 < \ell_2$ and
    \cref{lm:sa-prelim}\eqref{lm:sa-prelim-it-3}, we thus obtain $\Text[s
    \dd \Textlen] \prec \Pat'$ and $\lcp(\Pat', \Text[s \dd \Textlen]) \in [\ell_1 \dd
    \ell_2) = [\ell - |x| \dd 2\ell - |x|)$.  It remains to note that
    above we observed that $\Text[s - |x| \dd s) = \revstr{x}$.  Thus, we
    obtain $\Text[j \dd \Textlen] = \Text[s - |x| \dd \Textlen] \prec \revstr{x}\Pat' =
    \Pat$ and $\lcp(\Pat, \Text[j \dd \Textlen]) = \lcp(\revstr{x}\Pat',
    \revstr{x}\Text[s \dd \Textlen]) = |x| + \lcp(\Pat', \Text[s \dd \Textlen]) \in [\ell
    \dd 2\ell)$.
  \end{enumerate}

  2. The proof follows by first applying \cref{lm:equiv} and then
  symmetrically repeating all arguments above.
\end{proof}

\begin{lemma}\label{lm:sa-nonperiodic-posbeg-posend}
  Let $\ell \in [16 \dd \Textlen)$ and $\tau = \lfloor \tfrac{\ell}{3} \rfloor$.
  Let $\SSS$ be a $\tau$-synchronizing set of $\Text$. Let $j \in
  [1 \dd \Textlen - 3\tau + 2] \setminus \RTwo{\tau}{\Text}$, $c = \max \Sigma$,
  $s = \Successor{\SSS}{j}$, and $x, x', y, y'$ be such that:
  \begin{itemize}
  \item $\revstr{x} = \Text[j \dd s)$,
  \item $x' = x c^{\infty}$,
  \item $\revstr{x}y = \Text[j \dd \min(\Textlen + 1, j + \ell))$,
  \item $\revstr{x}y' = \Text[j \dd \min(\Textlen + 1, j + 2\ell))$.
  \end{itemize}
  Then:
  \begin{enumerate}
  \item\label{lm:sa-nonperiodic-posbeg-posend-it-1}
    $\PosBeg{\ell}{j}{\Text} \,{=}\, \{s \,{-}\, |x| : s \in \SSS,\
    x \preceq \revstr{\Textinf[s {-} 7\tau \dd s)} \prec x',
    \text{ and }
    y \preceq \Textinf[s \dd s {+} 7\tau) \prec y'\}$,
  \item\label{lm:sa-nonperiodic-posbeg-posend-it-2}
    $\PosEnd{\ell}{j}{\Text} \,{=}\, \{s \,{-}\, |x| : s \in \SSS,\
    x \preceq \revstr{\Textinf[s {-} 7\tau \dd s)} \prec x',
    \text{ and }
    y \preceq_{\rm inv} \Textinf[s \dd s {+} 7\tau) \prec_{\rm inv} y'\}$.
  \end{enumerate}
\end{lemma}
\begin{proof}

  1. Denote $\Pat = \Text[j \dd \Textlen]$ and $m = |\Pat| = \Textlen - j + 1$.  Note
  that $j \in [1 \dd \Textlen - 3\tau + 2]$ and $j \not\in \RTwo{\tau}{\Text}$ imply that $m
  \geq 3\tau - 1$ and $\per(\Pat[1 \dd 3\tau - 1]) >
  \tfrac{1}{3}\tau$. Thus, $\Pat$ is $\tau$-nonperiodic. Next,
  observe that since $\Text[\Textlen]$ does not occur in $\Text[1 \dd \Textlen)$, no
  nonempty suffix of $\Text$ is a proper prefix of $\Pat$. Next, note
  that by $j \in [1 \dd \Textlen - 3\tau + 2] \setminus \RTwo{\tau}{\Text}$, it holds $[j \dd
  j + \tau) \cap \SSS \neq \emptyset$. Thus, $s - j < \tau$, and hence
  letting $D = \Text[j \dd s + 2\tau)$ it holds $D \in \DistinguishingPrefs{\tau}{\Text}{\SSS}$.
  Thus, $\Pat$ has $D$ as a prefix. Observe that $\revstr{x}$ is a prefix of $D$
  and it holds $|\revstr{x}| + 2\tau = |D|$. Finally, observe that it
  holds $\min(\Textlen + 1, j + \ell) - j = \min(\Textlen + 1 - j, \ell) = \min(m,
  \ell)$ and $\min(\Textlen + 1, j + 2\ell) - j = \min(\Textlen + 1 - j, 2\ell) =
  \min(m, 2\ell)$. Thus, $\revstr{x}y = \Text[j \dd \min(\Textlen + 1, j +
  \ell)) = \Pat[1 \dd \min(m, \ell)]$ and $\revstr{x}y' = \Text[j \dd
  \min(\Textlen + 1, j + 2\ell)) = \Pat[1 \dd \min(m, 2\ell)]$. We have thus
  proved that all assumptions of
  \cref{lm:sa-nonperiodic-pat-posbeg-posend} hold for $\Pat$. The
  claim thus follows by combining the definition of
  $\PosBeg{\ell}{j}{\Text}$ and
  \cref{lm:sa-nonperiodic-pat-posbeg-posend}\eqref{lm:sa-nonperiodic-pat-posbeg-posend-it-1},
  i.e., $\PosBeg{\ell}{j}{\Text} = \PosBeg{\ell}{\Pat}{\Text} = \{s - |x| : s
  \in \SSS,\ x \preceq \revstr{\Textinf[s - 7\tau \dd s)} \prec x',
  \text{ and } y \preceq \Textinf[s \dd s + 7\tau) \prec y'\}$.

  2. The proof proceeds symmetrically, except we utilize
  \cref{lm:sa-nonperiodic-pat-posbeg-posend}\eqref{lm:sa-nonperiodic-pat-posbeg-posend-it-2}.
\end{proof}

\begin{lemma}\label{lm:sa-nonperiodic-posbeg-posend-size}
  Let $\ell \in [16 \dd \Textlen)$ and $\tau = \lfloor \tfrac{\ell}{3} \rfloor$.
  Let $\SSS$ be a $\tau$-synchronizing set of $\Text$ and
  $\SSScomp = \CompRepr{14\tau}{\SSS}{\Text}$.
  Let $j \in [1 \dd \Textlen - 3\tau + 2] \setminus \RTwo{\tau}{\Text}$, $c = \max \Sigma$,
  $s = \Successor{\SSS}{j}$, and $x, x', y, y'$ be such that:
  \begin{itemize}
  \item $\revstr{x} = \Text[j \dd s)$,
  \item $x' = x c^{\infty}$,
  \item $\revstr{x}y = \Text[j \dd \min(\Textlen + 1, j + \ell))$,
  \item $\revstr{x}y' = \Text[j \dd \min(\Textlen + 1, j + 2\ell))$.
  \end{itemize}
  Then, $|\PosBeg{\ell}{j}{\Text}| = \RangeCountFourSide{\Pts}{x}{x'}{y}{y'}$,
  where $\Pts = \StrStrPoints{7\tau}{\SSScomp}{\Text}$ (\cref{def:str-str}).
\end{lemma}
\begin{proof}

  Increasing all elements of a set by the same value does change
  its cardinality. By
  \cref{lm:sa-nonperiodic-posbeg-posend}\eqref{lm:sa-nonperiodic-posbeg-posend-it-1}
  and \cref{lm:sa-nonperiodic-count}, we thus obtain
  \begin{align*}
    |\PosBeg{\ell}{j}{\Text}|
      &= |\{s - |x| : s \in \SSS,\,
         x \preceq \revstr{\Textinf[s {-} 7\tau \dd s)} \prec x'
         \text{ and }
         y \preceq \Textinf[s \dd s {+} 7\tau) \prec y'\}|\\
      &= |\{s \in \SSS :
         x \preceq \revstr{\Textinf[s {-} 7\tau \dd s)} \prec x'
         \text{ and }
         y \preceq \Textinf[s \dd s {+} 7\tau) \prec y'\}|\\
      &= \RangeCountFourSide{\Pts}{x}{x'}{y}{y'}.
  \end{align*}
\end{proof}

\paragraph{Query Algorithms}

\begin{proposition}\label{pr:sa-nonperiodic-posbeg-posend-size}
  Let $k \in [4 \dd \lceil \log \Textlen \rceil)$, $\ell = 2^k$, $\tau
  = \lfloor \tfrac{\ell}{3} \rfloor$, and $j \in [1 \dd
  \Textlen] \setminus \RTwo{\tau}{\Text}$. Given the structure $\CompSaNonperiodic{\Text}$, the
  value $k$, the position $j$, and any $j' \in \OccThree{\ell}{j}{\Text}$
  satisfying $j' = \min \OccThree{2\ell}{j'}{\Text}$ as input, we can compute
  $|\PosBeg{\ell}{j}{\Text}|$ in $\bigO(\log^{2+\epsilon} \Textlen)$ time.
\end{proposition}
\begin{proof}

  If $j > \Textlen - 3\tau + 2$ then by $3\tau - 1 \leq \ell$ and the
  uniqueness of $\Text[\Textlen]$ in $\Text$, it holds $|\OccThree{\ell}{j}{\Text}| =
  1$. By $|\OccThree{2\ell}{j}{\Text}| > 0$ and
  $\OccThree{2\ell}{j}{\Text} \subseteq \OccThree{\ell}{j}{\Text}$ we return
  $|\PosBeg{\ell}{j}{\Text}| = 0$.

  Let us now assume $j \in [1 \dd \Textlen - 3\tau + 2]$. Denote $\Pts
  = \StrStrPoints{7\tau}{\SSScompgen{k}}{\Text}$. The algorithm consists of
  two steps:
  \begin{enumerate}
  \item First, using \cref{pr:sa-nonperiodic-succ}, we compute $s
    = \Successor{\SSS_k}{j}$ in $\bigO(\log \Textlen)$ time.
  \item Let $x, x', y, y'$ be such that $\Text[j \dd s) = \revstr{x}$,
    $x' = x c^{\infty}$, $\revstr{x}y = \Text[j \dd \min(\Textlen + 1, j
    + \ell))$, and $\revstr{x}y' = \Text[j \dd \min(\Textlen + 1, j + 2\ell))$.
    By \cref{lm:sa-nonperiodic-posbeg-posend-size},
    $|\PosBeg{\ell}{j}{\Text}| = \RangeCountFourSide{\Pts}{x}{x'}{y}{y'}
    = \RangeCountThreeSide{\Pts}{x}{x'}{y'} - \RangeCountThreeSide{\Pts}{x}{x'}{y}$
    (where the second equality holds by $y \preceq y'$). Observe also
    that using the structure $\CompSaNonperiodic{\Text}$, we can perform $\LCE_{\Text}$
    and $\LCE_{\revstr{\Text}}$ queries in $\bigO(\log \Textlen)$ time, and we
    can access any symbol of $\Text$ in $\bigO(\log \Textlen)$ time. Thus, we
    can compare any two substrings of $\Textinf$ or $\revstr{\Textinf}$
    (specified with their starting positions and lengths) in $t_{\rm
    cmp} = \bigO(\log \Textlen)$ time.  Thus, using the structure
    from \cref{pr:str-str} for $q = 7\tau$ and $P = \SSScompgen{k}$ (which
    is a component of $\CompSaNonperiodic{\Text}$) first with parameters
    $(i, q_l, q_r) = (s, s - j, \min(\Textlen + 1, j + 2\ell) - s)$ and then
    $(i, q_l, q_r) = (s, s - j, \min(\Textlen + 1, j + \ell) - s)$ we compute
    $|\PosBeg{\ell}{j}{\Text}|$ in $\bigO(\log^{2 + \epsilon} \Textlen +
    t_{\rm cmp} \log \Textlen) = \bigO(\log^{2 + \epsilon} \Textlen)$ time.
  \end{enumerate}
  In total, we spend $\bigO(\log^{2 + \epsilon} \Textlen)$ time.
\end{proof}

\subsubsection{Computing the Size of Occ}

\paragraph{Combinatorial Properties}

\begin{lemma}\label{lm:sa-nonperiodic-pat-occ}
  Let $\ell \in [16 \dd \Textlen)$ and $\tau = \lfloor \tfrac{\ell}{3} \rfloor$.
  Let $\SSS$ be a $\tau$-synchronizing set of $\Text$.
  Let $\Pat \in \Sigma^{m}$ be a $\tau$-nonperiodic pattern such that
  $m \geq 3\tau - 1$, no nonempty suffix of $\Text$ is a proper prefix of
  $\Pat$, and $\Pat$ is prefixed with some $D \in \DistinguishingPrefs{\tau}{\Text}{\SSS}$ (\cref{def:dist-prefix}).
  Let $d \in [\ell \dd 2\ell]$, $c = \max \Sigma$, and $x, x', y, y'$
  be such that:
  \begin{itemize}
  \item $\revstr{x}$ is a prefix of $D$ and $|\revstr{x}| + 2\tau =
    |D|$,
  \item $x' = x c^{\infty}$,
  \item $\revstr{x}y = \Pat[1 \dd \min(m, d)]$,
  \item $y' = y c^{\infty}$.
  \end{itemize}
  Then, $\OccThree{d}{\Pat}{\Text} = \{s - |x| : s \in \SSS,\
  x \preceq \revstr{\Textinf[s {-} 7\tau \dd s)} \prec x',\text{ and }
  y \preceq \Textinf[s \dd s {+} 7\tau) \prec y'\}$.
\end{lemma}
\begin{proof}

  Let $j \in \OccThree{d}{\Pat}{\Text}$. Denote $s = j + |x|$ We need to
  prove (a) $s \in \SSS$, (b) $x \preceq \revstr{\Textinf[s - 7\tau \dd
  s)} \prec x'$, and (c) $y \preceq \Textinf[s \dd s + 7\tau) \prec y'$.
  \begin{enumerate}[label=(\alph*)]
  \item By \cref{def:occ}, it holds $j \in [1 \dd \Textlen]$ and
    $\lcp(\Pat, \Text[j \dd \Textlen]) \geq \min(m, d)$. By $D \in \DistinguishingPrefs{\tau}{\Text}{\SSS}$
    there exists $j' \in [1 \dd \Textlen - 3\tau + 2] \setminus \RTwo{\tau}{\Text}$ such that
    $j' \in \OccTwo{D}{\Text}$ and $\Successor{\SSS}{j'} = j' + |D| - 2\tau
    = j' + |x|$. Thus, $j' + |x| \in \SSS$.  Since $j' \in [1 \dd \Textlen -
    3\tau + 2] \setminus \RTwo{\tau}{\Text}$ implies $[j' \dd j'
    + \tau) \cap \SSS \neq \emptyset$, we thus obtain $|x| < \tau$ and
    hence $|D| = |x| + 2\tau \leq 3\tau - 1 \leq \ell$. By
    $\lcp(\Pat, \Text[j \dd \Textlen]) \geq \min(m, d) \geq 3\tau - 1$ and by
    $D$ being a prefix of $\Pat$, we therefore have
    $j \in \OccTwo{D}{\Text}$. Consequently, $\Text[j' + |x| \dd j' + |x| +
    2\tau) = \Text[j + |x| \dd j + |x| + 2\tau)$, and hence by the
    consistency of $\SSS$, it holds $s = j + |x| \in \SSS$.
  \item By the above, $j \in \OccTwo{D}{\Text}$.  Since $\revstr{x}$ is a
    prefix of $D$, we thus have $\Text[j \dd j + |x|) = \revstr{x}$, or
    equivalently, $\revstr{\Text[j \dd j + |x|)} = x$, which we can write
    as $\revstr{\Text[s - |x| \dd s)} = x$.  Since $|x| < \tau$, this
    implies that $x$ is a prefix of $\revstr{\Textinf[s - 7\tau \dd
    s)}$. By $x' = x c^{\infty}$, we thus have
    $x \preceq \revstr{\Textinf[s {-} 7\tau \dd s)} \prec x'$.
  \item Recall that $\revstr{x}y = \Pat[1 \dd \min(m, d)]$. On the
    other hand, we assumed $\lcp(\Pat, \Text[j \dd \Textlen]) \geq \min(m,
    d)$. Thus, $\revstr{x}y$ is a prefix of $\Text[j \dd \Textlen]$. In
    particular, $y$ is a prefix of $\Text[s \dd \Textlen]$. Since $|y| \leq
    d \leq 2\ell \leq 7\tau$, we thus obtain that $y$ if a prefix of
    $\Textinf[s \dd s + 7\tau)$. By $y' = y c^{\infty}$, we thus have
    $y \preceq \Textinf[s \dd s + 7\tau) \prec y'$.
  \end{enumerate}

  Let us now consider any $s \in \SSS$ satisfying
  $x \preceq \revstr{\Textinf[s - 7\tau \dd s)} \prec x'$ and
  $y \preceq \Textinf[s \dd s + 7\tau) \prec y'$. Denote $j = s -
  |x|$. We will prove that $j \in \OccThree{d}{\Pat}{\Text}$, i.e., (a)
  $j \in [1 \dd \Textlen]$ and (b) $\lcp(\Pat, \Text[j \dd \Textlen]) \geq \min(m, d)$.
  \begin{enumerate}[label=(\alph*)]
  \item Recall that $x' = x c^{\infty}$. The inequality
    $x \preceq \revstr{\Textinf[s - 7\tau \dd s)} \prec x'$ thus implies
    that $x$ is a prefix of $\revstr{\Textinf[s - 7\tau \dd s)}$, or
    equivalently, $\Textinf[s - |x| \dd s) = \revstr{x}$. Suppose that $s
    - |x| < 1$. Then, there exists a nonempty prefix of $\revstr{x}$
    that is a suffix of $\Text$. Note, however, that $|x| < |D| \leq m$,
    i.e., $\revstr{x}$ is a proper prefix of $\Pat$. Thus, this
    contradicts the assumption about $\Text$ not having any nonempty
    suffix that is a proper prefix of $\Pat$. We thus have $s -
    |x| \geq 1$, and hence $j \in [1 \dd \Textlen]$.
  \item Recall that $y' = y c^{\infty}$. The assumption
    $y \preceq \Textinf[s \dd s + 7\tau) \prec y'$ thus implies that $y$
    is a prefix of $\Textinf[s \dd s + 7\tau)$. Combining with the above,
    we thus obtain that $\revstr{x}y$ is a prefix of $\Textinf[j \dd s +
    7\tau)$, i.e., $\Textinf[j \dd j + \min(m, d))
    = \revstr{x}y$. Suppose that $j + \min(m, d) > \Textlen + 1$. Then, there
    exists a nonempty suffix of $\Text$ that is a proper prefix of
    $\revstr{x}y$. But since $\revstr{x}y$ is a prefix of $\Pat$, this
    contradicts the assumption about $\Text$ not having a nonempty suffix
    that is a proper prefix of $\Pat$. Thus, we have $j + \min(\Textlen,
    d) \leq \Textlen + 1$, and hence $\revstr{x}y = \Text[j \dd j + \min(m,
    d))$. By $\revstr{x}y = \Pat[1 \dd \min(m, d)]$, this implies
    $\lcp(\Pat, \Text[j \dd \Textlen]) \geq |\revstr{x}y| = \min(m, d)$.
    \qedhere
  \end{enumerate}
\end{proof}

\begin{lemma}\label{lm:sa-nonperiodic-pos-occ}
  Let $\ell \in [16 \dd \Textlen)$ and $\tau = \lfloor \tfrac{\ell}{3} \rfloor$.
  Let $\SSS$ be a $\tau$-synchronizing set of $\Text$.
  Let $j \in [1 \dd \Textlen - 3\tau + 2] \setminus \RTwo{\tau}{\Text}$, $c = \max \Sigma$,
  $s = \Successor{\SSS}{j}$, $d \in [\ell \dd 2\ell]$, and $x, x', y,
  y'$ be such that:
  \begin{itemize}
  \item $\revstr{x} = \Text[j \dd s)$,
  \item $x' = x c^{\infty}$,
  \item $\revstr{x}y = \Text[j \dd \min(\Textlen + 1, j + d))$,
  \item $y' = y c^{\infty}$.
  \end{itemize}
  Then, $\OccThree{d}{j}{\Text} \,{=}\, \{s \,{-}\, |x| : s \in \SSS,\
  x \preceq \revstr{\Textinf[s {-} 7\tau \dd s)} \prec x', \text{ and }
  y \preceq \Textinf[s \dd s {+} 7\tau) \prec y'\}$.
\end{lemma}
\begin{proof}
  Denote $\Pat = \Text[j \dd \Textlen]$ and $m = |\Pat| = \Textlen - j + 1$.  Note that
  $j \in [1 \dd \Textlen - 3\tau + 2]$ and $j \not\in \RTwo{\tau}{\Text}$ imply that $m \geq
  3\tau - 1$ and $\per(\Pat[1 \dd 3\tau - 1]) > \tfrac{1}{3}\tau$.
  Thus, $\Pat$ is $\tau$-nonperiodic. Next, observe that since $\Text[\Textlen]$
  does not occur in $\Text[1 \dd \Textlen)$, no nonempty suffix of $\Text$ is a
  proper prefix of $\Pat$. Next, note that by $j \in [1 \dd \Textlen - 3\tau
  + 2] \setminus \RTwo{\tau}{\Text}$, it holds $[j \dd j
  + \tau) \cap \SSS \neq \emptyset$. Thus, $s - j < \tau$, and hence
  letting $D = \Text[j \dd s + 2\tau)$ it holds $D \in \DistinguishingPrefs{\tau}{\Text}{\SSS}$. Thus, $\Pat$
  has $D$ as a prefix. Observe that $\revstr{x}$ is a prefix of $D$
  and it holds $|\revstr{x}| + 2\tau = |D|$. Finally, observe that it
  holds $\min(\Textlen + 1, j + d) - j = \min(\Textlen + 1 - j, d) = \min(m,
  d)$. Thus, $\revstr{x}y = \Text[j \dd \min(\Textlen + 1, j + d))
  = \Pat[1 \dd \min(m, d)]$. We have thus proved that all assumptions
  of \cref{lm:sa-nonperiodic-pat-occ} hold for $\Pat$. The claim thus
  follows by combining the definition of $\OccThree{d}{j}{\Text}$
  and \cref{lm:sa-nonperiodic-pat-occ}, i.e., $\OccThree{d}{j}{\Text}
  = \OccThree{d}{\Pat}{\Text} = \{s - |x| : s \in \SSS,\
  x \preceq \revstr{\Textinf[s - 7\tau \dd s)} \prec x', \text{ and }
  y \preceq \Textinf[s \dd s + 7\tau) \prec y'\}$.
\end{proof}

\begin{lemma}\label{lm:sa-nonperiodic-pos-occ-size}
  Let $\ell \in [16 \dd \Textlen)$ and $\tau = \lfloor \tfrac{\ell}{3} \rfloor$.
  Let $\SSS$ be a $\tau$-synchronizing set of $\Text$ and
  $\SSScomp = \CompRepr{14\tau}{\SSS}{\Text}$.
  Let $j \in [1 \dd \Textlen - 3\tau + 2] \setminus \RTwo{\tau}{\Text}$, $c = \max \Sigma$,
  $s = \Successor{\SSS}{j}$, $d \in [\ell \dd 2\ell]$, and $x, x', y, y'$
  be such that
  \begin{itemize}
  \item $\revstr{x} = \Text[j \dd s)$,
  \item $x' = x c^{\infty}$,
  \item $\revstr{x}y = \Text[j \dd \min(\Textlen + 1, j + d))$,
  \item $y' = y c^{\infty}$.
  \end{itemize}
  Then, $|\OccThree{d}{j}{\Text}| = \RangeCountFourSide{\Pts}{x}{x'}{y}{y'}$,
  where $\Pts = \StrStrPoints{7\tau}{\SSScomp}{\Text}$ (\cref{def:str-str}).
\end{lemma}
\begin{proof}
  Increasing all elements of a set by the same value does change
  its cardinality. By \cref{lm:sa-nonperiodic-pos-occ} and
  \cref{lm:sa-nonperiodic-count}, we thus obtain:
  \begin{align*}
    |\OccThree{d}{j}{\Text}|
      &= |\{s - |x| : s \in \SSS,\,
         x \preceq \revstr{\Textinf[s {-} 7\tau \dd s)} \prec x'
         \text{ and }
         y \preceq \Textinf[s \dd s {+} 7\tau) \prec y'\}|\\
      &= |\{s \in \SSS :
         x \preceq \revstr{\Textinf[s {-} 7\tau \dd s)} \prec x'
         \text{ and }
         y \preceq \Textinf[s \dd s {+} 7\tau) \prec y'\}|\\
      &= \RangeCountFourSide{\Pts}{x}{x'}{y}{y'}.
      \qedhere
  \end{align*}
\end{proof}

\paragraph{Query Algorithms}

\begin{proposition}\label{pr:sa-nonperiodic-occ-size}
  Let $k \in [4 \dd \lceil \log \Textlen \rceil)$, $\ell = 2^k$, $\tau
  = \lfloor \tfrac{\ell}{3} \rfloor$, and $j \in [1 \dd
  \Textlen] \setminus \RTwo{\tau}{\Text}$. Given the structure $\CompSaNonperiodic{\Text}$, the
  value $k$, the position $j$, and any $j' \in \OccThree{\ell}{j}{\Text}$
  satisfying $j' = \min \OccThree{2\ell}{j'}{\Text}$ as input, we can compute
  $|\OccThree{2\ell}{j}{\Text}|$ in $\bigO(\log^{2+\epsilon} \Textlen)$ time.
\end{proposition}
\begin{proof}

  If $j > \Textlen - 3\tau + 2$ then by $3\tau - 1 \leq \ell$ and the
  uniqueness of $\Text[\Textlen]$ in $\Text$, it holds $|\OccThree{\ell}{j}{\Text}| =
  1$. By $|\OccThree{2\ell}{j}{\Text}| > 0$ and
  $\OccThree{2\ell}{j}{\Text} \subseteq \OccThree{\ell}{j}{\Text}$ we return
  $|\OccThree{2\ell}{j}{\Text}| = 1$.

  Let us now assume $j \in [1 \dd \Textlen - 3\tau + 2]$. Denote $\Pts
  = \StrStrPoints{7\tau}{\SSScompgen{k}}{\Text}$. The algorithm consists of two
  steps:
  \begin{enumerate}
  \item First, using \cref{pr:sa-nonperiodic-succ}, we compute $s
    = \Successor{\SSS_k}{j}$ in $\bigO(\log \Textlen)$ time.
  \item Let $x, x', y, y'$ be such that $\Text[j \dd s) = \revstr{x}$,
    $x' = x c^{\infty}$, $\revstr{x}y = \Text[j \dd \min(\Textlen + 1, j +
    2\ell))$, and $y' = y c^{\infty}$.
    By \cref{lm:sa-nonperiodic-pos-occ-size}, it then holds
    $|\OccThree{2\ell}{j}{\Text}| = \RangeCountFourSide{\Pts}{x}{x'}{y}{y'}
    = \RangeCountThreeSide{\Pts}{x}{x'}{y'} - \RangeCountThreeSide{\Pts}{x}{x'}{y}$
    (where the last equality follows by $y \preceq y'$). Observe also
    that using the structure $\CompSaNonperiodic{\Text}$, we can perform $\LCE_{\T}$
    and $\LCE_{\revstr{\T}}$ queries in $\bigO(\log \Textlen)$ time, and we
    can access any symbol of $\Text$ in $\bigO(\log \Textlen)$ time. Thus, we
    can compare any two substrings of $\Textinf$ or $\revstr{\Textinf}$
    (specified with their starting positions and lengths) in $t_{\rm
    cmp} = \bigO(\log \Textlen)$ time. Thus, using
    the structure from \cref{pr:str-str} for $q = 7\tau$ and $P
    = \SSScompgen{k}$ (which is a component of
    $\CompSaNonperiodic{\Text}$) with parameters $(i, q_l, q_r) =
    (s, s - j, \min(\Textlen + 1, j + 2\ell) - s)$ we compute
    $|\OccThree{2\ell}{j}{\Text}|$ in $\bigO(\log^{2 + \epsilon} \Textlen + t_{\rm
    cmp} \log \Textlen) = \bigO(\log^{2 + \epsilon} \Textlen)$ time.
  \end{enumerate}
  In total, we spend $\bigO(\log^{2 + \epsilon} \Textlen)$ time.
\end{proof}

\subsubsection{Computing a Position in Occ}

\paragraph{Combinatorial Properties}

\begin{lemma}\label{lm:sa-nonperiodic-pat-occ-elem}
  Let $\ell \in [16 \dd \Textlen)$ and $\tau = \lfloor \tfrac{\ell}{3} \rfloor$.
  Let $\SSS$ be a $\tau$-synchronizing set of $\Text$ and
  $\SSScomp = \CompRepr{14\tau}{\SSS}{\Text}$.
  Let $\Pat \in \Sigma^{m}$ be a $\tau$-nonperiodic pattern such that
  $m \geq 3\tau - 1$, no nonempty suffix of $\Text$ is a proper prefix of
  $\Pat$, and $\Pat$ is prefixed with $D \in \DistinguishingPrefs{\tau}{\Text}{\SSS}$ (\cref{def:dist-prefix}). Let $c
  = \max \Sigma$ and $x, x', y, y'$ be such that:
  \begin{itemize}
  \item $\revstr{x}$ is a prefix of $D$ and $|\revstr{x}| + 2\tau = |D|$,
  \item $x' = x c^{\infty}$,
  \item $\revstr{x}y = \Pat[1 \dd \min(m, \ell)]$.
  \end{itemize}
  Let also $\Pts \,{=}\, \StrStrPoints{7\tau}{\SSScomp}{\Text}$ (\cref{def:str-str}),
  $m = \RangeCountThreeSide{\Pts}{x}{x'}{y}$, and $b \,{=}\, \RangeBegThree{\ell}{\Pat}{\Text}$.
  Finally, let $i \in [1 \dd \Textlen]$ be such that $\SA[i] \in
  \OccThree{2\ell}{\Pat}{\Text}$. Then:
  \begin{itemize}
  \item It holds $m + (i - b) \in [1 \dd \RangeCountTwoSide{\Pts}{x}{x'}]$.
  \item Every $p \in \RangeSelect{\Pts}{x}{x'}{m + (i - b)}$ satisfies $p
    - |x| \in \OccThree{2\ell}{\Pat}{\Text}$.
  \end{itemize}
\end{lemma}
\begin{proof}

  The proof consists of three steps.

  1. First, we prove that it holds $m + (i - b) \in [1 \dd
  \RangeCountTwoSide{\Pts}{x}{x'}]$, i.e., the first claim.  First, observe
  that $\SA[i] \in \OccThree{2\ell}{\Pat}{\Text} \subseteq \OccThree{\ell}{\Pat}{\Text}$.
  Thus, $i > b$, which by
  $m \geq 0$ implies that $m + (i - b) \geq 1$.  Denote $m' =
  \RangeCountTwoSide{\Pts}{x}{x'}$. To prove $m + (i - b) \leq m'$, note that
  $i \leq \RangeEndThree{\ell}{\Pat}{\Text}$, i.e., letting $d =
  |\OccThree{\ell}{\Pat}{\Text}|$, we have $i - b \leq d$. Thus, it
  suffices to prove $m + d \leq m'$.
  \begin{itemize}
  \item First, we observe that by
    \cref{lm:sa-nonperiodic-count}\eqref{lm:sa-nonperiodic-count-it-1},
    letting $\SSS_{\rm low} = \{s \in \SSS : x \preceq \revstr{\Textinf[s
    - 7\tau \dd s)} \prec x'\text{ and } \Textinf[s \dd s + 7\tau) \prec
    y\}$, it holds $m = \RangeCountThreeSide{\Pts}{x}{x'}{y} = |\SSS_{\rm
    low}|$.
  \item By
    \cref{lm:sa-nonperiodic-pat-occ},
    it holds $\OccThree{\ell}{\Pat}{\Text} = \{s - |x| : s \in \SSS,\ x
    \preceq \revstr{\Textinf[s - 7\tau \dd s)} \prec x',\text{ and } y
    \preceq \Textinf[s \dd s + 7\tau) \prec y c^{\infty}\}$. Thus,
    letting $\SSS_{\rm mid} = \{s \in \SSS : x \preceq \revstr{\Textinf[s
    - 7\tau \dd s)} \prec x'\text{ and }y \preceq \Textinf[s \dd s +
    7\tau) \prec y c^{\infty}\}$, we have $|\SSS_{\rm mid}| =
    |\OccThree{\ell}{\Pat}{\Text}| = d$.
  \item Lastly, by
    \cref{lm:sa-nonperiodic-count}\eqref{lm:sa-nonperiodic-count-it-3},
    letting $\SSS_{\rm all} = \{s \in \SSS : x \preceq \revstr{\Textinf[s
    - 7\tau \dd s)} \prec x'\}$, we have $m' = |\SSS_{\rm all}|$.
  \end{itemize}
  Note now that $\SSS_{\rm low} \subseteq \SSS_{\rm all}$, $\SSS_{\rm
  mid} \subseteq \SSS_{\rm all}$, and $\SSS_{\rm low} \cap \SSS_{\rm
  mid} = \emptyset$. Thus, $m + d = |\SSS_{\rm low}| + |\SSS_{\rm
  mid}| \leq |\SSS_{\rm all}| = m'$.

  2. In the second step, we show that for every $k \in [1 \dd d]$, any
  $p \in \RangeSelect{\Pts}{x}{x'}{m + k}$ satisfies $\Textinf[p - |x| \dd p
  - |x| + 7\tau) = \Textinf[\SA[b + k] \dd \SA[b + k] + 7\tau)$.
  Consider any $k \in [1 \dd d]$.  First, recall (see above) that
  $\OccThree{\ell}{\Pat}{\Text} = \{s - |x| : s \in \SSS,\ x \preceq \Textinf[s
  - 7\tau \dd s) \prec x',\text{ and } y \preceq \Textinf[s \dd s +
  7\tau) \prec y c^{\infty}\}$.  This implies that $\SA[b + k] +
  |x| \in \SSS$. Denote $j = \SA[b + k] + |x|$ and let $y'
  = \Textinf[j \dd j + 7\tau)$. We claim that $m + k \in
  (\RangeCountThreeSide{\Pts}{x}{x'}{y'} \dd \IncRangeCountThreeSide{\Pts}{x}{x'}{y'}]$.
  \begin{itemize}
  \item We first show $\RangeCountThreeSide{\Pts}{x}{x'}{y'} < m +
    k$. By \cref{lm:sa-nonperiodic-count}\eqref{lm:sa-nonperiodic-count-it-1},
    $\RangeCountThreeSide{\Pts}{x}{x'}{y'} = |\{s \in \SSS :
    x \preceq \revstr{\Textinf[s - 7\tau \dd s)}\prec x'\text{ and
    } \Textinf[s \dd s + 7\tau) \prec y'\}|$.  Recall that $y$ satisfies
    $\Pat[1 \dd \min(m, \ell)] = \revstr{x}y$. On the other and,
    $\SA[b + k] \in \OccThree{\ell}{\Pat}{\Text}$ implies that
    $\lcp(\Pat, \Text[\SA[b + k] \dd \Textlen]) \geq \min(m, \ell)$, or
    equivalently, that $\Pat[1 \dd \min(m, \ell)]$ is a prefix of
    $\Text[\SA[b + k] \dd \Textlen]$.  Thus, $\revstr{x}y$ is a prefix of
    $\Textinf[\SA[b + k] \dd \SA[b + k] + \ell)$, and hence $y$ is a
    prefix of $\Textinf[j \dd \SA[b + k] + \ell) = \Textinf[j \dd j + \ell -
    |x|)$, which by $\ell - |x| \leq \ell \leq 7\tau$ implies that $y$
    is a prefix of $y'$.  Consequently, it holds
    $\RangeCountThreeSide{\Pts}{x}{x'}{y'} = |\SSS_{\rm low}| + |\SSS'_{\rm
    mid}|$, where $\SSS_{\rm low}$ is defined as above, and
    $\SSS'_{\rm mid} = \{s \in \SSS : x \preceq \revstr{\Textinf[s -
    7\tau \dd s)} \prec x'\text{ and } y \preceq \Textinf[s \dd s +
    7\tau) \prec y'\}$.  As noted above, it holds $|\SSS_{\rm low}| =
    m$. Thus, it remains to show $|\SSS'_{\rm mid}| < k$. Let ${\sf
    P}'_{\rm mid} = \{s - |x| : s \in \SSS'_{\rm mid}\}$. We will
    instead show that $|{\sf P}'_{\rm mid}| < k$ (this will prove the
    claim since $|{\sf P}'_{\rm mid}| = |\SSS'_{\rm mid}|$). Recall
    the definition of $\OccThree{\ell}{\Pat}{\Text}$ stated above and note
    that $y' \prec y c^{\infty}$, since $y$ is a prefix of $y'$.
    Thus, ${\sf P}'_{\rm mid} \subseteq \OccThree{\ell}{\Pat}{\Text}
    = \{\SA[b + k'] : k' \in [1 \dd d]\}$.  Note that for every
    $k' \in [k \dd d]$, letting $s' = \SA[b + k'] + |x|$ we have by
    definition of $\SA$ that $\Textinf[s' \dd s' +
    7\tau) \succeq \Textinf[\SA[b + k] + |x| \dd \SA[b + k] + |x| +
    7\tau) = \Textinf[j \dd j + 7\tau) = y'$. Thus, for such $k'$, we
    have $\SA[b + k'] \not\in {\sf P}'_{\rm mid}$, and hence $|{\sf
    P}'_{\rm mid}| < |\OccThree{\ell}{\SA[i]}{\Text}| - (d - k) = k$.
  \item We now show $m {+} k
    {\leq} \IncRangeCountThreeSide{\Pts}{x}{x'}{y'}$.
    By \cref{lm:sa-nonperiodic-count}\eqref{lm:sa-nonperiodic-count-it-2},
    $\IncRangeCountThreeSide{\Pts}{x}{x'}{y'} = |\{s \in \SSS :
    x \preceq \revstr{\Textinf[s - 7\tau \dd s)} \prec x'\text{ and
    } \Textinf[s \dd s + 7\tau) \preceq y'\}|$. Thus, similarly as above,
    $\IncRangeCountThreeSide{\Pts}{x}{x'}{y'} = |\SSS_{\rm low}| +
    |\SSS''_{\rm mid}| = |\SSS_{\rm low}| + |{\sf P}''_{\rm mid}|$,
    where $\SSS''_{\rm mid} = \{s \in \SSS : x \preceq \revstr{\Textinf[s
    - 7\tau \dd s)} \prec x'\text{ and } y \preceq \Textinf[s \dd s +
    7\tau) \preceq y'\}$ and ${\sf P}''_{\rm mid} = \{s - |x| :
    s \in \SSS''_{\rm mid}\}$. Thus, it remains to show $k \leq |{\sf
    P}''_{\rm mid}|$.  As before, we have ${\sf P}''_{\rm
    mid} \subseteq \OccThree{\ell}{\Pat}{\Text}$, except now we have that
    $\SA[b + k'] \in {\sf P}''_{\rm mid}$ for at least $k' \in [1 \dd
    k]$.  Thus, $k \leq |{\sf P}''_{\rm mid}|$.
  \end{itemize}
  By the above, for every $p \in \RangeSelect{\Pts}{x}{x'}{m + k}$, it
  holds $\Textinf[p \dd p + 7\tau) = y'$ and $x \preceq \revstr{\Textinf[p -
  7\tau \dd p)} \prec x'$, or equivalently, $\Textinf[p - |x| \dd p +
  7\tau) = \revstr{x}y'$.  On the other hand, we have $\Textinf[\SA[b +
  k] + |x| \dd \SA[b + k] + |x| + 7\tau) = y'$ and (by $\SA[b +
  k] \in \OccThree{\ell}{\Pat}{\Text}$) $\Textinf[\SA[b + k] \dd \SA[b + k] +
  |x|) = \revstr{x}$.  Thus, $\Textinf[p - |x| \dd p + 7\tau)
  = \Textinf[\SA[b + k] \dd \SA[b + k] + |x| + 7\tau)$.

  3. Applying the above step for $k = i - b$, we
  obtain that for every $p \in \RangeSelect{\Pts}{x}{x'}{m + (i - b)}$,
  it holds $\Textinf[p - |x| \dd p - |x| + 7\tau) = \Textinf[\SA[b +
  k] \dd \SA[b + k] + 7\tau) = \Textinf[\SA[i] \dd \SA[i] + 7\tau)$. By
  $2\ell \leq 7\tau$, we thus obtain $\Textinf[p - |x| \dd p - |x| +
  2\ell) = \Textinf[\SA[i] \dd \SA[i] + 2\ell)$, which
  by \cref{lm:pat-occ-equivalence} and
  $\SA[i] \in \OccThree{2\ell}{\Pat}{\Text}$ is equivalent to $p -
  |x| \in \OccThree{2\ell}{\Pat}{\Text}$.
\end{proof}

\begin{lemma}\label{lm:sa-nonperiodic-occ-elem}
  Let $\ell \in [16 \dd \Textlen)$ and $\tau = \lfloor \tfrac{\ell}{3} \rfloor$.
  Let $\SSS$ be a $\tau$-synchronizing set of $\Text$ and
  $\SSScomp = \CompRepr{14\tau}{\SSS}{\Text}$.
  Let $i \in [1 \dd \Textlen]$ be such that $\SA[i] \in [1 \dd \Textlen - 3\tau + 2]
  \sm \RTwo{\tau}{\Text}$. Let $j \in \OccThree{\ell}{\SA[i]}{\Text}$, $c = \max\Sigma$,
  $s = \Successor{\SSS}{j}$, and $x, x', y$ be such that:
  \begin{itemize}
  \item $\revstr{x} = \Text[j \dd s)$,
  \item $x' = x c^{\infty}$,
  \item $\revstr{x}y = \Text[j \dd \min(\Textlen + 1, j + \ell))$.
  \end{itemize}
  Let us also denote
  $\Pts = \StrStrPoints{7\tau}{\SSScomp}{\Text}$ (\cref{def:str-str}),
  $m = \RangeCountThreeSide{\Pts}{x}{x'}{y}$, and
  $b = \RangeBegThree{\ell}{\SA[i]}{\Text}$. Then:
  \begin{itemize}
  \item It holds $m + (i - b) \in [1 \dd \RangeCountTwoSide{\Pts}{x}{x'}]$.
  \item Every $p \in \RangeSelect{\Pts}{x}{x'}{m + (i - b)}$ satisfies $p
    - |x| \in \OccThree{2\ell}{\SA[i]}{\Text}$.
  \end{itemize}
\end{lemma}
\begin{proof}
  Denote $\Pat = \Text[\SA[i] \dd \Textlen]$ and $m = |\Pat| = \Textlen - \SA[i] + 1$.
  Note that $\SA[i] \in [1 \dd \Textlen - 3\tau + 2]$ and $\SA[i] \not\in \RTwo{\tau}{\Text}$
  imply that $m \geq 3\tau - 1$ and $\per(\Pat[1 \dd 3\tau - 1])
  > \tfrac{1}{3}\tau$. Thus, $\Pat$ is $\tau$-nonperiodic. Next,
  observe that since $\Text[\Textlen]$ does not occur in $\Text[1 \dd \Textlen)$, no
  nonempty suffix of $\Text$ is a proper prefix of $\Pat$.  Denote $D
  = \Text[j \dd s + 2\tau)$.
  By \cref{lm:nonperiodic-pos-lce}\eqref{lm:nonperiodic-pos-lce-it-1},
  $3\tau - 1 \leq \ell$, and $j \in \OccThree{\ell}{\SA[i]}{\Text}
  = \OccThree{\ell}{\Pat}{\Text}$, it follows that $j \in [1 \dd \Textlen - 3\tau +
  2] \sm \RTwo{\tau}{\Text}$ and $\Text[j \dd \Successor{\SSS}{j} + 2\tau)
  = \Text[j \dd s + 2\tau) = D$ is a prefix of $\Pat$. Note that
  $j \not\in \RTwo{\tau}{\Text}$ implies that $D \in \DistinguishingPrefs{\tau}{\Text}{\SSS}$.
  Next, note that the following properties hold for $x$ and $y$.
  \begin{itemize}
  \item First, we observe that by $\revstr{x} = \Text[j \dd s)$, the
     string $\revstr{x}$ is a prefix of $D$. Note also that by $|D| =
     s + 2\tau - j$ and $|\revstr{x}| = s - j$, it holds $|\revstr{x}|
     + 2\tau = |D|$.
  \item Second, we prove that it holds $\revstr{x}y
    = \Pat[1 \dd \min(m, \ell)]$.  Consider two cases:
    \begin{itemize}
    \item First, assume $j = \SA[i]$. By $\min(\Textlen + 1, j + \ell) - j
       = \min(\Textlen + 1, \SA[i] + \ell) - \SA[i] = \min(\Textlen + 1
       - \SA[i], \ell) = \min(m, \ell)$ we then obtain $\revstr{x}y
       = \Text[j \dd \min(\Textlen + 1, j + \ell)) = \Text[\SA[i] \dd \min(\Textlen +
       1, \SA[i] + \ell)) = \Pat[1 \dd \min(\Textlen + 1, \SA[i] + \ell)
       - \SA[i]] = \Pat[1 \dd \min(m, \ell)]$.
    \item Let us now thus assume $j \neq \SA[i]$. The assumption
       $j \in \OccThree{\ell}{\SA[i]}{\Text}$ and \cref{lm:occ-equivalence}
       then imply that $\LCE_{\Text}(j, \SA[i]) \geq \ell$, i.e.,
       $\Text[j \dd j + \ell) = \Text[\SA[i] \dd \SA[i] + \ell)$. This in
       turn implies $\min(\Textlen + 1, j + \ell) - j \geq \ell$,
       $m \geq \ell$, and $\revstr{x}y = \Text[j \dd \min(\Textlen + 1, j
       + \ell)) = \Text[j \dd j + \ell) = \Text[\SA[i] \dd \SA[i] + \ell)
       = \Pat[1 \dd \ell] = \Pat[1 \dd \min(m, \ell)]$.
    \end{itemize}
  \end{itemize}
  By definition, it follows that $b
  = \RangeBegThree{\ell}{\SA[i]}{\Text} = \RangeBegThree{\ell}{\Text[\SA[i] \dd \Textlen]}{\Text}
  = \RangeBegThree{\ell}{\Pat}{\Text}$ and $\SA[i] \in \OccThree{2\ell}{\SA[i]}{\Text}
  = \OccThree{2\ell}{\Text[\SA[i] \dd \Textlen]}{\Text}
  = \OccThree{2\ell}{\Pat}{\Text}$. It thus follows by
  \cref{lm:sa-nonperiodic-pat-occ-elem} that:
  \begin{itemize}
  \item It holds $m + (i - b) \in [1 \dd \RangeCountTwoSide{\Pts}{x}{x'}]$,
  \item Every $p \in \RangeSelect{\Pts}{x}{x'}{m + (i - b)}$ satisfies $p
    - |x| \in \OccThree{2\ell}{\Pat}{\Text}
    = \OccThree{2\ell}{\SA[i]}{\Text}$.
    \qedhere
  \end{itemize}
\end{proof}

\paragraph{Query Algorithms}

\begin{proposition}\label{pr:sa-nonperiodic-occ-elem}
  Let $k \in [4 \dd \lceil \log \Textlen \rceil)$, $\ell = 2^k$, $\tau
  = \lfloor \tfrac{\ell}{3} \rfloor$.  Let $i \in [1 \dd \Textlen]$ be such
  that $\SA[i] \in [1 \dd \Textlen] \sm \RTwo{\tau}{\Text}$.  Given
  $\CompSaNonperiodic{\Text}$ along with the values $k$, $i$,
  $\RangeBegThree{\ell}{\SA[i]}{\Text}$, $\RangeEndThree{\ell}{\SA[i]}{\Text}$, and some
  position $j \in \OccThree{\ell}{\SA[i]}{\Text}$ satisfying $j
  = \min \OccThree{2\ell}{j}{\Text}$ as input, we can compute some
  $j' \in \OccThree{2\ell}{\SA[i]}{\Text}$ in $\bigO(\log^{3 + \epsilon} \Textlen)$
  time.
\end{proposition}
\begin{proof}

  Denote $b = \RangeBegThree{\ell}{\SA[i]}{\Text}$ and $e
  = \RangeEndThree{\ell}{\SA[i]}{\Text}$, If $e - b = |\OccThree{\ell}{\SA[i]}{\Text}| =
  1$ then by
  $\OccThree{2\ell}{\SA[i]}{\Text} \subseteq \OccThree{\ell}{\SA[i]}{\Text}$, it
  holds $\OccThree{2\ell}{\SA[i]}{\Text} = \{j\}$, and hence we return $j' =
  j$.

  Let us thus assume $e - b > 1$. By the uniqueness of $\Text[\Textlen]$ in
  $\Text$, this implies that $\Textinf[j \dd j + \ell)$ occurs in $\Textinf$
  starting in at least two different positions in $[1 \dd \Textlen]$, and
  hence it does not contain $\Text[\Textlen]$. Thus, $\SA[i] \in [1 \dd \Textlen
  - \ell]$. By $3\tau - 1 \leq \ell$, we thus have $\SA[i] \in [1 \dd
  \Textlen - 3\tau + 2] \sm \RTwo{\tau}{\Text}$. Note that by
  $j \in \OccThree{\ell}{\SA[i]}{\Text} \subseteq \OccThree{3\tau - 1}{\SA[i]}{\Text}$
  and \cref{lm:nonperiodic-pos-lce}\eqref{lm:nonperiodic-pos-lce-it-2}, we then
  obtain $j \in [1 \dd \Textlen - 3\tau + 2] \sm \RTwo{\tau}{\Text}$.
  Observe that using
  $\CompSaNonperiodic{\Text}$, we can perform $\LCE_{\Text}$ and
  $\LCE_{\revstr{\Text}}$ queries in $\bigO(\log \Textlen)$ time, and we can
  access any symbol of $\Text$ in $\bigO(\log \Textlen)$ time. Thus, we can
  compare any two substrings of $\Textinf$ or $\revstr{\Textinf}$ (specified
  with their starting positions and lengths) in $t_{\rm cmp}
  = \bigO(\log \Textlen)$ time. The algorithm consists of three steps:
  \begin{enumerate}
  \item First, using \cref{pr:sa-nonperiodic-succ}, we compute $s
    = \Successor{\SSS_k}{j}$ in $\bigO(\log \Textlen)$ time (note that we
    utilize $j \in \OccThree{\ell}{j}{\Text}$).
  \item Denote $\Pts = \StrStrPoints{7\tau}{\SSScompgen{k}}{\Text}$ and let
    $x, x', y$ be such that $\revstr{x} = \Text[j \dd s)$, $x' = x
    c^{\infty}$, and $\revstr{x}y = \Text[j \dd \min(\Textlen + 1, j + \ell))$
    (where $c
    = \max \Sigma$). Using \cref{pr:str-str}\eqref{pr:str-str-it-1},
    in $\bigO(\log^{2 + \epsilon} \Textlen + t_{\rm cmp} \log \Textlen)
    = \bigO(\log^{2 + \epsilon} \Textlen)$ time we compute $m
    := \RangeCountThreeSide{\Pts}{x}{x'}{y}$ using parameters $(i, q_l, q_r)
    = (s, s - j, \min(\Textlen + 1, j + \ell) - s)$.
  \item Finally, using \cref{pr:str-str}\eqref{pr:str-str-it-2} with
    parameters $(i, q_l, r) = (s, s - j, m + (i - b))$ we compute some
    $p \in \RangeSelect{\Pts}{x}{x'}{r}$ in $\bigO(\log^{3 + \epsilon} \Textlen
    + t_{\rm cmp} \log \Textlen) = \bigO(\log^{3 + \epsilon} \Textlen)$ time.
    By \cref{lm:sa-nonperiodic-occ-elem}, it holds $p -
    |x| \in \OccThree{2\ell}{\SA[i]}{\Text}$. We thus return $j' := p - |x| =
    p - (s - j)$ as the answer.
  \end{enumerate}
  In total, we spend $\bigO(\log^{3 + \epsilon} \Textlen)$ time.
\end{proof}

\subsubsection{Computing a Position in a Cover}

\paragraph{Combinatorial Properties}

\begin{lemma}\label{lm:sa-nonperiodic-pat-occ-min}
  Let $\ell \in [16 \dd \Textlen)$ and $\tau = \lfloor \tfrac{\ell}{3} \rfloor$.
  Let $\SSS$ be a $\tau$-synchronizing set of $\Text$ and
  $\SSScomp = \CompRepr{14\tau}{\SSS}{\Text}$.
  Let $\Pat \in \Sigma^{m}$ be a $\tau$-nonperiodic pattern such that
  $m \geq 3\tau - 1$, no nonempty suffix of $\Text$ is a proper prefix of
  $\Pat$, and $\Pat$ is prefixed with some $D \in \DistinguishingPrefs{\tau}{\Text}{\SSS}$ (\cref{def:dist-prefix}).
  Let $d \in [\ell \dd 2\ell]$, $c = \max \Sigma$, and $x, x', y, y'$
  be such that:
  \begin{itemize}
  \item $\revstr{x}$ is a prefix of $D$ and $|\revstr{x}| + 2\tau =
    |D|$,
  \item $x' = x c^{\infty}$,
  \item $\revstr{x}y = \Pat[1 \dd \min(m, d)]$,
  \item $y' = y c^{\infty}$.
  \end{itemize}
  Then, $\min \OccThree{d}{\Pat}{\Text} = \RangeMinFourSide{\Pts}{x}{x'}{y}{y'} - |x|$,
  where $\Pts = \StrStrPoints{7\tau}{\SSScomp}{\Text}$ (\cref{def:str-str}).
\end{lemma}
\begin{proof}
  The result follows by putting together
  \cref{lm:sa-nonperiodic-pat-occ,lm:sa-nonperiodic-min}, i.e.,
  \begin{align*}
    \min \OccThree{d}{\Pat}{\Text}
      &= \min \{s - |x| : s \in \SSS,\
         x \preceq \revstr{\Textinf[s {-} 7\tau \dd s)} \prec x',
         \text{ and }
         y \preceq \Textinf[s \dd s {+} 7\tau) \prec y'\}\\
      &= \min \{s \in \SSS:
         x \preceq \revstr{\Textinf[s {-} 7\tau \dd s)} \prec x',
         \text{ and }
         y \preceq \Textinf[s \dd s {+} 7\tau) \prec y'\} - |x|\\
      &= \RangeMinFourSide{\Pts}{x}{x'}{y}{y'} - |x|.
      \qedhere
  \end{align*}
\end{proof}

\begin{lemma}\label{lm:sa-nonperiodic-pos-occ-min}
  Let $\ell \in [16 \dd \Textlen)$ and $\tau = \lfloor \tfrac{\ell}{3} \rfloor$.
  Let $\SSS$ be a $\tau$-synchronizing set of $\Text$ and
  $\SSScomp = \CompRepr{14\tau}{\SSS}{\Text}$.
  Let $j \in [1 \dd \Textlen - 3\tau + 2] \setminus \RTwo{\tau}{\Text}$, $c = \max \Sigma$,
  $s = \Successor{\SSS}{j}$, $d \in [\ell \dd 2\ell]$, and $x, x', y,
  y'$ be such that:
  \begin{itemize}
  \item $\revstr{x} = \Text[j \dd s)$,
  \item $x' = x c^{\infty}$,
  \item $\revstr{x}y = \Text[j \dd \min(\Textlen + 1, j + d))$,
  \item $y' = y c^{\infty}$.
  \end{itemize}
  Then, $\min \OccThree{d}{j}{\Text} = \RangeMinFourSide{\Pts}{x}{x'}{y}{y'} - |x|$,
  where $\Pts = \StrStrPoints{7\tau}{\SSScomp}{\Text}$ (\cref{def:str-str}).
\end{lemma}
\begin{proof}
  The result follows by \cref{lm:sa-nonperiodic-pat-occ-min} (see the
  proof of \cref{lm:sa-nonperiodic-pos-occ} for a similar argument).
\end{proof}

\paragraph{Query Algorithms}

\begin{proposition}\label{pr:sa-nonperiodic-occ-min}
  Let $k \in [4 \dd \lceil \log \Textlen \rceil)$, $\ell = 2^k$, $\tau
  = \lfloor \tfrac{\ell}{3} \rfloor$, and $j \in [1 \dd
  \Textlen] \setminus \RTwo{\tau}{\Text}$. Given the structure $\CompSaNonperiodic{\Text}$, the
  value $k$, the position $j$, and any $j' \in \OccThree{\ell}{j}{\Text}$
  satisfying $j' = \min \OccThree{2\ell}{j'}{\Text}$ as input, we can compute
  $\min \OccThree{2\ell}{j}{\Text}$ in $\bigO(\log^2 \Textlen)$ time.
\end{proposition}
\begin{proof}

  If $j > \Textlen - 3\tau + 2$, then we have $|\OccThree{\ell}{j}{\Text}| =
  1$. Thus, by $j \in \OccThree{\ell}{j}{\Text}$ and
  $\OccThree{2\ell}{j}{\Text} \subseteq \OccThree{\ell}{j}{\Text}$ we return that
  $\min \OccThree{2\ell}{j}{\Text} = j$.

  Let us thus assume $j \in [1 \dd \Textlen - 3\tau + 2]$. The algorithm
  consists of two steps:
  \begin{enumerate}
  \item Using \cref{pr:sa-nonperiodic-succ}, we compute $s
    = \Successor{\SSS_k}{j}$ in $\bigO(\log \Textlen)$ time.
  \item Let $x, x', y, y'$ be such that $\Textinf[j \dd s) = \revstr{x}$,
    $x' = x c^{\infty}$, $\Textinf[j \dd j + 2\ell) = \revstr{x}y$, and
    $y' = y c^{\infty}$ (where $c = \max \Sigma$).
    By \cref{lm:sa-nonperiodic-pos-occ-min}, it then holds
    $\min \OccThree{2\ell}{j}{\Text} = \RangeMinFourSide{\Pts}{x}{x'}{y}{y'} - |x|$,
    where $\Pts = \StrStrPoints{7\tau}{\SSScompgen{k}}{\Text}$
    (\cref{def:str-str}).  Observe also that using
    $\CompSaNonperiodic{\Text}$, we can perform $\LCE_{\T}$ and
    $\LCE_{\revstr{\T}}$ queries in $\bigO(\log \Textlen)$ time, and we can
    access any symbol of $\Text$ in $\bigO(\log \Textlen)$ time. Thus, we can
    compare any two substrings of $\Textinf$ or $\revstr{\Textinf}$
    (specified with their starting positions and lengths) in $t_{\rm
    cmp} = \bigO(\log \Textlen)$ time.  Thus, using the data structure
    from \cref{pr:str-str} for $q = 7\tau$ and $P = \SSScompgen{k}$
    (which is a component of $\CompSaNonperiodic{\Text}$) with parameters
    $(i, q_l, q_r) = (s, s - j, \min(\Textlen + 1, j + 2\ell) - s)$ we
    compute $\min \OccThree{2\ell}{j}{\Text}$ in $\bigO(\log^{1 + \epsilon} \Textlen
    + t_{\rm cmp} \log \Textlen) = \bigO(\log^2 \Textlen)$ time.
  \end{enumerate}
  In total, we spend $\bigO(\log^2 \Textlen)$ time.
\end{proof}

\subsubsection{Implementation of ISA Queries}\label{sec:sa-nonperiodic-isa}

\begin{proposition}\label{pr:sa-nonperiodic-isa}
  Let $k \in [4 \dd \lceil \log \Textlen \rceil)$. Denote $\ell = 2^k$ and
  $\tau = \lfloor \tfrac{\ell}{3} \rfloor$.  Let $j \in [1 \dd
  \Textlen] \setminus \RTwo{\tau}{\Text}$.  Given $\CompSaNonperiodic{\Text}$, the
  value $k$, the position $j$, any $j' \in \OccThree{\ell}{j}{\Text}$
  satisfying $j' = \min \OccThree{2\ell}{j'}{\Text}$, and
  $(\RangeBegThree{\ell}{j}{\Text}, \RangeEndThree{\ell}{j}{\Text})$ as input, we can compute
  $(\RangeBegThree{2\ell}{j}{\Text}, \RangeEndThree{2\ell}{j}{\Text})$ and some
  position $j'' \in \OccThree{2\ell}{j}{\Text}$ satisfying $j''
  = \min \OccThree{4\ell}{j''}{\Text}$ in $\bigO(\log^{2 + \epsilon} \Textlen)$
  time.
\end{proposition}
\begin{proof}
  Denote $b = \RangeBegThree{\ell}{j}{\Text}$. The algorithm proceeds in two
  steps (note that their order can be swapped):
  \begin{enumerate}
  \item First,
    using \cref{pr:sa-nonperiodic-posbeg-posend-size,pr:sa-nonperiodic-occ-size}
    in $\bigO(\log^{2 + \epsilon} \Textlen)$ time we compute $\delta =
    |\PosBeg{\ell}{j}{\Text}|$ and $m = |\OccThree{2\ell}{j}{\Text}|$.  We
    then have $(\RangeBegThree{2\ell}{j}{\Text}, \RangeEndThree{2\ell}{j}{\Text}) \allowbreak
    = \allowbreak (b + \delta, b + \delta + m)$.
  \item Second, using \cref{pr:sa-nonperiodic-occ-min} in
    $\bigO(\log^2 \Textlen)$ time we compute $j''
    = \min \OccThree{2\ell}{j}{\Text}$.  Such position clearly satisfies
    $j'' \in \OccThree{2\ell}{j}{\Text}$ and $j''
    = \min \OccThree{4\ell}{j''}{\Text}$.
  \end{enumerate}
  In total, we spend $\bigO(\log^{2 + \epsilon} \Textlen)$ time.
\end{proof}

\subsubsection{Implementation of SA Queries}\label{sec:sa-nonperiodic-sa}

\begin{proposition}\label{pr:sa-nonperiodic-sa}
  Let $k \in [4 \dd \lceil \log \Textlen \rceil)$. Denote $\ell = 2^k$,
  $\tau = \lfloor \tfrac{\ell}{3} \rfloor$. Let $i \in [1 \dd \Textlen]$ be
  such that $\SA[i] \in [1 \dd \Textlen] \sm \RTwo{\tau}{\Text}$. 
  Given $\CompSaNonperiodic{\Text}$ along with values $k$, $i$,
  $\RangeBegThree{\ell}{\SA[i]}{\Text}$, $\RangeEndThree{\ell}{\SA[i]}{\Text}$, and some $j \in
  \OccThree{\ell}{\SA[i]}{\Text}$ satisfying $j = \min \OccThree{2\ell}{j}{\Text}$
  as input, we can compute the pair
  $(\RangeBegThree{2\ell}{\SA[i]}{\Text},\allowbreak \RangeEndThree{2\ell}{\SA[i]}{\Text})$ along with some
  $j' \in \OccThree{2\ell}{\SA[i]}{\Text}$ satisfying $j' = \min
  \OccThree{4\ell}{j'}{\Text}$ in $\bigO(\log^{3 + \epsilon} \Textlen)$ time.
\end{proposition}
\begin{proof}
  Denote $b = \RangeBegThree{\ell}{\SA[i]}{\Text}$ and $e =
  \RangeEndThree{\ell}{\SA[i]}{\Text}$. The algorithm consists of two steps:
  \begin{enumerate}
  \item First, using \cref{pr:sa-nonperiodic-occ-elem} we compute
    some position $p \in \OccThree{2\ell}{\SA[i]}{\Text}$ in $\bigO(\log^{3 +
    \epsilon} \Textlen)$ time.  Note that then it holds
    $\OccThree{\ell}{p}{\Text} = \OccThree{\ell}{\SA[i]}{\Text}$ (and hence, in
    particular, $j \in \OccThree{\ell}{p}{\Text}$), $\RangeBegThree{\ell}{p}{\Text} = b$,
    and $\RangeEndThree{\ell}{p}{\Text} = e$. Moreover, by $3\tau - 1 \leq \ell$ and
    \cref{def:sss}, we have $p \in [1 \dd \Textlen] \setminus \RTwo{\tau}{\Text}$.
  \item Using
    \cref{pr:sa-nonperiodic-isa}, we compute $b' = \RangeBegThree{2\ell}{p}{\Text}$,
    $e' = \RangeEndThree{2\ell}{p}{\Text}$, and some $j' \in \OccThree{2\ell}{p}{\Text}$
    satisfying $j' = \min \OccThree{4\ell}{j'}{\Text}$ in $\bigO(\log^{2 +
    \epsilon} \Textlen)$ time.  By $p \in
    \OccThree{2\ell}{\SA[i]}{\Text}$, we then have $\RangeBegThree{2\ell}{\SA[i]}{\Text} =
    b'$, $\RangeEndThree{2\ell}{\SA[i]}{\Text} = e'$, and $j' \in
    \OccThree{2\ell}{\SA[i]}{\Text}$.
  \end{enumerate}
  In total, we spend $\bigO(\log^{3 + \epsilon} \Textlen)$ time.
\end{proof}

\subsubsection{Construction Algorithm}\label{sec:sa-nonperiodic-construction}

\begin{proposition}\label{pr:sa-nonperiodic-construction}
  Given the LZ77 parsing of $\Text$, we can construct
  $\CompSaNonperiodic{\Text}$ in $\bigO(\SubstringComplexity{\Text} \log^7 \Textlen)$ time.
\end{proposition}
\begin{proof}

  First, using~\cite[Theorem~6.11]{resolutionfull}
  (resp.\ \cite[Theorem~6.21]{resolutionfull}), in $\bigO(\LZSize{\Text} \log^4 \Textlen)$
  time we construct a structure that, given any substring $S$ of $\Text$
  (specified with its starting position and the length) in
  $\bigO(\log^3 \Textlen)$ time returns $\min \OccTwo{S}{\Text}$ (resp.\
  $|\OccTwo{S}{\Text}|$).

  We then construct the components of $\CompSaNonperiodic{\Text}$
  (\cref{sec:sa-nonperiodic-ds}) as follows:
  \begin{enumerate}

  \item In $\bigO(\SubstringComplexity{\Text} \log^7 \Textlen)$ time we construct
    $\CompSaCore{\Text}$ using \cref{pr:sa-core-construction}.

  \item Using \cref{pr:comp-sss-construction} with $c = 14$, in
    $\bigO(\SubstringComplexity{\Text} \log^7 \Textlen)$ time we construct
    $\{\SSScompgen{k}\}_{k \in [4 \dd \lceil \log \Textlen \rceil)}$ (as defined
    in \cref{sec:sa-nonperiodic-ds}). Then, for $k \in
    [4 \dd \lceil \log \Textlen \rceil)$, in $\bigO(n_k \log n_k)
    = \bigO(n_k \log \Textlen)$ time (where $n_k$ is defined
    in \cref{sec:sa-nonperiodic-ds}), we sort the elements of
    $\SSScompgen{k}$ and write to
    $\ArrSSSComp{k}$. By \cref{pr:comp-sss-construction}, we have
    $\sum_{k \in [4 \dd \lceil \log \Textlen \rceil)} n_k
    = \bigO(\SubstringComplexity{\Text} \log \tfrac{\Textlen \log \sigma}{\SubstringComplexity{\Text} \log \Textlen})$.  Thus,
    over all $k$, the sorting takes $\bigO(\SubstringComplexity{\Text} \log^2 \Textlen)$ time.

  \item For $k = 4, \dots, \lceil \log \Textlen \rceil - 1$, we construct the
    structure from \cref{pr:str-str} with $P = \SSScompgen{k}$
    and $q = 7\tau_k$ as input. Over all $k$, this takes
    $\bigO(\sum_{k \in [4 \dd \lceil \log \Textlen \rceil)} n_k \log^3 \Textlen)
    = \bigO(\SubstringComplexity{\Text} \log^4 \Textlen)$ time (see above).  Recall that using
    $\CompSaCore{\Text}$ we can perform $\LCE_{\Text}$ and
    $\LCE_{\revstr{\Text}}$ queries in $\bigO(\log \Textlen)$ time, and we can
    access any symbol of $\Text$ in $\bigO(\log \Textlen)$ time. Thus, we can
    compare any two substrings of $\Textinf$ or $\revstr{\Textinf}$
    (specified with their starting positions and lengths) in
    $\bigO(\log \Textlen)$ time. Consequently, all structures needed
    in \cref{pr:str-str} were either constructed above or are part of
    $\CompSaCore{\Text}$.
  \end{enumerate}
  In total, the construction takes $\bigO(\SubstringComplexity{\Text} \log^7 \Textlen + \LZSize{\Text} \log^4
  \Textlen) = \bigO(\SubstringComplexity{\Text} \log^7 \Textlen)$ time.
\end{proof}

\subsection{The Periodic Patterns and Positions}\label{sec:sa-periodic}

\subsubsection{Preliminaries}\label{sec:sa-periodic-prelim}

\paragraph{Notation and Definitions for Patterns}

\begin{definition}[Necklace-consistent function]\label{def:nc}
  A function $f: \Sigma^{+} \rightarrow \Sigma^{+}$ is said to be
  \emph{necklace-consistent} if it satisfies the following conditions
  for every $S, S'\in \Sigma^+$:
  \begin{enumerate}
  \item The strings $f(S)$ and $S$ are cyclically equivalent.
  \item If $S$ and $S'$ are cyclically equivalent, then $f(S)= f(S')$.
  \end{enumerate}
\end{definition}

Let $f$ be some necklace-consistent function. Let also $\tau \geq 1$
and $\Pat \in \Sigma^{m}$ be a $\tau$-periodic pattern. Denote $p =
\per(\Pat[1 \dd 3\tau - 1])$.  We define $\RootPat{f}{\tau}{\Pat}
:= f(\Pat[1 \dd p])$ and $\RunEndPat{\tau}{\Pat} := 1 + p +
\lcp(\Pat[1 \dd m], \Pat[1 + p \dd m])$.  Observe that then we can
write $\Pat[1 \dd \RunEndPat{\tau}{\Pat}) = H' H^{k} H''$, where $H =
\RootPat{f}{\tau}{\Pat}$, and $H'$ (resp.\ $H''$) is a proper
suffix (resp.\ prefix) of $H$. Note that there is always only one way
to write $\Pat[1 \dd \RunEndPat{\tau}{\Pat})$ in this way, since the
opposite would contradict the synchronization property of primitive
strings~\cite[Lemma~1.11]{AlgorithmsOnStrings}. We denote
$\HeadPat{f}{\tau}{\Pat} := |H'|$, $\ExpPat{f}{\tau}{\Pat} :=
k$, and $\TailPat{f}{\tau}{\Pat} := |H''|$. For any $t \geq 3\tau
- 1$, we define $\ExpCutPat{f}{\tau}{\Pat}{t} :=
\min(\ExpPat{f}{\tau}{\Pat}, \lfloor \tfrac{t - s}{|H|} \rfloor)$
and $\RunEndCutPat{f}{\tau}{\Pat}{t} := 1 + s +
\ExpCutPat{f}{\tau}{\Pat}{t}\cdot |H|$, where $s =
\HeadPat{f}{\tau}{\Pat}$ and $H = \RootPat{f}{\tau}{\Pat}$.
We denote $\RunEndFullPat{f}{\tau}{\Pat} :=
\RunEndCutPat{f}{\tau}{\Pat}{m}$. Note that
$\RunEndFullPat{f}{\tau}{\Pat} = j + s +
\ExpPat{f}{\tau}{\Pat}\cdot |H| = \RunEndPat{\tau}{\Pat} -
\TailPat{f}{\tau}{\Pat}$. Next, letting $p = |\per(\Pat[1 \dd
3\tau - 1])|$, we define $\TypePat{\tau}{\Pat} = +1$
if $\RunEndPat{\tau}{\Pat} \leq |\Pat|$ and
$\Pat[\RunEndPat{\tau}{\Pat}] \succ \Pat[\RunEndPat{\tau}{\Pat} - p]$,
and $\TypePat{\tau}{\Pat} = -1$ otherwise.

\begin{lemma}\label{lm:periodic-pat-lce}
  Let $\tau \geq 1$, $\Pat \in \Sigma^{+}$ be a $\tau$-periodic
  pattern, and $f$ be a necklace-consistent function. For every $\Pat'
  \in \Sigma^{+}$, $\lcp(\Pat, \Pat') \geq 3\tau - 1$ holds if and
  only if $\Pat'$ is $\tau$-periodic, $\RootPat{f}{\tau}{\Pat'} =
  \RootPat{f}{\tau}{\Pat}$, and $\HeadPat{f}{\tau}{\Pat'} =
  \HeadPat{f}{\tau}{\Pat}$.  Moreover, if $\RunEndPat{\tau}{\Pat}
  \leq |\Pat|$ and $\lcp(\Pat, \Pat') \geq \RunEndPat{\tau}{\Pat}$
  (which holds, in particular, when $\Pat$ is a prefix of $\Pat'$),
  then:
  \begin{itemize}
  \item $\RunEndPat{\tau}{\Pat'} = \RunEndPat{\tau}{\Pat}$,
  \item $\TailPat{f}{\tau}{\Pat'} = \TailPat{f}{\tau}{\Pat}$,
  \item $\RunEndFullPat{f}{\tau}{\Pat'} =
    \RunEndFullPat{f}{\tau}{\Pat}$,
  \item $\ExpPat{f}{\tau}{\Pat'} = \ExpPat{f}{\tau}{\Pat}$,
  \item $\TypePat{\tau}{\Pat'} = \TypePat{\tau}{\Pat}$.
  \end{itemize}
\end{lemma}
\begin{proof}

  Denote $s = \HeadPat{f}{\tau}{\Pat}$, $H =
  \RootPat{f}{\tau}{\Pat}$, and $p = |H|$.  Let us first assume
  $\lcp(\Pat, \Pat') \geq 3\tau - 1$. This implies $|\Pat'| \geq
  \lcp(\Pat, \Pat') \geq 3\tau - 1$ and $\Pat'[1 \dd 3\tau - 1] =
  \Pat[1 \dd 3\tau - 1]$. Thus, $\per(\Pat'[1 \dd 3\tau - 1]) =
  \per(\Pat[1 \dd 3\tau - 1]) = p \leq \tfrac{1}{3}\tau$, i.e.,
  $\Pat'$ is $\tau$-periodic. Moreover, $\RootPat{f}{\tau}{\Pat'}
  = f(\Pat'[1 \dd p]) = f(\Pat[1 \dd p]) = H$.  To show
  $\HeadPat{f}{\tau}{\Pat'} = s$, note that by $|H| \leq \tau$,
  the string $H' H^2$ (where $H'$ is a length-$s$ suffix of $H$) is a
  prefix of $\Pat[1 \dd 3\tau - 1] = \Pat'[1 \dd 3\tau - 1]$.  On the
  other hand, letting $s' = \HeadPat{f}{\tau}{\Pat'}$, it holds
  that $\widehat{H} H^2$ (where $\widehat{H}$ is a length-$s'$ suffix
  of $H$) is a prefix of $\Pat'[1 \dd 3\tau - 1]$. Thus, by the
  synchronization property of primitive
  strings~\cite[Lemma~1.11]{AlgorithmsOnStrings} applied to the two
  copies of $H$, we have $s' = s$, i.e., $\HeadPat{f}{\tau}{\Pat'}
  = \HeadPat{f}{\tau}{\Pat}$.  For the converse implication,
  assume that $\Pat'$ is $\tau$-periodic and it holds
  $\RootPat{f}{\tau}{\Pat'} = H$ and $\HeadPat{f}{\tau}{\Pat'}
  = s$.  This implies that both $\Pat[1 \dd \RunEndPat{\tau}{\Pat})$
  and $\Pat'[1 \dd \RunEndPat{\tau}{\Pat'})$ are prefixes of $H'
  \cdot H^{\infty}[1 \dd)$ (where $H'$ is as above). Thus, by
  $\RunEndPat{\tau}{\Pat} - 1, \RunEndPat{\tau}{\Pat'} - 1 \geq
  3\tau - 1$, we obtain $\lcp(\Pat, \Pat') \geq 3\tau - 1$.

  Let us now assume $\RunEndPat{\tau}{\Pat} \leq |\Pat|$ and
  $\lcp(\Pat, \Pat') \geq \RunEndPat{\tau}{\Pat}$.  Note that
  $\RunEndPat{\tau}{\Pat} > 3\tau - 1$ and the first part of the
  claim then imply that $\Pat'$ is $\tau$-periodic and it holds
  $\RootPat{f}{\tau}{\Pat'} = H$ and $\HeadPat{f}{\tau}{\Pat'}
  = s$.  By $\RunEndPat{\tau}{\Pat} \leq |\Pat|$, it follows that
  $\Pat[\RunEndPat{\tau}{\Pat}] \neq \Pat[\RunEndPat{\tau}{\Pat} -
  p]$.  Combining with $\lcp(\Pat, \Pat') \geq
  \RunEndPat{\tau}{\Pat}$, we thus have $\lcp(\Pat[1 \dd |\Pat|],
  \Pat[1 + p \dd |\Pat|]) = \lcp(\Pat'[1 \dd |\Pat'|], \Pat'[1 + p \dd
  |\Pat'|])$.  Consequently, $\RunEndPat{\tau}{\Pat'} = 1 + p +
  \lcp(\Pat'[1 \dd |\Pat'|], \Pat'[1 + p \dd |\Pat'|]) = 1 + p +
  \lcp(\Pat[1 \dd |\Pat|], \Pat[1 + p \dd |\Pat|]) =
  \RunEndPat{\tau}{\Pat}$.  We then also obtain
  $\TailPat{f}{\tau}{\Pat'} = (\RunEndPat{\tau}{\Pat'} - 1 -
  \HeadPat{f}{\tau}{\Pat'}) \bmod p = (\RunEndPat{\tau}{\Pat} - 1
  - \HeadPat{f}{\tau}{\Pat}) \bmod p =
  \TailPat{f}{\tau}{\Pat}$, $\RunEndFullPat{f}{\tau}{\Pat'} =
  \RunEndPat{\tau}{\Pat'} - \TailPat{f}{\tau}{\Pat'} =
  \RunEndPat{\tau}{\Pat} - \TailPat{f}{\tau}{\Pat} =
  \RunEndFullPat{f}{\tau}{\Pat}$, $\ExpPat{f}{\tau}{\Pat'} =
  \lfloor \tfrac{\RunEndFullPat{f}{\tau}{\Pat'} - 1}{p} \rfloor =
  \lfloor \tfrac{\RunEndFullPat{f}{\tau}{\Pat} - 1}{p} \rfloor =
  \ExpPat{f}{\tau}{\Pat}$. Finally, we note that by
  $\RunEndPat{\tau}{\Pat} \leq |\Pat|$ and $\RunEndPat{\tau}{\Pat'}
  = \RunEndPat{\tau}{\Pat} \leq \lcp(\Pat, \Pat') \leq |\Pat'|$, we
  obtain that $\TypePat{\tau}{\Pat} = -1$ holds if and only if
  $\TypePat{\tau}{\Pat} = -1$.  Thus, $\TypePat{\tau}{\Pat}
  = \TypePat{\tau}{\Pat'}$.
\end{proof}

\begin{lemma}\label{lm:pat-lex}
  Let $\tau \geq 1$, $f$ be a necklace-consistent function, and
  $\Pat_1, \Pat_2 \in \Sigma^{+}$ be $\tau$-periodic patterns such
  that $\RootPat{f}{\tau}{\Pat_1} = \RootPat{f}{\tau}{\Pat_2}$
  and $\HeadPat{f}{\tau}{\Pat_1} =
  \HeadPat{f}{\tau}{\Pat_2}$. Denote $t_1 =
  \RunEndPat{\tau}{\Pat_1} - 1$ and $t_2 = \RunEndPat{\tau}{\Pat_2}
  - 1$. Then, it holds $\lcp(\Pat_1, \Pat_2) \geq \min(t_1,
  t_2)$. Moreover:
  \begin{enumerate}
  \item\label{lm:pat-lex-it-1} If $\TypePat{\tau}{\Pat_1} \neq
    \TypePat{\tau}{\Pat_2}$ or $t_1 \neq t_2$, then $\Pat_1 \neq
    \Pat_2$ and $\lcp(\Pat_1, \Pat_2) = \min(t_1, t_2)$,
  \item\label{lm:pat-lex-it-2} If $\TypePat{\tau}{\Pat_1} \neq
    \TypePat{\tau}{\Pat_2}$, then $\Pat_1 \prec \Pat_2$ if and only if
    $\TypePat{\tau}{\Pat_1} < \TypePat{\tau}{\Pat_2}$,
  \item\label{lm:pat-lex-it-3} If $\TypePat{\tau}{\Pat_1} = -1$,
    then $t_1 < t_2$ implies $\Pat_1 \prec \Pat_2$,
  \item\label{lm:pat-lex-it-4} If $\TypePat{\tau}{\Pat_1} =
    +1$, then $t_1 < t_2$ implies $\Pat_1 \succ \Pat_2$,
  \item\label{lm:pat-lex-it-5} If $\TypePat{\tau}{\Pat_1} =
    \TypePat{\tau}{\Pat_2}  = -1$ and $t_1 \neq
    t_2$, then $t_1 < t_2$ if and only if $\Pat_1 \prec \Pat_2$,
  \item\label{lm:pat-lex-it-6} If $\TypePat{\tau}{\Pat_1} =
    \TypePat{\tau}{\Pat_2} = +1$ and $t_1 \neq t_2$, then
    $t_1 < t_2$ if and only if $\Pat_1 \succ \Pat_2$.
  \end{enumerate}
\end{lemma}
\begin{proof}

  Denote $H = \RootPat{f}{\tau}{\Pat_1} =
  \RootPat{f}{\tau}{\Pat_2}$, $s = \HeadPat{f}{\tau}{\Pat_1} =
  \HeadPat{f}{\tau}{\Pat_2}$, and $p = |H|$. Let $H'$ be a
  length-$s$ suffix of $H$ and let $Q = H' H^{\infty}$. We first
  observe that both $\Pat_1[1 \dd \RunEndPat{\tau}{\Pat_1})$ and
  $\Pat_2[1 \dd \RunEndPat{\tau}{\Pat_2})$ are prefixes of $Q$, i.e.,
  $\Pat_1[1 \dd \RunEndPat{\tau}{\Pat_1}) = \Pat_1[1 \dd t_1] = Q[1
  \dd t_1]$ and $\Pat_2[1 \dd \RunEndPat{\tau}{\Pat_2}) = \Pat_2[1
  \dd t_2] = Q[1 \dd t_2]$. Thus, $\lcp(\Pat_1, \Pat_2) \geq \min(t_1,
  t_2)$.

  We now prove the remaining claims:
  \begin{enumerate}

  \item Recall that $\Pat_1[1 \dd \RunEndPat{\tau}{\Pat_1}) =
    \Pat_1[1 \dd t_1] = Q[1 \dd t_1]$, $\Pat_2[1 \dd
    \RunEndPat{\tau}{\Pat_2}) = \Pat_2[1 \dd t_2] = Q[1 \dd t_2]$,
    and $\lcp(\Pat_1, \Pat_2) \geq \min(t_1, t_2)$.  Thus, it remains
    to prove $\Pat_1 \neq \Pat_2$ and $\lcp(\Pat_1, \Pat_2) \leq
    \min(t_1, t_2)$.  Let us first assume $t_1 \neq t_2$. Without the
    loss of generality, let $t_1 < t_2$. Consider two cases:
    \begin{itemize}
    \item First, assume $\RunEndPat{\tau}{\Pat_1} = |\Pat_1| +
      1$. Then, $\Pat_1 = Q[1 \dd t_1]$.  This implies $\lcp(\Pat_1,
      \Pat_2) \leq t_1 = \min(t_1, t_2)$.  On the other hand, from
      $|\Pat_2| \geq t_2 > t_1 = |\Pat_1|$, we then obtain $\Pat_1
      \neq \Pat_2$.
    \item Let us now assume $\RunEndPat{\tau}{\Pat_1} \leq
      |\Pat_1|$. By $t_1 < t_2$, this implies $\Pat_1[1 + t_1] =
      \Pat_1[\RunEndPat{\tau}{\Pat_1}] \neq
      \Pat_1[\RunEndPat{\tau}{\Pat_1} - p] = \Pat_1[1 + t_1 - p] =
      Q[1 + t_1 - p] = Q[1 + t_1] = \Pat_2[1 + t_1]$. Thus, $\Pat_1
      \neq \Pat_2$ and $\lcp(\Pat_1, \Pat_2) \leq t_1 = \min(t_1,
      t_2)$.
    \end{itemize}
    Let us now assume $t_1 = t_2$ and
    $\TypePat{\tau}{\Pat_1} \neq \TypePat{\tau}{\Pat_2}$. Without the
    loss of generality, let us assume $\TypePat{\tau}{\Pat_1} = -1$
    and $\TypePat{\tau}{\Pat_2} = +1$. Note that then
    $\RunEndPat{\tau}{\Pat_2} \leq |\Pat_2|$ and
    $\Pat_2[\RunEndPat{\tau}{\Pat_2}] \succ \Pat_2[\RunEndPat{\tau}{\Pat_2}
    - p]$.  Consider two cases:
    \begin{itemize}
    \item First, assume $\RunEndPat{\tau}{\Pat_1} = |\Pat_1| + 1$.
      Then, it holds $\Pat_1 = Q[1 \dd t_1]$. This immediately implies
      $\lcp(\Pat_1, \Pat_2) \leq t_1 = \min(t_1, t_2)$.  On the other
      hand, $\TypePat{\tau}{\Pat_2} = +1$ implies
      $|\Pat_2| \geq \RunEndPat{\tau}{\Pat_2} = t_2 + 1 > t_1 =
      |\Pat_1|$. Thus, $\Pat_1 \neq \Pat_2$.
    \item Let us now assume $\RunEndPat{\tau}{\Pat_1} \leq |\Pat_1|$.
      By $\TypePat{\tau}{\Pat_1} = -1$, we then must have
      $\Pat_1[\RunEndPat{\tau}{\Pat_1}] \prec \Pat_1[\RunEndPat{\tau}{\Pat_1}
      - p]$. Recall that in the proof
      of \cref{lm:pat-lex}\eqref{lm:pat-lex-it-1}, we observed that in
      this situation it holds $\Pat_1[1 + t_1] \prec Q[1 + t_1]$ and
      $Q[1 + t_2] \prec \Pat_2[1 + t_2]$.  By $t_1 = t_2$, this
      implies $\Pat_1 \neq \Pat_2$ and
      $\lcp(\Pat_1, \Pat_2) \leq \min(t_1, t_2)$.
    \end{itemize}

  \item Let us first assume $\TypePat{\tau}{\Pat_1} = -1$ and
    $\TypePat{\tau}{\Pat_2} = +1$. We will prove that $\Pat_1 \prec
    Q \prec \Pat_2$.  Recall that
    $\Pat_1[1 \dd \RunEndPat{\tau}{\Pat_1}) = \Pat_1[1 \dd t_1] =
    Q[1 \dd t_1]$.  If $t_1 = |\Pat_1|$ then we immediately obtain
    $\Pat_1 \prec Q$. Otherwise, $\TypePat{\tau}{\Pat_1} = -1$ implies
    that $t_1 < |\Pat_1|$ and $\Pat_1[1 + t_1] \prec \Pat_1[1 + t_1 -
    p] = Q[1 + t_1 - p] = Q[1 + t_1]$. Thus, we again have
    $\Pat_1 \prec Q$. Next, recall that
    $\Pat_2[1 \dd \RunEndPat{\tau}{\Pat_2}) = \Pat_2[1 \dd t_2] =
    Q[1 \dd t_2]$.  On the other hand, $\TypePat{\tau}{\Pat_2} = +1$
    implies that $\RunEndPat{\tau}{\Pat_2} \leq |\Pat_2|$ and
    $\Pat_2[1 + t_2] \succ \Pat_2[1 + t_2 - p] = Q[1 + t_2 - p] = Q[1
    + t_2]$. Hence, $\Pat_2 \succ Q$. We have thus proved
    $\Pat_1 \prec Q \prec \Pat_2$, which implies
    $\Pat_1 \prec \Pat_2$.  Let us now assume
    $\Pat_1 \prec \Pat_2$. Suppose
    $\TypePat{\tau}{\Pat_1} \geq \TypePat{\tau}{\Pat_2}$. By the
    assumption $\TypePat{\tau}{\Pat_1} \neq \TypePat{\tau}{\Pat_2}$,
    we then must have $\TypePat{\tau}{\Pat_1} = +1$ and
    $\TypePat{\tau}{\Pat_2} = -1$.  Analogously as above, it then
    follows $\Pat_2 \prec Q \prec \Pat_1$, which contradicts the
    assumption.  Thus, $\TypePat{\tau}{\Pat_1}
    < \TypePat{\tau}{\Pat_2}$.

  \item Recall that $\Pat_1[1 \dd \RunEndPat{\tau}{\Pat_1})
    = \Pat_1[1 \dd t_1] = Q[1 \dd t_1]$ and
    $\Pat_2[1 \dd \RunEndPat{\tau}{\Pat_2}) = \Pat_2[1 \dd t_2] =
    Q[1 \dd t_2]$.  If $\RunEndPat{\tau}{\Pat_1} = |\Pat_1| + 1$, then
    $\Pat_1 = \Pat_1[1 \dd \RunEndPat{\tau}{\Pat_1}) = Q[1 \dd t_1]$.
    By $t_1 < t_2$, we thus obtain $\Pat_1 = Q[1 \dd t_1] \prec
    Q[1 \dd t_2]
    = \Pat_2[1 \dd \RunEndPat{\tau}{\Pat_2}) \preceq \Pat_2$.  Let us
    thus assume $\RunEndPat{\tau}{\Pat_1} \leq |\Pat_1|$.  By $t_1 <
    t_2$, we have $\Pat_1[1 \dd t_1] = \Pat_2[1 \dd t_1]$. On the
    other hand, by $\TypePat{\tau}{\Pat_1} = -1$ and $t_1 < t_2$, we
    have $\Pat_1[1 + t_1]
    = \Pat_1[\RunEndPat{\tau}{\Pat_1}] \prec \Pat_1[\RunEndPat{\tau}{\Pat_1}
    - p] = \Pat_1[1 + t_1 - p] = Q[1 + t_1 - p] = Q[1 + t_1]
    = \Pat_2[1 + t_1]$.  We thus obtain $\Pat_1[1 \dd 1 +
    t_1] \prec \Pat_2[1 \dd 1 + t_1]$, which implies
    $\Pat_1 \prec \Pat_2$.

  \item Recall (as above) that
    $\Pat_1[1 \dd \RunEndPat{\tau}{\Pat_1}) =
    \Pat_1[1 \dd t_1] = Q[1 \dd t_1]$ and $\Pat_2[1 \dd
    \RunEndPat{\tau}{\Pat_2}) = \Pat_2[1 \dd t_2] = Q[1 \dd t_2]$.
    The assumptions $\TypePat{\tau}{\Pat_1} = +1$ and $t_1 < t_2$ imply
    $\Pat_1[1 + t_1] = \Pat_1[\RunEndPat{\tau}{\Pat_1}] \succ
    \Pat_1[\RunEndPat{\tau}{\Pat_1} - p] = \Pat_1[1 + t_1 - p] = Q[1
    + t_1 - p] = Q[1 + t_1] = \Pat_2[1 + t_1]$.  We thus obtain
    $\Pat_1[1 \dd 1 + t_1] \succ \Pat_2[1 \dd 1 + t_1]$, which implies
    $\Pat_1 \succ \Pat_2$.

  \item The first implication follows by
    \cref{lm:pat-lex}\eqref{lm:pat-lex-it-3}. Let us thus assume
    $\Pat_1 \prec \Pat_2$.  Suppose $t_1 \geq t_2$. The assumption
    $t_1 \neq t_2$ then implies $t_1 > t_2$. Recall that it holds
    $\Pat_1[1 \dd \RunEndPat{\tau}{\Pat_1}) = \Pat_1[1 \dd t_1] = Q[1
    \dd t_1]$ and $\Pat_2[1 \dd \RunEndPat{\tau}{\Pat_2}) = \Pat_2[1
    \dd t_2] = Q[1 \dd t_2]$.  By $t_1 > t_2$, we thus have $\Pat_1[1
    \dd t_2] = \Pat_2[1 \dd t_2]$. Consider two cases. If
    $\RunEndPat{\tau}{\Pat_2} + 1 = |\Pat_2|$, then $\Pat_2 = \Pat_2[1 \dd
    \RunEndPat{\tau}{\Pat_2}) = Q[1 \dd t_2] \prec Q[1 \dd t_1] =
    \Pat_1[1 \dd \RunEndPat{\tau}{\Pat_1}) \preceq \Pat_1$, a
    contradiction.  Assume now $\RunEndPat{\tau}{\Pat_2} \leq |\Pat_2|$.
    Then, $\Pat_2[1 + t_2]
    = \Pat_2[\RunEndPat{\tau}{\Pat_2}] \prec
    \Pat_2[\RunEndPat{\tau}{\Pat_2} - p] = \Pat_2[1 + t_2 - p] = Q[1
    + t_2 - p] = Q[1 + t_2] = \Pat_1[1 + t_2]$. Therefore, we obtain
    $\Pat_2[1 \dd 1 + t_2] \prec \Pat_1[1 \dd 1 + t_2]$, which again
    implies $\Pat_2 \prec \Pat_1$, a contradiction. Thus, we must have
    $t_1 < t_2$.

  \item The first implication follows by
    \cref{lm:pat-lex}\eqref{lm:pat-lex-it-4}. Let us thus assume
    $\Pat_1 \succ \Pat_2$.  Suppose $t_1 \geq t_2$. The assumption
    $t_1 \neq t_2$ then implies $t_1 > t_2$.
    The assumptions $\TypePat{\tau}{\Pat_2} =
    +1$ and $t_1 > t_2$ imply $\RunEndPat{\tau}{\Pat_2} \leq
    |\Pat_2|$ and $\Pat_2[1 + t_2] = \Pat_2[\RunEndPat{\tau}{\Pat_2}]
    \succ \Pat_2[\RunEndPat{\tau}{\Pat_2} - p] = \Pat_2[1 + t_2 - p]
    = Q[1 + t_2 - p] = Q[1 + t_2] = \Pat_1[1 + t_2]$.  We thus obtain
    $\Pat_2[1 \dd 1 + t_2] \succ \Pat_1[1 \dd 1 + t_2]$, which implies
    $\Pat_2 \succ \Pat_1$, contradicting the assumption. Thus, we must
    have $t_1 < t_2$.  \qedhere
  \end{enumerate}
\end{proof}

\begin{lemma}\label{lm:pat-expcut}
  Let $\tau \geq 1$, $\Pat \in \Sigma^{+}$ be a $\tau$-periodic
  pattern, and $f$ be a necklace-consistent function.  Let $t \geq t'
  \geq 3\tau - 1$ and $\Pat' = \Pat[1 \dd \min(|\Pat|, t)]$. Then,
  $\Pat'$ is $\tau$-periodic and it holds:
  \begin{enumerate}
  \item\label{lm:pat-expcut-it-1}
    $\ExpCutPat{f}{\tau}{\Pat}{t'} =
    \ExpCutPat{f}{\tau}{\Pat'}{t'}$,
  \item\label{lm:pat-expcut-it-2}
    $\ExpCutPat{f}{\tau}{\Pat}{t} =
    \ExpPat{f}{\tau}{\Pat'}$,
  \item\label{lm:pat-expcut-it-3}
    $\RunEndCutPat{f}{\tau}{\Pat}{t'} =
    \RunEndCutPat{f}{\tau}{\Pat'}{t'}$,
  \item\label{lm:pat-expcut-it-4}
    $\RunEndCutPat{f}{\tau}{\Pat}{t} =
    \RunEndFullPat{f}{\tau}{\Pat'}$.
  \end{enumerate}
\end{lemma}
\begin{proof}

  1. Denote $m = |\Pat|$, $s = \HeadPat{f}{\tau}{\Pat}$, $H =
  \RootPat{f}{\tau}{\Pat}$, $p = |H|$, and $k_1 =
  \ExpCutPat{f}{\tau}{\Pat}{t'} = \min(\ExpPat{f}{\tau}{\Pat},
  \lfloor \tfrac{t' - s}{p} \rfloor)$. Observe that by $m \geq 3\tau -
  1$ and $t \geq 3\tau - 1$, we have $|\Pat'| = \min(m, t) \geq 3\tau
  - 1$. Thus, by \cref{lm:periodic-pat-lce}, $\Pat'$ is $\tau$-periodic and it
  holds $\HeadPat{f}{\tau}{\Pat'} = s$ and
  $\RootPat{f}{\tau}{\Pat'} = H$. We first show
  $\ExpCutPat{f}{\tau}{\Pat'}{t'} = k_1$. Consider two cases:
  \begin{itemize}
  \item First, assume $t \leq \RunEndPat{\tau}{\Pat} - 1$.  This
    implies $\ExpPat{f}{\tau}{\Pat} = \lfloor
    \tfrac{\RunEndPat{\tau}{\Pat} - 1 - s}{p} \rfloor \geq \lfloor
    \tfrac{t - s}{p} \rfloor \geq \lfloor \tfrac{t' - s}{p}
    \rfloor$. Thus, $k_1 = \min(\ExpPat{f}{\tau}{\Pat}, \lfloor
    \tfrac{t' - s}{p} \rfloor) = \lfloor \tfrac{t' - s}{p} \rfloor$.
    On the other hand, we then have $m \geq \RunEndPat{\tau}{\Pat} -
    1 \geq t$ and hence $|\Pat'| = \min(m, t) = t$.  Observe that by
    $p + \lcp(\Pat[1 \dd m], \Pat[1 + p \dd m]) =
    \RunEndPat{\tau}{\Pat} - 1 \geq t$, we then also have $p +
    \lcp(\Pat[1 \dd t], \Pat[1 + p \dd t]) = t$.  Thus,
    $\RunEndPat{\tau}{\Pat'} - 1 = p + \lcp(\Pat'[1 \dd t], \Pat'[1 +
    p \dd t]) = p + \lcp(\Pat[1 \dd t], \Pat[1 + p \dd t]) =
    t$. This in turn implies $\ExpPat{f}{\tau}{\Pat'} = \lfloor
    \tfrac{\RunEndPat{\tau}{\Pat'} - 1 - s}{p} \rfloor = \lfloor
    \tfrac{t - s}{p} \rfloor \geq \lfloor \tfrac{t' - s}{p} \rfloor$.
    Consequently, $\ExpCutPat{f}{\tau}{\Pat'}{t'} =
    \min(\ExpPat{f}{\tau}{\Pat'}, \lfloor \tfrac{t' - s}{p}
    \rfloor) = \lfloor \tfrac{t' - s}{p} \rfloor = k_1$.
  \item Let us now assume $\RunEndPat{\tau}{\Pat} - 1 < t$.  Consider
    two subcases. If $m < t$, then $\Pat' = \Pat$ and we immediately
    obtain $\ExpCutPat{f}{\tau}{\Pat'}{t'} =
    \ExpCutPat{f}{\tau}{\Pat}{t'}$.  Otherwise (i.e., $m \geq t$),
    we have $\RunEndPat{\tau}{\Pat} \leq m$ and $\lcp(\Pat', \Pat) =
    t \geq \RunEndPat{\tau}{\Pat}$, and thus by \cref{lm:periodic-pat-lce}, it
    holds $\ExpPat{f}{\tau}{\Pat'} = \ExpPat{f}{\tau}{\Pat}$.
    This implies $\ExpCutPat{f}{\tau}{\Pat'}{t'} =
    \ExpCutPat{f}{\tau}{\Pat}{t'}$.
  \end{itemize}

  We prove the remaining claims as follows:
  \begin{itemize}
  \item To prove the second claim, observe that by $t \geq |\Pat'| \geq
    \RunEndPat{\tau}{\Pat'} - 1$ it follows that
    $\ExpCutPat{f}{\tau}{\Pat'}{t} =
    \min(\ExpPat{f}{\tau}{\Pat'}, \lfloor \tfrac{t - s}{p}
    \rfloor) = \min(\lfloor \tfrac{\RunEndPat{\tau}{\Pat'} - 1 -
    s}{p} \rfloor, \lfloor \tfrac{t - s}{p} \rfloor) = \lfloor
    \tfrac{\RunEndPat{\tau}{\Pat'} - 1 - s}{p} \rfloor =
    \ExpPat{f}{\tau}{\Pat'}$.  Thus, we obtain by
    \cref{lm:pat-expcut}\eqref{lm:pat-expcut-it-1} that
    $\ExpCutPat{f}{\tau}{\Pat}{t} =
    \ExpCutPat{f}{\tau}{\Pat'}{t} = \ExpPat{f}{\tau}{\Pat'}$.
  \item By \cref{lm:pat-expcut}\eqref{lm:pat-expcut-it-1},
    $\RunEndCutPat{f}{\tau}{\Pat}{t'} = s +
    \ExpCutPat{f}{\tau}{\Pat}{t'} p = s +
    \ExpCutPat{f}{\tau}{\Pat'}{t'} p =
    \RunEndCutPat{f}{\tau}{\Pat'}{t'}$.
  \item By \cref{lm:pat-expcut}\eqref{lm:pat-expcut-it-2},
    $\RunEndCutPat{f}{\tau}{\Pat}{t} = s +
    \ExpCutPat{f}{\tau}{\Pat}{t} p = s +
    \ExpPat{f}{\tau}{\Pat'} p = \RunEndFullPat{f}{\tau}{\Pat'}$.
    \qedhere
  \end{itemize}
\end{proof}

\paragraph{Notation and Definitions for Positions}

Let $\tau \in [1 \dd \floor{\frac{\Textlen}{2}}]$ and $f$ be some
necklace-consistent function. Observe that if $j \in \RTwo{\tau}{\Text}$,
then $\Text[j \dd \Textlen]$ is $\tau$-periodic
(\cref{def:periodic-pattern}). Letting $\Pat = \Text[j \dd \Textlen]$ and $p =
\per(\Text[j \dd j + 3\tau - 1))$, we denote $\RootPos{f}{\tau}{\Text}{j} :=
\RootPat{f}{\tau}{\Pat}$, $\RunEndPos{\tau}{\Text}{j} := j +
\RunEndPat{\tau}{\Pat} - 1 = j + p + \LCE_{\Text}(j, j + p)$,
$\HeadPos{f}{\tau}{\Text}{j} := \HeadPat{f}{\tau}{\Pat}$,
$\ExpPos{f}{\tau}{\Text}{j} := \ExpPat{f}{\tau}{\Pat}$,
$\TailPos{f}{\tau}{\Text}{j} \,{:=}\, \TailPat{f}{\tau}{\Pat}$,
$\ExpCutPos{f}{\tau}{\Text}{j}{t} \,{:=}\,
\ExpCutPat{f}{\tau}{\Pat}{t}$, $\RunEndCutPos{f}{\tau}{\Text}{j}{t} := j
+ \RunEndCutPat{f}{\tau}{\Pat}{t} - 1$ (with $t \geq 3\tau - 1$),
$\RunEndFullPos{f}{\tau}{\Text}{j} := j + \RunEndFullPat{f}{\tau}{\Pat} -
1$, and $\TypePos{\tau}{\Text}{j} := \TypePat{\tau}{\Pat}$.
Note that $\RunEndFullPos{f}{\tau}{\Text}{j} = j + s +
\ExpPos{f}{\tau}{\Text}{j}|H| = \RunEndPos{\tau}{\Text}{j} -
\TailPos{f}{\tau}{\Text}{j}$, where $s = \HeadPos{f}{\tau}{\Text}{j}$ and $H
= \RootPos{f}{\tau}{\Text}{j}$.  We also let $\RunBeg{\tau}{\Text}{j} = j -
\lcs(\Text[1 \dd j), \Text[1 \dd j + p))$.

Let $H \in \Sigma^{+}$ and $s \in \Zz$.
We will repeatedly refer to the following subsets of $\RTwo{\tau}{\Text}$:
\begin{itemize}[itemsep=0.5pt]
\item $\RMinusTwo{\tau}{\Text} := \{j \in \RTwo{\tau}{\Text} :
  \TypePos{\tau}{\Text}{j} = -1\}$,
\item $\RPlusTwo{\tau}{\Text} := \RTwo{\tau}{\Text} \sm \RMinusTwo{\tau}{\Text}$,
\item $\RFour{f}{H}{\tau}{\Text} := \{j \in \RTwo{\tau}{\Text} :
  \RootPos{f}{\tau}{\Text}{j} = H\}$,
\item $\RMinusFour{f}{H}{\tau}{\Text} := \RMinusTwo{\tau}{\Text} \cap \RFour{f}{H}{\tau}{\Text}$,
\item $\RPlusFour{f}{H}{\tau}{\Text} := \RPlusTwo{\tau}{\Text} \cap \RFour{f}{H}{\tau}{\Text}$,
\item $\RFive{f}{s}{H}{\tau}{\Text} :=
  \{j \in \RFour{f}{H}{\tau}{\Text} : \HeadPos{f}{\tau}{\Text}{j} = s\}$,
\item $\RMinusFive{f}{s}{H}{\tau}{\Text} := \RMinusTwo{\tau}{\Text} \cap \RFive{f}{s}{H}{\tau}{\Text}$,
\item $\RPlusFive{f}{s}{H}{\tau}{\Text} := \RPlusTwo{\tau}{\Text} \cap \RFive{f}{s}{H}{\tau}{\Text}$,
\item $\RSix{f}{s}{k}{H}{\tau}{\Text} := \{j \in \RFive{f}{s}{H}{\tau}{\Text} :
  \ExpPos{f}{\tau}{\Text}{j} = k\}$,
\item $\RMinusSix{f}{s}{k}{H}{\tau}{\Text} := \RMinusTwo{\tau}{\Text} \cap \RSix{f}{s}{k}{H}{\tau}{\Text}$,
\item $\RPlusSix{f}{s}{k}{H}{\tau}{\Text} := \RPlusTwo{\tau}{\Text} \cap \RSix{f}{s}{k}{H}{\tau}{\Text}$.
\end{itemize}

Maximal blocks of positions from $\RTwo{\tau}{\Text}$ play an important role
in our data structure. The starting positions of these blocks are
defined as
\begin{align*}
  \RPrimTwo{\tau}{\Text} := \{j\in \RTwo{\tau}{\Text} : j-1 \notin \RTwo{\tau}{\Text}\}.
\end{align*}
We also let:
\begin{itemize}[itemsep=0.5pt]
\item $\RPrimMinusTwo{\tau}{\Text} = \RPrimTwo{\tau}{\Text} \cap \RMinusTwo{\tau}{\Text}$,
\item $\RPrimPlusTwo{\tau}{\Text} = \RPrimTwo{\tau}{\Text} \cap \RPlusTwo{\tau}{\Text}$,
\item $\RPrimMinusFour{f}{H}{\tau}{\Text} = \RPrimTwo{\tau}{\Text} \cap \RMinusFour{f}{H}{\tau}{\Text}$,
\item $\RPrimPlusFour{f}{H}{\tau}{\Text} = \RPrimTwo{\tau}{\Text} \cap \RPlusFour{f}{H}{\tau}{\Text}$.
\end{itemize}

\begin{lemma}\label{lm:R-text-block}
  Let $\tau \in [1 \dd \floor{\frac{\Textlen}{2}}]$ and $f$ be any
  necklace-consistent function. For any position $j \in
  \RTwo{\tau}{\Text}$ such that $j-1 \in \RTwo{\tau}{\Text}$, it holds
  \begin{itemize}
  \item $\RootPos{f}{\tau}{\Text}{j {-} 1} = \RootPos{f}{\tau}{\Text}{j}$,
  \item $\RunEndPos{\tau}{\Text}{j {-} 1} = \RunEndPos{\tau}{\Text}{j}$,
  \item $\TailPos{f}{\tau}{\Text}{j {-} 1} = \TailPos{f}{\tau}{\Text}{j}$,
  \item $\RunEndFullPos{f}{\tau}{\Text}{j {-} 1} =
    \RunEndFullPos{f}{\tau}{\Text}{j}$,
  \item $\TypePos{\tau}{\Text}{j {-} 1} = \TypePos{\tau}{\Text}{j}$.
  \end{itemize}
\end{lemma}
\begin{proof}

  Denote $\Pat = \Text[j - 1 \dd j + 3\tau - 2)$, $\Pat' = \Text[j \dd j +
  3\tau - 1)$, $p = \per(\Pat)$ and $p' = \per(\Pat')$. We first prove
  that $p = p'$. Suppose $p \neq p'$. Denote $X = \Text[j \dd j + \tau)$.
  Observe that since $\Pat$ and $\Pat'$ overlap by $3\tau - 2 \geq
  \tau$ symbols, $X$ occurs in both $\Pat$ and $\Pat'$.  Thus, $X$ has
  periods $p$ and $p'$. Note that it is not possible that $p \mid p'$,
  since this implies that the length-$p'$ prefix of $X$ is not
  primitive (contradicting $p' = \per(\Pat')$).  By $p, p' \leq
  \tfrac{1}{3}\tau$ and the Weak Periodicity
  Lemma~\cite{periodicitylemma}, we therefore have that $X$ has period
  $p'' = \gcd(p, p')$. Since $p$ does not divide $p'$ we thus have
  $p'' < p'$. By definition, we also have $p'' \mid p$. This, however,
  again implies that $Y[1 \dd p']$ is not primitive, contradicting $p'
  = \per(\Pat')$. Thus, $p = p'$. We now show $\RootPos{f}{\tau}{\Text}{j
  - 1} = \RootPos{f}{\tau}{\Text}{j}$. Observe that $p, p' \leq
  \tfrac{1}{3}\tau$ implies that $\{\Text[j-1+\delta \dd j-1+\delta+p) :
  \delta \in [0 \dd p)\} = \{\Text[j+\delta \dd j + \delta + p) : \delta
  \in [0 \dd p)\}$.  Thus, the strings $\Text[j-1 \dd j-1+p)$ and $\Text[j \dd
  j+p)$ are cyclically equivalent. Consequently, since $f$ is
  necklace-consistent, we have $\RootPos{f}{\tau}{\Text}{j-1} = f(\Text[j - 1
  \dd j - 1 + p)) = f(\Text[j \dd j + p)) = \RootPos{f}{\tau}{\Text}{j}$.

  By the above, it holds $\Text[j - 1] = \Text[j - 1 + p]$. Thus,
  $\RunEndPos{\tau}{\Text}{j - 1} = j - 1 + p + \LCE_{\Text}(j - 1, j - 1 + p) =
  j + p + \LCE_{\Text}(j, j + p) = \RunEndPos{\tau}{\Text}{j}$.

  Let $H'$ (resp.\ $H''$) be a suffix (resp.\ prefix) of $H$ of length
  $\HeadPos{f}{\tau}{\Text}{j-1}$
  (resp.\ $\TailPos{f}{\tau}{\Text}{j-1}$). Then, $\Text[j-1 \dd
  \RunEndPos{\tau}{\Text}{j-1}) = H' H^{k} H''$, where $k =
  \ExpPos{f}{\tau}{\Text}{j-1}$. By $\RunEndPos{\tau}{\Text}{j-1} =
  \RunEndPos{\tau}{\Text}{j}$, and the uniqueness of this decomposition, it
  thus follows that we either have $\Text[j \dd \RunEndPos{\tau}{\Text}{j}) =
  H'[2 \dd |H'|] H^k H''$ (when $|H'| \geq 1$), or $\Text[j \dd
  \RunEndPos{\tau}{\Text}{j}) = H[2 \dd p] H^{k-1} H''$ (if $|H'| = 0$). In
  either case, we have $\TailPos{f}{\tau}{\Text}{j} = |H''| =
  \TailPos{f}{\tau}{\Text}{j-1}$.

  By the above, and the definition of $\RunEndFullPos{f}{\tau}{\Text}{j-1}$
  it follows that $\RunEndFullPos{f}{\tau}{\Text}{j-1} =
  \RunEndPos{\tau}{\Text}{j-1} - \TailPos{f}{\tau}{\Text}{j-1} =
  \RunEndPos{\tau}{\Text}{j} - \TailPos{f}{\tau}{\Text}{j} =
  \RunEndFullPos{f}{\tau}{\Text}{j}$.

  By $\RunEndPos{\tau}{\Text}{j-1} = \RunEndPos{\tau}{\Text}{j}$ and $p = p'$, we
  immediately obtain $\TypePos{\tau}{\Text}{j-1} = \TypePos{\tau}{\Text}{j}$.
\end{proof}

\begin{lemma}\label{lm:beg-end}
  Let $\tau \in [1 \dd \floor{\frac{\Textlen}{2}}]$, and $j \in
  \RTwo{\tau}{\Text}$. Then:
  \begin{enumerate}
  \item\label{lm:beg-end-it-1} It holds $\RunEndPos{\tau}{\Text}{j} =
    \max\{j' \in [j \dd \Textlen] : [j \dd j'] \subseteq \RTwo{\tau}{\Text}\} +
    3\tau - 1$,
  \item\label{lm:beg-end-it-2} If holds $\RunBeg{\tau}{\Text}{j} =
    \min\{j' \in [1 \dd j] : [j' \dd j] \subseteq \RTwo{\tau}{\Text}\}$.
  \end{enumerate}
\end{lemma}
\begin{proof}

  1. Let $p = \per(\Text[j \dd j + 3\tau - 1))$ and $j_e =
  \RunEndPos{\tau}{\Text}{j} - 3\tau + 1$. Recall that, by definition, $p
  \leq \tfrac{1}{3}\tau$. On the other hand, by $\RunEndPos{\tau}{\Text}{j}
  = j + p + \LCE_{\Text}(j, j + p)$, the string $\Text[j \dd
  \RunEndPos{\tau}{\Text}{j}) = \Text[j \dd j_e + 3\tau - 1)$ has period $p$.
  Thus, for every $j' \in [j \dd j_e]$, it holds $\per(\Text[j' \dd j' +
  3\tau - 1)) \leq p \leq \tfrac{1}{3}\tau$.  Hence $[j \dd j_e]
  \subseteq \RTwo{\tau}{\Text}$. Moreover, by \cref{lm:R-text-block}, for
  every $j' \in [j \dd j_e]$, it holds $\RunEndPos{\tau}{\Text}{j'} =
  \RunEndPos{\tau}{\Text}{j} = j_e$.  Suppose now that $j_e + 1 \in
  \RTwo{\tau}{\Text}$. By definition, we have $\RunEndPos{\tau}{\Text}{j_e + 1} \geq (j_e +
  1) + 3\tau - 1 = j_e + 3\tau$. By \cref{lm:R-text-block}, however,
  we would then also have $\RunEndPos{\tau}{\Text}{j_e + 1} =
  \RunEndPos{\tau}{\Text}{j_e} = j_e + 3\tau - 1$, a contradiction.  Thus,
  $j_e + 1 \not\in \RTwo{\tau}{\Text}$, and hence $j_e = \max\{j' \in [j \dd
  \Textlen] : [j \dd j'] \subseteq \RTwo{\tau}{\Text}\}$. Consequently,
  $\RunEndPos{\tau}{\Text}{j} = j_e + 3\tau - 1 = \max\{j' \in [j \dd \Textlen] :
  [j \dd j'] \subseteq \RTwo{\tau}{\Text}\} + 3\tau - 1$.

  2. Denote $j_b = \RunBeg{\tau}{\Text}{j}$. By $\RunBeg{\tau}{\Text}{j} = j
  - \lcs(\Text[1 \dd j), \Text[1 \dd j + p))$, it follows that the string
  $\Text[j_b \dd j + 3\tau - 1)$ has period $p$. Thus, for every $j' \in
  [j_b \dd j]$, it holds $\per(\Text[j' \dd j' + 3\tau - 1)) \leq p \leq
  \tfrac{1}{3}\tau$.  Hence, $[j_b \dd j] \subseteq
  \RTwo{\tau}{\Text}$. Suppose now that $j_b - 1 \in \RTwo{\tau}{\Text}$.  By
  \cref{lm:R-text-block}, we then have $\per(\Text[j_b - 1 \dd j_b + 3\tau
  - 2)) = |\RootPos{f}{\tau}{\Text}{j_b-1}| =
  |\RootPos{f}{\tau}{\Text}{j_b}| = \per(\Text[j_b \dd j_b + 3\tau - 1)) =
  p$. In particular, $\Text[j_b - 1] = \Text[j_b - 1 + p]$.  This implies $j -
  \lcs(\Text[1 \dd j), \Text[1 \dd j + p)) \leq j_b - 1$, a
  contradiction. Thus, $\RunBeg{\tau}{\Text}{j} = j_b = \min\{j' \in [1
  \dd j] : [j' \dd j] \subseteq \RTwo{\tau}{\Text}\}$.
\end{proof}

\begin{lemma}\label{lm:periodic-pos-lce}
  Let $\tau \in [1 \dd \lfloor \tfrac{\Textlen}{2} \rfloor]$, $f$ be
  any necklace-consistent function, and $j \in [1 \dd \Textlen]$.
  \begin{enumerate}
  \item\label{lm:periodic-pos-lce-it-1}
    Let $\Pat \in \Sigma^{+}$ be a $\tau$-periodic pattern. Then,
    the following conditions are equivalent:
    \begin{itemize}
    \item $j \in \OccThree{3\tau - 1}{\Pat}{\Text}$,
    \item $\lcp(\Pat, \Text[j \dd \Textlen]) \geq 3\tau - 1$,
    \item $j \in \RTwo{\tau}{\Text}$, $\RootPos{f}{\tau}{\Text}{j} =
      \RootPat{f}{\tau}{\Pat}$, and $\HeadPos{f}{\tau}{\Text}{j} =
      \HeadPat{f}{\tau}{\Pat}$.
    \end{itemize}
    Moreover, if, letting $t = \RunEndPat{\tau}{\Pat} - 1$, it holds
    $\lcp(\Pat, \Text[j \dd \Textlen]) > t$, then:
    \begin{itemize}
    \item $\RunEndPat{\tau}{\Pat} - 1 = \RunEndPos{\tau}{\Text}{j} - j$,
    \item $\TailPat{f}{\tau}{\Pat} = \TailPos{f}{\tau}{\Text}{j}$,
    \item $\RunEndFullPat{f}{\tau}{\Pat} - 1 =
      \RunEndFullPos{f}{\tau}{\Text}{j} - j$,
    \item $\ExpPat{f}{\tau}{\Pat} = \ExpPos{f}{\tau}{\Text}{j}$,
    \item $\TypePat{\tau}{\Pat} = \TypePos{\tau}{\Text}{j}$.
    \end{itemize}
  \item\label{lm:periodic-pos-lce-it-2}
    Let $j' \in \RTwo{\tau}{\Text}$. Then, the following conditions
    are equivalent:
    \begin{itemize}
    \item $j \in \OccThree{3\tau - 1}{j'}{\Text}$,
    \item $\LCE_{\Text}(j', j) \geq 3\tau - 1$,
    \item $j \in \RTwo{\tau}{\Text}$,
      $\RootPos{f}{\tau}{\Text}{j} = \RootPos{f}{\tau}{\Text}{j'}$, and
      $\HeadPos{f}{\tau}{\Text}{j'} = \HeadPos{f}{\tau}{\Text}{j}$.
    \end{itemize}
    Moreover, if letting $t = \RunEndPos{\tau}{\Text}{j} - j$, it holds
    $\LCE_{\Text}(j, j') > t$, then:
    \begin{itemize}
    \item $\RunEndPos{\tau}{\Text}{j'} - j' = \RunEndPos{\tau}{\Text}{j} - j$,
    \item $\TailPos{f}{\tau}{\Text}{j'} = \TailPos{f}{\tau}{\Text}{j}$,
    \item $\RunEndFullPos{f}{\tau}{\Text}{j'} - j' =
      \RunEndFullPos{f}{\tau}{\Text}{j} - j$,
    \item $\ExpPos{f}{\tau}{\Text}{j'} = \ExpPos{f}{\tau}{\Text}{j}$,
    \item $\TypePos{\tau}{\Text}{j'} = \TypePos{\tau}{\Text}{j}$.
    \end{itemize}
  \end{enumerate}
\end{lemma}
\begin{proof}

  1. We begin by showing the three implications:
  \begin{itemize}
  \item Assume $j \in \OccThree{3\tau - 1}{\Pat}{\Text}$.  Note that $\Pat$
    being $\tau$-periodic implies that $|\Pat| \geq 3\tau - 1$.  Thus,
    $j \in \OccThree{3\tau - 1}{\Pat}{\Text}$ by definition implies
    $\lcp(\Pat, \Text[j \dd \Textlen]) \geq \min(|\Pat|, 3\tau - 1) \geq 3\tau
    - 1$.
  \item Let us now assume $\lcp(\Pat, \Text[j \dd \Textlen]) \geq 3\tau -
    1$. This implies $j \in [1 \dd \Textlen - 3\tau + 2]$.  Moreover,
    denoting $\Pat' = \Text[j \dd \Textlen]$, it then follows by
    \cref{lm:periodic-pat-lce}, that $\Pat'$ is $\tau$-periodic and it holds
    $\RootPat{f}{\tau}{\Pat'} = \RootPat{f}{\tau}{\Pat}$ and
    $\HeadPat{f}{\tau}{\Pat'} =
    \HeadPat{f}{\tau}{\Pat}$. Consequently, $\per(\Text[j \dd j +
    3\tau - 1)) = \per(\Pat'[1 \dd 3\tau - 1]) \leq \tfrac{1}{3}\tau$,
    i.e., $j \in \RTwo{\tau}{\Text}$.  We then also immediately obtain
    $\RootPos{f}{\tau}{\Text}{j} = \RootPat{f}{\tau}{\Pat'} =
    \RootPat{f}{\tau}{\Pat}$ and $\HeadPos{f}{\tau}{\Text}{j} =
    \HeadPat{f}{\tau}{\Pat'} = \HeadPat{f}{\tau}{\Pat}$.
  \item Let us finally assume that it holds $j \in \RTwo{\tau}{\Text}$,
    $\RootPos{f}{\tau}{\Text}{j} = \RootPat{f}{\tau}{\Pat}$, and
    $\HeadPos{f}{\tau}{\Text}{j} = \HeadPat{f}{\tau}{\Pat}$. Denote
    $\Pat' = \Text[j \dd \Textlen]$. Note that $j \in \RTwo{\tau}{\Text}$ implies that
    $\Pat'$ is $\tau$-periodic. Moreover, by definition, we have
    $\RootPat{f}{\tau}{\Pat'} = \RootPos{f}{\tau}{\Text}{j}$ and
    $\HeadPat{f}{\tau}{\Pat'} = \HeadPos{f}{\tau}{\Text}{j}$. Thus,
    we obtain $\RootPat{f}{\tau}{\Pat'} =
    \RootPat{f}{\tau}{\Pat}$ and $\HeadPat{f}{\tau}{\Pat'} =
    \HeadPat{f}{\tau}{\Pat}$. Consequently, it follows by
    \cref{lm:periodic-pat-lce}, that $\lcp(\Pat, \Pat') \geq 3\tau - 1$. By
    definition of $\Pat'$, we thus obtain $j \in
    \OccThree{3\tau - 1}{\Pat}{\Text}$.
  \end{itemize}
  Let us now assume $\lcp(\Pat, \Text[j \dd \Textlen]) > t$, where $t =
  \RunEndPat{\tau}{\Pat} - 1$.  Note that this implies $|\Pat| \geq
  \lcp(\Pat, \Pat') \geq t + 1 = \RunEndPat{\tau}{\Pat}$.  Letting
  $\Pat' = \Text[j \dd \Textlen]$, we also have $\lcp(\Pat, \Pat') \geq t + 1 =
  \RunEndPat{\tau}{\Pat}$.  Thus, by combining \cref{lm:periodic-pat-lce} with
  the definitions of the values below, we obtain:
  \begin{itemize}
  \item $\RunEndPat{\tau}{\Pat} - 1 = \RunEndPat{\tau}{\Pat'} - 1 =
    \RunEndPos{\tau}{\Text}{j} - j$,
  \item $\TailPat{f}{\tau}{\Pat} = \TailPat{f}{\tau}{\Pat'} =
    \TailPos{f}{\tau}{\Text}{j}$,
  \item $\RunEndFullPat{f}{\tau}{\Pat} - 1 =
    \RunEndFullPat{f}{\tau}{\Pat'} - 1 = \RunEndFullPos{f}{\tau}{\Text}{j}
    - j$,
  \item $\ExpPat{f}{\tau}{\Pat} = \ExpPat{f}{\tau}{\Pat'} =
    \ExpPos{f}{\tau}{\Text}{j}$,
  \item $\TypePat{\tau}{\Pat} = \TypePat{\tau}{\Pat'} =
    \TypePos{\tau}{\Text}{j}$.
  \end{itemize}

  2. Denote $\Pat := \Text[j' \dd \Textlen]$. Observe that $j' \in \RTwo{\tau}{\Text}$
  implies that $\Pat$ is $\tau$-periodic and, by definition, it holds
  $\OccThree{3\tau - 1}{j'}{\Text} = \OccThree{3\tau - 1}{\Pat}{\Text}$, $\LCE_{\Text}(j', j) = \lcp(\Pat, \Text[j \dd
  \Textlen])$, $\RootPos{f}{\tau}{\Text}{j'} = \RootPat{f}{\tau}{\Pat}$,
  $\HeadPos{f}{\tau}{\Text}{j'} = \HeadPat{f}{\tau}{\Pat}$,
  $\RunEndPos{\tau}{\Text}{j'} - j' = \RunEndPat{\tau}{\Pat} - 1$,
  $\TailPos{f}{\tau}{\Text}{j'} = \TailPat{f}{\tau}{\Pat}$,
  $\RunEndFullPos{f}{\tau}{\Text}{j'} - j' = \RunEndFullPat{f}{\tau}{\Pat}
  - 1$, $\ExpPos{f}{\tau}{\Text}{j'} = \ExpPat{f}{\tau}{\Pat}$, and
  $\TypePos{\tau}{\Text}{j'} = \TypePat{\tau}{\Pat}$. Thus, all
  claims follow by \cref{lm:periodic-pos-lce}\eqref{lm:periodic-pos-lce-it-1}.
\end{proof}

\begin{lemma}\label{lm:R-lex-block-pat}
  Let $\tau \in [1 \dd \floor{\frac{\Textlen}{2}}]$, $f$ be any
  necklace-consistent function, and $j \in \RTwo{\tau}{\Text}$.
  Let $\Pat \in \Sigma^{m}$ be a $\tau$-periodic pattern such that
  $\RootPat{f}{\tau}{\Pat} = \RootPos{f}{\tau}{\Text}{j}$ and
  $\HeadPat{f}{\tau}{\Pat} = \HeadPos{f}{\tau}{\Text}{j}$.  Then,
  letting $t_1 = \RunEndPos{\tau}{\Text}{j} - j$ and $t_2 =
  \RunEndPat{\tau}{\Pat} - 1$, it holds $\lcp(\Text[j \dd \Textlen], \Pat)
  \geq \min(t_1, t_2)$ and:
  \begin{enumerate}
  \item\label{lm:R-lex-block-pat-it-1}
    If $\TypePos{\tau}{\Text}{j} \neq
    \TypePat{\tau}{\Pat}$ or $t_1 \neq t_2$, then
    $\Text[j {\dd} \Textlen] \neq \Pat$ and
    $\lcp(\Text[j {\dd} \Textlen], \Pat) = \min(t_1, t_2)$,
  \item\label{lm:R-lex-block-pat-it-2}
    If $\TypePos{\tau}{\Text}{j} \neq \TypePat{\tau}{\Pat}$,
    then $\Text[j \dd \Textlen] \prec \Pat$ if and only if
    $\TypePos{\tau}{\Text}{j} < \TypePat{\tau}{\Pat}$,
  \item\label{lm:R-lex-block-pat-it-3}
    If $\TypePos{\tau}{\Text}{j} = -1$, then $t_1 < t_2$ implies
    $\Text[j \dd \Textlen] \prec \Pat$,
  \item\label{lm:R-lex-block-pat-it-4}
    If $\TypePos{\tau}{\Text}{j} = +1$, then $t_1 < t_2$ implies
    $\Text[j \dd \Textlen] \succ \Pat$,
  \item\label{lm:R-lex-block-pat-it-5}
    If $\TypePos{\tau}{\Text}{j} = \TypePat{\tau}{\Pat} = -1$, and
    $t_1 \neq t_2$, then $t_1 < t_2$ if and only if
    $\Text[j \dd \Textlen] \prec \Pat$,
  \item\label{lm:R-lex-block-pat-it-6}
    If $\TypePos{\tau}{\Text}{j} = \TypePat{\tau}{\Pat} = +1$, and
    $t_1 \neq t_2$, then $t_1 < t_2$ if and only if
    $\Text[j \dd \Textlen] \succ \Pat$.
  \end{enumerate}
\end{lemma}
\begin{proof}
  Denote $\Pat_1 := \Text[j \dd \Textlen]$ and $\Pat_2 := \Pat$. Observe
  that $\Pat_1$ is $\tau$-periodic and it holds
  $\RootPat{f}{\tau}{\Pat_1} = \RootPos{f}{\tau}{\Text}{j} =
  \RootPat{f}{\tau}{\Pat_2}$ and $\HeadPat{f}{\tau}{\Pat_1} =
  \HeadPos{f}{\tau}{\Text}{j} = \HeadPat{f}{\tau}{\Pat_2}$.
  Moreover, we then have $\TypePat{\tau}{\Pat_1} =
  \TypePos{\tau}{\Text}{j}$ and $\RunEndPat{\tau}{\Pat_2} - 1 =
  \RunEndPos{\tau}{\Text}{j} - j = t_1$. All claims thus follow by
  \cref{lm:pat-lex}.
\end{proof}

\begin{lemma}\label{lm:R-lex-block-pos}
  Let $\tau \in [1 \dd \floor{\frac{\Textlen}{2}}]$, $f$ be any
  necklace-consistent function, and $j \in \RTwo{\tau}{\Text}$. Let
  $j' \in \RTwo{\tau}{\Text}$ be such that $\RootPos{f}{\tau}{\Text}{j'} =
  \RootPos{f}{\tau}{\Text}{j}$ and $\HeadPos{f}{\tau}{\Text}{j'} =
  \HeadPos{f}{\tau}{\Text}{j}$.  Then, letting $t_1 =
  \RunEndPos{\tau}{\Text}{j} - j$ and $t_2 = \RunEndPos{\tau}{\Text}{j'} - j'$,
  it holds $\LCE_{\Text}(j, j') \geq \min(t_1, t_2)$ and:
  \begin{enumerate}
  \item\label{lm:R-lex-block-pos-it-1}
    If $\TypePos{\tau}{\Text}{j} \neq \TypePos{\tau}{\Text}{j'}$ or $t_1
    \neq t_2$, then $\LCE_{\Text}(j, j') = \min(t_1, t_2)$,
  \item\label{lm:R-lex-block-pos-it-2}
    If $\TypePos{\tau}{\Text}{j} \neq \TypePos{\tau}{\Text}{j'}$, then
    $\Text[j \dd \Textlen] \prec \Text[j' \dd \Textlen]$ if and only if
    $\TypePos{\tau}{\Text}{j} <
    \TypePos{\tau}{\Text}{j'}$,
  \item\label{lm:R-lex-block-pos-it-3}
    If $\TypePos{\tau}{\Text}{j} = \TypePos{\tau}{\Text}{j'} = -1$ and
    $t_1 \neq t_2$, then $t_1 < t_2$ if and only if
    $\Text[j \dd \Textlen] \prec \Text[j'
    \dd \Textlen]$,
  \item\label{lm:R-lex-block-pos-it-4}
    If $\TypePos{\tau}{\Text}{j} = \TypePos{\tau}{\Text}{j'} = +1$ and
    $t_1 \neq t_2$, then $t_1 < t_2$ if and only if
    $\Text[j \dd \Textlen] \succ \Text[j'
    \dd \Textlen]$.
  \end{enumerate}
\end{lemma}
\begin{proof}
  Denote $\Pat_3 := \Text[j' \dd \Textlen]$. Observe that $\Pat_3$ is
  $\tau$-periodic and it holds $\RootPat{f}{\tau}{\Pat_3} =
  \RootPos{f}{\tau}{\Text}{j'} = \RootPos{f}{\tau}{\Text}{j}$ and
  $\HeadPat{f}{\tau}{\Pat_3} = \HeadPos{f}{\tau}{\Text}{j'} =
  \HeadPos{f}{\tau}{\Text}{j}$.  Moreover, we then have
  $\TypePat{\tau}{\Pat_3} = \TypePos{\tau}{\Text}{j'}$ and
  $\RunEndPat{\tau}{\Pat_3} - 1 = \RunEndPos{\tau}{\Text}{j'} - j' =
  t_2$. All claims thus follow by
  \cref{lm:R-lex-block-pat}.
\end{proof}

\begin{lemma}\label{lm:run-end}
  Let $\tau \in [1 \dd \floor{\frac{\Textlen}{2}}]$.  Let $j, j',
  j'' \in [1 \dd \Textlen]$ be such that $j, j'' \in \RTwo{\tau}{\Text}$, $j'
  \not\in \RTwo{\tau}{\Text}$, and $j < j' < j''$. Then, it holds
  $\RunEndPos{\tau}{\Text}{j} \leq j'' + \tau - 1$.
\end{lemma}
\begin{proof}

  Consider any $i \in \RTwo{\tau}{\Text}$ and let $b_i =
  \RunBeg{\tau}{\Text}{i}$, $e_i = \RunEndPos{\tau}{\Text}{i}$, and $p_i =
  \per(\Text[i \dd i + 3\tau - 1))$.  By definition, the fragment $\Text[b_e
  \dd e_i)$ has a period $p$.  Since, however, it contains $\Text[i \dd i
  + 3\tau - 1)$ as a substring, we must have $\per(\Text[b_i \dd e_i)) =
  p_i$. On the other hand, by definition of $\RunBeg{\tau}{\Text}{i}$
  and $\RunEndPos{\tau}{\Text}{i}$, the substring $\Text[b_i \dd e_i)$ cannot be
  extended in either left or right in $\Text$ without increasing its
  shortest period. Thus, by $p \leq \lfloor \tfrac{1}{3}\tau \rfloor$,
  $\Text[b_i \dd e_i)$ is a therefore a
  \emph{run}~\cite{KolpakovK99,michael1989detecting}.
  Any two distinct runs with periods
  at most $k$ must overlap by less than $2k$ symbols (see,
  e.g.,~\cite[Fact~2.2.4]{phdtomek}). Lastly, observe
  that by \cref{lm:beg-end}, we then also have $[b_i \dd e_i - 3\tau +
  1] \subseteq \RTwo{\tau}{\Text}$, $b_i - 1 \not\in \RTwo{\tau}{\Text}$, and $e_i -
  3\tau + 2 \not\in \RTwo{\tau}{\Text}$.

  Let us now consider substrings $\Text[\RunBeg{\tau}{\Text}{j} \dd
  \RunEndPos{\tau}{\Text}{j})$ and $\Text[\RunBeg{\tau}{\Text}{j''} \dd
  \RunEndPos{\tau}{\Text}{j''})$. By the above discussion we have
  $\RunEndPos{\tau}{\Text}{j} - \RunBeg{\tau}{\Text}{j''} < 2\lfloor
  \tfrac{1}{3}\tau \rfloor$.  By $\RunBeg{\tau}{\Text}{j''} \leq j''$, we
  thus obtain $\RunEndPos{\tau}{\Text}{j} \leq \RunBeg{\tau}{\Text}{j''} + 2
  \lfloor \tfrac{1}{3}\tau \rfloor \leq j'' + \tau - 1$.
\end{proof}

\begin{lemma}\label{lm:efull}
  Let $\tau \in [1 \dd \floor{\frac{\Textlen}{2}}]$ and $f$ be any necklace-consistent function.
  For any $j, j' \in
  \RPrimTwo{\tau}{\Text}$, $j \neq j'$ implies $\RunEndFullPos{f}{\tau}{\Text}{j} \neq
  \RunEndFullPos{f}{\tau}{\Text}{j'}$.
\end{lemma}
\begin{proof}
  Without the loss of generality, let us assume $j < j'$. We first
  derive an upper bound on $\RunEndFullPos{f}{\tau}{\Text}{j}$ and a lower
  bound on $\RunEndFullPos{f}{\tau}{\Text}{j'}$:
  \begin{itemize}
  \item By definition, it holds $j' - 1 \not\in \RTwo{\tau}{\Text}$. By
    \cref{lm:run-end} applied for $j$, $j' - 1$, and $j'$, we thus obtain
    $\RunEndPos{\tau}{\Text}{j} \leq j' + \tau - 1$. Consequently,
    $\RunEndFullPos{f}{\tau}{\Text}{j} \leq \RunEndPos{\tau}{\Text}{j} \leq j' +
    \tau - 1$.
  \item By definition, $\RunEndPos{\tau}{\Text}{j'} - j' \geq 3\tau - 1$, or
    equivalently, $\RunEndPos{\tau}{\Text}{j'} \geq j' + 3\tau - 1$.  On the
    other hand, $\RunEndPos{\tau}{\Text}{j'} - \RunEndFullPos{f}{\tau}{\Text}{j'}
    = \TailPos{f}{\tau}{\Text}{j'} \leq \lfloor \tfrac{1}{3}\tau
    \rfloor$. Combining the two inequalities, we thus obtain
    $\RunEndFullPos{f}{\tau}{\Text}{j'} \geq \RunEndPos{\tau}{\Text}{j'} - \lfloor
    \tfrac{1}{3}\tau \rfloor \geq j' + 3\tau - 1 - \lfloor
    \tfrac{1}{3}\tau \rfloor \geq j' + 2\tau - 1$.
  \end{itemize}
  By combining the two bounds, we obtain $\RunEndFullPos{f}{\tau}{\Text}{j}
  \leq j' + \tau - 1 < j' + 2\tau - 1 \leq
  \RunEndFullPos{f}{\tau}{\Text}{j'}$. Thus, $\RunEndFullPos{f}{\tau}{\Text}{j}
  \neq \RunEndFullPos{f}{\tau}{\Text}{j'}$.
\end{proof}

\begin{lemma}\label{lm:full-run-shifted}
  Let $\tau \in [1 \dd \lfloor \tfrac{\Textlen}{2} \rfloor]$ and
  $f$ be any necklace-consistent function. Let $i \in \RTwo{\tau}{\Text}$ and
  $\ell = \RunEndPos{\tau}{\Text}{i} - i$. For every $j' \in [1 \dd \Textlen]$,
  $\delta \in [0 \dd \ell]$, and $\ell' > \ell$, $\Textinf[j' -
  \delta \dd j' - \delta + \ell') = \Textinf[i \dd i + \ell')$
  implies that, letting $j = j' - \delta$, it holds $j \in
  \RTwo{\tau}{\Text}$ and:
  \begin{itemize}
  \item $\RootPos{f}{\tau}{\Text}{j} = \RootPos{f}{\tau}{\Text}{i}$,
  \item $\HeadPos{f}{\tau}{\Text}{j} = \HeadPos{f}{\tau}{\Text}{i}$,
  \item $\RunEndPos{\tau}{\Text}{j} - j = \RunEndPos{\tau}{\Text}{i} - i$,
  \item $\TailPos{f}{\tau}{\Text}{j} = \TailPos{f}{\tau}{\Text}{i}$,
  \item $\RunEndFullPos{f}{\tau}{\Text}{j} - j =
    \RunEndFullPos{f}{\tau}{\Text}{i} - i$,
  \item $\ExpPos{f}{\tau}{\Text}{j} = \ExpPos{f}{\tau}{\Text}{i}$,
  \item $\TypePos{\tau}{\Text}{j} = \TypePos{\tau}{\Text}{i}$.
  \end{itemize}
\end{lemma}
\begin{proof}
  First, observe that if $p \in [1 \dd \Textlen]$ and for $t \geq 0$
  (resp.\ $t' \geq 0$) the substring $\Textinf[p-t \dd p)$
  (resp.\ $\Textinf[p \dd p + t')$) does not contain symbol
  $\Text[\Textlen]$, then it holds $p-t \geq 1$ (resp.\ $p + t' \leq \Textlen$).
  Since for every $j \in \RTwo{\tau}{\Text}$, we have $\RunEndPos{\tau}{\Text}{j}
  \leq \Textlen$, then $i + \ell \leq \Textlen$ and by the uniqueness of
  $\Text[\Textlen]$ in $\Text$, the substring $\Textinf[i \dd i + \ell)$ does not
  contain $\Text[\Textlen]$.  Consequently, by $0 \leq \delta \leq \ell <
  \ell'$ and $\Textinf[j' - \delta \dd j' - \delta + \ell') =
  \Textinf[i \dd i + \ell')$, we have $\Textinf[i - \delta \dd i)
  = \Textinf[j' - \delta \dd j')$ and $\Textinf[i \dd i + \delta')
  = \Textinf[j' \dd j' + \delta')$ (where $\delta' = \ell -
  \delta$). Thus, $j' - \delta \geq 1$ and $j + \ell = j' + \delta'
  \leq \Textlen$, i.e., $[j \dd j + \ell] \subseteq [1 \dd \Textlen]$.
  Thus, $\LCE_{\Text}(i, j) \geq \ell + 1 > \RunEndPos{\tau}{\Text}{i} - i
  \geq 3\tau - 1$. The lemma thus follows by
  \cref{lm:periodic-pos-lce}\eqref{lm:periodic-pos-lce-it-2}.
\end{proof}

\begin{lemma}\label{lm:expcut}
  Let $\tau \in [1 \dd \lfloor \tfrac{\Textlen}{2} \rfloor]$ and
  $f$ be any necklace-consistent function. Let $j \in \RTwo{\tau}{\Text}$.
  Consider any $t \geq 3\tau - 1$.
  \begin{enumerate}
  \item\label{lm:expcut-it-1}
    Let $\Pat \in \Sigma^{+}$ be a $\tau$-periodic pattern satisfying
    $\RunEndPat{\tau}{\Pat} \leq |\Pat|$. If $j \in
    \OccThree{t}{\Pat}{\Text}$, then it holds:
    \begin{itemize}
    \item $\ExpCutPos{f}{\tau}{\Text}{j}{t} =
      \ExpCutPat{f}{\tau}{\Pat}{t}$,
    \item $\RunEndCutPos{f}{\tau}{\Text}{j}{t} - j =
      \RunEndCutPat{f}{\tau}{\Pat}{t} - 1$.
    \end{itemize}
  \item\label{lm:expcut-it-2}
    Let $j' \in \RTwo{\tau}{\Text}$. If $j \in \OccThree{t}{j'}{\Text}$, then it
    holds:
    \begin{itemize}
    \item $\ExpCutPos{f}{\tau}{\Text}{j}{t} =
      \ExpCutPos{f}{\tau}{\Text}{j'}{t}$,
    \item $\RunEndCutPos{f}{\tau}{\Text}{j}{t} - j =
      \RunEndCutPos{f}{\tau}{\Text}{j'}{t} - j'$.
    \end{itemize}
  \end{enumerate}
\end{lemma}
\begin{proof}
  Denote $H = \RootPos{f}{\tau}{\Text}{j}$, $s =
  \HeadPos{f}{\tau}{\Text}{j}$, and $p = |H|$.

  1. We first observe that $t \geq 3\tau - 1$ implies that $j \in
  \OccThree{t}{\Pat}{\Text} \subseteq \OccThree{3\tau - 1}{\Pat}{\Text}$.  Thus, by
  \cref{lm:periodic-pos-lce}\eqref{lm:periodic-pos-lce-it-1}, it follows that
  $\RootPat{f}{\tau}{\Pat} = H$ and $\HeadPat{f}{\tau}{\Pat} =
  s$. Denote $t' = \RunEndPat{\tau}{\Pat} - 1$. We consider two
  cases:
  \begin{itemize}
  \item Assume $t \leq t'$.  Then, it holds
    $\ExpCutPat{f}{\tau}{\Pat}{t} =
    \min(\ExpPat{f}{\tau}{\Pat}, \lfloor \tfrac{t - s}{p} \rfloor)
    = \min(\lfloor \tfrac{\RunEndPat{\tau}{\Pat} - 1 - s}{p} \rfloor,
    \lfloor \tfrac{t - s}{p} \rfloor) = \min(\lfloor \tfrac{t' - s}{p}
    \rfloor, \lfloor \tfrac{t - s}{p} \rfloor) = \lfloor \tfrac{t -
    s}{p} \rfloor$.  Note that we then also have $|\Pat| \geq
    \RunEndPat{\tau}{\Pat} - 1 = t' \geq t$.  By definition of
    $\OccThree{t}{\Pat}{\Text}$, this implies $\lcp(\Pat, \Text[j \dd \Textlen]) \geq
    \min(|\Pat|, t) = t$.  By definition of $\RunEndPat{\tau}{\Pat}$,
    we therefore have $p + \lcp(\Pat[1 \dd |\Pat|], \Pat[1 + p \dd
    |\Pat|]) = \RunEndPat{\tau}{\Pat} - 1 \geq t$.  Consequently,
    the assumption $\lcp(\Pat, \Text[j \dd \Textlen]) \geq t$ implies $p +
    \LCE_{\Text}(j, j + p) \geq t$, and hence $\RunEndPos{\tau}{\Text}{j} = j +
    p + \LCE_{\Text}(j, j + p) \geq j + t$, or equivalently,
    $\RunEndPos{\tau}{\Text}{j} - j \geq t$.  This implies
    $\ExpPos{f}{\tau}{\Text}{j} = \lfloor \tfrac{\RunEndPos{\tau}{\Text}{j} - j
    - s}{p} \rfloor \geq \lfloor \tfrac{t - s}{p} \rfloor$.  We thus
    obtain $\ExpCutPos{f}{\tau}{\Text}{j}{t} =
    \min(\ExpPos{f}{\tau}{\Text}{j}, \lfloor \tfrac{t - s}{p} \rfloor) =
    \lfloor \tfrac{t - s}{p} \rfloor =
    \ExpCutPat{f}{\tau}{\Pat}{t}$.
  \item Let us now assume $t > t'$. Recall that we assumed $|\Pat|
    \geq \RunEndPat{\tau}{\Pat} = t' + 1$.  Consequently, from $j \in
    \OccThree{t}{\Pat}{\Text}$, it follows that $\lcp(\Pat, \Text[j \dd \Textlen]) \geq
    \min(|\Pat|, t) > t'$.  By
    \cref{lm:periodic-pos-lce}\eqref{lm:periodic-pos-lce-it-1}, it follows that
    $\RunEndPos{\tau}{\Text}{j} - j = \RunEndPat{\tau}{\Pat} - 1 = t'$.
    This in turn implies that $\ExpPos{f}{\tau}{\Text}{j} = \lfloor
    \tfrac{\RunEndPos{\tau}{\Text}{j} - j - s}{p} \rfloor = \lfloor
    \tfrac{t' - s}{p} \rfloor = \lfloor \tfrac{\RunEndPat{\tau}{\Pat}
    - 1 - s}{p} \rfloor = \ExpPat{f}{\tau}{\Pat}$.  Hence, we obtain
    $\ExpCutPos{f}{\tau}{\Text}{j}{t} = \min(\ExpPos{f}{\tau}{\Text}{j},
    \lfloor \tfrac{t - s}{p} \rfloor) =
    \min(\ExpPat{f}{\tau}{\Pat}, \lfloor \tfrac{t - s}{p} \rfloor)
    = \ExpCutPat{f}{\tau}{\Pat}{t}$.
  \end{itemize}
  In both cases, we thus have $\ExpCutPos{f}{\tau}{\Text}{j}{t} =
  \ExpCutPat{f}{\tau}{\Pat}{t}$, i.e., the first claim.  This
  immediately gives the second claim, since,
  $\RunEndCutPos{f}{\tau}{\Text}{j}{t} - j = s +
  \ExpCutPos{f}{\tau}{\Text}{j}{t} \cdot p = s +
  \ExpCutPat{f}{\tau}{\Pat}{t} \cdot p =
  \RunEndCutPat{f}{\tau}{\Pat}{t} - 1$.

  2. Let $\Pat = \Text[j' \dd \Textlen]$. Note that $\Pat$ is $\tau$-periodic
  and, by the uniqueness of $\Text[\Textlen]$ in $\Text$, it holds $|\Pat| \geq
  \RunEndPat{\tau}{\Pat}$ and $j \in \OccThree{t}{j'}{\Text} =
  \OccThree{t}{\Pat}{\Text}$.  Note also that, by definition, it holds
  $\ExpCutPos{f}{\tau}{\Text}{j'}{t} = \ExpCutPat{f}{\tau}{\Pat}{t}$
  and $\RunEndCutPos{f}{\tau}{\Text}{j'}{t} - j' =
  \RunEndCutPat{f}{\tau}{\Pat}{t} - 1$. Thus, the claims follow by
  \cref{lm:expcut}\eqref{lm:expcut-it-1}.
\end{proof}

\begin{lemma}\label{lm:special-pat-properties}
  Let $\tau \in [1 \dd \lfloor \tfrac{\Textlen}{2} \rfloor]$ and
  $f$ be any necklace-consistent function. Let $j \in \RTwo{\tau}{\Text}$.
  Denote $s = \HeadPos{f}{\tau}{\Text}{j}$, $H =
  \RootPos{f}{\tau}{\Text}{j}$, $p = |H|$, $H' = H(p-s \dd p]$. Let $t,
  t'$ be such that $3\tau - 1 \leq t' \leq t$. Let $\Pat$ be a
  length-$t$ prefix of $H' H^{\infty}$. Denote $k = \lfloor \tfrac{t -
  s}{p} \rfloor$, $k' = \lfloor \tfrac{t' - s}{p} \rfloor$, $x_l = s
  + k' p$, and $y_u = \Text[j + s \dd j + s + t' - x_l)$. Then:
  \begin{itemize}
  \item $\Pat$ is $\tau$-periodic,
  \item $\Pat$ does not contain $\Text[\Textlen]$,
  \item $j \in \OccThree{3\tau - 1}{\Pat}{\Text}$,
  \item $\HeadPat{f}{\tau}{\Pat} = s$,
  \item $\RootPat{f}{\tau}{\Pat} = H$,
  \item $\RunEndPat{\tau}{\Pat} - 1 = t$,
  \item $\TypePat{\tau}{\Pat} = -1$,
  \item $\ExpPat{f}{\tau}{\Pat} = k$,
  \item $\ExpCutPat{f}{\tau}{\Pat}{t'} = k'$,
  \item $\RunEndCutPat{f}{\tau}{\Pat}{t'} - 1 = x_l$,
  \item $\Pat[\RunEndCutPat{f}{\tau}{\Pat}{t'} \dd t'] = y_u$.
  \end{itemize}
\end{lemma}
\begin{proof}
  By $|\Pat| = t \geq t' \geq 3\tau - 1$ and
  $|H| \leq \tfrac{1}{3}\tau$, it holds $\per(\Pat[1 \dd 3\tau -
  1]) \leq |H| \leq \tfrac{1}{3}\tau$.  Thus, $\Pat$ is
  $\tau$-periodic (\cref{def:periodic-pattern}). Since $\Pat$ is a
  substring of $H^{\infty}$, $\Pat$ does not contain symbol $\Text[\Textlen]$.
  Next, observe that $\RootPos{f}{\tau}{\Text}{j} = H$ and
  $\HeadPos{f}{\tau}{\Text}{j} = s$ implies that, letting $\Pat'$ be a
  prefix of $H' H^{\infty}$ of length $3\tau - 1$, $\Text[j \dd \Textlen]$ has
  $\Pat'$ as a prefix. Thus, $\lcp(\Pat, \Text[j \dd
  \Textlen]) \geq \lcp(\Pat, \Pat') = 3\tau - 1$, i.e., $j \in
  \OccThree{3\tau - 1}{\Pat}{\Text}$.
  By \cref{lm:periodic-pos-lce}\eqref{lm:periodic-pos-lce-it-1},
  this also implies $\HeadPat{f}{\tau}{\Pat} = s$ and
  $\RootPat{f}{\tau}{\Pat} = H$. This in turn implies
  $\RunEndPat{\tau}{\Pat} - 1 = p + \lcp(\Pat, \Pat[1 + p \dd t]) =
  t$.  Consequently, $\TypePat{\tau}{\Pat} = -1$.  Next,
  observe that $\ExpPat{f}{\tau}{\Pat}
  = \lfloor \tfrac{\RunEndPat{\tau}{\Pat} - 1 - s}{p} \rfloor
  = \lfloor \tfrac{t - s}{p} \rfloor = k$. We then also have
  $\ExpCutPat{f}{\tau}{\Pat}{t'}
  = \min(\ExpPat{f}{\tau}{\Pat}, \lfloor \tfrac{t' -
  s}{p} \rfloor) = \min(k, k') = k'$, and
  $\RunEndCutPat{f}{\tau}{\Pat}{t'} - 1 = s + k' p = x_l$. Lastly,
  observe that $\Pat[\RunEndCutPat{f}{\tau}{\Pat}{t'} \dd t']$ is a
  prefix of $H$ of length $t' - \RunEndCutPat{f}{\tau}{\Pat}{t'} + 1
  = t' - x_l$.  By $\Text[j + s \dd j + s + p) = H$ we thus have
  $\Pat[\RunEndCutPat{f}{\tau}{\Pat}{t'} \dd t'] = \Text[j + s \dd j + s
  + t' - x_l) = y_u$.
\end{proof}

\begin{lemma}\label{lm:RskH}
  Let $\tau \in [1 \dd \lfloor \tfrac{\Textlen}{2} \rfloor]$
  and $f$ be any necklace-consistent function.  Let $H \in \Sigma^{+}$,
  $p = |H|$, $s \in [0 \dd p)$, and $k_{\min} = \lceil \tfrac{3\tau -
  1 - s}{p} \rceil - 1$. Then:
  \begin{enumerate}
  \item For every $k \in [1 \dd k_{\min}]$,
    \[\RMinusSix{f}{s}{k}{H}{\tau}{\Text} \subseteq
      \{\RunEndFullPos{f}{\tau}{\Text}{j} - s - kp :
        j \in \RPrimMinusFour{f}{H}{\tau}{\Text}
        \text{ and }
        s + kp \leq \RunEndFullPos{f}{\tau}{\Text}{j} - j\},\]
  \item\label{lm:RskH-it-2} For every $k \in (k_{\min} \dd \Textlen]$,
    \[\RMinusSix{f}{s}{k}{H}{\tau}{\Text} =
      \{\RunEndFullPos{f}{\tau}{\Text}{j} - s - kp :
        j \in \RPrimMinusFour{f}{H}{\tau}{\Text}
        \text{ and }
        s + kp \leq \RunEndFullPos{f}{\tau}{\Text}{j} - j\}.\]
  \end{enumerate}
\end{lemma}
\begin{proof}

  For every $k \in [1 \dd \Textlen]$, denote $A_k
  := \{\RunEndFullPos{f}{\tau}{\Text}{j} - s - kp :
  j \in \RPrimMinusFour{f}{H}{\tau}{\Text} \text{ and }s +
  kp \leq \RunEndFullPos{f}{\tau}{\Text}{j} - j\}$.

  Consider any $k \in [1 \dd \Textlen]$. In the first step, we prove that
  $\RMinusSix{f}{s}{k}{H}{\tau}{\Text} \subseteq A_k$. Let
  $j \in \RMinusSix{f}{s}{k}{H}{\tau}{\Text}$, and $j' = \RunBeg{\tau}{\Text}{j}$. Then,
  by \cref{lm:beg-end}\eqref{lm:beg-end-it-2}, it holds $[j' \dd
  j] \subseteq \RTwo{\tau}{\Text}$.  By \cref{lm:R-text-block} we thus have
  $\RootPos{f}{\tau}{\Text}{j'} = \RootPos{f}{\tau}{\Text}{j} = H$ and
  $\TypePos{\tau}{\Text}{j'} = \TypePos{\tau}{\Text}{j} = -1$.  On the other
  hand, by \cref{lm:beg-end}\eqref{lm:beg-end-it-2}, $j' \in
  \RPrimTwo{\tau}{\Text}$. Hence, $j' \in \RPrimMinusFour{f}{H}{\tau}{\Text}$. Note
  that \cref{lm:R-text-block} also implies
  $\RunEndFullPos{f}{\tau}{\Text}{j'}
  = \RunEndFullPos{f}{\tau}{\Text}{j}$. Consequently, since $j' \leq j$, it
  holds $\RunEndFullPos{f}{\tau}{\Text}{j'} -
  j' \geq \RunEndFullPos{f}{\tau}{\Text}{j'} - j
  = \RunEndFullPos{f}{\tau}{\Text}{j} - j = \HeadPos{f}{\tau}{\Text}{j}
  + \ExpPos{f}{\tau}{\Text}{j}|\RootPos{f}{\tau}{\Text}{j}| = s + kp$. On
  the other hand, $\RunEndFullPos{f}{\tau}{\Text}{j} - j = s + kp$ and
  $\RunEndFullPos{f}{\tau}{\Text}{j} = \RunEndFullPos{f}{\tau}{\Text}{j'}$ imply
  $j = \RunEndFullPos{f}{\tau}{\Text}{j} - s - kp
  = \RunEndFullPos{f}{\tau}{\Text}{j'} - s - kp$. We have thus proved that
  $j \in A_k$.

  Let us now consider $k \in (k_{\min} \dd \Textlen]$. In the second step, we
  prove that for such $k$, it holds
  $A_k \subseteq \RMinusSix{f}{s}{k}{H}{\tau}{\Text}$.  Let $j \in A_k$. Then,
  there exists $j' \in \RPrimMinusFour{f}{H}{\tau}{\Text}$ such that $s +
  kp \leq \RunEndFullPos{f}{\tau}{\Text}{j'} - j'$ and $j
  = \RunEndFullPos{f}{\tau}{\Text}{j'} - s - kp$.  The assumptions
  immediately imply that $j' \leq j$. On the other hand, $k >
  k_{\min}$ implies that $\RunEndPos{\tau}{\Text}{j'} -
  j \geq \RunEndFullPos{f}{\tau}{\Text}{j'} - j = s + kp \geq s + (k_{\min}
  + 1) \cdot p = s + \lceil \tfrac{3\tau - 1 - s}{p} \rceil p \geq s
  + \tfrac{3\tau - 1 - s}{p} \cdot p = 3\tau - 1$. Thus, it follows
  by \cref{lm:beg-end}\eqref{lm:beg-end-it-1}, that $[j' \dd
  j] \subseteq \RTwo{\tau}{\Text}$.  By \cref{lm:R-text-block}, we thus have
  $j \in \RMinusFour{f}{H}{\tau}{\Text}$ and $\RunEndFullPos{f}{\tau}{\Text}{j}
  = \RunEndFullPos{f}{\tau}{\Text}{j'}$. Thus, $\HeadPos{f}{\tau}{\Text}{j} =
  (\RunEndFullPos{f}{\tau}{\Text}{j} - j) \bmod p =
  (\RunEndFullPos{f}{\tau}{\Text}{j'} - j) \bmod p = (s + kp) \bmod p = s$
  and $\ExpPos{f}{\tau}{\Text}{j}
  = \lfloor \tfrac{\RunEndFullPos{f}{\tau}{\Text}{j} - j}{p} \rfloor
  = \lfloor \tfrac{s + kp}{p} \rfloor = k$.  We have thus proved that
  $j \in \RMinusSix{f}{s}{k}{H}{\tau}{\Text}$.
\end{proof}

\begin{lemma}\label{lm:RskH-size}
  Let $\tau \in [1 \dd \lfloor \tfrac{\Textlen}{2} \rfloor]$
  and $f$ be any necklace-consistent function.  Let $H \in \Sigma^{+}$,
  $p = |H|$, $s \in [0 \dd p)$, and $k_{\min} = \lceil \tfrac{3\tau -
  1 - s}{p} \rceil - 1$. Then:
  \begin{itemize}
  \item For every $k \in [1 \dd k_{\min}]$,
    $|\RMinusSix{f}{s}{k}{H}{\tau}{\Text}| \leq |\{j \in \RPrimMinusFour{f}{H}{\tau}{\Text} :
    s + kp \leq \RunEndFullPos{f}{\tau}{\Text}{j} - j\}|$,
  \item For every $k \in (k_{\min} \dd \Textlen]$,
    $|\RMinusSix{f}{s}{k}{H}{\tau}{\Text}| = |\{j \in \RPrimMinusFour{f}{H}{\tau}{\Text} :
    s + kp \leq \RunEndFullPos{f}{\tau}{\Text}{j} - j\}|$.
  \end{itemize}
\end{lemma}
\begin{proof}
  First, observe that adding the same value to every element of a set
  does not change its cardinality. Moreover, by \cref{lm:efull}, for
  every $i, i' \in \RPrimMinusTwo{\tau}{\Text}$, $i \neq i'$ implies
  $\RunEndFullPos{f}{\tau}{\Text}{i} \neq \RunEndFullPos{f}{\tau}{\Text}{i'}$. It
  thus follows by \cref{lm:RskH} that for $k \in [1 \dd k_{\min}]$,
  \begin{align*}
    |\RMinusSix{f}{s}{k}{H}{\tau}{\Text}|
      &\leq |\{\RunEndFullPos{f}{\tau}{\Text}{j} - s - kp :
            j \in \RPrimMinusFour{f}{H}{\tau}{\Text}
            \text{ and }
            s + kp \leq \RunEndFullPos{f}{\tau}{\Text}{j} - j\}|\\
      &=    |\{\RunEndFullPos{f}{\tau}{\Text}{j} :
            j \in \RPrimMinusFour{f}{H}{\tau}{\Text}
            \text{ and }
            s + kp \leq \RunEndFullPos{f}{\tau}{\Text}{h} - j\}|\\
      &=    |\{j \in \RPrimMinusFour{f}{H}{\tau}{\Text} :
            s + kp \leq \RunEndFullPos{f}{\tau}{\Text}{h} - j\}|.
  \end{align*}
  The proof for $k \in (k_{\min} \dd \Textlen]$ follows analogously, except
  the first inequality is replaced with an equality.
\end{proof}

\begin{definition}\label{def:periodic-input}
  Let $\tau \in [1 \dd \lfloor \tfrac{\Textlen}{2} \rfloor]$ and $f$ be
  any necklace-consistent function.  Let $H \in \Sigma^{+}$.
  We define
  \[
    \PairsMinus{f}{H}{\tau}{\Text} :=
      \{
        (\RunEndFullPos{f}{\tau}{\Text}{p},
         \min(7\tau, \RunEndFullPos{f}{\tau}{\Text}{p} -
                     \RunBeg{\tau}{\Text}{p}))
        : p \in \CompRepr{14\tau}{\RMinusFour{f}{H}{\tau}{\Text}}{\Text}
      \}.
  \]
\end{definition}

\begin{lemma}\label{lm:periodic-input}
  Let $\tau \in [1 \dd \lfloor \tfrac{\Textlen}{2} \rfloor]$
  and $f$ be any necklace-consistent function. For every
  $H \in \Sigma^{+}$, it holds
  \[
    \PairsMinus{f}{H}{\tau}{\Text} =
      \{
        (\RunEndFullPos{f}{\tau}{\Text}{p_j},
         \min(7\tau, \RunEndFullPos{f}{\tau}{\Text}{p_j} -
                     \RunBeg{\tau}{\Text}{p_j}))
        : j \in [1 \dd m]\text{ and }p_j \in \RMinusFour{f}{H}{\tau}{\Text}
      \},
  \]
  where $\PairsMinus{f}{H}{\tau}{\Text}$ is as
  in \cref{def:periodic-input} and $(p_j, t_j)_{j \in [1 \dd m]}
  = \IntervalRepr{\CompRepr{14\tau}{\RTwo{\tau}{\Text}}{\Text}}$.
\end{lemma}
\begin{proof}

  Denote the set on the right-hand side by $A$. For every
  $p \in \RTwo{\tau}{\Text}$, let us also denote $\alpha(p)
  := \RunEndFullPos{f}{\tau}{\Text}{p}$ and $\beta(p)
  := \min(7\tau, \RunEndFullPos{f}{\tau}{\Text}{p}
  - \RunBeg{\tau}{\Text}{p})$.

  Let $a \in A$. Then, there exists $j \in [1 \dd m]$ such that
  $p_j \in \RMinusFour{f}{H}{\tau}{\Text}$ and $a =
  (\alpha(p_j), \beta(p_j))$. Recall that
  $\CompRepr{14\tau}{\RTwo{\tau}{\Text}}{\Text}
  = \RTwo{\tau}{\Text} \cap \Cover{14\tau}{\Text}$.
  By \cref{def:interval-representation}, it holds $[p_j \dd p_j +
  t_j) \subseteq \RTwo{\tau}{\Text} \cap \Cover{14\tau}{\Text}$.  In
  particular, $p_j \in \Cover{14\tau}{\Text}$. Combining with
  $p_j \in \RMinusFour{f}{H}{\tau}{\Text}$, we thus obtain
  $p_j \in \RMinusFour{f}{H}{\tau}{\Text} \cap \Cover{14\tau}{\Text}
  = \CompRepr{14\tau}{\RMinusFour{f}{H}{\tau}{\Text}}{\Text}$.
  By \cref{def:periodic-input}, we thus obtain
  $(\alpha(p_j), \beta(p_j)) \in \PairsMinus{f}{H}{\tau}{\Text}$. We
  have thus proved $a \in \PairsMinus{f}{H}{\tau}{\Text}$.

  We now show the second inclusion. Let
  $a \in \PairsMinus{f}{H}{\tau}{\Text}$.  Then, there exists
  $p \in \CompRepr{14\tau}{\RMinusFour{f}{H}{\tau}{\Text}}{\Text}
  = \RMinusFour{f}{H}{\tau}{\Text} \cap \Cover{14\tau}{\Text}$ such
  that $a = (\alpha(p), \beta(p))$.  Let $p' = \min\{t \in [1 \dd p] :
  [t \dd p] \subseteq \RTwo{\tau}{\Text} \cap \Cover{14\tau}{\Text}\}$.  We
  prove three properties of $p'$:
  \begin{itemize}
  \item First, we observe that by \cref{lm:R-text-block}, it follows
    that $\TypePos{\tau}{\Text}{p'} = \TypePos{\tau}{\Text}{p}$ and
    $\RootPos{f}{\tau}{\Text}{p'} = \RootPos{f}{\tau}{\Text}{p}$.
    Thus, $p' \in \RMinusFour{f}{H}{\tau}{\Text}$.
  \item Next, observe that, by definition, we have
    $p'-1 \not\in \RTwo{\tau}{\Text} \cap \Cover{14\tau}{\Text}$,
    i.e., $p'$ is the leftmost element of a maximal block in
    $\RTwo{\tau}{\Text} \cap \Cover{14\tau}{\Text}$.  Thus, there
    exists $j \in [1 \dd m]$ such that $p_j = p'$.
  \item Lastly, observe that by \cref{lm:R-text-block,lm:beg-end}, it
    holds $\RunEndFullPos{f}{\tau}{\Text}{p'}
    = \RunEndFullPos{f}{\tau}{\Text}{p}$ and $\RunBeg{\tau}{\Text}{p'}
    = \RunBeg{\tau}{\Text}{p}$. Consequently, $(\alpha(p'), \beta(p'))
    = (\alpha(p), \beta(p))$.
  \end{itemize}
  Putting everything together, we thus obtain that there exists $j \in
  [1 \dd m]$ such that $p' = p_j$ and
  $p_j \in \RMinusFour{f}{H}{\tau}{\Text}$. Thus,
  $(\alpha(p'), \beta(p')) \in A$.  By $(\alpha(p'), \beta(p')) =
  (\alpha(p), \beta(p)) = a$, we thus have $a \in A$.
\end{proof}

\begin{corollary}\label{cor:nruns}
  Let $\tau \in [1 \dd \lfloor \tfrac{\Textlen}{2} \rfloor]$
  and $f$ be any necklace-consistent function. Then, it holds
  \begin{align*}
    \sum_{H \in \Sigma^{+}} |\PairsMinus{f}{H}{\tau}{\Text}| \leq
    |\IntervalRepr{\CompRepr{14\tau}{\RTwo{\tau}{\Text}}{\Text}}|.
  \end{align*}
\end{corollary}
\begin{proof}
  The result follows immediately from \cref{lm:periodic-input}.
\end{proof}

\begin{lemma}\label{lm:unique-labels}
  Let $\tau \in [1 \dd \lfloor \tfrac{\Textlen}{2} \rfloor]$
  and $f$ be any necklace-consistent function.  For every
  $H \in \Sigma^{+}$, the labels in
  $\WInts{7\tau}{\PairsMinus{f}{H}{\tau}{\Text}}{\Text}$
  (\cref{def:intervals}) are unique.
\end{lemma}
\begin{proof}
  For every $j \in \RTwo{\tau}{\Text}$, denote $\alpha(j)
  = \RunEndFullPos{f}{\tau}{\Text}{j}$ and $\beta(j)
  = \min(7\tau, \RunEndFullPos{f}{\tau}{\Text}{j} - \RunBeg{\tau}{\Text}{j})$.
  Denote $\Ints = \WInts{7\tau}{\PairsMinus{f}{H}{\tau}{\Text}}{\Text}$.  Let $q_1,
  q_2 \in \Ints$.  For $k \in \{1, 2\}$, denote $q_k = (e_k,
  w_k, \ell_k)$, and assume $\ell_1 = \ell_2$.  We need to prove that
  $q_1 = q_2$.  By \cref{def:intervals} and the definition of
  $\PairsMinus{f}{H}{\tau}{\Text}$, for $k \in \{1, 2\}$ there exists
  $j_k \in \CompRepr{14\tau}{\RMinusFour{f}{H}{\tau}{\Text}}{\Text}
  \subseteq \RMinusFour{f}{H}{\tau}{\Text}$ such that, letting $i_k
  = \alpha(j_k) = \RunEndFullPos{f}{\tau}{\Text}{j_k}$, it holds $e_k
  = \beta(j_k) = \min(7\tau, \RunEndFullPos{f}{\tau}{\Text}{j_k}
  - \RunBeg{\tau}{\Text}{j_k})$, $w_k = |\{i' \in [1 \dd \Textlen] :
  \Textinf[i' - 7\tau \dd i' + 7\tau) = \Textinf[i_k - 7\tau \dd
  i_k + 7\tau)\}|$, and $\ell_k = \min \{i' \in [1 \dd \Textlen] :
  \Textinf[i' - 7\tau \dd i' + 7\tau) = \Textinf[i_k - 7\tau \dd
  i_k + 7\tau)\}$.  For $k \in \{1, 2\}$, denote $X_k = \Textinf[i_k
  - 7\tau \dd i_k + 7\tau)$.  Note that for $k \in \{1, 2\}$, it holds
  $\Textinf[\ell_k - 7\tau \dd \ell_k + 7\tau) = X_k$.  Thus, the
  assumption $\ell_1 = \ell_2$ implies $X_1 = \Textinf[\ell_1 -
  7\tau \dd \ell_1 + 7\tau) = \Textinf[\ell_2 - 7\tau \dd \ell_2 +
  7\tau) = X_2$.  This in turn implies that it holds $w_1 = |\{i' \in
  [1 \dd \Textlen] : \Textinf[i' - 7\tau \dd i' + 7\tau) = X_1\}| =
  |\{i' \in [1 \dd \Textlen] : \Textinf[i' - 7\tau \dd i' + 7\tau) = X_2\}|
  = w_2$.  Finally, observe that for every $i' \in \RFour{f}{H}{\tau}{\Text}$,
  letting $Y = \Textinf[\RunEndFullPos{f}{\tau}{\Text}{i'} -
  7\tau \dd \RunEndFullPos{f}{\tau}{\Text}{i'})$, it holds
  $\min(7\tau, \RunEndFullPos{f}{\tau}{\Text}{i'} - \RunBeg{\tau}{\Text}{i'})
  = \lcs(Y[1 \dd 7\tau], Y[1 \dd 7\tau - |H|])$.  Thus, by $X_1 =
  X_2$, we obtain $e_1 = \min(7\tau, \RunEndFullPos{f}{\tau}{\Text}{j_1}
  - \RunBeg{\tau}{\Text}{j_1}) = \lcs(X_1, X_1[1 \dd 7\tau - |H|])
  = \lcs(X_2, X_2[1 \dd 7\tau - |H|])
  = \min(7\tau, \RunEndFullPos{f}{\tau}{\Text}{j_2}
  - \RunBeg{\tau}{\Text}{j_2}) = e_2$. Putting everything together, we
  have thus proved $(e_1, w_1) = (e_2, w_2)$. Combining with the
  assumption $\ell_1 = \ell_2$, we thus obtain $q_1 = q_2$.
\end{proof}

\subsubsection{A Necklace-Consistent Function}

\begin{definition}\label{def:canonical-function}
  Let $\tau \in [1 \dd \lfloor \tfrac{\Textlen}{2}
  \rfloor]$. Let $(r_i)_{i \in [1 \dd q]}$ be a sequence
  containing all elements of $\RPrimTwo{\tau}{\Text}$ in sorted order, i.e, for
  any $i, i' \in [1 \dd q]$, $i < i'$ implies $r_i <
  r_{i'}$. Denote
  \begin{align*}
    \Text' &= \textstyle \prod_{i=1,2,\dots,q} \Text[r_i \dd r_i +
      2\tau {-} 1) \Text[\Textlen] \in \Sigma^{2\tau q},\\
    \mathcal{Q} &= \textstyle \bigcup_{i \in \RTwo{\tau}{\Text}}\{\Text[i {+}
      \delta \dd i {+} \delta {+} \per(\Text[i {\dd} i {+} 3\tau {-} 1)))
      : \delta \in [0 \dd \tau)\} \subseteq \Sigma^{\leq \tau/3},
  \end{align*}
  and let
  \begin{itemize}
  \item $g: \Z \rightarrow \Z$ be defined by $g(i) = 1 + 2\tau \lfloor
    \tfrac{i - 1}{2\tau} \rfloor$,
  \item $h: \mathcal{Q} \rightarrow \Sigma^{+}$ be defined by $h(X) =
    \Text'[g(s) \dd g(s) + |X|)$, where $s = \min
    \OccTwo{X^{\infty}[1 \dd \tau]}{\Text'}$,
  \item $h' : \Sigma^{+} \rightarrow \Sigma^{+}$ be defined by $h'(X)
    = \min\{\rot^{i}(X) : i \in [0 \dd |X|)\}$.
  \end{itemize}
  Then, by $f_{\tau,\Text} : \Sigma^{+} \rightarrow \Sigma^{+}$ we define
  the function such that for every $X \in \Sigma^{+}$,
  \vspace{2ex}
  \[
    f_{\tau,\Text}(X) =
    \begin{cases}
      h(X) & \text{if $X \in \mathcal{Q}$},\\
      h'(X) & \text{otherwise}.
    \end{cases}
    \vspace{2ex}
  \]
\end{definition}

\begin{lemma}\label{lm:canonical-function-1}
  Let $\tau \in [1 \dd \lfloor \tfrac{\Textlen}{2}
  \rfloor]$. Then, $f_{\tau,\Text}$ (\cref{def:canonical-function}) is
  well-defined.
\end{lemma}
\begin{proof}
  Recall that $\RPrimTwo{\tau}{\Text} \subseteq \RTwo{\tau}{\Text} \subseteq [1 \dd \Textlen
  - 3\tau + 2]$.  Thus, for $i \in [1 \dd q]$, it holds $r_i +
  2\tau - 1 \leq \Textlen - \tau + 1 \leq \Textlen$.  Hence, $\Text[r_i \dd
  r_i + 2\tau - 1)$ in the formula for $\Text'$ is well-defined.

  Next, we show that for every $X \in \mathcal{Q}$, the value $\min
  \OccTwo{X^{\infty}[1 \dd \tau]}{\Text'}$ is well-defined, i.e., that
  $\OccTwo{X^{\infty}[1 \dd \tau]}{\Text'} \neq \emptyset$.  Let $j \in
  \RTwo{\tau}{\Text}$ and $\delta \in [0 \dd \tau)$ be such that $X = \Text[j +
  \delta \dd j + \delta + p)$, where $p = \per(\Text[j \dd j + 3\tau -
  1))$. Observe, that $p$ being a period of $\Text[j \dd j + 3\tau - 1)$
  implies that $\Text[j + \delta \dd j + 3\tau - 1)$ is a prefix of
  $X^{\infty}$.  Thus, by $\delta < \tau$, it holds $X^{\infty}[1 \dd
  \tau] = \Text[j + \delta \dd j + \delta + \tau)$.  Note also that $p
  \leq \tfrac{1}{3}\tau$. Let $j' \in [1 \dd j]$ be the smallest
  integer such that $[j' \dd j] \subseteq \RTwo{\tau}{\Text}$.  Then, $j' \in
  \RPrimTwo{\tau}{\Text}$. Consider the substring $\Text[j' \dd
  \RunEndPos{\tau}{\Text}{j'})$. On the one hand, by $[j' \dd j] \subseteq
  \RTwo{\tau}{\Text}$ and \cref{lm:beg-end}\eqref{lm:beg-end-it-1}, it holds
  $\RunEndPos{\tau}{\Text}{j'} \geq j + 3\tau - 1$, and hence $j' \leq j +
  \delta < j + \delta + \tau < j + 3\tau - 1 \leq
  \RunEndPos{\tau}{\Text}{j'}$. Thus, $X^{\infty}[1 \dd \tau]$ is a
  substring of $\Text[j' \dd \RunEndPos{\tau}{\Text}{j'})$. On the other hand,
  by \cref{lm:R-text-block}, $\per(\Text[j' \dd j' + 3\tau - 1)) =
  \per(\Text[j \dd j + 3\tau - 1)) = p$. By definition of
  $\RunEndPos{\tau}{\Text}{j'}$, $\Text[j' \dd \RunEndPos{\tau}{\Text}{j'})$ thus has
  period $p$. Consequently, there exists $\delta' \in [0 \dd p)$ such
  that $\Text[j' + \delta' \dd j' + \delta' + \tau) = X^{\infty}[1 \dd
  \tau]$. Note that by $\delta' < p \leq \tfrac{1}{3}\tau \leq \tau$,
  we then have $j' + \delta' + \tau \leq j' + 2\tau -
  1$. Consequently, letting $i \in [1 \dd q]$ be such that $r_i =
  j'$, we have that $X^{\infty}[1 \dd \tau]$ is a substring of
  $\Text[r_{i} \dd r_{i} + 2\tau - 1)$, and hence
  $\OccTwo{X^{\infty}[1 \dd \tau]}{\Text'} \neq \emptyset$.
\end{proof}

\begin{lemma}\label{lm:canonical-function-2}
  Let $\tau \in [1 \dd \lfloor \tfrac{\Textlen}{2}
  \rfloor]$. Then, $f_{\tau,\Text}$ (\cref{def:canonical-function}) is
  necklace-consistent.
\end{lemma}
\begin{proof}
  First, we show that for any $X \in \Sigma^{+}$, the strings
  $f_{\tau, \Text}(X)$ and $X$ are cyclically equivalent.  Consider two
  cases:
  \begin{enumerate}
  \item Let us first assume $X \in \mathcal{Q}$.  Denote $s = \min
    \OccTwo{X^{\infty}[1 \dd \tau]}{\Text'}$ and $i = \lceil \tfrac{s}{2\tau}
    \rceil$. We prove the following three properties.
    \begin{enumerate}[label=(\alph*)]
    \item First, we show that $[s \dd s + \tau) \subseteq [1 +
      2\tau(i-1) \dd 2\tau i)$, i.e., $X^{\infty}[1 \dd \tau]$ is a
      substring of $\Text[r_{i} \dd r_{i} + 2\tau - 1)$.  By
      definition of $\mathcal{Q}$, there exists $j \in \RTwo{\tau}{\Text}$
      and $\delta \in [0 \dd \tau)$ such that $X = \Text[j + \delta \dd j
      + \delta + p)$, where $p = \per(\Text[j \dd j + 3\tau - 1)) \leq
      \tfrac{1}{3}\tau$. By $j \in \RTwo{\tau}{\Text}$, we have $j \in [1 \dd
      \Textlen - 3\tau + 2]$. Therefore, $j + \delta + p \leq j + 2\tau - 1
      \leq \Textlen - \tau + 1 \leq \Textlen$.  Consequently, $\Text[\Textlen]$ does not
      occur in $X$, and hence by definition of $\Text'$, we have $[s \dd s
      + \tau) \subseteq [1 + 2\tau(i-1) \dd 2\tau i)$. Note that for
      such $s$, we have $g(s) = 1 + 2\tau \lfloor \tfrac{s - 1}{2\tau}
      \rfloor = 1 + 2\tau(i-1)$.
    \item Second, we show that $\per(\Text[r_{i} \dd r_{i} +
      3\tau - 1)) = |X|$. Denote $p' = \per(\Text[r_{i} \dd
      r_{i} + 3\tau - 1))$ and let $\delta' \in [0 \dd \tau)$ be
      such that $\Text[r_{i} + \delta' \dd r_{i} + \delta' +
      \tau) = X^{\infty}[1 \dd \tau]$.  Then, $\Text[r_{i} + \delta'
      \dd r_{i} + \delta' + \tau)$ has periods $|X|$ and $p'$. By
      $|X|, p' \leq \tfrac{1}{3}\tau$ and the weak periodicity
      lemma~\cite{periodicitylemma}, $\gcd(|X|, p')$ is therefore a
      period of $\Text[r_{i} + \delta' \dd r_{i} + \delta' +
      \tau)$.  Since every length-$p'$ substring of $\Text[r_{i} \dd
      r_{i} + 3\tau - 1)$ is primitive, we thus have $\gcd(|X|,
      p') = p'$. Consequently, $p' \mid |X|$.  On the other hand, $X$
      being a substring of $\Text[j \dd j + 3\tau - 1)$ and $\per(\Text[j \dd
      j + 3\tau - 1)) = |X|$ imply that $X$ is primitive. Thus, $p' =
      |X|$.
    \item Third, we show that $\Text[r_{i} + t \dd r_{i} + t +
      |X|) = X$ holds for some $t \in [0 \dd |X|)$.  Since $X$ is a
      substring of $\Text[r_{i} \dd r_{i} + 3\tau - 1)$ and
      $\per(\Text[r_{i} \dd r_{i} + 3\tau - 1)) = |X|$, we can
      write $\Text[r_{i} \dd r_{i} + 3\tau - 1) = X' X^k X''$,
      where $k \geq 1$ and $X'$ (resp.\ $X''$) is a proper suffix
      (resp.\ prefix) of $X$. Letting $t = |X'|$, we obtain the claim.
      Note that we then also have $\Text[r_i \dd r_i + |X|) =
      \rot^{t}(X)$.
    \end{enumerate}
    Combining the above properties, we thus obtain
    \begin{align*}
      f_{\tau, \Text}(X)
        &= h(X)\\
        &= \Text'[g(s) \dd g(s) + |X|)\\
        &= \Text'[1 + 2\tau(i-1) \dd 1 + 2\tau(i-1) + |X|)\\
        &= \Text[r_{i} \dd r_{i} + |X|)\\
        &= \rot^{t}(X).
    \end{align*}
    In other words, $f_{\tau, \Text}(X)$ and $X$ are cyclically
    equivalent.
  \item Let us now assume $X \not\in \mathcal{Q}$. Then, $f_{\tau,
    \Text}(X) = \min \{\rot^{i}(X) : i \in [0 \dd |X|)\}$.  Since every
    element of $\{\rot^{i}(X) : i \in [0 \dd |X|)\}$ is cyclically
    equivalent with $X$ by definition, the claim follows.
  \end{enumerate}

  We now show the second condition in \cref{def:nc} for $f_{\tau, \Text}$.
  Let $X, X' \in \Sigma^{+}$ and assume that $X$ and $X'$ are
  cyclically equivalent. Consider two cases:
  \begin{enumerate}
  \item Let $X \in \mathcal{Q}$.  Denote $s = \min
    \OccTwo{X^{\infty}[1 \dd \tau]}{\Text'}$ and $i = \lceil \tfrac{s}{2\tau} \rceil$.  By the
    three properties of $\Text[r_{i} \dd r_{i} + 3\tau - 1)$
    proved above, it holds $\{\Text[r_{i} + t \dd r_{i} + t +
    |X|) : t \in [0 \dd |X|)\} = \{\rot^{i}(X) : i \in [0 \dd
    |X|)\}$. Since $X'$ is in the latter set, there exists
    $\delta' \in [0 \dd |X'|)$ such that $\Text[r_{i} + \delta' \dd
    r_{i} + \delta' + |X'|) = X'$.  This implies $X' \in
    \mathcal{Q}$. Moreover, since $\per(\Text[r_{i} \dd r_{i} +
    3\tau - 1)) = |X| = |X'|$, we then have $\Text[r_{i} + \delta'
    \dd r_{i} + \delta' + \tau) = \Text'[1 + 2\tau(i-1) + \delta' \dd
    1 + 2\tau(i-1) + \delta' + \tau) = X'^{\infty}[1 \dd
    \tau]$. Consequently, letting $s' = \min
    \OccTwo{X'^{\infty}[1 \dd \tau]}{\Text'}$, we have $s' < 1 + 2\tau i$ and hence $g(s) = 1 +
    2\tau(i-1) \geq 1 + 2\tau \lfloor \tfrac{s' - 1}{2\tau} \rfloor =
    g(s')$.  By $X' \in \mathcal{Q}$ and the symmetry of our argument,
    we analogously obtain $g(s) \leq g(s')$. Thus, $g(s) = g(s')$ and
    hence
    \begin{align*}
      f_{\tau, \Text}(X)
        &= h(X)\\
        &= \Text'[g(s) \dd g(s) + |X|)\\
        &= \Text'[g(s') \dd g(s') + |X'|)\\
        &= h(X')\\
        &= f_{\tau, \Text}(X').
    \end{align*}
  \item Let us now assume $X \not\in \mathcal{Q}$. By the above, we
    then have $X' \not\in \mathcal{Q}$ and thus $f_{\tau, \Text}(X) =
    h'(X) = \min\{\rot^{i}(X) : i \in [0 \dd |X|)\} =
    \min\{\rot^{i}(X') : i \in [0 \dd |X'|)\} = h'(X') = f_{\tau,
    \Text}(X')$.  \qedhere
  \end{enumerate}
\end{proof}

Next, we first describe an alternative construction of the
necklace-consistent function from \cref{def:canonical-function}, that
instead of the whole set $\RTwo{\tau}{\Text}$ represented using
$\RPrimTwo{\tau}{\Text}$, utilizes the set $\CompRepr{k}{\RTwo{\tau}{\Text}}{\Text}$. This will be used
during the construction algorithms. After introducing it in
\cref{def:canonical-function-comp}, we prove that it is well-defined
(\cref{lm:canonical-function-comp-1}) and indeed equal to $f_{\tau,
  \Text}$ (\cref{lm:canonical-function-comp-2}) for every $k \geq 3\tau -
1$.

\begin{definition}\label{def:canonical-function-comp}
  Let $\tau \in [1 \dd \lfloor \tfrac{\Textlen}{2} \rfloor]$ and $k
  \geq 3\tau - 1$. Let $\mathcal{Q}$, $g$, and $h'$ be as in
  \cref{def:canonical-function}. Let $(a_i, t_i)_{i \in [1 \dd q']}
  = \IntervalRepr{\CompRepr{k}{\RTwo{\tau}{\Text}}{\Text}}$. Denote
  \[
    \Text'_{\rm comp} = \textstyle \prod_{i=1,2,\dots,q'} \Text[a_i \dd a_i +
    2\tau - 1) \Text[\Textlen] \in \Sigma^{2\tau q'}.
  \]
  Let $h_{\rm comp}: \mathcal{Q} \rightarrow \Sigma^{+}$ be a function
  such that for every $X \in \mathcal{Q}$, it holds
  $h_{\rm comp}(X) = \Text'_{\rm comp}[g(s) \dd g(s) + |X|)$,
  where $s = \min \OccTwo{X^{\infty}[1 \dd \tau]}{\Text'_{\rm comp}}$.
  Finally, let $f_{k,\tau,\Text} : \Sigma^{+} \rightarrow \Sigma^{+}$ be a
  function such that for every $X \in \Sigma^{+}$,
  \vspace{2ex}
  \[
    f_{k,\tau,\Text}(X) =
    \begin{cases}
      h_{\rm comp}(X) & \text{if $X \in \mathcal{Q}$},\\
      h'(X) & \text{otherwise}.
    \end{cases}
    \vspace{2ex}
  \]
\end{definition}

\begin{lemma}\label{lm:canonical-function-comp-1}
  Let $\tau \in [1 \dd \lfloor \tfrac{\Textlen}{2} \rfloor]$ and $k
  \geq 3\tau - 1$.  Then, $f_{k,\tau,\Text}$
  (\cref{def:canonical-function-comp}) is well-defined.
\end{lemma}
\begin{proof}
  For every $i \in [1 \dd q']$, we have $a_i \in
  \CompRepr{k}{\RTwo{\tau}{\Text}}{\Text}$.  On the other hand, by \cref{def:comp}, we have
  $\CompRepr{k}{\RTwo{\tau}{\Text}}{\Text} = \RTwo{\tau}{\Text} \cap \Cover{k}{\Text} \subseteq
  \RTwo{\tau}{\Text} \subseteq [1 \dd \Textlen - 3\tau + 2]$.  Thus, for every $i
  \in [1 \dd q']$, it holds $a_i + 2\tau - 1 \leq \Textlen - \tau + 1 \leq
  \Textlen$, and hence $\Text[a_i \dd a_i + 2\tau - 1)$ in the formula for
  $\Text'_{\rm comp}$ is well-defined.

  We now show that for every $X \in \mathcal{Q}$, it holds
  $\OccTwo{X^{\infty}[1 \dd \tau]}{\Text'_{\rm comp}} \neq \emptyset$. By
  definition of $\mathcal{Q}$, there exists $j \in \RTwo{\tau}{\Text}$ and
  $\delta \in [0 \dd \tau)$ such that $X = \Text[j + \delta \dd j + \delta
  + p)$, where $p = \per(\Text[j \dd j + 3\tau - 1))$. In
  \cref{lm:canonical-function-1}, we proved that we then have $\Text[j +
  \delta \dd j + \delta + \tau) = X^{\infty}[1 \dd \tau]$.  Let $j' =
  \min \OccThree{k}{j}{\Text}$. Note that $j' \leq j \leq \Textlen - 3\tau + 2$
  (where the latter follows by $j \in \RTwo{\tau}{\Text} \subseteq [1 \dd \Textlen
  - 3\tau + 2]$).  Thus, by $j' \in \OccThree{k}{j}{\Text}$ and $k \geq 3\tau
  - 1$ we have $\Text[j' \dd j' + 3\tau - 1) = \Text[j \dd j + 3\tau - 1)$.
  In particular, $\Text[j' + \delta \dd j' + \delta + \tau) = X^{\infty}[1
  \dd \tau]$.  Note now that by \cref{lm:sa-core-nav}, we have $j' \in
  \CompRepr{k}{\RTwo{\tau}{\Text}}{\Text}$.  Let $j'' \in [1 \dd j']$ be the smallest
  integer such that $[j'' \dd j'] \subseteq \CompRepr{k}{\RTwo{\tau}{\Text}}{\Text}$.
  In particular, we then have $[j'' \dd j'] \subseteq
  \RTwo{\tau}{\Text}$. In the proof of \cref{lm:canonical-function-1} we observed that
  this, combined with $X^{\infty}[1 \dd \tau]$ being a substring of
  $\Text[j' \dd j' + 3\tau - 1)$, implies that $\Text[j'' \dd
  \RunEndPos{\tau}{\Text}{j''})$ has period $|X|$ and that $X^{\infty}[1 \dd
  \tau]$ is a substring of $\Text[j'' \dd \RunEndPos{\tau}{\Text}{j''})$.  Thus,
  there exists $\delta' \in [0 \dd |X|)$ such that $\Text[j'' + \delta'
  \dd j'' + \delta' + \tau) = X^{\infty}[1 \dd \tau]$.  Letting $i \in
  [1 \dd q']$ be such that $a_i = j''$, we obtain by $\delta' < \tau$,
  that $X^{\infty}[1 \dd \tau]$ is a substring of $\Text[a_i \dd a_i +
  2\tau - 1)$, and therefore
  $\OccTwo{X^{\infty}[1 \dd \tau]}{\Text'_{\rm comp}} \neq \emptyset$.
\end{proof}

\begin{lemma}\label{lm:canonical-function-comp-2}
  Let $\tau \in [1 \dd \lfloor \tfrac{\Textlen}{2} \rfloor]$ and $k
  \geq 3\tau - 1$.  Then, it holds $f_{k,\tau,\Text} = f_{\tau,\Text}$
  (\cref{def:canonical-function-comp,def:canonical-function}).
\end{lemma}
\begin{proof}
  By definition, we have $f_{k,\tau,\Text}(X) = f_{\tau,\Text}(X)$ for every
  $X \in \Sigma^{+} \sm \mathcal{Q}$. Thus, it remains to show that
  $h_{\rm comp} = h$ (where $h$ is as in
  \cref{def:canonical-function}), i.e., that for every $X \in
  \mathcal{Q}$, it holds $\Text'_{\rm comp}[g(s_{\rm comp}) \dd g(s_{\rm
  comp}) + |X|) = \Text'[g(s) \dd g(s) + |X|)$, where $s_{\rm comp} = \min
  \OccTwo{X^{\infty}[1 \dd \tau]}{\Text'_{\rm comp}}$, $s = \min
  \OccTwo{X^{\infty}[1 \dd \tau]}{\Text'}$, and $\Text'$ is as in
  \cref{def:canonical-function}. Let $(r_i)_{i \in [1 \dd q]}$
  and $(a_i, t_i)_{i \in [1 \dd q']}$ be as in
  \cref{def:canonical-function} and
  \cref{def:canonical-function-comp}, respectively. Denote $i = \lceil
  \tfrac{s}{2\tau} \rceil$ and $k = \lceil \tfrac{s_{\rm comp}}{2\tau}
  \rceil$.
  \begin{enumerate}
  \item First, we prove that it holds $a_{k} \leq r_{i}$.  To
    achieve this, we will show that $r_{i} \in \Cover{k}{\Text}$.
    Suppose that $r_{i} \not\in \Cover{k}{\Text}$.  Then, $\min
    \OccThree{k}{r_{i}}{\Text} < r_{i}$ (see \cref{lm:cover-equivalence}),
    i.e., there exists $j \in [1 \dd r_{i})$ such that
    $\Textinf[j \dd j + k) = \Textinf[r_{i} \dd r_{i} +
    k)$. Note that $j < r_{i} \leq \Textlen - 3\tau + 2$ (where the
    latter follows by $r_{i} \in \RTwo{\tau}{\Text} \subseteq [1 \dd \Textlen
    - 3\tau + 2]$).  Thus, $\Text[j \dd j + 3\tau - 1) = \Text[r_{i} \dd
    r_{i} + 3\tau - 1)$. In the proof of
    \cref{lm:canonical-function-2}, we showed that $X^{\infty}[1 \dd
    \tau]$ is a substring of $\Text[r_{i} \dd r_{i} + 2\tau -
    1)$ and $\per(\Text[r_{i} \dd r_{i} + 3\tau - 1)) = |X|$.
    Consequently, $\per(\Text[j \dd j + 3\tau - 1)) = \per(\Text[r_{i}
    \dd r_{i} + 3\tau - 1)) = |X| \leq \tfrac{1}{3}\tau$ and thus
    $j \in \RTwo{\tau}{\Text}$. Let $j' \in [1 \dd j]$ be the smallest
    integer such that $[j' \dd j] \subseteq \RTwo{\tau}{\Text}$. By
    \cref{lm:R-text-block}, $\per(\Text[j' \dd j' + 3\tau - 1)) = \per(\Text[j
    \dd j + 3\tau - 1)) = |X|$. Thus, by definition of
    $\RunEndPos{\tau}{\Text}{j'}$, the string $\Text[j' \dd
    \RunEndPos{\tau}{\Text}{j'})$ has period $|X|$. Since $X^{\infty}[1 \dd
    \tau]$ is the substring of $\Text[j' \dd \RunEndPos{\tau}{\Text}{j'})$, we
    therefore obtain $\delta \in [0 \dd |X|)$ satisfying $\Text[j' +
    \delta \dd j' + \delta + \tau) = X^{\infty}[1 \dd \tau]$, i.e.,
    $X^{\infty}[1 \dd \tau]$ is a substring of $\Text[j' \dd j' + 2\tau -
    1)$.  Equivalently, letting $t \in [1 \dd q]$ be such that
    $r_{t} = j'$ (such $t$ exists since $j' \in \RPrimTwo{\tau}{\Text}$),
    $X^{\infty}[1 \dd \tau]$ is a substring of $\Text'[1 + 2\tau(t-1) \dd
    2\tau t)$.  Since $j', r_{i} \in \RPrimTwo{\tau}{\Text}$ and $j' \leq j
    < r_{i}$, we have $t < i$.  Therefore, $\min
    \OccTwo{X^{\infty}[1 \dd \tau]}{\Text'} < 1 + 2\tau t \leq 1 + 2\tau(i-1)
    \leq s$, which contradicts the definition of $s$. We have thus
    proved $r_{i} \in \Cover{k}{\Text}$.  Combined with $r_{i}
    \in \RPrimTwo{\tau}{\Text}$, this implies that there exists $h \in [1 \dd
    q']$ such that $a_h = r_{i}$.  Since $X^{\infty}[1 \dd \tau]$
    is a substring of $\Text[r_{i} \dd r_{i} + 2\tau - 1)$, we
    thus obtain that the leftmost occurrence of $X^{\infty}[1 \dd
    \tau]$ in $\Text'_{\rm comp}$ occurs either inside or to the left of
    the block $\Text[a_h \dd a_h + 2\tau - 1)\Text[\Textlen]$. More precisely,
    $s_{\rm comp} < 1 + 2\tau h$. This implies $k = \lceil
    \tfrac{s_{\rm comp}}{2\tau} \rceil \leq h$ and hence $a_{k} \leq
    a_{h} = r_{i}$.
  \item Second, we prove that it holds $a_{k} \geq r_{i}$.
    Suppose that $a_{k} < r_{i}$.  From the definition of $k$, it
    follows that $X^{\infty}[1 \dd \tau]$ occurs inside $\Text'_{\rm
    comp}[1 + 2\tau(k-1) \dd 2\tau k) = \Text[a_k \dd a_k + 2\tau - 1)$.
    Recall that $a_k \in \CompRepr{k}{\RTwo{\tau}{\Text}}{\Text} \subseteq
    \RTwo{\tau}{\Text}$.  Let $j \in [1 \dd a_k]$ be the smallest integer such that
    $[j \dd a_k] \subseteq \RTwo{\tau}{\Text}$.  Consider the substring $\Text[j
    \dd \RunEndPos{\tau}{\Text}{j})$. In the proof of
    \cref{lm:canonical-function-1}, we showed that then $\Text[j \dd
    \RunEndPos{\tau}{\Text}{j})$ has period $|X|$ and $X^{\infty}[1 \dd
    \tau]$ is a substring of $\Text[j \dd \RunEndPos{\tau}{\Text}{j})$. Thus,
    there exists $\delta \in [0 \dd |X|)$ such that $\Text[j + \delta \dd
    j + \delta + \tau) = X^{\infty}[1 \dd \tau]$. Note also that $j
    \in \RPrimTwo{\tau}{\Text}$.  Letting $i' \in [1 \dd q]$ be such that
    $r_{i'} = j$, we have thus obtained an occurrence of
    $X^{\infty}[1 \dd \tau]$ in $\Text[r_{i'} \dd r_{i'} + 2\tau
    - 1)$.  Note that since $j \leq a_k < r_i$, we have $i' < i$.
    By $\Text[r_{i'} \dd r_{i'} + 2\tau - 1) = \Text'[1 + 2\tau
    (i' - 1) \dd 2\tau i')$, we therefore have $\min
    \OccTwo{X^{\infty}[1 \dd \tau]}{\Text'} < 2\tau i' < 1 + 2\tau (i-1) \leq s$, which
    contradicts the definition of $s$.
  \end{enumerate}
  By the above, we thus have $a_{k} = r_{i}$. Consequently,
  \begin{align*}
    h_{\rm comp}(X)
      &= \Text'_{\rm comp}[g(s_{\rm comp}) \dd g(s_{\rm comp}) + |X|)\\
      &= \Text'_{\rm comp}[1 + 2\tau(k - 1) \dd 1 + 2\tau(k - 1) + |X|)\\
      &= \Text[a_{k} \dd a_{k} + |X|)\\
      &= \Text[r_{i} \dd r_{i} + |X|)\\
      &= \Text'[1 + 2\tau(i-1) \dd 1 + 2\tau(i-1) + |X|)\\
      &= \Text'[g(s) \dd g(s) + |X|)\\
      &= h(X). \qedhere
  \end{align*}
\end{proof}

\subsubsection{Decomposition Lemmas}

\begin{definition}\label{def:pos-sets-for-pat}
  Let $f$ be a necklace-consistent function.
  Let $\ell \in [16 \dd \Textlen)$ and $\tau = \lfloor \tfrac{\ell}{3} \rfloor$.
  For every $\tau$-periodic pattern $\Pat \in \Sigma^{+}$, we define:
  \begin{align*}
    \PosLowMinus{f}{\Pat}{\Text} &:= \{j' \in \RMinusFive{f}{s}{H}{\tau}{\Text} :
      \ExpPos{f}{\tau}{\Text}{j'} = k_1\text{ and }(\Text[j' \dd \Textlen]
      \succeq \Pat \text{ or }\lcp(\Pat, \Text[j' \dd \Textlen]) \geq \ell)\},\\
    \PosMidMinus{f}{\Pat}{\Text} &:= \{j' \in \RMinusFive{f}{s}{H}{\tau}{\Text} :
      \ExpPos{f}{\tau}{\Text}{j'} \in (k_1 \dd k_2]\},\\
    \PosHighMinus{f}{\Pat}{\Text} &:= \{j' \in \RMinusFive{f}{s}{H}{\tau}{\Text} :
      \ExpPos{f}{\tau}{\Text}{j'} = k_2\text{ and }(\Text[j' \dd \Textlen]
      \succeq \Pat \text{ or }\lcp(\Pat, \Text[j' \dd \Textlen]) \geq 2\ell)\},
  \end{align*}
  where
    $s = \HeadPat{f}{\tau}{\Pat}$,
    $H = \RootPat{f}{\tau}{\Pat}$,
    $k_1 = \ExpCutPat{f}{\tau}{\Pat}{\ell}$, and
    $k_2 = \ExpCutPat{f}{\tau}{\Pat}{2\ell}$.
  We denote
    $\DeltaLowMinus{f}{\Pat}{\Text} := |\PosLowMinus{f}{\Pat}{\Text}|$,
    $\DeltaMidMinus{f}{\Pat}{\Text} := |\PosMidMinus{f}{\Pat}{\Text}|$, and
    $\DeltaHighMinus{f}{\Pat}{\Text} := |\PosHighMinus{f}{\Pat}{\Text}|$.
  Next, by
    $\PosLowPlus{f}{\Pat}{\Text}$,
    $\PosMidPlus{f}{\Pat}{\Text}$, and
    $\PosHighPlus{f}{\Pat}{\Text}$
  we denote the symmetric versions of the above sets (i.e., with
    $\RMinusFive{f}{s}{H}{\tau}{\Text}$ replaced by $\RPlusFive{f}{s}{H}{\tau}{\Text}$,
    and the condition $\Text[j' \dd \Textlen] \succeq \Pat$ replaced with
    $\Text[j' \dd \Textlen] \succeq_{\rm inv} \Pat$).
  Finally, we denote
    $\PosLowAll{f}{\Pat}{\Text} :=
      \PosLowMinus{f}{\Pat}{\Text} \cup \PosLowPlus{f}{\Pat}{\Text}$,
    $\PosMidAll{f}{\Pat}{\Text} :=
      \PosMidMinus{f}{\Pat}{\Text} \cup \PosMidPlus{f}{\Pat}{\Text}$, and
    $\PosHighAll{f}{\Pat}{\Text} :=
      \PosHighMinus{f}{\Pat}{\Text} \cup \PosHighPlus{f}{\Pat}{\Text}$.
\end{definition}

\begin{remark}\label{rm:poslow-posmid-poshigh-plus}
  Observe that, similarly as when characterizing
  $\PosBeg{\ell}{\Pat}{\Text}$ (\cref{lm:equiv}; see also \cref{rm:posbeg-posend-equiv}), it is not correct
  to simply change $\succeq$ to $\preceq$ when defining the symmetric
  versions of sets $\PosLowMinus{f}{\Pat}{\Text}$,
  $\PosMidMinus{f}{\Pat}{\Text}$, and $\PosHighMinus{f}{\Pat}{\Text}$.
  Instead, we replace $\succeq$ in \cref{def:pos-sets-for-pat} with $\succeq_{\rm inv}$,
  where the inverted lexicographic order is as in \cref{def:inv}.
  Observe that then $\RPlusFour{f}{H}{\tau}{\Text}$ from
  the standard order becomes $\RMinusFour{f}{H}{\tau}{\Text}$ in
  the inverted order. Thus, the
  combinatorial results and query algorithms concerning $\PosLowPlus{f}{\Pat}{\Text}$,
  $\PosMidPlus{f}{\Pat}{\Text}$, and $\PosHighPlus{f}{\Pat}{\Text}$
  are identical as for $\PosLowMinus{f}{\Pat}{\Text}$,
  $\PosMidMinus{f}{\Pat}{\Text}$, and $\PosHighMinus{f}{\Pat}{\Text}$ (with the inverted order).
  This is why the structure from \cref{sec:sa-periodic-ds}
  contains two parts: one for 
  the standard lexicographic order $\preceq$, and one for the inverted
  lexicographic order $\preceq_{\rm inv}$.
\end{remark}

\begin{lemma}\label{lm:pos-depends-on-prefix}
  Let $f$ be a necklace-consistent function.
  Let $\ell \in [16 \dd \Textlen)$ and $\tau = \lfloor \tfrac{\ell}{3} \rfloor$.
  Let $\Pat \in \Sigma^{+}$ be a $\tau$-periodic pattern and let $m = |\Pat|$.
  Then:
  \begin{enumerate}[itemsep=1pt]
  \item\label{lm:pos-depends-on-prefix-it-1}
    Let $\Pat' = \Pat[1 \dd \min(m, \ell)]$. Then, $\Pat'$ is
    $\tau$-periodic and $\PosLowMinus{f}{\Pat}{\Text} = \PosLowMinus{f}{\Pat'}{\Text}$,
  \item\label{lm:pos-depends-on-prefix-it-2}
    Let $\Pat' = \Pat[1 \dd \min(m, 2\ell)]$. Then, $\Pat'$ is
    $\tau$-periodic and $\PosMidMinus{f}{\Pat}{\Text} = \PosMidMinus{f}{\Pat'}{\Text}$,
  \item\label{lm:pos-depends-on-prefix-it-3}
    Let $\Pat' = \Pat[1 \dd \min(m, 2\ell)]$. Then, $\Pat'$ is
    $\tau$-periodic and $\PosHighMinus{f}{\Pat}{\Text} = \PosHighMinus{f}{\Pat'}{\Text}$.
  \end{enumerate}
\end{lemma}
\begin{proof}

  Denote $H = \RootPat{f}{\tau}{\Pat}$,
  $s = \HeadPat{f}{\tau}{\Pat}$, $p = |H|$,
  $k_1 = \ExpCutPat{f}{\tau}{\Pat}{\ell}$, and
  $k_2 = \ExpCutPat{f}{\tau}{\Pat}{2\ell}$.

  1. By $m \geq 3\tau - 1$ and $\ell \geq 3\tau - 1$, we obtain
  $|\Pat'| \geq 3\tau - 1$. Thus, it follows by \cref{lm:periodic-pat-lce} that
  $\Pat'$ is $\tau$-periodic. Moreover, by the same result, we also
  have $\RootPat{f}{\tau}{\Pat'} = H$ and $\HeadPat{f}{\tau}{\Pat'} = s$.
  Denote $k'_1 = \ExpCutPat{f}{\tau}{\Pat'}{\ell}$.  Note that by
  \cref{lm:pat-expcut}\eqref{lm:pat-expcut-it-1}, it follows that
  $k'_1 = \ExpCutPat{f}{\tau}{\Pat'}{\ell} = \ExpCutPat{f}{\tau}{\Pat}{\ell} = k_1$.
  Given the above, we thus prove the two inclusions as follows:
  \begin{itemize}
  \item Let $j \in \PosLowMinus{f}{\Pat}{\Text}$. Then, it holds $j \in \RMinusFive{f}{s}{H}{\tau}{\Text}$,
    $\ExpPos{f}{\tau}{\Text}{j} = k_1$, and $\Text[j \dd \Textlen] \succeq \Pat$ or $\lcp(\Pat,
    \Text[j \dd \Textlen]) \geq \ell$. By the above, to show $j \in
    \PosLowMinus{f}{\Pat'}{\Text}$, it thus remains to prove that either
    $\Text[j \dd \Textlen] \succeq \Pat'$ or $\lcp(\Pat', \Text[j \dd \Textlen]) \geq \ell$.  By
    \cref{lm:sa-prelim}\eqref{lm:sa-prelim-it-1}, the assumptions
    $\Text[j \dd \Textlen] \succeq \Pat$ or $\lcp(\Pat, \Text[j \dd \Textlen]) \geq \ell$
    imply $\Text[j \dd \Textlen] \succeq \Pat[1 \dd \min(m, \ell)] = \Pat'$. We
    have thus proved $j \in \PosLowMinus{f}{\Pat'}{\Text}$.
  \item Let $j' \in \PosLowMinus{f}{\Pat'}{\Text}$. Then, it holds $j \in \RMinusFive{f}{s}{H}{\tau}{\Text}$,
    $\ExpPos{f}{\tau}{\Text}{j} = k'_1$, and $\Text[j \dd \Textlen] \succeq \Pat'$ or $\lcp(\Pat',
    \Text[j \dd \Textlen]) \geq \ell$. By the above, to show $j \in
    \PosLowMinus{f}{\Pat}{\Text}$, it thus remains to prove that either $\Text[j \dd \Textlen]
    \succeq \Pat$ or $\lcp(\Pat, \Text[j \dd \Textlen]) \geq \ell$. Let us
    consider two cases. If $\Text[j \dd \Textlen] \succeq \Pat' = \Pat[1 \dd
    \min(m, \ell)]$, then the claim follows by
    \cref{lm:sa-prelim}\eqref{lm:sa-prelim-it-1}. Otherwise (i.e., if
    $\lcp(\Pat', \Text[j \dd \Textlen]) \geq \ell$), then $|\Pat'| = \ell$, and
    hence $\Pat' = \Pat[1 \dd \ell]$.  This immediately implies
    $\lcp(\Pat, \Text[j \dd \Textlen]) \geq \ell$. We have thus proved $j \in
    \PosLowMinus{f}{\Pat}{\Text}$.
  \end{itemize}

  2. By $m \geq 3\tau - 1$ and $2\ell \geq 3\tau - 1$, we obtain
  $|\Pat'| \geq 3\tau - 1$. Thus, it follows by \cref{lm:periodic-pat-lce} that
  $\Pat'$ is $\tau$-periodic. Moreover, by the same result, we also
  have $\RootPat{f}{\tau}{\Pat'} = H$ and $\HeadPat{f}{\tau}{\Pat'} = s$.
  Denote
  $k'_1 = \ExpCutPat{f}{\tau}{\Pat'}{\ell}$ and
  $k'_2 = \ExpCutPat{f}{\tau}{\Pat'}{2\ell}$.
  By
  \cref{lm:pat-expcut}\eqref{lm:pat-expcut-it-1}, it follows that
  $k'_1 = \ExpCutPat{f}{\tau}{\Pat'}{\ell} = \ExpCutPat{f}{\tau}{\Pat}{\ell} = k_1$ and
  $k'_2 = \ExpCutPat{f}{\tau}{\Pat'}{2\ell} = \ExpCutPat{f}{\tau}{\Pat}{2\ell} = k_2$.
  We thus have $\PosMidMinus{f}{\Pat}{\Text} = \PosMidMinus{f}{\Pat'}{\Text}$.

  3. The proof follows analogously as in
  \cref{lm:pos-depends-on-prefix}\eqref{lm:pos-depends-on-prefix-it-1}.
\end{proof}

\begin{remark}\label{rm:lm-decomposition}
  Note that the following lemma holds even when $\Pat$ does not occur
  in $\Text$ (and we indeed use it in that case).
  Note also that it is important (\cref{pr:pat-fully-periodic}) that the lemma below holds not only for
  patterns $\Pat$ satisfying $\RunEndPat{\tau}{\Pat} \leq |\Pat|$
  but also for the case $\RunEndPat{\tau}{\Pat} = |\Pat| + 1$
  (we have $\TypePat{\tau}{\Pat} = -1$ for such patterns).
\end{remark}

\begin{lemma}\label{lm:pat-decomposition}
  Let $f$ be a necklace-consistent function.
  Let $\ell \in [16 \dd \Textlen)$ and $\tau = \lfloor \tfrac{\ell}{3} \rfloor$.
  Let $\Pat \in \Sigma^{+}$ be a $\tau$-periodic pattern satisfying
  $\TypePat{\tau}{\Pat} = -1$. Then,
  \[\DeltaBeg{\ell}{\Pat}{\Text} = \DeltaLowMinus{f}{\Pat}{\Text} +
  \DeltaMidMinus{f}{\Pat}{\Text} - \DeltaHighMinus{f}{\Pat}{\Text}.\]
\end{lemma}
\begin{proof}

  First, observe that by definition it holds $\PosBeg{\ell}{\Pat}{\Text}
  \cap \PosHighMinus{f}{\Pat}{\Text} = \emptyset$ and $\PosLowMinus{f}{\Pat}{\Text} \cap
  \PosMidMinus{f}{\Pat}{\Text} = \emptyset$. We will prove that
  $\PosBeg{\ell}{\Pat}{\Text} \cup \PosHighMinus{f}{\Pat}{\Text} = \PosLowMinus{f}{\Pat}{\Text} \cup
  \PosMidMinus{f}{\Pat}{\Text}$, since then we have
  \begin{align*}
    \DeltaLowMinus{f}{\Pat}{\Text} + \DeltaMidMinus{f}{\Pat}{\Text}
      &= |\PosLowMinus{f}{\Pat}{\Text}| + |\PosMidMinus{f}{\Pat}{\Text}|\\
      &= |\PosLowMinus{f}{\Pat}{\Text} \cup \PosMidMinus{f}{\Pat}{\Text}|\\
      &= |\PosBeg{\ell}{\Pat}{\Text} \cup \PosHighMinus{f}{\Pat}{\Text}|\\
      &= |\PosBeg{\ell}{\Pat}{\Text}| + |\PosHighMinus{f}{\Pat}{\Text}|\\
      &= \DeltaBeg{\ell}{\Pat}{\Text} + \DeltaHighMinus{f}{\Pat}{\Text}
  \end{align*}
  which implies the claim. It therefore remains to show
  $\PosBeg{\ell}{\Pat}{\Text} \cup \PosHighMinus{f}{\Pat}{\Text} = \PosLowMinus{f}{\Pat}{\Text} \cup
  \PosMidMinus{f}{\Pat}{\Text}$.  Denote $s = \HeadPat{f}{\tau}{\Pat}$ and $H =
  \RootPat{f}{\tau}{\Pat}$. Observe that for every $j' \in
  \PosBeg{\ell}{\Pat}{\Text}$, it holds $\lcp(\Pat, \Text[j' \dd \Textlen]) \geq \ell
  \geq 3\tau - 1$.  Thus, by \cref{lm:periodic-pos-lce}\eqref{lm:periodic-pos-lce-it-1},
  we have $j' \in \RFive{f}{s}{H}{\tau}{\Text}$. We furthermore have $\TypePos{\tau}{\Text}{j'} = -1$,
  since otherwise by
  \cref{lm:R-lex-block-pat}\eqref{lm:R-lex-block-pat-it-1} and \cref{lm:R-lex-block-pat}\eqref{lm:R-lex-block-pat-it-2},
  we would
  have $\Pat \prec \Text[j' \dd \Textlen]$, which would contradict $j' \in
  \PosBeg{\ell}{\Pat}{\Text}$. Thus,  $\PosBeg{\ell}{\Pat}{\Text} \subseteq
  \RMinusFive{f}{s}{H}{\tau}{\Text}$.

  First first show the inclusion $\PosLowMinus{f}{\Pat}{\Text} \cup \PosMidMinus{f}{\Pat}{\Text}
  \subseteq \PosBeg{\ell}{\Pat}{\Text} \cup \PosHighMinus{f}{\Pat}{\Text}$. Let $j' \in
  \PosLowMinus{f}{\Pat}{\Text}$. Note that $k_1 \leq k_2$. Let us thus consider two
  cases:
  \begin{itemize}
  \item Let us first assume $k_1 < k_2$. This is equivalent to
    $\min(\ExpPat{f}{\tau}{\Pat}, \lfloor \tfrac{\ell - s}{|H|} \rfloor) <
    \min(\ExpPat{f}{\tau}{\Pat}, \lfloor \tfrac{2\ell - s}{|H|} \rfloor)$, which
    implies $\lfloor \tfrac{\ell - s}{|H|} \rfloor < \ExpPat{f}{\tau}{\Pat}$.
    Therefore, $k_1 = \lfloor \tfrac{\ell - s}{|H|} \rfloor <
    \ExpPat{f}{\tau}{\Pat}$.  Denote $t = \RunEndPos{\tau}{\Text}{j'} - j'$ and $t' = \RunEndPat{\tau}{\Pat} -
    1$.  Consequently, by combining $\TypePos{\tau}{\Text}{j'} = -1$ and $t =
    \RunEndPos{\tau}{\Text}{j'} - j' = s + \ExpPos{f}{\tau}{\Text}{j'}|H| + \TailPos{f}{\tau}{\Text}{j'} = s + k_1|H| +
    \TailPos{f}{\tau}{\Text}{j'} < s + (k_1 + 1)|H| \leq s + \ExpPat{f}{\tau}{\Pat}|H| \leq s +
    \ExpPat{f}{\tau}{\Pat}|H| + \TailPat{f}{\tau}{\Pat} = \RunEndPat{\tau}{\Pat} - 1 = t'$, we obtain
    from \cref{lm:R-lex-block-pat}\eqref{lm:R-lex-block-pat-it-3} that
    $\Text[j' \dd \Textlen] \prec \Pat$. By definition of $\PosLowMinus{f}{\Pat}{\Text}$, this
    implies $\lcp(\Pat, \Text[j' \dd \Textlen]) \geq \ell$. On the other hand,
    by $t < t'$ it follows from
    \cref{lm:R-lex-block-pat}\eqref{lm:R-lex-block-pat-it-1} that
    $\lcp(\Pat, \Text[j' \dd \Textlen]) = t = \RunEndPos{\tau}{\Text}{j'} - j' = s + \ExpPos{f}{\tau}{\Text}{j'}|H|
    + \TailPos{f}{\tau}{\Text}{j'} = s + k_1|H| + \TailPos{f}{\tau}{\Text}{j'} = s + \lfloor \tfrac{\ell -
    s}{|H|} \rfloor |H| + \TailPos{f}{\tau}{\Text}{j'} \leq \ell + \TailPos{f}{\tau}{\Text}{j'} < \ell +
    \tau < 2\ell$.  We have thus proved that $\Text[j' \dd \Textlen] \prec \Pat$
    and $\lcp(\Pat, \Text[j' \dd \Textlen]) \in [\ell \dd 2\ell)$, i.e., $j' \in
    \PosBeg{\ell}{\Pat}{\Text}$.
  \item Let us now assume $k_1 = k_2$. By $j' \in \PosLowMinus{f}{\Pat}{\Text}$,
    this immediately implies $\ExpPos{f}{\tau}{\Text}{j'} = k_2$. Consider two cases. If
    $\Text[j' \dd \Textlen] \succeq \Pat$ or $\lcp(\Pat, \Text[j' \dd \Textlen]) \geq
    2\ell$, then we immediately obtain $j' \in
    \PosHighMinus{f}{\Pat}{\Text}$. Otherwise, we have $\Text[j' \dd \Textlen] \prec \Pat$ and
    $\lcp(\Pat, \Text[j' \dd \Textlen]) < 2\ell$.  Combining this with
    $\lcp(\Pat, \Text[j' \dd \Textlen]) \geq \ell$ (following from $j' \in
    \PosLowMinus{f}{\Pat}{\Text}$), we thus have $j' \in \PosBeg{\ell}{\Pat}{\Text}$.
  \end{itemize}
  We have thus proved $\PosLowMinus{f}{\Pat}{\Text} \subseteq \PosBeg{\ell}{\Pat}{\Text}
  \cup \PosHighMinus{f}{\Pat}{\Text}$.  To show $\PosMidMinus{f}{\Pat}{\Text} \subseteq
  \PosBeg{\ell}{\Pat}{\Text} \cup \PosHighMinus{f}{\Pat}{\Text}$, consider $j' \in
  \PosMidMinus{f}{\Pat}{\Text}$, and note that $\PosMidMinus{f}{\Pat}{\Text} \neq \emptyset$
  implies $k_1 < k_2$. As noted above, this implies $k_1 = \lfloor
  \tfrac{\ell - s}{|H|} \rfloor < \ExpPat{f}{\tau}{\Pat}$.  Letting $t' =
  \RunEndPat{\tau}{\Pat} - 1$, we thus have $t' = s + \ExpPat{f}{\tau}{\Pat}|H| +
  \TailPat{f}{\tau}{\Pat} \geq (\lfloor \tfrac{\ell - s}{|H|} \rfloor + 1)|H|
  \geq \ell$.  On the other hand, $j' \in \PosMidMinus{f}{\Pat}{\Text}$ implies $k_1
  < \ExpPos{f}{\tau}{\Text}{j'}$.  Letting $t = \RunEndPos{\tau}{\Text}{j'} - j'$, we thus have $t = s +
  \ExpPos{f}{\tau}{\Text}{j'}|H| + \TailPos{f}{\tau}{\Text}{j'} \geq (k_1 + 1)|H| = (\lfloor \tfrac{\ell -
  s}{|H|} \rfloor + 1)|H| \geq \ell$.  Thus, by
  \cref{lm:R-lex-block-pat}, we obtain
  $\lcp(\Pat, \Text[j' \dd \Textlen]) \geq \min(t, t') \geq \ell$.  Recall now
  that $\ExpPos{f}{\tau}{\Text}{j'} \in (k_1 \dd k_2]$, and consider two cases:
  \begin{itemize}
  \item First, assume $\ExpPos{f}{\tau}{\Text}{j'} = k_2$. Then, either at least one of
    the following conditions: (1) $\Text[j' \dd \Textlen] \geq \Pat$, (2)
    $\lcp(\Pat, \Text[j' \dd \Textlen]) \geq 2\ell$, is true, or both are false
    (i.e., it holds $\Text[j' \dd \Textlen] \prec \Pat$ and $\lcp(\Pat, \Text[j'
    \dd \Textlen]) < 2\ell$). If the first possibility holds, then we
    immediately obtain $j' \in \PosHighMinus{f}{\Pat}{\Text}$. Otherwise, by
    combining with $\lcp(\Pat, \Text[j' \dd \Textlen]) \geq \ell$, we have $j'
    \in \PosBeg{\ell}{\Pat}{\Text}$.
  \item Next, assume $\ExpPos{f}{\tau}{\Text}{j'} \in (k_1 \dd k_2)$. In
    particular, $\ExpPos{f}{\tau}{\Text}{j'} < k_2 = \min(\ExpPat{f}{\tau}{\Pat}, \lfloor
    \tfrac{2\ell - s}{|H|} \rfloor)$, i.e., $\ExpPos{f}{\tau}{\Text}{j'} < \ExpPat{f}{\tau}{\Pat}$
    and $\ExpPos{f}{\tau}{\Text}{j'} < \lfloor \tfrac{2\ell - s}{|H|} \rfloor$.  The
    first inequality implies $\ExpPos{f}{\tau}{\Text}{j'} - j' = s + \ExpPos{f}{\tau}{\Text}{j'}|H| +
    \TailPos{f}{\tau}{\Text}{j'} < s + (\ExpPos{f}{\tau}{\Text}{j'} + 1)|H| \leq s + \ExpPat{f}{\tau}{\Pat}|H| \leq s
    + \ExpPat{f}{\tau}{\Pat}|H| + \TailPat{f}{\tau}{\Pat} = \RunEndPat{\tau}{\Pat} - 1$.  By the second
    inequality, on the other hand, $\ExpPos{f}{\tau}{\Text}{j'} - j' = s + \ExpPos{f}{\tau}{\Text}{j'}|H|
    + \TailPos{f}{\tau}{\Text}{j'} < s + (\ExpPos{f}{\tau}{\Text}{j'} + 1)|H| \leq s + \lfloor
    \tfrac{2\ell - s}{|H|} \rfloor |H| \leq 2\ell$. By
    \cref{lm:R-lex-block-pat}\eqref{lm:R-lex-block-pat-it-1}, we
    thus obtain $\lcp(\Pat, \Text[j' \dd \Textlen]) = \min(\RunEndPos{\tau}{\Text}{j'} - j',
    \RunEndPat{\tau}{\Pat} - 1) = \RunEndPos{\tau}{\Text}{j'} - j' < 2\ell$.  Since above we proved
    $\lcp(\Pat, \Text[j' \dd \Textlen]) \geq \ell$, we thus have $\lcp(\Pat,
    \Text[j' \dd \Textlen]) \in [\ell \dd 2\ell)$.  Lastly, it follows from
    $\RunEndPos{\tau}{\Text}{j'} - j' < \RunEndPat{\tau}{\Pat} - 1$ by applying
    \cref{lm:R-lex-block-pat}\eqref{lm:R-lex-block-pat-it-3} that
    $\Text[j' \dd \Textlen] \prec \Pat$. We have therefore proved $j' \in
    \PosBeg{\ell}{\Pat}{\Text}$.
  \end{itemize}
  We have thus proved $\PosMidMinus{f}{\Pat}{\Text} \subseteq \PosBeg{\ell}{\Pat}{\Text}
  \cup \PosHighMinus{f}{\Pat}{\Text}$. Combining this with the above proof of
  $\PosLowMinus{f}{\Pat}{\Text} \subseteq \PosBeg{\ell}{\Pat}{\Text} \cup
  \PosHighMinus{f}{\Pat}{\Text}$, we obtain $\PosLowMinus{f}{\Pat}{\Text} \cup \PosMidMinus{f}{\Pat}{\Text}
  \subseteq \PosBeg{\ell}{\Pat}{\Text} \cup \PosHighMinus{f}{\Pat}{\Text}$.

  We now prove the opposite inclusion, i.e., $\PosBeg{\ell}{\Pat}{\Text}
  \cup \PosHighMinus{f}{\Pat}{\Text} \subseteq \PosLowMinus{f}{\Pat}{\Text} \cup \PosMidMinus{f}{\Pat}{\Text}$.
  Consider $j' \in \PosBeg{\ell}{\Pat}{\Text}$. First, observe that we have
  $\RunEndPos{\tau}{\Text}{j'} - j' \leq \RunEndPat{\tau}{\Pat} - 1$, i.e., $\RunEndPos{\tau}{\Text}{j'} - j' =
  \min(\RunEndPos{\tau}{\Text}{j'} - j', \RunEndPat{\tau}{\Pat} - 1)$ since otherwise, by
  \cref{lm:R-lex-block-pat}\eqref{lm:R-lex-block-pat-it-5}
  we would have $\Text[j' \dd \Textlen] \succeq \Pat$, contradicting $j' \in
  \PosBeg{\ell}{\Pat}{\Text}$.  Second, note that by
  \cref{lm:R-lex-block-pat}, we have
  $\min(\RunEndPos{\tau}{\Text}{j'} - j', \RunEndPat{\tau}{\Pat} - 1) \leq \lcp(\Pat, \Text[j' \dd
  \Textlen])$.  Third, by $j' \in \PosBeg{\ell}{\Pat}{\Text}$, we have
  $\lcp(\Pat, \Text[j' \dd \Textlen]) < 2\ell$.  Combining these three
  inequalities, we obtain $\RunEndPos{\tau}{\Text}{j'} - j' = \min(\RunEndPos{\tau}{\Text}{j'} - j',
  \RunEndPat{\tau}{\Pat} - 1) \leq \lcp(\Pat, \Text[j' \dd \Textlen]) < 2\ell$, which
  implies $\ExpPos{f}{\tau}{\Text}{j'} = \lfloor \tfrac{\RunEndPos{\tau}{\Text}{j'} - j' - s}{|H|} \rfloor
  \leq \lfloor \tfrac{2\ell - s}{|H|} \rfloor$.  On the other hand,
  from $\RunEndPos{\tau}{\Text}{j'} - j' \leq \RunEndPat{\tau}{\Pat} - 1$ we also obtain $\ExpPos{f}{\tau}{\Text}{j'}
  = \lfloor \tfrac{\RunEndPos{\tau}{\Text}{j'} - j' - s}{|H|} \rfloor \leq \lfloor
  \tfrac{\RunEndPat{\tau}{\Pat} - 1 - s}{|H|} \rfloor = \ExpPat{f}{\tau}{\Pat}$. Combining
  the two upper bound on $\ExpPos{f}{\tau}{\Text}{j'}$, we thus have $\ExpPos{f}{\tau}{\Text}{j'} \leq
  \min(\ExpPat{f}{\tau}{\Pat}, \lfloor \tfrac{2\ell - s}{|H|} \rfloor) = k_2$.
  We next show that $\ExpPos{f}{\tau}{\Text}{j'} \geq k_1$. Consider two cases:
  \begin{itemize}
  \item First, assume $\RunEndPat{\tau}{\Pat} - 1 < \ell$. This implies
    $\ExpPat{f}{\tau}{\Pat} = \lfloor \tfrac{\RunEndPat{\tau}{\Pat} - 1 - s}{|H|} \rfloor
    \leq \lfloor \tfrac{\ell - s}{|H|} \rfloor$ and hence $k_1 =
    \min(\ExpPat{f}{\tau}{\Pat}, \lfloor \tfrac{\ell - s}{|H|} \rfloor) =
    \ExpPat{f}{\tau}{\Pat}$.  On the other hand, $j' \in
    \PosBeg{\ell}{\Pat}{\Text}$ implies $\lcp(\Pat, \Text[j' \dd \Textlen]) \geq
    \ell$. Thus, $\lcp(\Pat, \Text[j' \dd \Textlen]) > \RunEndPat{\tau}{\Pat} - 1$.
    By \cref{lm:periodic-pos-lce}\eqref{lm:periodic-pos-lce-it-1}, we therefore obtain
    $\RunEndPos{\tau}{\Text}{j'} - j' = \RunEndPat{\tau}{\Pat} - 1$. Consequently, $\ExpPos{f}{\tau}{\Text}{j'} =
    \lfloor \tfrac{\RunEndPos{\tau}{\Text}{j'} - j' - s}{|H|} \rfloor = \lfloor
    \tfrac{\RunEndPat{\tau}{\Pat} - 1 - s}{|H|} \rfloor = \ExpPat{f}{\tau}{\Pat} = k_1$.
  \item Let us now assume $\RunEndPat{\tau}{\Pat} - 1 \geq \ell$. Note that this
    implies $\RunEndPos{\tau}{\Text}{j'} - j' \geq \ell$, since otherwise by
    \cref{lm:R-lex-block-pat}\eqref{lm:R-lex-block-pat-it-1}, we would
    have $\lcp(\Pat, \Text[j' \dd \Textlen]) = \min(\RunEndPat{\tau}{\Pat} - 1, \RunEndPos{\tau}{\Text}{j'} -
    j') < \ell$, which contradicts $j' \in
    \PosBeg{\ell}{\Pat}{\Text}$. Consequently, $\ExpPos{f}{\tau}{\Text}{j'} = \lfloor
    \tfrac{\RunEndPos{\tau}{\Text}{j'} - j' - s}{|H|} \rfloor \geq \lfloor \tfrac{\ell -
    s}{|H|} \rfloor$, and hence $k_1 = \min(\ExpPos{f}{\tau}{\Text}{j'}, \lfloor
    \tfrac{\ell - s}{|H|} \rfloor) = \lfloor \tfrac{\ell - s}{|H|}
    \rfloor \leq \ExpPos{f}{\tau}{\Text}{j'}$.
  \end{itemize}
  We have thus proved that $\ExpPos{f}{\tau}{\Text}{j'} \in [k_1 \dd
  k_2]$. Consider now two cases. If $\ExpPos{f}{\tau}{\Text}{j'} > k_1$,
  then we have $\ExpPos{f}{\tau}{\Text}{j'} \in (k_1 \dd k_2]$, and hence
  $j' \in \PosMidMinus{f}{\Pat}{\Text}$ by definition. Otherwise (i.e.,
  $\ExpPos{f}{\tau}{\Text}{j'} = k_1$), combining with $\lcp(\Pat,
  \Text[j' \dd \Textlen]) \geq \ell$ (following from
  $j' \in \PosBeg{\ell}{\Pat}{\Text}$) we have
  $j' \in \PosLowMinus{f}{\Pat}{\Text}$. Thus,
  $j' \in \PosLowMinus{f}{\Pat}{\Text} \cup \PosMidMinus{f}{\Pat}{\Text}$, and
  hence $\PosBeg{\ell}{\Pat}{\Text}
  \subseteq \PosLowMinus{f}{\Pat}{\Text} \cup \PosMidMinus{f}{\Pat}{\Text}$. It
  therefore remains to show
  $\PosHighMinus{f}{\Pat}{\Text} \subseteq \PosLowMinus{f}{\Pat}{\Text} \cup \PosMidMinus{f}{\Pat}{\Text}$.
  Let $j' \in \PosHighMinus{f}{\Pat}{\Text}$. Consider two cases. If $k_1 <
  k_2$, then we immediately obtain $j' \in \PosMidMinus{f}{\Pat}{\Text}$,
  since $\ExpPos{f}{\tau}{\Text}{j'} \in (k_1 \dd k_2]$. Otherwise, we
  have $\ExpPos{f}{\tau}{\Text}{j'} = k_2 = k_1$, and either $\Text[j' \dd
  \Textlen] \succeq \Pat$ or $\lcp(\Pat, \Text[j' \dd \Textlen]) \geq
  2\ell \geq \ell$. This implies
  $j' \in \PosLowMinus{f}{\Pat}{\Text}$. Thus,
  $\PosHighMinus{f}{\Pat}{\Text} \subseteq \PosLowMinus{f}{\Pat}{\Text} \cup \PosMidMinus{f}{\Pat}{\Text}$.
  We have therefore proved that
  $\PosBeg{\ell}{\Pat}{\Text}
  \cup \PosHighMinus{f}{\Pat}{\Text} \subseteq \PosLowMinus{f}{\Pat}{\Text} \cup \PosMidMinus{f}{\Pat}{\Text}$.
\end{proof}

\begin{definition}\label{def:pos-sets-for-pos}
  Let $f$ be a necklace-consistent function.
  Let $\ell \in [16 \dd \Textlen)$ and $\tau = \lfloor \tfrac{\ell}{3} \rfloor$.
  For every $j \in \RTwo{\tau}{\Text}$, we define:
  \begin{itemize}[itemsep=1pt]
  \item $\PosLowMinus{f}{j}{\Text}  := \PosLowMinus{f}{\Text[j \dd \Textlen]}{\Text}$,
  \item $\PosMidMinus{f}{j}{\Text}  := \PosMidMinus{f}{\Text[j \dd \Textlen]}{\Text}$,
  \item $\PosHighMinus{f}{j}{\Text} := \PosHighMinus{f}{\Text[j \dd \Textlen]}{\Text}$.
  \end{itemize}
  The sets
  $\PosLowPlus{f}{j}{\Text}$, $\PosMidPlus{f}{j}{\Text}$, and
  $\PosHighPlus{f}{j}{\Text}$ are defined analogously. We also let
  \begin{itemize}
  \item $\PosLowAll{f}{j}{\Text}   := \PosLowAll{f}{\Text[j \dd \Textlen]}{\Text}$,
  \item $\PosMidAll{f}{j}{\Text}   := \PosMidAll{f}{\Text[j \dd \Textlen]}{\Text}$,
  \item $\PosHighAll{f}{j}{\Text}  := \PosHighAll{f}{\Text[j \dd \Textlen]}{\Text}$.
  \end{itemize}
  Finally, we denote
  \begin{itemize}[itemsep=0pt]
  \item $\DeltaLowMinus{f}{j}{\Text}  := |\PosLowMinus{f}{j}{\Text}|$,
  \item $\DeltaMidMinus{f}{j}{\Text}  := |\PosMidMinus{f}{j}{\Text}|$,
  \item $\DeltaHighMinus{f}{j}{\Text} := |\PosHighMinus{f}{j}{\Text}|$.
  \end{itemize}
\end{definition}

\begin{remark}\label{rm:pos-sets-for-pos}
  Note that all above sets are well-defined since $j \in \RTwo{\tau}{\Text}$
  implies that $\Text[j \dd \Textlen]$ is $\tau$-periodic.
\end{remark}

\begin{lemma}\label{lm:pos-sets-for-pos}
  Let $f$ be a necklace-consistent function.  Let
  $\ell \in [16 \dd \Textlen)$ and $\tau
  = \lfloor \tfrac{\ell}{3} \rfloor$.  For every $j \in \RTwo{\tau}{\Text}$,
  it holds:
  \begin{align*}
    \PosLowMinus{f}{j}{\Text} &= \{j' \in \RMinusFive{f}{s}{H}{\tau}{\Text} :
      \ExpPos{f}{\tau}{\Text}{j'} = k_1\text{ and }(\Text[j' \dd \Textlen] \succeq \Text[j \dd \Textlen]
      \text{ or }\LCE_{\Text}(j, j') \geq \ell)\},\\
    \PosMidMinus{f}{j}{\Text} &= \{j' \in \RMinusFive{f}{s}{H}{\tau}{\Text} :
      \ExpPos{f}{\tau}{\Text}{j'} \in (k_1 \dd k_2]\},\\
    \PosHighMinus{f}{j}{\Text} &= \{j' \in \RMinusFive{f}{s}{H}{\tau}{\Text} :
      \ExpPos{f}{\tau}{\Text}{j'} = k_2\text{ and }(\Text[j' \dd \Textlen] \succeq \Text[j \dd \Textlen]
      \text{ or }\LCE_{\Text}(j, j') \geq 2\ell)\},
  \end{align*}
  where
  $s = \HeadPos{f}{\tau}{\Text}{j}$,
  $H = \RootPos{f}{\tau}{\Text}{j}$,
  $k_1 = \ExpCutPos{f}{\tau}{\Text}{j}{\ell}$, and
  $k_2 = \ExpCutPos{f}{\tau}{\Text}{j}{2\ell}$.
\end{lemma}
\begin{proof}
  Denote $\Pat = \Text[j \dd \Textlen]$. The claim follows by putting
  together \cref{def:pos-sets-for-pat,def:pos-sets-for-pos}, and
  observing that (by definition), it holds
  $\RootPat{f}{\tau}{\Pat} = \RootPos{f}{\tau}{\Text}{j} = H$,
  $\HeadPat{f}{\tau}{\Pat} = \HeadPos{f}{\tau}{\Text}{j} = s$,
  $\ExpCutPat{f}{\tau}{\Pat}{\ell}
  = \ExpCutPos{f}{\tau}{\Text}{j}{\ell} = k_1$,
  $\ExpCutPat{f}{\tau}{\Pat}{2\ell}
  = \ExpCutPos{f}{\tau}{\Text}{j}{2\ell} = k_2$, and $\lcp(\Pat, \Text[j' \dd
  \Textlen]) \geq \ell$ (resp.\ $\lcp(\Pat, \Text[j' \dd \Textlen]) \geq 2\ell$)
  holds if and only if $\LCE_{\Text}(j, j') \geq \ell$ (resp.\
  $\LCE_{\Text}(j, j') \geq 2\ell$).
\end{proof}

\begin{lemma}\label{lm:pos-decomposition}
  Let $f$ be a necklace-consistent function.  Let
  $\ell \in [16 \dd \Textlen)$ and $\tau
  = \lfloor \tfrac{\ell}{3} \rfloor$.  For every $j \in \RMinusTwo{\tau}{\Text}$,
  it holds: \[\DeltaBeg{\ell}{j}{\Text} = \DeltaLowMinus{f}{j}{\Text}
  + \DeltaMidMinus{f}{j}{\Text} - \DeltaHighMinus{f}{j}{\Text}.\]
\end{lemma}
\begin{proof}
  Note that if we let $\Pat = \Text[j \dd \Textlen]$, then
  $\TypePat{\tau}{\Pat} = -1$.  Thus,
  by \cref{lm:pat-decomposition}, we obtain $\DeltaBeg{\ell}{j}{\Text}
  = \DeltaBeg{\ell}{\Pat}{\Text} = \DeltaLowMinus{f}{\Pat}{\Text}
  + \DeltaMidMinus{f}{\Pat}{\Text} - \DeltaHighMinus{f}{\Pat}{\Text}
  = \DeltaLowMinus{f}{j}{\Text} + \DeltaMidMinus{f}{j}{\Text}
  - \DeltaHighMinus{f}{j}{\Text}$.
\end{proof}

\begin{lemma}\label{lm:poshigh-equivalence}
  Let $f$ be a necklace-consistent function.  Let
  $\ell \in [16 \dd \Textlen)$, $\tau = \lfloor \tfrac{\ell}{3} \rfloor$,
  and $\Pat \in \Sigma^{+}$ be a $\tau$-periodic pattern. For every
  $i, j \in [1 \dd \Textlen]$ such that $\Textinf[i \dd i + 7\tau) =
  \Textinf[j \dd j + 7\tau)$, $i \in \PosHighMinus{f}{\tau}{\Pat}$
  holds if and only if $j \in \PosHighMinus{f}{\tau}{\Pat}$.
\end{lemma}
\begin{proof}
  Denote $H = \RootPat{f}{\tau}{\Pat}$, $s
  = \HeadPat{f}{\tau}{\Pat}$, $p = |H|$, and $k_2
  = \ExpCutPat{f}{\tau}{\Pat}{2\ell}$. Let
  $i \in \PosHighMinus{f}{\tau}{\Pat}$.  We will prove that
  $j \in \PosHighMinus{f}{\tau}{\Pat}$ (the proof of the opposite
  implication follows by symmetry). If $i = j$, then the claim follows
  immediately.  Let us thus assume $i \neq
  j$. By \cref{lm:occ-equivalence}, we then have $\LCE_{\Text}(i, j) \geq
  7\tau$. The assumption $i \in \PosHighMinus{f}{\tau}{\Pat}$ implies
  $i \in \RMinusFive{f}{s}{H}{\tau}{\Text}$, $\ExpPos{f}{\tau}{\Text}{i} = k_2$,
  and it holds either $\Text[i \dd \Textlen] \succeq \Pat$ or $\lcp(\Pat,
  \Text[i \dd \Textlen]) \geq 2\ell$. First, note that
  by \cref{lm:periodic-pos-lce}\eqref{lm:periodic-pos-lce-it-2}, we
  obtain $j \in \RFive{f}{s}{H}{\tau}{\Text}$. Next, recall that $k_2
  = \min(\ExpPat{f}{\tau}{\Pat}, \lfloor \tfrac{2\ell -
  s}{p} \rfloor) \leq \lfloor \tfrac{2\ell -
  s}{p} \rfloor$. Consequently, $\RunEndFullPos{f}{\tau}{\Text}{i} - i = s
  + \ExpPos{f}{\tau}{\Text}{i}p = s + k_2 p \leq 2\ell$.  This implies
  $\RunEndPos{\tau}{\Text}{i} - i = (\RunEndFullPos{f}{\tau}{\Text}{i} - i) +
  (\RunEndPos{\tau}{\Text}{i} - \RunEndFullPos{f}{\tau}{\Text}{i}) <
  (\RunEndFullPos{f}{\tau}{\Text}{i} - i) + p \leq 2\ell
  + \lfloor \tau/3 \rfloor \leq 7\tau$ (where $2\ell
  + \lfloor \tau/3 \rfloor \leq 7\tau$ follows by $\tau
  = \lfloor \tfrac{\ell}{3} \rfloor$ and $\ell \geq 16$).  Combining
  this with $\LCE_{\Text}(i, j) \geq 7\tau$ and
  applying \cref{lm:periodic-pos-lce}\eqref{lm:periodic-pos-lce-it-2},
  we thus have $\ExpPos{f}{\tau}{\Text}{j} = \ExpPos{f}{\tau}{\Text}{i} =
  k_2$ and $\TypePos{\tau}{\Text}{j} = \TypePos{\tau}{\Text}{i} = -1$, i.e.,
  $j \in \RMinusFive{f}{s}{H}{\tau}{\Text}$.  It remains to show that it holds
  $\Text[j \dd \Textlen] \succeq \Pat$ or $\lcp(\Pat, \Text[j \dd \Textlen]) \geq
  2\ell$.  Recall, that we assumed $\Text[i \dd \Textlen] \succeq \Pat$ or
  $\lcp(\Pat, \Text[i \dd \Textlen]) \geq 2\ell$.  Consider two cases:
  \begin{itemize}
  \item If $\lcp(\Pat, \Text[i \dd \Textlen]) \geq 2\ell$, then by combining
    with $\lcp(\Text[i \dd \Textlen], \Text[j \dd \Textlen]) \geq 7\tau$ and $2\ell \leq
    7\tau$, it follows that $\lcp(\Pat, \Text[j \dd
    \Textlen]) \geq \min(\lcp(\Pat, \Text[i \dd \Textlen]), \lcp(\Text[i \dd \Textlen],
    \Text[j \dd \Textlen])) \geq \min(2\ell, 7\tau) = 2\ell$.
  \item Let us now assume that $\lcp(\Pat, \Text[i \dd \Textlen]) < 2\ell$ and
    $\Text[i \dd \Textlen] \succeq \Pat$.  Note that we cannot have $\Pat =
    \Text[i \dd \Textlen]$, since that would imply $\Textlen - i + 1 = |\Pat|
    = \lcp(\Pat, \Text[i \dd \Textlen]) < 2\ell$ which by $2\ell \leq 7\tau$
    would contradict $\Textlen - i + 1 \geq 7\tau$ (following from
    $\LCE_{\Text}(i, j) \geq 7\tau$). We thus have $\Text[i \dd
    \Textlen] \succ \Pat$, which implies that either $\Pat$ is a proper
    prefix of $\Text[i \dd \Textlen]$, or, letting $\ell' = \lcp(\Pat, \Text[i \dd
    \Textlen])$, it holds $\ell' < |\Pat|$, $i + \ell' \leq \Textlen$, and $\Text[i
    + \ell'] \succ \Pat[1 + \ell']$.  Consider two subcases:
    \begin{itemize}
    \item If $\Pat$ is a proper prefix of $\Text[i \dd \Textlen]$, then $|\Pat|
      = \lcp(\Pat, \Text[i \dd \Textlen]) < 2\ell$. Consequently, by
      $2\ell \leq 7\tau$, the assumption $\LCE_{\Text}(i, j) \geq 7\tau$
      implies that $\Pat$ is also a proper prefix of $\Text[j \dd
      \Textlen]$. We thus have $\Text[j \dd \Textlen] \succ \Pat$.
    \item Let us now assume that $\ell' < |\Pat|$, $i + \ell' \leq
      \Textlen$, and $\Text[i + \ell'] \succ \Pat[1 + \ell']$.  By $\ell' <
      2\ell$, and $\LCE_{\Text}(i, j) \geq 7\tau$, this implies that
      $\lcp(\Pat, \Text[j \dd \Textlen]) = \ell'$ and $\Text[j + \ell'] = \Text[i
      + \ell'] \succ \Pat[1 + \ell']$. Thus, we again obtain $\Text[j \dd
      \Textlen] \succ \Pat$.  \qedhere
    \end{itemize}
  \end{itemize}
\end{proof}

\subsubsection{The Data Structure}\label{sec:sa-periodic-ds}

\paragraph{Definitions}

For every $k \in [4 \dd \lceil \log \Textlen \rceil)$, denote $\ell_k = 2^k$,
$\tau_k = \lfloor \tfrac{\ell_k}{3} \rfloor$, $f_k := f_{\tau_k, \Text}$
(\cref{def:canonical-function}), $n_{{\rm runs},k} =
|\IntervalRepr{\CompRepr{14\tau_k}{\RTwo{\tau_k}{\Text}}{\Text}}|$
(\cref{def:interval-representation,def:comp}), and let
$\ArrRoot{k}[1 \dd n_{{\rm runs},k}]$ be such that for $j \in
[1 \dd n_{{\rm runs},k}]$ it holds $\ArrRoot{k}[j] =
(\HeadPos{f_k}{\tau_k}{\Text}{p_j}, |\RootPos{f_k}{\tau_k}{\Text}{p_j}|)$,
where $(p_j,t_j)_{j \in [1 \dd n_{{\rm runs},k}]} =
\IntervalRepr{\CompRepr{14\tau_k}{\RTwo{\tau_k}{\Text}}{\Text}}$.

\paragraph{Components}

The data structure, denoted $\CompSaPeriodic{\Text}$, consists of
two parts. The first part consists of the following four components:

\begin{enumerate}
\item The index core $\CompSaCore{\Text}$ (\cref{sec:sa-core-ds}). It
  needs $\bigO(\SubstringComplexity{\Text} \log \tfrac{\Textlen \log \sigma}{\SubstringComplexity{\Text} \log \Textlen})$
  space.
\item For $k \in [4 \dd \lceil \log \Textlen \rceil)$, we store the array
  $\ArrRoot{k}[1 \dd n_{{\rm runs},k}]$ in plain form using
  $\bigO(n_{{\rm runs},k})$ space. By the same analysis as
  in \cref{sec:sa-core-ds}, the total space used by the arrays is
  bounded by
  \[
    \bigO(\textstyle\sum_{k \in [4 \dd \lceil \log \Textlen \rceil)} n_{{\rm runs},k})
    = \bigO(\SubstringComplexity{\Text} \log \tfrac{\Textlen \log \sigma}{\SubstringComplexity{\Text} \log \Textlen}).
  \]
\item For every $k \in [4 \dd \lceil \log \Textlen \rceil)$ and
  $H \in \Sigma^{+}$ such that
  $\PairsMinus{f_k}{H}{\tau_k}{\Text} \neq \emptyset$
  (\cref{def:periodic-input}), we store a data structure
  from \cref{pr:int-str} for $q = 7\tau_k$ and $P =
  \PairsMinus{f_k}{H}{\tau_k}{\Text}$.  By \cref{pr:int-str}, the size of a
  single structure is
  $\bigO(|\PairsMinus{f_k}{H}{\tau_k}{\Text}|)$. By \cref{cor:nruns}, the total
  size of all the structures is thus
  \begin{align*}
    \bigO(\textstyle\sum_{k \in [4 \dd \lceil \log \Textlen \rceil)}
          \textstyle\sum_{H \in \Sigma^{+}} |\PairsMinus{f_k}{H}{\tau_k}{\Text}|)
       &= \bigO(\textstyle\sum_{k \in [4 \dd \lceil \log \Textlen \rceil)} n_{{\rm runs},k})\\
       &= \bigO(\SubstringComplexity{\Text} \log \tfrac{\Textlen \log \sigma}{\SubstringComplexity{\Text} \log \Textlen}).
  \end{align*}
  To access the structures, for every $k \in [4 \dd \lceil \log
  \Textlen \rceil)$, we store an array $\ArrPtrFirst{k}[1 \dd n_{{\rm
  runs},k}]$ such that letting $(p_j,t_j)_{j \in [1 \dd n_{{\rm
  runs},k}]} = \IntervalRepr{\CompRepr{14\tau_k}{\RTwo{\tau_k}{\Text}}{\Text}}$,
  $\ArrPtrFirst{k}[j]$ stores the pointer to the structure for $P =
  \PairsMinus{f_k}{H}{\tau_k}{\Text}$, where $H = \RootPos{f_k}{\tau_k}{\Text}{p_j}$ (or a null
  pointer, if $\PairsMinus{f_k}{H}{\tau_k}{\Text} = \emptyset$). In total, the
  arrays need $\bigO(\sum_{k \in [4 \dd \lceil \log \Textlen \rceil)} n_{{\rm
  runs},k}) = \bigO(\SubstringComplexity{\Text} \log \tfrac{\Textlen \log \sigma}{\SubstringComplexity{\Text} \log \Textlen})$
  space.
\item For $k \in [4 \dd \lceil \log \Textlen \rceil)$ and
  $H \in \Sigma^{+}$ such that
  $\PairsMinus{f_k}{H}{\tau_k}{\Text} \neq \emptyset$, we store a data
  structure from \cref{pr:mod-queries} for $q = 7\tau_k$ and $P =
  \PairsMinus{f_k}{H}{\tau_k}{\Text}$, with $h = |H|$.  The assumptions
  from \cref{pr:mod-queries} are satisfied since the labels in
  $\WInts{7\tau}{\PairsMinus{f_k}{H}{\tau_k}{\Text}}{\Text}$ are unique
  by \cref{lm:unique-labels}. By \cref{pr:mod-queries}, the size of a
  single structure is $\bigO(|\PairsMinus{f_k}{H}{\tau_k}{\Text}|)$.  In total
  the structures need
  $\bigO(\SubstringComplexity{\Text} \log \tfrac{\Textlen \log \sigma}{\SubstringComplexity{\Text} \log \Textlen})$ (see
  above).  To access the structures, we store an array
  $\ArrPtrSecond{k}[1 \dd n_{{\rm runs},k}]$ such that, letting
  $(p_j,t_j)_{j \in [1 \dd n_{{\rm runs},k}]}
  = \IntervalRepr{\CompRepr{14\tau_k}{\RTwo{\tau_k}{\Text}}{\Text}}$,
  $\ArrPtrSecond{k}[j]$ stores the pointer to the structure for $P
  = \PairsMinus{f_k}{H}{\tau_k}{\Text}$, where $H = \RootPos{f_k}{\tau_k}{\Text}{p_j}$ (or a null
  pointer, if $\PairsMinus{f_k}{H}{\tau_k}{\Text} = \emptyset$). In total, the
  arrays need $\bigO(\SubstringComplexity{\Text} \log \tfrac{\Textlen \log \sigma}{\SubstringComplexity{\Text} \log
  \Textlen})$ space (see above).
\end{enumerate}

The second part of the structure consists of the same components as
above, except in all definitions the standard lexicographic order
$\preceq$ is replaced with the inverted lexicographic order
$\preceq_{\rm inv}$ (\cref{def:inv}). No other orders are changed. As
a result, e.g.,
\begin{enumerate*}[label={\rm (\arabic*)}] \item
In \cref{def:periodic-input}, $\RMinusFour{f}{H}{\tau}{\Text}$ is
replaced with $\RPlusFour{f}{H}{\tau}{\Text}$, and \item In the structures
for range queries (\cref{pr:int-str}) we use $\preceq_{\rm inv}$ instead of $\preceq$ as
the order on the second coordinate (while the order on the first
coordinate remains the same). \end{enumerate*}
See \cref{rm:poslow-posmid-poshigh-plus} for the justification.

In total, $\CompSaPeriodic{\Text}$ needs
$\bigO(\SubstringComplexity{\Text} \log \tfrac{\Textlen \log \sigma}{\SubstringComplexity{\Text} \log \Textlen})$ space.

\begin{remark}\label{rm:ptr}
  Note that in the definition of $\ArrPtrFirst{k}$ (resp.\
  $\ArrPtrSecond{k}$), we store in $\ArrPtrFirst{k}[j]$ (resp.\
  $\ArrPtrSecond{k}[j]$) the pointer to the structure
  from \cref{pr:int-str} (resp.\ \cref{pr:mod-queries}) for $P =
  \PairsMinus{f_k}{H}{\tau_k}{\Text}$, where $H = \RootPos{f_k}{\tau_k}{\Text}{p_j}$ and
  $\PairsMinus{f_k}{H}{\tau_k}{\Text} \neq \emptyset$, even if $\TypePos{\tau_k}{\Text}{p_j} =
  +1$. This is used, e.g.,
  in \cref{pr:sa-periodic-poslow-poshigh-pat}.
\end{remark}

\subsubsection{Basic Combinatorial Properties}\label{sec:sa-periodic-basic}

\paragraph{Weighted Range Queries}

\begin{lemma}\label{lm:sa-periodic-count}
  Let $\tau \in [1 \dd \lfloor \tfrac{\Textlen}{2} \rfloor]$
  and $f$ be any necklace-consistent function.
  Let $H \in \Sigma^{+}$ and $\Pts = \IntStrPoints{7\tau}{\PairsMinus{f}{H}{\tau}{\Text}}{\Text}$
  (\cref{def:int-str,def:periodic-input}). For any $x_l \in [0 \dd 7\tau]$ and $y_l, y_u
  \in \Sigma^{*}$, it holds:
  \begin{enumerate}[leftmargin=4.5ex,itemsep=1.5ex]
  \item\label{lm:sa-periodic-count-it-1}
    $\RangeCountFourSide{\Pts}{x_l}{\Textlen}{y_l}{y_u} =\\
    \hspace*{1em}
    |\{j \in \RPrimMinusFour{f}{H}{\tau}{\Text} :
      x_l \leq \RunEndFullPos{f}{\tau}{\Text}{j} - j,\text{and }
      y_l \preceq \Textinf[\RunEndFullPos{f}{\tau}{\Text}{j} \dd
                              \RunEndFullPos{f}{\tau}{\Text}{j} + 7\tau)
          \prec y_u\}|$,
  \item\label{lm:sa-periodic-count-it-2}
    $\IncRangeCountThreeSide{\Pts}{x_l}{\Textlen}{y_u} =\\
    \hspace*{1em}
    |\{j \in \RPrimMinusFour{f}{H}{\tau}{\Text} :
      x_l \leq \RunEndFullPos{f}{\tau}{\Text}{j} - j,\text{and }
      \Textinf[\RunEndFullPos{f}{\tau}{\Text}{j} \dd
                  \RunEndFullPos{f}{\tau}{\Text}{j} + 7\tau) \preceq y_u\}|$,
  \item\label{lm:sa-periodic-count-it-3}
    $\RangeCountTwoSide{\Pts}{x_l}{\Textlen} =\\
    \hspace*{1em}
    |\{j \in \RPrimMinusFour{f}{H}{\tau}{\Text} :
      x_l \leq \RunEndFullPos{f}{\tau}{\Text}{j} - j\}|$.
  \end{enumerate}
\end{lemma}
\begin{proof}
  1. Denote $\Rcomp = \CompRepr{14\tau}{\RMinusFour{f}{H}{\tau}{\Text}}{\Text}$, and
  recall that we have $\PairsMinus{f}{H}{\tau}{\Text} = \{(\RunEndFullPos{f}{\tau}{\Text}{j}, \allowbreak
  \min(\RunEndFullPos{f}{\tau}{\Text}{j} - \RunBeg{\tau}{\Text}{j}, 7\tau)) : j \in \Rcomp\}$
  (\cref{def:periodic-input}).
  The proof consists of four steps labeled (a) through (d).

  (a) Denote
  \begin{align*}
    Q &= \{\Textinf[\RunEndFullPos{f}{\tau}{\Text}{i} - 7\tau \dd \RunEndFullPos{f}{\tau}{\Text}{i} + 7\tau) :
         i \in \Rcomp, x_l \leq \RunEndFullPos{f}{\tau}{\Text}{i} -
         \RunBeg{\tau}{\Text}{i},\text{ and }\\
      &  \hspace{6.1cm}
         y_l \preceq \Textinf[\RunEndFullPos{f}{\tau}{\Text}{i} \dd \RunEndFullPos{f}{\tau}{\Text}{i} + 7\tau)
         \prec y_u\},\\
    A &= \{j \in [1 \dd \Textlen] : \Textinf[j - 7\tau \dd j + 7\tau) \in Q\}.\\
  \end{align*}
  In the first step, we prove that
  $\RangeCountFourSide{\Pts}{x_l}{\Textlen}{y_l}{y_u} = |A|$.  Let $\mathcal{R} =
  \{(x,y,w,\ell) \in \Pts : x_l \leq x < \Textlen\text{ and }y_l \preceq y
  \prec y_u\}$.  First, observe that for every $p =
  (x_p,y_p,w_p,\ell_p) \in \mathcal{R}$, it holds $\Textinf[\ell_p -
  7\tau \dd \ell_p + 7\tau) \in Q$.  To see this, note that $p \in
  \IntStrPoints{7\tau}{\PairsMinus{f}{H}{\tau}{\Text}}{\Text}$. By definition of $\PairsMinus{f}{H}{\tau}{\Text}$,
  there exists $i_p \in \Rcomp$ such that $x_p =
  \min(\RunEndFullPos{f}{\tau}{\Text}{i_p} - \RunBeg{\tau}{\Text}{i_p}, 7\tau)$, $y_p =
  \Textinf[\RunEndFullPos{f}{\tau}{\Text}{i_p} \dd \RunEndFullPos{f}{\tau}{\Text}{i_p} + 7\tau)$, and $\ell_p$
  satisfies $\Textinf[\ell_p - 7\tau \dd \ell_p +7\tau) =
  \Textinf[\RunEndFullPos{f}{\tau}{\Text}{i_p} - 7\tau \dd \RunEndFullPos{f}{\tau}{\Text}{i_p} + 7\tau)$.  Thus, $p
  \in \mathcal{R}$ implies $x_l \leq x_p = \min(\RunEndFullPos{f}{\tau}{\Text}{i_p} -
  \RunBeg{\tau}{\Text}{i_p}, 7\tau) \leq \RunEndFullPos{f}{\tau}{\Text}{i_p} - \RunBeg{\tau}{\Text}{i_p}$ and $y_l
  \preceq y_p = \Textinf[\RunEndFullPos{f}{\tau}{\Text}{i_p} \dd \RunEndFullPos{f}{\tau}{\Text}{i_p} + 7\tau) \prec
  y_u$. By definition we thus have $\Textinf[\RunEndFullPos{f}{\tau}{\Text}{i_p} - 7\tau \dd
  \RunEndFullPos{f}{\tau}{\Text}{i_p} + 7\tau) \in Q$. Since $\Textinf[\RunEndFullPos{f}{\tau}{\Text}{i_p} - 7\tau
  \dd \RunEndFullPos{f}{\tau}{\Text}{i_p} + 7\tau) = \Textinf[\ell_p - 7\tau \dd \ell_p +
  7\tau)$, we therefore obtain $\Textinf[\ell_p - 7\tau \dd \ell_p +
  7\tau) \in Q$.  Let $g: \mathcal{R} \rightarrow Q$ be defined so
  that for every $p = (x,y,w,\ell) \in \mathcal{R}$, we let $g(p) =
  \Textinf[\ell - 7\tau \dd \ell + 7\tau)$. We prove that $g$ is a
  bijection.
  \begin{itemize}
  \item Let $p_1, p_2 \in \mathcal{R}$ be such that $g(p_1) =
    g(p_2)$. We show that $p_1 = p_2$.  For $k \in \{1, 2\}$, denote
    $p_k = (x_k, y_k, w_k, \ell_k)$ and let $i_k \in \Rcomp$ be
    such that $x_k = \min(\RunEndFullPos{f}{\tau}{\Text}{i_k} - \RunBeg{\tau}{\Text}{i_k}, 7\tau)$, $y_k =
    \Textinf[\RunEndFullPos{f}{\tau}{\Text}{i_k} \dd \RunEndFullPos{f}{\tau}{\Text}{i_k} + 7\tau)$, and
    $\Textinf[\ell_k - 7\tau \dd \ell_k + 7\tau) =
    \Textinf[\RunEndFullPos{f}{\tau}{\Text}{i_k} - 7\tau \dd \RunEndFullPos{f}{\tau}{\Text}{i_k} + 7\tau)$.  Let
    also $S_k := \Textinf[\ell_k - 7\tau \dd \ell_k + 7\tau) =
    \Textinf[\RunEndFullPos{f}{\tau}{\Text}{i_k} - 7\tau \dd \RunEndFullPos{f}{\tau}{\Text}{i_k} + 7\tau)$.  By the
    assumption, $S_1 = g(p_1) = g(p_2) = S_2$. Consequently, it holds
    $\ell_1 = \min\{j \in [1 \dd \Textlen] : \Textinf[j - 7\tau \dd j + 7\tau) =
    S_1\} = \min\{j \in [1 \dd \Textlen] : \Textinf[j - 7\tau \dd j + 7\tau) =
    S_2\} = \ell_2$. Similarly, $w_1 = |\{j \in [1 \dd \Textlen] : \Textinf[j - 7\tau
    \dd j + 7\tau) = S_1\}| = |\{j \in [1 \dd \Textlen] : \Textinf[j - 7\tau \dd
    j + 7\tau) = S_2\}| = w_2$.  We also have $y_1 =
    \Textinf[\RunEndFullPos{f}{\tau}{\Text}{i_1} \dd \RunEndFullPos{f}{\tau}{\Text}{i_1} + 7\tau) = S_1(7\tau \dd
    14\tau] = S_2(7\tau \dd 14\tau] = \Textinf[\RunEndFullPos{f}{\tau}{\Text}{i_2} \dd
    \RunEndFullPos{f}{\tau}{\Text}{i_2} + 7\tau) = y_2$. Lastly, observe that due to $\Text[\Textlen]
    = \Textinf[0]$ being unique in $\Text$, for every $i \in \RFour{f}{H}{\tau}{\Text}$,
    letting $Z = \Textinf[\RunEndFullPos{f}{\tau}{\Text}{i} - 7\tau \dd
    \RunEndFullPos{f}{\tau}{\Text}{i})$, we have $\min(\RunEndFullPos{f}{\tau}{\Text}{i} - \RunBeg{\tau}{\Text}{i}, 7\tau) = |H| +
    \lcs(Z, Z[1 \dd 7\tau - |H|])$.  Thus, $x_1 = \min(\RunEndFullPos{f}{\tau}{\Text}{i_1}
    - \RunBeg{\tau}{\Text}{i_1}, 7\tau) = |H| + \lcs(S_1[1 \dd 7\tau], S_1[1 \dd 7\tau -
    |H|]) = |H| + \lcs(S_2[1 \dd 7\tau], S_2[1 \dd 7\tau - |H|]) =
    \min(\RunEndFullPos{f}{\tau}{\Text}{i_2} - \RunBeg{\tau}{\Text}{i_2}, 7\tau) = x_2$. We have thus proved
    $p_1 = p_2$.
  \item Let us now consider any $X \in Q$. Then, there exists $i \in \Rcomp$ such
    that $x_l \leq \RunEndFullPos{f}{\tau}{\Text}{i} - \RunBeg{\tau}{\Text}{i}$, $y_l \preceq
    \Textinf[\RunEndFullPos{f}{\tau}{\Text}{i} \dd \RunEndFullPos{f}{\tau}{\Text}{i} + 7\tau) \prec y_u$, and
    $\Textinf[\RunEndFullPos{f}{\tau}{\Text}{i} - 7\tau \dd \RunEndFullPos{f}{\tau}{\Text}{i} = 7\tau) = X$.
    Note that by definition of $\PairsMinus{f}{H}{\tau}{\Text}$, $i \in \Rcomp$ implies
    that $(\RunEndFullPos{f}{\tau}{\Text}{i}, \min(\RunEndFullPos{f}{\tau}{\Text}{i} - \RunBeg{\tau}{\Text}{i}, 7\tau)) \in
    \PairsMinus{f}{H}{\tau}{\Text}$. Consequently, there exists $p = (x, y, w, \ell) \in
    \Pts$ such that $x = \min(\RunEndFullPos{f}{\tau}{\Text}{i} - \RunBeg{\tau}{\Text}{i}, 7\tau)$, $y =
    \Textinf[\RunEndFullPos{f}{\tau}{\Text}{i} \dd \RunEndFullPos{f}{\tau}{\Text}{i} + 7\tau)$, and $\ell$
    satisfies $\Textinf[\ell - 7\tau \dd \ell + 7\tau) =
    \Textinf[\RunEndFullPos{f}{\tau}{\Text}{i} - 7\tau \dd \RunEndFullPos{f}{\tau}{\Text}{i} + 7\tau) = X$.  By
    $x_l \leq 7\tau$, the inequality $x_l \leq \RunEndFullPos{f}{\tau}{\Text}{i} -
    \RunBeg{\tau}{\Text}{i}$ implies $x_l \leq \min(\RunEndFullPos{f}{\tau}{\Text}{i} - \RunBeg{\tau}{\Text}{i}, 7\tau) =
    x$.  On the other hand, $x \leq \RunEndFullPos{f}{\tau}{\Text}{i} - \RunBeg{\tau}{\Text}{i} \leq \Textlen - 1
    < \Textlen$ holds by the uniqueness of $\Text[\Textlen]$ in $\Text$. Finally,note that we then also have $y_l \preceq
    y = \Textinf[\RunEndFullPos{f}{\tau}{\Text}{i} \dd \RunEndFullPos{f}{\tau}{\Text}{i} + 7\tau) \prec y_u$. Thus, it
    holds $p \in \mathcal{R}$.  Since, as noted above, $\ell$
    satisfies $\Textinf[\ell - 7\tau \dd \ell + 7\tau) = X$, we thus
    obtain $g(p) = \Textinf[\ell - 7\tau \dd \ell + 7\tau) = X$.
  \end{itemize}
  We have thus proved that $g$ is a bijection. Thus, for every $(x_1,
  y_1, w_1, \ell_1), (x_2, y_2, w_2, \ell_2) \in \mathcal{R}$, $(x_1,
  y_1, w_1, \ell_1) \neq (x_2, y_2, w_2, \ell_2)$ implies
  $\Textinf[\ell_1 - 7\tau \dd \ell_1 + 7\tau) \neq \Textinf[\ell_2 - 7\tau
  \dd \ell_2 + 7\tau)$. Observe now that in \cref{def:int-str}, we can
  equivalently set $c(i) = |\{i' \in [1 \dd \Textlen] : \Textinf[i' - q \dd i' +
  q) = \Textinf[m(i) - q \dd m(i) + q)\}|$.  On the other hand, by the
  above, letting $P_k = \{j \in [1 \dd \Textlen] : \Textinf[j - 7\tau \dd j +
  7\tau) = \Textinf[\ell_k - 7\tau \dd \ell_k + 7\tau)\}$ for $k \in \{1,
  2\}$, we have $P_1 \cap P_2 = \emptyset$. Combining that, we obtain
  $\RangeCountFourSide{\Pts}{x_l}{\Textlen}{y_l}{y_u} = \sum_{(x,y,w,\ell) \in \mathcal{R}}
  w = \sum_{(x,y,w,\ell) \in \mathcal{R}} |\{j \in [1 \dd \Textlen] : \Textinf[j
  - 7\tau \dd j + 7\tau) = \Textinf[\ell - 7\tau \dd \ell + 7\tau)\}| =
  |\{j \in [1 \dd \Textlen] : \Textinf[j - 7\tau \dd j + 7\tau) \in Q\}| = |A|$.

  (b) Denote
  \begin{align*}
    Q' &= \{\Textinf[\RunEndFullPos{f}{\tau}{\Text}{i} - 7\tau \dd \RunEndFullPos{f}{\tau}{\Text}{i} + 7\tau) :
          i \in \RMinusFour{f}{H}{\tau}{\Text}, x_l \leq \RunEndFullPos{f}{\tau}{\Text}{i} - i,\text{ and }\\
       &  \hspace{6.1cm} y_l \preceq
          \Textinf[\RunEndFullPos{f}{\tau}{\Text}{i} \dd \RunEndFullPos{f}{\tau}{\Text}{i} + 7\tau) \prec y_u\}.
  \end{align*}
  In the second step, we show that $Q = Q'$.
  \begin{itemize}
  \item First, consider any $X \in Q'$. Then, there exists $i \in
    \RMinusFour{f}{H}{\tau}{\Text}$ such that $X = \Textinf[\RunEndFullPos{f}{\tau}{\Text}{i} - 7\tau \dd
    \RunEndFullPos{f}{\tau}{\Text}{i} + 7\tau)$, $x_l \leq \RunEndFullPos{f}{\tau}{\Text}{i} - i$, and
    $y_l \preceq \Textinf[\RunEndFullPos{f}{\tau}{\Text}{i} \dd \RunEndFullPos{f}{\tau}{\Text}{i} + 7\tau) \prec y_u$.
    Let $i' = \max(i, \RunEndFullPos{f}{\tau}{\Text}{i} - 7\tau)$. Note that then $[i \dd i']
    \subseteq \RTwo{\tau}{\Text}$ and thus by \cref{lm:R-text-block}, $i' \in
    \RMinusFour{f}{H}{\tau}{\Text}$ and $\RunEndFullPos{f}{\tau}{\Text}{i'} = \RunEndFullPos{f}{\tau}{\Text}{i}$.  By $x_l \leq
    7\tau$, we then also have $\RunEndFullPos{f}{\tau}{\Text}{i'} - i' \in [x_l \dd
    7\tau]$.  Let $j = \min \{t \in [1 \dd \Textlen] : \Textinf[t - 7\tau \dd t
    + 7\tau) = X\}$ and let $j'$ be such that $\RunEndFullPos{f}{\tau}{\Text}{i'} - i' = j
    - j'$.  We prove the following properties of $j'$:
    \begin{itemize}[leftmargin=3.5ex]
    \item First, we show that $j' \in \RMinusFour{f}{H}{\tau}{\Text}$ and $\RunEndFullPos{f}{\tau}{\Text}{j'}
      = j$. Observe, that letting $\delta = j - j'$ and $\ell' = (j -
      j') + \tau$, we have $\delta \in [0 \dd \RunEndPos{\tau}{\Text}{i'} - i']$,
      $\RunEndPos{\tau}{\Text}{i'} - i' = (\RunEndPos{\tau}{\Text}{i'} - \RunEndFullPos{f}{\tau}{\Text}{i'}) + (\RunEndFullPos{f}{\tau}{\Text}{i'} -
      i') < \tau + (j - j') = \ell'$.  Moreover, $\Textinf[\RunEndFullPos{f}{\tau}{\Text}{i'}
      - 7\tau \dd \RunEndFullPos{f}{\tau}{\Text}{i'} + 7\tau) = \Textinf[j - 7\tau \dd j +
      7\tau)$ implies $\Textinf[j - \delta \dd j - \delta + \ell') =
      \Textinf[j' \dd j + \tau) = \Textinf[i' \dd \RunEndFullPos{f}{\tau}{\Text}{i'} + \tau) =
      \Textinf[i' \dd i' + \ell')$.  Finally, we also have $j \in [1 \dd
      \Textlen]$. By \cref{lm:full-run-shifted}, we thus have $j' \in \RMinusFour{f}{H}{\tau}{\Text}$,
      and $\RunEndFullPos{f}{\tau}{\Text}{j'} = j' + (\RunEndFullPos{f}{\tau}{\Text}{i'} - i') = j$.
    \item Second, we prove that $y_l \preceq \Textinf[\RunEndFullPos{f}{\tau}{\Text}{j'} \dd
      \RunEndFullPos{f}{\tau}{\Text}{j'} {+} 7\tau) \prec y_u$. For this, it suffices to
      note that $\Textinf[\RunEndFullPos{f}{\tau}{\Text}{j'} \dd \RunEndFullPos{f}{\tau}{\Text}{j'} {+} 7\tau) =
      \Textinf[j \dd j {+} 7\tau) = X(7\tau \dd 14\tau]$. On the other hand, $X(7\tau \dd 14\tau] =
      \Textinf[\RunEndFullPos{f}{\tau}{\Text}{i} \dd \RunEndFullPos{f}{\tau}{\Text}{i} {+} 7\tau)$. Thus, by the
      assumption about the string $\Textinf[\RunEndFullPos{f}{\tau}{\Text}{i} \dd \RunEndFullPos{f}{\tau}{\Text}{i} + 7\tau)$,
      we obtain
      $y_l \preceq \Textinf[\RunEndFullPos{f}{\tau}{\Text}{j'} \dd \RunEndFullPos{f}{\tau}{\Text}{j'} + 7\tau)
      \prec y_u$.
    \item Finally, we prove $j' \in \Cover{14\tau}{\Text}$
      (\cref{cons:cover}).  For this, by \cref{lm:cover-equivalence} it suffices
      to prove that there exists $p \in [1 \dd \Textlen]$ such that $p = \min
      \OccThree{14\tau}{p}{\Text}$ and $j' \in [p \dd p + 14\tau)$. If $j
      \leq 14\tau$, then by $j' \leq j$ it suffices to take $p =
      1$. Let us thus assume $j > 14\tau$ and let $p = j - 7\tau$. To
      show that $p = \min \OccThree{14\tau}{p}{\Text}$, note that by
      definition of $j$, for every $t \in [1 \dd j)$, we have $\Textinf[t
      - 7\tau \dd t + 7\tau) \neq X$. Thus, for every $p' \in [1 \dd
      p)$, it holds $\Textinf[p' \dd p' + 14\tau) \neq X$. By $\Textinf[p
      \dd p + 14\tau) = X$, we therefore obtain $p = \min
      \OccThree{14\tau}{p}{\Text}$. It remains to note that $j' < j$ and $j'
      = j - (\RunEndFullPos{f}{\tau}{\Text}{i'} - i') \geq j - 7\tau$. Thus, $j' \in [j -
      7\tau \dd j) \subseteq [p \dd p + 14\tau)$.
    \end{itemize}
    By combining the above properties, we obtain that $j' \in
    \RMinusFour{f}{H}{\tau}{\Text} \cap \Cover{14\tau}{\Text} = \Rcomp$,
    $\RunEndFullPos{f}{\tau}{\Text}{j'} - \RunBeg{\tau}{\Text}{j'} \geq \RunEndFullPos{f}{\tau}{\Text}{j'} - j' = j - j' =
    \RunEndFullPos{f}{\tau}{\Text}{i'} - i' \geq x_l$, and $y_l \preceq \Textinf[\RunEndFullPos{f}{\tau}{\Text}{j'}
    \dd \RunEndFullPos{f}{\tau}{\Text}{j'} + 7\tau) \prec y_u$. Thus, by definition of $Q$, we
    have $\Textinf[\RunEndFullPos{f}{\tau}{\Text}{j'} - 7\tau \dd \RunEndFullPos{f}{\tau}{\Text}{j'} + 7\tau) \in
    Q$. Recalling, that $\Textinf[\RunEndFullPos{f}{\tau}{\Text}{j'} - 7\tau \dd \RunEndFullPos{f}{\tau}{\Text}{j'}
    + 7\tau) = \Textinf[j - 7\tau \dd j + 7\tau) = X$, we thus obtain $X
    \in Q$.
  \item Let us now consider $X \in Q$. By definition, then there
    exists $i \in \Rcomp$ such that $x_l \leq \RunEndFullPos{f}{\tau}{\Text}{i} -
    \RunBeg{\tau}{\Text}{i}$, $y_l \preceq \Textinf[\RunEndFullPos{f}{\tau}{\Text}{i} \dd \RunEndFullPos{f}{\tau}{\Text}{i} + 7\tau)
    \prec y_u$, and $\Textinf[\RunEndFullPos{f}{\tau}{\Text}{i} - 7\tau \dd \RunEndFullPos{f}{\tau}{\Text}{i} + 7\tau)
    = X$. By $\Rcomp = \Cover{14\tau}{\Text} \cap \RMinusFour{f}{H}{\tau}{\Text}
    \subseteq \RMinusFour{f}{H}{\tau}{\Text}$, we have $i \in \RMinusFour{f}{H}{\tau}{\Text}$. Denote $i' =
    \RunBeg{\tau}{\Text}{i}$. By \cref{lm:R-text-block}, we have $i' \in
    \RMinusFour{f}{H}{\tau}{\Text}$ and $\RunEndFullPos{f}{\tau}{\Text}{i'} = \RunEndFullPos{f}{\tau}{\Text}{i}$. Thus, it holds
    $\Textinf[\RunEndFullPos{f}{\tau}{\Text}{i'} \dd \RunEndFullPos{f}{\tau}{\Text}{i'} + 7\tau) =
    \Textinf[\RunEndFullPos{f}{\tau}{\Text}{i} \dd \RunEndFullPos{f}{\tau}{\Text}{i} + 7\tau)$ and hence $y_l \preceq
    \Textinf[\RunEndFullPos{f}{\tau}{\Text}{i'} \dd \RunEndFullPos{f}{\tau}{\Text}{i'} + 7\tau) \prec y_u$.  By
    $\RunEndFullPos{f}{\tau}{\Text}{i} - \RunBeg{\tau}{\Text}{i} \geq x_l$, we also obtain $\RunEndFullPos{f}{\tau}{\Text}{i'}
    - i' = \RunEndFullPos{f}{\tau}{\Text}{i} - \RunBeg{\tau}{\Text}{i} \geq x_l$. We have thus proved that
    $X = \Textinf[\RunEndFullPos{f}{\tau}{\Text}{i} - 7\tau \dd \RunEndFullPos{f}{\tau}{\Text}{i} + 7\tau) =
    \Textinf[\RunEndFullPos{f}{\tau}{\Text}{i'} - 7\tau \dd \RunEndFullPos{f}{\tau}{\Text}{i'} + 7\tau) \in Q'$.
  \end{itemize}

  (c) Denote
  \begin{align*}
    A' &= \{\RunEndFullPos{f}{\tau}{\Text}{i} : i \in \RPrimMinusFour{f}{H}{\tau}{\Text}, x_l \leq \RunEndFullPos{f}{\tau}{\Text}{i} - i,
          \text{ and }y_l \preceq \Textinf[\RunEndFullPos{f}{\tau}{\Text}{i} \dd \RunEndFullPos{f}{\tau}{\Text}{i} +
          7\tau) \prec y_u\}.
  \end{align*}
  In the third step, we prove that $A = A'$.
  \begin{itemize}
  \item Consider any $j \in A'$. Then, there exists $j' \in
    \RPrimMinusFour{f}{H}{\tau}{\Text}$ such that $j = \RunEndFullPos{f}{\tau}{\Text}{j'}$, $x_l \leq \RunEndFullPos{f}{\tau}{\Text}{j'}
    - j'$, and $y_l \preceq \Textinf[\RunEndFullPos{f}{\tau}{\Text}{j'} \dd \RunEndFullPos{f}{\tau}{\Text}{j'} + 7\tau)
    \prec y_u$. To show that $j \in A$, we need to prove that $\Textinf[j
    - 7\tau \dd j + 7\tau) \in Q'$ (we use here that $Q = Q'$).  This
    in turn requires showing that there exists $t \in \RMinusFour{f}{H}{\tau}{\Text}$ such
    that $x_l \leq \RunEndFullPos{f}{\tau}{\Text}{t} - t$, $y_l \preceq \Textinf[\RunEndFullPos{f}{\tau}{\Text}{t} \dd
    \RunEndFullPos{f}{\tau}{\Text}{t} + 7\tau) \prec y_u$, and $\Textinf[\RunEndFullPos{f}{\tau}{\Text}{t} -
    7\tau \dd \RunEndFullPos{f}{\tau}{\Text}{t} + 7\tau) = \Textinf[j - 7\tau \dd j +
    7\tau)$. Let $t = j'$. Then, we indeed have $t \in \RPrimMinusFour{f}{H}{\tau}{\Text}
    \subseteq \RMinusFour{f}{H}{\tau}{\Text}$, $\RunEndFullPos{f}{\tau}{\Text}{t} - t = \RunEndFullPos{f}{\tau}{\Text}{j'} - j' \geq
    x_l$ and $\Textinf[\RunEndFullPos{f}{\tau}{\Text}{t} \dd \RunEndFullPos{f}{\tau}{\Text}{t} + 7\tau) =
    \Textinf[\RunEndFullPos{f}{\tau}{\Text}{j'} \dd \RunEndFullPos{f}{\tau}{\Text}{j'} + 7\tau)$ (which implies
    $y_l \preceq \Textinf[\RunEndFullPos{f}{\tau}{\Text}{t} \dd \RunEndFullPos{f}{\tau}{\Text}{t} + 7\tau) \prec y_u$).
    Lastly, $\Textinf[\RunEndFullPos{f}{\tau}{\Text}{t} - 7\tau \dd \RunEndFullPos{f}{\tau}{\Text}{t} + 7\tau) =
    \Textinf[j - 7\tau \dd j + 7\tau)$. Thus, $j \in A$.
  \item Let us now take $j \in A$. Then, $j \in [1 \dd \Textlen]$ and
    $\Textinf[j - 7\tau \dd j + 7\tau) \in Q'$ (we again use that $Q =
    Q'$), which in turn implies that there exists a position $i \in
    \RMinusFour{f}{H}{\tau}{\Text}$ such that $\Textinf[\RunEndFullPos{f}{\tau}{\Text}{i} - 7\tau \dd \RunEndFullPos{f}{\tau}{\Text}{i}
    + 7\tau) = \Textinf[j - 7\tau \dd j + 7\tau)$, $x_l \leq \RunEndFullPos{f}{\tau}{\Text}{i}
    - i$, and $y_l \preceq \Textinf[\RunEndFullPos{f}{\tau}{\Text}{i} \dd \RunEndFullPos{f}{\tau}{\Text}{i} + 7\tau)
    \prec y_u$. Let $i' = \max(i, \RunEndFullPos{f}{\tau}{\Text}{i} - 7\tau)$. We proceed
    analogously as in the second step. First, note that then $[i \dd
    i'] \subseteq \RTwo{\tau}{\Text}$ and thus by \cref{lm:R-text-block}, $i' \in
    \RMinusFour{f}{H}{\tau}{\Text}$ and $\RunEndFullPos{f}{\tau}{\Text}{i'} = \RunEndFullPos{f}{\tau}{\Text}{i}$. By $x_l \leq
    7\tau$, we then also have $\RunEndFullPos{f}{\tau}{\Text}{i'} - i' \in [x_l \dd
    7\tau]$. Let $j'$ be such that $\RunEndFullPos{f}{\tau}{\Text}{i'} - i' = j -
    j'$. Letting $\delta = j - j'$ and $\ell' = (j - j') + \tau$, we
    have $\delta \in [0 \dd \RunEndPos{\tau}{\Text}{i'} - i']$ and $\RunEndPos{\tau}{\Text}{i'} - i' <
    \ell'$.  Moreover, $\Textinf[\RunEndFullPos{f}{\tau}{\Text}{i'} - 7\tau \dd \RunEndFullPos{f}{\tau}{\Text}{i'}
    + 7\tau) = \Textinf[j - 7\tau \dd j + 7\tau)$ implies $\Textinf[j -
    \delta \dd j - \delta + \ell') = \Textinf[j' \dd j + \tau) = \Textinf[i'
    \dd \RunEndFullPos{f}{\tau}{\Text}{i'} + \tau) = \Textinf[i' \dd i' + \ell')$.  By
    \cref{lm:full-run-shifted}, we thus obtain $j' \in \RMinusFour{f}{H}{\tau}{\Text}$, and
    $\RunEndFullPos{f}{\tau}{\Text}{j'} = j' + (\RunEndFullPos{f}{\tau}{\Text}{i'} - i') = j$.  Let us now
    define $j'' = \RunBeg{\tau}{\Text}{j'}$. By definition, $j'' \in \RPrimTwo{\tau}{\Text}$. On the
    other hand, by \cref{lm:R-text-block}, $j'' \in
    \RMinusFour{f}{H}{\tau}{\Text}$. Thus, $j'' \in
    \RPrimMinusFour{f}{H}{\tau}{\Text}$. By \cref{lm:R-text-block}, we also have
    $\RunEndFullPos{f}{\tau}{\Text}{j''} = \RunEndFullPos{f}{\tau}{\Text}{j'} = j$. Next, note that $j'' =
    \RunBeg{\tau}{\Text}{j'} \leq j'$ and our earlier assumptions imply
    $\RunEndFullPos{f}{\tau}{\Text}{j''} - j'' = \RunEndFullPos{f}{\tau}{\Text}{j'} - j'' \geq \RunEndFullPos{f}{\tau}{\Text}{j'} -
    j' = j - j' = \RunEndFullPos{f}{\tau}{\Text}{i'} - i' \geq x_l$. Lastly, we also have
    $\Textinf[\RunEndFullPos{f}{\tau}{\Text}{j''} \dd \RunEndFullPos{f}{\tau}{\Text}{j''} + 7\tau) =
    \Textinf[\RunEndFullPos{f}{\tau}{\Text}{j'} \dd \RunEndFullPos{f}{\tau}{\Text}{j'} + 7\tau) = \Textinf[j \dd j +
    7\tau) = \Textinf[\RunEndFullPos{f}{\tau}{\Text}{i} \dd \RunEndFullPos{f}{\tau}{\Text}{i} + 7\tau)$. Consequently,
    $y_l \preceq \Textinf[\RunEndFullPos{f}{\tau}{\Text}{j''} \dd \RunEndFullPos{f}{\tau}{\Text}{j''} + 7\tau)
    \prec y_u$. We have thus proved that $j \in A'$ (with $j''$ as the
    position satisfying $\RunEndFullPos{f}{\tau}{\Text}{j''} = j$).
  \end{itemize}

  (d) Denote
  \begin{align*}
    A'' &= \{i \in \RPrimMinusFour{f}{H}{\tau}{\Text}: x_l \leq \RunEndFullPos{f}{\tau}{\Text}{i} - i\text{ and }
            y_l \preceq \Textinf[\RunEndFullPos{f}{\tau}{\Text}{i} \dd \RunEndFullPos{f}{\tau}{\Text}{i} + 7\tau)
            \prec y_u\}.
  \end{align*}
  In this step, we put everything together.
  By \cref{lm:efull}, for every $i, i' \in
  \RPrimTwo{\tau}{\Text}$, $i \neq i'$ implies $\RunEndFullPos{f}{\tau}{\Text}{i} \neq \RunEndFullPos{f}{\tau}{\Text}{i'}$.  This
  implies $|A'| = |A''|$. Combining this with (a) and (c), we thus
  obtain $\RangeCountFourSide{\Pts}{x_l}{\Textlen}{y_l}{y_u} = |A| = |A'| = |A''|$,
  i.e., the claim.

  2. The proof is analogous, except we replace the condition
  $y_l \preceq \Textinf[\RunEndFullPos{f}{\tau}{\Text}{i} \dd \RunEndFullPos{f}{\tau}{\Text}{i} + 7\tau) \prec y_u$
  in the definition of $Q$, $Q'$, $A'$, and $A''$ with the
  condition $\Textinf[\RunEndFullPos{f}{\tau}{\Text}{i} \dd \RunEndFullPos{f}{\tau}{\Text}{i} + 7\tau) \preceq y_u$.

  3. The proof is again analogous, except we remove the condition
  $y_l \preceq \Textinf[\RunEndFullPos{f}{\tau}{\Text}{i} \dd \RunEndFullPos{f}{\tau}{\Text}{i} + 7\tau) \prec y_u$
  in the definition of $Q$, $Q'$, $A'$, and $A''$.
\end{proof}

\begin{lemma}\label{lm:RskH-size-2}
  Let $\tau \in [1 \dd \lfloor \tfrac{\Textlen}{2} \rfloor]$
  and $f$ be any necklace-consistent function.
  Let $H \in \Sigma^{+}$, $p = |H|$, $s \in [0 \dd p)$, $k_{\min} =
  \lceil \tfrac{3\tau - 1 - s}{p} \rceil - 1$, $k_{\max} = \lfloor
  \tfrac{7\tau - s}{p} \rfloor$, and $\Pts =
  \IntStrPoints{7\tau}{\PairsMinus{f}{H}{\tau}{\Text}}{\Text}$. Then:
  \begin{itemize}
  \item For every $k \in [1 \dd k_{\min}]$, it holds
    $\RangeCountTwoSide{\Pts}{s + kp}{\Textlen} \geq |\RMinusSix{f}{s}{k}{H}{\tau}{\Text}|$,
  \item For every $k \in (k_{\min} \dd k_{\max}]$, it holds
    $\RangeCountTwoSide{\Pts}{s + kp}{\Textlen} = |\RMinusSix{f}{s}{k}{H}{\tau}{\Text}|$.
  \end{itemize}
\end{lemma}
\begin{proof}
  By combining \cref{lm:RskH-size} and
  \cref{lm:sa-periodic-count}, it holds for $k \in [1 \dd k_{\min}]$ that:
  \begin{align*}
    |\RMinusSix{f}{s}{k}{H}{\tau}{\Text}|
      &\leq |\{j \in \RPrimMinusFour{f}{H}{\tau}{\Text} : s + kp \leq \RunEndFullPos{f}{\tau}{\Text}{j} - j\}|\\
      &=    \RangeCountTwoSide{\Pts}{s + kp}{\Textlen}.
  \end{align*} 
  The proof for $k \in (k_{\min} \dd k_{\max}]$ follows analogously,
  except the inequality in the first line is replaced with an
  equality. Note that \cref{lm:sa-periodic-count} requires that
  $s + kp \leq 7\tau$, which holds here, since $k \leq k_{\max}$
  implies $s + kp \leq s + \lfloor \tfrac{7\tau - s}{p} \rfloor p \leq
  7\tau$.
\end{proof}

\begin{lemma}\label{lm:elem-of-RskH}
  Let $\tau \in [1 \dd \lfloor \tfrac{\Textlen}{2} \rfloor]$
  and $f$ be any necklace-consistent function.
  Let $H \in \Sigma^{+}$, $p = |H|$, $s \in [0 \dd p)$, $k_{\min} =
  \lceil \tfrac{3\tau - 1 - s}{p} \rceil - 1$, $k_{\max} = \lfloor
  \tfrac{7\tau - s}{p} \rfloor$, and $\Pts =
  \IntStrPoints{7\tau}{\PairsMinus{f}{H}{\tau}{\Text}}{\Text}$.  Let $k \in (k_{\min} \dd
  k_{\max}]$ be such that $\RMinusSix{f}{s}{k}{H}{\tau}{\Text} \neq \emptyset$. Let also $x
  = s + kp$, $c = \RangeCountTwoSide{\Pts}{x}{\Textlen}$, and $c' \in [1 \dd
  |\RMinusSix{f}{s}{k}{H}{\tau}{\Text}|]$. Then:
  \begin{itemize}
  \item It holds $c' \in [1 \dd c]$, i.e., $\RangeSelect{\Pts}{x}{\Textlen}{c'}$
    is well-defined,
  \item Every position $j \in \RangeSelect{\Pts}{x}{\Textlen}{c'}$ satisfies $j
    - x \in \RMinusSix{f}{s}{k}{H}{\tau}{\Text}$.
  \end{itemize}
\end{lemma}
\begin{proof}

  By \cref{lm:RskH-size-2}, it holds
  $|\RMinusSix{f}{s}{k}{H}{\tau}{\Text}| = \RangeCountTwoSide{\Pts}{s + kp}{\Textlen} =
  \RangeCountTwoSide{\Pts}{x}{\Textlen} = c$.  Thus, $c' \in [1 \dd c]$, i.e.,
  $\RangeSelect{\Pts}{x}{\Textlen}{c'}$ is well-defined.

  Let us now consider any $j \in \RangeSelect{\Pts}{x}{\Textlen}{c'}$. Then, by
  definition (see \cref{sec:range-queries}), there exists
  $(x',y',w',\ell') \in \Pts$ such that $x \leq x' < \Textlen$ and $j =
  \ell'$.  Since $\Pts = \IntStrPoints{7\tau}{\PairsMinus{f}{H}{\tau}{\Text}}{\Text}$, it follows
  by \cref{def:int-str} that there exists $(q,h) \in \PairsMinus{f}{H}{\tau}{\Text}$ such
  that $x' = h$ and $j = \ell'$ satisfies $\Textinf[j - 7\tau \dd j) =
  \Textinf[q - 7\tau \dd q)$. Furthermore, by definition of $\PairsMinus{f}{H}{\tau}{\Text}$
  (see \cref{sec:sa-periodic-ds}), there exists $q' \in
  \CompRepr{14\tau}{\RMinusFour{f}{H}{\tau}{\Text}}{\Text} \subseteq \RMinusFour{f}{H}{\tau}{\Text}$ such that $q =
  \RunEndFullPos{f}{\tau}{\Text}{q'}$ and $h = \min(\RunEndFullPos{f}{\tau}{\Text}{q'} - \RunBeg{\tau}{\Text}{q'},
  7\tau)$. Denote $q'' = \RunBeg{\tau}{\Text}{q'}$.  Putting everything together we
  thus obtain that $x \leq x' = h = \min(q - q'', 7\tau) \leq q - q''$
  and $\Textinf[j - 7\tau \dd j + 7\tau) = \Textinf[q - 7\tau \dd q +
  7\tau)$. Note also that since by definition, it holds $[q'' \dd
  q']$, it follows by $q' \in \RMinusFour{f}{H}{\tau}{\Text}$ and \cref{lm:R-text-block}
  that $q'' \in \RMinusFour{f}{H}{\tau}{\Text}$ and $\RunEndFullPos{f}{\tau}{\Text}{q''} = \RunEndFullPos{f}{\tau}{\Text}{q'}$.
  Denote $q''' = q - x$. We will prove that $q''' \in
  \RMinusSix{f}{s}{k}{H}{\tau}{\Text}$. First, note that by the assumption $q - q'' \geq
  x$, we immediately obtain $q'' \leq q'''$. On the other hand, $k
  > k_{\min}$ implies that $\RunEndPos{\tau}{\Text}{q''} - q''' \geq \RunEndFullPos{f}{\tau}{\Text}{q''} -
  q''' = \RunEndFullPos{f}{\tau}{\Text}{q'} - q''' = q - q''' = x = s + kp \geq s +
  (k_{\min} + 1) \cdot p = s + \lceil \tfrac{3\tau - 1 - s}{p} \rceil p \geq
  s + \tfrac{3\tau - 1 - s}{p} \cdot p \geq 3\tau - 1$. Thus, it
  follows by $q''' \geq q''$ and
  \cref{lm:beg-end}\eqref{lm:beg-end-it-1}, that $[q'' \dd q''']
  \subseteq \RTwo{\tau}{\Text}$.  By \cref{lm:R-text-block}, we thus obtain $q''' \in
  \RMinusFour{f}{H}{\tau}{\Text}$ and $\RunEndFullPos{f}{\tau}{\Text}{q'''} = \RunEndFullPos{f}{\tau}{\Text}{q''} = \RunEndFullPos{f}{\tau}{\Text}{q'} =
  q$.  Thus, $\HeadPos{f}{\tau}{\Text}{q'''} = (\RunEndFullPos{f}{\tau}{\Text}{q'''} - q''') \bmod p = (q -
  q''') \bmod p = x \bmod p = (s + pk) \bmod p = s$ and $\ExpPos{f}{\tau}{\Text}{q''} =
  \lfloor \tfrac{\RunEndFullPos{f}{\tau}{\Text}{q'''} - q'''}{p} \rfloor = \lfloor \tfrac{q
  - q'''}{p} \rfloor = \lfloor \tfrac{x}{p} \rfloor = \lfloor \tfrac{s
  + pk}{p} \rfloor = k$.  We have thus proved $q''' \in
  \RMinusSix{f}{s}{k}{H}{\tau}{\Text}$. Recall now that $\Textinf[j - 7\tau \dd j + 7\tau) =
  \Textinf[q - 7\tau \dd q + 7\tau)$.  By $k \leq k_{\max}$, we have $x = s
  + kp \leq s + \lfloor \tfrac{7\tau - s}{p} \rfloor p \leq 7\tau$.
  Thus, we obtain $\Textinf[j - x \dd j) = \Textinf[q - x \dd
  q)$.  On the other hand, by the assumption, we have $\Textinf[j \dd j +
  \tau) = \Textinf[q \dd q + \tau)$.  Thus, letting $\ell' = x + \tau$,
  it holds $\Textinf[j - x \dd j - x + \ell') = \Textinf[q - x \dd q - x +
  \ell') = \Textinf[q''' \dd q''' + \ell')$.  Combining this with $j \in
  [1 \dd \Textlen]$ and $\RunEndPos{\tau}{\Text}{q'''} - q''' = (\RunEndPos{\tau}{\Text}{q'''} - \RunEndFullPos{f}{\tau}{\Text}{q'''})
  + (\RunEndFullPos{f}{\tau}{\Text}{q'''} - q''') < \tau + (\RunEndFullPos{f}{\tau}{\Text}{q'''} - q''') = \tau
  + (q - q''') = \tau + x = \ell'$, by \cref{lm:full-run-shifted}
  implies that $j - x \in \RTwo{\tau}{\Text}$, $\HeadPos{f}{\tau}{\Text}{j - x} = \HeadPos{f}{\tau}{\Text}{q'''}$,
  $\RootPos{f}{\tau}{\Text}{j - x} = \RootPos{f}{\tau}{\Text}{q'''}$,
  $\TypePos{f}{\tau}{j - x} = \TypePos{f}{\tau}{q'''}$, and
  $\ExpPos{f}{\tau}{\Text}{j - x} = \ExpPos{f}{\tau}{\Text}{q'''}$. Thus, $j - x \in \RMinusSix{f}{s}{k}{H}{\tau}{\Text}$.
\end{proof}

\begin{lemma}\label{lm:elem-of-RskH-2}
  Let $\tau \in [1 \dd \lfloor \tfrac{\Textlen}{2} \rfloor]$
  and $f$ be any necklace-consistent function.
  Let $H \in \Sigma^{+}$, $p = |H|$, $s \in [0 \dd p)$, $k_{\min} =
  \lceil \tfrac{3\tau - 1 - s}{p} \rceil - 1$, $k_{\max} = \lfloor
  \tfrac{7\tau - s}{p} \rfloor$, and $\Pts =
  \IntStrPoints{7\tau}{\PairsMinus{f}{H}{\tau}{\Text}}{\Text}$.  Let $k \in (k_{\min} \dd k_{\max}]$ be such that
  $\RMinusSix{f}{s}{k}{H}{\tau}{\Text} \neq \emptyset$. Let also $x = s + kp$. Then,
  $c = \RangeCountTwoSide{\Pts}{x}{\Textlen}$ satisfies $c \geq 1$ and
  every position $j \in \RangeSelect{\Pts}{x}{\Textlen}{1}$ satisfies $j
  - x \in \RMinusSix{f}{s}{k}{H}{\tau}{\Text}$.
\end{lemma}
\begin{proof}
  The result follows by \cref{lm:elem-of-RskH} with $c' = 1$.
\end{proof}

\begin{lemma}\label{lm:sa-periodic-min}
  Let $\tau \in [1 \dd \lfloor \tfrac{\Textlen}{2} \rfloor]$
  and $f$ be any necklace-consistent function.
  Let $H \in \Sigma^{+}$ and $\Pts = \IntStrPoints{7\tau}{\PairsMinus{f}{H}{\tau}{\Text}}{\Text}$
  (\cref{def:int-str}).  For any integer $x_l \in [0 \dd 7\tau]$ and any $y_l, y_u
  \in \Sigma^{*}$ such that $\RangeCountFourSide{\Pts}{x_l}{\Textlen}{y_l}{y_u} >
  0$, it holds $\RangeMinFourSide{\Pts}{x_l}{\Textlen}{y_l}{y_u} = \min \{\RunEndFullPos{f}{\tau}{\Text}{j} :
  j \in \RPrimMinusFour{f}{H}{\tau}{\Text}, x_l \leq \RunEndFullPos{f}{\tau}{\Text}{j} - j,\text{ and } y_l
  \preceq \Textinf[\RunEndFullPos{f}{\tau}{\Text}{j} \dd \RunEndFullPos{f}{\tau}{\Text}{j} + 7\tau) \prec y_u\}$.
\end{lemma}
\begin{proof}
  Denote $\Rcomp = \CompRepr{14\tau}{\RMinusFour{f}{H}{\tau}{\Text}}{\Text}$, and recall
  that by \cref{def:periodic-input}, we have $\PairsMinus{f}{H}{\tau}{\Text} =
  \{(\RunEndFullPos{f}{\tau}{\Text}{j}, \min(\RunEndFullPos{f}{\tau}{\Text}{j} - \RunBeg{\tau}{\Text}{j},
  7\tau)) : j \in \Rcomp\}$. The
  proof consists of two steps labeled (a)-(b).

  (a) Let $Q$ and $A$ be defined as in the proof of
  \cref{lm:sa-periodic-count}\eqref{lm:sa-periodic-count-it-1}, i.e.,
  $Q = \{\Textinf[\RunEndFullPos{f}{\tau}{\Text}{i} - 7\tau \dd \RunEndFullPos{f}{\tau}{\Text}{i} + 7\tau) : i \in
  \Rcomp, x_l \leq \RunEndFullPos{f}{\tau}{\Text}{i} - \RunBeg{\tau}{\Text}{i},\text{ and }y_l
  \preceq \Textinf[\RunEndFullPos{f}{\tau}{\Text}{i} \dd \RunEndFullPos{f}{\tau}{\Text}{i} + 7\tau) \prec y_u\}$
  and $A = \{j \in [1 \dd \Textlen] : \Textinf[j - 7\tau \dd j + 7\tau) \in
  Q\}$. Denote $\mathcal{R} = \{(x,y,w,\ell) \in \Pts : x_l \leq x <
  \Textlen\text{ and } y_l \preceq y \prec y_u\}$.  In the first step, we
  prove that $\RangeMinFourSide{\Pts}{x_l}{\Textlen}{y_l}{y_u} = \min A$.
  \begin{itemize}
  \item Let $i = \min A$. By definition of $A$, we then have $\Textinf[i
    - 7\tau \dd i + 7\tau) \in Q$. Denote $S = \Textinf[i - 7\tau \dd i +
    7\tau)$.  Recall from the proof of
    \cref{lm:sa-periodic-count}\eqref{lm:sa-periodic-count-it-1}, that
    the function $g: \mathcal{R} \rightarrow Q$ defined such that for
    every $p = (x, y, w, \ell) \in \mathcal{R}$, $g(p) = \Textinf[\ell -
    7\tau \dd \ell + 7\tau)$, is a bijection.  Let us thus consider
    $p_S = g^{-1}(S) \in \mathcal{R}$. Denote $p_S = (x_S, y_S, w_S,
    \ell_S)$. Observe, that we then have $\Textinf[\ell_S - 7\tau \dd
    \ell_S + 7\tau) = g(p_S) = S = \Textinf[i - 7\tau \dd i + 7\tau)$. By
    \cref{def:int-str}, it then holds $\ell_S = \min\{i' \in [1 \dd \Textlen]
    : \Textinf[i' - 7\tau \dd i' + 7\tau) = S\}$. Consequently, $\ell_S
    \leq i$ and thus $\RangeMinFourSide{\Pts}{x_l}{\Textlen}{y_l}{y_u} =
    \min_{(x,y,w,\ell) \in \mathcal{R}} \ell \leq \ell_S \leq i = \min
    A$.
  \item Let $p_{\min} = (x_{\min}, y_{\min}, w_{\min}, \ell_{\min})
    \in \mathcal{R}$ be such that $\RangeMinFourSide{\Pts}{x_l}{\Textlen}{y_l}{y_u} =
    \min_{(x,y,w,\ell) \in \mathcal{R}} \ell = \ell_{\min}$.  Let $S =
    g(p_{\min}) = \Textinf[\ell_{\min} - 7\tau \dd \ell_{\min} + 7\tau)$,
    where $g : \mathcal{R} \rightarrow Q$ is the bijection defined
    above. We then have $S \in Q$. Consequently, by $\ell_{\min} \in
    [1 \dd \Textlen]$ and $\Textinf[\ell_{\min} - 7\tau \dd \ell_{\min} + 7\tau)
    = S$, we have $\ell_{\min} \in A$ Thus, $\min A \leq \ell_{\min} =
    \RangeMinFourSide{\Pts}{x_l}{\Textlen}{y_l}{y_u}$.
  \end{itemize}

  (b) Let $A'$ be defined as in the proof of
  \cref{lm:sa-periodic-count}\eqref{lm:sa-periodic-count-it-1}, i.e.,
  $A' = \{\RunEndFullPos{f}{\tau}{\Text}{j} : j \in \RPrimMinusFour{f}{H}{\tau}{\Text}, x_l \leq \RunEndFullPos{f}{\tau}{\Text}{j} - j,
  \text{ and }y_l \preceq \Textinf[\RunEndFullPos{f}{\tau}{\Text}{j} \dd \RunEndFullPos{f}{\tau}{\Text}{j} + 7\tau)
  \prec y_u\}$. In the proof of
  \cref{lm:sa-periodic-count}\eqref{lm:sa-periodic-count-it-1}, we
  showed that $A = A'$. Combining this with
  $\RangeMinFourSide{\Pts}{x_l}{\Textlen}{y_l}{y_u} = \min A$ shown above, we obtain
  $\RangeMinFourSide{\Pts}{x_l}{\Textlen}{y_l}{y_u} = \min A = \min A' = \min
  \{\RunEndFullPos{f}{\tau}{\Text}{j} : j \in \RPrimMinusFour{f}{H}{\tau}{\Text}, x_l \leq \RunEndFullPos{f}{\tau}{\Text}{j} - j,
  \text{ and }y_l \preceq \Textinf[\RunEndFullPos{f}{\tau}{\Text}{j} \dd \RunEndFullPos{f}{\tau}{\Text}{j} + 7\tau)
  \prec y_u\}$, i.e., the claim.
\end{proof}

\begin{lemma}\label{lm:RskH-min}
  Let $\tau \in [1 \dd \lfloor \tfrac{\Textlen}{2} \rfloor]$
  and $f$ be any necklace-consistent function.
  Let $H \in \Sigma^{+}$, $p = |H|$, $s \in [0 \dd p)$, $\Pts =
  \IntStrPoints{7\tau}{\PairsMinus{f}{H}{\tau}{\Text}}{\Text}$, $k_{\min} = \lceil \tfrac{3\tau -
  1 - s}{p} \rceil - 1$, and $k_{\max} = \lfloor \tfrac{7\tau - s}{p}
  \rfloor$.  Let $k \in (k_{\min} \dd k_{\max}]$ be such that
  $\RMinusSix{f}{s}{k}{H}{\tau}{\Text} \neq \emptyset$. Then:
  \begin{enumerate}
  \item\label{lm:RskH-min-it-1} It holds $\min \RMinusSix{f}{s}{k}{H}{\tau}{\Text} =
    \RangeMinTwoSide{\Pts}{s + kp}{\Textlen} - s - kp$.
  \item\label{lm:RskH-min-it-2} The position $j = \min \RMinusSix{f}{s}{k}{H}{\tau}{\Text}$
    satisfies $j = \min \OccThree{4\ell}{j}{\Text}$.
  \end{enumerate}
\end{lemma}
\begin{proof}
  1. For every $k \in (k_{\min} \dd k_{\max}]$, denote $A_k :=
  \{\RunEndFullPos{f}{\tau}{\Text}{j} - s - kp : j \in \RPrimMinusFour{f}{H}{\tau}{\Text} \text{ and }s + kp \leq
  \RunEndFullPos{f}{\tau}{\Text}{j} - j\}$ and $B_k := \{\RunEndFullPos{f}{\tau}{\Text}{j} : j \in
  \RPrimMinusFour{f}{H}{\tau}{\Text}\text{ and }s + kp \leq \RunEndFullPos{f}{\tau}{\Text}{j} - j\}$. Immediately
  from the definition, we have $\min A_k = \min B_k - s - kp$. By
  \cref{lm:RskH}, we have $\RMinusSix{f}{s}{k}{H}{\tau}{\Text} = A_k$.  Next,
  observe that by \cref{lm:RskH-size-2}, we have $|\RMinusSix{f}{s}{k}{H}{\tau}{\Text}| =
  \RangeCountTwoSide{\Pts}{s + kp}{\Textlen}$. The assumption $\RMinusSix{f}{s}{k}{H}{\tau}{\Text} \neq
  \emptyset$ thus implies that $\RangeCountFourSide{\Pts}{s +
  kp}{\Textlen}{\emptystring}{c^{\infty}} = \RangeCountTwoSide{\Pts}{s + kp}{\Textlen} >
  0$ (where $c = \max\Sigma$).  Consequently, by
  \cref{lm:sa-periodic-min}, it holds $\min B_k = \RangeMinFourSide{\Pts}{s +
  kp}{\Textlen}{\emptystring}{c^{\infty}} = \RangeMinTwoSide{\Pts}{s + kp}{\Textlen}$
  (note that \cref{lm:sa-periodic-min} requires that $s + kp
  \leq 7\tau$, which holds here since $s + kp \leq s + \lfloor
  \tfrac{7\tau - s}{p} \rfloor \cdot p \leq 7\tau$).  Putting everything
  together, we thus obtain $\min \RMinusSix{f}{s}{k}{H}{\tau}{\Text} = \min A_k =
  \min B_k - s - kp = \RangeMinTwoSide{\Pts}{s + kp}{\Textlen} - s - kp$.

  2. First, note that $j \in \RMinusSix{f}{s}{k}{H}{\tau}{\Text}$ implies that $\RunEndPos{\tau}{\Text}{j} -
  j = s + kp \leq s + \lfloor \tfrac{7\tau - s}{p} \rfloor p \leq
  7\tau \leq 2\ell + \tau \leq 3\ell$.  Suppose that $j \neq \min
  \OccThree{4\ell}{j}{\Text}$. Then, there exists $j' \in [1 \dd j)$ such
  that $j' \in \OccThree{4\ell}{j}{\Text}$, or equivalently (by
  \cref{lm:occ-equivalence}) $\Textinf[j \dd j + 4\ell) = \Textinf[j' \dd j'
  + 4\ell)$. By \cref{lm:full-run-shifted} (with $\delta = 0$), we
  thus obtain $j' \in \RMinusSix{f}{s}{k}{H}{\tau}{\Text}$, contradicting the definition of
  $j$.
\end{proof}

\paragraph{Weighted Modular Constraint Queries}

\begin{lemma}\label{lm:sa-periodic-modcount}
  Let $\tau \in [1 \dd \lfloor \tfrac{\Textlen}{2} \rfloor]$
  and $f$ be any necklace-consistent function.
  Let $H \in \Sigma^{+}$, $p = |H|$, $s \in [0 \dd p)$, and $\Ints =
  \WInts{7\tau}{\PairsMinus{f}{H}{\tau}{\Text}}{\Text}$ (\cref{def:intervals,def:periodic-input}). Denote
  $k_{\min} = \lceil \tfrac{3\tau - 1 - s}{p} \rceil - 1$ and
  $k_{\max} = \lfloor \tfrac{7\tau - s}{p} \rfloor$. For every $k_1$
  and $k_2$ such that $k_{\min} \leq k_1 \leq k_2 \leq k_{\max}$, it
  holds
  \[
    \ModCountTwoSide{\Ints}{p}{s}{k_1}{k_2} =
    |\{j \in \RMinusFive{f}{s}{H}{\tau}{\Text} : \ExpPos{f}{\tau}{\Text}{j} \in (k_1 \dd k_2]\}|.
  \]
\end{lemma}
\begin{proof}
  Denote $\Rcomp = \CompRepr{14\tau}{\RMinusFour{f}{H}{\tau}{\Text}}{\Text}$. Recall
  that by \cref{def:periodic-input}, it holds 
  $\PairsMinus{f}{H}{\tau}{\Text} = \{(\RunEndFullPos{f}{\tau}{\Text}{j},
  \min(7\tau, \RunEndFullPos{f}{\tau}{\Text}{j} - \RunBeg{\tau}{\Text}{j})) : j \in \Rcomp\}$. Let $q \in
  (k_{\min} \dd k_{\max}]$. The proof consists of four steps labeled
  (a) through (d).

  (a) Denote
  \begin{align*}
    Q &= \{\Textinf[\RunEndFullPos{f}{\tau}{\Text}{i} - 7\tau \dd \RunEndFullPos{f}{\tau}{\Text}{i} + 7\tau) :
      i \in \Rcomp\text{ and }s + pq \leq \RunEndFullPos{f}{\tau}{\Text}{i} - \RunBeg{\tau}{\Text}{i}\},\\
    A &= \{j \in [1 \dd \Textlen] : \Textinf[j - 7\tau \dd j + 7\tau) \in Q\}.
  \end{align*}
  In the first step, we prove that it holds
  $\ModCountTwoSide{\Ints}{p}{s}{q-1}{q} = |A|$.  Let $\Ints' =
  \{(e,w,\ell) \in \Ints : s + pq \leq e\}$. We begin by proving that
  for every $u = (e_u,w_u,\ell_u) \in \Ints'$, it holds $\Textinf[\ell_u
  - 7\tau \dd \ell_u + 7\tau) \in Q$. By $u \in \Ints$, there exists
  $i_u \in \Rcomp$ such that it holds $e_u = \min(7\tau,
  \RunEndFullPos{f}{\tau}{\Text}{i_u} - \RunBeg{\tau}{\Text}{i_u})$, and $\ell_u$ satisfies $\Textinf[\ell_u
  - 7\tau \dd \ell_u + 7\tau) = \Textinf[\RunEndFullPos{f}{\tau}{\Text}{i_u} - 7\tau \dd
  \RunEndFullPos{f}{\tau}{\Text}{i_u} + 7\tau)$.
  By $u \in \Ints'$,
  we obtain $s + pq \leq e_u = \min(7\tau, \RunEndFullPos{f}{\tau}{\Text}{i_u} - \RunBeg{\tau}{\Text}{i_u})
  \leq \RunEndFullPos{f}{\tau}{\Text}{i_u} - \RunBeg{\tau}{\Text}{i_u}$.
  We thus have $\Textinf[\RunEndFullPos{f}{\tau}{\Text}{i_u} -
  7\tau \dd \RunEndFullPos{f}{\tau}{\Text}{i_u} + 7\tau) \in Q$. By $\Textinf[\RunEndFullPos{f}{\tau}{\Text}{i_u} -
  7\tau \dd \RunEndFullPos{f}{\tau}{\Text}{i_u} + 7\tau) = \Textinf[\ell_u - 7\tau \dd \ell_u
  + 7\tau)$, we thus obtain $\Textinf[\ell_u - 7\tau \dd \ell_u + 7\tau)
  \in Q$.

  Let $g : \Ints' \rightarrow Q$ be defined so that for every $u =
  (e,w,\ell) \in \Ints'$, we let $g(u) = \Textinf[\ell - 7\tau \dd \ell
  + 7\tau)$. We prove that $g$ is a bijection.
  \begin{itemize}
  \item Let $u_1, u_2 \in \Ints'$ be such that $g(u_1) = g(u_2)$. We
    show that $u_1 = u_2$. For $x \in \{1, 2\}$, denote $u_x = (e_x,
    w_x, \ell_x)$ and let $i_x \in \Rcomp$ be such that $e_x =
    \min(7\tau, \RunEndFullPos{f}{\tau}{\Text}{i_x} - \RunBeg{\tau}{\Text}{i_x})$ and $\Textinf[\ell_x -
    7\tau \dd \ell_x + 7\tau) = \Textinf[\RunEndFullPos{f}{\tau}{\Text}{i_x} - 7\tau \dd
    \RunEndFullPos{f}{\tau}{\Text}{i_x} + 7\tau)$. Let also $S_x := \Textinf[\ell_x - 7\tau
    \dd \ell_x + 7\tau) = \Textinf[\RunEndFullPos{f}{\tau}{\Text}{i_x} - 7\tau \dd
    \RunEndFullPos{f}{\tau}{\Text}{i_x} + 7\tau)$.  By the assumption, $S_1 = g(u_1) =
    g(u_2) = S_2$. This implies $\ell_1 = \min\{j \in [1 \dd \Textlen] :
    \Textinf[j - 7\tau \dd j + 7\tau) = S_1\} = \min\{j \in [1 \dd \Textlen] :
    \Textinf[j - 7\tau \dd j + 7\tau) = S_2\} = \ell_2$ and $w_1 = |\{j
    \in [1 \dd \Textlen] : \Textinf[j - 7\tau \dd j - 7\tau) = S_1\}| = |\{j \in
    [1 \dd \Textlen] : \Textinf[j - 7\tau \dd j + 7\tau) = S_2\}| = w_2$.
    Observe that due to $\Text[\Textlen] = \Textinf[0]$ being unique in $\Text$, for
    every $j \in \RFour{f}{H}{\tau}{\Text}$, letting $Z = \Textinf[\RunEndFullPos{f}{\tau}{\Text}{j} - 7\tau \dd
    \RunEndFullPos{f}{\tau}{\Text}{j} + 7\tau)$, it holds $\min(7\tau, \RunEndFullPos{f}{\tau}{\Text}{j} -
    \RunBeg{\tau}{\Text}{j}) = |H| + \lcs(Z[1 \dd 7\tau], Z[1 \dd 7\tau-|H|])$.
    Thus, $e_1 = \min(7\tau, \RunEndFullPos{f}{\tau}{\Text}{i_1} - \RunBeg{\tau}{\Text}{i_1}) = |H| +
    \lcs(S_1[1 \dd 7\tau], S_1[1 \dd 7\tau - |H|]) = |H| + \lcs(S_2[1
    \dd 7\tau], S_2[1 \dd 7\tau - |H|]) = \min(7\tau, \RunEndFullPos{f}{\tau}{\Text}{i_2} -
    \RunBeg{\tau}{\Text}{i_2}) = e_2$.  Consequently, $u_1 = u_2$.
  \item Let $X \in Q$. Then, there exits $i \in \Rcomp$ such that
    $s + pq \leq \RunEndFullPos{f}{\tau}{\Text}{i} - \RunBeg{\tau}{\Text}{i}$
    and $\Textinf[\RunEndFullPos{f}{\tau}{\Text}{i} - 7\tau \dd \RunEndFullPos{f}{\tau}{\Text}{i} + 7\tau) = X$.
    Note that $i \in \Rcomp$ implies that we have $(\RunEndFullPos{f}{\tau}{\Text}{i}, \min(7\tau,
    \RunEndFullPos{f}{\tau}{\Text}{i} - \RunBeg{\tau}{\Text}{i})) \in \PairsMinus{f}{H}{\tau}{\Text}$. Hence, there
    exists $u = (e, w, \ell) \in \Ints$ such that $e = \min(7\tau,
    \RunEndFullPos{f}{\tau}{\Text}{i} - \RunBeg{\tau}{\Text}{i})$ and $\Textinf[\ell - 7\tau \dd \ell +
    7\tau) = \Textinf[\RunEndFullPos{f}{\tau}{\Text}{i} - 7\tau \dd \RunEndFullPos{f}{\tau}{\Text}{i} + 7\tau) =
    X$.  Observe now that $s + pq \leq s + pk_{\max} = s + p \lfloor
    \tfrac{7\tau-s}{p} \rfloor \leq 7\tau$ and $s + pq \leq \RunEndFullPos{f}{\tau}{\Text}{i} - \RunBeg{\tau}{\Text}{i}$
    imply that $s + pq \leq \min(7\tau, \RunEndFullPos{f}{\tau}{\Text}{i} -
    \RunBeg{\tau}{\Text}{i})$. Thus, $s + pq \leq e$.  Consequently, it holds $(e, w,
    \ell) \in \Ints'$. Since, as noted earlier, we have $\Textinf[\ell -
    7\tau \dd \ell + 7\tau) = X$, it follows that $g(p) = \Textinf[\ell -
    7\tau \dd \ell + 7\tau) = X$.
  \end{itemize}
  We have thus proved that $g$ is a bijection. Observe now that:
  \begin{itemize}
  \item By definition, $\ModCountTwoSide{\Ints}{p}{s}{q-1}{q} =
    \sum_{(e,w,\ell) \in \Ints} w \cdot |\{j \in [0 \dd e] : j \bmod p
    = s\text{ and }q-1 < \lfloor \tfrac{j}{p} \rfloor \leq q\}| =
    \sum_{(e,w,\ell) \in \Ints} w \cdot |\{j \in [0 \dd e] : j \bmod p
    = s\text{ and } \lfloor \tfrac{j}{p} \rfloor = q\}| =
    \sum_{(e,w,\ell) \in \Ints} w \cdot |\{j \in [0 \dd e] : j = s +
    pq\}| = \sum_{(e,w,\ell) \in \Ints: s+pq \leq e} w =
    \sum_{(e,w,\ell) \in \Ints'} w$.
  \item By \cref{def:intervals}, we thus have
    $\ModCountTwoSide{\Ints}{p}{s}{q-1}{q} = \sum_{(e,w,\ell) \in
    \Ints'} w = \sum_{(e,w,\ell) \in \Ints'} |\{j \in [1 \dd \Textlen] :
    \Textinf[j-7\tau \dd j+7\tau) = \Textinf[\ell-7\tau \dd
    \ell+7\tau)\}|$. Since $g$ is a bijection, for any
    $(e_1,w_1,\ell_1), (e_2,w_2,\ell_2) \in \Ints'$, $(e_1,w_1,\ell_1)
    \neq (e_2,w_2,\ell_2)$ implies $\Textinf[\ell_1 - 7\tau \dd \ell_1 +
    7\tau) \neq \Textinf[\ell_2 - 7\tau \dd \ell_2 + 7\tau)$. Thus, in
    the above expression, no position $j \in [1 \dd \Textlen]$ is accounted
    twice (i.e., for different elements of $\Ints'$). On the other
    hand, $g$ being a bijection also implies that for every $j \in [1
    \dd \Textlen]$, $\Textinf[j-7\tau \dd j+7\tau) \in Q$ implies that there
    exists some $u = (e,w,\ell) \in \Ints'$ such that $\Textinf[j-7\tau
    \dd j+7\tau) = \Textinf[\ell-7\tau \dd \ell+7\tau)$. Consequently,
    $\ModCountTwoSide{\Ints}{p}{s}{q-1}{q} = \sum_{(e,w,\ell) \in
    \Ints'} |\{j \in [1 \dd \Textlen] : \Textinf[j-7\tau \dd j+7\tau) =
    \Textinf[\ell-7\tau \dd \ell+7\tau)\}| = |\{j \in [1 \dd \Textlen] :
    \Textinf[j-7\tau \dd j+7\tau) \in Q\}| = |A|$.
  \end{itemize}

  (b) Denote
  \begin{align*}
    Q' &= \{\Textinf[\RunEndFullPos{f}{\tau}{\Text}{i} - 7\tau \dd \RunEndFullPos{f}{\tau}{\Text}{i} + 7\tau) :
       i \in \RMinusFour{f}{H}{\tau}{\Text}\text{ and }s + pq \leq \RunEndFullPos{f}{\tau}{\Text}{i} - i\}.
  \end{align*}
  In the second step, we prove that $Q = Q'$.
  \begin{itemize}
  \item Let $X \in Q$. Then, there exists $i \in \Rcomp$ such that
    $s + pq \leq \RunEndFullPos{f}{\tau}{\Text}{i} - \RunBeg{\tau}{\Text}{i}$
    and $\Textinf[\RunEndFullPos{f}{\tau}{\Text}{i} - 7\tau \dd \RunEndFullPos{f}{\tau}{\Text}{i} + 7\tau) = X$.
    By $\Rcomp \subseteq \RMinusFour{f}{H}{\tau}{\Text}$, we have $i \in
    \RMinusFour{f}{H}{\tau}{\Text}$. Denote $i' =
    \RunBeg{\tau}{\Text}{i}$. By \cref{lm:R-text-block}, we have $i' \in
    \RMinusFour{f}{H}{\tau}{\Text}$ and $\RunEndFullPos{f}{\tau}{\Text}{i'} = \RunEndFullPos{f}{\tau}{\Text}{i}$.  Consequently,
    $s + pq \leq \RunEndFullPos{f}{\tau}{\Text}{i} - \RunBeg{\tau}{\Text}{i} = \RunEndFullPos{f}{\tau}{\Text}{i'} - i'$.
    By definition of
    $Q'$, $\Textinf[\RunEndFullPos{f}{\tau}{\Text}{i'} - 7\tau \dd \RunEndFullPos{f}{\tau}{\Text}{i'} + 7\tau) \in
    Q'$.  Recalling that $\Textinf[\RunEndFullPos{f}{\tau}{\Text}{i'} - 7\tau \dd
    \RunEndFullPos{f}{\tau}{\Text}{i'} + 7\tau) = \Textinf[\RunEndFullPos{f}{\tau}{\Text}{i} - 7\tau \dd
    \RunEndFullPos{f}{\tau}{\Text}{i} + 7\tau) = X$, we thus obtain $X \in Q'$.
  \item Let $X \in Q'$.  Then, there
    exists $i \in \RMinusFour{f}{H}{\tau}{\Text}$ such that $s + pq \leq \RunEndFullPos{f}{\tau}{\Text}{i} - i$
    and $X = \Textinf[\RunEndFullPos{f}{\tau}{\Text}{i} - 7\tau \dd
    \RunEndFullPos{f}{\tau}{\Text}{i} + 7\tau)$. Let $i' = \max(i, \RunEndFullPos{f}{\tau}{\Text}{i} -
    7\tau)$. Then, $[i \dd i'] \subseteq \RTwo{\tau}{\Text}$.
    By \cref{lm:R-text-block}, we thus have $i' \in \RMinusFour{f}{H}{\tau}{\Text}$
    and $\RunEndFullPos{f}{\tau}{\Text}{i'} = \RunEndFullPos{f}{\tau}{\Text}{i}$. By $s + pq \leq s + pk_{\max}
    = s + p \lfloor \tfrac{7\tau - s}{p} \rfloor \leq s + (7\tau - s) =
    7\tau$, we also have $\RunEndFullPos{f}{\tau}{\Text}{i'} - i' \in [s + pq \dd 7\tau]$.
    Let $j = \min\{t \in [1 \dd \Textlen] : \Textinf[t - 7\tau \dd t + 7\tau) =
    X\}$ and let $j'$ be such that $\RunEndFullPos{f}{\tau}{\Text}{i'} - i' = j - j'$. We
    prove the following properties of $j'$:
    \begin{itemize}[leftmargin=3.5ex]
    \item First, as in the proof of \cref{lm:sa-periodic-count}, we
      obtain that $j' \in \RMinusFour{f}{H}{\tau}{\Text}$ and $\RunEndFullPos{f}{\tau}{\Text}{j'} = j$.
    \item Second, we show that $s + pq \leq \RunEndFullPos{f}{\tau}{\Text}{j'} - j'$.
      First, recall that $\RunEndFullPos{f}{\tau}{\Text}{j'} = j$. On the other hand, above
      we observed that $\RunEndFullPos{f}{\tau}{\Text}{i'} - i' \geq s + pq$. Thus, we obtain
      $s + pq \leq \RunEndFullPos{f}{\tau}{\Text}{i'} - i' = j - j' = \RunEndFullPos{f}{\tau}{\Text}{j'} - j'$.
    \item Lastly, using the argument as in the proof of
      \cref{lm:sa-periodic-count}, we obtain $j' \in
      \Cover{14\tau}{\Text}$.
    \end{itemize}
    We have thus proved $j' \in \RMinusFour{f}{H}{\tau}{\Text} \cap \Cover{14\tau}{\Text} =
    \Rcomp$ and $s + pq \leq \RunEndFullPos{f}{\tau}{\Text}{j'} - j' \leq
    \RunEndFullPos{f}{\tau}{\Text}{j'} - \RunBeg{\tau}{\Text}{j'}$.
    Consequently, by definition of $Q$, $\Textinf[\RunEndFullPos{f}{\tau}{\Text}{j'} - 7\tau
    \dd \RunEndFullPos{f}{\tau}{\Text}{j'} + 7\tau) \in Q$. Therefore, by
    $\Textinf[\RunEndFullPos{f}{\tau}{\Text}{j'} - 7\tau \dd \RunEndFullPos{f}{\tau}{\Text}{j'} + 7\tau) = \Textinf[j
    - 7\tau \dd j + 7\tau) = X$, we have $X \in Q$.
  \end{itemize}

  (c) Denote
  \begin{align*}
    A' &= \{\RunEndFullPos{f}{\tau}{\Text}{j} : j \in \RPrimMinusFour{f}{H}{\tau}{\Text}\text{ and }
           s + pq \leq \RunEndFullPos{f}{\tau}{\Text}{j} - j\}.
  \end{align*}
  In the third step, we prove that $A = A'$.
  \begin{itemize}
  \item Let $j \in A$. Then, $j \in [1 \dd \Textlen]$ and $\Textinf[j - 7\tau
    \dd j + 7\tau) \in Q'$ (recall, that $Q = Q'$).  Thus, there
    exists $i \in \RMinusFour{f}{H}{\tau}{\Text}$ satisfying $s + pq \leq \RunEndFullPos{f}{\tau}{\Text}{i} - i$
    and $\Textinf[\RunEndFullPos{f}{\tau}{\Text}{i} -
    7\tau \dd \RunEndFullPos{f}{\tau}{\Text}{i} + 7\tau) = \Textinf[j - 7\tau \dd j +
    7\tau)$. Let $i' = \max(i, \RunEndFullPos{f}{\tau}{\Text}{i} - 7\tau)$.  It holds $[i
    \dd i'] \subseteq \RTwo{\tau}{\Text}$ and hence by \cref{lm:R-text-block},
    we have $i' \in \RMinusFour{f}{H}{\tau}{\Text}$ and $\RunEndFullPos{f}{\tau}{\Text}{i'} = \RunEndFullPos{f}{\tau}{\Text}{i}$. By
    $s + pq \leq 7\tau$, we also have $\RunEndFullPos{f}{\tau}{\Text}{i'} - i' \in [s + pq
    \dd 7\tau]$. Let $j'$ be such that $\RunEndFullPos{f}{\tau}{\Text}{i'} - i' = j -
    j'$. By the same argument as in the second step of this proof, we
    then have $j' \in \RMinusFour{f}{H}{\tau}{\Text}$ and $\RunEndFullPos{f}{\tau}{\Text}{j'} = j$.  Let $j'' =
    \RunBeg{\tau}{\Text}{j'}$. By definition and \cref{lm:R-text-block}, we
    have $j'' \in \RPrimMinusFour{f}{H}{\tau}{\Text}$ and
    $\RunEndFullPos{f}{\tau}{\Text}{j''} = \RunEndFullPos{f}{\tau}{\Text}{j'} = j$. Observe now that by $j'' =
    \RunBeg{\tau}{\Text}{j'}$, we have $j'' \leq j'$. Thus,
    $s + pq \leq \RunEndFullPos{f}{\tau}{\Text}{i'} - i' = j - j' \leq j - j''= \RunEndFullPos{f}{\tau}{\Text}{j''} - j''$.
    We have thus proved that $j \in A'$ (with $j''$ as the position in
    $\RPrimMinusFour{f}{H}{\tau}{\Text}$ satisfying $\RunEndFullPos{f}{\tau}{\Text}{j''} = j$).
  \item Let $j \in A'$. Then, there exists $j' \in \RPrimMinusFour{f}{H}{\tau}{\Text}$ such
    that $s + pq \leq \RunEndFullPos{f}{\tau}{\Text}{j'} - j'$
    and $\RunEndFullPos{f}{\tau}{\Text}{j'} = j$. To show $j \in A$, we need to prove that
    $\Textinf[j - 7\tau \dd j + 7\tau) \in Q'$ (recall that $Q = Q'$),
    which in turn requires showing that there exists $t \in
    \RMinusFour{f}{H}{\tau}{\Text}$ such that $s + pq \leq \RunEndFullPos{f}{\tau}{\Text}{t} - t$
    and $\Textinf[\RunEndFullPos{f}{\tau}{\Text}{t} - 7\tau \dd \RunEndFullPos{f}{\tau}{\Text}{t} +
    7\tau) = \Textinf[j - 7\tau \dd j + 7\tau)$. Let $t = j'$. Then, we
    indeed have $t \in \RPrimMinusFour{f}{H}{\tau}{\Text} \subseteq \RMinusFour{f}{H}{\tau}{\Text}$,
    $s + pq \leq \RunEndFullPos{f}{\tau}{\Text}{t} - t$, and
    $\Textinf[\RunEndFullPos{f}{\tau}{\Text}{t} - 7\tau \dd \RunEndFullPos{f}{\tau}{\Text}{t} + 7\tau) =
    \Textinf[\RunEndFullPos{f}{\tau}{\Text}{j'} - 7\tau \dd \RunEndFullPos{f}{\tau}{\Text}{j'} + 7\tau) = \Textinf[j -
    7\tau \dd j + 7\tau)$. Hence, $j \in A$.
  \end{itemize}

  (d) We now put everything together. By the first three steps, it
  holds $\ModCountTwoSide{\Ints}{p}{s}{q-1}{q} = |A| = |A'|$. On the other
  hand, since adding a fixed value to every element of a set does not change its cardinality,
  it follows by \cref{lm:RskH}\eqref{lm:RskH-it-2}, that $|A'| = |\RMinusSix{f}{s}{q}{H}{\tau}{\Text}|$. By
  \cref{lm:mod-queries-properties}\eqref{lm:mod-queries-properties-it-2},
  we thus have:
  \begin{align*}
    \ModCountTwoSide{\Ints}{p}{s}{k_1}{k_2}
      &= \textstyle\sum_{q=k_1+1}^{k_2}
           \ModCountTwoSide{\Ints}{p}{s}{q-1}{q}\\
      &= \textstyle\sum_{q=k_1+1}^{k_2}
           |\RMinusSix{f}{s}{q}{H}{\tau}{\Text}|\\
      &= |\{j \in \RMinusFive{f}{s}{H}{\tau}{\Text} : \ExpPos{f}{\tau}{\Text}{j} \in (k_1 \dd k_2]\}|.
      \qedhere
  \end{align*}
\end{proof}

\subsubsection{Basic Navigation Primitives}\label{sec:sa-periodic-nav}

\begin{proposition}\label{pr:sa-periodic-nav-head-root}
  Let $k \in [4 \dd \lceil \log \Textlen \rceil)$, $\ell = 2^k$, $\tau
  = \lfloor \tfrac{\ell}{3} \rfloor$, and $f = f_{\tau,\Text}$
  (\cref{def:canonical-function}).  Given $\CompSaPeriodic{\Text}$, the
  value $k$, and any $j \in \RTwo{\tau}{\Text}$ such that $j
  = \min \OccThree{2\ell}{j}{\Text}$, we can in $\bigO(\log \Textlen)$ time compute
  $\HeadPos{f}{\tau}{\Text}{j}$ and $|\RootPos{f}{\tau}{\Text}{j}|$.
\end{proposition}
\begin{proof}
  Denote $\Rcomp := \CompRepr{14\tau}{\RTwo{\tau}{\Text}}{\Text}
  = \RTwo{\tau}{\Text} \cap \Cover{14\tau}{\Text}$ (\cref{def:comp}). Observe
  that $14\tau \geq 2\ell$ (following from $\tau
  = \lfloor \tfrac{\ell}{3} \rfloor$ and $\ell \geq 16$). Thus,
  by \cref{lm:cover}, $j
  = \min \OccThree{2\ell}{j}{\Text}$ implies that $j \in \Cover{14\tau}{\Text}$.
  Thus, $j \in \Rcomp$. Consequently, there exists $i \in
  [1 \dd n_{{\rm runs},k}]$ (\cref{sec:sa-core-ds}) such that, letting
  $\ArrRuns{k}[i] = (p_i, t_i)$,
  it holds $p_i \leq j < p_i + t_i$.  Using binary search, we compute
  the index $i$ in $\bigO(\log \Textlen)$ time.  We then retrieve
  $\ArrRoot{k}[i] = (s, p) = (\HeadPos{f}{\tau}{\Text}{p_i},
  |\RootPos{f}{\tau}{\Text}{p_i}|)$ (\cref{sec:sa-periodic-ds}).
  Observe now that by $[p_i \dd p_i +
  t_i) \subseteq \RTwo{\tau}{\Text}$ and the uniqueness of
  run-decomposition, the value $\HeadPos{f}{\tau}{\Text}{p_i}$
  determines $\HeadPos{f}{\tau}{\Text}{p_i + \delta}$ for every
  $\delta \in [0 \dd t_i)$.  More precisely,
  $\HeadPos{f}{\tau}{\Text}{p_i + \delta} = (s - \delta) \bmod
  p$. Therefore, we have $\HeadPos{f}{\tau}{\Text}{j} = (s - j +
  p_i) \bmod p$.  On the other hand, by \cref{lm:R-text-block}, we
  have $|\RootPos{f}{\tau}{\Text}{j}| = |\RootPos{f}{\tau}{\Text}{p_i}| =
  p$.
\end{proof}

\begin{proposition}\label{pr:sa-periodic-nav-run-end}
  Let $k \in [4 \dd \lceil \log \Textlen \rceil)$, $\ell = 2^k$, $\tau
  = \lfloor \tfrac{\ell}{3} \rfloor$, $f = f_{\tau,\Text}$
  (\cref{def:canonical-function}), and $j \in \RTwo{\tau}{\Text}$.  Given
  $\CompSaPeriodic{\Text}$ and the values $k$, $j$,
  $\HeadPos{f}{\tau}{\Text}{j}$, and $|\RootPos{f}{\tau}{\Text}{j}|$,
  we can in $\bigO(\log \Textlen)$ time compute:
  \begin{itemize}
  \item $\RunEndPos{\tau}{\Text}{j}$,
  \item $\ExpPos{f}{\tau}{\Text}{j}$,
  \item $\RunEndFullPos{f}{\tau}{\Text}{j}$,
  \item $\ExpCutPos{f}{\tau}{\Text}{j}{\ell}$,
  \item $\ExpCutPos{f}{\tau}{\Text}{j}{2\ell}$.
  \end{itemize}
\end{proposition}
\begin{proof}
  Denote $s = \HeadPos{f}{\tau}{\Text}{j}$, $H
  = \RootPos{f}{\tau}{\Text}{j}$, and $p = |H|$. Recall that using
  $\CompSaCore{\Text}$ (which is part of $\CompSaPeriodic{\Text}$),
  we can answer $\LCE_{\Text}$ queries in $\bigO(\log
  \Textlen)$ time.  All values are thus consecutively computed as follows:
  \begin{itemize}[itemsep=2pt]
  \item $t := \RunEndPos{\tau}{\Text}{j} = j + p + \LCE_{\Text}(j, j + p)$,
  \item $k := \ExpPos{f}{\tau}{\Text}{j} = \lfloor \tfrac{\RunEndPos{\tau}{\Text}{j} - j -
    \HeadPos{f}{\tau}{\Text}{j}}{|\RootPos{f}{\tau}{\Text}{j}|} \rfloor =
    \lfloor \tfrac{t - j - s}{p} \rfloor$,
  \item $t' := \RunEndFullPos{f}{\tau}{\Text}{j} =
    j + \ExpPos{f}{\tau}{\Text}{j} \cdot |\RootPos{f}{\tau}{\Text}{j}| =
    j + kp$,
  \item $k_1 := \ExpCutPos{f}{\tau}{\Text}{j}{\ell}
    = \min(\ExpPos{f}{\tau}{\Text}{j}, \lfloor \tfrac{\ell
    - \HeadPos{f}{\tau}{\Text}{j}}{|\RootPos{f}{\tau}{\Text}{j}|} \rfloor)
    = \min(k, \lfloor \tfrac{\ell - s}{p} \rfloor)$,
  \item $k_2 := \ExpCutPos{f}{\tau}{\Text}{j}{2\ell} =
    \min(k, \lfloor \tfrac{2\ell - s}{p} \rfloor)$.
  \end{itemize}
  In total, the query takes $\bigO(\log \Textlen)$ time.
\end{proof}

\begin{proposition}\label{pr:sa-periodic-pts-access}
  Let $k \in [4 \dd \lceil \log \Textlen \rceil)$, $\ell = 2^k$, $\tau
  = \lfloor \tfrac{\ell}{3} \rfloor$, and $f = f_{\tau,\Text}$
  (\cref{def:canonical-function}).  Given $\CompSaPeriodic{\Text}$, the
  value $k$, and any $j \in \RTwo{\tau}{\Text}$ such that $j
  = \min \OccThree{2\ell}{j}{\Text}$, we can in $\bigO(\log \Textlen)$ time
  determine if $\PairsMinus{f}{H}{\tau}{\Text} \neq \emptyset$ (where $H
  = \RootPos{f}{\tau}{\Text}{j}$), and if so, return the pointer to the
  data structure from \cref{pr:int-str} for weighted range queries on
  $\IntStrPoints{7\tau}{\PairsMinus{f}{H}{\tau}{\Text}}{\Text}$ (\cref{def:int-str,def:periodic-input}).
\end{proposition}
\begin{proof}
  First, recall that $2\ell \leq
  14\tau$. By \cref{lm:cover-equivalence} and \cref{lm:cover}, this
  implies that $j \in \CompRepr{14\tau}{\RTwo{\tau}{\Text}}{\Text}$.
  Consequently, there exists $i \in [1 \dd n_{{\rm runs},k}]$
  (\cref{sec:sa-core-ds}) such that, letting $\ArrRuns{k}[i] =
  (p_i, t_i)$, it holds $p_i \leq j < p_i + t_i$. Using binary search
  over the array $\ArrRuns{k}[1 \dd n_{{\rm runs},k}]$ (stored as
  part of $\CompSaCore{\Text}$; see \cref{sec:sa-core-ds}), we compute
  the index $i$ in $\bigO(\log \Textlen)$ time.  We then have $[p_i \dd
  j] \subseteq \RTwo{\tau}{\Text}$, and hence by \cref{lm:R-text-block},
  $\RootPos{f}{\tau}{\Text}{p_i} = \RootPos{f}{\tau}{\Text}{j} = H$. We
  then retrieve the pointer $\mu_{H}$ to the structure
  from \cref{pr:int-str} for $\PairsMinus{f}{H}{\tau}{\Text}$ (i.e., answering
  range queries on $\IntStrPoints{7\tau}{\PairsMinus{f}{H}{\tau}{\Text}}{\Text}$) from
  $\ArrPtrFirst{k}[i]$ (\cref{sec:sa-periodic-ds}).  If $\mu_{H}$ is a
  null pointer (note that this is possible, since we did not assume
  $\TypePos{\tau}{\Text}{j}$), then we return that $\PairsMinus{f}{H}{\tau}{\Text}
  = \emptyset$. Otherwise, we return $\mu_{H}$ as the answer.
\end{proof}

\begin{proposition}\label{pr:sa-periodic-ints-access}
  Let $k \in [4 \dd \lceil \log \Textlen \rceil)$, $\ell = 2^k$, $\tau
  = \lfloor \tfrac{\ell}{3} \rfloor$, and $f = f_{\tau,\Text}$
  (\cref{def:canonical-function}).  Given $\CompSaPeriodic{\Text}$, the
  value $k$, and any $j \in \RTwo{\tau}{\Text}$ such that $j
  = \min \OccThree{2\ell}{j}{\Text}$, we can in $\bigO(\log \Textlen)$ time
  determine if $\PairsMinus{f}{H}{\tau}{\Text} \neq \emptyset$ (where $H
  = \RootPos{f}{\tau}{\Text}{j}$), and if so, return the pointer to the
  data structure from \cref{pr:mod-queries} for modular constraint
  queries on $\WInts{7\tau}{\PairsMinus{f}{H}{\tau}{\Text}}{\Text}$
  (\cref{def:intervals,def:periodic-input}).
\end{proposition}
\begin{proof}
  Similarly as in the proof of \cref{pr:sa-periodic-pts-access}, using
  binary search over $\ArrRuns{k}[1 \dd n_{{\rm runs},k}]$, we
  first compute $i \in [1 \dd n_{{\rm runs},k}]$ such that, letting
  $\ArrRuns{k}[i] = (p_i, t_i)$, it holds $p_i \leq j < p_i + t_i$. We
  then have $[p_i \dd j] \subseteq \RTwo{\tau}{\Text}$, and hence
  by \cref{lm:R-text-block}, $\RootPos{f}{\tau}{\Text}{p_i}
  = \RootPos{f}{\tau}{\Text}{j} = H$.  We then retrieve the pointer
  $\mu_{H}$ to the structure from \cref{pr:mod-queries} for
  $\PairsMinus{f}{H}{\tau}{\Text}$ (i.e., answering weighted modular constraint
  queries on $\WInts{7\tau}{\PairsMinus{f}{H}{\tau}{\Text}}{\Text}$) from
  $\ArrPtrSecond{k}[i]$ (\cref{sec:sa-periodic-ds}).  If $\mu_H$ is
  a null pointer, we return that $\PairsMinus{f}{H}{\tau}{\Text}
  = \emptyset$. Otherwise, we return $\mu_{H}$ as the answer.
\end{proof}

\subsubsection{Computing the Size of Poslow
  and Poshigh}\label{sec:sa-periodic-poslow-poshigh}

\paragraph{Combinatorial Properties}

\begin{lemma}\label{lm:poslow-poshigh-run}
  Let $\ell \in [16 \dd \Textlen)$, $\tau = \lfloor \tfrac{\ell}{3} \rfloor$,
  and $f$ be any necklace-consistent function.
  Let $\Pat \in \Sigma^{m}$ be a $\tau$-periodic pattern such that
  $\TypePat{\tau}{\Pat} = -1$ and $\Text[\Textlen]$ does not occur in $\Pat[1 \dd m)$.  Let
  $i \in \RTwo{\tau}{\Text}$. Denote $H = \RootPat{f}{\tau}{\Pat}$ and $t = \RunEndPos{\tau}{\Text}{i} - i - 3\tau
  + 2$. Then, $|\PosLowMinus{f}{\Pat}{\Text} \cap [i \dd i + t)| \leq 1$ and
  $|\PosHighMinus{f}{\Pat}{\Text} \cap [i \dd i + t)| \leq 1$. Moreover,
  $|\PosLowMinus{f}{\Pat}{\Text} \cap [i \dd i + t)| = 1$ (resp.\ $|\PosHighMinus{f}{\Pat}{\Text}
  \cap [i \dd i + t)| = 1$) holds if and only if
  \begin{itemize}
  \item $i \in \RMinusFour{f}{H}{\tau}{\Text}$,
  \item $\RunEndFullPos{f}{\tau}{\Text}{i} - i \geq \RunEndCutPat{f}{\tau}{\Pat}{\ell} - 1$
    (resp.\ $\RunEndFullPos{f}{\tau}{\Text}{i} - i \geq \RunEndCutPat{f}{\tau}{\Pat}{2\ell} - 1$) and
  \item $\Pat[\RunEndCutPat{f}{\tau}{\Pat}{\ell} \dd \min(m,\ell)] \preceq
    \Textinf[\RunEndFullPos{f}{\tau}{\Text}{i} \dd \RunEndFullPos{f}{\tau}{\Text}{i} + 7\tau)$\\ (resp.\
    $\Pat[\RunEndCutPat{f}{\tau}{\Pat}{2\ell} \dd \min(m, 2\ell)] \preceq
    \Textinf[\RunEndFullPos{f}{\tau}{\Text}{i} \dd \RunEndFullPos{f}{\tau}{\Text}{i} + 7\tau)$).
  \end{itemize}
  Lastly, if $\PosLowMinus{f}{\Pat}{\Text} \cap [i \dd i + t) \neq \emptyset$
  (resp.\ $\PosHighMinus{f}{\Pat}{\Text} \cap [i \dd i + t) \neq \emptyset$), then
  $\PosLowMinus{f}{\Pat}{\Text} \cap [i \dd i + t) = \{\RunEndFullPos{f}{\tau}{\Text}{i} -
  (\RunEndCutPat{f}{\tau}{\Pat}{\ell} - 1)\}$ (resp.\ $\PosHighMinus{f}{\Pat}{\Text} \cap [i \dd i + t)
  = \{\RunEndFullPos{f}{\tau}{\Text}{i} - (\RunEndCutPat{f}{\tau}{\Pat}{2\ell} - 1)\}$).
\end{lemma}
\begin{proof}

  Below we only prove the lemma for $\PosLowMinus{f}{\Pat}{\Text}$. The version for
  $\PosHighMinus{f}{\Pat}{\Text}$ is identical, except we replace $\RunEndCutPat{f}{\tau}{\Pat}{\ell}$,
  $\ExpCutPat{f}{\tau}{\Pat}{\ell}$, and $\ell$ with $\RunEndCutPat{f}{\tau}{\Pat}{2\ell}$,
  $\ExpCutPat{f}{\tau}{\Pat}{2\ell}$, and $2\ell$, respectively. Note that below
  we also use the fact that $\ell \leq 7\tau$. Note that we also have
  $2\ell \leq 7\tau$.

  Denote $s = \HeadPat{f}{\tau}{\Pat}$ and $k_1 = \ExpCutPat{f}{\tau}{\Pat}{\ell} =
  \min(\ExpPat{f}{\tau}{\Pat}, \lfloor \tfrac{\ell - s}{|H|} \rfloor)$. First,
  note that by definition we have $\RunEndPos{\tau}{\Text}{i} - i \geq 3\tau - 1$. Thus,
  $t > 0$. By \cref{lm:beg-end}\eqref{lm:beg-end-it-1}, it holds $[i
  \dd \RunEndPos{\tau}{\Text}{i} - 3\tau + 1] = [i \dd i + t) \subseteq \RTwo{\tau}{\Text}$. From
  \cref{lm:R-text-block}, we thus obtain that for every $\delta \in [0
  \dd t)$, it holds $\RunEndFullPos{f}{\tau}{\Text}{i + \delta} = \RunEndFullPos{f}{\tau}{\Text}{i}$, which
  implies $\RunEndFullPos{f}{\tau}{\Text}{i + \delta} - (i + \delta) = \RunEndFullPos{f}{\tau}{\Text}{i} - i -
  \delta$. Consider any $j \in \PosLowMinus{f}{\Pat}{\Text}$. By definition, we then
  have $j \in \RMinusSix{f}{s}{k_1}{H}{\tau}{\Text}$. Thus, $\RunEndFullPos{f}{\tau}{\Text}{j} - j = s + k_1 p =
  \RunEndCutPat{f}{\tau}{\Pat}{\ell} - 1$. Consequently, $i + \delta \in \PosLowMinus{f}{\Pat}{\Text}$
  implies $\RunEndFullPos{f}{\tau}{\Text}{i + \delta} - (i + \delta) = \RunEndFullPos{f}{\tau}{\Text}{i} - i -
  \delta = \RunEndCutPat{f}{\tau}{\Pat}{\ell} - 1$, i.e., $\delta = (\RunEndFullPos{f}{\tau}{\Text}{i} - i) -
  (\RunEndCutPat{f}{\tau}{\Pat}{\ell} - 1)$, and hence $|\PosLowMinus{f}{\Pat}{\Text} \cap [i \dd i +
  t)| \leq 1$.

  We now prove the equivalence. Let us first assume $|\PosLowMinus{f}{\Pat}{\Text}
  \cap [i \dd i + t)| = 1$, i.e., that for some $\delta \in [0 \dd
  t)$, it holds $i + \delta \in \PosLowMinus{f}{\Pat}{\Text}$. As noted above, this
  implies $i + \delta \in \RMinusSix{f}{s}{k_1}{H}{\tau}{\Text}$.  By $[i \dd i + \delta]
  \subseteq \RTwo{\tau}{\Text}$ and \cref{lm:R-text-block}, we thus have $i \in
  \RMinusFour{f}{H}{\tau}{\Text}$, i.e., the first condition. Next, recall from the above
  that $i + \delta \in \PosLowMinus{f}{\Pat}{\Text}$ implies $\RunEndCutPat{f}{\tau}{\Pat}{\ell} - 1 =
  \RunEndFullPos{f}{\tau}{\Text}{i} - i - \delta \leq \RunEndFullPos{f}{\tau}{\Text}{i} - i$. This establishes
  the second condition. We now show the third condition. By $k_1 \leq
  \ExpPat{f}{\tau}{\Pat}$, it holds $\Pat[1 \dd \RunEndCutPat{f}{\tau}{\Pat}{\ell}) = H' H^{k_1}$,
  where $H'$ is a length-$s$ suffix of $H$. On the other hand, by
  definition, $i + \delta \in \RMinusSix{f}{s}{k_1}{H}{\tau}{\Text}$ implies $\Text[i + \delta
  \dd \RunEndFullPos{f}{\tau}{\Text}{i + \delta}) = \Text[i + \delta \dd \RunEndFullPos{f}{\tau}{\Text}{i}) = H'
  H^{k_1}$. Therefore, the assumption $\Text[i + \delta \dd \Textlen] \succeq
  \Pat$ or $\lcp(\Pat, \Text[i + \delta \dd \Textlen]) \geq \ell$ following from
  $i + \delta \in \PosLowMinus{f}{\Pat}{\Text}$ is equivalent to $\Text[\RunEndFullPos{f}{\tau}{\Text}{i}
  \dd \Textlen] \succeq \Pat[\RunEndCutPat{f}{\tau}{\Pat}{\ell} \dd m]$ or
  $\lcp(\Pat[\RunEndCutPat{f}{\tau}{\Pat}{\ell} \dd m], \Text[\RunEndFullPos{f}{\tau}{\Text}{i} \dd \Textlen]) \geq \ell
  - (\RunEndCutPat{f}{\tau}{\Pat}{\ell} - 1)$.  Consider now two cases:
  \begin{itemize}[leftmargin=3.8ex]
  \item First, assume $\lcp(\Pat[\RunEndCutPat{f}{\tau}{\Pat}{\ell} \dd m],
    \Text[\RunEndFullPos{f}{\tau}{\Text}{i} \dd \Textlen]) \geq \ell - (\RunEndCutPat{f}{\tau}{\Pat}{\ell} - 1)$.  Then,
    $m - \RunEndCutPat{f}{\tau}{\Pat}{\ell} + 1 \geq \ell - (\RunEndCutPat{f}{\tau}{\Pat}{\ell} - 1)$, or
    equivalently $m \geq \ell$.  By $\ell \leq 7\tau$, we thus
    have $\Textinf[\RunEndFullPos{f}{\tau}{\Text}{i} \dd \RunEndFullPos{f}{\tau}{\Text}{i} + 7\tau) \allowbreak \succeq
    \Textinf[\RunEndFullPos{f}{\tau}{\Text}{i} \dd \RunEndFullPos{f}{\tau}{\Text}{i} + \ell - (\RunEndCutPat{f}{\tau}{\Pat}{\ell} -
    1)) = \Pat[\RunEndCutPat{f}{\tau}{\Pat}{\ell} \dd \ell] = \Pat[\RunEndCutPat{f}{\tau}{\Pat}{\ell} \dd
    \min(\ell,m)]$.
  \item Let us now assume $\Text[\RunEndFullPos{f}{\tau}{\Text}{i} \dd \Textlen] \succeq
    \Pat[\RunEndCutPat{f}{\tau}{\Pat}{\ell} \dd m]$. This implies that for every $q \geq
    0$, it holds $\Textinf[\RunEndFullPos{f}{\tau}{\Text}{i} \dd \RunEndFullPos{f}{\tau}{\Text}{i} + q] \succeq
    \Pat[\RunEndCutPat{f}{\tau}{\Pat}{\ell} \dd \min(m, \RunEndCutPat{f}{\tau}{\Pat}{\ell} + q)]$.  In
    particular, for $q = 7\tau$, we have $\Textinf[\RunEndFullPos{f}{\tau}{\Text}{i} \dd
    \RunEndFullPos{f}{\tau}{\Text}{i} + 7\tau) \succeq \Pat[\RunEndCutPat{f}{\tau}{\Pat}{\ell} \dd \min(m,
    \RunEndCutPat{f}{\tau}{\Pat}{\ell} + 7\tau)] \succeq \Pat[\RunEndCutPat{f}{\tau}{\Pat}{\ell} \dd \min(m,
    \ell)]$, where the last inequality follows by $\RunEndCutPat{f}{\tau}{\Pat}{\ell} +
    7\tau \geq \ell$.
  \end{itemize}

  To prove the opposite implication, consider any $i \in \RTwo{\tau}{\Text}$ and
  assume that it holds $i \in \RMinusFour{f}{H}{\tau}{\Text}$, $\RunEndFullPos{f}{\tau}{\Text}{i} - i \geq
  \RunEndCutPat{f}{\tau}{\Pat}{\ell} - 1$, and $\Textinf[\RunEndFullPos{f}{\tau}{\Text}{i} \dd \RunEndFullPos{f}{\tau}{\Text}{i} +
  7\tau) \succeq \Pat[\RunEndCutPat{f}{\tau}{\Pat}{\ell} \dd \min(m, \ell)]$. Let $\delta
  = (\RunEndFullPos{f}{\tau}{\Text}{i} - i) - (\RunEndCutPat{f}{\tau}{\Pat}{\ell} - 1)$.  We will prove that
  $\delta \in [0 \dd t)$ and $i + \delta \in \PosLowMinus{f}{\Pat}{\Text}$. The
  inequality $\delta \geq 0$ follows from the definition of $\delta$
  and our assumptions. To show $\delta < t$, we consider two cases:
  \begin{itemize}
  \item First, let us assume that it holds $\RunEndPat{\tau}{\Pat} - 1 < \ell$.
    Then, $k_1 = \ExpCutPat{f}{\tau}{\Pat}{\ell} = \min(\ExpPat{f}{\tau}{\Pat}, \lfloor
    \tfrac{\ell - s}{|H|} \rfloor) = \min(\lfloor \tfrac{\RunEndPat{\tau}{\Pat} -
    1 - s}{|H|} \rfloor, \lfloor \tfrac{\ell - s}{|H|} \rfloor) =
    \lfloor \tfrac{\RunEndPat{\tau}{\Pat} - 1 - s}{|H} \rfloor = \ExpPat{f}{\tau}{\Pat}$.
    Thus, $\RunEndCutPat{f}{\tau}{\Pat}{\ell} = 1 + s + k_1|H| = 1 + s + \ExpPat{f}{\tau}{\Pat}|H| =
    \RunEndFullPat{f}{\tau}{\Pat}$.  Denote $q := \TailPat{f}{\tau}{\Pat} = \RunEndPat{\tau}{\Pat} -
    \RunEndFullPat{f}{\tau}{\Pat}$ and $q' := \TailPos{f}{\tau}{\Text}{i} = \RunEndPos{\tau}{\Text}{i} -
    \RunEndFullPos{f}{\tau}{\Text}{i}$. We show that $q' \geq q$. Suppose that $q' < q$.
    Note that by definition, $\Pat[\RunEndFullPat{f}{\tau}{\Pat} \dd \RunEndPat{\tau}{\Pat}) =
    H[1 \dd q]$ and $\Text[\RunEndFullPos{f}{\tau}{\Text}{i} \dd \RunEndPos{\tau}{\Text}{i}) = H[1 \dd
    q']$. We therefore obtain that $\lcp(\Pat[\RunEndFullPat{f}{\tau}{\Pat} \dd m], \Text[\RunEndFullPos{f}{\tau}{\Text}{i} \dd
    \Textlen]) \,{\geq}\, \min(q, q') \allowbreak = \allowbreak q'$.
    By the uniqueness of $\Text[\Textlen]$ in $\Text$, we have $\RunEndPos{\tau}{\Text}{i} \leq
    \Textlen$. Thus, $\RunEndFullPos{f}{\tau}{\Text}{i} + q' \leq \Textlen$. Therefore, by
    $\lcp(\Pat[\RunEndFullPat{f}{\tau}{\Pat} \dd m], \Text[\RunEndFullPos{f}{\tau}{\Text}{i} \dd \Textlen]) \geq
    q'$ and $\TypePos{\tau}{\Text}{i} = -1$, we have $\Text[\RunEndFullPos{f}{\tau}{\Text}{i} + q'] =
    \Text[\RunEndPos{\tau}{\Text}{i}] \prec \Text[\RunEndPos{\tau}{\Text}{i} - |H|] = \Text[\RunEndFullPos{f}{\tau}{\Text}{i} + q' -
    |H|]$ and hence $\Text[\RunEndFullPos{f}{\tau}{\Text}{i} + q'] \prec \Text[\RunEndFullPos{f}{\tau}{\Text}{i} + q'
    - |H|] = \Pat[\RunEndFullPat{f}{\tau}{\Pat} + q' - |H|] = \Pat[\RunEndFullPat{f}{\tau}{\Pat} +
    q']$.  Consequently, $\Text[\RunEndFullPos{f}{\tau}{\Text}{i} \dd \RunEndFullPos{f}{\tau}{\Text}{i} + q'] \prec
    \Pat[\RunEndFullPat{f}{\tau}{\Pat} \dd \RunEndFullPat{f}{\tau}{\Pat} + q']$.  Observe now that
    we have $\RunEndFullPat{f}{\tau}{\Pat} + q' \leq \RunEndFullPat{f}{\tau}{\Pat} + q - 1 =
    \RunEndPat{\tau}{\Pat} - 1 \leq m$ and $\RunEndFullPat{f}{\tau}{\Pat} + q' \leq
    \RunEndFullPat{f}{\tau}{\Pat} + q - 1 = \RunEndPat{\tau}{\Pat} - 1 < \ell$.  Thus,
    $\RunEndFullPat{f}{\tau}{\Pat} + q' \leq \min(m, \ell)$.  On the other hand, we
    also have $\RunEndFullPos{f}{\tau}{\Text}{i} + q' < \RunEndFullPos{f}{\tau}{\Text}{i} + |\RootPos{f}{\tau}{\Text}{i}| <
    \RunEndFullPos{f}{\tau}{\Text}{i} + \tau < \RunEndFullPos{f}{\tau}{\Text}{i} + 7\tau$.  We therefore obtain
    that $\Text[\RunEndFullPos{f}{\tau}{\Text}{i} \dd \RunEndFullPos{f}{\tau}{\Text}{i} + q'] \prec
    \Pat[\RunEndFullPat{f}{\tau}{\Pat} \dd \RunEndFullPat{f}{\tau}{\Pat} + q']$ implies
    $\Textinf[\RunEndFullPos{f}{\tau}{\Text}{i} \dd \RunEndFullPos{f}{\tau}{\Text}{i} + 7\tau) \prec
    \Pat[\RunEndFullPat{f}{\tau}{\Pat} \dd \min(m,\ell)]$, contradicting the main
    assumption, since $\Pat[\RunEndFullPat{f}{\tau}{\Pat} \dd \min(m,\ell)] =
    \Pat[\RunEndCutPat{f}{\tau}{\Pat}{\ell} \dd \min(m,\ell)]$.  We have thus proved $q'
    \geq q$, i.e., $\RunEndPos{\tau}{\Text}{i} - \RunEndFullPos{f}{\tau}{\Text}{i} \geq \RunEndPat{\tau}{\Pat} -
    \RunEndFullPat{f}{\tau}{\Pat}$. Observe now that $\RunEndFullPos{f}{\tau}{\Text}{i} - (i + \delta) =
    \RunEndCutPat{f}{\tau}{\Pat}{\ell} - 1 = \RunEndFullPat{f}{\tau}{\Pat} - 1$. Thus,
    \begin{align*}
      \RunEndPos{\tau}{\Text}{i} - (i + \delta)
        &=    (\RunEndFullPos{f}{\tau}{\Text}{i} - (i + \delta)) + (\RunEndPos{\tau}{\Text}{i} - \RunEndFullPos{f}{\tau}{\Text}{i})\\
        &=    (\RunEndFullPat{f}{\tau}{\Pat} - 1) + (\RunEndPos{\tau}{\Text}{i} - \RunEndFullPos{f}{\tau}{\Text}{i})\\
        &\geq (\RunEndFullPat{f}{\tau}{\Pat} - 1) + (\RunEndPat{\tau}{\Pat} - \RunEndFullPat{f}{\tau}{\Pat})\\
        &=    \RunEndPat{\tau}{\Pat} - 1
         \geq 3\tau - 1.
    \end{align*}
  \item Let us now assume $\RunEndPat{\tau}{\Pat} - 1 \geq \ell$.  Note that we
    then have $m \geq \RunEndPat{\tau}{\Pat} - 1 \geq \ell$.  Thus, $\min(m,\ell)
    = \ell$.  We then also have $k_1 = \min(\ExpPat{f}{\tau}{\Pat}, \lfloor
    \tfrac{\ell - s}{|H|} \rfloor) = \min(\lfloor \tfrac{\RunEndPat{\tau}{\Pat} -
    1 - s}{|H|} \rfloor, \lfloor \tfrac{\ell - s}{|H|} \rfloor) =
    \lfloor \tfrac{\ell - s}{|H|} \rfloor$ and hence $\RunEndCutPat{f}{\tau}{\Pat}{\ell}
    \geq 1 + s + \lfloor \tfrac{\ell - s}{|H|} \rfloor|H| > 1 + s +
    (\tfrac{\ell - s}{|H|} - 1)|H| = \ell - (|H| - 1)$.  Denote $q :=
    1 + \ell - \RunEndCutPat{f}{\tau}{\Pat}{\ell}$ and $q' := \TailPos{f}{\tau}{\Text}{i} = \RunEndPos{\tau}{\Text}{i} -
    \RunEndFullPos{f}{\tau}{\Text}{i}$. We prove that $q' \geq q$. Suppose $q' < q$.
    First, note that by the above, it holds $q < |H|$. By $\RunEndPat{\tau}{\Pat}
    - 1 \geq \ell$, we thus have $\Pat[\RunEndCutPat{f}{\tau}{\Pat}{\ell} \dd \ell] = H[1
    \dd q]$.  On the other hand, by definition we have
    $\Text[\RunEndFullPos{f}{\tau}{\Text}{i} \dd \RunEndPos{\tau}{\Text}{i}) = H[1 \dd q']$.  Thus,
    $\lcp(\Pat[\RunEndCutPat{f}{\tau}{\Pat}{\ell} \dd m], \Text[\RunEndFullPos{f}{\tau}{\Text}{i} \dd \Textlen]) \geq
    \min(q, q') = q'$. By the uniqueness of $\Text[\Textlen]$ in $\Text$, we have
    $\RunEndPos{\tau}{\Text}{i} \leq \Textlen$. Thus, $\RunEndFullPos{f}{\tau}{\Text}{i} + q' \leq \Textlen$. Therefore, by
    $\lcp(\Pat[\RunEndCutPat{f}{\tau}{\Pat}{\ell} \dd m], \Text[\RunEndFullPos{f}{\tau}{\Text}{i} \dd \Textlen]) \geq q'$
    and $\TypePos{\tau}{\Text}{i} = -1$, we have $\Text[\RunEndFullPos{f}{\tau}{\Text}{i} + q'] = \Text[\RunEndPos{\tau}{\Text}{i}]
    \prec \Text[\RunEndPos{\tau}{\Text}{i} - |H|] = \Text[\RunEndFullPos{f}{\tau}{\Text}{i} + q' - |H|]$, and hence
    $\Text[\RunEndFullPos{f}{\tau}{\Text}{i} + q'] \prec \Text[\RunEndFullPos{f}{\tau}{\Text}{i} + q' - |H|] =
    \Pat[\RunEndCutPat{f}{\tau}{\Pat}{\ell} + q' - |H|] = \Pat[\RunEndCutPat{f}{\tau}{\Pat}{\ell} + q']$; note
    that $\Pat[\RunEndCutPat{f}{\tau}{\Pat}{\ell} + q' - |H|]$ is well-defined since
    $\RunEndCutPat{f}{\tau}{\Pat}{\ell} > \ell - (|H| - 1) \geq 2\tau$ and $|H| < \tau$.
    Consequently, $\Text[\RunEndFullPos{f}{\tau}{\Text}{i} \dd \RunEndFullPos{f}{\tau}{\Text}{i} + q'] \prec
    \Pat[\RunEndCutPat{f}{\tau}{\Pat}{\ell} \dd \RunEndCutPat{f}{\tau}{\Pat}{\ell} + q']$.  Observe now that
    we have $\RunEndCutPat{f}{\tau}{\Pat}{\ell} + q' \leq \RunEndCutPat{f}{\tau}{\Pat}{\ell} + q - 1 =
    \ell$. On the other hand, $\RunEndFullPos{f}{\tau}{\Text}{i} + q' < \RunEndFullPos{f}{\tau}{\Text}{i} +
    |\RootPos{f}{\tau}{\Text}{i}| < \RunEndFullPos{f}{\tau}{\Text}{i} + \tau < \RunEndFullPos{f}{\tau}{\Text}{i} + 7\tau$.  Thus,
    we obtain $\Textinf[\RunEndFullPos{f}{\tau}{\Text}{i} \dd \RunEndFullPos{f}{\tau}{\Text}{i} + 7\tau) \prec
    \Pat[\RunEndCutPat{f}{\tau}{\Pat}{\ell} \dd \ell] = \Pat[\RunEndCutPat{f}{\tau}{\Pat}{\ell} \dd \min(m,
    \ell)]$, which contradicts our main assumption. We have thus
    proved $q' \geq q$, i.e., $\RunEndPos{\tau}{\Text}{i} - \RunEndFullPos{f}{\tau}{\Text}{i} \geq 1 + \ell -
    \RunEndCutPat{f}{\tau}{\Pat}{\ell}$. Thus,
    \begin{align*}
      \RunEndPos{\tau}{\Text}{i} - (i + \delta)
         &=    (\RunEndFullPos{f}{\tau}{\Text}{i} - (i + \delta)) + (\RunEndPos{\tau}{\Text}{i} - \RunEndFullPos{f}{\tau}{\Text}{i})\\
         &=    (\RunEndCutPat{f}{\tau}{\Pat}{\ell} - 1) + (\RunEndPos{\tau}{\Text}{i} - \RunEndFullPos{f}{\tau}{\Text}{i})\\
         &\geq (\RunEndCutPat{f}{\tau}{\Pat}{\ell} - 1) + (1 + \ell - \RunEndCutPat{f}{\tau}{\Pat}{\ell})\\
         &=    \ell
          \geq 3\tau - 1.
    \end{align*}
  \end{itemize}
  In both cases, we have shown $\RunEndPos{\tau}{\Text}{i} - (i + \delta) \geq 3\tau -
  1$, or equivalently, $\delta \leq \RunEndPos{\tau}{\Text}{i} - i - 3\tau + 1 < t$. It
  remains to show $i + \delta \in \PosLowMinus{f}{\Pat}{\Text}$. First, recall that
  by $\RunEndFullPos{f}{\tau}{\Text}{i + \delta} = \RunEndFullPos{f}{\tau}{\Text}{i}$ and the definition of
  $\delta$, we have $\RunEndFullPos{f}{\tau}{\Text}{i + \delta} - (i + \delta) =
  \RunEndFullPos{f}{\tau}{\Text}{i} - (i + \delta) = \RunEndCutPat{f}{\tau}{\Pat}{\ell} - 1$.  Thus, since
  above we noted that $[i \dd i + t) \subseteq \RMinusFour{f}{H}{\tau}{\Text}$ (and hence
  $i + \delta \in \RMinusFour{f}{H}{\tau}{\Text}$), we have $\HeadPos{f}{\tau}{\Text}{i + \delta} =
  (\RunEndFullPos{f}{\tau}{\Text}{i + \delta} - (i + \delta)) \bmod |H| = (\RunEndCutPat{f}{\tau}{\Pat}{\ell} -
  1) \bmod |H| = (\HeadPat{f}{\tau}{\Pat} + k_1|H|) \bmod |H| = \HeadPat{f}{\tau}{\Pat}$ and
  $\ExpPos{f}{\tau}{\Text}{i + \delta} = \lfloor \tfrac{\RunEndFullPos{f}{\tau}{\Text}{i + \delta} - (i +
  \delta)}{|H|} \rfloor = \lfloor \tfrac{\RunEndCutPat{f}{\tau}{\Pat}{\ell} - 1}{|H|}
  \rfloor = k_1$. We have therefore proved that $i + \delta \in
  \RMinusFive{f}{s}{H}{\tau}{\Text}$ and $\ExpPos{f}{\tau}{\Text}{i + \delta} = k_1$. Thus, to show $i +
  \delta \in \PosLowMinus{f}{\Pat}{\Text}$ it remains to prove that it holds
  $\Textinf[i + \delta \dd \Textlen] \succeq \Pat$ or $\lcp(\Pat, \Text[i + \delta
  \dd \Textlen]) \geq \ell$. To this end, first note that, letting $H'$ be a
  length-$s$ suffix of $H$, by the above we have $\Text[i + \delta \dd
  \RunEndFullPos{f}{\tau}{\Text}{i + \delta}) = \Text[i + \delta \dd \RunEndFullPos{f}{\tau}{\Text}{i}) = \Pat[1
  \dd \RunEndCutPat{f}{\tau}{\Pat}{\ell}) = H' H^{k_1}$. Consequently, $\lcp(\Pat, \Text[i +
  \delta \dd \Textlen]) = (\RunEndCutPat{f}{\tau}{\Pat}{\ell} - 1) + \lcp(\Pat[\RunEndCutPat{f}{\tau}{\Pat}{\ell} \dd
  m], \Text[\RunEndFullPos{f}{\tau}{\Text}{i} \dd \Textlen])$.  Denote $h = \lcp(\Pat[\RunEndCutPat{f}{\tau}{\Pat}{\ell}
  \dd m], \Text[\RunEndFullPos{f}{\tau}{\Text}{i} \dd \Textlen])$.  Consider two cases:
  \begin{itemize}
  \item First, assume $h \geq \ell - (\RunEndCutPat{f}{\tau}{\Pat}{\ell} - 1)$. Then, we
    immediately obtain $\lcp(\Pat, \Text[i + \delta \dd \Textlen]) =
    (\RunEndCutPat{f}{\tau}{\Pat}{\ell} - 1) + h \geq \ell$, and hence $i + \delta \in
    \PosLowMinus{f}{\Pat}{\Text}$.
  \item Second, assume $h < \ell - (\RunEndCutPat{f}{\tau}{\Pat}{\ell} - 1)$. We will
    prove that $\Text[\RunEndFullPos{f}{\tau}{\Text}{i} \dd \Textlen] \succeq \Pat[\RunEndCutPat{f}{\tau}{\Pat}{\ell} \dd
    m]$.  Consider two subcases:
    \begin{itemize}
    \item Let us first assume that $\RunEndCutPat{f}{\tau}{\Pat}{\ell} + h \leq m$.  Since
      $\Text[\Textlen]$ does not occur in $\Pat[1 \dd m)$, this implies
      $\RunEndFullPos{f}{\tau}{\Text}{i} + h \leq \Textlen$. Thus, we obtain $\Text[\RunEndFullPos{f}{\tau}{\Text}{i} + h] \neq
      \Pat[\RunEndCutPat{f}{\tau}{\Pat}{\ell} + h]$, and consequently $\Text[\RunEndFullPos{f}{\tau}{\Text}{i} \dd
      \RunEndFullPos{f}{\tau}{\Text}{i} + h] \neq \Pat[\RunEndCutPat{f}{\tau}{\Pat}{\ell} \dd \RunEndCutPat{f}{\tau}{\Pat}{\ell} +
      h]$.  We show that it holds $\Text[\RunEndFullPos{f}{\tau}{\Text}{i} \dd \RunEndFullPos{f}{\tau}{\Text}{i} + h] \succ
      \Pat[\RunEndCutPat{f}{\tau}{\Pat}{\ell} \dd \RunEndCutPat{f}{\tau}{\Pat}{\ell} + h]$. To see this,
      assume $\Text[\RunEndFullPos{f}{\tau}{\Text}{i} \dd \RunEndFullPos{f}{\tau}{\Text}{i} + h] \prec
      \Pat[\RunEndCutPat{f}{\tau}{\Pat}{\ell} \dd \RunEndCutPat{f}{\tau}{\Pat}{\ell} + h]$.  Note that $h \leq
      \ell - 1 \leq 7\tau - 1$ and $\RunEndCutPat{f}{\tau}{\Pat}{\ell} + h \leq \ell$.
      Thus, we also have $\Textinf[\RunEndFullPos{f}{\tau}{\Text}{i} \dd \RunEndFullPos{f}{\tau}{\Text}{i} + 7\tau
      - 1] \prec \Pat[\RunEndCutPat{f}{\tau}{\Pat}{\ell} \dd \min(m,\ell)]$, which
      contradicts our assumption. Hence, $\Text[\RunEndFullPos{f}{\tau}{\Text}{i} \dd
      \RunEndFullPos{f}{\tau}{\Text}{i} + h] \succ \Pat[\RunEndCutPat{f}{\tau}{\Pat}{\ell} \dd \RunEndCutPat{f}{\tau}{\Pat}{\ell} +
      h]$. This implies $\Text[\RunEndFullPos{f}{\tau}{\Text}{i} \dd \Textlen] \succ
      \Pat[\RunEndCutPat{f}{\tau}{\Pat}{\ell} \dd m]$.
    \item Let us now assume $\RunEndCutPat{f}{\tau}{\Pat}{\ell} + h = m + 1$. Since
      $\Text[\Textlen]$ does not occur in $\Pat[1 \dd m)$, the suffix $\Text[\RunEndFullPos{f}{\tau}{\Text}{i} \dd
      \Textlen]$ can thus match with $\Pat$ only up to symbol $\Text[\Textlen]$.
      Thus, $\RunEndFullPos{f}{\tau}{\Text}{i} + h - 1 \leq \Textlen$.  Consequently, we have
      $\Text[\RunEndFullPos{f}{\tau}{\Text}{i} \dd \Textlen] \succeq \Text[\RunEndFullPos{f}{\tau}{\Text}{i} \dd \RunEndFullPos{f}{\tau}{\Text}{i}
      + h - 1] = \Pat[\RunEndCutPat{f}{\tau}{\Pat}{\ell} \dd \RunEndCutPat{f}{\tau}{\Pat}{\ell} + h - 1] =
      \Pat[\RunEndCutPat{f}{\tau}{\Pat}{\ell} \dd m]$.
    \end{itemize}
    We have thus proved $\Text[\RunEndFullPos{f}{\tau}{\Text}{i} \dd \Textlen] \succeq
    \Pat[\RunEndCutPat{f}{\tau}{\Pat}{\ell} \dd m]$.  Combining with $\Text[i+\delta \dd
    \RunEndFullPos{f}{\tau}{\Text}{i}) = \Pat[1 \dd \RunEndCutPat{f}{\tau}{\Pat}{\ell})$, this implies
    $\Text[i+\delta \dd \Textlen] \succeq \Pat$. Thus, $i+\delta \in
    \PosLowMinus{f}{\Pat}{\Text}$.  \qedhere
  \end{itemize}

  To show the last implication, recall that we proved that $i + \delta
  \in \PosLowMinus{f}{\Pat}{\Text}$ (where $\delta \in [0 \dd t)$) implies $\delta =
  (\RunEndFullPos{f}{\tau}{\Text}{i} - i) - (\RunEndCutPat{f}{\tau}{\Pat}{\ell} - 1)$. Thus, $\PosLowMinus{f}{\Pat}{\Text}
  \cap [i \dd i + t) \neq \emptyset$ implies $\PosLowMinus{f}{\Pat}{\Text} \cap [i
  \dd i + t) = \{i + \delta\} = \{\RunEndFullPos{f}{\tau}{\Text}{i} - (\RunEndCutPat{f}{\tau}{\Pat}{\ell} -
  1)\}$.
\end{proof}

\begin{lemma}\label{lm:poslow-poshigh-single}
  Let $\ell \in [16 \dd \Textlen)$, $\tau = \lfloor \tfrac{\ell}{3} \rfloor$,
  and $f$ be any necklace-consistent function.
  Let $\Pat \in \Sigma^{m}$ be a $\tau$-periodic pattern such that
  $\TypePat{\tau}{\Pat} = -1$ and $\Text[\Textlen]$ does not occur in $\Pat[1 \dd m)$.
  Denote $H = \RootPat{f}{\tau}{\Pat}$, $s = \HeadPat{f}{\tau}{\Pat}$, $k_1 =
  \ExpCutPat{f}{\tau}{\Pat}{\ell}$ (resp.\ $k_2 = \ExpCutPat{f}{\tau}{\Pat}{2\ell}$). For
  every $i \in \RTwo{\tau}{\Text}$, $i \in \PosLowMinus{f}{\Pat}{\Text}$ (resp.\ $i \in
  \PosHighMinus{f}{\Pat}{\Text}$) holds if and only if
  \begin{itemize}
  \item $i \in \RMinusSix{f}{s}{k_1}{H}{\tau}{\Text}$ (resp.\ $i \in \RMinusSix{f}{s}{k_2}{H}{\tau}{\Text}$) and
  \item $\Pat[\RunEndCutPat{f}{\tau}{\Pat}{\ell} \dd \min(m, \ell)] \preceq
    \Textinf[\RunEndFullPos{f}{\tau}{\Text}{i} \dd \RunEndFullPos{f}{\tau}{\Text}{i} + 7\tau)$\\
    (resp.\ $\Pat[\RunEndCutPat{f}{\tau}{\Pat}{2\ell} \dd \min(m, 2\ell)] \preceq
    \Textinf[\RunEndFullPos{f}{\tau}{\Text}{i} \dd \RunEndFullPos{f}{\tau}{\Text}{i} + 7\tau)$).
  \end{itemize}
\end{lemma}
\begin{proof}

  Below we only prove the lemma for $\PosLowMinus{f}{\Pat}{\Text}$. The version for
  $\PosHighMinus{f}{\Pat}{\Text}$ is identical, except we replace $\PosLowMinus{f}{\Pat}{\Text}$,
  $k_1$, $\RunEndCutPat{f}{\tau}{\Pat}{\ell}$, and $\ell$ with $\PosHighMinus{f}{\Pat}{\Text}$, $k_2$,
  $\RunEndCutPat{f}{\tau}{\Pat}{2\ell}$, and $2\ell$, respectively.

  Let $i \in \RTwo{\tau}{\Text}$ and assume $i \in \PosLowMinus{f}{\Pat}{\Text}$.  By definition of
  $\PosLowMinus{f}{\Pat}{\Text}$, it then holds $i \in \RMinusFive{f}{s}{H}{\tau}{\Text}$ and $\ExpPos{f}{\tau}{\Text}{i} =
  k_1$, i.e., $i \in \RMinusSix{f}{s}{k_1}{H}{\tau}{\Text}$.  This proves the first
  claim. To show the second claim, note that by $\RunEndPos{\tau}{\Text}{i} - i \geq
  3\tau - 1$, letting $t = \RunEndPos{\tau}{\Text}{i} - i - 3\tau + 2$, we also have
  $\PosLowMinus{f}{\Pat}{\Text} \cap [i \dd i + t) \neq \emptyset$. By
  \cref{lm:poslow-poshigh-run}, we thus obtain
  $\Pat[\RunEndCutPat{f}{\tau}{\Pat}{\ell} \dd \min(m,\ell)] \preceq \Textinf[\RunEndFullPos{f}{\tau}{\Text}{i}
  \dd \RunEndFullPos{f}{\tau}{\Text}{i} + 7\tau)$.

  Let us consider $i \in \RTwo{\tau}{\Text}$ and assume that it holds $i \in
  \RMinusSix{f}{s}{k_1}{H}{\tau}{\Text}$ and $\Pat[\RunEndCutPat{f}{\tau}{\Pat}{\ell} \dd \min(m,\ell)] \preceq
  \Textinf[\RunEndFullPos{f}{\tau}{\Text}{i} \dd \RunEndFullPos{f}{\tau}{\Text}{i} + 7\tau)$.  The first
  assumption implies that $\RunEndFullPos{f}{\tau}{\Text}{i} - i = \HeadPos{f}{\tau}{\Text}{i} +
  \ExpPos{f}{\tau}{\Text}{i}\cdot |\RootPos{f}{\tau}{\Text}{i}| = s + k_1 \cdot |H| = \HeadPat{f}{\tau}{\Pat} +
  \ExpCutPat{f}{\tau}{\Pat}{\ell} \cdot |\RootPat{f}{\tau}{\Pat}| = \RunEndCutPat{f}{\tau}{\Pat}{\ell} - 1
  = \RunEndCutPat{f}{\tau}{\Pat}{\ell} - 1$. By \cref{lm:poslow-poshigh-run}, it thus
  follows that, letting $t = \RunEndPos{\tau}{\Text}{i} - i - 3\tau + 2 > 0$, we have
  $\PosLowMinus{f}{\Pat}{\Text} \cap [i \dd i + t) = \{\RunEndFullPos{f}{\tau}{\Text}{i} -
  (\RunEndCutPat{f}{\tau}{\Pat}{\ell} - 1)\} = \{\RunEndFullPos{f}{\tau}{\Text}{i} - (\RunEndFullPos{f}{\tau}{\Text}{i} - i)\} =
  \{i\}$. Thus, $i \in \PosLowMinus{f}{\Pat}{\Text}$.
\end{proof}

\begin{lemma}\label{lm:pat-poslow-poshigh-all}
  Let $\ell \in [16 \dd \Textlen)$, $\tau = \lfloor \tfrac{\ell}{3} \rfloor$,
  and $f$ be any necklace-consistent function.
  Let $\Pat \in \Sigma^m$ be a $\tau$-periodic pattern such that
  $\TypePat{\tau}{\Pat} = -1$ and $\Text[\Textlen]$ does not occur in $\Pat[1 \dd m)$.  Let
  $x_l = \RunEndCutPat{f}{\tau}{\Pat}{\ell} - 1$, $y_u = \Pat[\RunEndCutPat{f}{\tau}{\Pat}{\ell} \dd
  \min(m,\ell)]$, $x'_l = \RunEndCutPat{f}{\tau}{\Pat}{2\ell} - 1$, and $y'_u =
  \Pat[\RunEndCutPat{f}{\tau}{\Pat}{2\ell} \dd \min(m,2\ell)]$.  Then, letting $H =
  \RootPat{f}{\tau}{\Pat}$, it holds:
  \begin{itemize}[itemsep=1pt]
  \item $\PosLowMinus{f}{\Pat}{\Text} =
    \{\RunEndFullPos{f}{\tau}{\Text}{j} - x_l : j \in \RPrimMinusFour{f}{H}{\tau}{\Text},\,
    x_l \leq \RunEndFullPos{f}{\tau}{\Text}{j} - j,\text{ and }\\
    y_u \preceq \Textinf[\RunEndFullPos{f}{\tau}{\Text}{j} \dd \RunEndFullPos{f}{\tau}{\Text}{j} + 7\tau)\}$,
  \item $\PosHighMinus{f}{\Pat}{\Text} =
    \{\RunEndFullPos{f}{\tau}{\Text}{j} - x'_l : j \in \RPrimMinusFour{f}{H}{\tau}{\Text},\,
    x'_l \leq \RunEndFullPos{f}{\tau}{\Text}{j} - j,\text{ and }\\
    y'_u \preceq \Textinf[\RunEndFullPos{f}{\tau}{\Text}{j} \dd \RunEndFullPos{f}{\tau}{\Text}{j} + 7\tau)\}$.
  \end{itemize}
\end{lemma}
\begin{proof}
  Below we prove only the first formula. The proof for the second
  formula is analogous. By \cref{def:pos-sets-for-pat}, we have
  $\PosLowMinus{f}{\Pat}{\Text} \subseteq \RMinusFour{f}{H}{\tau}{\Text}$.  On the other hand, by
  \cref{lm:beg-end}\eqref{lm:beg-end-it-1} it holds $\RMinusFour{f}{H}{\tau}{\Text} =
  \bigcup_{j \in \RPrimMinusFour{f}{H}{\tau}{\Text}} [j \dd \RunEndPos{\tau}{\Text}{j} - 3\tau + 1]$.  Since
  this is a disjoint union, we thus have $\PosLowMinus{f}{\Pat}{\Text} = \bigcup_{j
  \in \RPrimMinusFour{f}{H}{\tau}{\Text}} \PosLowMinus{f}{\Pat}{\Text} \cap [j \dd \RunEndPos{\tau}{\Text}{j} - 3\tau + 1]$.
  By \cref{lm:poslow-poshigh-run}, for every $j \in \RPrimMinusFour{f}{H}{\tau}{\Text}$,
  $|\PosLowMinus{f}{\Pat}{\Text} \cap [j \dd \RunEndPos{\tau}{\Text}{j} - 3\tau + 1]| \leq
  1$. Moreover, $|\PosLowMinus{f}{\Pat}{\Text} \cap [j \dd \RunEndPos{\tau}{\Text}{j} - 3\tau + 1]| =
  1$ holds if and only if $x_l \leq \RunEndFullPos{f}{\tau}{\Text}{j} - j$ and $y_u \preceq
  \Textinf[\RunEndFullPos{f}{\tau}{\Text}{j} \dd \RunEndFullPos{f}{\tau}{\Text}{j} + 7\tau)$.  Furthermore, if $x_l
  \leq \RunEndFullPos{f}{\tau}{\Text}{j} - j$ and $y_u \preceq \Textinf[\RunEndFullPos{f}{\tau}{\Text}{j} \dd
  \RunEndFullPos{f}{\tau}{\Text}{j} + 7\tau)$, then $\PosLowMinus{f}{\Pat}{\Text} \cap [j \dd \RunEndPos{\tau}{\Text}{j} -
  3\tau + 1] = \{\RunEndFullPos{f}{\tau}{\Text}{j} - x_l\}$.  Consequently, letting
  $\mathcal{J} = \{j \in \RPrimMinusFour{f}{H}{\tau}{\Text} : x_l \leq \RunEndFullPos{f}{\tau}{\Text}{j} - j\text{
  and } y_u \preceq \Textinf[\RunEndFullPos{f}{\tau}{\Text}{j} \dd \RunEndFullPos{f}{\tau}{\Text}{j} + 7\tau)\}$ we
  thus obtain $\PosLowMinus{f}{\Pat}{\Text} = \bigcup_{j \in \RPrimMinusFour{f}{H}{\tau}{\Text}}
  \PosLowMinus{f}{\Pat}{\Text} \cap [j \dd \RunEndPos{\tau}{\Text}{j} - 3\tau + 1] = \bigcup_{j \in
  \mathcal{J}} \{\RunEndFullPos{f}{\tau}{\Text}{j} - x_l\}$, i.e., the claim.
\end{proof}

\begin{lemma}\label{lm:pat-poslow-poshigh-all-count}
  Let $\ell \in [16 \dd \Textlen)$, $\tau = \lfloor \tfrac{\ell}{3} \rfloor$,
  and $f$ be any necklace-consistent function.
  Let $\Pat \in \Sigma^m$ be a $\tau$-periodic pattern such that
  $\TypePat{\tau}{\Pat} = -1$ and $\Text[\Textlen]$ does not occur in $\Pat[1 \dd m)$.  Let
  $x_l = \RunEndCutPat{f}{\tau}{\Pat}{\ell} - 1$, $y_u = \Pat[\RunEndCutPat{f}{\tau}{\Pat}{\ell} \dd
  \min(m,\ell)]$, $x'_l = \RunEndCutPat{f}{\tau}{\Pat}{2\ell} - 1$, and $y'_u =
  \Pat[\RunEndCutPat{f}{\tau}{\Pat}{2\ell} \dd \min(m,2\ell)]$.  Then, letting $H =
  \RootPat{f}{\tau}{\Pat}$ and $\Pts = \IntStrPoints{7\tau}{\PairsMinus{f}{H}{\tau}{\Text}}{\Text}$,
  it holds:
  \begin{itemize}[itemsep=1pt]
  \item $|\PosLowMinus{f}{\Pat}{\Text}| = \RangeCountTwoSide{\Pts}{x_l}{\Textlen} -
    \RangeCountThreeSide{\Pts}{x_l}{\Textlen}{y_u}$,
  \item $|\PosHighMinus{f}{\Pat}{\Text}| = \RangeCountTwoSide{\Pts}{x'_l}{\Textlen} -
    \RangeCountThreeSide{\Pts}{x'_l}{\Textlen}{y'_u}$.
  \end{itemize}
\end{lemma}
\begin{proof}
  By \cref{lm:pat-poslow-poshigh-all,lm:sa-periodic-count}, it holds
  \begin{align*}
    |\PosLowMinus{f}{\Pat}{\Text}|
      &= |\{j' \in \RPrimMinusFour{f}{H}{\tau}{\Text} : x_l \leq \RunEndFullPos{f}{\tau}{\Text}{j'} - j'\text{ and }
         y_u \preceq \Textinf[\RunEndFullPos{f}{\tau}{\Text}{j'} \dd \RunEndFullPos{f}{\tau}{\Text}{j'} + 7\tau)\}|\\
      &= |\{j' \in \RPrimMinusFour{f}{H}{\tau}{\Text} : x_l \leq \RunEndFullPos{f}{\tau}{\Text}{j'} - j'\}|\,-\\
      &\quad\
         |\{j' \in \RPrimMinusFour{f}{H}{\tau}{\Text} : x_l \leq \RunEndFullPos{f}{\tau}{\Text}{j'} - j'\text{ and }
         \Textinf[\RunEndFullPos{f}{\tau}{\Text}{j'} \dd \RunEndFullPos{f}{\tau}{\Text}{j'} + 7\tau) \prec y_u\}|\\
      &= \RangeCountTwoSide{\Pts}{x_l}{\Textlen} - \RangeCountThreeSide{\Pts}{x_l}{\Textlen}{y_u}.
  \end{align*}
  Note that \cref{lm:sa-periodic-count} requires that $x_l, x'_l \in
  [0 \dd 7\tau]$, which holds here since $x_l = \RunEndCutPat{f}{\tau}{\Pat}{\ell} - 1
  \leq \ell$, and for $\tau = \lfloor
  \tfrac{\ell}{3} \rfloor$ and $\ell \geq 16$, it holds $\ell \leq
  7\tau$. The proof for $|\PosHighMinus{f}{\Pat}{\Text}|$ is analogous,
  except we use that $x'_l = \RunEndCutPat{f}{\tau}{\Pat}{2\ell} - j
  \leq 2\ell$, and for $\tau = \lfloor
  \tfrac{\ell}{3} \rfloor$ and $\ell \geq 16$, it holds $2\ell \leq
  7\tau$.
\end{proof}

\begin{lemma}\label{lm:pat-occ-poslow-poshigh-size}
  Let $\ell \in [16 \dd \Textlen)$, $\tau = \lfloor \tfrac{\ell}{3} \rfloor$,
  and $f$ be any necklace-consistent function.
  Let $\Pat \in \Sigma^m$ be a $\tau$-periodic pattern such that
  $\TypePat{\tau}{\Pat} = - 1$, $\RunEndPat{\tau}{\Pat} \leq |\Pat|$,
  and $\Text[\Textlen]$ does not occur in $\Pat[1 \dd m)$.
  Let also $j_1 \in \OccThree{\ell}{\Pat}{\Text}$ and $j_2 \in
  \OccThree{2\ell}{\Pat}{\Text}$.  Then, it holds $j_1, j_2 \in \RTwo{\tau}{\Text}$. Moreover, letting $x_l =
  \RunEndCutPos{f}{\tau}{\Text}{j_1}{\ell} - j_1$, $y_u = \Text[\RunEndCutPos{f}{\tau}{\Text}{j_1}{\ell} \dd j_1 + \min(m,
  \ell))$, $x'_l = \RunEndCutPos{f}{\tau}{\Text}{j_2}{2\ell} - j_2$, $y'_u = \Text[\RunEndCutPos{f}{\tau}{\Text}{j_2}{2\ell}
  \dd j_2 + \min(m, 2\ell))$, $H = \RootPat{f}{\tau}{\Pat}$, and $\Pts =
  \IntStrPoints{7\tau}{\PairsMinus{f}{H}{\tau}{\Text}}{\Text}$, it holds
  \begin{itemize}[itemsep=1pt]
  \item $|\PosLowMinus{f}{\Pat}{\Text}| = \RangeCountTwoSide{\Pts}{x_l}{\Textlen} -
    \RangeCountThreeSide{\Pts}{x_l}{\Textlen}{y_u}$,
  \item $|\PosHighMinus{f}{\Pat}{\Text}| = \RangeCountTwoSide{\Pts}{x'_l}{\Textlen} -
    \RangeCountThreeSide{\Pts}{x'_l}{\Textlen}{y'_u}$.
  \end{itemize}
\end{lemma}
\begin{proof}
  We show the claim only for $\PosLowMinus{f}{\Pat}{\Text}$ (the proof for
  $\PosHighMinus{f}{\Pat}{\Text}$ is analogous).  By $\ell \geq 3\tau - 1$
  and \cref{lm:periodic-pos-lce}\eqref{lm:periodic-pos-lce-it-1}, it holds
  $j_1 \in \RTwo{\tau}{\Text}$, $\HeadPos{f}{\tau}{\Text}{j_1} = \HeadPat{f}{\tau}{\Pat}$, $\RootPos{f}{\tau}{\Text}{j_1} =
  \RootPat{f}{\tau}{\Pat}$. By \cref{lm:expcut}\eqref{lm:expcut-it-1},
  we moreover have $\ExpCutPos{f}{\tau}{\Text}{j_1}{\ell} = \ExpCutPat{f}{\tau}{\Pat}{\ell}$.
  This implies that
  $\RunEndCutPat{f}{\tau}{\Pat}{\ell} - 1 = \HeadPat{f}{\tau}{\Pat} + \ExpCutPat{f}{\tau}{\Pat}{\ell} \cdot
  |\RootPat{f}{\tau}{\Pat}| = \HeadPos{f}{\tau}{\Text}{j_1} + \ExpCutPos{f}{\tau}{\Text}{j_1}{\ell} \cdot
  |\RootPos{f}{\tau}{\Text}{j_1}| = \RunEndCutPos{f}{\tau}{\Text}{j_1}{\ell} - j_1 = x_l$.  By definition of
  $\OccThree{\ell}{\Pat}{\Text}$, we have $\lcp(\Text[j_1 \dd \Textlen], \Pat) \geq
  \min(m, \ell)$. Consequently, $\Pat[\RunEndCutPat{f}{\tau}{\Pat}{\ell} \dd \min(m,
  \ell)] = \Text[j_1 + \RunEndCutPat{f}{\tau}{\Pat}{\ell} - 1 \dd j_1 + \min(m, \ell) - 1] =
  \Text[\RunEndCutPos{f}{\tau}{\Text}{j_1}{\ell} \dd j_1 + \min(m, \ell)) = y_u$.  Consequently, it
  follows by \cref{lm:pat-poslow-poshigh-all-count} that
  $|\PosLowMinus{f}{\Pat}{\Text}| = \RangeCountTwoSide{\Pts}{x_l}{\Textlen} -
  \RangeCountThreeSide{\Pts}{x_l}{\Textlen}{y_u}$.
\end{proof}

\begin{lemma}\label{lm:special-pat-poslow-poshigh-size}
  Let $\ell \in [16 \dd \Textlen)$, $\tau = \lfloor \tfrac{\ell}{3} \rfloor$,
  and $f$ be any necklace-consistent function.
  Let $j \in \RTwo{\tau}{\Text}$. Denote $s = \HeadPos{f}{\tau}{\Text}{j}$, $H = \RootPos{f}{\tau}{\Text}{j}$, $p = |H|$,
  $H' = H(p-s \dd p]$. Let $\Pat$ be a length-$2\ell$ prefix of $H'
  H^{\infty}$. Then, $\Pat$ is $\tau$-periodic. Moreover, letting $k_1
  = \lfloor \tfrac{\ell - s}{p} \rfloor$, $k_2 = \lfloor \tfrac{2\ell
  - s}{p} \rfloor$, $x_l = s + k_1 p$, $y_u = \Text[j + s \dd j + s +
  \ell - x_l)$, $x'_l = s + k_2 p$, $y'_u = \Text[j + s \dd j + s + 2\ell
  - x'_l)$, and $\Pts = \IntStrPoints{7\tau}{\PairsMinus{f}{H}{\tau}{\Text}}{\Text}$,
  it holds:
  \begin{itemize}[itemsep=1pt]
  \item $|\PosLowMinus{f}{\Pat}{\Text}| = \RangeCountTwoSide{\Pts}{x_l}{\Textlen} -
    \RangeCountThreeSide{\Pts}{x_l}{\Textlen}{y_u}$,
  \item $|\PosHighMinus{f}{\Pat}{\Text}| = \RangeCountTwoSide{\Pts}{x'_l}{\Textlen} -
    \RangeCountThreeSide{\Pts}{x'_l}{\Textlen}{y'_u}$.
  \end{itemize}
\end{lemma}
\begin{proof}
  By \cref{lm:special-pat-properties}, $\Pat$ is $\tau$-periodic,
  $\RootPat{f}{\tau}{\Pat} = H$, $\TypePat{\tau}{\Pat} = -1$, $\Text[\Textlen]$ does not occur in
  $\Pat[1 \dd 2\ell)$, $\RunEndCutPat{f}{\tau}{\Pat}{\ell} - 1 = x_l$, $\RunEndCutPat{f}{\tau}{\Pat}{2\ell} =
  x'_l$, $\Pat[\RunEndCutPat{f}{\tau}{\Pat}{\ell} \dd \ell] = y_u$, and
  $\Pat[\RunEndCutPat{f}{\tau}{\Pat}{2\ell} \dd 2\ell] = y'_u$. The claims therefore
  follow by \cref{lm:pat-poslow-poshigh-all-count}.
\end{proof}

\begin{lemma}\label{lm:pos-occ-poslow-poshigh-size}
  Let $\ell \in [16 \dd \Textlen)$, $\tau = \lfloor \tfrac{\ell}{3} \rfloor$,
  and $f$ be any necklace-consistent function.
  Let $j \in \RMinusTwo{\tau}{\Text}$, $j_1 \in \OccThree{\ell}{j}{\Text}$, and $j_2 \in
  \OccThree{2\ell}{j}{\Text}$. Then, it holds $j_1, j_2 \in \RTwo{\tau}{\Text}$.  Moreover,
  letting $x_l = \RunEndCutPos{f}{\tau}{\Text}{j_1}{\ell} - j_1$, $y_u = \Text[\RunEndCutPos{f}{\tau}{\Text}{j_1}{\ell} \dd
  \min(\Textlen + 1, j_1 + \ell))$, $x'_l = \RunEndCutPos{f}{\tau}{\Text}{j_2}{2\ell} - j_2$, $y'_u =
  \Text[\RunEndCutPos{f}{\tau}{\Text}{j_2}{2\ell} \dd \min(\Textlen + 1,j_2 + 2\ell))$, $H = \RootPos{f}{\tau}{\Text}{j}$,
  and $\Pts = \IntStrPoints{7\tau}{\PairsMinus{f}{H}{\tau}{\Text}}{\Text}$, it holds
  \begin{itemize}[itemsep=1pt]
  \item $|\PosLowMinus{f}{j}{\Text}| = \RangeCountTwoSide{\Pts}{x_l}{\Textlen} -
    \RangeCountThreeSide{\Pts}{x_l}{\Textlen}{y_u}$,
  \item $|\PosHighMinus{f}{j}{\Text}| = \RangeCountTwoSide{\Pts}{x'_l}{\Textlen} -
    \RangeCountThreeSide{\Pts}{x'_l}{\Textlen}{y'_u}$.
  \end{itemize}
\end{lemma}
\begin{proof}
  Denote $\Pat := \Text[j \dd \Textlen]$ and $m := |\Pat| = \Textlen - j + 1$.  By
  definition, $j \in \RMinusTwo{\tau}{\Text}$ implies that $\Pat$ is $\tau$-periodic,
  $\TypePat{\tau}{\Pat} = -1$, $\RunEndPat{\tau}{\Pat} \leq |\Pat|$, and
  that $\Text[\Textlen]$ does not occur in $\Pat[1 \dd
  m)$. We then also have $j_1 \in \OccThree{\ell}{j}{\Text} =
  \OccThree{\ell}{\Text[j \dd \Textlen]}{\Text} = \OccTwo{\Pat}{\Text}$ and $j_2 \in
  \OccThree{2\ell}{j}{\Text} = \OccThree{2\ell}{\Text[j \dd \Textlen]}{\Text} =
  \OccThree{2\ell}{\Pat}{\Text}$. Next, observe that $j_1 + \min(m, \ell) =
  \min(\Textlen + 1, j_1 + \ell)$. To see this, consider two cases:
  \begin{itemize}
  \item First, assume $m < \ell$.  The assumption $j_1 \in
    \OccThree{\ell}{j}{\Text}$ implies $\lcp(\Text[j_1 \dd \Textlen], \Pat) \geq
    \min(m, \ell) = m$.  Thus, $\Text[j_1 \dd j_1 + m) = \Pat$. By
    $\Pat[m] = \Text[\Textlen]$ and the uniqueness of $\Text[\Textlen]$ in $\Text$, we then
    have $j_1 + m - 1 = \Textlen$. Consequently, $j_1 + \min(m, \ell) = j_1 +
    \min(\Textlen - j_1 + 1, \ell) = \min(\Textlen + 1, j_1 + \ell)$.
  \item Second, assume $m \geq \ell$. The assumption $j_1 \in
    \OccThree{\ell}{j}{\Text}$ then implies $\lcp(\Text[j_1 \dd \Textlen], \Pat) \geq
    \min(m, \ell) = \ell$. Thus, $\Text[j_1 \dd j_1 + \ell) = \Pat[1 \dd
    \ell]$.  This implies that $\min(\Textlen + 1, j_1 + \ell) = j_1 + \ell$,
    and hence $j_1 + \min(m, \ell) = j_1 + \ell = \min(\Textlen + 1, j_1 +
    \ell)$.
  \end{itemize}
  We thus have $\Text[\RunEndCutPos{f}{\tau}{\Text}{j_1}{\ell} \dd j_1 + \min(m, \ell)) =
  \Text[\RunEndCutPos{f}{\tau}{\Text}{j_1}{\ell} \dd \min(\Textlen + 1, j_1 + \ell)) = y_u$.  Consequently,
  the claim follows by \cref{lm:pat-occ-poslow-poshigh-size}.
\end{proof}

\begin{lemma}\label{lm:pos-poslow-poshigh-size}
  Let $\ell \in [16 \dd \Textlen)$, $\tau = \lfloor \tfrac{\ell}{3} \rfloor$,
  and $f$ be any necklace-consistent function.
  Let $j \in \RMinusTwo{\tau}{\Text}$, $x_l = \RunEndCutPos{f}{\tau}{\Text}{j}{\ell} - j$, $y_u = \Text[\RunEndCutPos{f}{\tau}{\Text}{j}{\ell}
  \dd \min(\Textlen + 1, j + \ell))$, $x'_l = \RunEndCutPos{f}{\tau}{\Text}{j}{2\ell} - j$, and $y'_u =
  \Text[\RunEndCutPos{f}{\tau}{\Text}{j}{2\ell} \dd \min(\Textlen + 1,j + 2\ell))$.  Letting $H =
  \RootPos{f}{\tau}{\Text}{j}$ and $\Pts = \IntStrPoints{7\tau}{\PairsMinus{f}{H}{\tau}{\Text}}{\Text}$,
  it holds
  \begin{itemize}[itemsep=1pt]
  \item $|\PosLowMinus{f}{j}{\Text}| = \RangeCountTwoSide{\Pts}{x_l}{\Textlen} -
    \RangeCountThreeSide{\Pts}{x_l}{\Textlen}{y_u}$,
  \item $|\PosHighMinus{f}{j}{\Text}| = \RangeCountTwoSide{\Pts}{x'_l}{\Textlen} -
    \RangeCountThreeSide{\Pts}{x'_l}{\Textlen}{y'_u}$.
  \end{itemize}
\end{lemma}
\begin{proof}
  By definition, we have $j \in \OccThree{\ell}{j}{\Text}$ and $j \in
  \OccThree{2\ell}{j}{\Text}$.  Thus, the claim follows by
  \cref{lm:pos-occ-poslow-poshigh-size}.
\end{proof}

\paragraph{Query Algorithms}

\begin{proposition}\label{pr:pos-poslow-poshigh-size}
  Let $k \in [4 \dd \lceil \log \Textlen \rceil)$, $\ell = 2^k$, $\tau
  = \lfloor \tfrac{\ell}{3} \rfloor$, and $f = f_{\tau,\Text}$
  (\cref{def:canonical-function}).  Let $j \in \RMinusTwo{\tau}{\Text}$.
  Given $\CompSaPeriodic{\Text}$, the value $k$, any position
  $j' \in \OccThree{\ell}{j}{\Text}$ (resp.\ $j' \in \OccThree{2\ell}{j}{\Text}$),
  any $j'' \in \OccThree{3\tau - 1}{j}{\Text}$ satisfying $j''
  = \min \OccThree{2\ell}{j''}{\Text}$, and the values
  $\HeadPos{f}{\tau}{\Text}{j}$, $|\RootPos{f}{\tau}{\Text}{j}|$, and
  $\ExpCutPos{f}{\tau}{\Text}{j}{\ell}$ (resp.\
  $\ExpCutPos{f}{\tau}{\Text}{j}{2\ell}$) as input, we can compute
  $|\PosLowMinus{f}{j}{\Text}|$ (resp.\ $|\PosHighMinus{f}{j}{\Text}|$) in
  $\bigO(\log^{2 + \epsilon} \Textlen)$ time.
\end{proposition}
\begin{proof}
  Let $s = \HeadPos{f}{\tau}{\Text}{j}$, $H
  = \RootPos{f}{\tau}{\Text}{j}$, $p = |H|$, $k_1
  = \ExpCutPos{f}{\tau}{\Text}{j}{\ell}$, and $k_2
  = \ExpCutPos{f}{\tau}{\Text}{j}{2\ell}$.  In $\bigO(1)$ time we set
  $x_l := \RunEndCutPos{f}{\tau}{\Text}{j}{\ell} - j = s + k_1 p$ (resp.\
  $x_l := \RunEndCutPos{f}{\tau}{\Text}{j}{2\ell} - j = s + k_2 p$).
  Denote $y_u = \Text[\RunEndCutPos{f}{\tau}{\Text}{j}{\ell} \dd \min(\Textlen + 1, j
  + \ell))$ (resp.\ $y_u
  = \Text[\RunEndCutPos{f}{\tau}{\Text}{j}{2\ell} \dd \min(\Textlen + 1, j +
  2\ell))$). Using \cref{pr:sa-periodic-pts-access} and the position
  $j''$ as input, in $\bigO(\log \Textlen)$ time we retrieve the pointer to
  the structure from \cref{pr:int-str} for $\PairsMinus{f}{H}{\tau}{\Text}$
  (note that $j'' \in \RFour{f}{H}{\tau}{\Text}$ holds
  by \cref{lm:periodic-pos-lce}\eqref{lm:periodic-pos-lce-it-2}),
  i.e., performing weighted range queries on $\Pts
  = \IntStrPoints{7\tau}{\PairsMinus{f}{H}{\tau}{\Text}}{\Text}$.  Note that this
  pointer is not null, since we assumed $j \in \RMinusTwo{\tau}{\Text}$,
  which implies $\PairsMinus{f}{H}{\tau}{\Text} \neq \emptyset$.  Note also
  that using $\CompSaNonperiodic{\Text}$, we can perform $\LCE_{\Text}$ and
  $\LCE_{\revstr{\Text}}$ queries in $\bigO(\log \Textlen)$ time, and we can
  access any symbol of $\Text$ in $\bigO(\log \Textlen)$ time. Thus, we can
  compare any two substrings of $\Textinf$ or $\revstr{\Textinf}$ (specified
  with their starting positions and lengths) in $t_{\rm cmp}
  = \bigO(\log \Textlen)$ time.  In $\bigO(\log^{2 + \epsilon} \Textlen + t_{\rm
  cmp} \log \Textlen) = \bigO(\log^{2 + \epsilon} \Textlen)$ time we thus compute $q
  := \RangeCountTwoSide{\Pts}{x_l}{\Textlen} - \RangeCountThreeSide{\Pts}{x_l}{\Textlen}{y_u}$
  using \cref{pr:int-str} with $i = j' + x_l$ and $q_r = \min(\Textlen + 1,
  j' + \ell) - i$ (resp.\ $q_r = \min(\Textlen + 1, j' + 2\ell) -
  i$). By \cref{lm:pos-poslow-poshigh-size}, it holds
  $|\PosLowMinus{f}{j}{\Text}| = q$ (resp.\ $|\PosHighMinus{f}{j}{\Text}| =
  q$).  In total, we spend $\bigO(\log^{2 + \epsilon} \Textlen)$ time.
\end{proof}

\begin{proposition}\label{pr:sa-periodic-poslow-poshigh-pat}
  Let $k \in [4 \dd \lceil \log \Textlen \rceil)$, $\ell = 2^k$, $\tau
  = \lfloor \tfrac{\ell}{3} \rfloor$, and $f = f_{\tau,\Text}$
  (\cref{def:canonical-function}).  Let $j \in \RTwo{\tau}{\Text}$ be a
  position satisfying $j = \min \OccThree{2\ell}{j}{\Text}$.  Denote $s
  = \HeadPos{f}{\tau}{\Text}{j}$ and $H = \RootPos{f}{\tau}{\Text}{j}$.
  Let $H'$ be a length-$s$ suffix of $H$ and $\Pat$ be a
  length-$2\ell$ prefix of $H' H^{\infty}$. Then, $\Pat$ is
  $\tau$-periodic. Moreover, given $\CompSaPeriodic{\Text}$, the value
  $k$, the position $j$, and the values $s$ and $|H|$ as input, we can
  compute $|\PosLowMinus{f}{\Pat}{\Text}|$ and
  $|\PosHighMinus{f}{\Pat}{\Text}|$ in $\bigO(\log^{2 + \epsilon} \Textlen)$
  time.
\end{proposition}
\begin{proof} 
  Denote $p = |H|$. In $\bigO(1)$ time we compute $k_1
  := \lfloor \tfrac{\ell - s}{p} \rfloor$, $k_2
  := \lfloor \tfrac{2\ell - s}{p} \rfloor$, $x_l := s + k_1 p$, and
  $x'_l := s + k_2 p$. Denote $y_u = \Text[j + s \dd j + s + \ell - x_l)$
  and $y'_u = \Text[j + s \dd j + s + 2\ell -
  x'_l)$. Using \cref{pr:sa-periodic-pts-access} and the position $j$
  as input, in $\bigO(\log \Textlen)$ time we check if
  $\PairsMinus{f}{H}{\tau}{\Text} \neq \emptyset$, and if so, we retrieve the
  pointer $\mu_{H}$ to the structure from \cref{pr:int-str} for
  $\PairsMinus{f}{H}{\tau}{\Text}$, i.e., performing weighted range queries on
  $\Pts = \IntStrPoints{7\tau}{\PairsMinus{f}{H}{\tau}{\Text}}{\Text}$.  Note that
  using $\CompSaCore{\Text}$ (which is part of $\CompSaPeriodic{\Text}$), we
  can lexicographically compare any two substrings of $\Textinf$ or
  $\revstr{\Textinf}$ (specified with their starting positions and
  lengths) in $t_{\rm cmp} = \bigO(\log \Textlen)$ time.  If
  $\PairsMinus{f}{H}{\tau}{\Text} = \emptyset$ (observe that this is possible,
  since we did not assume anything about $\TypePos{\tau}{\Text}{j}$),
  then we return $|\PosLowMinus{f}{\Pat}{\Text}| =
  |\PosHighMinus{f}{\Pat}{\Text}| = 0$.  Otherwise, in $\bigO(\log^{2
  + \epsilon} \Textlen + t_{\rm cmp} \log \Textlen) = \bigO(\log^{2 + \epsilon} \Textlen)$
  time we compute $q_1 := \RangeCountTwoSide{\Pts}{x_l}{\Textlen}
  - \RangeCountThreeSide{\Pts}{x_l}{\Textlen}{y_u}$ and $q_2
  := \RangeCountTwoSide{\Pts}{x'_l}{\Textlen} - \RangeCountThreeSide{\Pts}{x'_l}{\Textlen}{y'_u}$
  using \cref{pr:int-str} first with $i = j + s$ and $q_r = \ell -
  x_l$, and then with $i = j + s$ and $q_r = 2\ell -
  x'_l$. By \cref{lm:special-pat-poslow-poshigh-size}, it holds
  $|\PosLowMinus{f}{\Pat}{\Text}| = q_1$ and $|\PosHighMinus{f}{\Pat}{\Text}|
  = q_2$.  In total, we spend $\bigO(\log^{2 + \epsilon} \Textlen)$ time.
\end{proof}

\subsubsection{Computing the Size of Posmid}\label{sec:sa-periodic-posmid}

\paragraph{Combinatorial Properties}

\begin{lemma}\label{lm:kmin-kmax-range}
  Let $\ell \geq 16$, $\tau = \lfloor \tfrac{\ell}{3} \rfloor$,
  and $f$ be any necklace-consistent function.
  Let $\Pat \in \Sigma^{+}$ be a $\tau$-periodic pattern and let $s =
  \HeadPat{f}{\tau}{\Pat}$, $H = \RootPat{f}{\tau}{\Pat}$, and $p = |H|$.  Letting $k_{\min}
  = \lceil \tfrac{3\tau - 1 - s}{p} \rceil - 1$ and $k_{\max} = \lfloor
  \tfrac{7\tau - s}{p} \rfloor$, it holds $k_{\min} \leq
  \ExpCutPat{f}{\tau}{\Pat}{\ell} \leq \ExpCutPat{f}{\tau}{\Pat}{2\ell} \leq k_{\max}$.
\end{lemma}
\begin{proof}
  Denote $k_1 = \ExpCutPat{f}{\tau}{\Pat}{\ell}$ and $k_2 =
  \ExpCutPat{f}{\tau}{\Pat}{2\ell}$.  We prove the three inequalities as follows.
  \begin{itemize}
  \item First, note that $\Pat$ being $\tau$-periodic implies that
    $\RunEndPat{\tau}{\Pat} - 1 \geq 3\tau - 1$, which in turn implies
    $\ExpPat{f}{\tau}{\Pat} = \lfloor \tfrac{\RunEndPat{\tau}{\Pat} - 1 - s}{p} \rfloor \geq
    \lfloor \tfrac{3\tau - 1 - s}{p} \rfloor \geq \lceil \tfrac{3\tau
    - 1 - s}{p} \rceil - 1$. On the other hand, by $\tau = \lfloor
    \tfrac{\ell}{3} \rfloor$ it follows that $3\tau - 1 \leq \ell$,
    and hence $\lfloor \tfrac{\ell - s}{p} \rfloor \geq \lfloor
    \tfrac{3\tau - 1 - s}{p} \rfloor \geq \lceil \tfrac{3\tau - 1 -
    s}{p} \rceil - 1$. Putting the two together, we thus obtain
    $k_{\min} = \lceil \tfrac{3\tau - 1 - s}{p} \rceil - 1 \leq
    \min(\ExpPat{f}{\tau}{\Pat}, \lfloor \tfrac{\ell - s}{p} \rfloor) = k_1$.
  \item Second, note that $k_1 = \min(\ExpPat{f}{\tau}{\Pat}, \lfloor \tfrac{\ell
    - s}{p} \rfloor) \leq \min(\ExpPat{f}{\tau}{\Pat}, \lfloor \tfrac{2\ell -
    s}{p} \rfloor) = k_2$.
  \item Lastly, $k_2 = \min(\ExpPat{f}{\tau}{\Pat}, \lfloor \tfrac{2\ell - s}{p}
    \rfloor) \leq \lfloor \tfrac{2\ell - s}{p} \rfloor \leq \lfloor
    \tfrac{7\tau - s}{p} \rfloor = k_{\max}$, where $2\ell \leq 7\tau$
    follows by $\tau = \lfloor \tfrac{\ell}{3} \rfloor$ and $\ell \geq
    16$.  \qedhere
  \end{itemize}
\end{proof}

\begin{lemma}\label{lm:pat-posmid-size}
  Let $\ell \in [16 \dd \Textlen)$, $\tau = \lfloor \tfrac{\ell}{3} \rfloor$,
  and $f$ be any necklace-consistent function.
  Let $\Pat \in \Sigma^{+}$ be a $\tau$-periodic pattern.  Denote $s =
  \HeadPat{f}{\tau}{\Pat}$, $H = \RootPat{f}{\tau}{\Pat}$, $p = |H|$, $k_1 =
  \ExpCutPat{f}{\tau}{\Pat}{\ell}$, $k_2 = \ExpCutPat{f}{\tau}{\Pat}{2\ell}$, and $\Ints =
  \WInts{7\tau}{\PairsMinus{f}{H}{\tau}{\Text}}{\Text}$. Then, it holds
  \[|\PosMidMinus{f}{\Pat}{\Text}| = \ModCountTwoSide{\Ints}{p}{s}{k_1}{k_2}.\]
\end{lemma}
\begin{proof}
  Denote $k_{\min} = \lceil \tfrac{3\tau - 1 - s}{p} \rceil - 1$ and
  $k_{\max} = \lfloor \tfrac{7\tau - s}{p} \rfloor$. By
  \cref{lm:kmin-kmax-range}, it holds $k_{\min} \leq k_1 \leq k_2 \leq
  k_{\max}$. Thus, by applying \cref{def:pos-sets-for-pat} and
  \cref{lm:sa-periodic-modcount}, we obtain $|\PosMidMinus{f}{\Pat}{\Text}| = |\{j'
  \in \RMinusFive{f}{s}{H}{\tau}{\Text} : \ExpPos{f}{\tau}{\Text}{j'} \in (k_1 \dd k_2]\}| =
  \ModCountTwoSide{\Ints}{p}{s}{k_1}{k_2}$.
\end{proof}

\paragraph{Query Algorithms}

\begin{proposition}\label{pr:sa-periodic-posmid}
  Let $k \in [4 \dd \lceil \log \Textlen \rceil)$, $\ell = 2^k$, $\tau
  = \lfloor \tfrac{\ell}{3} \rfloor$, and $f = f_{\tau,\Text}$
  (\cref{def:canonical-function}). Let $\Pat \in \Sigma^{+}$ be a
  $\tau$-periodic pattern. Given $\CompSaPeriodic{\Text}$, the values
  $k$, $\HeadPat{f}{\tau}{\Pat}$, $|\RootPat{f}{\tau}{\Pat}|$,
  $\ExpCutPat{f}{\tau}{\Pat}{\ell}$,
  $\ExpCutPat{f}{\tau}{\Pat}{2\ell}$, and some
  $j \in \OccTwo{\Pat[1 \dd 3\tau - 1]}{\Text}$ satisfying $j
  = \min \OccThree{2\ell}{j}{\Text}$, we can compute
  $|\PosMidMinus{f}{\Pat}{\Text}|$ in $\bigO(\log^{2+\epsilon} \Textlen)$ time.
\end{proposition}
\begin{proof}
  Denote $s = \HeadPat{f}{\tau}{\Pat}$, $H
  = \RootPat{f}{\tau}{\Pat}$, $p = |H|$, $k_1
  = \ExpCutPat{f}{\tau}{\Pat}{\ell}$, and $k_2
  = \ExpCutPat{f}{\tau}{\Pat}{2\ell}$.
  Using \cref{pr:sa-periodic-ints-access} and the position $j$ as
  input, in $\bigO(\log \Textlen)$ time we check if
  $\PairsMinus{f}{H}{\tau}{\Text} \neq \emptyset$ (note that
  $j \in \RFour{f}{H}{\tau}{\Text}$ follows
  by \cref{lm:periodic-pos-lce}\eqref{lm:periodic-pos-lce-it-1}), and
  if so, retrieve the pointer to the structure
  from \cref{pr:mod-queries} for $\PairsMinus{f}{H}{\tau}{\Text}$, i.e.,
  performing weighted modular constraint queries on $\Ints
  = \WInts{7\tau}{\PairsMinus{f}{H}{\tau}{\Text}}{\Text}$. If
  $\PairsMinus{f}{H}{\tau}{\Text} = \emptyset$, then
  by \cref{lm:pat-posmid-size}, it holds $|\PosMidMinus{f}{\Pat}{\Text}|
  = \ModCountTwoSide{\Ints}{p}{s}{k_1}{k_2} = 0$, and hence we return
  $0$.  Otherwise, in $\bigO(\log^{2 + \epsilon} \Textlen)$ time we compute
  $q := \ModCountTwoSide{\Ints}{p}{s}{k_1}{k_2}$
  using \cref{pr:mod-queries}.  By \cref{lm:pat-posmid-size}, it holds
  $|\PosMidMinus{f}{\Pat}{\Text}| = q$.  In total, we spend $\bigO(\log^{2
  + \epsilon} \Textlen)$ time.
\end{proof}

\begin{proposition}\label{pr:sa-periodic-posmid-pos}
  Let $k \in [4 \dd \lceil \log \Textlen \rceil)$, $\ell = 2^k$, $\tau
  = \lfloor \tfrac{\ell}{3} \rfloor$, and $f = f_{\tau,\Text}$
  (\cref{def:canonical-function}).  Let $j \in \RTwo{\tau}{\Text}$.  Given
  $\CompSaPeriodic{\Text}$, the values $k$, $\HeadPos{f}{\tau}{\Text}{j}$,
  $|\RootPos{f}{\tau}{\Text}{j}|$, $\ExpCutPos{f}{\tau}{\Text}{j}{\ell}$,
  $\ExpCutPos{f}{\tau}{\Text}{j}{2\ell}$, and some $j' \in
  \OccThree{3\tau - 1}{j}{\Text}$ satisfying $j' = \min \OccThree{2\ell}{j'}{\Text}$ as input, we
  can compute $|\PosMidMinus{f}{j}{\Text}|$ in $\bigO(\log^{2 + \epsilon}
  \Textlen)$ time.
\end{proposition}
\begin{proof}
  Denote $s = \HeadPos{f}{\tau}{\Text}{j}$, $H
  = \RootPos{f}{\tau}{\Text}{j}$, $p = |H|$, $k_1
  = \ExpCutPos{f}{\tau}{\Text}{j}{\ell}$, $k_2
  = \ExpCutPos{f}{\tau}{\Text}{j}{2\ell}$, and $\Pat = \Text[j \dd \Textlen]$.  By
  $j \in \RTwo{\tau}{\Text}$, $\Pat$ is $\tau$-periodic.  We then also have
  $j' \in \OccThree{3\tau - 1}{j}{\Text} = \OccTwo{\Pat[1 \dd 3\tau - 1]}{\Text}$
  and $\HeadPat{f}{\tau}{\Pat} = s$, $\RootPat{f}{\tau}{\Pat}
  = H$, $\ExpCutPat{f}{\tau}{\Pat}{\ell} = k_1$, and
  $\ExpCutPat{f}{\tau}{\Pat}{2\ell} = k_2$.  Thus, using $s$, $p$,
  $k_1$, $k_2$, and $j'$ as input to \cref{pr:sa-periodic-posmid}, we
  can compute $|\PosMidMinus{f}{j}{\Text}| = |\PosMidMinus{f}{\Text[j \dd
  \Textlen]}{\Text}| = |\PosMidMinus{f}{\Pat}{\Text}|$ in $\bigO(\log^{2 + \epsilon}
  \Textlen)$ time.
\end{proof}

\begin{proposition}\label{pr:sa-periodic-posmid-pat}
  Let $k \in [4 \dd \lceil \log \Textlen \rceil)$, $\ell = 2^k$, $\tau
  = \lfloor \tfrac{\ell}{3} \rfloor$, and $f = f_{\tau,\Text}$
  (\cref{def:canonical-function}).  Let $j \in \RTwo{\tau}{\Text}$ be a
  position satisfying $j = \min \OccThree{2\ell}{j}{\Text}$.  Denote $s
  = \HeadPos{f}{\tau}{\Text}{j}$ and $H = \RootPos{f}{\tau}{\Text}{j}$.
  Let $H'$ be a length-$s$ suffix of $H$ and $\Pat$ be a
  length-$2\ell$ prefix of $H' H^{\infty}$. Then, $\Pat$ is
  $\tau$-periodic.  Moreover, given $\CompSaPeriodic{\Text}$, the value
  $k$, the position $j$, and the values $\HeadPos{f}{\tau}{\Text}{j}$
  and $|\RootPos{f}{\tau}{\Text}{j}|$ as input, we compute
  $|\PosMidMinus{f}{\Pat}{\Text}|$ in $\bigO(\log^{2+\epsilon} \Textlen)$ time.
\end{proposition}
\begin{proof}
  Denote $p = |H|$. In $\bigO(1)$ time, we compute $k_1
  = \lfloor \tfrac{\ell - s}{p} \rfloor$ and $k_2
  = \lfloor \tfrac{2\ell -
  s}{p} \rfloor$. By \cref{lm:special-pat-properties}, $\Pat$ is
  $\tau$-periodic and it holds $\HeadPat{f}{\tau}{\Pat} = s$,
  $|\RootPat{f}{\tau}{\Pat}| = p$,
  $\ExpCutPat{f}{\tau}{\Pat}{\ell} = k_1$, and
  $\ExpCutPat{f}{\tau}{\Pat}{2\ell} = k_2$.  It also implies that
  $j \in \OccThree{3\tau - 1}{\Pat}{\Text}$. Thus, using $j$, $s$, $p$,
  $k_1$, and $k_2$ as input to \cref{pr:sa-periodic-posmid}, we
  compute $|\PosMidMinus{f}{\tau}{\Pat}|$ in $\bigO(\log^{2 + \epsilon}
  \Textlen)$ time.
\end{proof}

\subsubsection{Computing the Exponent}

\paragraph{Combinatorial Properties}

\begin{lemma}\label{lm:sa-periodic-poslow-poshigh-range}
  Let $\ell \in [16 \dd \Textlen)$, $\tau = \lfloor \tfrac{\ell}{3} \rfloor$,
  and $f$ be any necklace-consistent function.
  Let $\Pat \in \Sigma^{+}$ be a $\tau$-periodic pattern.
  Then, it holds $\PosLowMinus{f}{\Pat}{\Text} = \{\SA[i] : i
  \in (b \dd e]\}$ (resp.\ $\PosHighMinus{f}{\Pat}{\Text} = \{\SA[i] : i \in
  (b \dd e]\}$), where $b = \RangeBegThree{\ell}{\Pat}{\Text}$ (resp.\ $b =
  \RangeBegThree{2\ell}{\Pat}{\Text}$) and $e = b + \DeltaLowMinus{f}{\Pat}{\Text}$ (resp.\ 
  $e = b + \DeltaHighMinus{f}{\Pat}{\Text}$).
\end{lemma}
\begin{proof}

  We prove the claim only for $\PosLowMinus{f}{\Pat}{\Text}$. The proof for
  $\PosHighMinus{f}{\Pat}{\Text}$ is nearly identical, except we replace
  $\PosLowMinus{f}{\Pat}{\Text}$, $\DeltaLowMinus{f}{\Pat}{\Text}$, $\ell$, $k_1$, and
  $\ExpCutPat{f}{\tau}{\Pat}{\ell}$ with $\PosHighMinus{f}{\Pat}{\Text}$, $\DeltaHighMinus{f}{\Pat}{\Text}$,
  $2\ell$, $k_2$, and $\ExpCutPat{f}{\tau}{\Pat}{2\ell}$, respectively.

  Denote $m = |\Pat|$ and $\Pat' = \Pat[1 \dd \min(m,\ell)]$. First,
  we prove that $\PosLowMinus{f}{\Pat}{\Text} \subseteq \{\SA[i] : i \in (b \dd
  \Textlen]\}$.  Consider any $j' \in \PosLowMinus{f}{\Pat}{\Text}$ and let $i' \in [1 \dd
  \Textlen]$ be such that $\SA[i'] = j'$. By definition of $\PosLowMinus{f}{\Pat}{\Text}$,
  it holds $\Text[j' \dd \Textlen] \succeq \Pat$ or $\lcp(\Pat, \Text[j' \dd \Textlen])
  \geq \ell$.  By \cref{lm:sa-prelim}\eqref{lm:sa-prelim-it-1},
  this implies that $\Text[j' \dd \Textlen] \succeq \Pat'$. On
  the other hand, by \cref{lm:sa-prelim}\eqref{lm:sa-prelim-it-2},
  for every $j \in [1 \dd
  \Textlen]$, $\Text[j \dd \Textlen] \prec \Pat$ and $\lcp(\Pat, \Text[j \dd \Textlen]) < \ell$
  holds it and only if $\Text[j \dd \Textlen] \prec \Pat'$. Thus, applying the
  definition of $b$, we obtain $b = |\{j \in [1 \dd \Textlen] : \Text[j \dd \Textlen]
  \prec \Pat\text{ and } \lcp(\Pat, \Text[j \dd \Textlen]) < \ell\}| = |\{j \in
  [1 \dd \Textlen] : \Text[j \dd \Textlen] \prec \Pat'\}|$.  By definition of suffix
  array, positions in the last set must occupy its initial segment.
  We thus have $\{\SA[i] : i \in [1 \dd b]\} = \{j \in [1 \dd \Textlen] :
  \Text[j \dd \Textlen] \prec \Pat'\}$.  Consequently, $j' \not\in \{\SA[i] : i
  \in [1 \dd b]\}$, i.e., $i' \in (b \dd \Textlen]$. We have thus proved
  $\PosLowMinus{f}{\Pat}{\Text} \subseteq \{\SA[i] : i \in (b \dd \Textlen]\}$.

  Next, we show that for every $i' \in (b + 1 \dd \Textlen]$, $\SA[i'] \in
  \PosLowMinus{f}{\Pat}{\Text}$ implies $\SA[i'-1] \in \PosLowMinus{f}{\Pat}{\Text}$.  Denote $s
  = \HeadPat{f}{\tau}{\Pat}$, $H = \RootPat{f}{\tau}{\Pat}$, and $p = |H|$.
  \begin{itemize}
  \item As noted above $i' - 1 > b$ implies that $\Text[\SA[i'-1] \dd \Textlen]
    \succeq \Pat'$. By \cref{lm:sa-prelim}\eqref{lm:sa-prelim-it-1},
    this implies $\Text[\SA[i' - 1] \dd \Textlen] \succeq \Pat$ or $\lcp(\Pat,
    \Text[\SA[i' - 1] \dd \Textlen]) \geq \ell$.
  \item First, note that $\SA[i'] \in \PosLowMinus{f}{\Pat}{\Text}$ implies that
    $\SA[i'] \in \RMinusFive{f}{s}{H}{\tau}{\Text}$. By
    \cref{lm:periodic-pos-lce}\eqref{lm:periodic-pos-lce-it-1}, we thus have $\lcp(\Pat,
    \Text[\SA[i'] \dd \Textlen]) \geq 3\tau - 1$.  Denote $h = \lcp(\Text[\SA[i' -
    1] \dd \Textlen], \Text[\SA[i'] \dd \Textlen])$. We show that $h \geq 3\tau -
    1$. Assume the opposite and consider two cases:
    \begin{itemize}
    \item First, assume $\Textlen - \SA[i' - 1] + 1 = h$. Then, $\Text[\SA[i'-1]
      \dd \Textlen]$ is a proper prefix of $\Text[\SA[i'] \dd \SA[i'] + 3\tau -
      1) = \Pat[1 \dd 3\tau - 1]$. Note, however, that $|\Pat'| =
      \min(m, \ell) \geq 3\tau - 1$, since $\ell \geq 3\tau - 1$
      follows by the definition, and $m \geq 3\tau - 1$ follows since
      we assumed that $\Pat$ is $\tau$-periodic
      (\cref{def:periodic-pattern}). Since $\Pat'$ is a prefix of
      $\Pat$, we thus obtain that $\Text[\SA[i'-1] \dd \Textlen]$ is a proper
      prefix of $\Pat'$. Thus, $\Text[\SA[i'-1] \dd \Textlen] \prec \Pat'$.
      However, as observed above, this implies $i' - 1 \in [1 \dd b]$,
      contradicting the assumption $i' \in (b + 1 \dd \Textlen]$.
    \item Let us now assume $\Textlen - \SA[i' - 1] + 1 > h$. Then,
      $\SA[i'-1]$ being in the suffix array earlier than $\SA[i']$
      implies that $\Textlen - \SA[i'] + 1 > h$ and $\Text[\SA[i' - 1] + h]
      \prec \Text[\SA[i'] + h]$. Note, however, since by $h < 3\tau - 1$,
      we have $\Text[\SA[i'] \dd \SA[i'] + h] = \Pat'[1 \dd h + 1]$. We
      therefore again obtain $\Text[\SA[i' - 1] \dd \Textlen] \prec \Pat'$,
      which by the above implies $i' - 1 \in [1 \dd b]$, contradicting
      the assumption $i' \in (b + 1 \dd \Textlen]$.
    \end{itemize}
    We thus obtain $h \geq 3\tau - 1$. By
    \cref{lm:periodic-pos-lce}\eqref{lm:periodic-pos-lce-it-2}, this implies
    $\HeadPos{f}{\tau}{\Text}{\SA[i'-1]} = \HeadPos{f}{\tau}{\Text}{\SA[i']} = s$ and $\RootPos{f}{\tau}{\Text}{\SA[i'-1]} =
    \RootPos{f}{\tau}{\Text}{\SA[i']} = H$. Lastly, observe that we must also have
    $\TypePos{\tau}{\Text}{\SA[i'-1]} = -1$, since otherwise by
    \cref{lm:R-lex-block-pos}\eqref{lm:R-lex-block-pos-it-2}, we would
    have $\Text[\SA[i'-1] \dd \Textlen] \succ \Text[\SA[i'] \dd \Textlen]$, contradicting
    the definition of suffix array. We have thus proved $\SA[i'-1] \in
    \RMinusFive{f}{s}{H}{\tau}{\Text}$.
  \item Denote $k_1 = \ExpCutPat{f}{\tau}{\Pat}{\ell}$. We show that
    $\ExpPos{f}{\tau}{\Text}{\SA[i'-1]} = k_1$. To this end, we first prove that $\Pat'$
    is $\tau$-periodic and it holds $\HeadPat{f}{\tau}{\Pat'} = s$,
    $\RootPat{f}{\tau}{\Pat'} = H$, and $\ExpPat{f}{\tau}{\Pat'} = k_1$.  As noted above, it
    holds $|\Pat'| \geq 3\tau - 1$. Moreover, $\Pat'$ is a prefix of
    $\Pat$.  By \cref{lm:periodic-pat-lce}, we thus obtain that $\Pat'$ is
    $\tau$-periodic and it holds $\HeadPat{f}{\tau}{\Pat'} = \HeadPat{f}{\tau}{\Pat} = s$
    and $\RootPat{f}{\tau}{\Pat'} = \RootPat{f}{\tau}{\Pat} = H$. Lastly, by
    \cref{lm:pat-expcut}, we have $k_1 = \ExpCutPat{f}{\tau}{\Pat}{\ell} =
    \ExpPat{f}{\tau}{\Pat[1 \dd \min(m, \ell)]} = \ExpPat{f}{\tau}{\Pat'}$.  We have thus
    proved that $\Pat'$ is $\tau$-periodic and it holds $\HeadPat{f}{\tau}{\Pat'}
    = s$, $\RootPat{f}{\tau}{\Pat'} = H$, and $\ExpPat{f}{\tau}{\Pat'} = k_1$.  Observe now
    that by \cref{lm:R-lex-block-pos}\eqref{lm:R-lex-block-pos-it-3},
    we have $\RunEndPos{\tau}{\Text}{\SA[i'-1]} - \SA[i'-1] \leq \RunEndPos{\tau}{\Text}{\SA[i']}-\SA[i']$.
    Thus, $\ExpPos{f}{\tau}{\Text}{\SA[i'-1]} = \lfloor
    \tfrac{\RunEndPos{\tau}{\Text}{\SA[i'-1]}-\SA[i'-1]-s}{p} \rfloor \leq \lfloor
    \tfrac{\RunEndPos{\tau}{\Text}{\SA[i']}-\SA[i']-s}{p} \rfloor = \ExpPos{f}{\tau}{\Text}{\SA[i']} =
    k_1$ (the last equality follows by $\SA[i'] \in \PosLowMinus{f}{\Pat}{\Text}$).
    Consequently, to show $\ExpPos{f}{\tau}{\Text}{\SA[i'-1]} = k_1$, it remains to
    prove that we cannot have $\ExpPos{f}{\tau}{\Text}{\SA[i'-1]} < k_1$. Let us thus
    assume that this holds.  Then, $\RunEndPos{\tau}{\Text}{\SA[i'-1]} - \SA[i'-1] =
    \HeadPos{f}{\tau}{\Text}{\SA[i'-1]} + \ExpPos{f}{\tau}{\Text}{\SA[i'-1]}|\RootPos{f}{\tau}{\Text}{\SA[i'-1]}| +
    \TailPos{f}{\tau}{\Text}{\SA[i'-1]} \leq s + (k_1-1)p + \TailPos{f}{\tau}{\Text}{\SA[i'-1]} < s + k_1
    p \leq s + k_1 p + \TailPat{f}{\tau}{\Pat'} = \HeadPat{f}{\tau}{\Pat'} +
    \ExpPat{f}{\tau}{\Pat'}|\RootPat{f}{\tau}{\Pat'}| + \TailPat{f}{\tau}{\Pat'} = \RunEndPat{\tau}{\Pat'} - 1$.
    By \cref{lm:R-lex-block-pat}\eqref{lm:R-lex-block-pat-it-3}, we then
    have $\Text[\SA[i'-1] \dd \Textlen] \prec \Pat'$. As noted above, this implies
    $i'-1 \in [1 \dd b]$, contradicting $i' \in (b + 1 \dd \Textlen]$. We
    have thus proved that $\ExpPos{f}{\tau}{\Text}{\SA[i'-1]} = k_1$.
  \end{itemize}
  Combining the above conditions implies $\SA[i'-1] \in
  \PosLowMinus{f}{\Pat}{\Text}$. Consequently, $\PosLowMinus{f}{\Pat}{\Text} = \{\SA[i] : i \in (b
  \dd b + |\PosLowMinus{f}{\Pat}{\Text}|]\} = \{\SA[i] : i \in (b \dd e]\}$.
\end{proof}

\begin{corollary}\label{cor:sa-periodic-poslow-poshigh-range}
  Let $\ell \in [16 \dd \Textlen)$, $\tau = \lfloor \tfrac{\ell}{3} \rfloor$,
  and $f$ be any necklace-consistent function.
  Let $\Pat \in \Sigma^{+}$ be a $\tau$-periodic pattern.
  For any $i \in [1 \dd \Textlen]$, $\SA[i] \in
  \PosLowMinus{f}{\Pat}{\Text}$ (resp.\ $\SA[i] \in \PosHighMinus{f}{\Pat}{\Text}$) holds if and
  only if $\RangeBegThree{\ell}{\Pat}{\Text} < i \leq \RangeBegThree{\ell}{\Pat}{\Text} +
  \DeltaLowMinus{f}{\Pat}{\Text}$ (resp.\ $\RangeBegThree{2\ell}{\Pat}{\Text} < i \leq
  \RangeBegThree{2\ell}{\Pat}{\Text} + \DeltaHighMinus{f}{\Pat}{\Text}$).
\end{corollary}

\begin{lemma}\label{lm:sa-periodic-exp-sets}
  Let $\ell \in [16 \dd \Textlen)$, $\tau = \lfloor \tfrac{\ell}{3} \rfloor$,
  and $f$ be any necklace-consistent function.
  Let $\Pat \in \Sigma^{+}$ be a $\tau$-periodic pattern satisfying
  $\TypePat{\tau}{\Pat} = -1$.  Denote $s = \HeadPat{f}{\tau}{\Pat}$, $H =
  \RootPat{f}{\tau}{\Pat}$, $k_1 = \ExpCutPat{f}{\tau}{\Pat}{\ell}$,
  $b = \RangeBegThree{\ell}{\Pat}{\Text}$, and $\delta = \DeltaLowMinus{f}{\Pat}{\Text}$. For any $k \geq k_1$, let
  $E_{k} := \bigcup_{t \in (k_1 \dd k]}\RMinusSix{f}{s}{t}{H}{\tau}{\Text}$ and $m_k =
  |E_k|$.  It holds $E_k = \{\SA[i] : i \in (b + \delta \dd b +
  \delta + m_k]\}$.
\end{lemma}
\begin{proof}

  Denote $p = |H|$, $i' = b + \delta$, and $B = \{j \in [1 \dd \Textlen] :
  \Text[j \dd \Textlen] \prec \Pat\text{ and }\lcp(\Pat, \Text[j \dd \Textlen]) < \ell\}$.
  In the proof of \cref{lm:sa-periodic-poslow-poshigh-range}, we
  observed that $B = \{\SA[i] : i \in [1 \dd b]\}$. We proceed in
  two steps.

  First, we show that $E_{k} \subseteq \{\SA[i''] : i'' \in (i' \dd
  \Textlen]\}$. The claim holds trivially for $k = k_1$ since $E_{k_1} =
  \emptyset$. Let us thus assume $k > k_1$. Let $j \in E_{k}$ and let
  $i \in [1 \dd \Textlen]$ be such that $\SA[i] = j$. Note that by
  definition, we have $\ExpPat{f}{\tau}{\Pat} \geq k_1$. Consider two cases:
  \begin{itemize}
  \item First, assume that $\ExpPat{f}{\tau}{\Pat} = k_1$. Then, $\RunEndPat{\tau}{\Pat} - 1 =
    \HeadPat{f}{\tau}{\Pat} + \ExpPat{f}{\tau}{\Pat}|\RootPat{f}{\tau}{\Pat}| + \TailPat{f}{\tau}{\Pat} = s + k_1
    p + \TailPat{f}{\tau}{\Pat} < s + (k_1 + 1)p \leq s + (k_1 + 1)p + \TailPos{f}{\tau}{\Text}{j}
    \leq \HeadPos{f}{\tau}{\Text}{j} + \ExpPos{f}{\tau}{\Text}{j}|\RootPos{f}{\tau}{\Text}{j}| + \TailPos{f}{\tau}{\Text}{j} = \RunEndPos{\tau}{\Text}{j} -
    j$. By \cref{lm:R-lex-block-pat}\eqref{lm:R-lex-block-pat-it-1} and \cref{lm:R-lex-block-pat}\eqref{lm:R-lex-block-pat-it-5},
    we thus have $\Pat \prec \Text[j \dd \Textlen]$. Consequently, $j \not\in B$,
    and hence $i \in (b \dd \Textlen]$. Note, however, that since for every $j'
    \in \PosLowMinus{f}{\Pat}{\Text}$ we have $\ExpPos{f}{\tau}{\Text}{j'} = k_1 < \ExpPos{f}{\tau}{\Text}{j}$, it also
    holds $j \not\in \PosLowMinus{f}{\Pat}{\Text}$. By
    \cref{cor:sa-periodic-poslow-poshigh-range}, we thus have
    $i \not\in (b \dd b + \delta] = (b \dd i']$. Consequently,
    $i \in (i' \dd \Textlen]$. Thus, $j \in \{\SA[i''] : i'' \in (i' \dd \Textlen]\}$.
  \item Let us now assume $\ExpPat{f}{\tau}{\Pat} > k_1$.  We begin by observing
    that the assumption $\ExpPat{f}{\tau}{\Pat} > k_1$ implies that, letting $H'$
    be a length-$s$ suffix of $H$, the string $H' H^{k_1 + 1}$ is a
    prefix of $\Pat$. On the other hand, by definition of $E_{k}$, we
    have $\ExpPos{f}{\tau}{\Text}{j} > k_1$. Thus, $H' H^{k_1 + 1}$ is also a prefix of
    $\Text[j \dd \Textlen]$, and hence $\lcp(\Pat, \Text[j \dd \Textlen]) \geq s + (k_1 +
    1)p$. Observe now that the assumption $k_1 = \min(\ExpPat{f}{\tau}{\Pat},
    \lfloor \tfrac{\ell - s}{p} \rfloor) < \ExpPat{f}{\tau}{\Pat}$ implies that
    $k_1 = \lfloor \tfrac{\ell - s}{p} \rfloor$.  Consequently, $s +
    (k_1 + 1)p = s + (\lfloor \tfrac{\ell - s}{p} \rfloor + 1)p \geq s
    + (\tfrac{\ell - s}{p})p = \ell$. Thus, $\lcp(\Pat, \Text[j \dd \Textlen])
    \geq \ell$, and hence $j \not\in B$. Similarly, as in the first
    case, we also have $j \not\in \PosLowMinus{f}{\Pat}{\Text}$.  Thus, again $i \in
    (i' \dd \Textlen]$ and hence $j \in \{\SA[i''] : i'' \in (i' \dd \Textlen]\}$.
  \end{itemize}

  Second, we prove that for every $i'' \in (i' + 1 \dd \Textlen]$, $\SA[i'']
  \in E_{k}$ implies $\SA[i''-1] \in E_{k}$.  We begin by proving that
  $\SA[i''-1] \in \RFive{f}{s}{H}{\tau}{\Text}$ and $\ExpPos{f}{\tau}{\Text}{\SA[i''-1]} \geq k_1$.  Observe
  that $i'' - 1 > i' \geq b$ implies that $\SA[i''-1] \not\in B$,
  i.e., that it holds $\Text[\SA[i''-1] \dd \Textlen] \succeq \Pat$ or
  $\lcp(\Pat, \Text[\SA[i''-1] \dd \Textlen]) \geq \ell$. Consider two cases:
  \begin{itemize}
  \item First, assume $\lcp(\Pat, \Text[\SA[i''-1] \dd \Textlen]) \geq \ell$.
    By $\ell \geq 3\tau - 1$ and
    \cref{lm:periodic-pos-lce}\eqref{lm:periodic-pos-lce-it-1}, we then obtain
    $\SA[i''-1] \in \RFive{f}{s}{H}{\tau}{\Text}$.  Denote $\Pat' = \Pat[1 \dd
    \ell]$. Note that by \cref{lm:periodic-pat-lce}, $\Pat'$ is
    $\tau$-periodic and we have $\HeadPat{f}{\tau}{\Pat'} = s$ and $\RootPat{f}{\tau}{\Pat'}
    = H$.  Observe that $\RunEndPos{\tau}{\Text}{\SA[i''-1]} - \SA[i''-1] = p +
    \lcp(\Text[\SA[i''-1] \dd \Textlen], \Text[\SA[i''-1] + p \dd \Textlen]) \geq p +
    \lcp(\Text[\SA[i''-1] \dd \SA[i''-1] + \ell), \Text[\SA[i''-1] + p \dd
    \SA[i''-1] + \ell)) = p + \lcp(\Pat'[1 \dd \ell], \Pat'[1 + p \dd
    \ell]) = \RunEndPat{\tau}{\Pat'} - 1$. Consequently, $\ExpPos{f}{\tau}{\Text}{\SA[i''-1]} =
    \lfloor \tfrac{\RunEndPos{\tau}{\Text}{\SA[i''-1]} - \SA[i''-1] - s}{p} \rfloor \geq
    \lfloor \tfrac{\RunEndPat{\tau}{\Pat'} - 1 - s}{p} \rfloor =
    \ExpPat{f}{\tau}{\Pat'}$. It remains to note that by \cref{lm:pat-expcut},
    $\ExpPat{f}{\tau}{\Pat'} = \ExpCutPat{f}{\tau}{\Pat}{\ell} = k_1$. Hence,
    $\ExpPos{f}{\tau}{\Text}{\SA[i''-1]} \geq k_1$.
  \item Let us now assume $\Text[\SA[i''-1] \dd \Textlen] \succeq \Pat$. If
    $\Text[\SA[i''-1] \dd \Textlen] = \Pat$, then by
    \cref{lm:periodic-pos-lce}\eqref{lm:periodic-pos-lce-it-1}, we immediately obtain
    that $\SA[i''-1] \in \RFive{f}{s}{H}{\tau}{\Text}$ and also by definition
    $\ExpPos{f}{\tau}{\Text}{\SA[i''-1]} = \ExpPat{f}{\tau}{\Text[\SA[i''-1] \dd \Textlen]} = \ExpPat{f}{\tau}{\Pat}
    \geq k_1$. Let us thus assume $\Text[\SA[i''-1] \dd \Textlen] \succ \Pat$.
    Denote $X = \Pat$, $Y = \Text[\SA[i''-1] \dd \Textlen]$, and $Z =
    \Text[\SA[i''] \dd \Textlen]$. We then have $X \prec Y \prec Z$. Note now
    that both $X$ and $Z$ are $\tau$-periodic and it holds $\HeadPat{f}{\tau}{X}
    = \HeadPat{f}{\tau}{Z}$ and $\RootPat{f}{\tau}{X} = \RootPat{f}{\tau}{Z}$.  Thus, by
    \cref{lm:periodic-pat-lce}, we have $\lcp(X, Z) \geq 3\tau - 1$.  Observe
    now that since $\lcp(X, Z) = \min(\lcp(X, Y), \lcp(Y, Z))$ holds
    for any strings satisfying $X \prec Y \prec Z$, we thus must have
    that $\lcp(X, Y) = \lcp(\Pat, \Text[\SA[i''-1] \dd \Textlen]) \geq 3\tau -
    1$.  Thus, by \cref{lm:periodic-pos-lce}\eqref{lm:periodic-pos-lce-it-1}, we have
    $\SA[i''-1] \in \RFive{f}{s}{H}{\tau}{\Text}$.  Let $H'$ be a length-$s$ suffix of
    $H$. By $\ExpPat{f}{\tau}{\Pat} \geq \ExpCutPat{f}{\tau}{\Pat}{\ell} = k_1$, $H'
    H^{k_1}$ is a prefix of $\Pat$.  On the other hand, $\SA[i''] \in
    E_{k}$ implies $\ExpPos{f}{\tau}{\Text}{\SA[i'']} > k_1$. Thus, $H' H^{k_1}$ is also
    a prefix of $\Text[\SA[i''] \dd \Textlen]$.  Consequently, $\lcp(X, Z) \geq
    s + k_1 p$. By the same argument as above, we thus also have
    $\lcp(X, Y) = \lcp(\Pat, \Text[\SA[i''-1] \dd \Textlen]) \geq s + k_1 p$,
    i.e., $H' H^{k_1}$ is also a prefix of $\Text[\SA[i''-1] \dd
    \Textlen]$. Since $\SA[i''-1] \in \RFive{f}{s}{H}{\tau}{\Text}$, this implies
    $\RunEndPos{\tau}{\Text}{\SA[i''-1]} - \SA[i''-1] = p + \lcp(\Text[\SA[i''-1] \dd \Textlen],
    \Text[\SA[i''-1] + p \dd \Textlen]) \geq s + k_1 p$, and consequently,
    $\ExpPos{f}{\tau}{\Text}{\SA[i''-1]} = \lfloor \tfrac{\RunEndPos{\tau}{\Text}{\SA[i''-1]} - \SA[i''-1]
    - s}{p} \rfloor \geq \lfloor \tfrac{k_1 p}{p} \rfloor = k_1$.
  \end{itemize}
  We have thus proved that $\SA[i''-1] \in \RFive{f}{s}{H}{\tau}{\Text}$ and
  $\ExpPos{f}{\tau}{\Text}{\SA[i''-1]} \geq k_1$.  Next, observe that we must have
  $\TypePos{\tau}{\Text}{\SA[i''-1]} = -1$, since otherwise by $\TypePos{\tau}{\Text}{\SA[i'']} = -1$
  and \cref{lm:R-lex-block-pos}\eqref{lm:R-lex-block-pos-it-2}, we
  would have $\Text[\SA[i''-1] \dd \Textlen] \succ \Text[\SA[i''] \dd \Textlen]$,
  contradicting the definition of the suffix array. We have thus
  proved that $\SA[i''-1] \in \RMinusFive{f}{s}{H}{\tau}{\Text}$. Recall now the definition
  of $\PosLowMinus{f}{\Pat}{\Text}$. Observe that almost all conditions are
  satisfied for $\SA[i''-1]$, except possibly the condition on the
  exponent. Since, however, we assumed $i'' - 1 > i'$, by
  \cref{lm:sa-periodic-poslow-poshigh-range} we have $\SA[i''-1]
  \not\in \PosLowMinus{f}{\Pat}{\Text}$. Consequently, we must have
  $\ExpPos{f}{\tau}{\Text}{\SA[i''-1]} \neq k_1$, and hence $\ExpPos{f}{\tau}{\Text}{\SA[i''-1]} > k_1$.
  Lastly, we prove that $\ExpPos{f}{\tau}{\Text}{\SA[i''-1]} \leq k$.  Suppose
  $\ExpPos{f}{\tau}{\Text}{\SA[i''-1]} \geq k+1$. Then, it holds $\RunEndPos{\tau}{\Text}{\SA[i''-1]} -
  \SA[i''-1] = \HeadPos{f}{\tau}{\Text}{\SA[i''-1]} +
  \ExpPos{f}{\tau}{\Text}{\SA[i''-1]}|\RootPos{f}{\tau}{\Text}{\SA[i''-1]}| + \TailPos{f}{\tau}{\Text}{\SA[i''-1]} \geq s +
  (k+1)p > s + kp + \TailPos{f}{\tau}{\Text}{\SA[i'']} \geq s + \ExpPos{f}{\tau}{\Text}{\SA[i'']}p +
  \TailPos{f}{\tau}{\Text}{\SA[i'']} = \HeadPos{f}{\tau}{\Text}{\SA[i'']} +
  \ExpPos{f}{\tau}{\Text}{\SA[i'']}|\RootPos{f}{\tau}{\Text}{\SA[i'']}| + \TailPos{f}{\tau}{\Text}{\SA[i'']} =
  \RunEndPos{\tau}{\Text}{\SA[i'']} - \SA[i'']$. By $\SA[i''-1], \SA[i''] \in
  \RMinusFive{f}{s}{H}{\tau}{\Text}$ and
  \cref{lm:R-lex-block-pos}\eqref{lm:R-lex-block-pos-it-3} this,
  however, implies $\Text[\SA[i''-1] \dd \Textlen] \succ \Text[\SA[i''] \dd \Textlen]$,
  contradicting the definition of the suffix array. Thus, it holds
  $\ExpPos{f}{\tau}{\Text}{\SA[i''-1]} \leq k$.  We have therefore proved
  $\ExpPos{f}{\tau}{\Text}{\SA[i''-1]} \in (k_1 \dd k]$, which concludes the proof of
  $\SA[i''-1] \in E_{k}$.

  Putting the above steps together implies that
  $E_k = \{\SA[i''] : i'' \in (i' \dd i' + m_k]\}$.
\end{proof}

\begin{lemma}\label{lm:sa-periodic-exp-pat}
  Let $\ell \in [16 \dd \Textlen)$, $\tau = \lfloor \tfrac{\ell}{3} \rfloor$,
  and $f$ be any necklace-consistent function.
  Let $\Pat \in \Sigma^{+}$ be a $\tau$-periodic pattern such that it
  holds $\TypePat{\tau}{\Pat} = -1$, $\RunEndPat{\tau}{\Pat} \leq |\Pat|$,
  and $\ExpCutPat{f}{\tau}{\Pat}{\ell} < \ExpPat{f}{\tau}{\Pat}
  \leq \lfloor \tfrac{7\tau - s}{p} \rfloor$, where $s =
  \HeadPat{f}{\tau}{\Pat}$, $H = \RootPat{f}{\tau}{\Pat}$, and $p = |H|$. Denote $\Ints =
  \WInts{7\tau}{\PairsMinus{f}{H}{\tau}{\Text}}{\Text}$, $k_1 = \ExpCutPat{f}{\tau}{\Pat}{\ell}$, $c =
  \ModCountOneSide{\Ints}{p}{s}{k_1}$, and $i' = \RangeBegThree{\ell}{\Pat}{\Text}
  + \DeltaLowMinus{f}{\Pat}{\Text}$.  Let $i \in [1 \dd \Textlen]$ be such that $\SA[i] \in
  \OccTwo{\Pat}{\Text}$. Then, it holds:
  \begin{itemize}
  \item $c + (i - i') \in [1 \dd \ModCountZeroSide{\Ints}{p}{s}]$,
  \item $\ExpPat{f}{\tau}{\Pat} = \ModSelect{\Ints}{p}{s}{c + (i-i')}$.
  \end{itemize}
\end{lemma}
\begin{proof}

  Denote $k = \ExpPat{f}{\tau}{\Pat}$. The proof consists of two parts.

  In the first step, we prove that for every $j' \in \OccTwo{\Pat}{\Text}$,
  it holds $j' \in \RMinusSix{f}{s}{k}{H}{\tau}{\Text}$.
  Recall that $\RunEndPat{\tau}{\Pat} \leq |\Pat|$,
  or equivalently, $|\Pat| > \RunEndPat{\tau}{\Pat} - 1$. Consequently, $j' \in
  \OccTwo{\Pat}{\Text}$ implies $\lcp(\Pat, \Text[j' \dd \Textlen]) = |\Pat| >
  \RunEndPat{\tau}{\Pat} - 1$. From \cref{lm:periodic-pos-lce}\eqref{lm:periodic-pos-lce-it-1} we
  therefore obtain that $\HeadPos{f}{\tau}{\Text}{j'} = \HeadPat{f}{\tau}{\Pat}$, $\RootPos{f}{\tau}{\Text}{j'} =
  \RootPat{f}{\tau}{\Pat}$, $\TypePos{\tau}{\Text}{j'} = \TypePat{\tau}{\Pat} = -1$, and
  $\ExpPos{f}{\tau}{\Text}{j'} = \ExpPat{f}{\tau}{\Pat} = k$. In other words, $j' \in
  \RMinusSix{f}{s}{k}{H}{\tau}{\Text}$.

  In the second step, we prove that it holds $c + (i-i') \in [1 \dd
  \ModCountZeroSide{\Ints}{p}{s}]$ (i.e., that
  $\ModSelect{\Ints}{p}{s}{c + (i-i')}$ is well-defined) and $k =
  \ModSelect{\Ints}{p}{s}{c + (i-i')}$ (i.e., the main claim). Recall
  that $k_1 < k \leq \lfloor \tfrac{7\tau - s}{p} \rfloor$. For any
  $k' \geq k_1$, let $E_{k'} = \bigcup_{t \in (k_1 \dd k']}
  \RMinusSix{f}{s}{t}{H}{\tau}{\Text}$ and $m_{k'} = |E_{k'}|$.  Note that by $k_1 < k$,
  the sets $E_{k}$ and $E_{k-1}$ are well-defined and it holds
  $\RMinusSix{f}{s}{k}{H}{\tau}{\Text} = E_{k} \setminus E_{k-1}$.  Thus, by the first
  step, we have $\SA[i] \in E_{k} \setminus E_{k-1}$.  On the other
  hand, by \cref{lm:sa-periodic-exp-sets}, it holds $E_{k-1} =
  \{\SA[i''] : i'' \in (i' \dd i' + m_{k-1}]\}$ and $E_{k} =
  \{\SA[i''] : i'' \in (i' \dd i' + m_{k}]\}$.  Thus, we obtain $i \in
  (i' + m_{k-1} \dd i' + m_{k}]$. Observe now that by
  \cref{lm:sa-periodic-modcount,lm:kmin-kmax-range} we have $m_{k-1}
  = |E_{k-1}| = |\{j' \in \RMinusFive{f}{s}{H}{\tau}{\Text} : \ExpPos{f}{\tau}{\Text}{j'} \in (k_1 \dd
  k-1]\}| = \ModCountTwoSide{\Ints}{p}{s}{k_1}{k-1}$.  By
  \cref{lm:mod-queries-properties}\eqref{lm:mod-queries-properties-it-1},
  we thus obtain $m_{k-1} = \ModCountTwoSide{\Ints}{p}{s}{k_1}{k-1} =
  \ModCountOneSide{\Ints}{p}{s}{k-1} -
  \ModCountOneSide{\Ints}{p}{s}{k_1} =
  \ModCountOneSide{\Ints}{p}{s}{k-1} - c$.  Analogously, $m_{k} =
  \ModCountOneSide{\Ints}{p}{s}{k} - c$. We thus obtain $i \in (i' +
  \ModCountOneSide{\Ints}{p}{s}{k-1} - c \dd i' +
  \ModCountOneSide{\Ints}{p}{s}{k} - c]$, or equivalently, $c + (i -
  i') \in (\ModCountOneSide{\Ints}{p}{s}{k-1} \dd
  \ModCountOneSide{\Ints}{p}{s}{k}]$. This implies $c + (i-i') \in [1
  \dd \ModCountZeroSide{\Ints}{p}{s}]$ and, by definition of the
  weighted modular constraint selection queries (see
  \cref{sec:mod-queries}), $k = \ModSelect{\Ints}{p}{s}{c + (i-i')}$.
\end{proof}

\begin{lemma}\label{lm:sa-periodic-exp-pos}
  Let $\ell \in [16 \dd \Textlen)$, $\tau = \lfloor \tfrac{\ell}{3} \rfloor$,
  and $f$ be any necklace-consistent function.
  Let $i \in [1 \dd \Textlen]$ be such that $\SA[i] \in \RMinusTwo{\tau}{\Text}$ and
  $\RunEndPos{\tau}{\Text}{\SA[i]} - \SA[i] < 2\ell$.  Denote $b =
  \RangeBegThree{\ell}{\SA[i]}{\Text}$, $\delta = \DeltaLowMinus{f}{\SA[i]}{\Text}$, and $k_1 =
  \ExpCutPos{f}{\tau}{\Text}{\SA[i]}{\ell}$. If $i \leq b + \delta$, then
  $\ExpPos{f}{\tau}{\Text}{\SA[i]} = k_1$. Otherwise, letting $H = \RootPos{f}{\tau}{\Text}{\SA[i]}$, $p =
  |H|$, $s = \HeadPos{f}{\tau}{\Text}{\SA[i]}$, $\Ints = \WInts{7\tau}{\PairsMinus{f}{H}{\tau}{\Text}}{\Text}$,
  and $c = \ModCountOneSide{\Ints}{p}{s}{k_1}$, it holds:
  \begin{itemize}
  \item $c + (i - (b + \delta)) \in [1 \dd
    \ModCountZeroSide{\Ints}{p}{s}]$,
  \item $\ExpPos{f}{\tau}{\Text}{\SA[i]} = \ModSelect{\Ints}{p}{s}{c + (i - (b +
    \delta))}$.
  \end{itemize}
\end{lemma}
\begin{proof}

  Denote $\Pat = \Text[\SA[i] \dd \Textlen]$. First, observe that since we
  assumed $\SA[i] \in \RMinusTwo{\tau}{\Text}$, it holds that $\Pat$ is $\tau$-periodic,
  $\TypePat{\tau}{\Pat} = -1$, and $\RunEndPat{\tau}{\Pat} \leq |\Pat|$. By
  \cref{lm:periodic-pos-lce}\eqref{lm:periodic-pos-lce-it-1}, we also have $\HeadPat{f}{\tau}{\Pat}
  = s$ and $\RootPat{f}{\tau}{\Pat} = H$.  Next, note that by definition it
  holds:
  \begin{itemize}
  \item $\RangeBegThree{\ell}{\SA[i]}{\Text} = \RangeBegThree{\ell}{\Text[\SA[i] \dd \Textlen]}{\Text} =
    \RangeBegThree{\ell}{\Pat}{\Text}$,
  \item $\DeltaLowMinus{f}{\SA[i]}{\Text} = \DeltaLowMinus{f}{\Text[\SA[i], \dd \Textlen]}{\Text} =
    \DeltaLowMinus{f}{\Pat}{\Text}$, and
  \item $\ExpCutPos{f}{\tau}{\Text}{\SA[i]}{\ell} = \ExpCutPat{f}{\tau}{\Text[\SA[i] \dd \Textlen]}{\ell} =
    \ExpCutPat{f}{\tau}{\Pat}{\ell}$.
  \end{itemize}
  Consequently, we have $\RangeBegThree{\ell}{\Pat}{\Text} = b$, $\DeltaLowMinus{f}{\Pat}{\Text}
  = \delta$, and $\ExpCutPat{f}{\tau}{\Pat}{\ell} = k_1$.  By
  \cref{lm:sa-periodic-poslow-poshigh-range}, we thus have
  $\PosLowMinus{f}{\Pat}{\Text} = \{\SA[t] : t \in (b \dd b + \delta]\}$.

  Let us first assume that it holds $i \leq b + \delta$.  By
  definition, it holds $\SA[i] \in \OccThree{\ell}{\SA[i]}{\Text}$. Thus, $b
  < i$. On the other hand, we have $i \leq b + \delta$ by the
  assumption. Thus, by the above, it holds $\SA[i] \in
  \PosLowMinus{f}{\Pat}{\Text}$. By definition of $\PosLowMinus{f}{\Pat}{\Text}$, this implies
  $\ExpPos{f}{\tau}{\Text}{\SA[i]} = \ExpCutPat{f}{\tau}{\Pat}{\ell} = k_1$.

  Let us now assume $i > b + \delta$. By the above characterization of
  $\PosLowMinus{f}{\Pat}{\Text}$, this implies $\SA[i] \not\in \PosLowMinus{f}{\Pat}{\Text}$.  We
  now show $k_1 < \ExpPat{f}{\tau}{\Pat}$. By definition,
  $\ExpCutPat{f}{\tau}{\Pat}{\ell} = \min(\ExpPat{f}{\tau}{\Pat}, \lfloor \tfrac{\ell -
  s}{p} \rfloor) \leq \ExpPat{f}{\tau}{\Pat}$. Thus, $k_1 \leq
  \ExpPat{f}{\tau}{\Pat}$. Suppose that it holds $k_1 = \ExpPat{f}{\tau}{\Pat}$. Then,
  $\SA[i] \in \RMinusFive{f}{s}{H}{\tau}{\Text}$ and $\ExpPos{f}{\tau}{\Text}{\SA[i]} = \ExpPat{f}{\tau}{\Text[\SA[i] \dd \Textlen]} =
  \ExpPat{f}{\tau}{\Pat} = k_1 = \ExpCutPat{f}{\tau}{\Pat}{\ell}$.  Moreover, we
  then have $\Text[\SA[i] \dd \Textlen] = \Pat$. Thus, by definition of
  $\PosLowMinus{f}{\Pat}{\Text}$, we have $\SA[i] \in \PosLowMinus{f}{\Pat}{\Text}$, which
  contradicts our earlier observation. Consequently, we have $k_1 <
  \ExpPat{f}{\tau}{\Pat}$. On the other hand, observe that $\RunEndPos{\tau}{\Text}{\SA[i]} -
  \SA[i] < 2\ell$ implies $\ExpPat{f}{\tau}{\Pat} = \ExpPat{f}{\tau}{\Text[\SA[i] \dd \Textlen]} =
  \ExpPos{f}{\tau}{\Text}{\SA[i]} = \lfloor \tfrac{\RunEndPos{\tau}{\Text}{\SA[i]} - \SA[i] - s}{p}
  \rfloor \leq \lfloor \tfrac{2\ell - s}{p} \rfloor \leq \lfloor
  \tfrac{7\tau - s}{p} \rfloor$, where $2\ell \leq 7\tau$ follows by
  $\tau = \lfloor \tfrac{\ell}{3} \rfloor$ and $\ell \geq 16$.  Thus,
  it holds $k_1 < \ExpPat{f}{\tau}{\Pat} \leq \lfloor \tfrac{7\tau - s}{p}
  \rfloor$. Observe also that $\Pat = \Text[\SA[i] \dd \Textlen]$ implies that
  $\SA[i] \in \OccTwo{\Pat}{\Text}$. In particular, $\OccTwo{\Pat}{\Text} \neq
  \emptyset$. Lastly, recall that $\TypePat{\tau}{\Pat} = -1$. By
  \cref{lm:sa-periodic-exp-pat}, we thus have that, letting $i' =
  \RangeBegThree{\ell}{\Pat}{\Text} + \DeltaLowMinus{f}{\Pat}{\Text}$, it holds $c + (i - i') \in
  [1 \dd \ModCountZeroSide{\Ints}{p}{s}]$ and $\ExpPat{f}{\tau}{\Pat} =
  \ModSelect{\Ints}{p}{s}{c + (i - i')}$.  By $i' = b + \delta$, we
  thus obtain $c + (i - (b + \delta)) \in [1 \dd
  \ModCountZeroSide{\Ints}{p}{s}]$ and $\ExpPos{f}{\tau}{\Text}{\SA[i]} =
  \ExpPat{f}{\tau}{\Text[\SA[i] \dd \Textlen]} = \ExpPat{f}{\tau}{\Pat} = \ModSelect{\Ints}{p}{s}{c +
  (i - (b + \delta))}$.
\end{proof}

\paragraph{Query Algorithms}

\begin{proposition}\label{pr:sa-periodic-exp}
  Let $k \in [4 \dd \lceil \log \Textlen \rceil)$, $\ell = 2^k$, $\tau
  = \lfloor \tfrac{\ell}{3} \rfloor$, and $f = f_{\tau,\Text}$
  (\cref{def:canonical-function}). Let $i \in [1 \dd \Textlen]$ be such that
  $\SA[i] \in \RMinusTwo{\tau}{\Text}$ and $\RunEndPos{\tau}{\Text}{\SA[i]}
  - \SA[i] < 2\ell$. Given $\CompSaPeriodic{\Text}$, the value $k$, the
  position $i$, some $j \in \OccThree{3\tau - 1}{\SA[i]}{\Text}$ satisfying
  $j = \min \OccThree{2\ell}{j}{\Text}$, and the values
  $\HeadPos{f}{\tau}{\Text}{\SA[i]}$,
  $|\RootPos{f}{\tau}{\Text}{\SA[i]}|$, $\RangeBegThree{\ell}{\SA[i]}{\Text}$,
  $\ExpCutPos{f}{\tau}{\Text}{\SA[i]}{\ell}$, and
  $\DeltaLowMinus{f}{\SA[i]}{\Text}$ as input, we can compute
  $\ExpPos{f}{\tau}{\Text}{\SA[i]}$ in $\bigO(\log^{3 + \epsilon} \Textlen)$
  time.
\end{proposition}
\begin{proof}
  Denote $s = \HeadPos{f}{\tau}{\Text}{\SA[i]}$, $H
  = \RootPos{f}{\tau}{\Text}{\SA[i]}$, $p = |H|$, $b
  = \RangeBegThree{\ell}{\SA[i]}{\Text}$, $k_1
  = \ExpCutPos{f}{\tau}{\Text}{\SA[i]}{\ell}$, $\delta
  = \DeltaLowMinus{f}{\SA[i]}{\Text}$, and $\Ints
  = \WInts{7\tau}{\PairsMinus{f}{H}{\tau}{\Text}}{\Text}$.  First, we check if
  $i \leq b + \delta$. If so, then by \cref{lm:sa-periodic-exp-pos},
  it holds $\ExpPos{f}{\tau}{\Text}{\SA[i]}
  = \ExpCutPos{f}{\tau}{\Text}{\SA[i]}{\ell}$, and thus we return $k_1$
  as the output. Let us now assume $i > b + \delta$.
  Using \cref{pr:sa-periodic-ints-access} and the position $j$ as
  input, in $\bigO(\log \Textlen)$ time we retrieve the pointer to the
  structure from \cref{pr:mod-queries} for $\PairsMinus{f}{H}{\tau}{\Text}$
  ($j \in \RFour{f}{H}{\tau}{\Text}$ follows
  by \cref{lm:periodic-pos-lce}\eqref{lm:periodic-pos-lce-it-2}),
  i.e., performing weighted modular constraint queries on $\Ints
  = \WInts{7\tau}{\PairsMinus{f}{H}{\tau}{\Text}}{\Text}$.  Note that the pointer
  is not null, since we assumed $\SA[i] \in \RMinusTwo{\tau}{\Text}$.  Thus,
  $\RMinusFour{f}{H}{\tau}{\Text} \neq \emptyset$, which implies
  $\PairsMinus{f}{H}{\tau}{\Text} \neq \emptyset$.  Next,
  using \cref{pr:mod-queries} in $\bigO(\log^{2 + \epsilon} \Textlen)$ time
  we compute $c = \ModCountOneSide{\Ints}{p}{s}{k_1}$. In
  $\bigO(\log^{3 + \epsilon} \Textlen)$ time, we then compute $k
  = \ModSelect{\Ints}{p}{s}{c + (i - (b
  + \delta))}$. By \cref{lm:sa-periodic-exp-pos}, it holds $k
  = \ExpPos{f}{\tau}{\Text}{\SA[i]}$. We thus return $k$ as the
  answer. In total, we spend $\bigO(\log^{3 + \epsilon} \Textlen)$ time.
\end{proof}

\subsubsection{Computing the Size of Occ}\label{sec:sa-periodic-occ}

\paragraph{Combinatorial Properties}

\begin{lemma}\label{lm:sa-periodic-occ-run}
  Let $\ell \in [16 \dd \Textlen)$, $\tau = \lfloor \tfrac{\ell}{3} \rfloor$,
  and $f$ be any necklace-consistent function.
  Let $\Pat \in \Sigma^{m}$ be a $\tau$-periodic pattern such that
  $\TypePat{\tau}{\Pat} = -1$ and $\Text[\Textlen]$ does not occur in $\Pat[1 \dd m)$. Let
  $i \in \RTwo{\tau}{\Text}$. Denote $H = \RootPat{f}{\tau}{\Pat}$ and $t = \RunEndPos{\tau}{\Text}{i} - i - 3\tau
  + 2$. Then, $|\OccThree{2\ell}{\Pat}{\Text} \cap \PosHighMinus{f}{\Pat}{\Text} \cap [i
  \dd i + t)| \leq 1$. Moreover, $|\OccThree{2\ell}{\Pat}{\Text} \cap
  \PosHighMinus{f}{\Pat}{\Text} \cap [i \dd i + t)| = 1$ holds if and only if
  \begin{itemize}
  \item $i \in \RMinusFour{f}{H}{\tau}{\Text}$,
  \item $\RunEndFullPos{f}{\tau}{\Text}{i} - i \geq \RunEndCutPat{f}{\tau}{\Pat}{2\ell} - 1$, and
  \item $\Pat[\RunEndCutPat{f}{\tau}{\Pat}{2\ell} \dd \min(m,2\ell)]$ is a prefix of
    $\Textinf[\RunEndFullPos{f}{\tau}{\Text}{i} \dd \RunEndFullPos{f}{\tau}{\Text}{i} + 7\tau)$.
  \end{itemize}
  Lastly, if $\OccThree{2\ell}{\Pat}{\Text} \cap \PosHighMinus{f}{\Pat}{\Text} \cap [i \dd
  i + t) \neq \emptyset$, then $\OccThree{2\ell}{\Pat}{\Text} \cap
  \PosHighMinus{f}{\Pat}{\Text} \cap [i \dd i + t) = \{\RunEndFullPos{f}{\tau}{\Text}{i} -
  (\RunEndCutPat{f}{\tau}{\Pat}{2\ell} - 1)\}$.
\end{lemma}
\begin{proof}

  Denote $s = \HeadPat{f}{\tau}{\Pat}$, $p = |H|$, and $k_2 =
  \ExpCutPat{f}{\tau}{\Pat}{2\ell}$. By definition, $\RunEndPos{\tau}{\Text}{i} - i \geq 3\tau -
  1$. Thus, $t > 0$. By \cref{lm:beg-end}\eqref{lm:beg-end-it-1}, it
  holds $[i \dd \RunEndPos{\tau}{\Text}{i} - 3\tau + 1] = [i \dd i + t) \subseteq \RTwo{\tau}{\Text}$.
  From \cref{lm:R-text-block}, we thus obtain that for every $\delta
  \in [0 \dd t)$, it holds $\RunEndFullPos{f}{\tau}{\Text}{i + \delta} = \RunEndFullPos{f}{\tau}{\Text}{i}$,
  which implies $\RunEndFullPos{f}{\tau}{\Text}{i + \delta} - (i + \delta) = \RunEndFullPos{f}{\tau}{\Text}{i} -
  i - \delta$.  Consider any $j \in \OccThree{2\ell}{\Pat}{\Text} \cap
  \PosHighMinus{f}{\Pat}{\Text}$. By definition of $\PosHighMinus{f}{\Pat}{\Text}$, we then have
  $j \in \RMinusSix{f}{s}{k_2}{H}{\tau}{\Text}$.  Thus, $\RunEndFullPos{f}{\tau}{\Text}{j} - j = s + k_2 p =
  \RunEndCutPat{f}{\tau}{\Pat}{2\ell} - 1$.  Consequently, $i + \delta \in
  \OccThree{2\ell}{\Pat}{\Text} \cap \PosHighMinus{f}{\Pat}{\Text}$ implies
  $\RunEndFullPos{f}{\tau}{\Text}{i + \delta} - (i + \delta) = \RunEndFullPos{f}{\tau}{\Text}{i} - i - \delta = \RunEndCutPat{f}{\tau}{\Pat}{2\ell}
  - 1$, i.e., $\delta = (\RunEndFullPos{f}{\tau}{\Text}{i} - i) - (\RunEndCutPat{f}{\tau}{\Pat}{2\ell} - 1)$,
  and hence $|\OccThree{2\ell}{\Pat}{\Text} \cap \PosHighMinus{f}{\Pat}{\Text} \cap [i \dd
  i + t)| \leq 1$.

  We now prove the equivalence. Let us first assume that
  $|\OccThree{2\ell}{\Pat}{\Text} \cap \PosHighMinus{f}{\Pat}{\Text} \cap [i \dd i + t)| =
  1$, i.e., that for some $\delta \in [0 \dd t)$, it holds $i + \delta
  \in \OccThree{2\ell}{\Pat}{\Text} \cap \PosHighMinus{f}{\Pat}{\Text}$. As noted above,
  this implies $i + \delta \in \RMinusSix{f}{s}{k_2}{H}{\tau}{\Text}$. By $[i \dd i +
  \delta] \subseteq \RTwo{\tau}{\Text}$ and \cref{lm:R-text-block}, this implies $i
  \in \RMinusFour{f}{H}{\tau}{\Text}$, i.e., the first condition.  Second, recall from the
  first paragraph that $i + \delta \in \OccThree{2\ell}{\Pat}{\Text} \cap
  \PosHighMinus{f}{\Pat}{\Text}$ implies $\RunEndCutPat{f}{\tau}{\Pat}{2\ell} - 1 = \RunEndFullPos{f}{\tau}{\Text}{i} - i -
  \delta \leq \RunEndFullPos{f}{\tau}{\Text}{i} - i$.  This establishes the second
  condition. We now show the third condition.  As noted above, $i +
  \delta \in \OccThree{2\ell}{\Pat}{\Text} \cap \PosHighMinus{f}{\Pat}{\Text}$ implies that
  $i + \delta \in \RSix{f}{s}{k_2}{H}{\tau}{\Text}$. Consequently, $\Text[i + \delta \dd
  \RunEndFullPos{f}{\tau}{\Text}{i + \delta}) = \Text[i + \delta \dd \RunEndFullPos{f}{\tau}{\Text}{i}) = \Pat[1
  \dd \RunEndCutPat{f}{\tau}{\Pat}{2\ell}) = H' H^{k_2}$, where $H'$ is a length-$s$
  suffix of $H$. Thus, $\Pat[1 \dd \min(m,2\ell)]$ being a prefix of
  $\Text[i + \delta \dd \Textlen]$ (which follows from $i + \delta \in
  \OccThree{2\ell}{\Pat}{\Text}$) implies that $\Pat[\RunEndCutPat{f}{\tau}{\Pat}{2\ell} \dd
  \min(m,2\ell)]$ is a prefix of $\Text[\RunEndFullPos{f}{\tau}{\Text}{i} \dd \Textlen]$. By
  $|\Pat[\RunEndCutPat{f}{\tau}{\Pat}{2\ell} \dd \min(m,2\ell)]| = \min(m,2\ell) -
  \RunEndCutPat{f}{\tau}{\Pat}{2\ell} + 1 \leq 2\ell \leq 7\tau$ (where the last
  inequality follows by $\tau = \lfloor \tfrac{\ell}{3} \rfloor$ and
  $\ell \geq 16$), we thus obtain that $\Pat[\RunEndCutPat{f}{\tau}{\Pat}{2\ell} \dd
  \min(m,2\ell)]$ is a prefix of $\Textinf[\RunEndFullPos{f}{\tau}{\Text}{i} \dd \RunEndFullPos{f}{\tau}{\Text}{i}
  + 7\tau)$.

  We now prove the opposite implication. Let $i \in \RTwo{\tau}{\Text}$ and assume
  $i \in \RMinusFour{f}{H}{\tau}{\Text}$, $\RunEndFullPos{f}{\tau}{\Text}{i} - i \geq
  \RunEndCutPat{f}{\tau}{\Pat}{2\ell} - 1$, and $\Pat[\RunEndCutPat{f}{\tau}{\Pat}{2\ell} \dd \min(m,2\ell)]$
  is a prefix of $\Textinf[\RunEndFullPos{f}{\tau}{\Text}{i} \dd \RunEndFullPos{f}{\tau}{\Text}{i} + 7\tau)$.
  Observe that this implies that $\Pat[\RunEndCutPat{f}{\tau}{\Pat}{2\ell} \dd \min(m,
  2\ell)] \preceq \Textinf[\RunEndFullPos{f}{\tau}{\Text}{i} \dd \RunEndFullPos{f}{\tau}{\Text}{i} + 7\tau)$ and
  hence by \cref{lm:poslow-poshigh-run}, letting $\delta =
  (\RunEndFullPos{f}{\tau}{\Text}{i} - i) - (\RunEndCutPat{f}{\tau}{\Pat}{2\ell} - 1)$, we have $i + \delta \in
  [0 \dd t)$ and $i + \delta \in \PosHighMinus{f}{\Pat}{\Text}$.  It thus remains to
  prove that $i + \delta \in \OccThree{2\ell}{\Pat}{\Text}$, i.e., that
  $\lcp(\Pat, \Text[i + \delta \dd \Textlen]) \geq \min(m,2\ell)$, or
  equivalently, that $i + \delta + \min(m,2\ell) - 1 \leq \Textlen$ and $\Text[i
  + \delta \dd i + \delta + \min(m, 2\ell) - 1] = \Pat[1 \dd
  \min(m,2\ell)]$.  We proceed in two steps:
  \begin{itemize}
  \item First, we prove that $\Text[i+\delta \dd \RunEndFullPos{f}{\tau}{\Text}{i}) = \Pat[1
    \dd \RunEndCutPat{f}{\tau}{\Pat}{2\ell})$. First, note that $[i \dd i + \delta]
    \subseteq \RTwo{\tau}{\Text}$, $i \in \RFour{f}{H}{\tau}{\Text}$ and \cref{lm:R-text-block} imply
    that $i + \delta \in \RFour{f}{H}{\tau}{\Text}$ and $\RunEndFullPos{f}{\tau}{\Text}{i + \delta} =
    \RunEndFullPos{f}{\tau}{\Text}{i}$.  On the other hand, by definition of $\delta$, we
    have $\RunEndFullPos{f}{\tau}{\Text}{i} - (i + \delta) = \RunEndCutPat{f}{\tau}{\Pat}{2\ell} - 1$.
    Consequently, $\HeadPos{f}{\tau}{\Text}{i + \delta} = (\RunEndFullPos{f}{\tau}{\Text}{i + \delta} - (i +
    \delta)) \bmod |\RootPos{f}{\tau}{\Text}{i + \delta}| = (\RunEndFullPos{f}{\tau}{\Text}{i} - (i +
    \delta)) \bmod p = (\RunEndCutPat{f}{\tau}{\Pat}{2\ell} - 1) \bmod p = \HeadPat{f}{\tau}{\Pat} =
    s$.  We also have $\ExpPos{f}{\tau}{\Text}{i + \delta} = \lfloor
    \tfrac{\RunEndFullPos{f}{\tau}{\Text}{i + \delta} - (i + \delta)}{|\RootPos{f}{\tau}{\Text}{i + \delta}|} \rfloor =
    \lfloor \tfrac{\RunEndFullPos{f}{\tau}{\Text}{i} - (i + \delta)}{p} \rfloor = \lfloor
    \tfrac{\RunEndCutPat{f}{\tau}{\Pat}{2\ell} - 1}{p} \rfloor = \lfloor \tfrac{s + k_2
    p}{p} \rfloor = k_2$.  Consequently, we obtain that $\Text[i + \delta
    \dd \RunEndFullPos{f}{\tau}{\Text}{i}) = \Text[i + \delta \dd \RunEndFullPos{f}{\tau}{\Text}{i + \delta}) = H'
    H^{k_2}$, where $H'$ is a length-$s$ suffix of $H$.  On the other
    hand, we also have by definition $\Pat[1 \dd \RunEndCutPat{f}{\tau}{\Pat}{2\ell}) = H'
    H^{k_2}$. Thus, $\Text[i + \delta \dd \RunEndFullPos{f}{\tau}{\Text}{i}) = \Pat[1 \dd
    \RunEndCutPat{f}{\tau}{\Pat}{2\ell})$.
  \item Second, we prove that it holds $i + \delta + \min(m, 2\ell) - 1\leq \Textlen$
    and $\Text[\RunEndFullPos{f}{\tau}{\Text}{i} \dd i + \delta + \min(m, 2\ell) - 1] =
    \Pat[\RunEndCutPat{f}{\tau}{\Pat}{2\ell} \dd \min(m,2\ell)]$.  Recall, that we assumed
    that $\Pat[\RunEndCutPat{f}{\tau}{\Pat}{2\ell} \dd \min(m,2\ell)]$ is a prefix of
    $\Textinf[\RunEndFullPos{f}{\tau}{\Text}{i} \dd \RunEndFullPos{f}{\tau}{\Text}{i} + 7\tau)$.  Note, however,
    that we also assumed that $\Text[\Textlen]$ does not occur in $\Pat[1 \dd
    m)$. Consequently, $\Text[\Textlen]$ also does not occur in
    $\Pat[\RunEndCutPat{f}{\tau}{\Pat}{2\ell} \dd \min(m,2\ell))$, and thus, $\RunEndFullPos{f}{\tau}{\Text}{i}
    + (\min(m,2\ell) - \RunEndCutPat{f}{\tau}{\Pat}{2\ell}) \leq \Textlen$. By recalling that $i
    + \delta = i + (\RunEndFullPos{f}{\tau}{\Text}{i} - i) - (\RunEndCutPat{f}{\tau}{\Pat}{2\ell} - 1) =
    \RunEndFullPos{f}{\tau}{\Text}{i} - \RunEndCutPat{f}{\tau}{\Pat}{2\ell} + 1$, we equivalently obtain
    $\RunEndFullPos{f}{\tau}{\Text}{i} + \min(m,2\ell) - \RunEndCutPat{f}{\tau}{\Pat}{2\ell} = i + \delta +
    \min(m,2\ell) - 1 \leq \Textlen$, i.e., the first part of the claim.
    Moreover, observe that $\Pat[\RunEndCutPat{f}{\tau}{\Pat}{2\ell} \dd \min(m,2\ell)]$ being a prefix
    of $\Text[\RunEndFullPos{f}{\tau}{\Text}{i} \dd \RunEndFullPos{f}{\tau}{\Text}{i} + 7\tau)$ then implies that
    it holds $\Pat[\RunEndCutPat{f}{\tau}{\Pat}{2\ell} \dd \min(m,2\ell)] =
    \Text[\RunEndFullPos{f}{\tau}{\Text}{i} \dd \RunEndFullPos{f}{\tau}{\Text}{i} + \min(m,2\ell) -
    \RunEndCutPat{f}{\tau}{\Pat}{2\ell}] = \Text[\RunEndFullPos{f}{\tau}{\Text}{i} \dd i + \delta + \min(m,2\ell)
    - 1]$, i.e., the second part of the claim.
  \end{itemize}

  To show the last implication, recall that we proved that $i + \delta
  \in \OccThree{2\ell}{\Pat}{\Text} \cap \PosHighMinus{f}{\Pat}{\Text}$ (where $\delta \in
  [0 \dd t)$) implies $\delta = (\RunEndFullPos{f}{\tau}{\Text}{i} - i) - (\RunEndCutPat{f}{\tau}{\Pat}{2\ell}
  - 1)$. Thus, $\OccThree{2\ell}{\Pat}{\Text} \cap \PosHighMinus{f}{\Pat}{\Text} \cap [i
  \dd i + t) \neq \emptyset$ implies $\OccThree{2\ell}{\Pat}{\Text} \cap
  \PosHighMinus{f}{\Pat}{\Text} \cap [i \dd i + t) = \{i + \delta\} = \{\RunEndFullPos{f}{\tau}{\Text}{i}
  - (\RunEndCutPat{f}{\tau}{\Pat}{2\ell} - 1)\}$.
\end{proof}

\begin{lemma}\label{lm:sa-periodic-occ-poshigh-single}
  Let $\ell \in [16 \dd \Textlen)$, $\tau = \lfloor \tfrac{\ell}{3} \rfloor$,
  and $f$ be any necklace-consistent function.
  Let $\Pat \in \Sigma^{m}$ be a $\tau$-periodic pattern such that
  $\TypePat{\tau}{\Pat} = -1$ and $\Text[\Textlen]$ does not occur in $\Pat[1 \dd
  m)$. Denote $H = \RootPat{f}{\tau}{\Pat}$, $s = \HeadPat{f}{\tau}{\Pat}$, and $k_2 =
  \ExpCutPat{f}{\tau}{\Pat}{2\ell}$. For every $i \in \RTwo{\tau}{\Text}$, $i \in
  \OccThree{2\ell}{\Pat}{\Text} \cap \PosHighMinus{f}{\Pat}{\Text}$ holds if and only if
  \begin{itemize}
  \item $i \in \RMinusSix{f}{s}{k_2}{H}{\tau}{\Text}$ and
  \item $\Pat[\RunEndCutPat{f}{\tau}{\Pat}{2\ell} \dd \min(m,2\ell)]$ is a prefix of
    $\Textinf[\RunEndFullPos{f}{\tau}{\Text}{i} \dd \RunEndFullPos{f}{\tau}{\Text}{i} + 7\tau)$.
  \end{itemize}
\end{lemma}
\begin{proof}

  Consider any $i \in \RTwo{\tau}{\Text}$ and assume $i \in \OccThree{2\ell}{\Pat}{\Text}
  \cap \PosHighMinus{f}{\Pat}{\Text}$. By \cref{def:pos-sets-for-pat}, this implies
  $i \in \RMinusSix{f}{s}{k_2}{H}{\tau}{\Text}$.  To show the second claim, note that by
  $\RunEndPos{\tau}{\Text}{i} - i \geq 3\tau - 1$, letting $t = \RunEndPos{\tau}{\Text}{i} - i - 3\tau +
  2$, we have $\OccThree{2\ell}{\Pat}{\Text} \cap \PosHighMinus{f}{\Pat}{\Text} \cap [i \dd
  i + t) \neq \emptyset$. By \cref{lm:sa-periodic-occ-run}, we thus
  obtain that $\Pat[\RunEndCutPat{f}{\tau}{\Pat}{2\ell} \dd \min(m,2\ell)]$ is a prefix of
  $\Textinf[\RunEndFullPos{f}{\tau}{\Text}{i} \dd \RunEndFullPos{f}{\tau}{\Text}{i} + 7\tau)$.

  Let us now consider $i \in \RTwo{\tau}{\Text}$ and assume $i \in
  \RMinusSix{f}{s}{k_2}{H}{\tau}{\Text}$ and that $\Pat[\RunEndCutPat{f}{\tau}{\Pat}{2\ell} \dd \min(m,2\ell)]$
  is a prefix of $\Textinf[\RunEndFullPos{f}{\tau}{\Text}{i} \dd \RunEndFullPos{f}{\tau}{\Text}{i} + 7\tau)$. The
  first assumption implies $\RunEndFullPos{f}{\tau}{\Text}{i} - i = s + k_2 \cdot |H| =
  \HeadPat{f}{\tau}{\Pat} + \ExpCutPat{f}{\tau}{\Pat}{2\ell} \cdot |\RootPat{f}{\tau}{\Pat}| =
  \RunEndCutPat{f}{\tau}{\Pat}{2\ell} - 1$. By \cref{lm:sa-periodic-occ-run}, it thus
  follows that, letting $t = \RunEndPos{\tau}{\Text}{i} - i - 3\tau + 2 > 0$, we have
  $\OccThree{2\ell}{\Pat}{\Text} \cap \PosHighMinus{f}{\Pat}{\Text} \cap [i \dd i + t) =
  \{\RunEndFullPos{f}{\tau}{\Text}{i} - (\RunEndCutPat{f}{\tau}{\Pat}{2\ell} - 1)\} = \{\RunEndFullPos{f}{\tau}{\Text}{i} -
  (\RunEndFullPos{f}{\tau}{\Text}{i} - i)\} = \{i\}$. Thus, $i \in \OccThree{2\ell}{\Pat}{\Text}
  \cap \PosHighMinus{f}{\Pat}{\Text}$.
\end{proof}

\begin{lemma}\label{lm:sa-periodic-occ-poshigh-size-pat}
  Let $\ell \in [16 \dd \Textlen)$, $\tau = \lfloor \tfrac{\ell}{3} \rfloor$,
  and $f$ be any necklace-consistent function.
  Let $\Pat \in \Sigma^{m}$ be a $\tau$-periodic pattern such that
  $\TypePat{\tau}{\Pat} = -1$ and $\Text[\Textlen]$ does not occur in $\Pat[1 \dd m)$. Let
  $c = \max \Sigma$, $x_l = \RunEndCutPat{f}{\tau}{\Pat}{2\ell} - 1$, $y_l =
  \Pat[\RunEndCutPat{f}{\tau}{\Pat}{2\ell} \dd \min(m,2\ell)]$, and $y_u = y_l
  c^{\infty}$.  Then, letting $H = \RootPat{f}{\tau}{\Pat}$ and $\Pts =
  \IntStrPoints{7\tau}{\PairsMinus{f}{H}{\tau}{\Text}}{\Text}$, it holds:
  \begin{itemize}
  \item $\OccThree{2\ell}{\Pat}{\Text} \cap \PosHighMinus{f}{\Pat}{\Text} = \{\RunEndFullPos{f}{\tau}{\Text}{j}
    - x_l : j \in \RPrimMinusFour{f}{H}{\tau}{\Text},\\x_l \leq \RunEndFullPos{f}{\tau}{\Text}{j} - j,\text{ and
    }y_l \preceq \Textinf[\RunEndFullPos{f}{\tau}{\Text}{j} \dd \RunEndFullPos{f}{\tau}{\Text}{j} + 7\tau) \prec
    y_u\}$,
  \item $|\OccThree{2\ell}{\Pat}{\Text} \cap \PosHighMinus{f}{\Pat}{\Text}| =
    \RangeCountFourSide{\Pts}{x_l}{\Textlen}{y_l}{y_u}$.
  \end{itemize}
\end{lemma}
\begin{proof}

  By \cref{def:pos-sets-for-pat}, we have $\PosHighMinus{f}{\Pat}{\Text} \subseteq
  \RMinusFour{f}{H}{\tau}{\Text}$.  On the other hand, by
  \cref{lm:beg-end}\eqref{lm:beg-end-it-1} it holds $\RMinusFour{f}{H}{\tau}{\Text} =
  \bigcup_{j \in \RPrimMinusFour{f}{H}{\tau}{\Text}} [j \dd \RunEndPos{\tau}{\Text}{j} - 3\tau + 1]$.  Since
  this is a disjoint union, we thus have $\OccThree{2\ell}{\Pat}{\Text} \cap
  \PosHighMinus{f}{\Pat}{\Text} = \bigcup_{j \in \RPrimMinusFour{f}{H}{\tau}{\Text}} \OccThree{2\ell}{\Pat}{\Text}
  \cap \PosHighMinus{f}{\Pat}{\Text} \cap [j \dd \RunEndPos{\tau}{\Text}{j} - 3\tau + 1]$.  By
  \cref{lm:sa-periodic-occ-run}, for every $j \in \RPrimMinusFour{f}{H}{\tau}{\Text}$,
  $|\OccThree{2\ell}{\Pat}{\Text} \cap \PosHighMinus{f}{\Pat}{\Text} \cap [j \dd \RunEndPos{\tau}{\Text}{j} -
  3\tau + 1]| \leq 1$. Moreover, $|\OccThree{2\ell}{\Pat}{\Text} \cap
  \PosHighMinus{f}{\Pat}{\Text} \cap [j \dd \RunEndPos{\tau}{\Text}{j} - 3\tau + 1]| = 1$ holds if and
  only if $x_l \leq \RunEndFullPos{f}{\tau}{\Text}{j} - j$ and $y_l$ is a prefix of
  $\Textinf[\RunEndFullPos{f}{\tau}{\Text}{j} \dd \RunEndFullPos{f}{\tau}{\Text}{j} + 7\tau)$.  Furthermore, if
  $x_l \leq \RunEndFullPos{f}{\tau}{\Text}{j} - j$ and $y_l$ is a prefix of
  $\Textinf[\RunEndFullPos{f}{\tau}{\Text}{j} \dd \RunEndFullPos{f}{\tau}{\Text}{j} + 7\tau)$, then
  $\OccThree{2\ell}{\Pat}{\Text} \cap \PosHighMinus{f}{\Pat}{\Text} \cap [j \dd \RunEndPos{\tau}{\Text}{j} -
  3\tau + 1] = \{\RunEndFullPos{f}{\tau}{\Text}{j} - x_l\}$. Consequently, letting
  \begin{align*}
    \mathcal{J}
      &= \{j \in \RPrimMinusFour{f}{H}{\tau}{\Text} : x_l \leq \RunEndFullPos{f}{\tau}{\Text}{j} - j
          \text{ and } y_u\text{ is a prefix of }
          \Textinf[\RunEndFullPos{f}{\tau}{\Text}{j} \dd \RunEndFullPos{f}{\tau}{\Text}{j} + 7\tau)\}\\
      &=  \{j \in \RPrimMinusFour{f}{H}{\tau}{\Text} : x_l \leq \RunEndFullPos{f}{\tau}{\Text}{j} - j
          \text{ and }y_l \preceq \Textinf[\RunEndFullPos{f}{\tau}{\Text}{j} \dd
          \RunEndFullPos{f}{\tau}{\Text}{j} + 7\tau) \prec y_u\},
  \end{align*}
  we have $\OccThree{2\ell}{\Pat}{\Text} \cap \PosHighMinus{f}{\Pat}{\Text} = \bigcup_{j \in
  \mathcal{J}} \{\RunEndFullPos{f}{\tau}{\Text}{j} - x_l\}$, i.e., the first claim.

  To show the second claim, note that for every $j \in \mathcal{J}$, we
  have $\RunEndFullPos{f}{\tau}{\Text}{j} - x_l \in [j \dd \RunEndPos{\tau}{\Text}{j} - 3\tau + 1]$.  Since
  by \cref{lm:beg-end}\eqref{lm:beg-end-it-1}, for every $j_1, j_2 \in
  \RPrimTwo{\tau}{\Text}$, $j_1 \neq j_2$ implies that $[j_1 \dd \RunEndPos{\tau}{\Text}{j_1} - 3\tau + 1]$
  and $[j_2 \dd \RunEndPos{\tau}{\Text}{j_2} - 3\tau + 1]$ are disjoint, we thus obtain
  that $|\OccThree{2\ell}{\Pat}{\Text} \cap \PosHighMinus{f}{\Pat}{\Text}| = |\bigcup_{j
  \in \mathcal{J}} \{\RunEndFullPos{f}{\tau}{\Text}{j} - x_l\}| =
  |\mathcal{J}|$. Consequently, it follows by
  \cref{lm:sa-periodic-count}\eqref{lm:sa-periodic-count-it-1} that
  $|\OccThree{2\ell}{\Pat}{\Text} \cap \PosHighMinus{f}{\Pat}{\Text}| =
  \RangeCountFourSide{\Pts}{x_l}{\Textlen}{y_l}{y_u}$.  Note that
  \cref{lm:sa-periodic-count}\eqref{lm:sa-periodic-count-it-1}
  requires $x_l \in [0 \dd 7\tau]$, which holds here since $x_l =
  \RunEndCutPat{f}{\tau}{\Pat}{2\ell} - 1 \leq 2\ell \leq 7\tau$, where the last
  inequality holds by $\tau = \lfloor \tfrac{\ell}{3} \rfloor$ and
  $\ell \geq 16$.
\end{proof}

\begin{lemma}\label{lm:occ-sub-poshigh}
  Let $\ell \in [16 \dd \Textlen)$, $\tau = \lfloor \tfrac{\ell}{3} \rfloor$,
  and $f$ be any necklace-consistent function.
  Let $\Pat \in \Sigma^{m}$ be a $\tau$-periodic pattern such that
  $\TypePat{\tau}{\Pat} = -1$, $\RunEndPat{\tau}{\Pat} - 1 < \min(m, 2\ell)$, and $\Text[\Textlen]$
  does not occur in $\Pat[1 \dd m)$. Then, it holds $\ExpPat{f}{\tau}{\Pat} =
  \ExpCutPat{f}{\tau}{\Pat}{2\ell}$, $\RunEndFullPat{f}{\tau}{\Pat} = \RunEndCutPat{f}{\tau}{\Pat}{2\ell}$, and
  $\OccThree{2\ell}{\Pat}{\Text} \subseteq \PosHighMinus{f}{\Pat}{\Text}$.
\end{lemma}
\begin{proof}

  Denote $s = \HeadPat{f}{\tau}{\Pat}$, $p = |H|$, $k = \ExpPat{f}{\tau}{\Pat}$, and $k_2 =
  \ExpCutPat{f}{\tau}{\Pat}{2\ell}$.  We start by observing that $\RunEndPat{\tau}{\Pat} - 1
  < 2\ell$ implies that $k_2 = \min(\ExpPat{f}{\tau}{\Pat}, \lfloor \tfrac{2\ell
  - s}{p} \rfloor) = \min(\lfloor \tfrac{\RunEndPat{\tau}{\Pat} - 1 - s}{p}
  \rfloor, \lfloor \tfrac{2\ell - s}{p} \rfloor) = \lfloor
  \tfrac{\RunEndPat{\tau}{\Pat} - 1 - s}{p} \rfloor = \ExpPat{f}{\tau}{\Pat} = k$. This
  implies $\RunEndFullPat{f}{\tau}{\Pat} = 1 + s + k p = 1 + s + k_2 p =
  \RunEndCutPat{f}{\tau}{\Pat}{2\ell} = \RunEndCutPat{f}{\tau}{\Pat}{2\ell}$.

  Next, we prove $\OccThree{2\ell}{\Pat}{\Text} \subseteq
  \PosHighMinus{f}{\Pat}{\Text}$.  Let $t \in \OccThree{2\ell}{\Pat}{\Text}$. By
  \cref{lm:sa-periodic-occ-single}, it follows that $t \in
  \RMinusSix{f}{s}{k}{H}{\tau}{\Text} = \RMinusSix{f}{s}{k_2}{H}{\tau}{\Text}$ and, letting $\Pat_{\rm sub} =
  \Pat[\RunEndFullPat{f}{\tau}{\Pat} \dd \min(m,2\ell)] = \Pat[\RunEndCutPat{f}{\tau}{\Pat}{2\ell} \dd
  \min(m,2\ell)]$, the string $\Pat_{\rm sub}$ is a prefix of
  $\Textinf[\RunEndFullPos{f}{\tau}{\Text}{t} \dd \RunEndFullPos{f}{\tau}{\Text}{t} + 7\tau)$.  Thus,
  $\Textinf[\RunEndFullPos{f}{\tau}{\Text}{t} \dd \RunEndFullPos{f}{\tau}{\Text}{t} + 7\tau) \succeq \Pat_{\rm
  sub}$. By \cref{lm:poslow-poshigh-single}, we thus have $t \in
  \PosHighMinus{f}{\Pat}{\Text}$.
\end{proof}

\begin{lemma}\label{lm:sa-periodic-occ-single}
  Let $\ell \in [16 \dd \Textlen)$, $\tau = \lfloor \tfrac{\ell}{3} \rfloor$,
  and $f$ be any necklace-consistent function.
  Let $\Pat \in \Sigma^{m}$ be a $\tau$-periodic pattern such that
  $\TypePat{\tau}{\Pat} = -1$, $\RunEndPat{\tau}{\Pat} - 1 < \min(m,2\ell)$, and $\Text[\Textlen]$
  does not occur in $\Pat[1 \dd m)$. Denote $H = \RootPat{f}{\tau}{\Pat}$, $s =
  \HeadPat{f}{\tau}{\Pat}$, and $k = \ExpPat{f}{\tau}{\Pat}$. For every $i \in \RTwo{\tau}{\Text}$, $i \in
  \OccThree{2\ell}{\Pat}{\Text}$ holds if and only if
  \begin{itemize}
  \item $i \in \RMinusSix{f}{s}{k}{H}{\tau}{\Text}$ and
  \item $\Pat[\RunEndFullPat{f}{\tau}{\Pat} \dd \min(m,2\ell)]$ is a prefix of
    $\Textinf[\RunEndFullPos{f}{\tau}{\Text}{i} \dd \RunEndFullPos{f}{\tau}{\Text}{i} + 7\tau)$.
  \end{itemize}
\end{lemma}
\begin{proof}
  Let us denote $k_2 = \ExpCutPat{f}{\tau}{\Pat}{2\ell}$.  By \cref{lm:occ-sub-poshigh},
  it holds $k_2 = k$, $\RunEndCutPat{f}{\tau}{\Pat}{2\ell} = \RunEndFullPat{f}{\tau}{\Pat}$, and
  $\OccThree{2\ell}{\Pat}{\Text} \subseteq \PosHighMinus{f}{\Pat}{\Text}$. Thus, it follows
  by \cref{lm:sa-periodic-occ-poshigh-single} that for every $i \in
  \RTwo{\tau}{\Text}$, $i \in \OccThree{2\ell}{\Pat}{\Text} \cap \PosHighMinus{f}{\Pat}{\Text} =
  \OccThree{2\ell}{\Pat}{\Text}$ holds if and only if $i \in \RMinusSix{f}{s}{k_2}{H}{\tau}{\Text}
  = \RMinusSix{f}{s}{k}{H}{\tau}{\Text}$ and $\Pat[\RunEndCutPat{f}{\tau}{\Pat}{2\ell} \dd \min(m, 2\ell)] =
  \Pat[\RunEndFullPat{f}{\tau}{\Pat} \dd \min(m, 2\ell)]$ is a prefix of
  $\Textinf[\RunEndFullPos{f}{\tau}{\Text}{i} \dd \RunEndFullPos{f}{\tau}{\Text}{i} + 7\tau)$.
\end{proof}

\begin{lemma}\label{lm:sa-periodic-occ-size-pat}
  Let $\ell \in [16 \dd \Textlen)$, $\tau = \lfloor \tfrac{\ell}{3} \rfloor$,
  and $f$ be any necklace-consistent function.
  Let $\Pat \in \Sigma^{m}$ be a $\tau$-periodic pattern such that
  $\TypePat{\tau}{\Pat} = -1$, $\RunEndPat{\tau}{\Pat} - 1 < \min(m,2\ell)$, and $\Text[\Textlen]$
  does not occur in $\Pat[1 \dd m)$.  Let $c = \max \Sigma$, $x_l =
  \RunEndFullPat{f}{\tau}{\Pat} - 1$, $y_l = \Pat[\RunEndFullPat{f}{\tau}{\Pat} \dd
  \min(m,2\ell)]$, and $y_u = y_l c^{\infty}$.  Then, letting $H
  \,{=}\, \RootPat{f}{\tau}{\Pat}$ and $\Pts = \IntStrPoints{7\tau}{\PairsMinus{f}{H}{\tau}{\Text}}{\Text}$,
  it holds:
  \begin{itemize}[leftmargin=3.5ex]
  \item $\OccThree{2\ell}{\Pat}{\Text} = \{\RunEndFullPos{f}{\tau}{\Text}{j} - x_l : j \in
    \RPrimMinusFour{f}{H}{\tau}{\Text},\ x_l \leq \RunEndFullPos{f}{\tau}{\Text}{j} - j,\text{ and }\\y_l \preceq
    \Textinf[\RunEndFullPos{f}{\tau}{\Text}{j} \dd \RunEndFullPos{f}{\tau}{\Text}{j} + 7\tau) \prec y_u\}$,
  \item $|\OccThree{2\ell}{\Pat}{\Text}| =
    \RangeCountFourSide{\Pts}{x_l}{\Textlen}{y_l}{y_u}$.
  \end{itemize}
\end{lemma}
\begin{proof}
  By \cref{lm:occ-sub-poshigh}, it holds $\RunEndCutPat{f}{\tau}{\Pat}{2\ell} - 1 =
  \RunEndFullPat{f}{\tau}{\Pat} - 1 = x_l$, and $\OccThree{2\ell}{\Pat}{\Text} \subseteq
  \PosHighMinus{f}{\Pat}{\Text}$.  Observe that we then also have
  $\Pat[\RunEndCutPat{f}{\tau}{\Pat}{2\ell} \dd \min(m,2\ell)] = \Pat[\RunEndFullPat{f}{\tau}{\Pat} \dd
  \min(m,2\ell)] = y_l$.  Thus, it follows by
  \cref{lm:sa-periodic-occ-poshigh-size-pat} that $\OccThree{2\ell}{\Pat}{\Text}
  \cap \PosHighMinus{f}{\Pat}{\Text} = \OccThree{2\ell}{\Pat}{\Text} = \{\RunEndFullPos{f}{\tau}{\Text}{j} - x_l : j
  \in \RPrimMinusFour{f}{H}{\tau}{\Text},\ x_l \leq \RunEndFullPos{f}{\tau}{\Text}{j} - j,\text{ and }y_l \preceq
  \Textinf[\RunEndFullPos{f}{\tau}{\Text}{j} \dd \RunEndFullPos{f}{\tau}{\Text}{j} + 7\tau) \prec y_u\}$
  and $|\OccThree{2\ell}{\Pat}{\Text} \cap \PosHighMinus{f}{\Pat}{\Text}|
  = |\OccThree{2\ell}{\Pat}{\Text}| = \RangeCountFourSide{\Pts}{x_l}{\Textlen}{y_l}{y_u}$.
\end{proof}

\begin{lemma}\label{lm:sa-periodic-occ-size-pos}
  Let $\ell \in [16 \dd \Textlen)$, $\tau = \lfloor \tfrac{\ell}{3} \rfloor$,
  and $f$ be any necklace-consistent function.
  Let $j \in \RMinusTwo{\tau}{\Text}$ be such that $\RunEndPos{\tau}{\Text}{j} - j < 2\ell$. Let $c =
  \max \Sigma$, $x_l = \RunEndFullPos{f}{\tau}{\Text}{j} - j$, $y_l = \Text[\RunEndFullPos{f}{\tau}{\Text}{j} \dd
  \min(\Textlen + 1, j + 2\ell))$, and $y_u = y_l c^{\infty}$.  Then, letting
  $H = \RootPos{f}{\tau}{\Text}{j}$ and $\Pts = \IntStrPoints{7\tau}{\PairsMinus{f}{H}{\tau}{\Text}}{\Text}$, it
  holds $|\OccThree{2\ell}{j}{\Text}| =
  \RangeCountFourSide{\Pts}{x_l}{\Textlen}{y_l}{y_u}$.
\end{lemma}
\begin{proof}
  Denote $\Pat = \Text[j \dd \Textlen]$ and $m = |\Pat|$. First, observe that by
  $j \in \RMinusTwo{\tau}{\Text}$, $\Pat$ is $\tau$-periodic and it holds
  $\TypePat{\tau}{\Pat} = -1$. Note also that $\Text[\Textlen]$ does not occur in $\Pat[1 \dd
  m)$. Next, recall that, by definition, $\RunEndPos{\tau}{\Text}{j} = j + \RunEndPat{\tau}{\Pat} -
  1$. Thus, we have $\RunEndPat{\tau}{\Pat} - 1 = \RunEndPos{\tau}{\Text}{j} - j < 2\ell$. Since by
  the uniqueness of $\Text[\Textlen]$ in $\Text$ it holds $\RunEndPos{\tau}{\Text}{j} \leq \Textlen$, we
  also have $\RunEndPat{\tau}{\Pat} - 1 = \RunEndPos{\tau}{\Text}{j} - j \leq \Textlen - j < \Textlen - j + 1 =
  m$. We have thus proved that $\RunEndPat{\tau}{\Pat} - 1 < \min(m,
  2\ell)$. Next, observe that $\RunEndFullPos{f}{\tau}{\Text}{j} = j + \RunEndFullPat{f}{\tau}{\Pat} -
  1$. Thus, $\RunEndFullPat{f}{\tau}{\Pat} - 1 = \RunEndFullPos{f}{\tau}{\Text}{j} - j = x_l$.  Moreover,
  by $\Pat = \Text[j \dd \Textlen]$,
  \begin{align*}
    \Pat[\RunEndFullPat{f}{\tau}{\Pat} \dd \min(m,d)]
      &= \Text[j + \RunEndFullPat{f}{\tau}{\Pat} - 1 \dd j + \min(m, 2\ell) - 1]\\
      &= \Text[\RunEndFullPos{f}{\tau}{\Text}{j} \dd (j - 1) + \min(\Textlen - (j - 1), 2\ell)]\\
      &= \Text[\RunEndFullPos{f}{\tau}{\Text}{j} \dd \min(\Textlen, j + 2\ell - 1)]\\
      &= \Text[\RunEndFullPos{f}{\tau}{\Text}{j} \dd \min(\Textlen + 1, j + 2\ell))\\
      &= y_l.
  \end{align*}
  Lastly, by definition we have $\RootPos{f}{\tau}{\Text}{j} =
  \RootPat{f}{\tau}{\Text[j \dd \Textlen]} = \RootPat{f}{\tau}{\Pat}$, and hence $\RootPat{f}{\tau}{\Pat} = H$.  By
  \cref{lm:sa-periodic-occ-size-pat}, we therefore obtain that
  $|\OccThree{2\ell}{\Pat}{\Text}| = \RangeCountFourSide{\Pts}{x_l}{\Textlen}{y_l}{y_u}$.
  Since by definition we have $\OccThree{2\ell}{j}{\Text} =
  \OccThree{2\ell}{\Text[j \dd \Textlen]}{\Text} = \OccThree{2\ell}{\Pat}{\Text}$, we thus obtain the claim.
\end{proof}

\paragraph{Query Algorithms}

\begin{proposition}\label{pr:sa-periodic-occ-size}
  Let $k \in [4 \dd \lceil \log \Textlen \rceil)$, $\ell = 2^k$, $\tau
  = \lfloor \tfrac{\ell}{3} \rfloor$, and $f = f_{\tau,\Text}$
  (\cref{def:canonical-function}). Let $j \in \RMinusTwo{\tau}{\Text}$ be
  such that $\RunEndPos{\tau}{\Text}{j} - j < 2\ell$. Given
  $\CompSaPeriodic{\Text}$, the value $k$, the position $j$, some
  $j' \in \OccThree{3\tau - 1}{j}{\Text}$ satisfying $j'
  = \min \OccThree{2\ell}{j'}{\Text}$, and the values
  $\HeadPos{f}{\tau}{\Text}{j}$, $|\RootPos{f}{\tau}{\Text}{j}|$, and
  $\ExpPos{f}{\tau}{\Text}{j}$ as input, we can compute
  $|\OccThree{2\ell}{j}{\Text}|$ in $\bigO(\log^{2+\epsilon} \Textlen)$ time.
\end{proposition}
\begin{proof}
  Let us denote $s = \HeadPos{f}{\tau}{\Text}{j}$, $H
  = \RootPos{f}{\tau}{\Text}{j}$, $p = |H|$, $k
  = \ExpPos{f}{\tau}{\Text}{j}$, $c = \max\Sigma$, $y_l
  = \Text[\RunEndFullPos{f}{\tau}{\Text}{j} \dd \min(\Textlen + 1, j + 2\ell))$, $y_u
  = y_l c^{\infty}$, and $\Pts
  = \IntStrPoints{7\tau}{\PairsMinus{f}{H}{\tau}{\Text}}{\Text}$.  In $\bigO(1)$ time
  we set $x_l := \RunEndFullPos{f}{\tau}{\Text}{j} - j = s + kp$.
  Using \cref{pr:sa-periodic-pts-access} and the position $j'$ as
  input, in $\bigO(\log \Textlen)$ time we retrieve the pointer to the
  structure from \cref{pr:int-str} for $\PairsMinus{f}{H}{\tau}{\Text}$ (note
  that $j' \in \RFour{f}{H}{\tau}{\Text}$ holds
  by \cref{lm:periodic-pos-lce}\eqref{lm:periodic-pos-lce-it-2}),
  i.e., performing weighted range queries on $\Pts
  = \IntStrPoints{7\tau}{\PairsMinus{f}{H}{\tau}{\Text}}{\Text}$.  Note that this
  pointer is not null, since we assumed $j \in \RMinusTwo{\tau}{\Text}$,
  which implies $\PairsMinus{f}{H}{\tau}{\Text} \neq \emptyset$.  Note that
  using $\CompSaCore{\Text}$ (which is part of $\CompSaPeriodic{\Text}$), we
  can lexicographically compare any two substrings of $\Textinf$ or
  $\revstr{\Textinf}$ (specified with their starting positions and
  lengths) in $t_{\rm cmp} = \bigO(\log \Textlen)$ time.  In $\bigO(\log^{2
  + \epsilon} \Textlen + t_{\rm cmp} \log \Textlen) = \bigO(\log^{2 + \epsilon} \Textlen)$
  time we therefore compute $q := \RangeCountFourSide{\Pts}{x_l}{\Textlen}{y_l}{y_u}
  = \RangeCountThreeSide{\Pts}{x_l}{\Textlen}{y_u}
  - \RangeCountThreeSide{\Pts}{x_l}{\Textlen}{y_l}$ using \cref{pr:int-str} with $i
  = j + x_l$ and $q_r = \min(\Textlen + 1, j + 2\ell) - i$.
  By \cref{lm:sa-periodic-occ-size-pos}, it holds $q =
  |\OccThree{2\ell}{j}{\Text}|$.  In total, we spend $\bigO(\log^{2
  + \epsilon} \Textlen)$ time.
\end{proof}

\subsubsection{Computing a Position in Occ}\label{sec:sa-periodic-occ-elem}

\paragraph{Combinatorial Properties}

\begin{lemma}\label{lm:sa-periodic-occ-elem}
  Let $\ell \in [16 \dd \Textlen)$, $\tau = \lfloor \tfrac{\ell}{3} \rfloor$,
  and $f$ be any necklace-consistent function.
  Let $\Pat \in \Sigma^{m}$ be a $\tau$-periodic pattern such that
  $\TypePat{\tau}{\Pat} = -1$ and $\Text[\Textlen]$ does not occur in $\Pat[1 \dd m)$.
  Denote $H = \RootPat{f}{\tau}{\Pat}$, $\Pts = \IntStrPoints{7\tau}{\PairsMinus{f}{H}{\tau}{\Text}}{\Text}$,
  $i' = \RangeBegThree{\ell}{\Pat}{\Text} + \DeltaLowMinus{f}{\Pat}{\Text} + \DeltaMidMinus{f}{\Pat}{\Text}$,
  $x = \RunEndCutPat{f}{\tau}{\Pat}{2\ell} - 1$, and $c = \RangeCountTwoSide{\Pts}{x}{\Textlen}$.  Let
  $i \in [1 \dd \Textlen]$ be such that $\SA[i] \in \OccThree{2\ell}{\Pat}{\Text}
  \cap \PosHighMinus{f}{\Pat}{\Text}$. Then:
  \begin{itemize}
  \item It holds $c - (i' - i) \in [1 \dd c]$, i.e.,
    $\RangeSelect{\Pts}{x}{\Textlen}{c - (i'-i)}$ is well-defined,
  \item Every position $j \in \RangeSelect{\Pts}{x}{\Textlen}{c - (i'-i)}$
    satisfies $j - x \in \OccThree{2\ell}{\Pat}{\Text} \cap \PosHighMinus{f}{\Pat}{\Text}$.
  \end{itemize}
\end{lemma}
\begin{proof}

  Denote $s = \HeadPat{f}{\tau}{\Pat}$, $p = |H|$, and $k_2 =
  \ExpCutPat{f}{\tau}{\Pat}{2\ell}$.  The proof consists of three steps.

  1. First, we prove $c - (i' - i) \in [1 \dd c]$, i.e., the first
  claim.  Recall that by \cref{lm:sa-periodic-poslow-poshigh-range},
  it holds $\PosHighMinus{f}{\Pat}{\Text} = \{\SA[i''] : i'' \in
  (\RangeBegThree{2\ell}{\Pat}{\Text} \dd \RangeBegThree{2\ell}{\Pat}{\Text} + \DeltaHighMinus{f}{\Pat}{\Text}]\}$, which due to
  \cref{lm:pat-decomposition} and
  \cref{lm:pat-posless}, we
  can rewrite as $\PosHighMinus{f}{\Pat}{\Text} = \{\SA[i''] : i'' \in (i' -
  \DeltaHighMinus{f}{\Pat}{\Text} \dd i']\}$.  Consequently, $\SA[i] \in
  \PosHighMinus{f}{\Pat}{\Text}$ implies $i' - \DeltaHighMinus{f}{\Pat}{\Text} < i \leq i'$. On
  the one hand, we thus obtain $i' - i < \DeltaHighMinus{f}{\Pat}{\Text} \,{=}\,
  |\PosHighMinus{f}{\Pat}{\Text}| \leq |\RMinusSix{f}{s}{k_2}{H}{\tau}{\Text}| \leq \RangeCountTwoSide{\Pts}{s +
  k_2 p}{\Textlen} \,{=}\, \RangeCountTwoSide{\Pts}{\RunEndCutPat{f}{\tau}{\Pat}{2\ell} - 1}{\Textlen} =
  \RangeCountTwoSide{\Pts}{x}{\Textlen} = c$, where $|\PosHighMinus{f}{\Pat}{\Text}| \leq
  |\RMinusSix{f}{s}{k_2}{H}{\tau}{\Text}|$ follows by $\PosHighMinus{f}{\Pat}{\Text} \subseteq
  \RMinusSix{f}{s}{k_2}{H}{\tau}{\Text}$ and $|\RMinusSix{f}{s}{k_2}{H}{\tau}{\Text}| \leq \RangeCountTwoSide{\Pts}{s +
  k_2 p}{\Textlen}$ follows by \cref{lm:RskH-size-2}. Note that
  \cref{lm:RskH-size-2} requires $k_2 \leq \lfloor \tfrac{7\tau - s}{p}
  \rfloor$, which holds here since $k_2 = \min(\ExpPat{f}{\tau}{\Pat}, \lfloor
  \tfrac{2\ell - s}{p} \rfloor) \leq \lfloor \tfrac{2\ell - s}{p} \rfloor
  \leq \lfloor \tfrac{7\tau - s}{p} \rfloor$, where $2\ell \leq 7\tau$
  follows by $\tau = \lfloor \tfrac{\ell}{3} \rfloor$ and $\ell \geq 16$.
  We thus have $i' - i < c$, which is equivalent to $c - (i' - i) \geq 1$.
  On the other hand, above we proved that $i \leq i'$, which is
  equivalent to $c - (i' - i) \leq c$.  We thus obtain $c - (i' - i)
  \in [1 \dd c]$.

  2. Next, we show that for every $\delta \in [0 \dd
  \DeltaHighMinus{f}{\Pat}{\Text})$, any $j \in \RangeSelect{\Pts}{x}{\Textlen}{c - \delta}$
  satisfies $j - x \in [1 \dd \Textlen]$ and $\Textinf[j - x \dd j - x + 7\tau)
  = \Textinf[\SA[i' - \delta] \dd \SA[i' - \delta] + 7\tau)$.  Consider
  any $\delta \in [0 \dd \DeltaHighMinus{f}{\Pat}{\Text})$ and let $j' = \SA[i' -
  \delta]$. Denote $j'' = j' + x$ and let $y_u = \Textinf[j'' \dd j'' +
  7\tau)$. We first prove that it holds $c - \delta \in
  (\RangeCountThreeSide{\Pts}{x}{\Textlen}{y_u} \dd
  \IncRangeCountThreeSide{\Pts}{x}{\Textlen}{y_u}]$.
  \begin{enumerate}[label=(\alph*)]
    \item First, we show that it holds $c - \delta >
    \RangeCountThreeSide{\Pts}{x}{\Textlen}{y_u}$.  We start by observing that by
    \cref{lm:sa-periodic-count}\eqref{lm:sa-periodic-count-it-1}, we
    have $\RangeCountThreeSide{\Pts}{x}{\Textlen}{y_u} =
    \RangeCountFourSide{\Pts}{x}{\Textlen}{\emptystring}{y_u} = |A|$, where $A = \{t
    \in \RPrimMinusFour{f}{H}{\tau}{\Text} : x \leq \RunEndFullPos{f}{\tau}{\Text}{t} - t\text{ and }
    \Textinf[\RunEndFullPos{f}{\tau}{\Text}{t} \dd \RunEndFullPos{f}{\tau}{\Text}{t} + 7\tau) \prec y_u\}$. Note
    that \cref{lm:sa-periodic-count} requires $x \leq 7\tau$, which
    holds here by $x = \RunEndCutPat{f}{\tau}{\Pat}{2\ell} - 1 = s + k_2 p = s +
    \min(\ExpPat{f}{\tau}{\Pat}, \lfloor \tfrac{2\ell - s}{p} \rfloor) p \leq s +
    \lfloor \tfrac{2\ell - s}{p} \rfloor p \leq 2\ell \leq 7\tau$.  On
    the other hand, by
    \cref{lm:sa-periodic-count}\eqref{lm:sa-periodic-count-it-3}, we
    have $c = |A'|$, where $A' = \{t \in \RPrimMinusFour{f}{H}{\tau}{\Text} : x \leq
    \RunEndFullPos{f}{\tau}{\Text}{t} - t\}$.  Our claim is thus equivalent to $|A'| - |A|
    > \delta$.  Observe that $A \subseteq A'$. This implies $|A'| -
    |A| = |A' \setminus A|$.  Consequently, our goal is to prove that
    $|A' \setminus A| > \delta$.  To this end, we will show that,
    letting $B = \{\RunBeg{\tau}{\Text}{\SA[i'']} : i'' \in [i' - \delta \dd i']\}$,
    it holds $|B| = \delta + 1$ and $B \subseteq A' \setminus A$. We
    proceed in two steps:
    \begin{itemize}
    \item First, we show that $|B| = \delta + 1$. Let $q, q' \in [i' -
      \delta \dd i']$ and assume $q \neq q'$. We will prove that
      $\RunBeg{\tau}{\Text}{\SA[q]} \neq \RunBeg{\tau}{\Text}{\SA[q']}$. Above we proved that
      $\PosHighMinus{f}{\Pat}{\Text} = \{\SA[i''] : i'' \in (i' - \DeltaHighMinus{f}{\Pat}{\Text}
      \dd i']\}$. By $\delta \in [0 \dd \DeltaHighMinus{f}{\Pat}{\Text})$, we thus
      have $\SA[q], \SA[q'] \in \PosHighMinus{f}{\Pat}{\Text}$. Consequently,
      $\SA[q], \SA[q'] \in \RMinusFive{f}{s}{H}{\tau}{\Text}$ and $\ExpPos{f}{\tau}{\Text}{\SA[q]} =
      \ExpPos{f}{\tau}{\Text}{\SA[q']} = k_2$.  Since for every $t \in \RTwo{\tau}{\Text}$, it holds
      $\RunEndFullPos{f}{\tau}{\Text}{t} - t = \HeadPos{f}{\tau}{\Text}{t} + |\RootPos{f}{\tau}{\Text}{t}| \cdot \ExpPos{f}{\tau}{\Text}{t}$, we
      thus have $\RunEndFullPos{f}{\tau}{\Text}{\SA[q]} - \SA[q] = \RunEndFullPos{f}{\tau}{\Text}{\SA[q']} -
      \SA[q']$.  By definition of $\SA$, we have $\SA[q] \neq
      \SA[q']$. Assume without the loss of generality that $\SA[q] <
      \SA[q']$, and suppose that $[\SA[q] \dd \SA[q']] \subseteq \RTwo{\tau}{\Text}$.
      By \cref{lm:R-text-block}, we then have $\RunEndFullPos{f}{\tau}{\Text}{\SA[q]} =
      \RunEndFullPos{f}{\tau}{\Text}{\SA[q']}$, which implies $\SA[q] = \SA[q']$, a
      contradiction.  We thus cannot have $[\SA[q] \dd \SA[q']]
      \subseteq \RTwo{\tau}{\Text}$, i.e., there exists $t \in (\SA[q] \dd \SA[q'])$
      satisfying $t \not\in \RTwo{\tau}{\Text}$. By
      \cref{lm:beg-end}\eqref{lm:beg-end-it-2}, we then obtain
      $\RunBeg{\tau}{\Text}{\SA[q]} \leq \SA[q] < t < \RunBeg{\tau}{\Text}{\SA[q']}$. In particular,
      $\RunBeg{\tau}{\Text}{\SA[q]} \neq \RunBeg{\tau}{\Text}{\SA[q']}$.
    \item Second, we show that $B \subseteq A' \setminus A$. Let $i''
      \in [i' - \delta \dd i']$ and $t = \RunBeg{\tau}{\Text}{\SA[i'']}$.  To show $t
      \in A' \setminus A$, we need to prove that $t \in \RPrimMinusFour{f}{H}{\tau}{\Text}$,
      $x \leq \RunEndFullPos{f}{\tau}{\Text}{t} - t$, and $\Textinf[\RunEndFullPos{f}{\tau}{\Text}{t} \dd
      \RunEndFullPos{f}{\tau}{\Text}{t} + 7\tau) \succeq y_u$. Above, we observed that it
      holds $\SA[i''] \in \PosHighMinus{f}{\Pat}{\Text}$.  This implies $\SA[i'']
      \in \RMinusFive{f}{s}{H}{\tau}{\Text}$. On the other hand, by
      \cref{lm:beg-end}\eqref{lm:beg-end-it-2}, we have $[t \dd
      \SA[i'']] \subseteq \RTwo{\tau}{\Text}$. Thus, by \cref{lm:R-text-block}, we
      obtain $t \in \RMinusFour{f}{H}{\tau}{\Text}$.  Since
      \cref{lm:beg-end}\eqref{lm:beg-end-it-2} also implies $t \in
      \RPrimTwo{\tau}{\Text}$, we thus obtain $t \in \RPrimMinusFour{f}{H}{\tau}{\Text}$.  Next, observe that
      $\SA[i''] \in \PosHighMinus{f}{\Pat}{\Text}$ also implies $\ExpPos{f}{\tau}{\Text}{\SA[i'']} =
      k_2$.  Combining this with $\SA[i''] \in \RFive{f}{s}{H}{\tau}{\Text}$ yields
      $\RunEndFullPos{f}{\tau}{\Text}{\SA[i'']} - \SA[i''] = \HeadPos{f}{\tau}{\Text}{\SA[i'']} +
      \ExpPos{f}{\tau}{\Text}{\SA[i'']}|\RootPos{f}{\tau}{\Text}{\SA[i'']}| = s + k_2 p = \HeadPat{f}{\tau}{\Pat} +
      \ExpCutPat{f}{\tau}{\Pat}{2\ell}|\RootPat{f}{\tau}{\Pat}| = \RunEndCutPat{f}{\tau}{\Pat}{2\ell} - 1
      = x$.  Note now that we have $t \leq \SA[i'']$.  On the other
      hand, by $[t \dd \SA[i'']] \subseteq \RTwo{\tau}{\Text}$ and
      \cref{lm:R-text-block}, it holds $\RunEndFullPos{f}{\tau}{\Text}{t} =
      \RunEndFullPos{f}{\tau}{\Text}{\SA[i'']}$.  Thus, $x = \RunEndFullPos{f}{\tau}{\Text}{\SA[i'']} - \SA[i'']
      = \RunEndFullPos{f}{\tau}{\Text}{t} - \SA[i''] \leq \RunEndFullPos{f}{\tau}{\Text}{t} - t$.  It remains to
      prove $\Textinf[\RunEndFullPos{f}{\tau}{\Text}{t} \dd \RunEndFullPos{f}{\tau}{\Text}{t} + 7\tau) \succeq
      y_u$.  Recall that by $\delta \in [0 \dd \DeltaHighMinus{f}{\Pat}{\Text})$, we
      have $\SA[i'-\delta] \in \PosHighMinus{f}{\Pat}{\Text}$.  On the other hand,
      we also have $\SA[i''] \in \PosHighMinus{f}{\Pat}{\Text}$.  Consequently,
      $\SA[i'-\delta], \SA[i''] \in \RSix{f}{s}{k_2}{H}{\tau}{\Text}$. This implies
      $\LCE_{\Text}(\SA[i'-\delta], \SA[i'']) \geq s + k_2 p = x$. Note
      that by the uniqueness of $\Text[\Textlen]$ in $\Text$ and $i' - \delta \leq
      i''$, for every $x' \geq 0$, it holds $\Textinf[\SA[i' - \delta]
      \dd \SA[i' - \delta] + x') \preceq \Textinf[\SA[i''] \dd \SA[i''] +
      x')$. With $\LCE_{\Text}(\SA[i'-\delta], \SA[i'']) \geq x$, this is
      equivalent to $\Textinf[\SA[i'-\delta] + x \dd \SA[i'-\delta] + x +
      x') \preceq \Textinf[\SA[i''] + x \dd \SA[i''] + x + x')$.  In
      particular, for $x' = 7\tau$, we thus have $y_u = \Textinf[j'' \dd
      j'' + 7\tau) = \Textinf[\SA[i' - \delta] + x \dd \SA[i' - \delta] +
      x + 7\tau) \preceq \Textinf[\SA[i''] + x \dd \SA[i''] + x + 7\tau)
      = \Textinf[\RunEndFullPos{f}{\tau}{\Text}{\SA[i'']} \dd \RunEndFullPos{f}{\tau}{\Text}{\SA[i'']} + 7\tau)$.
    \end{itemize}
  \item Next, we show $c - \delta \leq
    \IncRangeCountThreeSide{\Pts}{x}{\Textlen}{y_u}$. By
    \cref{lm:sa-periodic-count}\eqref{lm:sa-periodic-count-it-2}, we
    have $\IncRangeCountThreeSide{\Pts}{x}{\Textlen}{y_u} = |A|$, where $A = \{t \in
    \RPrimMinusFour{f}{H}{\tau}{\Text} : x \leq \RunEndFullPos{f}{\tau}{\Text}{t} - t\text{ and }
    \Textinf[\RunEndFullPos{f}{\tau}{\Text}{t} \dd \RunEndFullPos{f}{\tau}{\Text}{t} + 7\tau) \preceq y_u\}$ (the
    requirement $x \leq 7\tau$ is proved above). On the other hand, by
    \cref{lm:sa-periodic-count}\eqref{lm:sa-periodic-count-it-3}, we
    have $c = |A'|$, where $A' = \{t \in \RPrimMinusFour{f}{H}{\tau}{\Text} : x \leq
    \RunEndFullPos{f}{\tau}{\Text}{t} - t\}$.  Our claim is thus equivalent to $|A'| - |A|
    \leq \delta$. By $A \subseteq A'$ our goal is to prove that $|A'
    \setminus A| \leq \delta$. We prove this in two steps.
    \begin{itemize}
    \item Let $B = \{\RunBeg{\tau}{\Text}{t'} : t' \in \PosHighMinus{f}{\Pat}{\Text}\text{ and }
      \Textinf[\RunEndFullPos{f}{\tau}{\Text}{t'} \dd \RunEndFullPos{f}{\tau}{\Text}{t'} + 7\tau) \succ y_u\}$. In
      the first step, we show that $A' \setminus A \subseteq
      B$.  Let $q \in A' \setminus A$, i.e., $q \in \RPrimMinusFour{f}{H}{\tau}{\Text}$,
      $\RunEndFullPos{f}{\tau}{\Text}{q} - q \geq x = \RunEndCutPat{f}{\tau}{\Pat}{2\ell} - 1$, and
      $\Textinf[\RunEndFullPos{f}{\tau}{\Text}{q} \dd \RunEndFullPos{f}{\tau}{\Text}{q} + 7\tau) \succ y_u$.
      Recall that $\delta \in [0 \dd \DeltaHighMinus{f}{\Pat}{\Text})$ implies that
      $\SA[i'-\delta] \in \PosHighMinus{f}{\Pat}{\Text}$. By
      \cref{lm:poslow-poshigh-single} we thus obtain
      $\Textinf[\RunEndFullPos{f}{\tau}{\Text}{\SA[i'-\delta]} \dd \RunEndFullPos{f}{\tau}{\Text}{\SA[i'-\delta]} +
      7\tau) \succeq \Pat_{\rm sub}$, where $P_{\rm sub} =
      \Pat[\RunEndCutPat{f}{\tau}{\Pat}{2\ell} \dd \min(m,2\ell)]$.  Consequently,
      $\Textinf[\RunEndFullPos{f}{\tau}{\Text}{q} \dd \RunEndFullPos{f}{\tau}{\Text}{q} + 7\tau) \succ y_u =
      \Textinf[\RunEndFullPos{f}{\tau}{\Text}{\SA[i'-\delta]} \dd \RunEndFullPos{f}{\tau}{\Text}{\SA[i'-\delta]} +
      7\tau) \succeq \Pat_{\rm sub}$.  It therefore follows by
      \cref{lm:poslow-poshigh-run} that, letting $t = \RunEndPos{\tau}{\Text}{q} - q -
      3\tau + 2$, it holds $\PosHighMinus{f}{\Pat}{\Text} \cap [q \dd q + t) =
      \{q'\}$, where $q' = \RunEndFullPos{f}{\tau}{\Text}{q} - (\RunEndCutPat{f}{\tau}{\Pat}{2\ell} - 1)$.
      Observe that $\RunEndFullPos{f}{\tau}{\Text}{q} - q \geq \RunEndCutPat{f}{\tau}{\Pat}{2\ell} - 1$
      implies that $q' \geq q$. Combining this with $q' < q + t =
      \RunEndPos{\tau}{\Text}{q} - 3\tau + 2$ and
      \cref{lm:beg-end}\eqref{lm:beg-end-it-1}, implies that $[q \dd
      q'] \subseteq \RTwo{\tau}{\Text}$.  Thus, by $q \in \RPrimTwo{\tau}{\Text}$ and
      \cref{lm:beg-end}\eqref{lm:beg-end-it-2}, we have $\RunBeg{\tau}{\Text}{q'} =
      q$. Lastly, note that $[q \dd q'] \subseteq \RTwo{\tau}{\Text}$ and
      \cref{lm:R-text-block} imply $\RunEndFullPos{f}{\tau}{\Text}{q'} =
      \RunEndFullPos{f}{\tau}{\Text}{q}$. Thus, $\Textinf[\RunEndFullPos{f}{\tau}{\Text}{q'} \dd
      \RunEndFullPos{f}{\tau}{\Text}{q'} + 7\tau) = \Textinf[\RunEndFullPos{f}{\tau}{\Text}{q} \dd \RunEndFullPos{f}{\tau}{\Text}{q} +
      7\tau) \succ y_u$. Hence, we obtain $q \in B$.
    \item Let us now denote $B' = \{\RunBeg{\tau}{\Text}{\SA[i'']} : i'' \in (i' - \delta \dd
      i']\}$.  In the second step we show that $B \subseteq B'$.
      Consider any $i_1, i_2 \in (i' - \DeltaHighMinus{f}{\Pat}{\Text} \dd i']$ and
      assume that it holds $i_1 < i_2$. As noted above, we then have $\SA[i_1],
      \SA[i_2] \in \PosHighMinus{f}{\Pat}{\Text}$. Consequently, $\SA[i_1], \SA[i_2]
      \in \RMinusSix{f}{s}{k_2}{H}{\tau}{\Text}$, and thus $\RunEndFullPos{f}{\tau}{\Text}{\SA[i_1]} - \SA[i_1]
      = \RunEndFullPos{f}{\tau}{\Text}{\SA[i_2]} - \SA[i_2] = s + k_2 p = x$ and
      $\Text[\SA[i_1] \dd \RunEndFullPos{f}{\tau}{\Text}{\SA[i_1]}) \allowbreak = \Text[\SA[i_2] \dd
      \RunEndFullPos{f}{\tau}{\Text}{\SA[i_2]})$. By definition of the suffix array and the
      uniqueness of $\Text[\Textlen]$ in $\Text$, this implies
      $\Textinf[\RunEndFullPos{f}{\tau}{\Text}{\SA[i_1]} \dd \RunEndFullPos{f}{\tau}{\Text}{\SA[i_1]} + 7\tau)
      \prec \Text[\RunEndFullPos{f}{\tau}{\Text}{\SA[i_2]} \dd \RunEndFullPos{f}{\tau}{\Text}{\SA[i_2]} +
      7\tau)$. Consequently, for every $i'' \in (i' -
      \DeltaHighMinus{f}{\Pat}{\Text} \dd i']$, $\Textinf[\RunEndFullPos{f}{\tau}{\Text}{\SA[i'']} \dd
      \RunEndFullPos{f}{\tau}{\Text}{\SA[i'']} + 7\tau) \succ y_u =
      \Textinf[\RunEndFullPos{f}{\tau}{\Text}{\SA[i'-\delta]} \dd \RunEndFullPos{f}{\tau}{\Text}{\SA[i'-\delta]} +
      7\tau)$ implies $i'' > i' - \delta$. Let us now consider any $t
      \in B$. Then, by definition, there exists $t' \in
      \PosHighMinus{f}{\Pat}{\Text}$ satisfying $\Textinf[\RunEndFullPos{f}{\tau}{\Text}{t'} \dd
      \RunEndFullPos{f}{\tau}{\Text}{t'} + 7\tau) \succ y_u$ and $\RunBeg{\tau}{\Text}{t'} = t$.  The
      assumption $t' \in \PosHighMinus{f}{\Pat}{\Text}$ implies that there exists
      $i'' \in (i' - \DeltaHighMinus{f}{\Pat}{\Text} \dd i']$ such that $t' =
      \SA[i'']$. By the above, the assumption
      $\Textinf[\RunEndFullPos{f}{\tau}{\Text}{\SA[i'']} \dd \RunEndFullPos{f}{\tau}{\Text}{\SA[i'']} + 7\tau)
      \succ y_u$ implies that $i'' \in (i' - \delta \dd i']$. We thus
      have $\RunBeg{\tau}{\Text}{\SA[i'']} = \RunBeg{\tau}{\Text}{t'} = t \in B'$, which concludes
      the proof of $B \subseteq B'$.
    \end{itemize}
  \item Observe now that $|B'| \leq \delta$. By the above, we thus
    have $|A' \setminus A| \leq |B| \leq |B'| \leq \delta$.
  \end{enumerate}
  We have thus proved that it holds $c - \delta \in
  (\RangeCountThreeSide{\Pts}{x}{\Textlen}{y_u} \dd
  \IncRangeCountThreeSide{\Pts}{x}{\Textlen}{y_u}]$.  By definition, this implies
  that $\RangeSelect{\Pts}{x}{\Textlen}{c - \delta} = \{j \in \Z : (x',y',w',j)
  \in \Pts, x \leq x' < \Textlen,\text{ and }y' = y_u\}$.  Let us thus
  consider some $j \in \RangeSelect{\Pts}{x}{\Textlen}{c - \delta}$.  By the
  above, there exists $(x',y',w',j) \in \Pts$ satisfying $x \leq x' <
  \Textlen$ and $y' = y_u$. By \cref{def:int-str}, the position $j$ satisfies
  $j \in [1 \dd \Textlen]$ and $\Textinf[j \dd j + 7\tau) = y' = y_u =
  \Textinf[\RunEndFullPos{f}{\tau}{\Text}{\SA[i'-\delta]} \dd \RunEndFullPos{f}{\tau}{\Text}{\SA[i'-\delta]} +
  7\tau) = \Textinf[\SA[i'-\delta] + x \dd \SA[i'-\delta] + x +
  7\tau)$. Consequently,
  $\Textinf[j \dd j + 7\tau - x) = \Textinf[\SA[i'-\delta] + x \dd
  \SA[i'-\delta] + 7\tau)$.  Next, we prove that $j - x \in [1 \dd \Textlen]$
  and $\Text[j - x \dd j) = \Text[\SA[i'-\delta] \dd \SA[i'-\delta] + x)$.
  We proceed in three steps:
  \begin{enumerate}[label=(\alph*)]
  \item By definition of $\Pts = \IntStrPoints{7\tau}{\PairsMinus{f}{H}{\tau}{\Text}}{\Text}$ (see
    \cref{def:int-str}), there exists $(q,h) \in \PairsMinus{f}{H}{\tau}{\Text}$ such that
    $x' = h$ and $\Textinf[q \dd q + 7\tau) = y_u$. Furthermore, by
    \cref{def:int-str}, position $j$ satisfies $\Textinf[j - 7\tau \dd j
    + 7\tau) = \Textinf[q - 7\tau \dd q + 7\tau)$. Recall also (see
    \cref{sec:sa-periodic-ds}), that $(q, h) \in \PairsMinus{f}{H}{\tau}{\Text}$ in turn
    implies that there exists $q' \in \CompRepr{14\tau}{\RMinusFour{f}{H}{\tau}{\Text}}{\Text}
    \subseteq \RMinusFour{f}{H}{\tau}{\Text}$ such that $q = \RunEndFullPos{f}{\tau}{\Text}{q'}$ and $h =
    \min(\RunEndFullPos{f}{\tau}{\Text}{q'} - \RunBeg{\tau}{\Text}{q'}, 7\tau)$.  Let us denote $q'' = q -
    x$.
  \item Next, we prove that $j - x \in [1 \dd \Textlen]$ and $\Textinf[j - x \dd
    j) = H' H^{k_2}$, where $H'$ is a length-$s$ suffix of $H$.
    Above, we observed that $x'$ satisfies $x \leq x' < \Textlen$. By $x' = h
    = \min(\RunEndFullPos{f}{\tau}{\Text}{q'} - \RunBeg{\tau}{\Text}{q'}, 7\tau)$, this implies $x \leq
    \RunEndFullPos{f}{\tau}{\Text}{q'} - \RunBeg{\tau}{\Text}{q'}$, or equivalently, $q'' = q - x =
    \RunEndFullPos{f}{\tau}{\Text}{q'} - x \geq \RunBeg{\tau}{\Text}{q'}$. On the other hand, by $q' \in
    \RFour{f}{H}{\tau}{\Text}$, \cref{lm:R-text-block}, and
    \cref{lm:beg-end}\eqref{lm:beg-end-it-2}, we have $\RunBeg{\tau}{\Text}{q'} \in
    \RFour{f}{H}{\tau}{\Text}$. Thus, $\RunBeg{\tau}{\Text}{q'} \leq q'' \leq \RunEndFullPos{f}{\tau}{\Text}{q'}$ and
    $\RunEndFullPos{f}{\tau}{\Text}{q'} - q'' = x = s + k_2 p$ imply that $\Text[q'' \dd q) =
    \Text[q'' \dd \RunEndFullPos{f}{\tau}{\Text}{q'}) = H' H^{k_2}$.  By $x \leq 2\ell \leq
    7\tau$ and $\Textinf[q - 7\tau \dd q) = \Textinf[j - 7\tau \dd j)$, we
    thus obtain $\Textinf[j - x \dd j) = H' H^{k_2}$. Since $H$ does not
    contain $\Text[\Textlen]$, we thus obtain that $\Textinf[j - x \dd j)$ also
    does not contain $\Text[\Textlen]$. Combining this with $j \in [1 \dd \Textlen]$,
    we thus obtain $j - x \in [1 \dd \Textlen]$.
  \item Lastly, observe that $\RunEndFullPos{f}{\tau}{\Text}{\SA[i'-\delta]} -
    \SA[i'-\delta] = x$ and $\SA[i'-\delta] \in \RFive{f}{s}{H}{\tau}{\Text}$ implies that
    $\Text[\SA[i'-\delta] \dd \SA[i'-\delta] + x) = \Text[\SA[i'-\delta] \dd
    \RunEndFullPos{f}{\tau}{\Text}{\SA[i'-\delta]}) = H' H^{k_2}$. By $\Textinf[j - x \dd j) =
    H' H^{k_2}$, we thus obtain $\Textinf[j - x \dd j) =
    \Textinf[\SA[i'-\delta] \dd \SA[i'-\delta] + x)$.
  \end{enumerate}
  We have thus proved that $j - x \in [1 \dd \Textlen]$ and $\Textinf[j - x \dd
  j) = \Textinf[\SA[i'-\delta] \dd \SA[i'-\delta] + x)$.  Combining this
  with $\Textinf[j \dd j - x + 7\tau) = \Textinf[\SA[i' - \delta] + x \dd
  \SA[i' - \delta] + 7\tau)$, we thus obtain $\Textinf[j - x \dd j - x +
  7\tau) = \Textinf[\SA[i' - \delta] \dd \SA[i' - \delta] + 7\tau)$.

  3. Applying the claim from the second step for $\delta = i' - i$
  (note that we have $\delta \in [0 \dd \DeltaHighMinus{f}{\Pat}{\Text})$, since in
  the first step we observed that $\SA[i] \in \PosHighMinus{f}{\Pat}{\Text}$ implies
  $i' - \DeltaHighMinus{f}{\Pat}{\Text} < i \leq i'$), we obtain that any $j \in
  \RangeSelect{\Pts}{x}{\Textlen}{c - (i' - i)}$ satisfies $j - x \in [1 \dd \Textlen]$
  and $\Textinf[j - x \dd j - x + 7\tau) = \Textinf[\SA[i' - \delta] \dd
  \SA[i' - \delta] + 7\tau) = \Textinf[\SA[i] \dd \SA[i] + 7\tau)$.  By
  $\SA[i] \in \OccThree{2\ell}{\Pat}{\Text}$, $2\ell \leq 7\tau$, and
  \cref{lm:pat-occ-equivalence}, we thus have $j - x \in
  \OccThree{2\ell}{\Pat}{\Text}$. On the other hand, $\SA[i] \in
  \PosHighMinus{f}{\Pat}{\Text}$ and \cref{lm:poshigh-equivalence} imply
  that $j - x \in \PosHighMinus{f}{\Pat}{\Text}$. We thus obtain
  $j - x \in \OccThree{2\ell}{\Pat}{\Text} \cap \PosHighMinus{f}{\Pat}{\Text}$.
\end{proof}

\begin{lemma}\label{lm:pat-simple-occ-elem}
  Let $\ell \in [16 \dd \Textlen)$, $\tau = \lfloor \tfrac{\ell}{3} \rfloor$,
  and $f$ be any necklace-consistent function.
  Let $\Pat \in \Sigma^{m}$ be a $\tau$-periodic pattern such that
  $\TypePat{\tau}{\Pat} = -1$, $\RunEndPat{\tau}{\Pat} - 1 < \min(m, 2\ell)$, and $\Text[\Textlen]$
  does not occur in $\Pat[1 \dd m)$. Denote $H = \RootPat{f}{\tau}{\Pat}$, $\Pts
  = \IntStrPoints{7\tau}{\PairsMinus{f}{H}{\tau}{\Text}}{\Text}$, $i' = \RangeBegThree{\ell}{\Pat}{\Text} +
  \DeltaLowMinus{f}{\Pat}{\Text} + \DeltaMidMinus{f}{\Pat}{\Text}$, $x = \RunEndFullPat{f}{\tau}{\Pat} - 1$, and
  $c = \RangeCountTwoSide{\Pts}{x}{\Textlen}$. Let $i \in [1 \dd \Textlen]$ be such that
  $\SA[i] \in \OccThree{2\ell}{\Pat}{\Text}$. Then:
  \begin{itemize}
  \item It holds $c - (i' - i) \in [1 \dd c]$, i.e.,
    $\RangeSelect{\Pts}{x}{\Textlen}{c - (i'-i)}$ is well-defined,
  \item Every position $j \in \RangeSelect{\Pts}{x}{\Textlen}{c - (i'-i)}$
    satisfies $j - x \in \OccThree{2\ell}{\Pat}{\Text}$.
  \end{itemize}
\end{lemma}
\begin{proof}
  By \cref{lm:occ-sub-poshigh}, it holds $x = \RunEndFullPat{f}{\tau}{\Pat} - 1 =
  \RunEndCutPat{f}{\tau}{\Pat}{2\ell} - 1$ and $\SA[i] \in \PosHighMinus{f}{\Pat}{\Text}$. The claim
  thus follows by \cref{lm:sa-periodic-occ-elem}.
\end{proof}

\begin{lemma}\label{lm:pat-fully-periodic-two-subsets-of-occ}
  Let $\ell \in [16 \dd \Textlen)$, $\tau = \lfloor \tfrac{\ell}{3} \rfloor$,
  and $f$ be any necklace-consistent function.
  Let $\Pat \in \Sigma^{+}$ be a $\tau$-periodic pattern satisfying
  $\RunEndPat{\tau}{\Pat} - 1 \geq 2\ell$. Denote $s = \HeadPat{f}{\tau}{\Pat}$, $H =
  \RootPat{f}{\tau}{\Pat}$, $p = |H|$, and $k_2 = \lfloor \tfrac{2\ell - s}{p}
  \rfloor$. Then:
  \begin{enumerate}
  \item It holds $\RSix{f}{s}{k_2+1}{H}{\tau}{\Text} \cup \PosHighAll{f}{\Pat}{\Text} \subseteq
    \OccThree{2\ell}{\Pat}{\Text}$.
  \item If $\OccThree{2\ell}{\Pat}{\Text} \neq \emptyset$, then
    $\RSix{f}{s}{k_2+1}{H}{\tau}{\Text} \cup \PosHighAll{f}{\Pat}{\Text} \neq
    \emptyset$.
  \end{enumerate}
\end{lemma}
\begin{proof}

  Denote $m = |\Pat|$ and $t = \TailPat{f}{\tau}{\Pat}$.  Observe that the
  assumption $\RunEndPat{\tau}{\Pat} - 1 \geq 2\ell$ implies that
  $\ExpCutPat{f}{\tau}{\Pat}{2\ell} = \min(\ExpPat{f}{\tau}{\Pat}, \lfloor \tfrac{2\ell -
  s}{p} \rfloor) = \min(\lfloor \tfrac{\RunEndPat{\tau}{\Pat} - 1 - s}{p}
  \rfloor, \lfloor \tfrac{2\ell - s}{p} \rfloor) = \lfloor
  \tfrac{2\ell - s}{p} \rfloor = k_2$ and $m \geq \RunEndPat{\tau}{\Pat} - 1 \geq
  2\ell$. Let us thus denote $\Pat' := \Pat[1 \dd 2\ell]$. The
  assumption $\RunEndPat{\tau}{\Pat} - 1 \geq 2\ell$ then implies that, letting
  $H'$ (resp.\ $H''$) be a length-$s$ suffix (resp.\ length-$t$
  prefix) of $H$, we have $\Pat' = H' H^{k_2} H''$. Consequently, by
  definition of $\OccThree{2\ell}{\Pat}{\Text}$, $j \in
  \OccThree{2\ell}{\Pat}{\Text}$ holds if and only if $\Pat'$ is a prefix of $\Text[j \dd \Textlen]$,
  i.e., $j \in \OccTwo{\Pat'}{\Text}$. Thus, $\OccThree{2\ell}{\Pat}{\Text} =
  \OccTwo{\Pat'}{\Text}$.  Note also that $\Pat'$ is $\tau$-periodic,
  $\HeadPat{f}{\tau}{\Pat'} = s$, $\RootPat{f}{\tau}{\Pat'} = H$, $\RunEndPat{\tau}{\Pat'} - 1 =
  2\ell$, and $\TypePat{\tau}{\Pat'} = -1$.

  1. We show each of the two inclusions separately:
  \begin{itemize}
  \item Let $j \in \RSix{f}{s}{k_2+1}{H}{\tau}{\Text}$. By definition, this implies that
    $H' H^{k_2+1}$ is a prefix of $\Text[j \dd \Textlen]$.  Thus, $j \in
    \OccTwo{\Pat'}{\Text} = \OccThree{2\ell}{\Pat}{\Text}$.
  \item Let $j \in \PosHighAll{f}{\Pat}{\Text} = \PosHighMinus{f}{\Pat}{\Text} \cup
    \PosHighPlus{f}{\Pat}{\Text}$.  Consider first the case $j \in
    \PosHighMinus{f}{\Pat}{\Text}$, i.e., $j \in \RMinusSix{f}{s}{k_2}{H}{\tau}{\Text}$ and either $\Text[j
    \dd \Textlen] \succeq \Pat$ or $\lcp(\Pat, \Text[j \dd \Textlen]) \geq 2\ell$. If
    the latter condition holds, then we immediately obtain $j \in
    \OccThree{2\ell}{\Pat}{\Text}$. Let us thus assume $\Text[j \dd \Textlen] \succeq
    \Pat$. Note that $\Pat \succeq \Pat'$, and hence $\Text[j \dd \Textlen]
    \succeq \Pat'$. We consider two cases. If $\Text[j \dd \Textlen] = \Pat'$,
    then $j \in \OccTwo{\Pat'}{\Text} = \OccThree{2\ell}{\Pat}{\Text}$.  Let us
    thus assume $\Text[j \dd \Textlen] \succ \Pat'$. Observe that we then have
    $\RunEndPos{\tau}{\Text}{j} - j \geq \RunEndPat{\tau}{\Pat'} - 1$, since otherwise, by
    \cref{lm:R-lex-block-pat}\eqref{lm:R-lex-block-pat-it-3},
    we would have $\Text[j \dd \Textlen] \prec \Pat'$, contradicting our
    assumption. Consequently, $\RunEndPos{\tau}{\Text}{j} - j \geq \RunEndPat{\tau}{\Pat'} - 1 =
    2\ell$. By $j \in \RFive{f}{s}{H}{\tau}{\Text}$, this implies that $H' H^{k_2} H''$ is
    a prefix of $\Text[j \dd \Textlen]$, i.e., $j \in \OccTwo{\Pat'}{\Text} =
    \OccThree{2\ell}{\Pat}{\Text}$. The proof of the inclusion
    $\PosHighPlus{f}{\Pat}{\Text} \subseteq \OccThree{2\ell}{\Pat}{\Text}$ is symmetric to
    the proof for $\PosHighMinus{f}{\Pat}{\Text}$.
  \end{itemize}

  2. Let us now assume $\OccThree{2\ell}{\Pat}{\Text} \neq
  \emptyset$. Consider any $j \in \OccThree{2\ell}{\Pat}{\Text} =
  \OccTwo{\Pat'}{\Text}$.  This implies that $\lcp(\Text[j \dd \Textlen], \Pat') \geq 3\tau -
  1$. Hence, by \cref{lm:periodic-pos-lce}\eqref{lm:periodic-pos-lce-it-1}, we obtain $j
  \in \RFive{f}{s}{H}{\tau}{\Text}$. Furthermore, observe that we must have $\RunEndPos{\tau}{\Text}{j} - j
  \geq \RunEndPat{\tau}{\Pat'} - 1$, since otherwise, by
  \cref{lm:R-lex-block-pat}\eqref{lm:R-lex-block-pat-it-1}, we would
  have $\lcp(\Pat', \Text[j \dd \Textlen]) = \min(\RunEndPos{\tau}{\Text}{j} - j, \RunEndPat{\tau}{\Pat'} - 1)
  = \RunEndPos{\tau}{\Text}{j} - j < \RunEndPat{\tau}{\Pat'} - 1 = 2\ell$, contradicting $\lcp(\Pat',
  \Text[j \dd \Textlen]) = 2\ell$ (implied by $j \in \OccTwo{\Pat'}{\Text}$).
  Consequently, $\ExpPos{f}{\tau}{\Text}{j} = \lfloor \tfrac{\RunEndPos{\tau}{\Text}{j} - j - s}{p}
  \rfloor \geq \lfloor \tfrac{\RunEndPat{\tau}{\Pat'} - 1 - s}{p} \rfloor =
  \lfloor \tfrac{2\ell - s}{p} \rfloor = k_2$. Let us thus consider
  two cases:
  \begin{itemize}
  \item First, assume that $\ExpPos{f}{\tau}{\Text}{j} = k_2$. Since we also have
    $\ExpCutPat{f}{\tau}{\Pat}{2\ell} = k_2$, $j \in \RFive{f}{s}{H}{\tau}{\Text}$, and $\lcp(\Pat,
    \Text[j \dd \Textlen]) \geq \lcp(\Pat', \Text[j \dd \Textlen]) = 2\ell$, we
    immediately obtain that either $j \in \PosHighMinus{f}{\Pat}{\Text}$ or $j \in
    \PosHighPlus{f}{\Pat}{\Text}$.  Thus, $\PosHighAll{f}{\Pat}{\Text} \neq \emptyset$.
  \item Let us now assume $\ExpPos{f}{\tau}{\Text}{j} > k_2$. Let $j' = \RunEndFullPos{f}{\tau}{\Text}{j} -
    (k_2 + 1)\cdot p - s$. We will prove that $j' \in \RSix{f}{s}{k_2+1}{H}{\tau}{\Text}$.
    First, observe that $\ExpPos{f}{\tau}{\Text}{j} > k_2$ implies that $\RunEndFullPos{f}{\tau}{\Text}{j} -
    j = \HeadPos{f}{\tau}{\Text}{j} + \ExpPos{f}{\tau}{\Text}{j} |\RootPos{f}{\tau}{\Text}{j}| \geq s + (k_1 + 1) \cdot p$.
    In other words, $j' \geq j$. On the other hand, note that $k_2 =
    \lfloor \tfrac{2\ell - s}{p} \rfloor$ implies that $\RunEndPos{\tau}{\Text}{j} - j'
    \geq \RunEndFullPos{f}{\tau}{\Text}{j} - j' = s + (k_2 + 1) \cdot p = s + (\lfloor
    \tfrac{2\ell - s}{p} \rfloor + 1) \cdot p \geq s + \lceil
    \tfrac{2\ell - s}{p} \rceil \cdot p \geq s + (2\ell - s) = 2\ell
    \geq 3\tau - 1$. We thus have $j' \in [j \dd \RunEndPos{\tau}{\Text}{j} - 3\tau -
    1]$. By \cref{lm:beg-end}\eqref{lm:beg-end-it-1}, we therefore
    obtain $[j \dd j'] \subseteq \RTwo{\tau}{\Text}$. This in turn, by
    \cref{lm:R-text-block}, implies $\RootPos{f}{\tau}{\Text}{j'} = \RootPos{f}{\tau}{\Text}{j} = H$ and
    $\RunEndFullPos{f}{\tau}{\Text}{j'} = \RunEndFullPos{f}{\tau}{\Text}{j}$. Consequently, $\HeadPos{f}{\tau}{\Text}{j'} =
    (\RunEndFullPos{f}{\tau}{\Text}{j'} - j') \bmod p = (\RunEndFullPos{f}{\tau}{\Text}{j} - j') \bmod p = (s +
    (k_2 + 1) \cdot p) \bmod p = s$ and $\ExpPos{f}{\tau}{\Text}{j'} = \lfloor
    \tfrac{\RunEndFullPos{f}{\tau}{\Text}{j'} - j'}{p} \rfloor = \lfloor \tfrac{s + (k_2 +
    1) \cdot p}{p} \rfloor = k_2 + 1$. We have thus proved that $j'
    \in \RSix{f}{s}{k_2+1}{H}{\tau}{\Text}$. Thus, $\RSix{f}{s}{k_2+1}{H}{\tau}{\Text} \neq \emptyset$.
    \qedhere
  \end{itemize}
\end{proof}

\begin{lemma}\label{lm:pat-fully-periodic-poshigh-elem}
  Let $\ell \in [16 \dd \Textlen)$, $\tau = \lfloor \tfrac{\ell}{3} \rfloor$,
  and $f$ be any necklace-consistent function.
  Let $\Pat \in \Sigma^{m}$ be a $\tau$-periodic pattern such that
  $\Text[\Textlen]$ does not occur in $\Pat[1 \dd m)$, $\RunEndPat{\tau}{\Pat} - 1 \geq
  2\ell$, and $\PosHighMinus{f}{\Pat}{\Text} \neq \emptyset$.  Denote $s =
  \HeadPat{f}{\tau}{\Pat}$, $H = \RootPat{f}{\tau}{\Pat}$, $p = |H|$, $k_2 = \lfloor
  \tfrac{2\ell - s}{p} \rfloor$, $\Pts =
  \IntStrPoints{7\tau}{\PairsMinus{f}{H}{\tau}{\Text}}{\Text}$,
  $x = s + k_2 p$, and $c = \RangeCountTwoSide{\Pts}{x}{\Textlen}$. Let $c'
  \in [0 \dd \DeltaHighMinus{f}{\Pat}{\Text})$.  Then:
  \begin{itemize}
  \item It holds $c - c' \in [1 \dd c]$, i.e.,
    $\RangeSelect{\Pts}{x}{\Textlen}{c - c'}$ is well-defined.
  \item Every position $j \in \RangeSelect{\Pts}{x}{\Textlen}{c - c'}$ satisfies
    $j - x \in \PosHighMinus{f}{\Pat}{\Text}$.
  \end{itemize}
\end{lemma}
\begin{proof}
  First, observe that $m \geq \RunEndPat{\tau}{\Pat} - 1 \geq 2\ell$. Let us thus
  denote $\Pat' := \Pat[1 \dd 2\ell]$. It holds $\lcp(\Pat', \Pat) =
  2\ell \geq 3\tau - 1$. By \cref{lm:periodic-pat-lce}, we thus obtain that
  $\Pat'$ is $\tau$-periodic and it holds $\HeadPat{f}{\tau}{\Pat'} =
  \HeadPat{f}{\tau}{\Pat} = s$ and $\RootPat{f}{\tau}{\Pat'} = \RootPat{f}{\tau}{\Pat} = H$.  Observe
  also that by definition of $\RunEndPat{\tau}{\Pat}$ and the assumption
  $\RunEndPat{\tau}{\Pat} - 1 \geq 2\ell$, we have $\RunEndPat{\tau}{\Pat} - 1 = p +
  \lcp(\Pat[1 \dd m], \Pat[1 + p \dd m]) \geq 2\ell$. Thus, $p +
  \lcp(\Pat[1 \dd 2\ell], \Pat[1 + p \dd 2\ell]) = 2\ell$, and hence
  $\RunEndPat{\tau}{\Pat'} - 1 = p + \lcp(\Pat'[1 \dd 2\ell], \Pat'[1 + p \dd
  2\ell]) = p + \lcp(\Pat[1 \dd 2\ell], \Pat[1 + p \dd 2\ell]) =
  2\ell$.  Consequently, $\TypePat{\tau}{\Pat'} = -1$ and $\ExpPat{f}{\tau}{\Pat'}
  = \lfloor \tfrac{\RunEndPat{\tau}{\Pat'} - 1 - \HeadPat{f}{\tau}{\Pat'}}{|\RootPat{f}{\tau}{\Pat'}|}
  \rfloor = \lfloor \tfrac{2\ell - s}{p} \rfloor = k_2$. Note also
  that then $\ExpCutPat{f}{\tau}{\Pat'}{2\ell} = \min(\ExpPat{f}{\tau}{\Pat'}, \lfloor
  \tfrac{2\ell - s}{p} \rfloor) = k_2$, which in turn implies
  $\RunEndCutPat{f}{\tau}{\Pat'}{2\ell} - 1 = s + \ExpCutPat{f}{\tau}{\Pat'}{2\ell} \cdot p = s + k_2
  p = x$. Next, observe that by
  \cref{lm:pos-depends-on-prefix}\eqref{lm:pos-depends-on-prefix-it-3},
  it holds $\DeltaHighMinus{f}{\Pat'}{\Text} = \DeltaHighMinus{f}{\Pat'[1 \dd 2\ell]}{\Text} =
  \DeltaHighMinus{f}{\Pat[1 \dd 2\ell]}{\Text} = \DeltaHighMinus{f}{\Pat}{\Text}$.  Denote $i' =
  \RangeBegThree{\ell}{\Pat'}{\Text} + \DeltaLowMinus{f}{\Pat'}{\Text} + \DeltaMidMinus{f}{\Pat'}{\Text}$.  By
  \cref{lm:pat-posless} and
  \cref{lm:pat-decomposition}, we then have $\RangeBegThree{2\ell}{\Pat'}{\Text} =
  \RangeBegThree{\ell}{\Pat'}{\Text} + \DeltaBeg{\ell}{\Pat'}{\Text} = \RangeBegThree{\ell}{\Pat'}{\Text} +
  \DeltaLowMinus{f}{\Pat'}{\Text} + \DeltaMidMinus{f}{\Pat'}{\Text} - \DeltaHighMinus{f}{\Pat'}{\Text} = i' -
  \DeltaHighMinus{f}{\Pat'}{\Text}$.  Consequently, by
  \cref{lm:sa-periodic-poslow-poshigh-range}, $\PosHighMinus{f}{\Pat'}{\Text} =
  \{\SA[t] : t \in (i' - \DeltaHighMinus{f}{\Pat'}{\Text} \dd i']\} = \{\SA[t] : t
  \in (i' - \DeltaHighMinus{f}{\Pat}{\Text} \dd i']\}$.  By $c' \in [0 \dd
  \DeltaHighMinus{f}{\Pat}{\Text})$, this implies $\SA[i' - c'] \in
  \PosHighMinus{f}{\Pat'}{\Text}$.  Since by
  \cref{lm:pat-fully-periodic-two-subsets-of-occ}, we have
  $\PosHighMinus{f}{\Pat'}{\Text} \subseteq \OccThree{2\ell}{\Pat'}{\Text}$, we thus also
  have $\SA[i'-c'] \in \OccThree{2\ell}{\Pat'}{\Text}$.  Combining all the
  facts established so far and applying
  \cref{lm:sa-periodic-occ-elem}, we thus obtain that $c - (i' - (i' -
  c')) = c - c' \in [1 \dd c]$, i.e., $\RangeSelect{\Pts}{x}{\Textlen}{c - c'}$
  is well-defined, and that every position $j' \in
  \RangeSelect{\Pts}{x}{\Textlen}{c - c'}$ satisfies $j' - x \in
  \PosHighMinus{f}{\Pat'}{\Text}$.  It remains to observe that by
  \cref{lm:pos-depends-on-prefix}\eqref{lm:pos-depends-on-prefix-it-3},
  it holds $\PosHighMinus{f}{\Pat'}{\Text} = \PosHighMinus{f}{\Pat'[1 \dd 2\ell]}{\Text} =
  \PosHighMinus{f}{\Pat[1 \dd 2\ell]}{\Text} = \PosHighMinus{f}{\Pat}{\Text}$. Consequently, $j' -
  x \in \PosHighMinus{f}{\Pat}{\Text}$.
\end{proof}

\begin{lemma}\label{lm:pat-fully-periodic-poshigh-elem-2}
  Let $\ell \in [16 \dd \Textlen)$, $\tau = \lfloor \tfrac{\ell}{3} \rfloor$,
  and $f$ be any necklace-consistent function.
  Let $\Pat \in \Sigma^{m}$ be a $\tau$-periodic pattern such that
  $\Text[\Textlen]$ does not occur in $\Pat[1 \dd m)$, $\RunEndPat{\tau}{\Pat} - 1 \geq
  2\ell$, and $\PosHighMinus{f}{\Pat}{\Text} \neq \emptyset$.  Denote $s =
  \HeadPat{f}{\tau}{\Pat}$, $H = \RootPat{f}{\tau}{\Pat}$, $p = |H|$, $k_2 = \lfloor
  \tfrac{2\ell - s}{p} \rfloor$, $\Pts =
  \IntStrPoints{7\tau}{\PairsMinus{f}{H}{\tau}{\Text}}{\Text}$,
  $x = s + k_2 p$, and $c = \RangeCountTwoSide{\Pts}{x}{\Textlen}$. Then, $c
  \geq 1$ and every position $j \in \RangeSelect{\Pts}{x}{\Textlen}{c}$
  satisfies $j - x \in \PosHighMinus{f}{\Pat}{\Text}$.
\end{lemma}
\begin{proof}
  The result follows by \cref{lm:pat-fully-periodic-poshigh-elem}
  with $c' = 0$.
\end{proof}

\begin{lemma}\label{lm:pos-simple-occ-elem}
  Let $\ell \in [16 \dd \Textlen)$, $\tau = \lfloor \tfrac{\ell}{3} \rfloor$,
  and $f$ be any necklace-consistent function.
  Let $i \in [1 \dd \Textlen]$ be such that $\SA[i] \in \RMinusTwo{\tau}{\Text}$ and
  $\RunEndPos{\tau}{\Text}{\SA[i]} - \SA[i] < 2\ell$. Denote $H = \RootPos{f}{\tau}{\Text}{\SA[i]}$, $\Pts
  = \IntStrPoints{7\tau}{\PairsMinus{f}{H}{\tau}{\Text}}{\Text}$ (\cref{def:int-str}),
  $i' \,{=}\, \RangeBegThree{\ell}{\SA[i]}{\Text} +
  \DeltaLowMinus{f}{\SA[i]}{\Text} + \DeltaMidMinus{f}{\SA[i]}{\Text}$, $x = \RunEndFullPos{f}{\tau}{\Text}{\SA[i]} -
  \SA[i]$, and $c = \RangeCountTwoSide{\Pts}{x}{\Textlen}$. Then:
  \begin{itemize}
  \item It holds $c - (i' - i) \in [1 \dd c]$, i.e.,
    $\RangeSelect{\Pts}{x}{\Textlen}{c - (i'-i)}$ is well-defined,
  \item Every position $j \in \RangeSelect{\Pts}{x}{\Textlen}{c - (i'-i)}$
    satisfies $j - x \in \OccThree{2\ell}{\SA[i]}{\Text}$.
  \end{itemize}
\end{lemma}
\begin{proof}
  Denote $\Pat := \Text[\SA[i] \dd \Textlen]$ and $m := |\Pat| = \Textlen - \SA[i] +
  1$.  First, observe that $\SA[i] \in \RMinusTwo{\tau}{\Text}$ implies that $\Pat$ is
  $\tau$-periodic and $\TypePat{\tau}{\Pat} = -1$.
  Next, we prove that $\RunEndPat{\tau}{\Pat} - 1 < \min(m,
  2\ell)$.  By the uniqueness of $\Text[\Textlen]$ in $\Text$, we have
  $\RunEndPos{\tau}{\Text}{\SA[i]} \leq \Textlen$. Thus, $\RunEndPat{\tau}{\Pat} - 1 =
  \RunEndPat{\tau}{\Text[\SA[i] \dd \Textlen]} - 1 = \RunEndPos{\tau}{\Text}{\SA[i]} - \SA[i] \leq \Textlen - \SA[i] < m$.  On the
  other hand, the assumption $\RunEndPos{\tau}{\Text}{\SA[i]} - \SA[i] < 2\ell$ implies
  $\RunEndPat{\tau}{\Pat} - 1 = \RunEndPos{\tau}{\Text}{\SA[i]} - \SA[i] < 2\ell$.  We thus have
  $\RunEndPat{\tau}{\Pat} - 1 < \min(m, 2\ell)$.  Next, note that by the
  uniqueness of $\Text[\Textlen]$ in $\Text$, the symbol $\Text[\Textlen]$ does not occur in
  $\Pat[1 \dd m)$. Observe also that, by definition, we have:
  \begin{itemize}[itemsep=1pt]
  \item $\RangeBegThree{\ell}{\Pat}{\Text} = \RangeBegThree{\ell}{\Text[\SA[i] \dd \Textlen]}{\Text} =
    \RangeBegThree{\ell}{\SA[i]}{\Text}$,
  \item $\DeltaLowMinus{f}{\Pat}{\Text} = \DeltaLowMinus{f}{\Text[\SA[i] \dd \Textlen]}{\Text} =
    \DeltaLowMinus{f}{\SA[i]}{\Text}$,
  \item $\DeltaMidMinus{f}{\Pat}{\Text} = \DeltaMidMinus{f}{\Text[\SA[i] \dd \Textlen]}{\Text} =
    \DeltaMidMinus{f}{\SA[i]}{\Text}$,
  \item $\RootPat{f}{\tau}{\Pat} = \RootPat{f}{\tau}{\Text[\SA[i] \dd \Textlen]} = \RootPos{f}{\tau}{\Text}{\SA[i]} =
    H$, and
  \item $\RunEndFullPat{f}{\tau}{\Pat} - 1 = \RunEndFullPat{f}{\tau}{\Text[\SA[i] \dd \Textlen]} - 1 =
    \RunEndFullPos{f}{\tau}{\Text}{\SA[i]} - \SA[i] = x$.
  \end{itemize}
  Lastly, note that since $\lcp(\Text[\SA[i] \dd \Textlen], \Pat) = m \geq
  \min(m, 2\ell)$, it holds $\SA[i] \in \OccThree{2\ell}{\Pat}{\Text}$.
  The claim thus follows by \cref{lm:pat-simple-occ-elem} (recall that
  $\OccThree{2\ell}{\Pat}{\Text} = \OccThree{2\ell}{\Text[\SA[i] \dd \Textlen]}{\Text} =
  \OccThree{2\ell}{\SA[i]}{\Text}$).
\end{proof}

\begin{lemma}\label{lm:pos-fully-periodic-two-subsets-of-occ}
  Let $\ell \in [16 \dd \Textlen)$, $\tau = \lfloor \tfrac{\ell}{3} \rfloor$,
  and $f$ be any necklace-consistent function.
  Let $j \in \RTwo{\tau}{\Text}$ be such that $\RunEndPos{\tau}{\Text}{j} - j \geq 2\ell$. Denote $s =
  \HeadPos{f}{\tau}{\Text}{j}$, $H = \RootPos{f}{\tau}{\Text}{j}$, $p = |H|$, and $k_2 = \lfloor
  \tfrac{2\ell - s}{p} \rfloor$. Then, it holds:
  \begin{enumerate}[itemsep=1pt]
  \item $\RSix{f}{s}{k_2+1}{H}{\tau}{\Text} \cup \PosHighAll{f}{j}{\Text} \subseteq \OccThree{2\ell}{j}{\Text}$,
  \item $\RSix{f}{s}{k_2+1}{H}{\tau}{\Text} \cup \PosHighAll{f}{j}{\Text} \neq \emptyset$.
  \end{enumerate}
\end{lemma}
\begin{proof}
  Denote $\Pat := \Text[j \dd \Textlen]$.  First, observe that, by definition,
  $j \in \RTwo{\tau}{\Text}$ implies that $\Pat$ is $\tau$-periodic and it holds
  $\HeadPat{f}{\tau}{\Pat} = \HeadPat{f}{\tau}{\Text[j \dd \Textlen]} = \HeadPos{f}{\tau}{\Text}{j} = s$ and
  $\RootPat{f}{\tau}{\Pat} = \RootPat{f}{\tau}{\Text[j \dd \Textlen]} = \RootPos{f}{\tau}{\Text}{j} = H$.  Next, note
  that $\RunEndPat{\tau}{\Pat} - 1 = \RunEndPat{\tau}{\Text[j \dd \Textlen]} - 1 = \RunEndPos{\tau}{\Text}{j} - j \geq
  2\ell$. Lastly, note that by definition, for every $j' \in [1 \dd
  \Textlen]$ and $k \geq 0$, it holds $j' \in
  \OccThree{k}{\Text[j' \dd \Textlen]}{\Text}$. In particular, $j \in \OccThree{2\ell}{\Text[j \dd \Textlen]}{\Text} =
  \OccThree{2\ell}{\Pat}{\Text}$.  Thus, $\OccThree{2\ell}{\Pat}{\Text} \neq
  \emptyset$. Combining the above, it follows by
  \cref{lm:pat-fully-periodic-two-subsets-of-occ} that
  $\RSix{f}{s}{k_2+1}{H}{\tau}{\Text} \cup \PosHighAll{f}{\Pat}{\Text} \subseteq \OccThree{2\ell}{\Pat}{\Text}$
  and $\RSix{f}{s}{k_2+1}{H}{\tau}{\Text} \cup \PosHighAll{f}{\Pat}{\Text} \neq \emptyset$.
  To obtain the claim, it remains to observe that $\PosHighAll{f}{\Pat}{\Text} =
  \PosHighAll{f}{\Text[j \dd \Textlen]}{\Text} = \PosHighAll{f}{j}{\Text}$ and
  $\OccThree{2\ell}{\Pat}{\Text} = \OccThree{2\ell}{\Text[j \dd \Textlen]}{\Text}
  = \OccThree{2\ell}{j}{\Text}$.
\end{proof}

\paragraph{Query Algorithms}

\begin{proposition}\label{pr:pos-partially-periodic-occ-elem}
  Let $k \in [4 \dd \lceil \log \Textlen \rceil)$, $\ell = 2^k$, $\tau
  = \lfloor \tfrac{\ell}{3} \rfloor$, and $f = f_{\tau,\Text}$
  (\cref{def:canonical-function}).  Let $i \in [1 \dd \Textlen]$ be such that
  $\SA[i] \in \RMinusTwo{\tau}{\Text}$ and $\RunEndPos{\tau}{\Text}{\SA[i]}
  - \SA[i] < 2\ell$. Given $\CompSaPeriodic{\Text}$, the value $k$, the
  position $i$, some $j \in \OccThree{3\tau - 1}{\SA[i]}{\Text}$ satisfying
  $j = \min \OccThree{2\ell}{j}{\Text}$, and the values
  $\HeadPos{f}{\tau}{\Text}{\SA[i]}$,
  $|\RootPos{f}{\tau}{\Text}{\SA[i]}|$, $\RangeBegThree{\ell}{\SA[i]}{\Text}$,
  $\DeltaLowMinus{f}{\SA[i]}{\Text}$, $\DeltaMidMinus{f}{\SA[i]}{\Text}$, and
  $\ExpPos{f}{\tau}{\Text}{\SA[i]}$ as input, we can compute a position
  $j' \in \OccThree{2\ell}{\SA[i]}{\Text}$ in $\bigO(\log^{3 + \epsilon} \Textlen)$
  time.
\end{proposition}
\begin{proof}
  Let us denote $s = \HeadPos{f}{\tau}{\Text}{\SA[i]}$, $H
  = \RootPos{f}{\tau}{\Text}{\SA[i]}$, $p = |H|$, $k
  = \ExpPos{f}{\tau}{\Text}{\SA[i]}$, $b = \RangeBegThree{\ell}{\SA[i]}{\Text}$,
  $\delta_1 = \DeltaLowMinus{f}{\SA[i]}{\Text}$, $\delta_2
  = \DeltaMidMinus{f}{\SA[i]}{\Text}$, and $\Pts
  = \IntStrPoints{7\tau}{\PairsMinus{f}{H}{\tau}{\Text}}{\Text}$.
  Using \cref{pr:sa-periodic-pts-access} and the position $j$ as
  input, in $\bigO(\log \Textlen)$ time we retrieve the pointer to the
  structure from \cref{pr:int-str} for $\PairsMinus{f}{H}{\tau}{\Text}$ (note
  that $j \in \RFour{f}{H}{\tau}{\Text}$ holds
  by \cref{lm:periodic-pos-lce}\eqref{lm:periodic-pos-lce-it-2}),
  i.e., performing weighted range queries on $\Pts$. Note that the
  pointer is not null, since we assumed $\SA[i] \in \RMinusTwo{\tau}{\Text}$.
  Thus, $\RMinusFour{f}{H}{\tau}{\Text} \neq \emptyset$, which implies
  $\PairsMinus{f}{H}{\tau}{\Text} \neq \emptyset$. In $\bigO(1)$ time we now
  calculate $x := \RunEndFullPos{f}{\tau}{\Text}{\SA[i]} - \SA[i] = s +
  kp$.  Then, using \cref{pr:int-str}, in $\bigO(\log^{2 + \epsilon}
  \Textlen)$ time we compute $c = \RangeCountTwoSide{\Pts}{x}{\Textlen}$. Next, in
  $\bigO(1)$ time we set $i' = \RangeBegThree{\ell}{\SA[i]}{\Text}
  + \DeltaLowMinus{f}{\SA[i]}{\Text} + \DeltaMidMinus{f}{\SA[i]}{\Text} = b
  + \delta_1 + \delta_2$. Finally, using \cref{pr:int-str}, in
  $\bigO(\log^{3 + \epsilon} \Textlen)$ time we compute a position
  $j \in \RangeSelect{\Pts}{x}{\Textlen}{c - (i' -
  i)}$. By \cref{lm:pos-simple-occ-elem}, it holds $j -
  x \in \OccThree{2\ell}{\SA[i]}{\Text}$. We thus return $j' := j - x$ as the
  answer. In total, we spend $\bigO(\log^{3 + \epsilon} \Textlen)$ time.
\end{proof}

\begin{proposition}\label{pr:pat-fully-periodic-occ-elem}
  Let $k \in [4 \dd \lceil \log \Textlen \rceil)$, $\ell = 2^k$, $\tau
  = \lfloor \tfrac{\ell}{3} \rfloor$, and $f = f_{\tau,\Text}$
  (\cref{def:canonical-function}).  Let $j \in \RTwo{\tau}{\Text}$ be a
  position satisfying $j = \min \OccThree{2\ell}{j}{\Text}$.  Denote $s
  = \HeadPos{f}{\tau}{\Text}{j}$ and $H = \RootPos{f}{\tau}{\Text}{j}$.
  Let $H'$ be a length-$s$ suffix of $H$ and $\Pat$ be a
  length-$2\ell$ prefix of $H' H^{\infty}$. Then, $\Pat$ is
  $\tau$-periodic. Moreover, assuming $\OccTwo{\Pat}{\Text} \neq \emptyset$,
  given $\CompSaPeriodic{\Text}$, value $k$, the position $j$, and the
  values $\HeadPos{f}{\tau}{\Text}{j}$, $|\RootPos{f}{\tau}{\Text}{j}|$,
  $\DeltaHighMinus{f}{\Pat}{\Text}$, and $\DeltaHighPlus{f}{\Pat}{\Text}$ as
  input, we can compute a position $j' \in \OccTwo{\Pat}{\Text}$ in
  $\bigO(\log^{3 + \epsilon} \Textlen)$ time.
\end{proposition}
\begin{proof}
  Denote $s = \HeadPos{f}{\tau}{\Text}{j}$ and $p =
  |\RootPos{f}{\tau}{\Text}{j}|$. By \cref{lm:special-pat-properties},
  $\Pat$ is $\tau$-periodic, $\Text[\Textlen]$ does not occur in $\Pat[1 \dd
  2\ell)$, $\RunEndPat{\tau}{\Pat} - 1 = 2\ell$,
  $\HeadPat{f}{\tau}{\Pat} = s$, and $\RootPat{f}{\tau}{\Pat}
  = H$ (hence in particular, $|\RootPat{f}{\tau}{\Pat}| = p$).  In
  $\bigO(1)$ time we compute the value $k_2 = \lfloor \tfrac{2\ell -
  s}{p} \rfloor$.  We then initialize the set $\mathcal{C}
  := \emptyset$, and perform the following two steps:
  \begin{enumerate}
  \item Denote $\Pts = \IntStrPoints{7\tau}{\PairsMinus{f}{H}{\tau}{\Text}}{\Text}$.
    First, using \cref{pr:sa-periodic-pts-access} and the position $j$
    as input, in $\bigO(\log \Textlen)$ time we retrieve the pointer
    $\mu^{-}_{H}$ to the structure from \cref{pr:int-str} for
    $\PairsMinus{f}{H}{\tau}{\Text}$, i.e., performing weighted range queries
    on $\Pts$. If $\mu^{-}_{H}$ is a null pointer, this step is
    finished (note that this implies $\RMinusFour{f}{H}{\tau}{\Text}
    = \emptyset$, and hence in particular $\PosHighMinus{f}{\Pat}{\Text}
    = \RMinusSix{f}{s}{k_2+1}{H}{\tau}{\Text} = \emptyset$).  Let us thus
    assume that $\mu^{-}_{H}$ is not null. We then perform the
    following two substeps:
    \begin{itemize}
    \item First, we check if $\DeltaHighMinus{f}{\Pat}{\Text} > 0$. If
      not, we finish this substep and move to the second substep.
      Otherwise, in $\bigO(1)$ time we compute the value $x = s + k_2
      p$. Next, using \cref{pr:int-str} in $\bigO(\log^{2 + \epsilon}
      \Textlen)$ time, we compute $c = \RangeCountTwoSide{\Pts}{x}{\Textlen}$. Then, using
      again \cref{pr:int-str}, in $\bigO(\log^{3 + \epsilon} \Textlen)$ time
      we compute a position $q \in \RangeSelect{\Pts}{x}{\Textlen}{c}$. Finally,
      we add the position $q - x$ to the set
      $\mathcal{C}$. By \cref{lm:pat-fully-periodic-poshigh-elem-2},
      it holds $q - x \in \PosHighMinus{f}{\Pat}{\Text}$.
    \item In the second substep, we first let $k = k_2 + 1$ and $x = s
      + kp$. Using \cref{pr:int-str} in $\bigO(\log^{2 + \epsilon} \Textlen)$
      time we compute $h = \RangeCountTwoSide{\Pts}{x}{\Textlen}$. Denote $k_{\min}
      = \lceil \tfrac{3\tau - 1 - s}{p} \rceil - 1$ and $k_{\max}
      = \lfloor \tfrac{7\tau - s}{p} \rfloor$.  Note that by
      $2\ell \geq \ell \geq 3\tau - 1$ we then have $k = k_2 + 1
      = \lfloor \tfrac{2\ell - s}{p} \rfloor +
      1 \geq \lceil \tfrac{2\ell -
      s}{p} \rceil \geq \lceil \tfrac{3\tau - 1 - s}{p} \rceil >
      k_{\min}$. On the other hand, $k = k_2 + 1
      = \lfloor \tfrac{2\ell - s +
      p}{p} \rfloor \leq \lfloor \tfrac{2\ell + \lfloor \tau/3 \rfloor
      - s}{p} \rfloor \leq \lfloor \tfrac{7\tau - s}{p} \rfloor =
      k_{\max}$, where $2\ell + \lfloor \tau/3 \rfloor \leq 7\tau$
      follows by $\tau = \lfloor \tfrac{\ell}{3} \rfloor$ and
      $\ell \geq 16$.  Consequently, by \cref{lm:RskH-size-2}, it
      holds $h = |\RMinusSix{f}{s}{k_2+1}{H}{\tau}{\Text}|$. If $h = 0$, we
      finish this step. Let us thus assume $h > 0$.
      By \cref{lm:elem-of-RskH-2}, we then have
      $\RangeCountTwoSide{\Pts}{x}{\Textlen} \geq 1$. Using \cref{pr:int-str} in
      $\bigO(\log^{3 + \epsilon} \Textlen)$ time we compute a position
      $q \in \RangeSelect{\Pts}{x}{\Textlen}{1}$. We add $q - x$ to the set
      $\mathcal{C}$.  By \cref{lm:elem-of-RskH-2}, it holds $q -
      x \in \RMinusSix{f}{s}{k_2+1}{H}{\tau}{\Text}$.
    \end{itemize}
  \item In the second step, we perform the symmetric computation using
    the structure from \cref{pr:int-str} for $P^{+}_{f,H}(\tau, \Text)$
    (see \cref{sec:sa-periodic-ds}). This results in potentially
    inserting elements of $\PosHighPlus{f}{\Pat}{\Text}$ and
    $\RPlusSix{f}{s}{k_2+1}{H}{\tau}{\Text}$ into $\mathcal{C}$.
  \end{enumerate}
  Observe that during the above computation, for each of the four sets
  $\PosHighMinus{f}{\Pat}{\Text}$, $\PosHighPlus{f}{\Pat}{\Text}$,
  $\RMinusSix{f}{s}{k_2+1}{H}{\tau}{\Text}$, and
  $\RPlusSix{f}{s}{k_2+1}{H}{\tau}{\Text}$, we first check if the set is
  nonempty, and if so, we insert into $\mathcal{C}$ one of its
  elements.  On the one hand,
  by \cref{lm:pat-fully-periodic-two-subsets-of-occ}, this implies
  that
  $\mathcal{C} \subseteq \PosHighMinus{f}{\Pat}{\Text} \cup
  \PosHighPlus{f}{\Pat}{\Text} \cup \RMinusSix{f}{s}{k_2+1}{H}{\tau}{\Text} \cup
  \RPlusSix{f}{s}{k_2+1}{H}{\tau}{\Text}
  = \PosHighAll{f}{\Pat}{\Text} \cup \RSix{f}{s}{k_2+1}{H}{\tau}{\Text} \subseteq
  \OccThree{2\ell}{\Pat}{\Text}
  = \OccTwo{\Pat}{\Text}$.  On the other hand, observe that the same lemma
  implies that
  $\PosHighAll{f}{\Pat}{\Text} \cup \RSix{f}{s}{k_2+1}{H}{\tau}{\Text} \neq
  \emptyset$. Thus,
  by the observation about the above procedure, we have
  $\mathcal{C} \neq \emptyset$. Consequently, to finish the algorithm,
  we pick an arbitrary element from $\mathcal{C}$ and return as the
  answer.  In total, we spend $\bigO(\log^{3 + \epsilon} \Textlen)$ time.
\end{proof}

\begin{remark}
  Note that in \cref{pr:pat-fully-periodic-occ-elem}, it is possible
  that $\mathcal{C} = \emptyset$ holds after the first step, even when
  the returned pointer $\mu^{-}_{H}$ is not null.  This is
  because \cref{lm:pat-fully-periodic-two-subsets-of-occ} guarantees
  only that
  $\PosHighAll{f}{\Pat}{\Text} \cup \RSix{f}{s}{k_2+1}{H}{\tau}{\Text} \neq \emptyset$
  (i.e.,
  $\PosHighMinus{f}{\Pat}{\Text} \cup \PosHighPlus{f}{\Pat}{\Text} \cup
  \RMinusSix{f}{s}{k_2+1}{H}{\tau}{\Text} \cup \RPlusSix{f}{s}{k_2+1}{H}{\tau}{\Text} \neq
  \emptyset$). It
  is, however, possible that
  $\PosHighMinus{f}{\Pat}{\Text} \cup \RMinusSix{f}{s}{k_2+1}{H}{\tau}{\Text}
  = \emptyset$, even when $\RMinusFour{f}{H}{\tau}{\Text} \neq \emptyset$.
\end{remark}

\subsubsection{Computing a Position in a Cover}\label{sec:sa-periodic-cover}

\paragraph{Combinatorial Properties}

\begin{lemma}\label{lm:min-of-occ-poshigh}
  Let $\ell \in [16 \dd \Textlen)$, $\tau = \lfloor \tfrac{\ell}{3} \rfloor$,
  and $f$ be a necklace-consistent function.
  Let $\Pat \in \Sigma^{m}$ be a $\tau$-periodic pattern such that
  $\TypePat{\tau}{\Pat} = -1$ and $\Text[\Textlen]$ does not occur in $\Pat[1 \dd m)$.
  Assume $\OccThree{2\ell}{\Pat}{\Text} \cap \PosHighMinus{f}{\Pat}{\Text} \neq
  \emptyset$. Denote $c = \max\Sigma$, $x_l = \RunEndCutPat{f}{\tau}{\Pat}{2\ell} - 1$,
  $y_l = \Pat[\RunEndCutPat{f}{\tau}{\Pat}{2\ell} \dd \min(m,2\ell)]$, $y_u = y_l
  c^{\infty}$, $H = \RootPat{f}{\tau}{\Pat}$, and $\Pts =
  \IntStrPoints{7\tau}{\PairsMinus{f}{H}{\tau}{\Text}}{\Text}$. Then:
  \begin{enumerate}
  \item\label{lm:min-of-occ-poshigh-it-1} It holds $\min
    \OccThree{2\ell}{\Pat}{\Text} \cap \PosHighMinus{f}{\Pat}{\Text} =
    \RangeMinFourSide{\Pts}{x_l}{\Textlen}{y_l}{y_u} - x_l$.
  \item\label{lm:min-of-occ-poshigh-it-2} The position $j = \min
    \OccThree{2\ell}{\Pat}{\Text} \cap \PosHighMinus{f}{\Pat}{\Text}$ satisfies $j = \min
    \OccThree{4\ell}{j}{\Text}$.
  \end{enumerate}
\end{lemma}
\begin{proof}

  1. By \cref{lm:sa-periodic-occ-poshigh-size-pat}, it holds
  $\OccThree{2\ell}{\Pat}{\Text} \cap \PosHighMinus{f}{\Pat}{\Text} = \{\RunEndFullPos{f}{\tau}{\Text}{j} - x_l
  : j \in \RPrimMinusFour{f}{H}{\tau}{\Text},\ x_l \leq \RunEndFullPos{f}{\tau}{\Text}{j} - j,\text{ and }y_l
  \preceq \Textinf[\RunEndFullPos{f}{\tau}{\Text}{j} \dd \RunEndFullPos{f}{\tau}{\Text}{j} + 7\tau) \prec y_u\}$.
  The same lemma also implies $|\OccThree{2\ell}{\Pat}{\Text} \cap
  \PosHighMinus{f}{\Pat}{\Text}| = \RangeCountFourSide{\Pts}{x_l}{\Textlen}{y_l}{y_u}$.
  Thus, $\OccThree{2\ell}{\Pat}{\Text} \cap \PosHighMinus{f}{\Pat}{\Text} \neq
  \emptyset$ implies $\RangeCountFourSide{\Pts}{x_l}{\Textlen}{y_l}{y_u} > 0$.
  Consequently, by \cref{lm:sa-periodic-min}, we have $\min
  \OccThree{2\ell}{\Pat}{\Text} \cap \PosHighMinus{f}{\Pat}{\Text} =
  \RangeMinFourSide{\Pts}{x_l}{\Textlen}{y_l}{y_u} - x_l$. Note that
  \cref{lm:sa-periodic-min} requires $x_l \in [0 \dd 7\tau]$, which
  holds here since $x_l = \RunEndCutPat{f}{\tau}{\Pat}{2\ell} - 1 \leq 2\ell$, and for
  $\tau = \lfloor \tfrac{\ell}{3} \rfloor$ and $\ell \geq 16$, it
  holds $2\ell \leq 7\tau$.

  2. Denote $s = \HeadPat{f}{\tau}{\Pat}$, $H = \RootPat{f}{\tau}{\Pat}$, $p = |H|$, $k =
  \ExpPat{f}{\tau}{\Pat}$, and $k_2 = \ExpCutPat{f}{\tau}{\Pat}{2\ell}$. By definition of
  $\PosHighMinus{f}{\Pat}{\Text}$, we have $j \in \RMinusSix{f}{s}{k_2}{H}{\tau}{\Text}$. This implies
  that $\RunEndPos{\tau}{\Text}{j} - j = s + k_2 p + \TailPos{f}{\tau}{\Text}{j} = s + \min(k, \lfloor
  \tfrac{2\ell - s}{p} \rfloor) p + \TailPos{f}{\tau}{\Text}{j} \leq s + \lfloor
  \tfrac{2\ell - s}{p} \rfloor p + \TailPos{f}{\tau}{\Text}{j} \leq 2\ell + \TailPos{f}{\tau}{\Text}{j} <
  2\ell + p < 2\ell + \tau \leq 3\ell$.  Suppose now that $j \neq \min
  \OccThree{4\ell}{j}{\Text}$.  Then, there exists $j' \in [1 \dd j)$ such
  that $j' \in \OccThree{4\ell}{j}{\Text}$, or equivalently (by
  \cref{lm:occ-equivalence}) $\Textinf[j \dd j + 4\ell) = \Textinf[j'
  \dd j' + 4\ell)$.  By \cref{lm:full-run-shifted} (with $\delta =
  0$), we thus obtain $j' \in \RMinusSix{f}{s}{k_2}{H}{\tau}{\Text}$.  Observe now that by
  $j \in \OccThree{2\ell}{\Pat}{\Text} \cap \PosHighMinus{f}{\Pat}{\Text}$ and
  \cref{lm:sa-periodic-occ-poshigh-single}, the pattern $\Pat' :=
  \Pat[\RunEndCutPat{f}{\tau}{\Pat}{2\ell} \dd \min(m, 2\ell)]$ is a prefix of
  $\Textinf[\RunEndFullPos{f}{\tau}{\Text}{j} \dd \RunEndFullPos{f}{\tau}{\Text}{j} + 7\tau)$. Note that
  $\RunEndFullPos{f}{\tau}{\Text}{j} + |\Pat'| - j = s + k_2 p + |\Pat'| = \RunEndCutPat{f}{\tau}{\Pat}{2\ell}
  - 1 + (\min(m, 2\ell) - \RunEndCutPat{f}{\tau}{\Pat}{2\ell} + 1) = \min(m, 2\ell) \leq
  2\ell$.  This occurrence of $\Pat'$ is therefore contained in
  $\Textinf[j \dd j + 2\ell)$.  Recall now that $\RunEndFullPos{f}{\tau}{\Text}{j'} - j' = s +
  k_2 p = \RunEndFullPos{f}{\tau}{\Text}{j} - j$.  Thus we obtain from $\Textinf[j \dd j +
  4\ell) = \Textinf[j' \dd j' + 4\ell)$ that $\Textinf[\RunEndFullPos{f}{\tau}{\Text}{j'} \dd
  \RunEndFullPos{f}{\tau}{\Text}{j'} + |\Pat'|) = \Pat'$. By $|\Pat'| \leq 2\ell \leq
  7\tau$ (where $2\ell \leq 7\tau$ follows by $\tau = \lfloor
  \tfrac{\ell}{3} \rfloor$ and $\ell \geq 16$), we thus obtain that
  $\Pat'$ is a prefix of $\Textinf[\RunEndFullPos{f}{\tau}{\Text}{j'} \dd \RunEndFullPos{f}{\tau}{\Text}{j'} +
  7\tau)$.  Combining this with $j' \in \RMinusSix{f}{s}{k_2}{H}{\tau}{\Text}$ implies by
  \cref{lm:sa-periodic-occ-poshigh-single} that $j' \in
  \OccThree{2\ell}{\Pat}{\Text} \cap \PosHighMinus{f}{\Pat}{\Text}$, contradicting the
  definition of $j$.
\end{proof}

\begin{lemma}\label{lm:partially-periodic-pat-min-occ}
  Let $\ell \in [16 \dd \Textlen)$, $\tau = \lfloor \tfrac{\ell}{3} \rfloor$,
  and $f$ be a necklace-consistent function.
  Let $\Pat \in \Sigma^{m}$ be a $\tau$-periodic pattern such that
  $\TypePat{\tau}{\Pat} = -1$, $\RunEndPat{\tau}{\Pat} - 1 < \min(m, 2\ell)$, and $\Text[\Textlen]$
  does not occur in $\Pat[1 \dd m)$.  Assume $\OccThree{2\ell}{\Pat}{\Text}
  \neq \emptyset$.  Denote $c = \max\Sigma$, $x_l = \RunEndFullPat{f}{\tau}{\Pat} -
  1$, $y_l = \Pat[\RunEndFullPat{f}{\tau}{\Pat} \dd \min(m,2\ell)]$, $y_u = y_l
  c^{\infty}$, $H = \RootPat{f}{\tau}{\Pat}$, and $\Pts =
  \IntStrPoints{7\tau}{\PairsMinus{f}{H}{\tau}{\Text}}{\Text}$. Then:
  \begin{enumerate}
  \item\label{lm:partially-periodic-pat-min-occ-it-1} It holds $\min
    \OccThree{2\ell}{\Pat}{\Text} = \RangeMinFourSide{\Pts}{x_l}{\Textlen}{y_l}{y_u} - x_l$.
  \item\label{lm:partially-periodic-pat-min-occ-it-2} The position $j
    = \min \OccThree{2\ell}{\Pat}{\Text}$ satisfies $j = \min
    \OccThree{4\ell}{j}{\Text}$.
  \end{enumerate}
\end{lemma}
\begin{proof}

  1. By \cref{lm:occ-sub-poshigh}, it holds $x_l = \RunEndFullPat{f}{\tau}{\Pat} - 1
  = \RunEndCutPat{f}{\tau}{\Pat}{2\ell} - 1$ and $\OccThree{2\ell}{\Pat}{\Text} \subseteq
  \PosHighMinus{f}{\Pat}{\Text}$.  Thus, we have $y_l = \Pat[\RunEndFullPat{f}{\tau}{\Pat} \dd
  \min(m,2\ell)] = \Pat[\RunEndCutPat{f}{\tau}{\Pat}{2\ell} \dd
  \min(m,2\ell)]$. Consequently, it follows by
  \cref{lm:min-of-occ-poshigh}\eqref{lm:min-of-occ-poshigh-it-1} that
  $\min \OccThree{2\ell}{\Pat}{\Text} = \min \OccThree{2\ell}{\Pat}{\Text} \cap
  \PosHighAll{f}{\Pat}{\Text} = \RangeMinFourSide{\Pts}{x_l}{\Textlen}{y_l}{y_u} - x_l$.

  2. By \cref{lm:occ-sub-poshigh}, it holds $\OccThree{2\ell}{\Pat}{\Text}
  \subseteq \PosHighMinus{f}{\Pat}{\Text}$. Therefore, by
  \cref{lm:min-of-occ-poshigh}\eqref{lm:min-of-occ-poshigh-it-2}, the
  position $j = \min \OccThree{2\ell}{\Pat}{\Text} = \min
  \OccThree{2\ell}{\Pat}{\Text} \cap \PosHighMinus{f}{\Pat}{\Text}$ satisfies $j = \min \OccThree{4\ell}{j}{\Text}$.
\end{proof}

\begin{lemma}\label{lm:fully-periodic-pat-min-pos}
  Let $\ell \in [16 \dd \Textlen)$, $\tau = \lfloor \tfrac{\ell}{3} \rfloor$,
  and $f$ be a necklace-consistent function.
  Let $\Pat \in \Sigma^m$ be a $\tau$-periodic pattern such that
  $\TypePat{\tau}{\Pat} = -1$, $\RunEndPat{\tau}{\Pat} - 1 \geq 2\ell$, and $\Text[\Textlen]$ does not
  occur in $\Pat[1 \dd m)$. Assume $\PosHighMinus{f}{\Pat}{\Text} \neq
  \emptyset$. Denote $c = \max\Sigma$, $x_l = \RunEndCutPat{f}{\tau}{\Pat}{2\ell} - 1$,
  $y_l = \Pat[\RunEndCutPat{f}{\tau}{\Pat}{2\ell} \dd \min(m, 2\ell)]$, and $y_u = y_l
  c^{\infty}$, $H = \RootPat{f}{\tau}{\Pat}$, and $\Pts =
  \IntStrPoints{7\tau}{\PairsMinus{f}{H}{\tau}{\Text}}{\Text}$.  Then:
  \begin{enumerate}
  \item\label{lm:fully-periodic-pat-min-pos-it-1} It holds $\min
    \PosHighMinus{f}{\Pat}{\Text} = \RangeMinFourSide{\Pts}{x_l}{\Textlen}{y_l}{y_u} - x_l$.
  \item\label{lm:fully-periodic-pat-min-pos-it-2} The position $j =
    \min \PosHighMinus{f}{\Pat}{\Text}$ satisfies $j = \min \OccThree{4\ell}{j}{\Text}$.
  \end{enumerate}
\end{lemma}
\begin{proof}

  1. By \cref{lm:pat-fully-periodic-two-subsets-of-occ}, $\RunEndPat{\tau}{\Pat}
  - 1 \geq 2\ell$ implies that $\PosHighMinus{f}{\Pat}{\Text} \subseteq
  \OccThree{2\ell}{\Pat}{\Text}$. Consequently, $\PosHighMinus{f}{\Pat}{\Text} \neq
  \emptyset$ implies that $\PosHighMinus{f}{\Pat}{\Text} = \PosHighMinus{f}{\Pat}{\Text} \cap
  \OccThree{2\ell}{\Pat}{\Text} \neq \emptyset$. Thus, it follows by
  \cref{lm:min-of-occ-poshigh}\eqref{lm:min-of-occ-poshigh-it-1}, that
  $\min \PosHighMinus{f}{\Pat}{\Text} = \min \PosHighMinus{f}{\Pat}{\Text} \cap
  \OccThree{2\ell}{\Pat}{\Text} = \RangeMinFourSide{\Pts}{x_l}{\Textlen}{y_l}{y_u} - x_l$.

  2. Combining $\PosHighMinus{f}{\Pat}{\Text} \subseteq \OccThree{2\ell}{\Pat}{\Text}$,
  $\PosHighMinus{f}{\Pat}{\Text} = \PosHighMinus{f}{\Pat}{\Text} \cap \OccThree{2\ell}{\Pat}{\Text} \neq
  \emptyset$, it follows by
  \cref{lm:min-of-occ-poshigh}\eqref{lm:min-of-occ-poshigh-it-2}, that
  the position $j = \min \PosHighMinus{f}{\Pat}{\Text} = \min \PosHighMinus{f}{\Pat}{\Text} \cap
  \OccThree{2\ell}{\Pat}{\Text}$ satisfies $j = \min \OccThree{4\ell}{j}{\Text}$.
\end{proof}

\begin{lemma}\label{lm:partially-periodic-pos-min-occ}
  Let $\ell \in [16 \dd \Textlen)$, $\tau = \lfloor \tfrac{\ell}{3} \rfloor$,
  and $f$ be a necklace-consistent function.
  Let $j \in \RMinusTwo{\tau}{\Text}$ be such that $\RunEndPos{\tau}{\Text}{j} - j < 2\ell$. Denote $c =
  \max\Sigma$, $x_l = \RunEndFullPos{f}{\tau}{\Text}{j} - j$, $y_l = \Text[\RunEndFullPos{f}{\tau}{\Text}{j} \dd
  \min(\Textlen + 1, j + 2\ell))$, $y_u = y_l c^{\infty}$, $H = \RootPos{f}{\tau}{\Text}{j}$,
  and $\Pts = \IntStrPoints{7\tau}{\PairsMinus{f}{H}{\tau}{\Text}}{\Text}$. Then, we have
  $\OccThree{2\ell}{j}{\Text} \neq \emptyset$ and:
  \begin{enumerate}
  \item\label{lm:partially-periodic-pos-min-occ-it-1} It holds $\min
    \OccThree{2\ell}{j}{\Text} = \RangeMinFourSide{\Pts}{x_l}{\Textlen}{y_l}{y_u} - x_l$.
  \item\label{lm:partially-periodic-pos-min-occ-it-2} The position $j
    = \min \OccThree{2\ell}{j}{\Text}$ satisfies $j = \min
    \OccThree{4\ell}{j}{\Text}$.
  \end{enumerate}
\end{lemma}
\begin{proof}

  First, note that $j \in \OccThree{2\ell}{j}{\Text} =
  \OccThree{2\ell}{\Text[j \dd \Textlen]}{\Text} = \OccThree{2\ell}{\Pat}{\Text}$. Thus, $\OccThree{2\ell}{\Pat}{\Text}
  \neq \emptyset$.

  1. Denote $\Pat := \Text[j \dd \Textlen]$ and $m := |\Pat| = \Textlen - j + 1$.
  First, observe that $j \in \RMinusTwo{\tau}{\Text}$ implies that $\Pat$ is
  $\tau$-periodic and $\TypePat{\tau}{\Pat} = -1$.  By the uniqueness of
  $\Text[\Textlen]$ in $\Text$, $\RunEndPos{\tau}{\Text}{j} \leq \Textlen$.  Thus, $\RunEndPat{\tau}{\Pat} - 1 =
  \RunEndPat{\tau}{\Text[j \dd \Textlen]} - 1 = \RunEndPos{\tau}{\Text}{j} - j \leq \Textlen - j < m$.  On the other
  hand, $\RunEndPos{\tau}{\Text}{j} - j < 2\ell$ implies $\RunEndPat{\tau}{\Pat} - 1 = \RunEndPos{\tau}{\Text}{j} - j
  < 2\ell$. Consequently, $\RunEndPat{\tau}{\Pat} - 1 < \min(m, 2\ell)$.  Next,
  note that by the uniqueness of $\Text[\Textlen]$ in $\Text$, the symbol $\Text[\Textlen]$
  does not occur in $\Pat[1 \dd m)$. Observe also that, by definition,
  we have:
  \begin{itemize}
  \item $\RootPat{f}{\tau}{\Pat} = \RootPat{f}{\tau}{\Text[j \dd \Textlen]} = \RootPos{f}{\tau}{\Text}{j} = H$, and
  \item $\RunEndFullPat{f}{\tau}{\Pat} - 1 = \RunEndFullPat{f}{\tau}{\Text[j \dd \Textlen]} - 1 =
    \RunEndFullPos{f}{\tau}{\Text}{j} - j = x_l$,
  \item $\Pat[\RunEndFullPat{f}{\tau}{\Pat} \dd \min(m,2\ell)] = \Text[j +
    \RunEndFullPat{f}{\tau}{\Pat} - 1 \dd j + \min(m,2\ell) - 1] = \Text[\RunEndFullPos{f}{\tau}{\Text}{j}
    \dd (j-1) + \min(\Textlen - (j-1), 2\ell)] = \Text[\RunEndFullPos{f}{\tau}{\Text}{j} \dd \min(\Textlen,
    j - 1 + 2\ell)] = \Text[\RunEndFullPos{f}{\tau}{\Text}{j} \dd \min(\Textlen + 1, j + 2\ell)) =
    y_l$.
  \end{itemize}
  By the above, it follows by
  \cref{lm:partially-periodic-pat-min-occ}\eqref{lm:partially-periodic-pat-min-occ-it-1},
  that $\min \OccThree{2\ell}{\Pat}{\Text} = \RangeMinFourSide{\Pts}{x_l}{\Textlen}{y_l}{y_u} -
  x_l$.  It remains to observe that $\OccThree{2\ell}{\Pat}{\Text} =
  \OccThree{2\ell}{\Text[j \dd \Textlen]}{\Text} = \OccThree{2\ell}{j}{\Text}$.

  2. Using the above properties of $\Pat$, by
  \cref{lm:partially-periodic-pat-min-occ}\eqref{lm:partially-periodic-pat-min-occ-it-2}
  it follows that the position $j = \min \OccThree{2\ell}{\Pat}{\Text} = \min
  \OccThree{2\ell}{j}{\Text}$ satisfies $j = \min \OccThree{4\ell}{j}{\Text}$.
\end{proof}

\begin{lemma}\label{lm:fully-periodic-pos-min-occ}
  Let $\ell \in [16 \dd \Textlen)$, $\tau = \lfloor \tfrac{\ell}{3} \rfloor$,
  and $f$ be a necklace-consistent function.
  Let $j \in \RMinusTwo{\tau}{\Text}$ be such that $\RunEndPos{\tau}{\Text}{j} - j \geq 2\ell$ and $j \in
  \PosHighMinus{f}{j}{\Text}$. Denote $c = \max\Sigma$, $x_l = \RunEndFullPos{f}{\tau}{\Text}{j} - j$,
  $y_l = \Text[\RunEndFullPos{f}{\tau}{\Text}{j} \dd \min(\Textlen + 1, j + 2\ell))$, $y_u = y_l
  c^{\infty}$, $H = \RootPos{f}{\tau}{\Text}{j}$, and $\Pts =
  \IntStrPoints{7\tau}{\PairsMinus{f}{H}{\tau}{\Text}}{\Text}$. Then:
  \begin{enumerate}
  \item\label{lm:full-periodic-pos-min-occ-it-1} It holds $\min
    \PosHighMinus{f}{j}{\Text} = \RangeMinFourSide{\Pts}{x_l}{\Textlen}{y_l}{y_u} - x_l$.
  \item\label{lm:full-periodic-pos-min-occ-it-2} The position $j =
    \min \PosHighMinus{f}{j}{\Text}$ satisfies $j = \min \OccThree{4\ell}{j}{\Text}$.
  \end{enumerate}
\end{lemma}
\begin{proof}
  1. Let $\Pat := \Text[j \dd \Textlen]$ and $m := |\Pat| = \Textlen - j + 1$.  First,
  observe that $j \in \RMinusTwo{\tau}{\Text}$ implies that $\Pat$ is $\tau$-periodic
  and $\TypePat{\tau}{\Pat} = -1$. By the uniqueness of $\Text[\Textlen]$ in $\Text$,
  the symbol $\Text[\Textlen]$ does not occur in $\Pat[1 \dd m)$.  Observe also
  that, by definition, we have $\RootPat{f}{\tau}{\Pat} = \RootPat{f}{\tau}{\Text[j \dd \Textlen]} =
  \RootPos{f}{\tau}{\Text}{j} = H$.  Next, observe that $\RunEndPat{\tau}{\Pat} - 1 =
  \RunEndPat{\tau}{\Text[j \dd \Textlen]} - 1 = \RunEndPos{\tau}{\Text}{j} - j \geq 2\ell$.  Denoting $s = \HeadPat{f}{\tau}{\Pat}$
  and $p = |\RootPat{f}{\tau}{\Pat}|$, this implies $\ExpPat{f}{\tau}{\Pat} = \lfloor
  \tfrac{\RunEndPat{\tau}{\Pat} - 1 - s}{p} \rfloor \geq \lfloor \tfrac{2\ell -
  s}{p} \rfloor$.  Consequently, $\ExpCutPat{f}{\tau}{\Pat}{2\ell} =
  \min(\ExpPat{f}{\tau}{\Pat}, \lfloor \tfrac{2\ell - s}{p} \rfloor) =
  \ExpCutPat{f}{\tau}{\Pat}{2\ell}$.  The assumption $j \in \PosHighMinus{f}{j}{\Text} =
  \PosHighMinus{f}{\Text[j \dd \Textlen]}{\Text} = \PosHighMinus{f}{\Pat}{\Text}$ by
  \cref{lm:poslow-poshigh-single}, thus implies that $\HeadPos{f}{\tau}{\Text}{j} = s$
  and $\ExpPos{f}{\tau}{\Text}{j} = \ExpCutPat{f}{\tau}{\Pat}{2\ell}$.  Thus, $\RunEndCutPat{f}{\tau}{\Pat}{2\ell} - 1
  = s + \ExpCutPat{f}{\tau}{\Pat}{2\ell} p = s + \ExpPos{f}{\tau}{\Text}{j} p = \RunEndFullPos{f}{\tau}{\Text}{j} -
  j$. This in turn implies that $\Pat[\RunEndCutPat{f}{\tau}{\Pat}{2\ell} \dd
  \min(m,2\ell)] = \Text[j + \RunEndCutPat{f}{\tau}{\Pat}{2\ell} - 1 \dd j + \min(m,2\ell) -
  1] = \Text[\RunEndFullPos{f}{\tau}{\Text}{j} \dd (j-1) + \min(\Textlen - (j-1), 2\ell)] =
  \Text[\RunEndFullPos{f}{\tau}{\Text}{j} \dd \min(\Textlen, j - 1 + 2\ell)] = \Text[\RunEndFullPos{f}{\tau}{\Text}{j} \dd
  \min(\Textlen + 1, j + 2\ell)) = y_l$. Lastly, observe that $j \in
  \PosHighMinus{f}{j}{\Text}$ implies that $\PosHighMinus{f}{\Pat}{\Text} = \PosHighMinus{f}{\Text[j \dd \Textlen]}{\Text}
  = \PosHighMinus{f}{j}{\Text} \neq \emptyset$.  Putting all this together, by
  \cref{lm:fully-periodic-pat-min-pos}\eqref{lm:fully-periodic-pat-min-pos-it-1}
  we thus obtain $\min \PosHighMinus{f}{j}{\Text} = \min \PosHighMinus{f}{\Text[j \dd \Textlen]}{\Text} =
  \min \PosHighMinus{f}{\Pat}{\Text} = \RangeMinFourSide{\Pts}{x_l}{\Textlen}{y_l}{y_u} - x_l$.

  2. By the above properties of $\Pat$, combined with
  \cref{lm:fully-periodic-pat-min-pos}\eqref{lm:fully-periodic-pat-min-pos-it-2}, we have that
  $j = \min \PosHighMinus{f}{j}{\Text} = \min \PosHighMinus{f}{\Text[j \dd \Textlen]}{\Text} =
  \min \PosHighMinus{f}{\Pat}{\Text}$ satisfies $j = \min \OccThree{4\ell}{j}{\Text}$.
\end{proof}

\begin{lemma}\label{lm:fully-periodic-pos-two-cases}
  Let $\ell \in [16 \dd \Textlen)$, $\tau = \lfloor \tfrac{\ell}{3} \rfloor$,
  and $f$ be a necklace-consistent function.
  Let $j \in \RMinusTwo{\tau}{\Text}$ be such that $\RunEndPos{\tau}{\Text}{j} - j \geq 2\ell$. Denote $s
  = \HeadPos{f}{\tau}{\Text}{j}$, $H = \RootPos{f}{\tau}{\Text}{j}$, $p = |H|$, and $k_2 = \lfloor
  \tfrac{2\ell - s}{p} \rfloor$. If $\ExpPos{f}{\tau}{\Text}{j} = k_2$, then $j \in
  \PosHighMinus{f}{j}{\Text}$. Otherwise, it holds $\RMinusSix{f}{s}{k_2 + 1}{H}{\tau}{\Text} \neq
  \emptyset$.
\end{lemma}
\begin{proof}

  First, observe that $\RunEndPos{\tau}{\Text}{j} - j \geq 2\ell$ and $s = \HeadPos{f}{\tau}{\Text}{j}$
  imply that $\ExpPos{f}{\tau}{\Text}{j} = \lfloor \tfrac{\RunEndPos{\tau}{\Text}{j} - j - s}{p} \rfloor
  \geq \lfloor \tfrac{2\ell - s}{p} \rfloor = k_2$.  This implies that
  $\ExpCutPos{f}{\tau}{\Text}{j}{2\ell} = \min(\ExpPos{f}{\tau}{\Text}{j}, k_2) = k_2$.

  Assume $\ExpPos{f}{\tau}{\Text}{j} = k_2 = \ExpCutPos{f}{\tau}{\Text}{j}{2\ell}$. By
  \cref{lm:pos-sets-for-pos}, we then obtain $j \in \PosHighMinus{f}{j}{\Text}$.

  Let us now assume $\ExpPos{f}{\tau}{\Text}{j} \neq k_2$, i.e., denoting $k =
  \ExpPos{f}{\tau}{\Text}{j}$, we then have $k \geq k_2 + 1$. Let $j' := \RunEndFullPos{f}{\tau}{\Text}{j} -
  s - (k_2 + 1)p$.  We will prove that $j' \in \RMinusSix{f}{s}{k_2+1}{H}{\tau}{\Text}$.
  On the one hand, the assumption $k \geq k_2 + 1$ implies that
  $\RunEndFullPos{f}{\tau}{\Text}{j} - j = s + kp \geq s + (k_2 + 1)p$.  Thus, $j' =
  \RunEndFullPos{f}{\tau}{\Text}{j} - s - (k_2 + 1)p \geq j$. On the other hand, we obtain
  from $k_2 = \lfloor \tfrac{2\ell - s}{p} \rfloor$ that $\RunEndPos{\tau}{\Text}{j} -
  j' \geq \RunEndFullPos{f}{\tau}{\Text}{j} - j' = s + (k_2 + 1)p = s + (\lfloor
  \tfrac{2\ell - s}{p} \rfloor + 1)p \geq s + \lceil \tfrac{2\ell -
  s}{p} \rceil p \geq 2\ell \geq 3\tau - 1$.  We thus have $j' \in [j
  \dd \RunEndPos{\tau}{\Text}{j} - 3\tau + 1]$. By
  \cref{lm:beg-end}\eqref{lm:beg-end-it-1}, this implies that $[j \dd
  j'] \subseteq \RTwo{\tau}{\Text}$. Observe now that by \cref{lm:R-text-block}, it
  holds $j' \in \RMinusFour{f}{H}{\tau}{\Text}$ and $\RunEndFullPos{f}{\tau}{\Text}{j'} = \RunEndFullPos{f}{\tau}{\Text}{j}$.  Thus,
  $\HeadPos{f}{\tau}{\Text}{j'} = (\RunEndFullPos{f}{\tau}{\Text}{j'} - j') \bmod p = (\RunEndFullPos{f}{\tau}{\Text}{j} - j')
  \bmod p = (s + (k_2 + 1)p) \bmod p = s$ and $\ExpPos{f}{\tau}{\Text}{j'} = \lfloor
  \tfrac{\RunEndFullPos{f}{\tau}{\Text}{j'} - j'}{p} \rfloor = \lfloor \tfrac{s + (k_2 +
  1)p}{p} \rfloor = k_2 + 1$. We have thus proved that $j' \in
  \RMinusSix{f}{s}{k_2+1}{H}{\tau}{\Text}$. Hence, $\RMinusSix{f}{s}{k_2+1}{H}{\tau}{\Text} \neq \emptyset$.
\end{proof}

\paragraph{Query Algorithms}

\begin{proposition}\label{pr:partially-periodic-pos-occ-min}
  Let $k \in [4 \dd \lceil \log \Textlen \rceil)$, $\ell = 2^k$, $\tau
  = \lfloor \tfrac{\ell}{3} \rfloor$, and $f = f_{\tau,\Text}$
  (\cref{def:canonical-function}). Let $j \in \RMinusTwo{\tau}{\Text}$ be
  such that $\RunEndPos{\tau}{\Text}{j} - j < 2\ell$.  Given
  $\CompSaPeriodic{\Text}$, the value $k$, the position $j$, some
  $j' \in \OccThree{3\tau - 1}{j}{\Text}$ satisfying $j'
  = \min \OccThree{2\ell}{j'}{\Text}$, and the values
  $\HeadPos{f}{\tau}{\Text}{j}$, $|\RootPos{f}{\tau}{\Text}{j}|$, and
  $\RunEndPos{\tau}{\Text}{j}$ as input, we can compute a position
  $j'' \in \OccThree{2\ell}{j}{\Text}$ satisfying $j''
  = \min \OccThree{4\ell}{j''}{\Text}$ in $\bigO(\log^2 \Textlen)$ time.
\end{proposition}
\begin{proof}
  Let us denote $s = \HeadPos{f}{\tau}{\Text}{j}$, $H
  = \RootPos{f}{\tau}{\Text}{j}$, $p = |H|$, $k
  = \ExpPos{f}{\tau}{\Text}{j}$, and $\Pts
  = \IntStrPoints{7\tau}{\PairsMinus{f}{H}{\tau}{\Text}}{\Text}$
  (\cref{def:int-str}). Furthermore, let $c = \max\Sigma$, $x_l
  = \RunEndFullPos{f}{\tau}{\Text}{j} - j$, $y_l
  = \Text[\RunEndFullPos{f}{\tau}{\Text}{j} \dd \min(\Textlen + 1, j + 2\ell))$, and
  $y_u = y_l c^{\infty}$.  Finally, let $j''
  = \RangeMinFourSide{\Pts}{x_l}{\Textlen}{y_l}{y_u}$. On the one hand,
  by \cref{lm:partially-periodic-pos-min-occ}\eqref{lm:partially-periodic-pos-min-occ-it-1},
  it holds $j'' = \min \OccThree{2\ell}{j}{\Text}$. On the other hand,
  by \cref{lm:partially-periodic-pos-min-occ}\eqref{lm:partially-periodic-pos-min-occ-it-2},
  position $j''$ satisfies $j'' = \min \OccThree{4\ell}{j''}{\Text}$. Thus,
  our goal is to compute $\RangeMinFourSide{\Pts}{x_l}{\Textlen}{y_l}{y_u}$.
  Using \cref{pr:sa-periodic-pts-access} and the position $j'$ as
  input, in $\bigO(\log \Textlen)$ time we retrieve the pointer to the
  structure from \cref{pr:int-str} for $\PairsMinus{f}{H}{\tau}{\Text}$ (note
  that $j' \in \RFour{f}{H}{\tau}{\Text}$ holds
  by \cref{lm:periodic-pos-lce}\eqref{lm:periodic-pos-lce-it-2}),
  i.e., performing weighted range queries on $\Pts$. Note that the
  pointer is not null, since we assumed $j \in \RMinusTwo{\tau}{\Text}$.
  Thus, it holds $\RMinusFour{f}{H}{\tau}{\Text} \neq \emptyset$, which in
  turn implies $\PairsMinus{f}{H}{\tau}{\Text} \neq \emptyset$.  In $\bigO(1)$
  time we now calculate $k := \ExpPos{f}{\tau}{\Text}{j}
  = \lfloor \tfrac{\RunEndPos{\tau}{\Text}{j} - s}{p} \rfloor$ and $x_l
  := \RunEndFullPos{f}{\tau}{\Text}{j} - j = \HeadPos{f}{\tau}{\Text}{j}
  + \ExpPos{f}{\tau}{\Text}{j} \cdot |\RootPos{f}{\tau}{\Text}{j}| = s +
  kp$.  Note that using $\CompSaCore{\Text}$ (which is part of
  $\CompSaPeriodic{\Text}$), we can lexicographically compare any two
  substrings of $\Textinf$ or $\revstr{\Textinf}$ (specified with their
  starting positions and lengths) in $t_{\rm cmp} = \bigO(\log \Textlen)$
  time.  In $\bigO(\log^{1 + \epsilon} \Textlen + t_{\rm cmp} \log \Textlen)
  = \bigO(\log^2 \Textlen)$ time, we thus compute and return
  $\RangeMinFourSide{\Pts}{x_l}{\Textlen}{y_l}{y_u} - x_l$ using \cref{pr:int-str} with
  arguments $i = j + x_l$ and $q_r = \min(\Textlen + 1, j + 2\ell) - i$.  In
  total, we spend $\bigO(\log^2 \Textlen)$ time.
\end{proof}

\begin{proposition}\label{pr:fully-periodic-pos-occ-min}
  Let $k \in [4 \dd \lceil \log \Textlen \rceil)$, $\ell = 2^k$, $\tau
  = \lfloor \tfrac{\ell}{3} \rfloor$, and $f = f_{\tau,\Text}$
  (\cref{def:canonical-function}). Let $j \in \RMinusTwo{\tau}{\Text}$ be
  such that $\RunEndPos{\tau}{\Text}{j} - j \geq 2\ell$.  Given
  $\CompSaPeriodic{\Text}$, the value $k$, the position $j$, some
  $j' \in \OccThree{3\tau - 1}{j}{\Text}$ satisfying $j'
  = \min \OccThree{2\ell}{j'}{\Text}$, and the values
  $\HeadPos{f}{\tau}{\Text}{j}$, $|\RootPos{f}{\tau}{\Text}{j}|$, and
  $\RunEndPos{\tau}{\Text}{j}$ as input, we can compute a position
  $j'' \in \OccThree{2\ell}{j}{\Text}$ satisfying $j''
  = \min \OccThree{4\ell}{j''}{\Text}$ in $\bigO(\log^2 \Textlen)$ time.
\end{proposition}
\begin{proof}

  Denote $s = \HeadPos{f}{\tau}{\Text}{j}$, $H
  = \RootPos{f}{\tau}{\Text}{j}$, $p = |H|$, $\Pts
  = \IntStrPoints{7\tau}{\PairsMinus{f}{H}{\tau}{\Text}}{\Text}$, $k
  = \ExpPos{f}{\tau}{\Text}{j}$, and $k_2 = \lfloor \tfrac{2\ell -
  s}{p} \rfloor$. We have two cases:
  \begin{enumerate}
  \item Let us first assume $k =
    k_2$. By \cref{lm:fully-periodic-pos-two-cases}, we then have
    $j \in \PosHighMinus{f}{\Pat}{\Text}$. Consequently, letting $c
    = \max\Sigma$, $x_l = \RunEndFullPos{f}{\tau}{\Text}{j} - j$, $y_l
    = \Text[\RunEndFullPos{f}{\tau}{\Text}{j} \dd \min(\Textlen + 1, j + 2\ell))$,
    $y_u = y_l c^{\infty}$, and $j'' = \RangeMinFourSide{\Pts}{x_l}{\Textlen}{y_l}{y_u} -
    x_l$, by \cref{lm:fully-periodic-pos-min-occ}, the position $j''$
    satisfies $j'' = \min \PosHighMinus{f}{j}{\Text}$ and it holds $j''
    = \min \OccThree{4\ell}{j''}{\Text}$. Finally, note that
    by \cref{lm:pos-fully-periodic-two-subsets-of-occ}, we have
    $\PosHighMinus{f}{j}{\Text} \subseteq \OccThree{2\ell}{j}{\Text}$.  Thus,
    $j'' \in \OccThree{2\ell}{j}{\Text}$.
  \item Let us now assume $k \neq
    k_2$. By \cref{lm:fully-periodic-pos-two-cases}, we then have
    $\RMinusSix{f}{s}{k_2+1}{H}{\tau}{\Text} \neq \emptyset$. Then, letting
    $j'' = \RangeMinTwoSide{\Pts}{s + (k_2+1)p}{\Textlen} - s - (k_2+1)p$,
    by \cref{lm:RskH-min}, the position $j''$ satisfies $j''
    = \min \RMinusSix{f}{s}{k_2+1}{H}{\tau}{\Text}$, and moreover, it holds
    $j'' = \min \OccThree{4\ell}{j''}{\Text}$. Note that \cref{lm:RskH-min}
    requires that $\lceil \tfrac{3\tau - 1 - s}{p} \rceil - 1 < k_2 +
    1 \leq \lfloor \tfrac{7\tau - s}{p} \rfloor$.  The first
    inequality holds since $3\tau - 1 \leq 2\ell$ implies that $k_2 +
    1 = \lfloor \tfrac{2\ell - s}{p} \rfloor +
    1 \geq \lceil \tfrac{2\ell - s}{p} \rceil \geq \lceil \tfrac{3\tau
    - 1 - s}{p} \rceil > \lceil \tfrac{3\tau - 1 - s}{p} \rceil -
    1$. For the second inequality, we note that $k_2 + 1
    = \lfloor \tfrac{2\ell - s}{p} \rfloor + 1 = \lfloor \tfrac{2\ell
    - s + p}{p} \rfloor \leq \lfloor \tfrac{2\ell - s
    + \lfloor \tau/3 \rfloor}{p} \rfloor \leq \lfloor \tfrac{7\tau -
    s}{p} \rfloor$, where $2\ell + \lfloor \tau/3 \rfloor \leq 7\tau$
    follows by $\tau = \lfloor \tfrac{\ell}{3} \rfloor$ and $\ell \geq
    16$. It remains to observe that
    by \cref{lm:pos-fully-periodic-two-subsets-of-occ}, we have
    $\RMinusSix{f}{s}{k_2+1}{H}{\tau}{\Text} \subseteq \OccThree{2\ell}{j}{\Text}$.
    Thus, $j'' \in \OccThree{2\ell}{j}{\Text}$.
  \end{enumerate}

  The query algorithm thus proceeds as follows.  By the uniqueness of
  $\Text[\Textlen]$ in $\Text$, $j \in \RTwo{\tau}{\Text}$ and $j' \in
  \OccThree{3\tau - 1}{j}{\Text}$ imply that $\LCE_{\Text}(j, j') \geq 3\tau -
  1$. By \cref{lm:periodic-pos-lce}\eqref{lm:periodic-pos-lce-it-2},
  we thus obtain $\RootPos{f}{\tau}{\Text}{j'}
  = \RootPos{f}{\tau}{\Text}{j} =
  H$. Using \cref{pr:sa-periodic-pts-access} and the position $j'$ as
  input, in $\bigO(\log \Textlen)$ time we retrieve the pointer to the
  structure from \cref{pr:int-str} for $\PairsMinus{f}{H}{\tau}{\Text}$, i.e.,
  performing weighted range queries on $\Pts$. Note that the pointer
  is not null, since we assumed $j \in \RMinusTwo{\tau}{\Text}$.  Thus,
  $\RMinusFour{f}{H}{\tau}{\Text} \neq \emptyset$, which implies
  $\PairsMinus{f}{H}{\tau}{\Text} \neq \emptyset$.  In $\bigO(1)$ time we now
  calculate $k := \ExpPos{f}{\tau}{\Text}{j} = \lfloor \tfrac{\RunEndPos{\tau}{\Text}{j} -
  s}{p} \rfloor$, $k_2 = \lfloor \tfrac{2\ell - s}{p} \rfloor$, and
  $x_l := \RunEndFullPos{f}{\tau}{\Text}{j} - j = s + kp$.  Note that using
  $\CompSaCore{\Text}$ (which is part of $\CompSaPeriodic{\Text}$), we can
  lexicographically compare any two substrings of $\Textinf$ or
  $\revstr{\Textinf}$ (specified with their starting positions and
  lengths) in $t_{\rm cmp} = \bigO(\log \Textlen)$ time.  We consider two
  cases:
  \begin{enumerate}
  \item If $k = k_2$, then in $\bigO(\log^{1 + \epsilon} \Textlen + t_{\rm
    cmp} \log \Textlen) = \bigO(\log^2 \Textlen)$ time we compute and return
    $\RangeMinFourSide{\Pts}{x_l}{\Textlen}{y_l}{y_u} - x_l$ using \cref{pr:int-str} with
    arguments $i = j + x_l$ and $q_r = \min(\Textlen + 1, j + 2\ell) - i$.
  \item Otherwise, in $\bigO(\log^{1 + \epsilon} \Textlen)$ time we compute
    and return $\RangeMinTwoSide{\Pts}{s + (k_2 + 1)p}{\Textlen} - s - (k_2 + 1)p$
    using \cref{pr:int-str}.
  \end{enumerate}
  In total, we spend $\bigO(\log^2 \Textlen)$ time.
\end{proof}

\begin{proposition}\label{pr:periodic-pos-occ-min}
  Let $k \in [4 \dd \lceil \log \Textlen \rceil)$, $\ell = 2^k$, and $\tau
  = \lfloor \tfrac{\ell}{3} \rfloor$.  Let $j \in \RTwo{\tau}{\Text}$. Given
  $\CompSaPeriodic{\Text}$, the value $k$, the position $j$, and any
  $j' \in \OccThree{3\tau - 1}{j}{\Text}$ satisfying $j'
  = \min \OccThree{2\ell}{j'}{\Text}$ input, we can compute a position
  $j'' \in \OccThree{2\ell}{j}{\Text}$ satisfying $j''
  = \min \OccThree{4\ell}{j''}{\Text}$ in $\bigO(\log^2 \Textlen)$ time.
\end{proposition}
\begin{proof}
  Let $f = f_{\tau,\Text}$ (\cref{def:canonical-function}).
  By the uniqueness of $\Text[\Textlen]$ in $\Text$, $j \in \RTwo{\tau}{\Text}$ and
  $j' \in \OccThree{3\tau - 1}{j}{\Text}$ imply that $\LCE_{\Text}(j, j') \geq
  3\tau -
  1$. By \cref{lm:periodic-pos-lce}\eqref{lm:periodic-pos-lce-it-2},
  we thus obtain $\RootPos{f}{\tau}{\Text}{j'}
  = \RootPos{f}{\tau}{\Text}{j}$ and $\HeadPos{f}{\tau}{\Text}{j'}
  = \HeadPos{f}{\tau}{\Text}{j}$.  First, in $\bigO(\log \Textlen)$ time we
  compute $s = \HeadPos{f}{\tau}{\Text}{j'} = \HeadPos{f}{\tau}{\Text}{j}$
  and $p = |\RootPos{f}{\tau}{\Text}{j'}| = |\RootPos{f}{\tau}{\Text}{j}|$
  using \cref{pr:sa-periodic-nav-head-root}. Next, in $\bigO(\log \Textlen)$
  time we compute $t := \RunEndPos{\tau}{\Text}{j}$
  using \cref{pr:sa-periodic-nav-run-end}.  Recall that using
  $\CompSaCore{\Text}$ (which is part of $\CompSaPeriodic{\Text}$), we can
  access any symbol of $\Text$ in $\bigO(\log \Textlen)$ time.  In $\bigO(\log
  \Textlen)$ time we thus determine $\TypePos{\tau}{\Text}{j}$ by comparing
  $\Text[\RunEndPos{\tau}{\Text}{j}]$ and $\Text[\RunEndPos{\tau}{\Text}{j} - p]$.
  Let us assume that $\TypePos{\tau}{\Text}{j} = -1$ (the case of
  $\TypePos{\tau}{\Text}{j} = +1$ is processed symmetrically). We then
  consider two cases.  If $\RunEndPos{\tau}{\Text}{j} - j < 2\ell$, we
  apply \cref{pr:partially-periodic-pos-occ-min}.  Otherwise (i.e.,
  $\RunEndPos{\tau}{\Text}{j} - j \geq 2\ell$), we
  apply \cref{pr:fully-periodic-pos-occ-min}.  In either case, we
  obtain a position $j'' \in \OccThree{2\ell}{j}{\Text}$ satisfying $j''
  = \min \OccThree{4\ell}{j''}{\Text}$. In total, we spend $\bigO(\log^2 \Textlen)$
  time.
\end{proof}

\subsubsection{Implementation of ISA Queries}\label{sec:sa-periodic-isa}

\begin{proposition}\label{pr:pat-fully-periodic}
  Let $k \in [4 \dd \lceil \log \Textlen \rceil)$. Denote $\ell = 2^k$ and
  $\tau = \lfloor \tfrac{\ell}{3} \rfloor$. Let $j \in \RTwo{\tau}{\Text}$
  be such that $j = \min \OccThree{2\ell}{j}{\Text}$.  Let $s
  = \HeadPos{f}{\tau}{\Text}{j}$, $H = \RootPos{f}{\tau}{\Text}{j}$, $H'$
  be a length-$s$ suffix of $H$, and $\Pat$ be a length-$2\ell$ prefix
  of $H' H^{\infty}$.  Given $\CompSaPeriodic{\Text}$ and the values $k$,
  $j$, $s$, $|H|$, $\RangeBegThree{\ell}{\Pat}{\Text}$, and $\RangeEndThree{\ell}{\Pat}{\Text}$
  as input, in $\bigO(\log^{3+\epsilon} \Textlen)$ time we can compute $b
  = \RangeBegTwo{\Pat}{\Text}$ and $e = \RangeBegTwo{\Pat}{\Text}$. Moreover, if $b \neq e$
  (i.e., $\OccTwo{\Pat}{\Text} \neq \emptyset$), we also return a position
  $j' \in \OccTwo{\Pat}{\Text}$ satisfying $j' = \min \OccThree{4\ell}{j'}{\Text}$.
\end{proposition}
\begin{proof}

  By \cref{lm:special-pat-properties}, it follows that $\Pat$ is
  $\tau$-periodic, $\HeadPat{f}{\tau}{\Pat} = s$,
  $\RootPat{f}{\tau}{\Pat} = H$, $\RunEndPat{\tau}{\Pat} - 1 =
  2\ell$, and $\TypePat{\tau}{\Pat} = -1$.  Denote $b'
  = \RangeBegThree{\ell}{\Pat}{\Text}$ and $e' = \RangeEndThree{\ell}{\Pat}{\Text}$. The query
  algorithm proceeds in four steps:

  \begin{enumerate}

  \item\label{pr:pat-fully-periodic-step-1}
    First, using \cref{pr:sa-periodic-poslow-poshigh-pat}, in
    $\bigO(\log^{2+\epsilon} \Textlen)$ time we compute $\delta^{-}_{1}
    := \DeltaLowMinus{f}{\Pat}{\Text}$ and $\delta^{-}_{3}
    := \DeltaHighMinus{f}{\Pat}{\Text}$. Using the symmetric variant
    of \cref{pr:sa-periodic-poslow-poshigh-pat}, in the same time we
    also compute $\delta^{+}_{1} := \DeltaLowPlus{f}{\Pat}{\Text}$ and
    $\delta^{+}_{3} := \DeltaHighPlus{f}{\Pat}{\Text}$.

  \item\label{pr:pat-fully-periodic-step-2}
    Next, using \cref{pr:sa-periodic-posmid-pat} and its symmetric
    version, in $\bigO(\log^{2+\epsilon} \Textlen)$ time we compute
    $\delta^{-}_{2} := \DeltaMidMinus{f}{\Pat}{\Text}$ and $\delta^{+}_{2}
    := \DeltaMidPlus{f}{\Pat}{\Text}$.  By $\TypePat{\tau}{\Pat} = -1$
    and \cref{lm:pat-decomposition}, we then have $b = b'
    + \delta^{-}_{1} + \delta^{-}_{2} - \delta^{-}_{3}$. By
    the symmetric version of \cref{lm:pat-decomposition},
    we also have $e = e' - \delta^{+}_{1} - \delta^{+}_{2} +
    \delta^{+}_{3}$.~\footnote{Note
    that after inverting the lexicographic order, the type of
    $\Pat$ is still $-1$, because $\RunEndPat{\tau}{\Pat} = |\Pat| + 1$.
    Thus, we can indeed use \cref{lm:pat-decomposition} without any changes
    (see also \cref{rm:lm-decomposition}).}
    Thus, we can determine $b$ and $e$ in $\bigO(1)$ time.  If $b = e$, this
    completes the algorithm.

  \item\label{pr:pat-fully-periodic-step-3}
    Let us thus assume $b \neq e$, i.e.,
    $\OccTwo{\Pat}{\Text} \neq \emptyset$. Using \cref{pr:pat-fully-periodic-occ-elem},
    in $\bigO(\log^{3 + \epsilon} \Textlen)$ time we compute a position
    $j_2 \in \OccTwo{\Pat}{\Text}$.

  \item\label{pr:pat-fully-periodic-step-4}
    Finally, using \cref{pr:periodic-pos-occ-min} and the position
    $j_2$ as input, in $\bigO(\log^2 \Textlen)$ time we compute a position
    $j' \in \OccThree{2\ell}{j_2}{\Text}$ satisfying $j'
    = \min \OccThree{4\ell}{j'}{\Text}$.  Note that since $\Pat$ does not
    contain symbol $\Text[\Textlen]$, it holds $j_2 + 2\ell \leq \Textlen$ and
    $\Text[j_2 \dd j_2 + 2\ell) = \Pat$.  Thus, $\OccTwo{\Pat}{\Text}
    = \OccThree{2\ell}{j_2}{\Text}$, and hence $j' \in \OccTwo{\Pat}{\Text}$.
  \end{enumerate}
  In total, we spend $\bigO(\log^{3 + \epsilon} \Textlen)$ time.
\end{proof}

\begin{proposition}\label{pr:pos-partially-periodic-isa}
  Let $k \in [4 \dd \lceil \log \Textlen \rceil)$, $\ell = 2^k$, $\tau
  = \lfloor \tfrac{\ell}{3} \rfloor$, and $f = f_{\tau,\Text}$
  (\cref{def:canonical-function}).  Let $j \in \RMinusTwo{\tau}{\Text}$ be
  such that $\RunEndPos{\tau}{\Text}{j} - j < 2\ell$. Given
  $\CompSaPeriodic{\Text}$, the value $k$, the position $j$, the values
  $\RangeBegThree{\ell}{j}{\Text}$, $\RangeEndThree{\ell}{j}{\Text}$,
  $\HeadPos{f}{\tau}{\Text}{j}$, $|\RootPos{f}{\tau}{\Text}{j}|$,
  $\RunEndPos{\tau}{\Text}{j}$, and some position
  $j' \in \OccThree{\ell}{j}{\Text}$ satisfying $j'
  = \min \OccThree{2\ell}{j'}{\Text}$ as input, in $\bigO(\log^{2 + \epsilon}
  \Textlen)$ time we can compute $\RangeBegThree{2\ell}{j}{\Text}$, $\RangeEndThree{2\ell}{j}{\Text}$,
  and a position $j'' \in \OccThree{2\ell}{j}{\Text}$ satisfying $j''
  = \min \OccThree{4\ell}{j''}{\Text}$.
\end{proposition}
\begin{proof}

  Denote $s = \HeadPos{f}{\tau}{\Text}{j}$, $H
  = \RootPos{f}{\tau}{\Text}{j}$, $p = |H|$, $t
  = \RunEndPos{\tau}{\Text}{j}$, $b = \RangeBegThree{\ell}{j}{\Text}$, and $e
  = \RangeEndThree{\ell}{j}{\Text}$. We begin by computing $k
  := \ExpPos{f}{\tau}{\Text}{j} = \lfloor \tfrac{t - j - s}{p} \rfloor$,
  $k_1 := \ExpCutPos{f}{\tau}{\Text}{j}{\ell}
  = \min(k, \lfloor \tfrac{\ell - s}{p} \rfloor)$, $k_2
  := \ExpCutPos{f}{\tau}{\Text}{j}{\ell} = \min(k, \lfloor \tfrac{2\ell
  - s}{p} \rfloor)$, and $q := \RunEndFullPos{f}{\tau}{\Text}{j} = j + s +
  kp$ in $\bigO(1)$ time.  The rest of the query algorithm proceeds in
  four steps: \begin{enumerate}

  \item\label{pr:sa-periodic-isa-step-1}
    First, using \cref{pr:pos-poslow-poshigh-size} (note that
    $j \in \OccThree{\ell}{j}{\Text}$ and $j \in \OccThree{2\ell}{j}{\Text}$), in
    $\bigO(\log^{2 + \epsilon} \Textlen)$ time we compute the values
    $\delta_1 := \DeltaLowMinus{f}{j}{\Text}$ and $\delta_3
    = \DeltaHighMinus{f}{j}{\Text}$.

  \item\label{pr:sa-periodic-isa-step-2}
    Next, using \cref{pr:sa-periodic-posmid-pos}, in $\bigO(\log^{2
    + \epsilon} \Textlen)$ time we compute $\delta_2
    := \DeltaMidMinus{f}{j}{\Text}$.  By \cref{lm:pos-decomposition}, we
    can now in $\bigO(1)$ time determine the value $b'
    := \RangeBegThree{2\ell}{j}{\Text} = \RangeBegThree{\ell}{j}{\Text} + \DeltaLowMinus{f}{j}{\Text}
    + \DeltaMidMinus{f}{j}{\Text} - \DeltaHighMinus{f}{j}{\Text} = b
    + \delta_1 + \delta_2 - \delta_3$.

  \item\label{pr:sa-periodic-isa-step-3}
    Next, using \cref{pr:sa-periodic-occ-size}, in $\bigO(\log^{2
    + \epsilon} \Textlen)$ time we compute $t' := |\OccThree{2\ell}{j}{\Text}|$. In
    $\bigO(1)$ time we then set $e' := \RangeEndThree{2\ell}{j}{\Text}
    = \RangeBegThree{2\ell}{j}{\Text} + |\OccThree{2\ell}{j}{\Text}| = e' + t'$.

  \item\label{pr:sa-periodic-isa-step-4}
    Lastly, using \cref{pr:partially-periodic-pos-occ-min}, in
    $\bigO(\log^2 \Textlen)$ time we compute a position
    $j'' \in \OccThree{2\ell}{j}{\Text}$ satisfying $j''
    = \min \OccThree{4\ell}{j''}{\Text}$.
  \end{enumerate}
  In total, we spend $\bigO(\log^{2 + \epsilon} \Textlen)$ time.
\end{proof}

\begin{proposition}\label{pr:sa-periodic-isa}
  Let $k \in [4 \dd \lceil \log \Textlen \rceil)$, $\ell = 2^k$, and $\tau
  = \lfloor \tfrac{\ell}{3} \rfloor$.  Let $j \in \RTwo{\tau}{\Text}$.
  Given $\CompSaPeriodic{\Text}$, the value $k$, the position $j$, any
  $j' \in \OccThree{\ell}{j}{\Text}$ satisfying $j'
  = \min \OccThree{2\ell}{j'}{\Text}$, and the values $\RangeBegThree{\ell}{j}{\Text}$ and
  $\RangeEndThree{\ell}{j}{\Text}$ as input, we can compute
  $(\RangeBegThree{2\ell}{j}{\Text}, \allowbreak \RangeEndThree{2\ell}{j}{\Text})$ and some
  position $j'' \in \OccThree{2\ell}{j}{\Text}$ satisfying $j''
  = \min \OccThree{4\ell}{j''}{\Text}$ in $\bigO(\log^{3 + \epsilon} \Textlen)$
  time.
\end{proposition}
\begin{proof}

  Let $f = f_{\tau,\Text}$ (\cref{def:canonical-function}).
  By \cref{lm:periodic-pos-lce}\eqref{lm:periodic-pos-lce-it-2}, it
  follows that $j' \in \RTwo{\tau}{\Text}$, $\HeadPos{f}{\tau}{\Text}{j'}
  = \HeadPos{f}{\tau}{\Text}{j}$, and $\RootPos{f}{\tau}{\Text}{j'}
  = \RootPos{f}{\tau}{\Text}{j}$.  First, in $\bigO(\log \Textlen)$ time we
  compute $s := \HeadPos{f}{\tau}{\Text}{j'}
  = \HeadPos{f}{\tau}{\Text}{j}$ and $p := |\RootPos{f}{\tau}{\Text}{j'}|
  = |\RootPos{f}{\tau}{\Text}{j}|$
  using \cref{pr:sa-periodic-nav-head-root}. Next, in $\bigO(\log \Textlen)$
  time we compute $t := \RunEndPos{\tau}{\Text}{j}$
  using \cref{pr:sa-periodic-nav-run-end}. We then consider two
  cases: \begin{itemize}

  \item
    Let us first assume $\RunEndPos{\tau}{\Text}{j} - j \geq 2\ell$. Let
    $\Pat$ be a length-$2\ell$ prefix of $H' H^{\infty}$, Note that
    then $\RangeBegThree{\ell}{j}{\Text} = \RangeBegThree{\ell}{\Pat}{\Text}$, and
    $\RangeEndThree{\ell}{j}{\Text}
    = \RangeEndThree{\ell}{\Pat}{\Text}$. Using \cref{pr:pat-fully-periodic}, in
    $\bigO(\log^{3 + \epsilon} \Textlen)$ time we compute $\RangeBegTwo{\Pat}{\Text}
    = \RangeBegThree{2\ell}{j}{\Text}$, $\RangeEndTwo{\Pat}{\Text} = \RangeEndThree{2\ell}{j}{\Text}$, and
    some position $j'' \in \OccTwo{\Pat}{\Text} = \OccThree{2\ell}{j}{\Text}$
    satisfying $j'' = \min \OccThree{4\ell}{j''}{\Text}$.

  \item
    Let us now assume $\RunEndPos{\tau}{\Text}{j} - j < 2\ell$.  Recall
    that using $\CompSaCore{\Text}$ (which is a component of
    $\CompSaPeriodic{\Text}$), we can access any symbol of $\Text$ in
    $\bigO(\log \Textlen)$ time.  In $\bigO(\log \Textlen)$ time we thus compute
    $\TypePos{\tau}{\Text}{j}$ by comparing $\Text[\RunEndPos{\tau}{\Text}{j}]
    = \Text[t]$ with $\Text[\RunEndPos{\tau}{\Text}{j} -
    |\RootPos{f}{\tau}{\Text}{j}|] = \Text[t - p]$.  Assume
    $\TypePos{\tau}{\Text}{j} = -1$ (the case $\TypePos{\tau}{\Text}{j} =
    +1$ is processed symmetrically).
    Using \cref{pr:pos-partially-periodic-isa}, in $\bigO(\log^{2
    + \epsilon} \Textlen)$ time we then compute $\RangeBegThree{2\ell}{j}{\Text}$,
    $\RangeEndThree{2\ell}{j}{\Text}$, and some $j'' \in \OccThree{2\ell}{j}{\Text}$
    satisfying $j'' = \min \OccThree{4\ell}{j''}{\Text}$.
  \end{itemize}
  In total, we spend $\bigO(\log^{3 + \epsilon} \Textlen)$ time.
\end{proof}

\subsubsection{Implementation of SA Queries}\label{sec:sa-periodic-sa}

\begin{proposition}\label{pr:pos-partially-periodic-sa}
  Let $k \in [4 \dd \lceil \log \Textlen \rceil)$, $\ell = 2^k$, $\tau
  = \lfloor \tfrac{\ell}{3} \rfloor$, and $f = f_{\tau,\Text}$
  (\cref{def:canonical-function}).  Let $i \in [1 \dd \Textlen]$ be such that
  $\SA[i] \in \RMinusTwo{\tau}{\Text}$ and $\RunEndPos{\tau}{\Text}{\SA[i]}
  - \SA[i] < 2\ell$.  Given $\CompSaPeriodic{\Text}$, the value $k$, the
  position $i$, the values $\HeadPos{f}{\tau}{\Text}{\SA[i]}$,
  $|\RootPos{f}{\tau}{\Text}{\SA[i]}|$, $\RangeBegThree{\ell}{\SA[i]}{\Text}$,
  $\RangeEndThree{\ell}{\SA[i]}{\Text}$, and some position
  $j \in \OccThree{\ell}{\SA[i]}{\Text}$ satisfying $j
  = \min \OccThree{2\ell}{j}{\Text}$ as input, in $\bigO(\log^{3 + \epsilon}
  \Textlen)$ time we can compute $\RangeBegThree{2\ell}{\SA[i]}{\Text}$,
  $\RangeEndThree{2\ell}{\SA[i]}{\Text}$, and a position
  $j' \in \OccThree{2\ell}{\SA[i]}{\Text}$ satisfying $j'
  = \min \OccThree{4\ell}{j'}{\Text}$.
\end{proposition}
\begin{proof}

  Let $b = \RangeBegThree{\ell}{\SA[i]}{\Text}$, $e = \RangeEndThree{\ell}{\SA[i]}{\Text}$, $s
  = \HeadPos{f}{\tau}{\Text}{\SA[i]}$, $H
  = \RootPos{f}{\tau}{\Text}{\SA[i]}$, and $p = |H|$.  Observe that
  by \cref{lm:periodic-pos-lce}\eqref{lm:periodic-pos-lce-it-2}, it
  holds $j \in \RTwo{\tau}{\Text}$, $\HeadPos{f}{\tau}{\Text}{j}
  = \HeadPos{f}{\tau}{\Text}{\SA[i]} = s$, and
  $\RootPos{f}{\tau}{\Text}{j} = \RootPos{f}{\tau}{\Text}{\SA[i]} = H$.
  The query algorithm consists of eight steps:
  \begin{enumerate}

  \item\label{pr:pos-partially-periodic-sa-step-1}
    Observe that $j \in \RFive{f}{s}{H}{\tau}{\Text}$ and
    $j \in \OccThree{\ell}{\SA[i]}{\Text}$
    by \cref{lm:expcut}\eqref{lm:expcut-it-2} imply that
    $\ExpCutPos{f}{\tau}{\Text}{j}{\ell}
    = \ExpCutPos{f}{\tau}{\Text}{\SA[i]}{\ell}$. First,
    using \cref{pr:sa-periodic-nav-run-end} in $\bigO(\log \Textlen)$ time we
    compute $k_1 := \ExpCutPos{f}{\tau}{\Text}{j}{\ell}
    = \ExpCutPos{f}{\tau}{\Text}{\SA[i]}{\ell}$.

  \item\label{pr:pos-partially-periodic-sa-step-2}
    We apply \cref{pr:pos-poslow-poshigh-size} to position $\SA[i]$
    and using the fact that $s = \HeadPos{f}{\tau}{\Text}{\SA[i]}$, $p =
    |\RootPos{f}{\tau}{\Text}{\SA[i]}|$,
    $j \in \OccThree{\ell}{\SA[i]}{\Text}$, and $k_1
    = \ExpCutPos{f}{\tau}{\Text}{\SA[i]}{\ell}$, in $\bigO(\log^{2
    + \epsilon} \Textlen)$ time we compute $\delta_1
    := \DeltaLowMinus{f}{\SA[i]}{\Text}$.

  \item\label{pr:pos-partially-periodic-sa-step-3}
    Using \cref{pr:sa-periodic-exp}, in $\bigO(\log^{3 + \epsilon} \Textlen)$
    time, we compute $k := \ExpPos{f}{\tau}{\Text}{\SA[i]}$. Observe
    that $\RunEndPos{\tau}{\Text}{\SA[i]} - \SA[i] < 2\ell$ implies
    $\ExpPos{f}{\tau}{\Text}{\SA[i]}
    = \lfloor \tfrac{\RunEndPos{\tau}{\Text}{\SA[i]} - \SA[i] -
    s}{p} \rfloor \leq \lfloor \tfrac{2\ell - s}{p} \rfloor$, and
    hence it holds $k_2 := \ExpCutPos{f}{\tau}{\Text}{\SA[i]}{2\ell}
    = \min(\ExpPos{f}{\tau}{\Text}{\SA[i]}, \lfloor \tfrac{2\ell -
    s}{p} \rfloor) = \ExpPos{f}{\tau}{\Text}{\SA[i]} = k$.

  \item\label{pr:pos-partially-periodic-sa-step-4}
    We apply \cref{pr:sa-periodic-posmid-pos} to $\SA[i]$, utilizing
    that $s = \HeadPos{f}{\tau}{\Text}{\SA[i]}$, $p =
    |\RootPos{f}{\tau}{\Text}{\SA[i]}|$, $k_1
    = \ExpCutPos{f}{\tau}{\Text}{\SA[i]}{\ell}$, and $k_2
    = \ExpCutPos{f}{\tau}{\Text}{\SA[i]}{2\ell}$, to compute $\delta_2
    := \DeltaMidMinus{f}{\SA[i]}{\Text}$ in $\bigO(\log^{2 + \epsilon} \Textlen)$
    time.

  \item\label{pr:sa-partially-periodic-sa-step-5}
    Using \cref{pr:pos-partially-periodic-occ-elem}, in $\bigO(\log^{3
    + \epsilon} \Textlen)$ time we compute a position
    $j_2 \in \OccThree{2\ell}{\SA[i]}{\Text}$.

  \item\label{pr:sa-partially-periodic-sa-step-6}
    Using \cref{pr:pos-poslow-poshigh-size} for position $\SA[i]$, we
    compute $\delta_3 := \DeltaHighMinus{f}{\SA[i]}{\Text}$ in
    $\bigO(\log^{2 + \epsilon} \Textlen)$ time.  Note that this time we
    utilize the fact that we have a position
    $j_2 \in \OccThree{2\ell}{\SA[i]}{\Text}$ and we know that $k_2
    = \ExpCutPos{f}{\tau}{\Text}{\SA[i]}{2\ell}$.  After computing
    $\delta_3$, by \cref{lm:pos-posless,lm:pos-decomposition}, we
    calculate $b' := \RangeBegThree{2\ell}{\SA[i]}{\Text} = \RangeBegThree{\ell}{\SA[i]}{\Text}
    + \DeltaLowMinus{f}{\SA[i]}{\Text} + \DeltaMidMinus{f}{\SA[i]}{\Text}
    - \DeltaHighMinus{f}{\SA[i]}{\Text} = b + \delta_1 + \delta_2
    - \delta_3$.

  \item\label{pr:sa-partially-periodic-sa-step-7}
    Observe that by \cref{lm:occ-equivalence},
    $j_2 \in \OccThree{2\ell}{\SA[i]}{\Text}$ implies $\OccThree{2\ell}{j_2}{\Text}
    = \OccThree{2\ell}{\SA[i]}{\Text}$ (and hence, in particular,
    $|\OccThree{2\ell}{j_2}{\Text}| =
    |\OccThree{2\ell}{\SA[i]}{\Text}|$). By \cref{lm:full-run-shifted}, we
    then have $\TypePos{\tau}{\Text}{j_2} = \TypePos{\tau}{\Text}{\SA[i]} =
    -1$, $\RunEndPos{\tau}{\Text}{j_2} - j_2 = \RunEndPos{\tau}{\Text}{\SA[i]}
    - \SA[i] < 2\ell$, $\HeadPos{f}{\tau}{\Text}{j_2}
    = \HeadPos{f}{\tau}{\Text}{\SA[i]} \allowbreak=\allowbreak s$,
    $\RootPos{f}{\tau}{\Text}{j_2} = \RootPos{f}{\tau}{\Text}{\SA[i]} = H$
    (in particular, $|\RootPos{f}{\tau}{\Text}{j_2}| = p$), and
    $\ExpPos{f}{\tau}{\Text}{j_2} = \ExpPos{f}{\tau}{\Text}{\SA[i]} = k$.
    We then also have $j \in \OccThree{3\tau-1}{j_2}{\Text}$.  Thus, using
    $j_2$, $j$, $s$, $p$, and $k$ as input
    to \cref{pr:sa-periodic-occ-size}, in $\bigO(\log^{2 + \epsilon}
    \Textlen)$ time we compute $q := |\OccThree{2\ell}{j_2}{\Text}| =
    |\OccThree{2\ell}{\SA[i]}{\Text}|$.  In $\bigO(1)$ time we then calculate
    $e' := \RangeEndThree{2\ell}{\SA[i]}{\Text} = \RangeBegThree{2\ell}{\SA[i]}{\Text} +
    |\OccThree{2\ell}{\SA[i]}{\Text}| = b' + q$.

  \item\label{pr:sa-partially-periodic-sa-step-8}
    Using \cref{pr:periodic-pos-occ-min} for position $j_2$ and $j$,
    in $\bigO(\log^2 \Textlen)$ time we compute a position
    $j' \in \OccThree{2\ell}{j_2}{\Text} = \OccThree{2\ell}{\SA[i]}{\Text}$
    satisfying $j' = \min \OccThree{4\ell}{j'}{\Text}$.

  \end{enumerate}
  In total, we spend $\bigO(\log^{3 + \epsilon} \Textlen)$ time.
\end{proof}

\begin{proposition}\label{pr:sa-periodic-sa}
  Let $k \in [4 \dd \lceil \log \Textlen \rceil)$, $\ell = 2^k$, and $\tau
  = \lfloor \tfrac{\ell}{3} \rfloor$.  Let $i \in [1 \dd \Textlen]$ be such
  that $\SA[i] \in \RTwo{\tau}{\Text}$.  Given $\CompSaPeriodic{\Text}$, the
  value $k$, the position $i$, the values $\RangeBegThree{\ell}{\SA[i]}{\Text}$,
  $\RangeEndThree{\ell}{\SA[i]}{\Text}$, and some position
  $j \in \OccThree{\ell}{\SA[i]}{\Text}$ satisfying $j
  = \min \OccThree{2\ell}{j}{\Text}$ as input, in $\bigO(\log^{3 + \epsilon}
  \Textlen)$ time we can compute $\RangeBegThree{2\ell}{\SA[i]}{\Text}$,
  $\RangeEndThree{2\ell}{\SA[i]}{\Text}$, and a position
  $j' \in \OccThree{2\ell}{\SA[i]}{\Text}$ satisfying $j'
  = \min \OccThree{4\ell}{j'}{\Text}$.
\end{proposition}
\begin{proof}

  Let $f = f_{\tau,\Text}$ (\cref{def:canonical-function}).  Denote $b
  = \RangeBegThree{\ell}{\SA[i]}{\Text}$, and $e = \RangeEndThree{\ell}{\SA[i]}{\Text}$, and $H
  = \RootPos{f}{\tau}{\Text}{\SA[i]}$.
  By \cref{lm:periodic-pos-lce}\eqref{lm:periodic-pos-lce-it-2}, we
  have $j \in \RTwo{\tau}{\Text}$, $\HeadPos{f}{\tau}{\Text}{j}
  = \HeadPos{f}{\tau}{\Text}{\SA[i]}$ and $\RootPos{f}{\tau}{\Text}{j}
  = \RootPos{f}{\tau}{\Text}{\SA[i]}$.  First, in $\bigO(\log \Textlen)$ time
  we compute $s := \HeadPos{f}{\tau}{\Text}{j}
  = \HeadPos{f}{\tau}{\Text}{\SA[i]}$ and $p :=
  |\RootPos{f}{\tau}{\Text}{j}| = |\RootPos{f}{\tau}{\Text}{\SA[i]}|$
  using \cref{pr:sa-periodic-nav-head-root}.  Let $H' = H(p - s \dd
  p]$ and $\Pat$ be a length-$2\ell$ prefix of $H' H^{\infty}$.
  Using \cref{pr:pat-fully-periodic} and the position $j$ as input, in
  $\bigO(\log^{3 + \epsilon} \Textlen)$ time we compute $b_{\Pat}
  := \RangeBegTwo{\Pat}{\Text}$ and $e_{\Pat} := \RangeEndTwo{\Pat}{\Text}$.  If
  $b_{\Pat} \neq e_{\Pat}$, we also obtain a position
  $j_{\Pat} \in \OccTwo{\Pat}{\Text}$ satisfying $j_{\Pat}
  = \min \OccThree{4\ell}{j_{\Pat}}{\Text}$.  We then consider two
  cases: \begin{itemize}

  \item
    If $i \in (b_{\Pat} \dd e_{\Pat}]$, then
    $\SA[i] \in \OccTwo{\Pat}{\Text}$.  This implies
    $\RangeBegThree{2\ell}{\SA[i]}{\Text} = \RangeBegTwo{\Pat}{\Text}$,
    $\RangeEndThree{2\ell}{\SA[i]}{\Text} = \RangeEndTwo{\Pat}{\Text}$, and $\OccTwo{\Pat}{\Text}
    = \OccThree{2\ell}{\SA[i]}{\Text}$.  We thus return $b_{\Pat}$,
    $e_{\Pat}$, and $j_{\Pat}$ as the answer.

  \item
    Let us now assume $i \not\in (b_{\Pat} \dd e_{\Pat}]$. Moreover,
    let us assume $i \leq b_{\Pat}$ (the case $i > e_{\Pat}$ is
    handled symmetrically). This implies that
    $\RunEndPos{\tau}{\Text}{\SA[i]} - \SA[i] < 2\ell$ and
    $\SA[i] \in \RMinusTwo{\tau}{\Text}$.  Consequently, we then compute
    $\RangeBegThree{2\ell}{\SA[i]}{\Text}$, $\RangeEndThree{2\ell}{\SA[i]}{\Text}$, and a
    position $j' \in \OccThree{2\ell}{\SA[i]}{\Text}$ satisfying $j'
    = \min \OccThree{4\ell}{j'}{\Text}$
    using \cref{pr:pos-partially-periodic-sa} in $\bigO(\log^{3
    + \epsilon} \Textlen)$ time.

  \end{itemize}
  In total, we spend $\bigO(\log^{3 + \epsilon} \Textlen)$ time.
\end{proof}

\subsubsection{Construction Algorithm}\label{sec:sa-periodic-construction}

\begin{proposition}\label{pr:sa-periodic-construction}
  Given the LZ77 parsing of $\Text$, we can construct
  $\CompSaPeriodic{\Text}$ in $\bigO(\SubstringComplexity{\Text} \log^7 \Textlen)$ time.
\end{proposition}
\begin{proof}

  First, using~\cite[Theorem~6.11]{resolutionfull}
  (resp.\ \cite[Theorem~6.21]{resolutionfull}), in $\bigO(\LZSize{\Text} \log^4 \Textlen)$
  time we construct a structure that, given any substring $S$ of $\Text$
  (specified with its starting position and the length) in
  $\bigO(\log^3 \Textlen)$ time returns $\min \OccTwo{S}{\Text}$ (resp.\
  $|\OccTwo{S}{\Text}|$). Using~\cite[Theorem~6.7]{resolutionfull}, in
  $\bigO(\LZSize{\Text} \log^2 \Textlen)$ time we also construct a structure that, given
  any substring $S$ of $\Text$ (specified as above), in $\bigO(\log^3 \Textlen)$
  time checks if $\per(S) \leq \tfrac{|S|}{2}$, and if so, returns
  $\per(S)$.

  We then construct the components of the first part of
  $\CompSaPeriodic{\Text}$ (see \cref{sec:sa-periodic-ds}) as follows:
  \begin{enumerate}

  \item In $\bigO(\SubstringComplexity{\Text} \log^7 \Textlen)$ time we construct
    $\CompSaCore{\Text}$ using \cref{pr:sa-core-construction}.

  \item Next, we construct the arrays $\ArrRoot{i}$.  Let $i \in
    [4 \dd \lceil \log \Textlen \rceil)$.  Denote $k = 14\tau_i$,
    $(p_j, t_j)_{j \in [1 \dd n_{{\rm runs},i}]}
    = \IntervalRepr{\CompRepr{k}{\RTwo{\tau_i}{\Text}}{\Text}}$, and
    \begin{align*}
      \Text'_{\rm comp}
        &= \textstyle \prod_{j = 1, 2, \dots, n_{{\rm runs},i}}
           \Text[p_j \dd p_j + 2\tau_i - 1)\Text[\Textlen],\\
      \Text''_{\rm comp}
        &= \textstyle\prod_{j = n_{{\rm runs},i}, \dots, 2, 1}
           \Text[p_j \dd p_j + 2\tau_i - 1)\Text[\Textlen],\\
      Q &= \textstyle\bigcup_{j \in \RTwo{\tau_i}{\Text}}
           \{\Text[j + t \dd j + t + \per(\Text[j \dd j + 3\tau_i - 1))) :
             t \in [0 \dd \tau_i)\} \subseteq \Sigma^{\leq \tau_i/3}.
    \end{align*}
    Let also $g : \Z \rightarrow \Z$ be defined by $g(j) = 1 +
    2\tau_i \lfloor \tfrac{j-1}{2\tau_i} \rfloor$.  Consider any
    $j \in \RTwo{\tau_i}{\Text}$. Let $p = \per(\Text[j \dd j + 3\tau_i - 1))$,
    $X = \Text[j \dd j + p)$, and $s
    = \min \OccTwo{X^{\infty}[1 \dd \tau_i]}{\Text'_{\rm comp}}$. Note that
    $X \in Q$, and hence by \cref{def:canonical-function-comp}, it
    holds $f_{k,\tau_i,\T}(X) = \Text'_{\rm comp}[g(s) \dd g(s) + |X|)$.
    Let us now denote $s'
    = \max \OccTwo{X^{\infty}[1 \dd \tau_i]}{\Text''_{\rm comp}}$. Observe
    that due to length-$(2\tau_i-1)$ substrings of $\Text'_{\rm comp}$
    and $\Text''_{\rm comp}$ being separated by the sentinel symbol
    $\Text[\Textlen]$, it holds $g(s') + g(s) = 2\tau_i(n_{{\rm runs},i} - 1) +
    2$. This implies that $\Text'_{\rm comp}[g(s) \dd g(s) + |X|)
    = \Text''_{\rm comp}[g(s') \dd g(s') + |X|)$.~\footnote{Note that the
    rightmost occurrence of $X^{\infty}[1 \dd \tau_i]$ in $\Text''_{\rm
    comp}$ may not correspond to the leftmost occurrence of
    $X^{\infty}[1 \dd \tau_i]$ in $\Text'_{\rm comp}$.}  Since we denoted
    $f_i = f_{\tau_i, \T}$ (see \cref{sec:sa-periodic-ds}), and since
    by $k = 14\tau_i \geq 3\tau_i - 1$
    and \cref{lm:canonical-function-comp-2}, it holds
    $f_{k, \tau_i, \T} = f_{\tau_i, \T}$, we thus obtain
    \begin{align*}
      \RootPos{f_i}{\tau_i}{\Text}{j}
        &= f_i(\Text[j \dd j + p))\\
        &= f_i(X)\\
        &= f_{k, \tau_i, \T}(X)\\
        &= \Text'_{\rm comp}[g(s) \dd g(s) + p)\\
        &= \Text''_{\rm comp}[g(s') \dd g(s') + p).
    \end{align*}
    By the synchronization property of primitive
    strings~\cite[Lemma~1.11]{AlgorithmsOnStrings}, we therefore obtain
    $\HeadPos{f_i}{\tau_i}{\Text}{j} = (g(s') - s') \bmod p$.  The key
    difficulty is thus computing $s'$.  Observe that $\Text[j \dd j
    + \tau_i) = X^{\infty}[1 \dd \tau_i]$.  Thus, if we define
    $\Text_{\rm aux} := \Text \cdot \Text''_{\rm comp}$, then, letting $s''
    = \max \OccThree{\tau_i}{j}{\Text_{\rm aux}}$, we have $s'' = \Textlen +
    s'$. Consequently, we have $g(s') = g(s'' - \Textlen)$, and it holds
    \begin{align*}
      \HeadPos{f_i}{\tau_i}{\Text}{j}
        &= (g(s') - s') \bmod p\\
        &= (g(s'' - \Textlen) - (s'' - \Textlen)) \bmod p\\
        &= (1 + 2\tau_i\lfloor
           \tfrac{s'' - n - 1}{2\tau_i} \rfloor - (s'' - \Textlen)) \bmod p.
    \end{align*}
    To compute $\ArrRoot{i}[1 \dd n_{{\rm runs},i}]$ we thus
    proceed as follows:
    \begin{enumerate}
    \item First, we construct an LZ77-like parsing of $\Text_{\rm
      aux}$. For this we first take the input LZ77 parsing of $\Text$. We
      then append the phrases encoding the remaining suffix, i.e.,
      $\Text''_{\rm comp}$. For $j = n_{{\rm runs},i}, \dots, 2, 1$ we
      encode the substring $\Text[p_j \dd p_j + 2\tau_i - 1)\Text[\Textlen]$ using
      two phrases: the first is of length $2\tau_i - 1$ and has a
      source at position $p_j$ (obtained from $\ArrRuns{i}[j]$,
      which is a component of $\CompSaCore{\Text}$;
      see \cref{sec:sa-core-ds}).  The second phrase is the single
      symbol. The resulting LZ77-like parsing has $\LZSize{\Text} + 2n_{{\rm
      runs},i} = \bigO(\LZSize{\Text} + \SubstringComplexity{\Text}) = \bigO(\LZSize{\Text})$ phrases, where the
      subsequent inequalities follow by \cref{lm:IR-comp-R-size}
      and~\cite{delta}. Note also that $|\Text_{\rm aux}| = |\Text| +
      |\Text''_{\rm comp}| = \Textlen + 2\tau_i \cdot n_{{\rm runs},i}
      = \bigO(\Textlen)$, where $\tau_i \cdot n_{{\rm runs},i} = \bigO(\Textlen)$
      follows by \cref{lm:IR-comp-R-size}.
    \item Using~\cite[Theorem~6.12]{resolution}, in $\bigO(\LZSize{\Text} \log^4
      \Textlen)$ time we then construct a data structure that lets us find
      the rightmost occurrence of every substring of $\Text_{\rm aux}$
      (represented using its starting position and the length). More
      precisely, given any $x \in [1 \dd |\Text_{\rm aux}|]$ and $t > 0$,
      we can compute $\max \OccThree{t}{x}{\Text_{\rm aux}}$ in $\bigO(\log^3
      \Textlen)$ time.
    \item We now compute $\ArrRoot{i}[1 \dd n_{{\rm runs},i}]$. For
      $j = 1, \dots, n_{{\rm runs},i}$ we perform the following steps:
      \begin{enumerate}
      \item First, from $\CompSaCore{\Text}$, we obtain $(p_j,t_j)
        = \ArrRuns{i}[j]$.
      \item In $\bigO(\log^3 \Textlen)$ time we compute $p = \per(\Text[p_j \dd
        p_j + 3\tau_i - 1))$ (note that $\per(\Text[p_j \dd p_j + 3\tau_i
        - 1)) \leq \tfrac{3\tau_i - 1}{2}$ holds by
        $p_j \in \RTwo{\tau_i}{\Text}$).  We then have
        $|\RootPos{f_i}{\tau_i}{\Text}{p_j}| = p$.
      \item In $\bigO(\log^3 \Textlen)$ time we compute $s''
        = \max \OccThree{\tau_i}{p_j}{\Text_{\rm aux}}$.  Using the equation
        above, we then have $\HeadPos{f_i}{\tau_i}{\Text}{p_j} = (1 +
        2\tau_i\lfloor \tfrac{s'' - n - 1}{2\tau_i} \rfloor - (s'' -
        \Textlen)) \bmod p$.
      \end{enumerate}
      Over all $j \in [1 \dd n_{{\rm runs},i}]$, we spend
      $\bigO(n_{{\rm runs},i} \cdot \log^3 \Textlen) = \bigO(\SubstringComplexity{\Text} \log^3
      \Textlen)$ time.
    \end{enumerate}
    The whole construction of $\ArrRoot{i}$ takes $\bigO(\LZSize{\Text} \log^4
    \Textlen) = \bigO(\SubstringComplexity{\Text} \log^5 \Textlen)$ time. Over all $i \in
    [4 \dd \lceil \log \Textlen \rceil)$, this sums up to
    $\bigO(\SubstringComplexity{\Text} \log^6 \Textlen)$ time.

  \item Next, we construct the structures for range queries. Let
    $i \in [4 \dd \lceil \log \Textlen \rceil)$.  Let $g$ and $\Text_{\rm aux}$
    be defined as above. Recall that above we observed that for every
    $j_1, j_2 \in \RTwo{\tau_i}{\Text}$, letting $s''_1
    = \max \OccThree{\tau_i}{j_1}{\Text_{\rm aux}}$ and $s''_2
    = \max \OccThree{\tau_i}{j_2}{\Text_{\rm aux}}$, $g(s''_1 - \Textlen) = g(s''_2
    - \Textlen)$ holds if and only if $\RootPos{f_i}{\tau_i}{\Text}{j_1}
    = \RootPos{f_i}{\tau_i}{\Text}{j_2}$.  Since $g(x)$ is of the form
    $1 + 2\tau_i \cdot x$, and for every $j \in \RTwo{\tau_i}{\Text}$,
    letting $s'' = \max \OccThree{\tau_i}{j}{\Text_{\rm aux}}$ it holds
    $g(s'' - \Textlen) \in [1 \dd 2\tau_i \cdot n_{{\rm runs},i}]$, we thus
    obtain that the function $g' : \RTwo{\tau_i}{\Text} \rightarrow [1 \dd
    n_{{\rm runs},i}]$ defined by $g'(j) = \tfrac{g(s'' - \Textlen) -
    1}{2\tau_i} + 1$ (where $s''$ is defined as above) also uniquely
    identifies $\RootPos{f_i}{\tau_i}{\Text}{j}$, but returns a smaller
    integer.  We thus proceed as follows:
    \begin{enumerate}
    \item First, we repeat the computation from the previous step,
      except this time after we compute $s''
      = \max \OccThree{\tau_i}{p_j}{\Text_{\rm aux}}$ (where $j \in [1 \dd
      n_{{\rm runs},i}]$), we also store $g'(p_j) = \tfrac{g(s'' - \Textlen)
      - 1}{2\tau_i} + 1$ in an auxiliary array at index $j$. This
      takes $\bigO(\SubstringComplexity{\Text} \log^5 \Textlen)$ time.
    \item Next, we compute the collections $\PairsMinus{f_i}{H}{\tau_i}{\Text}$
      (see \cref{def:periodic-input}). To this end, we iterate over
      $\ArrRoot{i}[1 \dd n_{{\rm runs},i}]$. Let $(p, s)
      = \ArrRoot{i}[j] = (\HeadPos{f_i}{\tau_i}{\Text}{p_j},
      |\RootPos{f_i}{\tau_i}{\Text}{p_j}|)$. Using $\CompSaCore{\Text}$, in
      $\bigO(\log \Textlen)$ time we compute $\RunEndPos{\tau_i}{\Text}{p_j} = p
      + \LCE_{\Text}(p_j, p_j + p)$.  In $\bigO(1)$ time we then
      calculate $\ExpPos{f_i}{\tau_i}{\Text}{p_j}
      = \lfloor \tfrac{\RunEndPos{\tau_i}{\Text}{p_j} - p_j -
      s}{p} \rfloor$ and $\RunEndFullPos{f_i}{\tau_i}{\Text}{p_j} = p_j + s
      + \ExpPos{f_i}{\tau_i}{\Text}{p_j} \cdot p$.  Next, using again
      $\CompSaCore{\Text}$ in $\bigO(\log \Textlen)$ time we compute
      $\RunBeg{\tau_i}{\Text}{p_j} = p_j - \lcs(\Text[1 \dd p_j), \Text[1 \dd
      p_j + p))$. Lastly, we determine $\TypePos{\tau_i}{\Text}{p_j}$ by
      comparing $\Text[\RunEndPos{\tau_i}{\Text}{p_j}]$ with
      $\Text[\RunEndPos{\tau_i}{\Text}{p_j} - p]$ in $\bigO(\log \Textlen)$ time.
      Using the above values we can now calculate the values
      $\alpha(p_j) = \RunEndFullPos{f_i}{\tau_i}{\Text}{p_j}$ and
      $\beta(p_j) = \min(7\tau_i, \RunEndFullPos{f_i}{\tau_i}{\Text}{p_j}
      - \RunBeg{\tau_i}{\Text}{p_j})$.  If $\TypePos{\tau_i}{\Text}{p_j} =
      -1$, we store the resulting pair $(\alpha(p_j), \beta(p_j))$ in
      a linked list associated with position $g'(p_j)$ (which is
      retrieved from the auxiliary array computed above).  Observe
      that since $g'(p_j)$ uniquely identifies
      $\RootPos{f_i}{\tau_i}{\Text}{p_j}$, it follows by
      \cref{lm:periodic-input} that the resulting linked lists
      contain precisely the collection
      $\{\PairsMinus{f_i}{H}{\tau_i}{\Text}\}_{H \in \Sigma^{+}}$
      (see \cref{def:periodic-input}).  Over all $j \in [1 \dd n_{{\rm
      runs},i}]$, the above process takes $\bigO(n_{{\rm
      runs},i} \cdot \log \Textlen) = \bigO(\SubstringComplexity{\Text} \log \Textlen)$ time.
    \item For each list we apply \cref{pr:int-str}. Note that all
      queries needed in \cref{pr:int-str} are either supported using
      $\CompSaCore{\Text}$ or using structures constructed above (in
      particular, we can lexicographically compare substrings using
      $\LCE$ and random access queries).  The pointer to each
      structure is stored in a temporary array along with the pointer
      to the linked list containing each collection
      $\PairsMinus{f_i}{H}{\tau_i}{\Text}$.  Since the total length of lists is
      $n_{{\rm runs},i}$, in total we thus spend $\bigO(n_{\rm
      runs} \cdot \log^3 \Textlen) = \bigO(\SubstringComplexity{\Text} \log^3 \Textlen)$ time.
    \item We perform a pass over the sequence $(p_j,t_j)_{j \in [1 \dd
      n_{{\rm runs},i}]}$.  For each $j$, we retrieve the integer
      $g'(p_j)$ identifying $H = \RootPos{f_i}{\tau_i}{\Text}{p_j}$,
      then obtain the pointer to the structure from \cref{pr:int-str}
      for $\PairsMinus{f_i}{H}{\tau_i}{\Text}$, and store it in
      $\ArrPtrFirst{i}[j]$. Note that this time we deliberately ignore
      $\TypePos{\tau_i}{\Text}{p_j}$ (see \cref{rm:ptr}). This takes
      $\bigO(n_{{\rm runs},i}) = \bigO(\SubstringComplexity{\Text})$ time.
    \end{enumerate}
    In total the above computation takes $\bigO(\SubstringComplexity{\Text} \log^5 \Textlen)$
    time. Over all $i \in [4 \dd \lceil \log \Textlen \rceil)$, this sums up
    to $\bigO(\SubstringComplexity{\Text} \log^6 \Textlen)$ time.

  \item We now compute the structures for weighted modular queries.
    Let $i \in [4 \dd \lceil \log \Textlen \rceil)$. Above we already
    constructed the collection of sets
    $\{\PairsMinus{f_i}{H}{\tau_i}{\Text}\}_{H \in \Sigma^{+}}$, each
    represented using a list. For each set $\PairsMinus{f_i}{H}{\tau_i}{\Text}$,
    we apply \cref{pr:mod-queries} with $q = 7\tau_i$ and $h = |H|$.
    Note again that all queries required by \cref{pr:mod-queries} are
    either supported using $\CompSaCore{\Text}$ or using structures
    constructed above.  Since the total length of lists is $n_{{\rm
    runs},i}$, in total we thus spend $\bigO(n_{{\rm
    runs},i} \cdot \log^3 \Textlen) = \bigO(\SubstringComplexity{\Text} \log^3 \Textlen)$ time.  The
    pointer to each structure is stored in a temporary array along
    with the pointer to the linked list containing
    $\PairsMinus{f_i}{H}{\tau_i}{\Text}$.  After all structures are constructed,
    we again perform a pass over the sequence $(p_j,t_j)_{j \in [1 \dd
    n_{{\rm runs},i}]}$. For each $j$, we retrieve the integer
    $g'(p_j)$ identifying $H = \RootPos{f_i}{\tau_i}{\Text}{p_j}$, then
    obtain the pointer to the structure from \cref{pr:mod-queries} for
    $\PairsMinus{f_i}{H}{\tau_i}{\Text}$, and store it in
    $\ArrPtrSecond{i}[j]$. We again deliberately ignore
    $\TypePos{\tau_i}{\Text}{p_j}$ (see \cref{rm:ptr}).  In total the
    above computation takes $\bigO(\SubstringComplexity{\Text} \log^3 \Textlen)$ time. Over all
    $i \in [4 \dd \lceil \log \Textlen \rceil)$, this sums up to
    $\bigO(\SubstringComplexity{\Text} \log^4 \Textlen)$ time.
  \end{enumerate}
  In total, the construction of the first part of
  $\CompSaPeriodic{\Text}$ takes $\bigO(\SubstringComplexity{\Text} \log^7 \Textlen)$ time. We then
  construct the second part analogously. In total, the construction
  takes $\bigO(\SubstringComplexity{\Text} \log^7 \Textlen)$ time.
\end{proof}

\subsection{The Final Data Structure}\label{sec:sa-final}

\subsubsection{The Data Structure}\label{sec:sa-final-ds}

\paragraph{Definitions}

Let $n_{\rm short}$ be the number of distinct length-$16$ substrings of $\Textinf$,
and let  Let $\ArrStr[1 \dd n_{\rm short}]$ be an array containing all length-$16$ substring of $\Textinf$
sorted in lexicographic order. Let $\ArrRange[1 \dd
n_{\rm short}]$ be an array such that for every $i \in [1 \dd n_{\rm
short}]$, it holds $\ArrRange[i] = (\RangeBegThree{16}{\ArrStr[i]}{\Text}, \allowbreak
\RangeEndThree{16}{\ArrStr[i]}{\Text})$. Finally, let $\ArrMinOcc[1 \dd n_{\rm
short}]$ be such that for every $i \in [1 \dd n_{\rm short}]$, it
holds $\ArrMinOcc[i] = \min \OccTwo{\ArrStr[i]}{\Text}$.

\paragraph{Components}

The data structure, denoted $\CompSa{\Text}$, consists of five parts:

\begin{enumerate}
\item The array $\ArrStr[1 \dd n_{\rm short}]$ stored in plan form.
  Note that, by definition of $\SubstringComplexity{\Text}$, it holds $n_{\rm short} <
  \SubstrCount{16}{\Text}+16 \leq 16\SubstringComplexity{\Text}+16 \le 32\SubstringComplexity{\Text}$.  Thus, the array needs
  $\bigO(\SubstringComplexity{\Text})$ space.
\item The array $\ArrRange[1 \dd n_{\rm short}]$ stored in plain
  form.  It needs $\bigO(\SubstringComplexity{\Text})$ space.
\item The array $\ArrMinOcc[1 \dd n_{\rm short}]$ stored in plain
  form.  It needs $\bigO(\SubstringComplexity{\Text})$ space.
\item The structure $\CompSaNonperiodic{\Text}$ (\cref{sec:sa-nonperiodic-ds}).
  It needs $\bigO(\SubstringComplexity{\Text} \log \tfrac{\Textlen \log \sigma}{\SubstringComplexity{\Text} \log \Textlen})$ space.
\item The structure $\CompSaPeriodic{\Text}$ (\cref{sec:sa-periodic-ds}).
  It needs $\bigO(\SubstringComplexity{\Text} \log \tfrac{\Textlen \log \sigma}{\SubstringComplexity{\Text} \log \Textlen})$ space.
\end{enumerate}

In total, $\CompSa{\Text}$ needs $\bigO(\SubstringComplexity{\Text} \log \tfrac{\Textlen \log
\sigma}{\SubstringComplexity{\Text} \log \Textlen})$ space.

\subsubsection{Implementation of ISA Queries}\label{sec:sa-final-isa}

\begin{proposition}\label{pr:sa-final-isa-short}
  Let $j \in [1 \dd \Textlen]$. Given $\CompSa{\Text}$ and the position $j$, we
  can compute the pair $(\RangeBegThree{16}{j}{\Text}, \RangeEndThree{16}{j}{\Text})$ and a
  position $j' \in \OccThree{16}{j}{\Text}$ satisfying $j' = \min
  \OccThree{32}{j'}{\Text}$ in $\bigO(\log \Textlen)$ time.
\end{proposition}
\begin{proof}
  In $\bigO(\log \Textlen)$ time we compute $X = \Textinf[j \dd j + 16)$ using
  random access queries. Using binary search, in $\bigO(\log \Textlen)$ time
  we then compute $i \in [1 \dd n_{\rm short}]$ such that $\ArrStr[i]
  = X$. Then,
  \begin{align*}
    \ArrRange[i]
      &= (\RangeBegThree{16}{\ArrStr[i]}{\Text}, \RangeEndThree{16}{\ArrStr[i]}{\Text})\\
      &= (\RangeBegThree{16}{X}{\Text}, \RangeEndThree{16}{X}{\Text})\\
      &= (\RangeBegThree{16}{j}{\Text}, \RangeEndThree{16}{j}{\Text})
  \end{align*}
  and $\ArrMinOcc[i] = \min \OccThree{16}{\ArrStr[i]}{\Text} = \min
  \OccThree{16}{X}{\Text} = \min \OccThree{16}{j}{\Text}$.  Letting $j' =
  \ArrMinOcc[i]$, we thus have $j' \in \OccThree{16}{j}{\Text}$. This in turn
  implies that $\OccThree{16}{j}{\Text} = \OccThree{16}{j'}{\Text}$, and hence $j' =
  \min \OccThree{16}{j}{\Text} = \min \OccThree{16}{j'}{\Text}$. It remains to
  observe that $\OccThree{32}{j'}{\Text} \subseteq \OccThree{16}{j'}{\Text}$.  Thus,
  $j' = \min \OccThree{32}{j'}{\Text}$.
\end{proof}

\begin{proposition}\label{pr:sa-final-isa-step}
  Let $k \in [4 \dd \lceil \log \Textlen \rceil)$, $\ell = 2^k$, and
  $j \in [1 \dd \Textlen]$. Given
  $\CompSa{\Text}$, the value $k$, the position $j$, the pair
  $(\RangeBegThree{\ell}{j}{\Text}, \RangeEndThree{\ell}{j}{\Text})$, and a position
  $j' \in \OccThree{\ell}{j}{\Text}$ satisfying $j'
  = \min \OccThree{2\ell}{j'}{\Text}$ as input, we can in $\bigO(\log^{3
  + \epsilon} \Textlen)$ time compute the pair
  $(\RangeBegThree{2\ell}{j}{\Text},\allowbreak \RangeEndThree{2\ell}{j}{\Text})$ and a position
  $j'' \in \OccThree{2\ell}{j}{\Text}$ satisfying $j''
  = \min \OccThree{4\ell}{j''}{\Text}$.
\end{proposition}
\begin{proof}
  Denote $\tau = \lfloor \tfrac{\ell}{3} \rfloor$.
  First, using $\CompSaCore{\Text}$ and \cref{pr:sa-core-nav} in
  $\bigO(\log \Textlen)$ time we check if $j \in \RTwo{\tau}{\Text}$. We then
  consider two cases:
  \begin{itemize}
  \item If $j \not\in \RTwo{\tau}{\Text}$, then we obtain the output in
    $\bigO(\log^{2 + \epsilon} \Textlen)$ time by
    applying \cref{pr:sa-nonperiodic-isa}.
  \item Otherwise ($j \in \RTwo{\tau}{\Text}$), we obtain the output in
    $\bigO(\log^{3 + \epsilon} \Textlen)$ time by
    applying \cref{pr:sa-periodic-isa}.
  \end{itemize}
  In total, we spend $\bigO(\log^{3 + \epsilon} \Textlen)$ time.
\end{proof}

\begin{proposition}\label{pr:sa-final-isa}
  Let $j \in [1 \dd \Textlen]$. Given $\CompSa{\Text}$ and the position $j$ as
  input, we can in $\bigO(\log^{4 + \epsilon} \Textlen)$ time compute
  $\ISA[j]$.
\end{proposition}
\begin{proof}
  First, using \cref{pr:sa-final-isa-short}, in $\bigO(\log \Textlen)$ time
  we compute $(\RangeBegThree{16}{j}{\Text}, \allowbreak \RangeEndThree{16}{j}{\Text})$
  and a position $j' \in \OccThree{16}{j}{\Text}$ satisfying $j' = \min
  \OccThree{32}{j'}{\Text}$. Then, for $k = 4, \dots, \lceil \log \Textlen \rceil -
  1$, we use \cref{pr:sa-final-isa-step} to compute in $\bigO(\log^{3
  + \epsilon} \Textlen)$ time $(\RangeBegThree{2^{k+1}}{j}{\Text}, \allowbreak
  \RangeEndThree{2^{k+1}}{j}{\Text})$ and a position $j'' \in
  \OccThree{2^{k+1}}{j}{\Text}$ satisfying $j'' = \min \OccThree{2^{k+2}}{j''}{\Text}$, using position
  $j$ along with the pair $(\RangeBegThree{2^{k}}{j}{\Text}, \RangeEndThree{2^k}{j}{\Text})$ and
  $j' \in \OccThree{2^k}{j}{\Text}$ satisfying $j' = \min
  \OccThree{2^{k+1}}{j'}{\Text}$ as input.  After executing all steps, we obtain
  $(\RangeBegThree{\ell}{j}{\Text}, \RangeEndThree{\ell}{j}{\Text})$ and a position $j'' \in
  \OccThree{\ell}{j}{\Text}$, where $\ell = 2^{\lceil \log \Textlen \rceil} \geq \Textlen$. Since for
  every $\ell' \geq \Textlen$, it holds $\OccThree{\ell'}{j}{\Text} = \{j\}$, we
  thus return $\ISA[j] = \RangeEndThree{\ell}{j}{\Text}$ as the answer. In total,
  the query takes $\bigO(\log^{4 + \epsilon} \Textlen)$ time.
\end{proof}

\subsubsection{Implementation of SA Queries}\label{sec:sa-final-sa}

\begin{proposition}\label{pr:sa-final-sa-short}
  Let $i \in [1 \dd \Textlen]$. Given $\CompSa{\Text}$ and the position $i$, we
  can compute the pair $(\RangeBegThree{16}{\SA[i]}{\Text}, \RangeEndThree{16}{\SA[i]}{\Text})$
  and $j' \in \OccThree{16}{\SA[i]}{\Text}$ satisfying $j' = \min
  \OccThree{32}{j'}{\Text}$ in $\bigO(\log \Textlen)$ time.
\end{proposition}
\begin{proof}
  Using binary search, in $\bigO(\log \Textlen)$ time we compute $i' \in [1
  \dd n_{\rm short}]$ such that, letting $\ArrRange[t] = (b_t, e_t)$
  (where $t \in [1 \dd n_{\rm short}]$), it holds $i \in (b_{i'} \dd
  e_{i'}]$.  Denote $X = \ArrStr[i']$. We then have $\Textinf[\SA[i] \dd
  \SA[i] + 16) = X$ and hence
  \begin{align*}
    \ArrRange[i']
      &= (\RangeBegThree{16}{\ArrStr[i']}{\Text}, \RangeEndThree{16}{\ArrStr[i']}{\Text})\\
      &= (\RangeBegThree{16}{X}{\Text}, \RangeEndThree{16}{X}{\Text})\\
      &= (\RangeBegThree{16}{\SA[i]}{\Text}, \RangeEndThree{16}{\SA[i]}{\Text}).
  \end{align*}
  It also holds $\ArrMinOcc[i'] = \min \OccThree{16}{\ArrStr[i']}{\Text} =
  \min \OccThree{16}{X}{\Text} = \min \OccThree{16}{\SA[i]}{\Text}$.  Letting $j' =
  \ArrMinOcc[i']$, we thus have $j' \in \OccThree{16}{\SA[i]}{\Text}$. This
  in turn implies $\OccThree{16}{\SA[i]}{\Text} = \OccThree{16}{j'}{\Text}$, and
  hence $j' = \min \OccThree{16}{\SA[i]}{\Text} = \min \OccThree{16}{j'}{\Text}$. It
  remains to observe that $\OccThree{32}{j'}{\Text} \subseteq
  \OccThree{16}{j'}{\Text}$.  Thus, $j' = \min \OccThree{32}{j'}{\Text}$.
\end{proof}

\begin{proposition}\label{pr:sa-final-sa-step}
  Let $k \in [4 \dd \lceil \log \Textlen \rceil)$, $\ell = 2^k$, and $i \in
  [1 \dd \Textlen]$. Given $\CompSa{\Text}$, the value $k$, the position $i$,
  the pair $(\RangeBegThree{\ell}{\SA[i]}{\Text}, \RangeEndThree{\ell}{\SA[i]}{\Text})$, and
  $j' \in \OccThree{\ell}{\SA[i]}{\Text}$ satisfying $j'
  = \min \OccThree{2\ell}{j'}{\Text}$ as input, we can in $\bigO(\log^{3
  + \epsilon} \Textlen)$ time compute
  $(\RangeBegThree{2\ell}{\SA[i]}{\Text},\allowbreak \RangeEndThree{2\ell}{\SA[i]}{\Text})$ and
  a position $j' \in \OccThree{2\ell}{\SA[i]}{\Text}$ satisfying $j'
  = \min \OccThree{4\ell}{j'}{\Text}$.
\end{proposition}
\begin{proof}
  Denote $\tau = \lfloor \tfrac{\ell}{3} \rfloor$. First, using
  $\CompSaCore{\Text}$ and \cref{pr:sa-core-nav} in $\bigO(\log \Textlen)$ time
  we check if $j' \in \RTwo{\tau}{\Text}$. Observe that by $3\tau -
  1 \leq \ell$ and $j' \in \OccThree{\ell}{\SA[i]}{\Text}$,
  $\SA[i] \in \RTwo{\tau}{\Text}$ holds if and only if
  $j' \in \RTwo{\tau}{\Text}$.  Thus, this test reveals also whether
  $\SA[i] \in \RTwo{\tau}{\Text}$. We then consider two cases:
  \begin{itemize}
  \item If $\SA[i] \not\in \RTwo{\tau}{\Text}$, then we compute the output
    using \cref{pr:sa-nonperiodic-sa} in $\bigO(\log^{3 + \epsilon}
    \Textlen)$ time.
  \item Otherwise ($\SA[i] \in \RTwo{\tau}{\Text}$), we compute the output
    using \cref{pr:sa-periodic-sa} in $\bigO(\log^{3 + \epsilon} \Textlen)$
    time.
  \end{itemize}
  In total, we spend $\bigO(\log^{3 + \epsilon} \Textlen)$ time.
\end{proof}

\begin{proposition}\label{pr:sa-final-sa}
  Let $i \in [1 \dd \Textlen]$. Given $\CompSa{\Text}$ and the position $i$ as
  input, we can in $\bigO(\log^{4 + \epsilon} \Textlen)$ time compute
  $\SA[i]$.
\end{proposition}
\begin{proof}
  First, using \cref{pr:sa-final-sa-short}, in $\bigO(\log \Textlen)$ time we
  compute the initial pair
  $(\RangeBegThree{16}{\SA[i]}{\Text}, \allowbreak \RangeEndThree{16}{\SA[i]}{\Text})$ and a
  position $j' \in \OccThree{16}{\SA[i]}{\Text}$ satisfying $j'
  = \min \OccThree{32}{j'}{\Text}$. Then, for $k = 4, \dots, \lceil
  \log \Textlen \rceil - 1$, we use \cref{pr:sa-final-sa-step} to compute in
  $\bigO(\log^{3 + \epsilon} \Textlen)$ time the pair
  $(\RangeBegThree{2^{k+1}}{\SA[i]}{\Text}, \allowbreak \RangeEndThree{2^{k+1}}{\SA[i]}{\Text})$
  and a position $j'' \in \OccThree{2^{k+1}}{\SA[i]}{\Text}$ satisfying $j''
  = \min \OccThree{2^{k+2}}{j''}{\Text}$, using index $i$ along with
  $(\RangeBegThree{2^{k}}{\SA[i]}{\Text}, \RangeEndThree{2^k}{\SA[i]}{\Text})$ and
  $j' \in \OccThree{2^k}{\SA[i]}{\Text}$ satisfying $j'
  = \min \OccThree{2^{k+1}}{j'}{\Text}$ as input.  After executing all steps,
  we obtain $(\RangeBegThree{\ell}{\SA[i]}{\Text}, \RangeEndThree{\ell}{\SA[i]}{\Text})$ and a
  position $j'' \in \OccThree{\ell}{\SA[i]}{\Text}$, where $\ell =
  2^{\lceil \log \Textlen \rceil} \geq \Textlen$. Since for every $\ell' \geq \Textlen$, it
  holds $\OccThree{\ell'}{\SA[i]}{\Text} = \{\SA[i]\}$, we thus return
  $\SA[i] = j''$ as the answer. In total, the query takes
  $\bigO(\log^{4 + \epsilon} \Textlen)$ time.
\end{proof}

\subsubsection{Construction Algorithm}\label{sec:sa-final-construction}

\begin{proposition}\label{pr:sa-final-construction}
  Given the LZ77 parsing of $\Text$, we can construct $\CompSa{\Text}$ in
  $\bigO(\SubstringComplexity{\Text} \log^7 \Textlen)$ time.
\end{proposition}
\begin{proof}

  First, using~\cite[Theorem~6.11]{resolutionfull}
  (resp.\ \cite[Theorem~6.21]{resolutionfull}), in $\bigO(\LZSize{\Text} \log^4 \Textlen)$
  time we construct a structure that, given any substring $S$ of $\Text$
  (specified with its starting position and the length) in
  $\bigO(\log^3 \Textlen)$ time returns $\min \OccTwo{S}{\Text}$ (resp.\
  $|\OccTwo{S}{\Text}|$).

  We then construct the components of $\CompSa{\Text}$ as follows:
  \begin{enumerate}

  \item Observe that every length-$16$ substring of $\Textinf$ overlaps a
    phrase boundary in the LZ77 parsing of $\Text$.  In $\bigO(\LZSize{\Text} \log \Textlen)$
    time we thus first construct an array containing all length-$16$
    substrings of $\Textinf$ overlapping phrase boundaries.  In
    $\bigO(\LZSize{\Text} \log \LZSize{\Text})$ time we then sort the array, and remove
    duplicates.  The resulting array is precisely $\ArrStr$.

  \item To compute $\ArrRange$, we then proceed as follows.  For $i =
    1, \dots, n_{\rm short}$, letting $X_i = \ArrStr[i]$, we compute
    $q_i = |\OccTwo{X_i}{\Text}|$ in $\bigO(\log^3 \Textlen)$ time. If $i = 1$, we
    set $\ArrRange[i] = (0, q_1)$.  Otherwise, letting $(b,e)
    = \ArrRange[i-1]$, we set $\ArrRange[i] = (e, e + q_i)$.  In
    total, we spend $\bigO(\LZSize{\Text} \log^3 \Textlen)$ time.

  \item To compute $\ArrMinOcc$, we proceed as follows. For $i =
    1, \dots, n_{\rm short}$, letting $X_i = \ArrStr[i]$, we compute
    $m_i = \min \OccTwo{X_i}{\Text}$ in $\bigO(\log^3 \Textlen)$ time, and then set
    $\ArrMinOcc[i] = m_i$. In total, we spend $\bigO(\LZSize{\Text} \log^3 \Textlen)$
    time.

  \item In $\bigO(\SubstringComplexity{\Text} \log^7 \Textlen)$ time we construct
    $\CompSaNonperiodic{\Text}$
    using \cref{pr:sa-nonperiodic-construction}.

  \item In $\bigO(\SubstringComplexity{\Text} \log^7 \Textlen)$ time we construct
    $\CompSaPeriodic{\Text}$ using \cref{pr:sa-periodic-construction}.
  \end{enumerate}
  In total, the construction takes $\bigO(\LZSize{\Text} \log^3 \Textlen + \SubstringComplexity{\Text} \log^7
  \Textlen) = \bigO(\SubstringComplexity{\Text} \log^7 \Textlen)$ time.
\end{proof}

\subsection{Summary}

By combining \cref{pr:sa-final-isa,pr:sa-final-sa,pr:sa-final-construction}
and the size upper bound from \cref{sec:sa-final-ds},
we obtain the following final result.

\thmain*

\bibliographystyle{alphaurl}
\bibliography{paper}

\newcommand{\etalchar}[1]{$^{#1}$}
\begin{thebibliography}{KRRW23}

\bibitem[ABBK17]{AbboudBBK17}
Amir Abboud, Arturs Backurs, Karl Bringmann, and Marvin K{\"{u}}nnemann.
\newblock Fine-grained complexity of analyzing compressed data: Quantifying
  improvements over decompress-and-solve.
\newblock In Chris Umans, editor, {\em 58th {IEEE} Annual Symposium on
  Foundations of Computer Science, {FOCS} 2017}, pages 192--203. {IEEE}
  Computer Society, 2017.
\newblock \href {https://doi.org/10.1109/FOCS.2017.26}
  {\path{doi:10.1109/FOCS.2017.26}}.

\bibitem[ABBK20]{AbboudBBK20}
Amir Abboud, Arturs Backurs, Karl Bringmann, and Marvin K{\"{u}}nnemann.
\newblock Impossibility results for grammar-compressed linear algebra.
\newblock In Hugo Larochelle, Marc'Aurelio Ranzato, Raia Hadsell,
  Maria{-}Florina Balcan, and Hsuan{-}Tien Lin, editors, {\em 34th Conference
  on Neural Information Processing System, NeurIPS 2020}, volume~33, pages
  8810--8823. Curran Associates, Inc., 2020.
\newblock URL:
  \url{https://proceedings.neurips.cc/paper/2020/file/645e6bfdd05d1a69c5e47b20f0a91d46-Paper.pdf}.

\bibitem[ABM08]{bwtbook}
Donald Adjeroh, Tim Bell, and Amar Mukherjee.
\newblock {\em The {B}urrows-{W}heeler Transform: Data Compression, Suffix
  Arrays, and Pattern Matching}.
\newblock Springer, Boston, MA, USA, 2008.
\newblock \href {https://doi.org/10.1007/978-0-387-78909-5}
  {\path{doi:10.1007/978-0-387-78909-5}}.

\bibitem[AJ22]{AkmalJ22}
Shyan Akmal and Ce~Jin.
\newblock Near-optimal quantum algorithms for string problems.
\newblock In Joseph~(Seffi) Naor and Niv Buchbinder, editors, {\em 33rd Annual
  {ACM-SIAM} Symposium on Discrete Algorithms, {SODA} 2022}, pages 2791--2832.
  {SIAM}, 2022.
\newblock \href {https://doi.org/10.1137/1.9781611977073.109}
  {\path{doi:10.1137/1.9781611977073.109}}.

\bibitem[BCG{\etalchar{+}}21]{blocktree}
Djamal Belazzougui, Manuel C{\'{a}}ceres, Travis Gagie, Paweł Gawrychowski,
  Juha K{\"{a}}rkk{\"{a}}inen, Gonzalo Navarro, Alberto~Ord{\'{o}}{\~{n}}ez
  Pereira, Simon~J. Puglisi, and Yasuo Tabei.
\newblock Block trees.
\newblock {\em Journal of Computer and System Sciences}, 117:1--22, 2021.
\newblock \href {https://doi.org/10.1016/j.jcss.2020.11.002}
  {\path{doi:10.1016/j.jcss.2020.11.002}}.

\bibitem[BDY16]{BergerDY16}
Bonnie Berger, Noah~M. Daniels, and Y.~William Yu.
\newblock Computational biology in the 21st century: Scaling with compressive
  algorithms.
\newblock {\em Communications of the {ACM}}, 59(8):72--80, 2016.
\newblock \href {https://doi.org/10.1145/2957324} {\path{doi:10.1145/2957324}}.

\bibitem[Bel14]{Belazzougui14}
Djamal Belazzougui.
\newblock Linear time construction of compressed text indices in compact space.
\newblock In David~B. Shmoys, editor, {\em 46th Annual {ACM} Symposium on
  Theory of Computing, {STOC} 2014}, pages 148--193. {ACM}, 2014.
\newblock \href {https://doi.org/10.1145/2591796.2591885}
  {\path{doi:10.1145/2591796.2591885}}.

\bibitem[BKW19]{BringmannWK19}
Karl Bringmann, Marvin K{\"{u}}nnemann, and Philip Wellnitz.
\newblock Few matches or almost periodicity: Faster pattern matching with
  mismatches in compressed texts.
\newblock In Timothy~M. Chan, editor, {\em 30th Annual {ACM-SIAM} Symposium on
  Discrete Algorithms, {SODA} 2019}, pages 1126--1145. {SIAM}, 2019.
\newblock \href {https://doi.org/10.1137/1.9781611975482.69}
  {\path{doi:10.1137/1.9781611975482.69}}.

\bibitem[BLR{\etalchar{+}}15]{BLRSRW15}
Philip Bille, Gad~M. Landau, Rajeev Raman, Kunihiko Sadakane, Srinivasa~Rao
  Satti, and Oren Weimann.
\newblock Random access to grammar-compressed strings and trees.
\newblock {\em SIAM Journal on Computing}, 44(3):513--539, 2015.
\newblock \href {https://doi.org/10.1137/130936889}
  {\path{doi:10.1137/130936889}}.

\bibitem[BN14]{BelazzouguiN14}
Djamal Belazzougui and Gonzalo Navarro.
\newblock Alphabet-independent compressed text indexing.
\newblock {\em ACM Transactions on Algorithms}, 10(4):23:1--23:19, 2014.
\newblock \href {https://doi.org/10.1145/2635816} {\path{doi:10.1145/2635816}}.

\bibitem[BW94]{bwt}
Michael Burrows and David~J. Wheeler.
\newblock A block-sorting lossless data compression algorithm.
\newblock Technical Report 124, Digital Equipment Corporation, Palo Alto,
  California, 1994.
\newblock URL:
  \url{https://www.hpl.hp.com/techreports/Compaq-DEC/SRC-RR-124.pdf}.

\bibitem[CEK{\etalchar{+}}21]{ChristiansenEKN21}
Anders~Roy Christiansen, Mikko~Berggren Ettienne, Tomasz Kociumaka, Gonzalo
  Navarro, and Nicola Prezza.
\newblock Optimal-time dictionary-compressed indexes.
\newblock {\em ACM Transactions on Algorithms}, 17(1):8:1--8:39, 2021.
\newblock \href {https://doi.org/10.1145/3426473} {\path{doi:10.1145/3426473}}.

\bibitem[Cha88]{chazelle}
Bernard Chazelle.
\newblock A functional approach to data structures and its use in
  multidimensional searching.
\newblock {\em SIAM Journal on Computing}, 17(3):427--462, 1988.
\newblock \href {https://doi.org/10.1137/0217026} {\path{doi:10.1137/0217026}}.

\bibitem[CHL07]{AlgorithmsOnStrings}
Maxime Crochemore, Christophe Hancart, and Thierry Lecroq.
\newblock {\em Algorithms on strings}.
\newblock Cambridge University Press, Cambridge, UK, 2007.
\newblock \href {https://doi.org/10.1017/cbo9780511546853}
  {\path{doi:10.1017/cbo9780511546853}}.

\bibitem[CKPR21]{Charalampopoulos21}
Panagiotis Charalampopoulos, Tomasz Kociumaka, Solon~P. Pissis, and Jakub
  Radoszewski.
\newblock Faster algorithms for longest common substring.
\newblock In Petra Mutzel, Rasmus Pagh, and Grzegorz Herman, editors, {\em 29th
  Annual European Symposium on Algorithms, {ESA} 2021}, volume 204 of {\em
  LIPIcs}, pages 30:1--30:17. Schloss Dagstuhl--Leibniz-Zentrum f{\"{u}}r
  Informatik, 2021.
\newblock \href {https://doi.org/10.4230/LIPIcs.ESA.2021.30}
  {\path{doi:10.4230/LIPIcs.ESA.2021.30}}.

\bibitem[CKW20]{CKW20}
Panagiotis Charalampopoulos, Tomasz Kociumaka, and Philip Wellnitz.
\newblock Faster approximate pattern matching: {A} unified approach.
\newblock In Sandy Irani, editor, {\em 61st {IEEE} Annual Symposium on
  Foundations of Computer Science, {FOCS} 2020}, pages 978--989. {IEEE}
  Computer Society, 2020.
\newblock \href {https://doi.org/10.1109/FOCS46700.2020.00095}
  {\path{doi:10.1109/FOCS46700.2020.00095}}.

\bibitem[CLL{\etalchar{+}}05]{charikar}
Moses Charikar, Eric Lehman, Ding Liu, Rina Panigrahy, Manoj Prabhakaran, Amit
  Sahai, and Abhi Shelat.
\newblock The smallest grammar problem.
\newblock {\em IEEE Transactions on Information Theory}, 51(7):2554--2576,
  2005.
\newblock \href {https://doi.org/10.1109/TIT.2005.850116}
  {\path{doi:10.1109/TIT.2005.850116}}.

\bibitem[CN11]{ClaudeN11}
Francisco Claude and Gonzalo Navarro.
\newblock Self-indexed grammar-based compression.
\newblock {\em Fundamenta Informaticae}, 111(3):313--337, 2011.
\newblock \href {https://doi.org/10.3233/FI-2011-565}
  {\path{doi:10.3233/FI-2011-565}}.

\bibitem[CNP21]{ClaudeNP21}
Francisco Claude, Gonzalo Navarro, and Alejandro Pacheco.
\newblock Grammar-compressed indexes with logarithmic search time.
\newblock {\em Journal of Computer and System Sciences}, 118:53--74, 2021.
\newblock \href {https://doi.org/10.1016/j.jcss.2020.12.001}
  {\path{doi:10.1016/j.jcss.2020.12.001}}.

\bibitem[DFH{\etalchar{+}}20]{Dinklage0HKK20}
Patrick Dinklage, Johannes Fischer, Alexander Herlez, Tomasz Kociumaka, and
  Florian Kurpicz.
\newblock Practical performance of space efficient data structures for longest
  common extensions.
\newblock In Fabrizio Grandoni, Grzegorz Herman, and Peter Sanders, editors,
  {\em 28th Annual European Symposium on Algorithms, {ESA} 2020}, volume 173 of
  {\em LIPIcs}, pages 39:1--39:20. Schloss Dagstuhl--Leibniz-Zentrum f{\"{u}}r
  Informatik, 2020.
\newblock \href {https://doi.org/10.4230/LIPIcs.ESA.2020.39}
  {\path{doi:10.4230/LIPIcs.ESA.2020.39}}.

\bibitem[DNP21]{Diaz-DominguezN21}
Diego D{\'{\i}}az{-}Dom{\'{\i}}nguez, Gonzalo Navarro, and Alejandro Pacheco.
\newblock An {LMS}-based grammar self-index with local consistency properties.
\newblock In Thierry Lecroq and H{\'{e}}l{\`{e}}ne Touzet, editors, {\em 28th
  International Symposium on String Processing and Information Retrieval
  {SPIRE} 2021}, volume 12944 of {\em LNCS}, pages 100--113. Springer, 2021.
\newblock \href {https://doi.org/10.1007/978-3-030-86692-1_9}
  {\path{doi:10.1007/978-3-030-86692-1_9}}.

\bibitem[FGK{\etalchar{+}}22]{FranciscoGKLN22}
Alexandre~P. Francisco, Travis Gagie, Dominik K{\"{o}}ppl, Susana Ladra, and
  Gonzalo Navarro.
\newblock Graph compression for adjacency-matrix multiplication.
\newblock {\em {SN} Computer Science}, 3(3):193, 2022.
\newblock \href {https://doi.org/10.1007/s42979-022-01084-2}
  {\path{doi:10.1007/s42979-022-01084-2}}.

\bibitem[FM05]{FerraginaM05}
Paolo Ferragina and Giovanni Manzini.
\newblock Indexing compressed text.
\newblock {\em Journal of the ACM}, 52(4):552--581, 2005.
\newblock \href {https://doi.org/10.1145/1082036.1082039}
  {\path{doi:10.1145/1082036.1082039}}.

\bibitem[FM10]{fm2010}
Paolo Ferragina and Giovanni Manzini.
\newblock On compressing the textual web.
\newblock In Brian~D. Davison, Torsten Suel, Nick Craswell, and Bing Liu,
  editors, {\em 3rd International Conference on Web Search and Web Data Mining,
  {WSDM} 2010}, pages 391--400. {ACM}, 2010.
\newblock \href {https://doi.org/10.1145/1718487.1718536}
  {\path{doi:10.1145/1718487.1718536}}.

\bibitem[FMG{\etalchar{+}}22]{FerraginaMGKNST22}
Paolo Ferragina, Giovanni Manzini, Travis Gagie, Dominik K{\"{o}}ppl, Gonzalo
  Navarro, Manuel Striani, and Francesco Tosoni.
\newblock Improving matrix-vector multiplication via lossless
  grammar-compressed matrices.
\newblock {\em Proceedings of the {VLDB} Endowment}, 15(10):2175--2187, 2022.
\newblock URL: \url{https://www.vldb.org/pvldb/vol15/p2175-tosoni.pdf}.

\bibitem[FW65]{periodicitylemma}
Nathan~J. Fine and Herbert~S. Wilf.
\newblock Uniqueness theorems for periodic functions.
\newblock {\em Proceedings of the American Mathematical Society},
  16(1):109--114, 1965.
\newblock \href {https://doi.org/10.2307/2034009} {\path{doi:10.2307/2034009}}.

\bibitem[Gaw11]{Gaw11}
Paweł Gawrychowski.
\newblock Pattern matching in {L}empel-{Z}iv compressed strings: Fast, simple,
  and deterministic.
\newblock In Camil Demetrescu and Magn{\'{u}}s~M. Halld{\'{o}}rsson, editors,
  {\em 19th Annual European Symposium on Algorithms, {ESA} 2011}, volume 6942
  of {\em LNCS}, pages 421--432. Springer, 2011.
\newblock \href {https://doi.org/10.1007/978-3-642-23719-5_36}
  {\path{doi:10.1007/978-3-642-23719-5_36}}.

\bibitem[Gaw13]{Gaw13}
Paweł Gawrychowski.
\newblock Optimal pattern matching in {LZW} compressed strings.
\newblock {\em ACM Transactions on Algorithms}, 9(3):25:1--25:17, 2013.
\newblock \href {https://doi.org/10.1145/2483699.2483705}
  {\path{doi:10.1145/2483699.2483705}}.

\bibitem[GBMP14]{sdsl}
Simon Gog, Timo Beller, Alistair Moffat, and Matthias Petri.
\newblock From theory to practice: Plug and play with succinct data structures.
\newblock In Joachim Gudmundsson and Jyrki Katajainen, editors, {\em 13th
  International Symposium on Experimental Algorithms, SEA 2014}, volume 8504 of
  {\em LNCS}, pages 326--337. Springer, 2014.
\newblock \href {https://doi.org/10.1007/978-3-319-07959-2_28}
  {\path{doi:10.1007/978-3-319-07959-2_28}}.

\bibitem[GG22]{GanardiG22}
Moses Ganardi and Paweł Gawrychowski.
\newblock Pattern matching on grammar-compressed strings in linear time.
\newblock In Joseph~(Seffi) Naor and Niv Buchbinder, editors, {\em 33rd Annual
  {ACM-SIAM} Symposium on Discrete Algorithms, {SODA} 2022}, pages 2833--2846.
  {SIAM}, 2022.
\newblock \href {https://doi.org/10.1137/1.9781611977073.110}
  {\path{doi:10.1137/1.9781611977073.110}}.

\bibitem[GGK{\etalchar{+}}12]{GagieGKNP12}
Travis Gagie, Paweł Gawrychowski, Juha K{\"{a}}rkk{\"{a}}inen, Yakov Nekrich,
  and Simon~J. Puglisi.
\newblock A faster grammar-based self-index.
\newblock In Adrian{-}Horia Dediu and Carlos Mart{\'{\i}}n{-}Vide, editors,
  {\em 6th International Conference on Language and Automata Theory and
  Applications, {LATA} 2012}, volume 7183 of {\em LNCS}, pages 240--251.
  Springer, 2012.
\newblock \href {https://doi.org/10.1007/978-3-642-28332-1_21}
  {\path{doi:10.1007/978-3-642-28332-1_21}}.

\bibitem[GGK{\etalchar{+}}14]{GagieGKNP14}
Travis Gagie, Paweł Gawrychowski, Juha K{\"{a}}rkk{\"{a}}inen, Yakov Nekrich,
  and Simon~J. Puglisi.
\newblock {LZ77}-based self-indexing with faster pattern matching.
\newblock In Alberto Pardo and Alfredo Viola, editors, {\em 11th Latin American
  Symposium onTheoretical Informatics, {LATIN} 2014}, volume 8392 of {\em
  LNCS}, pages 731--742. Springer, 2014.
\newblock \href {https://doi.org/10.1007/978-3-642-54423-1_63}
  {\path{doi:10.1007/978-3-642-54423-1_63}}.

\bibitem[GGV03]{wt}
Roberto Grossi, Ankur Gupta, and Jeffrey~Scott Vitter.
\newblock High-order entropy-compressed text indexes.
\newblock In {\em 14th Annual {ACM-SIAM} Symposium on Discrete Algorithms,
  {SODA} 2003}, pages 841--850. {ACM/SIAM}, 2003.
\newblock URL: \url{http://dl.acm.org/citation.cfm?id=644108.644250}.

\bibitem[GJL21]{balancing}
Moses Ganardi, Artur Jeż, and Markus Lohrey.
\newblock Balancing straight-line programs.
\newblock {\em Journal of the {ACM}}, 68(4):27:1--27:40, 2021.
\newblock \href {https://doi.org/10.1145/3457389} {\path{doi:10.1145/3457389}}.

\bibitem[GKK{\etalchar{+}}18]{dynstr}
Paweł Gawrychowski, Adam Karczmarz, Tomasz Kociumaka, Jakub Łącki, and Piotr
  Sankowski.
\newblock Optimal dynamic strings.
\newblock In Artur Czumaj, editor, {\em 29th Annual {ACM-SIAM} Symposium on
  Discrete Algorithms, {SODA} 2018}, pages 1509--1528. {SIAM}, 2018.
\newblock \href {https://doi.org/10.1137/1.9781611975031.99}
  {\path{doi:10.1137/1.9781611975031.99}}.

\bibitem[GKLS22]{GaneshKLS22}
Arun Ganesh, Tomasz Kociumaka, Andrea Lincoln, and Barna Saha.
\newblock How compression and approximation affect efficiency in string
  distance measures.
\newblock In Joseph~(Seffi) Naor and Niv Buchbinder, editors, {\em 33rd Annual
  {ACM-SIAM} Symposium on Discrete Algorithms, {SODA} 2022}, pages 2867--2919.
  {SIAM}, 2022.
\newblock \href {https://doi.org/10.1137/1.9781611977073.112}
  {\path{doi:10.1137/1.9781611977073.112}}.

\bibitem[GNP18]{GNPlatin18}
Travis Gagie, Gonzalo Navarro, and Nicola Prezza.
\newblock On the approximation ratio of {L}empel-{Z}iv parsing.
\newblock In Michael~A. Bender, Martin Farach{-}Colton, and Miguel~A. Mosteiro,
  editors, {\em 13th Latin American Symposium on Theoretical Informatics,
  {LATIN} 2018}, volume 10807 of {\em LNCS}, pages 490--503. Springer, 2018.
\newblock \href {https://doi.org/10.1007/978-3-319-77404-6_36}
  {\path{doi:10.1007/978-3-319-77404-6_36}}.

\bibitem[GNP20]{Gagie2020}
Travis Gagie, Gonzalo Navarro, and Nicola Prezza.
\newblock Fully functional suffix trees and optimal text searching in
  {BWT}-runs bounded space.
\newblock {\em Journal of the ACM}, 67(1):2:1--2:54, 2020.
\newblock \href {https://doi.org/10.1145/3375890} {\path{doi:10.1145/3375890}}.

\bibitem[GT23]{GT23}
Daniel Gibney and Sharma~V. Thankachan.
\newblock Compressibility-aware quantum algorithms on strings, 2023.
\newblock \href {https://doi.org/10.48550/ARXIV.2302.07235}
  {\path{doi:10.48550/ARXIV.2302.07235}}.

\bibitem[Gus97]{gusfield}
Dan Gusfield.
\newblock {\em Algorithms on Strings, Trees, and Sequences: {C}omputer Science
  and Computational Biology}.
\newblock Cambridge University Press, Cambridge, UK, 1997.
\newblock \href {https://doi.org/10.1017/cbo9780511574931}
  {\path{doi:10.1017/cbo9780511574931}}.

\bibitem[GV05]{GrossiV05}
Roberto Grossi and Jeffrey~Scott Vitter.
\newblock Compressed suffix arrays and suffix trees with applications to text
  indexing and string matching.
\newblock {\em SIAM Journal on Computing}, 35(2):378--407, 2005.
\newblock \href {https://doi.org/10.1137/S0097539702402354}
  {\path{doi:10.1137/S0097539702402354}}.

\bibitem[Hag98]{Hagerup98}
Torben Hagerup.
\newblock Sorting and searching on the word {RAM}.
\newblock In Michel Morvan, Christoph Meinel, and Daniel Krob, editors, {\em
  15th Annual Symposium on Theoretical Aspects of Computer Science, {STACS}
  1998}, volume 1373 of {\em LNCS}, pages 366--398. Springer, 1998.
\newblock \href {https://doi.org/10.1007/BFb0028575}
  {\path{doi:10.1007/BFb0028575}}.

\bibitem[HLLW13]{HermelinLLW13}
Danny Hermelin, Gad~M. Landau, Shir Landau, and Oren Weimann.
\newblock Unified compression-based acceleration of edit-distance computation.
\newblock {\em Algorithmica}, 65(2):339--353, 2013.
\newblock \href {https://doi.org/10.1007/s00453-011-9590-6}
  {\path{doi:10.1007/s00453-011-9590-6}}.

\bibitem[HPWO19]{hernaez2019genomic}
Mikel Hernaez, Dmitri Pavlichin, Tsachy Weissman, and Idoia Ochoa.
\newblock Genomic data compression.
\newblock {\em Annual Review of Biomedical Data Science}, 2(1):19--37, 2019.
\newblock \href {https://doi.org/10.1146/annurev-biodatasci-072018-021229}
  {\path{doi:10.1146/annurev-biodatasci-072018-021229}}.

\bibitem[HSS09]{HonSS03}
Wing{-}Kai Hon, Kunihiko Sadakane, and Wing{-}Kin Sung.
\newblock Breaking a time-and-space barrier in constructing full-text indices.
\newblock {\em SIAM Journal on Computing}, 38(6):2162--2178, 2009.
\newblock \href {https://doi.org/10.1137/070685373}
  {\path{doi:10.1137/070685373}}.

\bibitem[I17]{tomohiro-lce}
Tomohiro I.
\newblock Longest common extensions with recompression.
\newblock In Juha K{\"{a}}rkk{\"{a}}inen, Jakub Radoszewski, and Wojciech
  Rytter, editors, {\em 28th Annual Symposium on Combinatorial Pattern
  Matching, {CPM} 2017}, volume~78 of {\em LIPIcs}, pages 18:1--18:15. Schloss
  Dagstuhl--Leibniz-Zentrum f{\"{u}}r Informatik, 2017.
\newblock \href {https://doi.org/10.4230/LIPIcs.CPM.2017.18}
  {\path{doi:10.4230/LIPIcs.CPM.2017.18}}.

\bibitem[Je{\.{z}}15]{Jez2015}
Artur Je{\.{z}}.
\newblock Faster fully compressed pattern matching by recompression.
\newblock {\em ACM Transactions on Algorithms}, 11(3):20:1--20:43, 2015.
\newblock \href {https://doi.org/10.1145/2631920} {\path{doi:10.1145/2631920}}.

\bibitem[JN23]{JinN23}
Ce~Jin and Jakob Nogler.
\newblock Quantum speed-ups for string synchronizing sets, longest common
  substring, and $k$-mismatch matching.
\newblock In Nikhil Bansal and Viswanath Nagarajan, editors, {\em 34th Annual
  {ACM-SIAM} Symposium on Discrete Algorithms, {SODA} 2023}, pages 5090--5121.
  {SIAM}, 2023.
\newblock \href {https://doi.org/10.1137/1.9781611977554.ch186}
  {\path{doi:10.1137/1.9781611977554.ch186}}.

\bibitem[Kem19]{Kempa19}
Dominik Kempa.
\newblock Optimal construction of compressed indexes for highly repetitive
  texts.
\newblock In Timothy~M. Chan, editor, {\em 30th Annual {ACM-SIAM} Symposium on
  Discrete Algorithms, {SODA} 2019}, pages 1344--1357. {SIAM}, 2019.
\newblock \href {https://doi.org/10.1137/1.9781611975482.82}
  {\path{doi:10.1137/1.9781611975482.82}}.

\bibitem[KK99]{KolpakovK99}
Roman~M. Kolpakov and Gregory Kucherov.
\newblock Finding maximal repetitions in a word in linear time.
\newblock In {\em 40th {IEEE} Annual Symposium on Foundations of Computer
  Science, {FOCS} 1999}, pages 596--604. {IEEE} Computer Society, 1999.
\newblock \href {https://doi.org/10.1109/SFFCS.1999.814634}
  {\path{doi:10.1109/SFFCS.1999.814634}}.

\bibitem[KK19]{sss}
Dominik Kempa and Tomasz Kociumaka.
\newblock String synchronizing sets: Sublinear-time {BWT} construction and
  optimal {LCE} data structure.
\newblock In Moses Charikar and Edith Cohen, editors, {\em 51st Annual {ACM}
  {SIGACT} Symposium on Theory of Computing, {STOC} 2019}, pages 756--767.
  {ACM}, 2019.
\newblock \href {https://doi.org/10.1145/3313276.3316368}
  {\path{doi:10.1145/3313276.3316368}}.

\bibitem[KK20a]{resolution}
Dominik Kempa and Tomasz Kociumaka.
\newblock Resolution of the {B}urrows-{W}heeler {T}ransform conjecture.
\newblock In Sandy Irani, editor, {\em 61st {IEEE} Annual Symposium on
  Foundations of Computer Science, {FOCS} 2020}, pages 1002--1013. {IEEE}
  Computer Society, 2020.
\newblock \href {https://doi.org/10.1109/FOCS46700.2020.00097}
  {\path{doi:10.1109/FOCS46700.2020.00097}}.

\bibitem[KK20b]{resolutionfull}
Dominik Kempa and Tomasz Kociumaka.
\newblock Resolution of the {B}urrows-{W}heeler {T}ransform conjecture, 2020.
\newblock Full version of \cite{resolution}.
\newblock \href {http://arxiv.org/abs/1910.10631v3}
  {\path{arXiv:1910.10631v3}}.

\bibitem[KK22]{dynsa}
Dominik Kempa and Tomasz Kociumaka.
\newblock Dynamic suffix array with polylogarithmic queries and updates.
\newblock In Stefano Leonardi and Anupam Gupta, editors, {\em 54th Annual {ACM}
  {SIGACT} Symposium on Theory of Computing, STOC 2022}, pages 1657--1670.
  {ACM}, 2022.
\newblock \href {https://doi.org/10.1145/3519935.3520061}
  {\path{doi:10.1145/3519935.3520061}}.

\bibitem[KK23]{breaking}
Dominik Kempa and Tomasz Kociumaka.
\newblock Breaking the ${O(n)}$-barrier in the construction of compressed
  suffix arrays and suffix trees.
\newblock In Nikhil Bansal and Viswanath Nagarajan, editors, {\em 34th Annual
  {ACM-SIAM} Symposium on Discrete Algorithms, SODA 2023}, pages 5122--5202.
  {SIAM}, 2023.
\newblock \href {https://doi.org/10.1137/1.9781611977554.ch187}
  {\path{doi:10.1137/1.9781611977554.ch187}}.

\bibitem[KMS{\etalchar{+}}03]{collage}
Takuya Kida, Tetsuya Matsumoto, Yusuke Shibata, Masayuki Takeda, Ayumi
  Shinohara, and Setsuo Arikawa.
\newblock Collage system: A unifying framework for compressed pattern matching.
\newblock {\em Theoretical Computer Science}, 298(1):253--272, 2003.
\newblock \href {https://doi.org/10.1016/S0304-3975(02)00426-7}
  {\path{doi:10.1016/S0304-3975(02)00426-7}}.

\bibitem[KN10]{kreft2010navarro}
Sebastian Kreft and Gonzalo Navarro.
\newblock {LZ}77-like compression with fast random access.
\newblock In {\em 2010 Data Compression Conference}, pages 239--248. {IEEE}
  Computer Society, 2010.
\newblock \href {https://doi.org/10.1109/DCC.2010.29}
  {\path{doi:10.1109/DCC.2010.29}}.

\bibitem[KNO22]{KociumakaNO22}
Tomasz Kociumaka, Gonzalo Navarro, and Francisco Olivares.
\newblock Near-optimal search time in {\(\delta\)}-optimal space.
\newblock In Armando Casta{\~{n}}eda and Francisco
  Rodr{\'{\i}}guez{-}Henr{\'{\i}}quez, editors, {\em 15th Latin American
  Symposium on Theoretical Informatics, {LATIN} 2022}, volume 13568 of {\em
  LNCS}, pages 88--103. Springer, 2022.
\newblock \href {https://doi.org/10.1007/978-3-031-20624-5_6}
  {\path{doi:10.1007/978-3-031-20624-5_6}}.

\bibitem[KNP23]{delta}
Tomasz Kociumaka, Gonzalo Navarro, and Nicola Prezza.
\newblock Towards a definitive compressibility measure for repetitive
  sequences.
\newblock {\em IEEE Transactions on Information Theory}, 69(4):2074--2092,
  2023.
\newblock \href {https://doi.org/10.1109/TIT.2022.3224382}
  {\path{doi:10.1109/TIT.2022.3224382}}.

\bibitem[Koc18]{phdtomek}
Tomasz Kociumaka.
\newblock {\em Efficient Data Structures for Internal Queries in Texts}.
\newblock PhD thesis, University of Warsaw, 2018.
\newblock URL: \url{https://www.mimuw.edu.pl/~kociumaka/files/phd.pdf}.

\bibitem[KP18]{attractors}
Dominik Kempa and Nicola Prezza.
\newblock At the roots of dictionary compression: String attractors.
\newblock In Ilias Diakonikolas, David Kempe, and Monika Henzinger, editors,
  {\em 50th Annual {ACM} {SIGACT} Symposium on Theory of Computing, {STOC}
  2018}, pages 827--840. {ACM}, 2018.
\newblock \href {https://doi.org/10.1145/3188745.3188814}
  {\path{doi:10.1145/3188745.3188814}}.

\bibitem[KRRW23]{IPM}
Tomasz Kociumaka, Jakub Radoszewski, Wojciech Rytter, and Tomasz Waleń.
\newblock Internal pattern matching queries in a text and applications, 2023.
\newblock \href {http://arxiv.org/abs/1311.6235v5} {\path{arXiv:1311.6235v5}}.

\bibitem[KS22]{KempaS22}
Dominik Kempa and Barna Saha.
\newblock An upper bound and linear-space queries on the {LZ}-end parsing.
\newblock In Joseph~(Seffi) Naor and Niv Buchbinder, editors, {\em 33rd Annual
  {ACM-SIAM} Symposium on Discrete Algorithms, {SODA} 2022}, pages 2847--2866.
  {SIAM}, 2022.
\newblock \href {https://doi.org/10.1137/1.9781611977073.111}
  {\path{doi:10.1137/1.9781611977073.111}}.

\bibitem[KVNP20]{KosolobovVNP20}
Dmitry Kosolobov, Daniel Valenzuela, Gonzalo Navarro, and Simon~J. Puglisi.
\newblock {L}empel-{Z}iv-like parsing in small space.
\newblock {\em Algorithmica}, 82(11):3195--3215, 2020.
\newblock \href {https://doi.org/10.1007/s00453-020-00722-6}
  {\path{doi:10.1007/s00453-020-00722-6}}.

\bibitem[KY00]{KiefferY00}
John~C. Kieffer and En{-}Hui Yang.
\newblock Grammar-based codes: {A} new class of universal lossless source
  codes.
\newblock {\em {IEEE} Transactions on Information Theory}, 46(3):737--754,
  2000.
\newblock \href {https://doi.org/10.1109/18.841160}
  {\path{doi:10.1109/18.841160}}.

\bibitem[LZ76]{LZ76}
Abraham Lempel and Jacob Ziv.
\newblock On the complexity of finite sequences.
\newblock {\em IEEE Transactions on Information Theory}, 22(1):75--81, 1976.
\newblock \href {https://doi.org/10.1109/TIT.1976.1055501}
  {\path{doi:10.1109/TIT.1976.1055501}}.

\bibitem[Mai89]{michael1989detecting}
Michael~G. Main.
\newblock Detecting leftmost maximal periodicities.
\newblock {\em Discrete Applied Mathematics}, 25(1-2):145--153, 1989.
\newblock \href {https://doi.org/10.1016/0166-218X(89)90051-6}
  {\path{doi:10.1016/0166-218X(89)90051-6}}.

\bibitem[MNN17]{MunroNN17}
J.~Ian Munro, Gonzalo Navarro, and Yakov Nekrich.
\newblock Space-efficient construction of compressed indexes in deterministic
  linear time.
\newblock In Philip~N. Klein, editor, {\em 28th Annual {ACM-SIAM} Symposium on
  Discrete Algorithms, {SODA} 2017}, pages 408--424. {SIAM}, 2017.
\newblock \href {https://doi.org/10.1137/1.9781611974782.26}
  {\path{doi:10.1137/1.9781611974782.26}}.

\bibitem[{Nat}22]{estimate}
{National Human Genome Research Institute (NIH)}.
\newblock Genomic data science, 2022.
\newblock Accessed March 30, 2023.
\newblock URL:
  \url{https://www.genome.gov/about-genomics/fact-sheets/Genomic-Data-Science}.

\bibitem[Nav14]{Navarro14}
Gonzalo Navarro.
\newblock Wavelet trees for all.
\newblock {\em Journal of Discrete Algorithms}, 25:2--20, 2014.
\newblock \href {https://doi.org/10.1016/j.jda.2013.07.004}
  {\path{doi:10.1016/j.jda.2013.07.004}}.

\bibitem[Nav16]{navarrobook}
Gonzalo Navarro.
\newblock {\em Compact data structures: A practical approach}.
\newblock Cambridge University Press, Cambridge, UK, 2016.
\newblock \href {https://doi.org/10.1017/cbo9781316588284}
  {\path{doi:10.1017/cbo9781316588284}}.

\bibitem[Nav21a]{NavarroMeasures}
Gonzalo Navarro.
\newblock Indexing highly repetitive string collections, part {I}:
  Repetitiveness measures.
\newblock {\em ACM Computing Surveys}, 54(2):29:1--29:31, 2021.
\newblock \href {https://doi.org/10.1145/3434399} {\path{doi:10.1145/3434399}}.

\bibitem[Nav21b]{NavarroIndexes}
Gonzalo Navarro.
\newblock Indexing highly repetitive string collections, part {II}: Compressed
  indexes.
\newblock {\em ACM Computing Surveys}, 54(2):26:1--26:32, 2021.
\newblock \href {https://doi.org/10.1145/3432999} {\path{doi:10.1145/3432999}}.

\bibitem[NII{\etalchar{+}}16]{NishimotoMFCS}
Takaaki Nishimoto, Tomohiro I, Shunsuke Inenaga, Hideo Bannai, and Masayuki
  Takeda.
\newblock Fully dynamic data structure for {LCE} queries in compressed space.
\newblock In Piotr Faliszewski, Anca Muscholl, and Rolf Niedermeier, editors,
  {\em 41st International Symposium on Mathematical Foundations of Computer
  Science, {MFCS} 2016}, volume~58 of {\em LIPIcs}, pages 72:1--72:15. Schloss
  Dagstuhl--Leibniz-Zentrum f{\"{u}}r Informatik, 2016.
\newblock \href {https://doi.org/10.4230/LIPIcs.MFCS.2016.72}
  {\path{doi:10.4230/LIPIcs.MFCS.2016.72}}.

\bibitem[NM07]{NavarroM07}
Gonzalo Navarro and Veli M{\"{a}}kinen.
\newblock Compressed full-text indexes.
\newblock {\em ACM Computing Surveys}, 39(1):2:1--2:61, 2007.
\newblock \href {https://doi.org/10.1145/1216370.1216372}
  {\path{doi:10.1145/1216370.1216372}}.

\bibitem[PHR00]{Przeworski2000}
Molly Przeworski, Richard~R. Hudson, and Anna~Di Rienzo.
\newblock Adjusting the focus on human variation.
\newblock {\em Trends in Genetics}, 16(7):296--302, 2000.
\newblock \href {https://doi.org/10.1016/S0168-9525(00)02030-8}
  {\path{doi:10.1016/S0168-9525(00)02030-8}}.

\bibitem[PNB17]{PereiraNB17}
Alberto~Ord{\'{o}}{\~{n}}ez Pereira, Gonzalo Navarro, and Nieves~R. Brisaboa.
\newblock Grammar compressed sequences with rank/select support.
\newblock {\em Journal of Discrete Algorithms}, 43:54--71, 2017.
\newblock \href {https://doi.org/10.1016/j.jda.2016.10.001}
  {\path{doi:10.1016/j.jda.2016.10.001}}.

\bibitem[Pre19]{Prezza19}
Nicola Prezza.
\newblock Optimal rank and select queries on dictionary-compressed text.
\newblock In Nadia Pisanti and Solon~P. Pissis, editors, {\em 30th Annual
  Symposium on Combinatorial Pattern Matching, {CPM} 2019}, volume 128 of {\em
  LIPIcs}, pages 4:1--4:12. Schloss Dagstuhl--Leibniz-Zentrum f{\"{u}}r
  Informatik, 2019.
\newblock \href {https://doi.org/10.4230/LIPIcs.CPM.2019.4}
  {\path{doi:10.4230/LIPIcs.CPM.2019.4}}.

\bibitem[Ryt03]{Rytter03}
Wojciech Rytter.
\newblock Application of {L}empel--{Z}iv factorization to the approximation of
  grammar-based compression.
\newblock {\em Theoretical Computer Science}, 302(1--3):211--222, 2003.
\newblock \href {https://doi.org/10.1016/S0304-3975(02)00777-6}
  {\path{doi:10.1016/S0304-3975(02)00777-6}}.

\bibitem[Sad02]{cst}
Kunihiko Sadakane.
\newblock Succinct representations of {LCP} information and improvements in the
  compressed suffix arrays.
\newblock In {\em 13th Annual {ACM-SIAM} Symposium on Discrete Algorithms, SODA
  2002}, pages 225--232. {ACM/SIAM}, 2002.
\newblock URL: \url{http://dl.acm.org/citation.cfm?id=545381.545410}.

\bibitem[SS82]{macro}
James~A. Storer and Thomas~G. Szymanski.
\newblock Data compression via textual substitution.
\newblock {\em Journal of the ACM}, 29(4):928--951, 1982.
\newblock \href {https://doi.org/10.1145/322344.322346}
  {\path{doi:10.1145/322344.322346}}.

\bibitem[Tis15]{Tiskin15}
Alexander Tiskin.
\newblock Fast distance multiplication of unit-{M}onge matrices.
\newblock {\em Algorithmica}, 71(4):859--888, 2015.
\newblock \href {https://doi.org/10.1007/s00453-013-9830-z}
  {\path{doi:10.1007/s00453-013-9830-z}}.

\bibitem[ZL77]{LZ77}
Jacob Ziv and Abraham Lempel.
\newblock A universal algorithm for sequential data compression.
\newblock {\em IEEE Transactions on Information Theory}, 23(3):337--343, 1977.
\newblock \href {https://doi.org/10.1109/TIT.1977.1055714}
  {\path{doi:10.1109/TIT.1977.1055714}}.

\end{thebibliography}

\end{document}